\documentclass[a4paper,11pt,oneside]{book}
\usepackage[utf8]{inputenc}

\usepackage{amsmath,amsfonts,amsthm,amssymb,dsfont}
\usepackage{graphicx,wrapfig,lipsum}
\usepackage[section]{placeins}

\usepackage[numbers,sort&compress]{natbib}
\usepackage{xcolor}
\usepackage{esvect}
\usepackage{wrapfig}
\usepackage{tikz}
\usetikzlibrary{shapes.geometric,decorations.markings}
\usepackage{subfigure}

\usepackage{slashed}

\newcommand{\Vol}{\text{Vol}}
\newcommand{\beq}{\begin{equation}}
\newcommand{\eeq}{\end{equation}}
\newcommand{\bea}{\begin{eqnarray}}
\newcommand{\eea}{\end{eqnarray}}
\newcommand{\nn}{\nonumber}

 \allowdisplaybreaks

 \usepackage{tikz}
\usetikzlibrary{3d}
\usepackage{rotating}
\usepackage{comment}
\usetikzlibrary{positioning}

\usetikzlibrary{decorations.pathreplacing}

\usepackage{ytableau}

\usepackage[export]{adjustbox}
\usetikzlibrary{positioning,shapes}

\tikzset{
    process1/.style={rectangle, minimum width=1cm, minimum height=1cm, text width=1cm,text centered, draw=black},
    process2/.style={rectangle, minimum width=6cm, minimum height=1cm, text width=5cm,text centered, draw=black},
    decision/.style={diamond, text centered, draw=black, aspect=2, inner xsep=-2mm},
    stop/.style={rectangle, rounded corners, minimum width=3cm, minimum height=1cm,text centered, draw=black},
    arr/.style={thick,-stealth}
    }
    
\usetikzlibrary{arrows,decorations.markings}
\usetikzlibrary{shapes.geometric}
\usetikzlibrary{arrows.meta}
\usetikzlibrary{decorations.pathreplacing,calligraphy}
\newcommand{\midarrow}{\tikz \draw[-triangle 90] (0,0) -- +(.1,0);}

 \usepackage{subfigure}
 \usepackage{float}
            \usepackage{changepage}
            \usepackage{enumitem}  
            \usepackage{slashed}

\usepackage[hidelinks]{hyperref}

\hypersetup{
    colorlinks=true,
    linktocpage=true,
    linkcolor=blue,
    urlcolor=blue,
    citecolor=blue,
}

\numberwithin{equation}{section}
\begin{document}


\begin{titlepage}
\begin{center}

\includegraphics[scale=0.35]{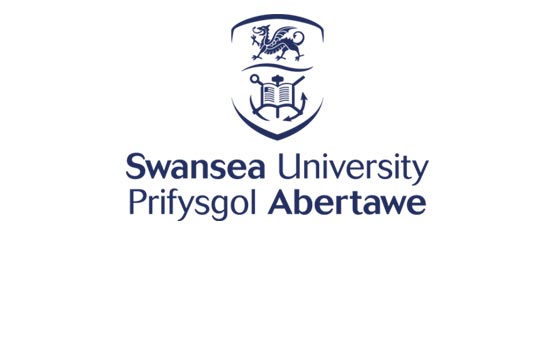} \vspace{0.5cm}

\textsc{\Large Doctoral Thesis } \vspace{0.5cm} 

\rule{15cm}{0.05cm} \vspace{0.4cm} 

\Large{\textbf{ 
Marginally deformed AdS$_{5}$/CFT$_{4}$ and spindle-like orbifolds
  }}\vspace{0.4cm} 

\rule{15cm}{0.05cm} \vspace{1.5cm} 
 
\large{\textit{by:}} \\
\Large{Paul Merrikin}  

\vspace{2cm}

\large \textit{
A thesis submitted in fulfilment of the requirements\\ 
for the Doctor of Philosophy} 

\vspace{0.3cm} 

\textit{of the}

\vspace{0.4cm}

Department of Physics,\\ 
Faculty of Science and Engineering,\\ 
Swansea University

\vspace{1.0cm} 

 September 11, 2025

\end{center}
\end{titlepage}

\newpage 
\pagenumbering{gobble}

\chapter*{Declaration of Authorship}
\noindent I, Paul Merrikin, declare that this thesis titled,``Marginally deformed AdS$_{5}$/CFT$_{4}$ and spindle-like orbifolds
" and the work presented in it are my own. I confirm that:

\begin{itemize} 
\item This work has not previously been accepted in substance for any degree and is not being concurrently submitted in candidature for any degree.

\item This thesis is the result of my own investigations, except where otherwise stated. Other sources are acknowledged by footnotes giving explicit references. A bibliography is appended. In particular, the thesis is based on the following publications:
    \begin{enumerate}
        \item N.~T.~Macpherson, P.~Merrikin and C.~Nunez,
``Marginally deformed AdS$_{5}$/CFT$_{4}$ and spindle-like orbifolds,''
JHEP \textbf{07} (2024), 042
\href{https://link.springer.com/article/10.1007/JHEP07(2024)042}{doi:10.1007/JHEP07(2024)042}
[arXiv:\href{https://arxiv.org/abs/2403.02380}{2403.02380 [hep-th]}].
\item P.~Merrikin,
``Marginally deformed AdS$_{5}$/CFT$_{4}$ backgrounds in Type IIB,''
JHEP \textbf{08} (2024), 181
\href{https://link.springer.com/article/10.1007/JHEP08(2024)181}{doi:10.1007/JHEP08(2024)181}
[arXiv:\href{https://arxiv.org/abs/2403.15326}{2403.15326 [hep-th]}].
    \end{enumerate}
    These papers are incorporated into chapters \ref{chap:setup} - \ref{chap:gamma}, and supplemented with additional results.
    \item I hereby give consent for my thesis, if accepted, to be available for photocopying and for inter-library loan, and for the title and summary to be made available to outside organisations.

\item The University’s ethical procedures have been followed and, where appropriate, that ethical approval has been granted.
\end{itemize} 

\vspace{0.5cm}
 
\noindent Signed: Paul Robert Greenwood Merrikin\\
\rule[0.5em]{25em}{0.5pt} 
 
\noindent Date: 
 September 11, 2025\\
\rule[0.5em]{25em}{0.5pt} 





\newpage
\pagenumbering{roman}
\setcounter{page}{6}

\chapter*{Abstract}
In this thesis, the AdS/CFT correspondence is used as a tool to explore novel AdS$_5$ Supergravity backgrounds 
(containing five-dimensional Anti-de Sitter spacetime) and their dual (four dimensional) Conformal Field Theory descriptions. In order to obtain precise results, both conformal symmetry and supersymmetry play an important role. However, in order to align with experimental observation, supersymmetry must be broken at low energies. 
In the absence of supersymmetry, finding deformations of a CFT which are marginal in nature (preserving conformal symmetry) is currently not well understood. Nevertheless, the solutions presented in this work may offer the best evidence to date of such deformations.

Multi-parameter families of non-supersymmetric type IIA and type IIB AdS$_5$ solutions are presented, promoting to $\mathcal{N}=1$ supersymmetry in some special cases. Contained within these solutions is an existing class of $\mathcal{N}=2$ type IIA solutions, recovered in one example when both deformation parameters are fixed to zero. The supersymmetry is studied using the method of G-structures, with the boundaries of the solutions carefully investigated - uncovering orbifold singularities within some solutions. In the type IIA backgrounds, both the spindle and its higher dimensional analogue play an important role, giving rise to rational quantization of charge. All parameters drop out of a quantity called the holographic central charge, pointing to marginal deformations of the existing $d=4$ $\mathcal{N}=2$ long linear quiver CFT. These marginal deformations are proposed to correspond to soft-SUSY breaking, with the Lagrangian nature of the CFT broken in some cases.


\newpage 


\chapter*{Acknowledgements}
My first debt of gratitude is owed to my supervisor, Carlos Nunez, whose capability as a Physicist is outweighed only by his outstanding character. I thank him for taking me on as a student, and I am truly grateful for his guidance, motivation, intellect, mentorship, encouragement, patience, never ending appetite for knowledge, and for his extraordinary generosity with his time over the past four years. In addition, I owe a huge debt of gratitude to Niall T. Macpherson for his generosity, guidance, intellect and very valuable tutelage (particularly in the realm of G-structures). I will be forever grateful to both, and I thank them for their collaboration on the work presented within this thesis. In addition, I would like to thank Vincent Menet and Alessandro Tomasiello for their generous invitation to Milan, and for two very informative and useful discussions. I also thank Christopher Couzens for a very insightful discussion. Lastly, I would like to thank all of my fellow postgraduate students, along with two previous academic tutors of particular value, Mark Dennis and Caroline Roelich. 

Outside academia, I am forever grateful to my family for their endless and overwhelming support and generosity - without whom this thesis would not have been possible. Particularly: my parents (Cheryl and Andy), grandparents (Marion, Derrick and Morvyth), brother (Michael and wife Tash), nieces (Melissa and Annabelle),  uncle (Phillip and Jeannie) and aunty (Elaine). Alongside my family, I would also like to thank my three oldest friends for their continued and very valuable friendship: Andrew Smith, Jack Worthington and Andrew Cutler. I would also like to thank Father Gorran Chapman for his many book vouchers and overall support. 

Last but not least, I would like to thank Bridie. 
In short, I thank her for her overwhelming kindness, support, patience, laughter, company, love, friendship, 
and for ultimately giving me a life outside of academia. I thank her entire family too, particularly Cath and Liz, for making me feel part of their family here in Wales.\\\\
 I dedicate this thesis to the loving memory of my grandfather, Albert, and uncle, Nigel.


\newpage
\tableofcontents 
\newpage

\pagenumbering{arabic}

\addcontentsline{toc}{chapter}{Introduction}
\chapter*{Introduction}
\label{Introduction}

Over the past two centuries, historical developments in physics have largely followed the theme of unification - in which seemingly diverse phenomena were recognised to be related to one another by some underlying principle. In 1865, the Maxwell equations of Electromagnetism were constructed \cite{Maxwell}, unifying electricity and magnetism (along with optics, by predicting electromagnetic waves) \cite{FirstCourseTextbook}. This led to Einstein's Special theory of relativity (SR) in 1905 \cite{Einstein:1905ve} (unifying Electromagnetism with Newtonian mechanics), which he further unified with gravity in 1915 \cite{Einstein:1915by} to develop the general theory of relativity (GR)- which, according to the curvature of `space-time', describes physics of the macroscopic scale. 

Then came along the development of Quantum Mechanics (QM) in the mid 1920s, describing physics of the microscopic scale. QM was then unified with SR in the 1940s, forming Quantum Field Theory (QFT), in which physics is described according to the exchange of fundamental field quanta. By the 1960s, it was shown that this framework could describe three of the four fundamental forces of nature: the electromagnetic force, 
the strong nuclear force, 
and the weak nuclear force
.
 The Standard Model (SM) of particle physics was then constructed, which is the best experimentally tested theory to date
 . However, there is one fundamental force missing from this framework- gravity.

 
Unfortunately, the usual quantization procedure of QFT does not work in the case of gravity, rendering the two great pillars of physics, GR and the SM, somewhat incompatible. The unification into one `Quantum Theory of Gravity' has since become one of the biggest problems of modern physics. 
Many attempts at unification have been made, but the most dominant approach has become `String Theory'.

\paragraph{String Theory }
The primary concept of string theory is that all the elementary `point' particles which we observe in nature are not point-like at all, but instead equivalent to different resonant patterns (or musical notes) of some tiny one dimensional elementary vibrating string
. Collectively, these vibrating strings then form one huge orchestra, composing and performing the elaborate symphony which we call the `universe'. These strings extend on the order of the Planck length, $10^{-35}$m, requiring experiments to reach the Planck energy, $10^{19}$GeV, to actually `see' the string directly. Hence, from our our feeble low energy `zoomed out' perspective, these strings would simply appear point like. This very simple and elegant idea has proven very powerful, with the new extended objects evading the usual quantization troubles of the point particle. 

 The theory already has a rich history of developments, originating in the 1960s as an attempt to explain the experimentally observed properties of the strong nuclear force. However, it was shown in 1974 by Schwarz and Scherk \cite{Scherk:1974ca} that the theory includes as part of the spectrum of string excitations a massless spin-2 particle - corresponding to the hypothesized graviton (the messenger particle for gravity). Subsequent work then demonstrated that string theory had scope to not just include all four fundamental forces, but all of matter too. This sparked the `first superstring revolution', between 1984 and 1994, during which time numerous properties of the standard model were shown to completely drop out naturally from the theory \cite{Chen}. However, in order for the theory to include both bosons (force carrying particles) and fermions (matter particles), `supersymmetry' needed to be incorporated. 
 
 \paragraph{Supersymmetry} Supersymmetry (or `SUSY') is a symmetry relating bosons (of integer spin) and fermions (of half-integer spin) of the same mass, in which each boson (e.g. the photon) comes with a fermionic superpartner (the `photino'), and vice versa. These are then related via supersymmetry transformations. Supersymmetric theories come with `supercharge', and are classified in terms of the number of supersymmetries, 
 $\mathcal{N}$, the theory has. For $\mathcal{N}>1$, the supercharges are rotated amongst themselves by an `R-symmetry'. In the case of an $\mathcal{N}=1$ theory, the R-symmetry is simply $U(1)$, which extends to $SU(2)\times U(1)$ in the case of an $\mathcal{N}=2$ theory \cite{Tong2}.

As it turns out, there are multiple ways in which supersymmetry can be incorporated into string theory, leading to five alternative versions: Type I, two `Heterotic' theories, and  two type II String theories (Type IIA and Type IIB). These theories are however forced to live in ten space-time dimensions, with the additional directions to our usual four space-time dimensions considered to be `compactified' to a very small size (evading our detection). The existence of five alternative theories was considered a major hurdle up until 1995, when Edward Witten \cite{Witten:1995ex} sparked what is now called the `second superstring revolution' -  outlining how all five theories are related to one another via a web of `dualities', and regarded as special cases of a single eleven-dimensional theory, called `M-Theory'.

\paragraph{Duality }
The concept of a duality is very important in physics, equating two seemingly disconnected theories via an Isomorphism - allowing the same phenomena to be viewed in multiple ways. Examples include: T-duality, S-duality, dimensional reductions from M-theory to type IIA (in which the size of the additional eleventh dimension is reduced to zero), and the AdS/CFT correspondence - which is an important example of the Holographic Principle, stating that the amount of information which can be stored within a volume $V_{d+1}$ is actually encoded on the boundary of that volume, with area $A_d$. All of these dualities will play an important role throughout this thesis. Most notable however is the AdS/CFT correspondence, which we will use as a tool to explore novel `Supergravity' backgrounds and their dual Conformal Field Theory (CFT) descriptions - field theories which are invariant under conformal transformations, meaning the physics is independent of length scale.

\paragraph{Supergravity} Supergravity (or `SUGRA') is a low energy approximation of string theory, where one has effectively `zoomed out' of the theory. Here, the extended nature of the string becomes less significant, allowing them to be approximated as point particles. The theory still contains supersymmetry and gravity, hence the name, but corresponds to a low energy effective theory - now considered a special type of supersymmetric field theory. See e.g. \cite{DallAgata:2022nfr}. These theories prove very useful, and play a pivotal role in the AdS/CFT correspondence. 

\paragraph{The AdS/CFT Correspondence}

The AdS/CFT correspondence, introduced by Maldacena in 1997 \cite{Maldacena:1997re}, elegantly relates gravitational theories in Anti-de Sitter (AdS) spacetime (of negative curvature) to non-gravitational Conformal Field Theories in one fewer dimension. This groundbreaking idea enables insights into the high energy (or ultra-violet, `UV') regime of dual field theories via supergravity, especially in theories with conformal and supersymmetry - symmetries essential for obtaining precise results.

Since its inception, the AdS/CFT correspondence has proven very fruitful in the study of supergravity backgrounds (with an AdS$_{d+1}$ factor) and their dual CFTs (of $d$ spatial dimensions). Some examples 
 include, for $d=1$ \cite{Lozano:2020txg,Lozano:2020sae, Lozano:2021rmk}, for $d=2$ \cite{Lozano:2019emq,Lozano:2019zvg,Legramandi:2020txf,Legramandi:2019xqd,Macpherson:2018mif,Lozano:2020bxo}, and for $d=3$ \cite{Akhond:2021ffz,Lozano:2016wrs,Assel:2012cj,Assel:2011xz,Merrikin:2021smb}. In the case of $d=4$, which will be the focus of this thesis, many different types of dual CFTs have been studied, but those of particular relevance include \cite{Gaiotto:2009gz,Lin:2004nb,Kaste:2003zd,Gauntlett:2004zh,Bah:2019jts, Bah:2022yjf,Macpherson:2016xwk}. For $d=5$, dual AdS$_6$ geometries were constructed in \cite{Legramandi:2021uds,Uhlemann:2019ypp,Gutperle:2018vdd,Apruzzi:2018cvq,DHoker:2017zwj,DHoker:2017mds}, with $d=6$ studied in \cite{Cremonesi:2015bld,Apruzzi:2015wna,Apruzzi:2014qva, Gaiotto:2014lca,Filippas:2019puw}.
All these cases preserve supersymmetry. For $d=7$, CFTs are not compatible with supersymmetry, however non-supersymmetric  AdS$_8$ backgrounds were found in \cite{Cordova:2018eba}.

Naturally, given that we see $d=4$ spacetime dimensions in our universe, AdS$_5$/CFT$_4$ solutions are potentially the most phenomenologically interesting - offering insights into high energy regimes of four dimensional field theories, with gravity naturally included in the dual framework. These field theories in general include both supersymmetry and conformal symmetry (with the latter stemming from the AdS space of the dual description). However, it is possible to consider non-AdS/non-CFT correspondence, see for example \cite{Aharony:2002up}.

\paragraph{Supersymmetry breaking}
Given that we haven't yet observed supersymmetric partners in particle physics experiments (such as the Large Hadron Collider), in order for string theory to be an accurate description of high energy physics, supersymmetry must be broken at an energy somewhere between the electroweak energy and the Planck energy. This then raises the natural question of supersymmetry breaking, of which there are a number of theoretical mechanisms - see \cite{Luty:2005sn} for a nice review. Once supersymmetry is broken, the masses of the superpartners can differ from the particles we observe, and must be higher than the energies of our current particle physics experiments (to avoid our detection).

A simple method of breaking supersymmetry is called `soft-SUSY breaking', in which supersymmetry breaking terms are added to the effective Lagrangian (terms which avoid quadratic divergences). If supersymmetry is then broken spontaneously at some high energy scale, the effective Lagrangian below such a scale would  be considered a `softly' broken supersymmetric theory. This method would then provide a natural extension to the SM.

From a supergravity perspective, supersymmetry is very useful in deriving solutions to the equations of motion. Work has been conducted into supersymmetry breaking solutions, the most notable for this thesis are \cite{Lust:2008zd,Legramandi:2019ulq,Menet:2023rnt,Menet:2023rml,Menet:2024afb,Held:2010az}. Unfortunately, when supersymmetry is broken within supergravity, the stability of the solution is no longer guaranteed. Various mechanisms of instability must be considered, including among others, the stability of D-brane probes
 - see \cite{Basile:2021vxh,Apruzzi:2021nle,Guarino:2020jwv} for useful discussions. Unstable solutions are then considered as part of the `Swampland' of supergravity solutions.  Breaking supersymmetry is then not so simple, and considered by some to be necessarily unstable for AdS vacua supported by fluxes - following a conjecture by Ooguri and Vafa in \cite{Ooguri:2016pdq} . However, more recent investigations have been refined by the techniques of Exceptional field theory, with stable non-supersymmetric AdS vacua presented in the works of \cite{Guarino:2020flh,Eloy:2023acy,Giambrone:2021wsm}- see also \cite{Malek:2020yue,Eloy:2020uix}.  
 

Hence, investigating supersymmetry breaking within the context of the AdS/CFT correspondence is still a natural and interesting line of research, particularly in the case of AdS$_5$/CFT$_4$ solutions. We will attempt to shed some new light on this within this work.

\paragraph{Overview of thesis} 
Given that CFTs are independent of length scale, they form a fixed point of the renormalization group (RG).
Operators then control the RG flow away from the conformal fixed point, and are classified as irrelevant, relevant or marginal. Marginal deformations do not break the conformality of a theory, and give rise to a family of CFTs. However, finding such deformations without the presence of supersymmetry is currently not very well understood. 

The AdS$_5$ supergravity backgrounds which are presented in this work may well be the best evidence to date of such solutions - finding supersymmetry breaking deformations of a supersymmetric solution (in both type IIA and type IIB) which are marginal in the dual $d=4$ CFT description! We will propose that these marginal deformations correspond to  ‘soft-SUSY’ breaking of the CFT (adding supersymmetry breaking terms to the Lagrangian). We will find that these deformations fall into two main categories: 
\begin{enumerate}
\item Some deformations preserve the Lagrangian nature of the original CFT, and are considered as marginal Lagrangian deformations 
- corresponding to integer quantization of charge in the dual supergravity description.
\item Other deformations break the Lagrangian nature of the original theory altogether, and are considered marginal non-Lagrangian deformations 
- corresponding to rational quantization of charge in the dual supergravity description.
\end{enumerate}
Interestingly, this rational quantization of charge is a consequence of additional orbifold singularities in the supergravity background - introduced into the solution by the marginal non-Lagrangian deformations. The most general solutions in fact contain both types of deformation, with the original $\mathcal{N}=2$ supersymmetry completely broken in general. In a particular type IIA subset, we find solutions containing varies stacks of branes, each orthogonal to its own \textit{spindle}, labelled $\mathbb{WCP}^1_{[n_-,n_+]}$. These spaces have the topology of a two-sphere with additional orbifold singularities at the poles $\mathbb{R}^2/\mathbb{Z}_{n_-,n_+}$ (producing a `lemon-like' object), with $n_-\neq n_+$ and gcd$(n_-,n_+)=1$. It is important to note however, when supersymmetry is completely broken, the stability of source D-branes is no longer guaranteed. 
There is no sign of instability as yet, but a more thorough investigation should one day be carried out. 

Additionally, fine-tuning the deformations appropriately gives rise to $\mathcal{N}=1$ preserving Lagrangian, and non-Lagrangian deformations 
(dual to both type IIA and type IIB supergravity). In the IIA supergravity description, this non-Lagrangian deformation is a consequence of a four-dimensional analogue of the spindle. These spaces then become more complicated under T-duality to type IIB.

In recent years, spindles have been the subject of much study within the context of supergravity - involving the near horizon limits of D-branes wrapping spindles. Examples in which spindles play a central role in the holographic duals include \cite{Ferrero:2020twa,Ferrero:2020laf,Couzens:2021cpk,Hosseini:2021fge,Boido:2021szx,Ferrero:2021wvk,Bah:2021hei,Ferrero:2021ovq,Couzens:2021rlk,Faedo:2021nub,Ferrero:2021etw,Couzens:2021tnv,Couzens:2022yjl,Arav:2022lzo,Couzens:2022yiv,Suh:2022pkg,Amariti:2023gcx,Inglese:2023tyc,Faedo:2024upq,Bomans:2024mrf}. In the solutions presented within this thesis however, we find branes which are instead orthogonal to the spindle - and in fact find multiple neighbouring spindles within the same solutions. In addition, our solutions include multiple versions of higher dimensional analogues of the spindle, labelled in one example $\mathbb{WCP}^2_{[l_{k-1},l_k,\frac{\zeta}{\xi}(l_{k-1}-l_k)]}$. These spaces are a generalization of other explicit examples appearing in \cite{Gauntlett:2004zh,Bianchi:2021uhn}, namely $\mathbb{WCP}^2_{[1,1,2]}$. To our knowledge, the generalized forms we uncover in this work are the first to appear in solutions of supergravity.

In summary, for the examples presented throughout this thesis, we find marginal deformations which both preserve and break the Lagrangian nature of the original CFT - corresponding to additional orbifold singularities in the dual supergravity description, with the latter giving rise to rational quantization of charge. In general, these deformations break the original $\mathcal{N}=2$ supersymmetry altogether, and are interpreted as soft-SUSY breaking in the dual CFT - with special subclasses enhancing to $\mathcal{N}=1$.


\paragraph{Plan of thesis:} This work is split into two parts:  Part \ref{Part:Preliminaries} focuses on some preliminary background material, while Part \ref{Part:SUSYdefs} contains new results. 
\\\\The content of Part \ref{Part:Preliminaries} is as follows:
\begin{itemize}
\item In Chapter \ref{chap:AdS/CFT}, we review some relevant aspects of eleven dimensional and ten dimensional (type II) supergravity, 
CFTs and the AdS/CFT correspondence. We then review in some depth the Gaiotto-Maldacena (GM) background \cite{Gaiotto:2009gz}, which is an AdS$_5$ $\mathcal{N}=2$ solution in $d=11$
. This is the `parent' solution to the results presented in Part \ref{Part:SUSYdefs}, and a primary focus of the chapter. We then note some existing deformations of the class, before concluding the chapter with some comments on 
spindles.
\item In Chapter \ref{sec:Gstructures}, we review some material on the method of G-structures, which recasts the usual supersymmetry conditions (in terms of a spinor and a metric) into a more elegant and easier to use approach (in terms of non spinorial and geometrical objects - forms). This framework is what facilitates the supersymmetry analysis presented throughout Part \ref{Part:SUSYdefs}. 
\end{itemize}
Part \ref{Part:SUSYdefs} begins with a brief overview, and contains the following content:
 \begin{itemize}
\item In Chapter \ref{chap:setup}, we 
provide further analysis on the GM class of solutions. We 
derive the appropriate G-Structure forms for the $d=11$ GM solution (written in terms of complex vielbeins), following an appropriate coordinate transformation (known as the `B$\ddot{\text{a}}$cklund' transformation) on the results given in \cite{Macpherson:2016xwk} for the Lin-Lunin-Maldacena (LLM) \cite{Lin:2004nb} solutions. This allows us to identify the $U(1)_R$ component of the original $SU(2)_R\times U(1)_R$ R-symmetry of the GM solution in $d=11$, allowing us to keep track of the supersymmetry under dimensional reduction and subsequent T-duality. We then perform an $SL(3,\mathds{R})$ transformation amongst three U(1) directions of the solution, requiring the G-structure forms to be rotated to a frame which takes into account both the $SL(3,\mathds{R})$ transformation and the specific $U(1)$ direction being reduced. A straightforward technique to do this is developed, noticing that the G-Structure conditions require a particular real two-form, $J$, to remain intact under the frame rotation. Using results from the literature, we then derive the type IIA G-Structure forms for the ten-dimensional $\mathcal{N}=2$ GM solution. Analysis at the boundary for this solution is then provided. The chapter then concludes with a brief discussion on supersymmetry breaking reductions to type IIA. 
\item In Chapter \ref{chap:typeIIA}, 
following the $SL(3,\mathds{R})$ transformation, we perform dimensional reductions along each of the three $U(1)$ directions of $d=11$ GM - leading to three two-parameter families of type IIA solutions, which are in general $\mathcal{N}=0$. Each solution is presented
, with the reduction along the $\beta$ coordinate forming the primary focus of the chapter (as this solution recovers the zero-parameter family of $\mathcal{N}=2$ GM)
. In all three solutions, the parameters introduce orbifold singularities into the supergravity background (including in some cases a set of neighbouring spindles), which then lead to rational quantization of D-brane charge. 
Fixing the two parameters appropriately, the backgrounds enhance to one-parameter families of $\mathcal{N}=1$ solutions. G-structure analysis is then provided, including a more general discussion on the supersymmetry breaking backgrounds - which are the first examples explicitly shown to break all three supersymmetry conditions, including the `gauge BPSness' condition (along with the solutions presented in Chapters \ref{chap:typeIIB} and \ref{chap:gamma}). 

The method of G-Structures also allows the higher form fluxes for the supersymmetric solutions to be derived very easily (such as $C_5$ and $C_7$). After some work, the results for the non-supersymmetric solutions were derived too, allowing the stability of a probe D6 brane to be studied - which was conducted for the $S^2$ preserved $\beta$ reduction case, showing no sign of instability.

Various mappings between the three general solutions can be performed. 
However, due to an integer condition on the parameters, these mathematically equivalent solutions are proposed to describe different physical backgrounds (with different rational charge
). These solutions are interesting in their own right, allowing one to pick the supersymmetry of a background by fine-tuning the parameters appropriately. 

We then discuss some aspects of the CFTs dual to our supergravity backgrounds. Calculating the holographic central charge (which is an effective weighted volume of the internal manifold of the supergravity, corresponding to the number of degrees of freedom of the dual CFT description), one finds that all parameters drop out completely. We conclude that the deformation parameters (introduced via an $SL(3,\mathds{R})$ transformation in $d=11$), which in general break the supersymmetry, correspond to marginal deformations in the dual CFT description. We propose that these deformations then correspond to soft-SUSY breaking of the CFT. 

Interestingly, dimensionally reducing in a more general fashion allows one to force integer quantization upon the system, which has the effect of changing the holographic central charge. However, given the preservation of AdS$_5$ (and hence conformality in the dual CFT), we propose that integer quantization is simply too much to ask of the system. We instead interpret the deformations which effect the quantization 
as  marginal non-Lagrangian deformations - which then break the Lagrangian nature of the dual theory.

\item In Chapter \ref{chap:typeIIB}, we perform an abelian T-duality on the three type IIA solutions - deriving multiple three-parameter families of type IIB backgrounds. These solutions are again $\mathcal{N}=0$ in general, but enhance to a one-parameter family of $\mathcal{N}=1$ solutions by an appropriate choice of parameters. These multiple (infinite) families have a zero five form flux, which could be of particular interest given that not many SUSY solutions are known with this property. The first two examples of such backgrounds were found in \cite{Macpherson:2014eza}, evading the prior classification of AdS$_5$ solutions considered in \cite{Gauntlett:2005ww} (which had a non-vanishing five-form flux). This led to the work of \cite{Couzens:2016iot}, where the classification was completed. In type IIB, we find no $\mathcal{N}=2$ backgrounds present. 

In order to verify that the one-parameter families of $\mathcal{N}=1$ type IIB solutions indeed preserve some supersymmetry, the IIB G-Structure forms and conditions needed to be derived from the IIA solution (via T-duality). This then becomes the focus of the opening section of the chapter. The G-structure description is then derived, with more general supersymmetry breaking G-structure forms included for one example. 

Peculiarities arise within these type IIB solutions, stemming from T-dualising within the orbifold structure, including backgrounds with rational quantization without the presence of orbifold singularities. 
\item In Chapter \ref{chap:gamma}, we provide the G-structure description for the $\gamma$-deformed GM solutions, first derived by N\'u\~nez, Roychowdhury, Speziali and Zacar\'\i{}as (NRSZ) \cite{Nunez:2019gbg}. These deformations are also marginal in the dual CFT, but until now, it was unclear whether supersymmetry was preserved in these backgrounds. Here we shed light on this question, demonstrating that the deformations in both the M-theory and type IIA solutions break supersymmetry completely in all cases. The type IIB solution however contains a special zero-parameter family of $\mathcal{N}=1$ solutions, derived by fixing $\gamma=-1$. This solution is actually a sub-class of the more general type IIB solutions derived in Chapter \ref{chap:typeIIB}, where we re-derive the solution. Interestingly, the $\gamma=-1$ deformation of NRSZ is the only $\mathcal{N}=1$ solution which preserves the Lagrangian nature of the dual CFT (due to integer charge). All other $\mathcal{N}=1$ solutions derived in this work break the Lagrangian nature due to rational quantization in the supergravity description. Hence, this type IIB solution seems the most likely dual description for the $\mathcal{N}=1$ CFT presented in the NRSZ paper.
\end{itemize}
We conclude with a 
summary of results, and outline some future directions. Most calculations are included within the text, but complimented with additional Appendices where required. A general $d$-dimensional form of the holographic central charge is included.


\part{Preliminaries}\label{Part:Preliminaries}
\chapter{The AdS/CFT correspondence}\label{chap:AdS/CFT}

\section{Introduction to Supergravity}

For further information on the topics reviewed in this section, see for example \cite{TomTextbook,DbraneTextbook,Legramandi:2020des}.
\subsection{Supergravity in Eleven Dimensions}
In eleven dimensions, supergravity contains the metric, $g_{MN}$, a three form potential, $A_3=A_{MNP}$, with four form field strength $G_4=dA_3$, and a single gravitino, $\psi_M$, where $M=0,...,10$. The bosonic action reads
\beq\label{eqn:11daction}
S_{11}=\frac{1}{2\kappa_{11}^2}\int \bigg(d^{11}x \sqrt{-g_{11}}\Big(R_{11}-\frac{1}{2(4!)}G_4^2\Big)-\frac{1}{6}A_3\wedge G_4\wedge G_4\bigg).
\eeq
Varying this action derives the following two equations of motion, which must be solved along with the Bianchi identity $dG_4=0$, namely
\beq
R_{MN}-\frac{1}{12}(G_4^2)_{MN}+\frac{1}{12^2}g_{MN}G_4^2=0,
\eeq
and
\beq
d\star G_4+\frac{1}{2}G_4\wedge G_4=0,~~~\Rightarrow~~~d\Big(\star G_4+\frac{1}{2}A_3\wedge G_4\Big)=0,
\eeq
with the last condition defining a six-form potential, $A_6$, as follows
\beq
dA_6=\star G_4+\frac{1}{2}A_3\wedge G_4,
\eeq
which is a magnetic potential when $A_3$ is electric, and vice versa.

Note the $\star$ notation means taking the `Hodge dual', which maps a $k$- form to a $(d-k)$- form
\beq
\star dx^{1}\wedge ...\wedge dx^{k} \rightarrow (-1)^{k(d-k)}dx^{k+1}\wedge ...\wedge dx^d,
\eeq
with $d=11$ in the present case. Note also that the wedge product is antisymmetric, with the properties
\beq\label{eqn:Leibniz}
\alpha_k\wedge \alpha_{k'}=(-1)^{kk'}\alpha_{k'}\wedge \alpha_k,~~~~~~~~~~d(\alpha_k\wedge \alpha_{k'}) = d\alpha_k\wedge \alpha_{k'}+(-1)^k\alpha_k\wedge d\alpha_{k'},
\eeq
with the latter called the `Leibniz identity'.

Eleven dimensional supergravity contains extended objects called membranes, a primary inspiration for the name `M-theory'. These objects are classified as elementary or solitonic, depending on whether they correspond to singular or non singular solutions, respectively. The potentials $A_3$ and $A_6$ then act as sources for M2- and M5- branes, with two and five spatial dimensions, respectively. 

 A supergravity solution is supersymmetric if the supersymmetry variations vanish, rendering the fields invariant under supersymmetry transformations. BPS branes are a special class of supersymmetric solutions in which the mass density and charge of the brane are equal. This property is known as the Bogomol'nyi-Prasad-Sommerfield (BPS) bound, and shields the branes from quantum corrections - allowing classical results to be extrapolated to the quantum level of string theory. See for example \cite{Passias}.
\subsection{Type II Supergravity in Ten Dimensions}\label{sec:typeII}
We will now review both type IIA and type IIB supergravity.  Here `type II' refers to the supersymmetry parameter, which is a ten dimensional Majorana spinor, $\varepsilon$, splitting into two 16 dimensional Majorana-Weyl Killing spinors, $\varepsilon^{1,2}$. These then generate a maximum of 32 supercharges, and have $(+,-)$ and $(+,+)$ chirality in IIA and IIB, respectively
\begin{align}\label{eqn:Chiralities}
\varepsilon=\varepsilon^1+\varepsilon^2,~~~~~~~~~~~~~~&\nn\\[2mm]
\Gamma_{(10)}\varepsilon^1=\varepsilon^1,~~~~~\Gamma_{(10)}\varepsilon^2=-\varepsilon^2,~~~~~&\text{Type IIA}\nn\\[2mm]
\Gamma_{(10)}\varepsilon^1=\varepsilon^1,~~~~~~~\Gamma_{(10)}\varepsilon^2=\varepsilon^2,~~~~~&\text{Type IIB}
\end{align}
noting $\Gamma_M$ are ten dimensional gamma matrices with $\Gamma_{(10)}=\Gamma^{0...9}$ and $M=0,...,9$.  

The matter fields of these theories are given by the fermions, made up of two gravitini and two dilatini
\beq
\psi_M^{1,2}~\text{(gravitino)},~~~~\lambda^{1,2}~\text{(dilatino)}.
\eeq
In IIA, $(\psi_M^{1},\lambda^1)$ and $(\psi_M^{2},\lambda^2)$ have positive and negative chiralities, respectively, and in IIB they all have positive chirality.

%
The bosonic fields consist of: a Neveu-Schwarz-Neveu-Schwarz (NSNS) sector, comprising of the metric $g_{MN}$ (of signature $(-,+,..,+)$), the dilaton $\Phi$ and two-form potential $B_{MN}$; and a Ramond-Ramond (RR) sector, with RR fields defined in terms of differential forms, $C_p$, of degree $p$. The RR forms are restricted to odd and even degrees in type IIA and IIB respectively, whereas the NSNS content is common to both. One can then define
\beq\label{eqn:formnotation}
B= \frac{1}{2}B_{MN}dx^M\wedge dx^N,~~~~~~C_p=\frac{1}{p!}C_{M_1...M_p}dx^{M_1}\wedge ...\wedge dx^{M_p},
\eeq
which allows for the following RR poly-forms to be built in each case
\beq\label{eqn:Cdefs}
C=
\begin{cases}
C_1+C_3+C_5+C_7+C_9&(IIA)\\
C_0+C_2+C_4+C_6+C_8&(IIB)
\end{cases}
.
\eeq
The field strengths are then defined in the following manner
\beq\label{eqn:origfieldstengths}
H=dB,~~~~~~~F_p=dC_{p-1}-H\wedge C_{p-3}
,~~~~~~~\text{where}~~~~~~d\equiv dx^M\frac{\partial}{\partial x^M}.
\eeq
This describes the `democratic formalism' of \cite{Bergshoeff:2001pv}, in which the number of RR fields has been doubled. To compensate, using ten-dimensional Hodge-duality, the `twisted field strengths', $F_p$, have the following duality condition
\beq\label{eqn:fieldrelation}
F_p=(-1)^{[\frac{p}{2}]}\star F_{10-p},
\eeq
with $[\frac{p}{2}]$ the integer part of $p$, and noting the special self-dual condition $F_5=\star F_5$. Typically then, the `electric basis' is used to define the fundamental fields as the fields with the smallest indices, namely $(F_0,F_2,F_4)$ in IIA and $(F_1,F_3,F_5)$ in IIB. 
In a similar manner, the dual NSNS potential, $B_6$, is defined as follows
\beq
dB_6=-\star H.
\eeq
One can then build a poly-form for the RR field strength
\beq
F=
\begin{cases}
 F_0+F_2+F_4+F_6+F_8+F_{10}&(IIA)\\
F_1+F_3+F_5+F_7+F_9&(IIB)
\end{cases}
, 
\eeq
with
\beq\label{eqn:Fdef}
F=dC-H\wedge C\equiv d_HC,~~~~~~\text{where}~~~~~d_H\equiv d-H\wedge,
\eeq
where $d_H$ is known as the `twisted exterior derivative'.
Type IIA can have a field of zero-form, $F_0$, which has no corresponding potential, and is called the `Romans mass'. When $F_0\neq0$, one has `massive' type IIA supergravity, for which \eqref{eqn:Fdef} must be modified to
\beq
F=  d_HC +e^{B\wedge}F_0,
\eeq
where $F_2=dC_1+F_0B$, $F_4=dC_3-H\wedge C_1+\frac{F_0}{2}B\wedge B$ and so on.
For the type IIA backgrounds discussed in this thesis, $F_0=0$, allowing us to ignore this additional term. In massless IIA, as a consequence of 
 \eqref{eqn:fieldrelation}, one has $F_{10}=dC_9-H_3\wedge C_7=0$.

We can now re-write \eqref{eqn:fieldrelation} in terms of the polyform by introducing an operator $\lambda$ in the following manner
\beq
F=\star \lambda F=\lambda * F,~~~~~~\text{where}~~~~~~\lambda \,\alpha_k =(-1)^{[\frac{p}{2}]}\alpha_k,
\eeq
for some differential form $\alpha_k$, of degree $k$. 

We note that $B_{MN}$ and $C_p$ are antisymmetric gauge potentials, generalising the $A_M$ potential from Electromagnetism to more dimensions, with the field strengths $H$ and $F_p$ playing a similar role to $F_{MN}$. We observe from \eqref{eqn:origfieldstengths} that one has the freedom to introduce the following gauge transformations
\beq\label{eqn:gaugetran}
B=B+d\Lambda_1,~~~~~~C_p=C_p+d\tilde{\Lambda}_{p-1}-H\wedge \tilde{\Lambda}_{p-3},
\eeq
with $\Lambda_1$ a 1-form and $\tilde{\Lambda}_k$ some $k$- form. As an example, one can perform the following gauge transformations for $F_5$: $C_2\rightarrow C_2+d\tilde{\Lambda}_1,~C_4\rightarrow C_4-H\wedge \tilde{\Lambda}_1$.

The action for the democratic formalism in `string frame' (and without sources) \cite{Koerber:2010bx} reads 
\beq\label{eqn:SUGRAactionsDem}
S=\frac{1}{2\kappa_{10}^2}\int d^{10}x\sqrt{-g_{10}} \bigg[e^{-2\Phi}\bigg(R+4\,d\Phi.d\Phi-\frac{1}{2}H.H\bigg)-\frac{1}{4}F.F\bigg],
\eeq
which when written in the usual formalism, take two alternative forms for (massive) type IIA and type IIB backgrounds \cite{Itsios:2013wd}\cite{Barranco:2013fza},
\begin{align}\label{eqn:SUGRAactions}
S_{mIIA}&=\frac{1}{2\kappa_{10}^2}\int
d^{10}x\sqrt{-g_{10}} \bigg[e^{-2\Phi}\bigg(R+4(\partial\Phi)^2-\frac{1}{12}H^2\bigg)-\frac{1}{2}\bigg(F_0^2+\frac{1}{2}F_2^2+\frac{1}{4!}F_4^2\bigg)\bigg]\nn\\[2mm]
&~~~~-\frac{1}{2}\bigg(dC_3\wedge dC_3\wedge B+\frac{1}{3}F_0 dC_3\wedge B^3 +\frac{1}{20}F_0^2B^5\bigg),\nn\\[2mm]
S_{IIB}&=\frac{1}{2\kappa_{10}^2}\int
d^{10}x\sqrt{-g_{10}} \bigg[e^{-2\Phi}\bigg(R+4(\partial\Phi)^2-\frac{1}{12}H^2\bigg)-\frac{1}{2}\bigg(F_1^2+\frac{1}{3!}F_3^2+
\frac{1}{2(5!)}F_5^2\bigg)\bigg]\nn\\[2mm]
&~~~~-\frac{1}{2}\Big(C_4\wedge H\wedge dC_2\Big),
\end{align}
where $F_k^2=F_{M_1...M_k}F^{M_1...M_k}$, $B^k\equiv B\wedge...\wedge B$ and $2\kappa_{10}^2 = (2\pi)^7l_s^8$, with $l_s$ the string length. For our purposes, we will fix $F_0=0$. These two actions include additional `Chern-Simons' terms compared to \eqref{eqn:SUGRAactionsDem}, however both forms lead to the same equations of motion. In the case of \eqref{eqn:SUGRAactionsDem}, one must impose the duality condition \eqref{eqn:fieldrelation} by hand. Given that this condition doesn't follow from the action, \eqref{eqn:SUGRAactionsDem} is known a pseudo-action - first discussed in \cite{Bergshoeff:2001pv}.  
With localized sources added to the background, such as D$p$-branes for example, one must include additional contributions to these actions - we will review the D$p$-brane contribution in \eqref{eq:branactions}.

The results above are written in string frame, signifying that the scalar curvature, $R$, includes an $e^{-2\Phi}$ factor. To recover the standard Einstein-Hilbert term in the action, one must use the alternative metric $g^E_{MN}\equiv e^{-\frac{\Phi}{2}}g_{MN}$. This is then referred to as the `Einstein frame'.

By varying the action, one derives the equations of motion to be satisfied (see 
\cite{Itsios:2013wd,Barranco:2013fza,Koerber:2010bx,Macpherson:2024qfi} for useful summaries). One then finds the `dilaton equation'
\beq\label{eqn:EOMD}
R+4D^2\Phi-4(\partial\Phi)^2-\frac{1}{12}H^2=0,
\eeq
the `Einstein equations'
\beq\label{eqn:EOME}
\hspace{-1cm}
R_{MN}+2D_MD_N\Phi-\frac{1}{4}H_{MN}^2 = 
e^{2\Phi} \begin{cases}
\frac{1}{2}(F_2^2)_{MN}+\frac{1}{12}(F_4^2)_{MN}-\frac{1}{4}g_{MN}\Big(F_0^2+\frac{1}{2}F_2^2 +\frac{1}{4!}F_4^2\Big)&IIA\\
 \frac{1}{2}(F_1^2)_{MN}+\frac{1}{4}(F_3^2)_{MN}+ \frac{1}{96}(F_5^2)_{MN}-\frac{1}{4}g_{MN}\Big( F_1^2 +\frac{1}{3!}F_3^2\Big) &IIB
\end{cases},
\eeq
the equations of motion for $H$,
\beq\label{eqn:EOMH}
d\Big(e^{-2\Phi}\star H\Big)= \begin{cases}
- F_0\,\star F_2+  F_2  \wedge \star F_4+\frac{1}{2}F_4\wedge F_4 &IIA\\
 F_1 \wedge \star F_3  +F_5\wedge F_3 &IIB
\end{cases},
\eeq
along with the following (source-free) Bianchi identities 
\beq\label{eqn:Bianchi}
dH=0,~~~~~d_HF=dF_p-H\wedge F_{p-2} = 0.
\eeq
 It is this Bianchi identity for $H$ which allows for the twisted exterior derivative, $d_H$. Fields satisfying these Bianchi identities (with no sources) are known as a `flux'. Additional contributions to these equations then arise from localized sources. For the equations of motion written in the democratic formalism, including the source terms, see \cite{Koerber:2010bx}.
 
The supersymmetry conditions for type II supergravity take the following approximate form
\begin{align}
&\delta_\varepsilon (\text{Bosons}) \sim \text{Fermions}=0,~~~~~~~~~~~~~~~~\delta_\varepsilon(\text{Fermions}) = \begin{cases}
  \delta\psi_M=\nabla_M\varepsilon +(\text{flux})\,\varepsilon &\\
    \delta \lambda = \Gamma^M \nabla_M\varepsilon +(\text{flux})\,\varepsilon & 
    \end{cases},\nn\\[2mm]
&~~~~~~~~~~~~~~~~~~~\Rightarrow ~~~~~~~~~~~~~\delta_\varepsilon\psi_M=0,~~~~~~~~~~~~~\delta_\varepsilon \lambda=0,
\end{align}
which must be satisfied, along with the Bianchi identities, to find supersymmetric solutions. The variation of the Bosonic fields are satisfied automatically due to the fermions vanishing, namely $\psi_M^a=\lambda^a=0$, as a consequence of the isometries of the backgrounds - see for example \cite{TomTextbook}. 
More concretely, the supersymmetry variations which must vanish for supersymmetry to be preserved, have the following forms \cite{Koerber:2010bx}
\begin{adjustwidth}{-1cm}{}
\vspace{-0.7cm}
\begin{align}\label{eqn:SUSYcondsSp}
&\delta \psi_M^1=\Big(\nabla_M+\frac{1}{4}\slashed{H}_M\Big)\varepsilon^1+\frac{e^\Phi}{16}\slashed{F}\,\Gamma_M \Gamma \varepsilon^2,~~~~~~~\delta \lambda^1= \Big(\Gamma^M\partial_M\Phi+\frac{1}{2}\slashed{H}\Big)\varepsilon^1    +\frac{1}{16}e^\Phi \Gamma^M \slashed{F}  \,\Gamma_M\,\Gamma\,\varepsilon^2 ,
\nn\\[2mm]
&\delta \psi_M^2=\Big(\nabla_M-\frac{1}{4}\slashed{H}_M\Big)\varepsilon^2-\frac{e^\Phi}{16}\lambda(\slashed{F})\,\Gamma_M \Gamma \varepsilon^1,~~~
\delta \lambda^2= \Big(\Gamma^M\partial_M\Phi-\frac{1}{2}\slashed{H}\Big)\varepsilon^2    -\frac{1}{16}e^\Phi \Gamma^M \lambda(\slashed{F})  \,\Gamma_M\,\Gamma\,\varepsilon^1,
\end{align}
\end{adjustwidth}
with the `modified dilatino equation' 
\begin{align}
\Gamma^M\delta \psi_M^1  - \delta \lambda^1&=\Big(\Gamma^M\nabla_M-\Gamma^M\partial_M\Phi+\frac{1}{4}\slashed{H}\Big)\varepsilon^1,~~~~~
\Gamma^M\delta \psi_M^2-\delta \lambda^2 =  \Big(\Gamma^M\nabla_M-\Gamma^M\partial_M\Phi-\frac{1}{4}\slashed{H}\Big)\varepsilon^2,
\end{align}
independent of any RR fields. Here $\nabla_M$ is the spin covariant derivative and the slash notation denotes that these fluxes come with gamma matrices in the following manner
\beq
\slashed{F}_k\equiv \frac{1}{k!}F_{M_1...M_k}\Gamma^{M_1}...\,\Gamma^{M_k},~~~~~~~~\slashed{H}_M\equiv \frac{1}{2}H_{MNP}\Gamma^{NP},~~~~~\slashed{H}\equiv \frac{1}{6}H_{MNP}\Gamma^{MNP}.
\eeq 
Comparisons with the forms given in \eqref{eqn:formnotation} leads to the following `Clifford map'
\beq\label{eqn:Clifford}
\alpha\equiv \sum_k\frac{1}{k!}\alpha_{i_1...i_k}dx^{i_1}\wedge ...\wedge dx^{i_k}~~~\leftrightarrow~~~\slashed{\alpha}\equiv \sum_k \frac{1}{k!}\alpha_{i_1...i_k}\gamma_{\alpha\beta}^{i_1...i_k},
\eeq
where the $k$-form $\slashed{\alpha}$ has two spinorial indices, $(\alpha,\beta)$, and hence called a `bispinor'.
A further discussion on supersymmetry is given in Section \ref{sec:Gstructures}, where a reformulation into the language of forms is reviewed.


\subsubsection{Branes}
Supergravity in ten dimensions contains a wider array of objects compared to its eleven dimensional counterpart, the most famous being `D-branes'. The `D' stands for `Dirichlet', and refers to the boundary conditions of open strings ending on the brane. This is in contrast to the M-branes of the eleven dimensional theory, which don't play the same role - as there are no open strings present in the theory. 

A D$p$-brane, of spatial dimension, $p$, is charged under $C_{p+1}$, and acts as a source for the RR potential. Given that the RR potentials have odd and even degree in IIA and IIB respectively, see \eqref{eqn:Cdefs}, one finds only $p$ even D-branes in IIA and $p$ odd D-branes in IIB. 

Due to the condition \eqref{eqn:fieldrelation}, if a D$p$-brane is electrically charged under $C_{p+1}$, it is magnetically charged under $C_{7-p}$, and vice versa. Given that $F_5=\star F_5$, one finds a self-dual D3 brane in type IIB. 

The smallest objects in the two theories are then a D$0$-brane in IIA, which is a point like object moving through time, and a D$(-1)$-brane in IIB, which is a point like object localized to only a single instant in time - consequently named the `Instanton'. The largest objects are a D$8$-brane in IIA, charged under $F_{10}=\star F_0$, and a space filling D$9$-brane in IIB (with a pure gauge potential).

From the NSNS sector, one concludes that the fundamental string, F1, must be electrically charged under $B_2$ and magnetically charged under $B_6$. The magnetic dual of F1 must then be some five-brane which is electrically charged under $B_6$ and magnetically charged under $B_2$ - this object is then called the `NS5-brane'. In IIA, these branes can be derived from dimensional reduction from the M5 brane of eleven dimensions, and in IIB can be derived from a D5 brane via an `S-duality'. This duality also relates the F1 with D1 branes.
 
An important characteristic of a brane is that it can end on another brane. This follows, via a chain of dualities, 
directly from the defining property of a D-brane - that a fundamental string, F1, ends on it. In the case of an F1 ending on a D3, performing an S-duality leads to a configuration in which a D1- brane ends on the D3-brane. Then, T-dualising along $(p-1)$ directions which are transverse to both branes, one derives a configuration in which a D$p$- brane ends on a D$(p+2)$- brane. Then, applying an S-duality to the configuration in which a D3-brane ends on a D5-brane, one finds a D3-brane which ends on an NS5- brane. Finally, performing a further T-duality along the worldvolume of the NS5 brane will lead to a D$(p\leq 6)$- brane ending on the NS5-brane. A very nice review is given in \cite{Giveon:1998sr}.

It is worth noting that there are additional (non-dynamical) extended objects called `Orientifold' (O)-planes, with an opposite tension to the D-brane (and hence seen as a source for anti-gravity). These objects will not be relevant to the solutions presented in this work, so we refrain from discussing them further here. See for example the discussion given in \cite{Legramandi:2020des} for further details.

\subsubsection{Brane solutions}
The presence of $p$-branes in the theory, extended along $(p+1)$- dimensions and localised in $(9-p)$- dimensions, are said to be `back-reacted' - as they have the effect of distorting flat space (and the flat space metric).  
 The solution then has $(p+1)$- dimensional Poincar\'e symmetry and $(9-p)$- dimensional rotational symmetry, allowing one to use polar coordinates for the transverse directions - with an $S^{8-p}$ sphere and radial coordinate, $r$. The $p$-brane solution is then derived using the symmetries of the brane, $ISO(p+1)\times SO(d-p)$, along with the equations of motion for the closed string.

For $N$ superimposed D$p$-branes (carrying $N$ units of RR charge $T_p$) extended along (or `wrapping') a flat manifold, $M_{||}^{1,p}$, transverse to $M_\perp^{9-p}$, the solution reads
\begin{align}\label{eqn:DpMetric}
&ds_{Dp}^2=h(r)^{-\frac{1}{2}}ds^2(M_{||}^{1,p})+h(r)^{\frac{1}{2}}\,ds^2(M_\perp^{9-p}),~~~~~~~e^\Phi=g_s\,h(r)^{\frac{3-p}{4}},~~~~~F=f_{8-p} \text{vol}(S^{8-p}), 
\nn\\[2mm]
&ds^2(M_{||}^{1,p})=-dt^2 +\sum_{a=1}^p (dx^a)^2,~~~~~~~ds^2(M_\perp^{9-p}) = \sum_{i=p+1}^9 (dx^i)^2  = dr^2+ r^2 ds^2(S^{8-p}),\nn\\[2mm]
&f_{8-p}\equiv \frac{N(2\pi l_s)^{7-p}}{\text{V}(S^{8-p})},~~~~~~~\text{V}(S^d)=\frac{2\pi^{(d+1)/2}}{\Gamma((d+1)/2)},~~~~~~r=\sqrt{\sum_{i=p+1}^9(x^i)^2},
\end{align}
with all remaining fields equal to zero, including $H=dB$. Here $\text{V}(S^{8-p})$ is the volume of the $S^{8-p}$ sphere (of unit radius), $\text{vol}(S^{8-p})$ is its volume form and $h(r)$ is a function of the radial direction, $r$, which for $p<7$, reads
\beq
h(r)=1+N\frac{r_0^{7-p}}{r^{7-p}},~~~~~~r_0^{7-p}\equiv \frac{g_s(2\pi\,l_s)^{7-p}}{(7-p)\text{V}(S^{8-p})},
\eeq
with $r_0$ the region of strong gravitational field. Considering the appropriate $h(r)$ for $p>7$ then leads to the conclusion that parallel D$p$-branes do not exert forces on one another. The $p=7$ case also requires more care - see \cite{TomTextbook} for details.

As one would expect, the solution tends to flat space as $r\rightarrow\infty$, where $h(r)\sim 1$. One can then approach the D$p$-brane stack by sending $r\rightarrow 0$, where $h(r)\sim r^{-7+p}$ to first order, which in most cases (with $p\neq 3$) gives rise to singularities, with the horizon at $r=0$ singular - which is expected to be resolved in the full string theory description. The behaviour in this limit was made more precise in \cite{Lozano:2019emq}, where for a  D$p$-brane wrapping some $M^{1,p}$ manifold, the metric and dilaton should be diffeomorphic to one of the following forms (to leading order)
\begin{adjustwidth}{-0.3cm}{}
\vspace{-0.6cm}
\begin{align}\label{eqn:Dbranemetrics}
\text{Dp brane: }&~ds^2 \sim r^{\frac{7-p}{2}}ds^2(M^{1,p})+r^{\frac{-7+p}{2}}\Big(dr^2 + r^2 ds^2(B^{8-p})\Big),~~~~e^\Phi\sim r^{\frac{(3-p)(-7+p)}{4}},\nn\\[2mm]
\begin{gathered}
\text{Dp smeared}\\ \text{on $\tilde{B}^s$ }
\end{gathered}:&~ds^2 \sim r^{\frac{7-p-s}{2}}ds^2(M^{1,p})+r^{\frac{-7+p+s}{2}}\Big(dr^2 +ds^2(\tilde{B}^s)+ r^2 ds^2(B^{8-p-s})\Big),~~e^\Phi\sim r^{\frac{(3-p)(-7+p+s)}{4}},
\end{align}
\end{adjustwidth}
where $B^{8-p}$ is a compact base one integrates over to get the D-brane charge and $\tilde{B}^s$ is some manifold the brane is smeared over (i.e. where it is not localized). 
The authors of \cite{Lozano:2019emq} also include analogous results for O-planes, which we omit here.

Notice that when $p=3$, the dilaton is constant, and the D3 solution describes an AdS$_5\times S^5$ geometry
\begin{align}\label{eqn:D3metric}
\text{D3 brane: }&~~ds^2 \sim ds^2(\text{AdS}_5) +ds^2(S^{5}),~~~~~~e^\Phi\sim 1,~~~~~~F_5=f_5 \text{vol}(S^5) \\
&~~ds^2(\text{AdS}_5) = \tilde{r}^{-2} \Big(ds^2(M^{1,p})+ d\tilde{r}^2   \Big),~~~~~\tilde{r}=r^{-1},\nn
\end{align}
with the D3 brane wrapping five-dimensional Anti-de Sitter space, AdS$_5$ (which has constant negative curvature). In this case, the horizon has a finite size. This example is then particularly pertinent in the AdS/CFT correspondence.


In the case of NS5 branes, one can use S-duality to deduce the correct solution from \eqref{eqn:DpMetric}, given by
\beq\label{eqn:NS5action}
ds_{NS5}^2=ds^2(M_{||}^{1,5})+h(r)\,(M_\perp^{4}),~~~~e^\Phi=g_s\,h(r)^{\frac{1}{2}},~~~~~H=2N \text{vol}(S^{3}),~~~~~h(r)=1+N\frac{l_s^2}{r^2}.
\eeq
 As in the D-brane case, the solution tends to flat space in the $r\rightarrow \infty$ limit. Approaching the NS5 brane stack, taking $r\rightarrow 0$, one finds $h(r)\sim r^{-2}$ to first order. One can then propose the NS5 branes have analogous behaviour to \eqref{eqn:Dbranemetrics}, wrapping some $M^{1,5}$ manifold in the following manner, 
 \begin{align}\label{eqn:NS5metrics}
\text{NS5 brane: }&~~ds^2 \sim ds^2(M^{1,5})+r^{-2}\Big(dr^2 + r^2 ds^2(B^{3})\Big),~~~~e^\Phi\sim r^{-1},\\[2mm]
\begin{gathered}
\text{NS5 smeared}\\ \text{on $\tilde{B}^s$ }
\end{gathered}:&~~ds^2 \sim ds^2(M^{1,p})+r^{ s-2 }\Big(dr^2 +ds^2(\tilde{B}^s)+ r^2 ds^2(B^{3-s})\Big),~~~e^\Phi\sim r^{\frac{s -2}{2}},\nn
\end{align}
 with $B^3$ the compact base integrated over to find the NS5 charge and $\tilde{B}^{3-s}$ some manifold over which the brane is smeared.
Comparing these results explicitly with the D5 brane, governed by \eqref{eqn:Dbranemetrics}, 
one can observe that the dilaton in each description behaves inversely, and the D5 brane has an additional multiplicative factor of $r^{\frac{2-s}{2}}\sim e^\Phi$ out the front of the whole metric with respect to the NS5 case. This demonstrates explicitly that NS5 branes are not D-branes, we will see further evidence of this shortly.

\subsubsection{ Effective action}
D$p$-branes are dynamical objects, with a mass per unit volume, feeling the force of gravity and responding to the various background fields present in the theory. The shape and location of the D-brane then change accordingly. One can view such dynamics as being controlled by the open strings that end on the D-brane, whose dynamics are themselves dependent on the background fields. One can then derive an action on the world-volume of the D$p$-brane which describe these dynamics (and added to \eqref{eqn:SUGRAactions} as a source term).

The bosonic part for the effective world-volume action of a D$p$-brane has two components, the `Dirac-Born-Infeld' (DBI) term, $S_{DBI}$, and the `Wess-Zumino' (WZ) (or `Chern-Simons') term, $S_{WZ}$, \cite{TomTextbook,Bachas:1995kx,Green:1996bh,Callan:1995xx,DbraneTextbook}
\begin{align} 
S_{\text{Dp}}&= S_{\text{DBI}}+S_{\text{WZ}},~~~~S_{\text{DBI}}= T_p\int e^{-\Phi}\sqrt{\det(g|_{D_p}+{\cal F})}\,d^{p+1}w,\nn\\[2mm]
S_{\text{WZ}}&= \mp T_p\int_{Dp} C|_{D_p}\wedge e^{-{\cal F}},~~~~~~~{\cal F}=B_2|_{D_p}+2\pi \tilde{f}_2,~~~~~~~T_p=\frac{1}{(2\pi)^pl_s^{p+1}},\label{eq:branactions}
\end{align}
where $\tilde{f}_2$ is the fields strength of a $U(1)$ gauge field living on the D-brane world-volume, $T_p$ is the $p$ volume tension (controlling the D-branes response to influences trying to change its shape, energy etc), and $C$ is the poly-form potential \eqref{eqn:Cdefs}. Following the notation used in \cite{TomTextbook}, the operation $|_{D_p}$ is included to make clear that the metric, $g_{MN}$, and all forms in space-time have a `pull back' (a restriction) to the $(p+1)$- dimensional D$p$-brane world volume (labelled $w^\mu$). This involves contracting each index $M$ with $\partial_\mu x^M$, where $\mu=0,...,p$ runs over the D$p$ world-volume. In the case of the metric, we have
\beq
g|_{D_p}=g_{\mu\nu} = g_{MN}\partial_\mu x^M \partial_\nu x^N.
\eeq
Fixing all fields to zero, except the background metric, $g_{MN}$, the action reduces to
\beq\label{eqn:onlygMNaction}
S_{\text{Dp}}= T_p\int e^{-\Phi}\sqrt{\det(g|_{D_p} )}\,d^{p+1}w,
\eeq
which measures the D$p$- brane volume in spacetime. 

The DBI action 
describes the coupling of the D$p$-brane to the NSNS fields. It requires the additional component, $\mathcal{F}$, describing the Electromagnetic fields on the fluctuating D$p$-brane. The need for this additional term becomes clear when considering T-duality, under which the $B_{2}$ potential mixes with the metric. 
One then requires a richer action to \eqref{eqn:onlygMNaction} in order to account for the dependence on $B_{2}$ when T-dualising to either a D$(p+1)$- or D$(p-1)$- brane. The preservation of spacetime gauge invariance then requires the $\mathcal{F}$ component to be a gauge invariant field strength on the D$p$ world-volume, with $d\mathcal{F}=H|_{D_p}$. Because one has the freedom to perform the gauge transformation \eqref{eqn:gaugetran}, $B_2$ is not itself gauge invariant. Hence, $\mathcal{F}$ requires an additional world-volume gauge field, $\tilde{f}_2 = -\frac{1}{2\pi}d\Lambda_1|_{D_p}$ in order to account for this gauge transformation. The exact form of the DBI action can be derived either by considering the $\sigma$- model (as it was originally derived), where it is essentially fixed by the D$p$-branes behaving like relativistic particles (see \cite{Callan:1995xx}), or by using T-duality to build up the action (see for example \cite{DbraneTextbook}\cite{Bachas:1995kx}). The full DBI action then obeys the rules of T-duality, as it should.

The WZ term describes the coupling of the D$p$-brane to the RR poly-form potentials. When $\mathcal{F}=0$, the only relevant potential is $C_{p+1}$, where
\beq
S_{\text{WZ}}= \mp T_p\int_{Dp} C|_{D_p}= \mp T_p\int_{w_{p+1}} C_{p+1}|_{D_p}= \mp T_p\int C_{M_0..M_p}\partial_0 x^{M_0}...\partial_p x^{M_p},
\eeq
which is simply the integral over the components of $C_{p+1}$ which span the $(p+1)$- dimensional world-volume, $w_{p+1}$, of the D$p$-brane. This term then calculates the charge of the D$p$ brane, with the brane tension, $T_p$, corresponding to a charge density. Hence, the elementary charge of the RR $C_{p+1}$ field is the D$p$-brane - analogous to the electron for the electric field.

When $\mathcal{F}\neq 0$ (with $F_0=0$), the D$p$ brane couples to additional RR potentials, $C_k$, with $k< p$, as follows
\beq
S_{\text{WZ}}= \mp T_p\int_{Dp}C|_{D_p}\wedge e^{-{\cal F}}= \mp T_p\sum_{k=0}^{\frac{p+1}{2}}\frac{(-1)^k}{k!} \int_{w_{p+1}}C_{p+1-2k}\wedge {\cal F}^k.
\eeq
For example, for a D2- and D4- brane, one finds the following WZ actions
\beq
S_{\text{WZ},D2}= \mp T_p\int_{D_2}\Big(C_3-\mathcal{F}\wedge C_1\Big),~~~~~~~~~S_{\text{WZ},D4}= \mp T_p\int_{D_4}\Big(C_5-\mathcal{F}\wedge C_3+\frac{1}{2}\mathcal{F}\wedge \mathcal{F}\wedge C_1\Big).
\eeq
In the D2 case, since $C_1$ couples to a D0 brane, the additional term should be interpreted as the presence of some D$0$-brane distribution on the D2. In fact, the D$0$-brane cannot be localized to a particular position within the D2, due to the integral over the whole D2 world-volume - the D$0$-brane is then said to be `smeared' over the worldvolume of the D2. This is a manifestation of the `Myers effect'. Consequently, the D2 charge contains a D0 charge - referred to as a `D2/D0 bound state'. The same arguments hold for the second example, where the D4 charge includes in general D2 and D0 charge. For the purposes of solutions considered in this work, where 
$\mathcal{F}\wedge \mathcal{F}=0$, the WZ actions clearly simplify.  We will discuss the D$p$-brane charge in more detail in the next subsection.

It is worth noting that when $F_0\neq0$, one must use an alternative form for the WZ action. This will not be necessary for our purposes, so the interested reader is directed to the discussion given in Section 1.3.3 of \cite{TomTextbook}. 

The $\mp$ factor out the front of the WZ action refers to the freedom to embed the D$p$ world-volume differently, by either flipping the sign of one of the coordinates or by exchanging the order of two coordinates in the wedge product - this is then related to the orientation of the D$p$-brane. Since the sign of the overall charge density is swapped, one interprets this as an anti-D$p$-brane. By convention, one uses $-$ for branes and $+$ for anti-branes.

One can include the $S_{Dp}$ action in the full ten-dimensional supergravity actions \eqref{eqn:SUGRAactions}, by introducing delta functions via the form
\beq
\delta_{D_p}\equiv \delta(x^1)...\delta(x^{9-p})dx^1\wedge ...\wedge dx^{9-p},
\eeq
such that, for example
\beq
\int_{Dp}C|_{D_p}=\int_{w_{p+1}} C_{p+1}|_{D_p}=\int C_{p+1} \wedge \delta_{D_{9-p}}.
\eeq
Varying the full action with respect to $C_{p+1}$, the Bianchi identity gains an additional source term on the right-hand side. In the $\mathcal{F}=0$ case, this reads
\beq\label{eqn:Bianchiwithsource}
d_HF_{8-p}=\mp 2\kappa_{10}^2T_{p}\delta_{D_{9-p}},
\eeq
and for the NS5 brane, one has a similar effective action
, with
\beq
dH=-2\kappa^2 T_{NS5}\delta_{NS5},~~~~~~~T_{NS5} = \frac{1}{(2\pi)^5l_s^6}.
\eeq
The right hand side of these Bianchi identities then correspond to the hodge duals of brane source currents, which lead to deriving the charge of a D$p$-brane. This is where our attention will now turn.




\subsubsection{Flux quantization and the Page charge}
In field configurations with only one non-trivial potential, which we will denote with a tilde, one has 
\beq
\tilde{F}_{p+2}=dC_{p+1},~~~~~~~~~~d\tilde{F}_{8-p}=0,
\eeq
in which the Bianchi identity, in the presence of a source, is dual to some current
\beq\label{eqn:simpleconslaw}
d\tilde{F}_{8-p}=\star j_{D_p},~~~~~~\Rightarrow~~~~d( \star j_{D_p})=0.
\eeq
From \eqref{eqn:fieldrelation}, one finds that the field strengths which couple to a D$p$- and D$(6-p)$- brane are dual to one another in the following manner
 \beq
F_{p+2}=(-1)^{[\frac{p+2}{2}]}\star F_{8-p}.
\eeq
With the conservation law given in \eqref{eqn:simpleconslaw}, and using Stoke's theorem,
\beq
\int_{S_k}d\alpha_{k-1} = \int_{\partial S_k}\alpha_{k-1},
\eeq
the electric (and magnetic) charges associated with the D$p$- and D$(6-p)$- brane then read,
\begin{align}
T_{p}=T_{6-p}^M&=  \int_{V^{9-p}}\star j_{D_p} = \int_{S^{8-p}}\star \tilde{F}_{p+2}, 
\nn\\[2mm]
T_{6-p}=T_{p}^M &=  \int_{V^{p+2}}\star j_{D_{(6-p)}}=  \int_{S^{p+2}} \star\tilde{F}_{8-p},
\end{align}
with the right hand side of each relation equivalent to the usual definition of charge, defined using a generalization of Gauss' law - surrounding the D$p$- and D$(6-p)$- branes by an $S^{8-p}$ and $S^{p+2}$ sphere, respectively.

One can check that the D$p$-brane charge is consistent with the generalization of the Dirac quantization argument of Electromagnetism. The wave function $\psi$ of an electrically charged D$p$-brane in the field of the magnetic D$(6-p)$- brane, takes the form 
\beq
\psi=\text{exp}\Big(i\,T_{6-p} \int_{E} C_{7-p}\Big)\psi,
\eeq
where $E$ is the equator of the $S^{8-p}$ sphere, and topologically equivalent to the $S^{7-p}$ sphere. The $S^{8-p}$ is divided into two semi-spheres, $U_N^{8-p},~U_S^{8-p}$, with the integral over each defined by Stoke's theorem, and in fact differing by an integral over the full $S^{8-p}$,
\beq\label{eqn:gluing}
 \int_{E} C_{7-p} =  \int_{U_{N,S}^{8-p}} \tilde{F}_{8-p},~~~~~~~\text{with}~~~~~~~~~ \int_{U_N^{8-p}} \tilde{F}_{8-p} -   \int_{U_S^{8-p}} \tilde{F}_{8-p} =   \int_{S^{8-p}} \tilde{F}_{8-p} = (-1)^{[\frac{p+2}{2}]}T_{p}.
\eeq
Using two different $C_{7-p}$ for each semi-sphere, which differ by a large gauge transformation,
\beq
C_{7-p}\rightarrow C_{7-p}+\Lambda_{7-p},~~~~~~~\frac{T_{6-p}}{2\pi}\int_{S^{7-p}}\Lambda_{7-p} \in \mathds{Z},
\eeq
the inconsistency between the two semi-spheres vanish. With a well defined wave function, 
\beq
\text{exp}\Big(i\,T_{6-p} \int_{E} C_{7-p}\Big)=1,~~~~~\Rightarrow~~~~~~T_{6-p} \int_{E} C_{7-p}=2\pi\,n~~~~~~~  n \in \mathds{Z},
\eeq
one then finds
\beq\label{eqn:Quantizedcharge}
(-1)^{[\frac{p+2}{2}]}T_{6-p}T_p=T_{6-p}\int_{S^{8-p}} \tilde{F}_{8-p}=2\pi\,n,~~~~~\Rightarrow~~~~~~Q_{D_p}=\frac{1}{(2\pi\,l_s)^{7-p}}\int_{S^{8-p}} \tilde{F}_{8-p} \in \mathds{Z} ,
\eeq
which gives the quantization condition for a $p$-brane charge, first presented in \cite{Deser:1997se}. Although the above arguments were presented using spheres, one is free to use a more general $\Sigma_{8-p}$ subspace. Of course, as we are surrounding the D$p$- brane with $\Sigma_{8-p}$ and using a generalization of Gauss' law, this subspace needs to be closed. For further discussions, see \cite{TomTextbook,DbraneTextbook,GravStringsTextbook,StringandMTextbook,StringandBranesTextbook,PolchinskiVol2Textbook}.


In the case of a general field configuration,  the twisted field strengths, $F_p$, and its `modified' Bianchi identity, $dF_p-H\wedge F_{p-2}=0$, complicate the usual notion of charge - giving rise to the three alternative options, outlined in \cite{Marolf:2000cb}. One can define: a `brane source charge', which is gauge invariant, localized but not quantized or conserved; a `Maxwell charge', which is gauge invariant, conserved but not quantized or localized; and a `Page Charge', which is conserved, localized and quantized, but not large gauge invariant.
This third option will become the most important, 
first considered by Page in \cite{Page:1983mke}, before being introduced to type IIA in \cite{Marolf:2000cb} (see also \cite{Zhou:2001rq}) and to type IIB in \cite{Zhou:2001cu}. 
 Let us review these different notions of charge a little further. 

The first option is the `Brane source charge', in which the Bianchi identity \eqref{eqn:Bianchiwithsource} is the dual to some current $\star j^{BS}_{Dp}$
\beq
d_HF_{8-p}=dF_{8-p}-H\wedge F_{6-p}=\star j^{BS}_{Dp},~~~~~~~~~Q^{BS}_{Dp}\sim \int_{V^{9-p}} \star j^{BS}_{Dp},
\eeq
each associated with a brane. The left hand side tells us that the current is both gauge invariant and localized (as it vanishes away from a source), so is associated with the external brane sources of the supergravity. However, this charge is expected to be non-quantized - as it is not conserved. That is, taking the external derivative using \eqref{eqn:Leibniz}, 
\beq
d(\star j^{BS}_{Dp}) =- dH\wedge F_{6-p}+H \wedge dF_{6-p} = - \star  j^{BS}_{NS5} \wedge F_{6-p}+H \wedge \Big(\star j^{BS}_{D(p+2)}+H\wedge F_{4-p}\Big),
\eeq
means that both NS5 branes and D$(p+2)$-branes are sources for the charge, with 
\beq
dH= \star  j^{BS}_{NS5}.
\eeq

The second option is the `Maxwell charge', which is defined to be simply the exterior derivative of $F_p$
\beq\label{eqn:maxwellcurrent1}
dF_{8-p}=\star j_{Dp}^{\text{Max}} = \star j^{BS}_{p}-H \wedge F_{6-p},
~~~~~~~~~Q^{\text{Max}}_{Dp}\sim \int_{V^{9-p}} \star j^{\text{Max}}_{Dp} =  \int_{\Sigma^{8-p}}  F_{8-p} ,
\eeq
in which the $H\wedge F_{6-p}$ component from the Bianchi identity is considered as a source for the field strength (so carries charge). This current is both gauge invariant and conserved. However, as it is carried by bulk fields, it is not localized to a brane. 

The final option is the `Page Charge', defined through writing the Bianchi identity as an exterior derivative. First, in the simple cases of a D4- and NS5- brane, one can write
\beq\label{eqn:D4page}
d(F_{4}-B \wedge F_{2} )=\star j_{D4}^{Page},~~~~~~~~~dH=\star j_{NS5}^{Page} \equiv  \star  j^{BS}_{NS5}, 
\eeq
which is identified as a current in the presence of a source, and otherwise corresponds to the source free Bianchi identities
\beq
dF_4-H\wedge F_2=0,~~~~dF_2=0~~~~~~\text{and}~~~~~dH=0,
\eeq
with
\beq
Q^{Page}_{D4}\sim \int_{V^{9-p}}\star j_{D4}^{Page} =  \int_{\Sigma^{8-p}} (F_{4}-B \wedge F_{2}),~~~~~~~~Q^{Page}_{NS5}\sim \int_{V^{9-p}}\star j_{NS5}^{Page} =  \int_{\Sigma^{8-p}}H.
\eeq
Notice from \eqref{eqn:D4page}, one could have replaced $B\wedge F_2$ with $H\wedge C_{1}$. It was shown that both options lead to the same Page charge in \cite{Marolf:2000cb}, as long as $\Sigma^{8-p}$ does not intersect any NS5- or D6- branes. 

This approach needs to be generalized a little further. In the case of a D2-brane for instance, in order to reproduce the source free Bianchi identities, the left hand side requires an additional component
\begin{align}
d\Big(F_6-B\wedge F_4+\frac{1}{2}B\wedge B\wedge F_2\Big) &= \Big(dF_6-H\wedge F_4\Big)-B\wedge \Big(dF_4-H\wedge F_2\Big)+\frac{1}{2}B\wedge B\wedge dF_2,\nn
\end{align}
with each term corresponding to a different Bianchi identity, leading to
\begin{align}\label{eqn:D2page}
 d(F_6\wedge e^{-B})& = (d_HF_6)\wedge e^{-B}=\star j_{D2}^{Page},~~~~~~~~~\text{with}~~~~~F_6\wedge e^{-B} = d(C_5\wedge e^{-B}),
\end{align}
where the second relation is easy to show. These results now naturally extend to a general D$p$-brane, defining the `Page flux', $\hat{F}^{\text{Page}}$, in the following manner
 \begin{align}\label{eqn:PageFlux}
&\hat{F}^{\text{Page}}=F \wedge e^{-B} = d(C\wedge e^{-B} ),
\end{align}
with
\beq
d\big(\hat{F}_{8-p}^{\text{Page}}\big) = \big(d_HF_{8-p}
\big)\wedge e^{-B}=\star j_{Dp}^{Page},
\eeq
reproducing \eqref{eqn:D4page} and \eqref{eqn:D2page} for $p=4$ and $p=2$, respectively. It then follows that the Page current is conserved and localized, however it is not gauge invariant. 

Due to the Maxwell current defined in \eqref{eqn:maxwellcurrent1}, one can now associate the twisted field strength $F_{8-p}$, defined in \eqref{eqn:origfieldstengths}, as a `Maxwell flux', where
\beq
F_{8-p}^{\text{Max}} \equiv   F_{8-p} =dC_{7-p}-H_3\wedge C_{5-p}.
\eeq
In summary then, in terms of the Maxwell flux, the three currents read
\beq
\star j_{Dp}^{BS} = d_HF_{8-p}^{\text{Max}},~~~~~~~~\star j_{Dp}^{\text{Max}} = dF_{8-p}^{\text{Max}},~~~~~~~~~\star j_{Dp}^{Page} = \big(d_HF_{8-p}^{\text{Max}}\big)\wedge e^{-B},
\eeq
where one can observe
\beq
\star j_{p}^{Page}= (\star j_{p}^{BS})\wedge e^{-B},
\eeq
which reproduces the various relations given in \cite{Zhou:2001rq} \cite{Zhou:2001cu} in a more general manner. 

It is this Page charge which will become important when deriving the charge of D$p$- branes, as it is both localized and conserved. Actually, the Page charge is conserved only in the absence of $B$ field sources, namely in the absence of NS5- branes (where the B field  is `topologically trivial' and can be smoothly contracted to zero throughout space). To calculate the quantized Page Charge, one must integrate over the Page current,
 \begin{align}
Q_{Dp}&=\frac{1}{(2\pi \,l_s)^{7-p}}\int_{V_{9-p}}\star j_{p}^{Page} = \frac{1}{(2\pi  \,l_s)^{7-p}}\int_{\Sigma_{8-p}}\hat{F}_{8-p}^{\text{Page}},
\end{align}
with the overall factor giving integer quantization, as in \eqref{eqn:Quantizedcharge}. 
The quantized Page charge for a D$p$- and NS5- brane is then defined by
 \begin{align}\label{eqn:Pagecharge}
Q_{Dp }&=\frac{1}{(2\pi  \,l_s)^{7-p}}\int_{\Sigma_{8-p}}d(C_{7-p}\wedge e^{-B} ),~~~~~~~~~Q_{NS5} = \frac{1}{(2\pi  \,l_s)^2}\int_{\Sigma_{3}}dB,
\end{align}
which will then identify the number of each brane appearing in the solution. 

In the presence of NS5 brane sources (with a `topologically non-trivial' $B$-field), the Page charge is no longer conserved throughout the entire space due to the large gauge transformation of $B$. In \cite{Zhou:2001cu}, a D3- brane probe was inserted into the region of $N$ coincident NS5 branes, defined by \eqref{eqn:NS5action}.
The action for this D3 is then given by \eqref{eq:branactions} for $p=3$ and $\mathcal{F}\wedge\mathcal{F}=0$, which include D1-branes living on the D3 probe. One can then vary the action to derive the equations of motion. A D3 brane initially located to the left of the NS5 brane stack, has $(n-N)$ D1-branes connecting the D3 brane to the $N$ coincident NS5 branes. When the D3 brane is moved across the NS5 branes however, one finds $n$ D1- branes connecting the NS5 stack with the D3 brane. That is, $N$ D1 branes are created by passing the D3 across the stack of $N$ NS5 branes. Hence, the D3 Page charge defined in two neighbouring regions, separated by NS5 branes, exhibits a jump in charge corresponding to the creation of D1 branes. This set of arguments then reproduce the `Hanany-Witten effect', first described in \cite{Hanany:1996ie}.  Further details are given in \cite{Zhou:2001cu} - see also \cite{Marolf:2001se,Nakatsu:1997ty}.

One is then able to construct a pictorial representation of the various branes present in the solution, with NS5 branes splitting the space into regions of varying D-brane charge. These set-ups are known as a `Hanany-Witten diagram', and play an important role in the AdS/CFT correspondence. 







\subsection{Dimensional reduction of M-Theory to Type IIA}
Let us now discuss dimensional reductions of M-Theory to type IIA, which is only possible if the M-Theory contains a compact $U(1)$ direction, $S^1$, of period $2\pi$. One can ignore any dependence on this circle by taking it to be small, allowing the theory to be described by ten dimensional fields. This procedure gives rise to a massless type IIA theory, as one is unable to switch on $F_0$ under the reduction. 

The eleven dimensional metric, $g_{MN}^{11}$, generates under the reduction a ten dimensional metric, $g_{MN}^{10}$, a vector field, $g_{M\,10}=C_M$ (identified with $C_1$), and a scalar, $g_{10\,10}=\Phi$ (identified with the dilaton). The eleven-dimensional three-form potential, $A_{MNP}$, then generates a three-form potential, $C_{MNP}$ (identified with $C_3$), and a two-form potential, $A_{MN\,10}=B_{MN}$ (identified with $B_2$). The eleven dimensional gravitino, $\psi_{M\alpha}$, generates a pair of spinors with both chiralities in ten dimensions, $\psi_{M\alpha}=\psi_{M}^1,\,\psi_{M\dot{\alpha}}=\psi_{M}^2,\psi_{10\alpha}=\lambda^1,\psi_{10\dot{\alpha}}=\lambda^2$ (identified with the IIA gravitinos and dilatinos). A summary of this discussion is as follows
\beq\label{eqn:11to10stuff}
g_{MN}^{11}\rightarrow 
\begin{cases}
g_{MN}^{10} \\
g_{M\,10}=C_M\\
g_{10\,10}=\Phi
\end{cases}~~~~~~~~~
A_{MNP}\rightarrow 
\begin{cases}
C_{MNP} \\
A_{MN\,10}=B_{MN}
\end{cases}~~~~~
\psi_{M\alpha}\rightarrow 
\begin{cases}
\psi_{M\alpha}=\psi_{M}^1\\
\psi_{M\dot{\alpha}}=\psi_{M}^2\\
\psi_{10\alpha}=\lambda^1\\
\psi_{10\dot{\alpha}}=\lambda^2
\end{cases}.
\eeq
For a dimensional reduction along a $U(1)$ direction, $\psi$ (which is not to be confused with the spinor notation), the formula reads
\begin{align}\label{eqn:reductionformula}
e^{-\frac{2}{3}\Phi
}ds_{IIA}^2 = ds^2 -  e^{\frac{4}{3}\Phi
}( d\psi+C_1)^2,~~~~~~~~A_3=C_3+B_2\wedge d\psi.
\end{align}
Under these identifications, the eleven dimensional action \eqref{eqn:11daction} becomes the massless IIA action given in \eqref{eqn:SUGRAactions} (with $F_0=0$). In addition, one needs to make the identification $\kappa_{11}^2=\kappa_{10}^2L_{\psi}$, with $L_{\psi}$ is the size of the $S^1$ direction coming from the integral over $d\psi$. \\
For a constant $g_s=e^\Phi$, the tension of the action \eqref{eq:branactions} effectively becomes $\tilde{T}_p=e^{-\Phi}T_p=g_s^{-1}T_p$. Hence, as $g_s$ becomes small, the D-branes become heavy. We then see that the mass of D0-branes follow from the form of $\tilde{T}_p$, where for $k$ D0-branes 
\beq\label{eqn:D0masses}
m_{D0}=\frac{k}{l_sg_s}~~~~\Rightarrow~~~~L_\psi=2\pi\,l_sg_s,
\eeq
which implies the size of the eleventh $U(1)$ dimension, $L_\psi$. For a constant dilaton, the Planck length becomes $l_P=g_s^{\frac{1}{4}}l_s$, with
\beq
L_\psi \ll l_{P}~~(g_s \ll 1),~~~~~~~L_\psi \gg l_{P}~~(g_s \gg 1),
\eeq
hence, for weak coupling the size of the $S^1$ is sub-Planckian (with an unclear meaning physically) and for strong coupling the $S^1$ size becomes macroscopic. Hence, one should view these eleven- and ten- dimensional effective actions as useful in the strong coupling ($g_s \gg 1$) and weak coupling ($g_s \ll 1$) regimes, respectively. Given that the string coupling, $g_s$, is dynamically generated by the string, one can conclude from $L_\psi$ that the dimension of spacetime itself is dynamically generated.

The concept of compactifying a theory on some $S^1$ was first considered by Kaluza \cite{Kaluza:1921tu} (and refined by Klein \cite{Klein:1926tv}) decades before string theory, in an attempt to unify gravity with Electromagnetism, by adding an additional $U(1)$ dimension to the usual four-dimensional GR. In that case, in an analogous manner to \eqref{eqn:11to10stuff}, the five-dimensional metric, with topology Mink$_4\times S^1$, became the usual four-dimensional metric, $g_{\mu\nu}$ (describing gravity), a vector field, $g_{\mu 4}$ (proportional to the $A_\mu$ of Electromagnetism in four dimensions) and an additional scalar, $g_{4\,4}$ (giving rise to massive spin two fields). The spin two fields are known as Kaluza-Klein (KK) modes and have a tower of masses, 
\beq\label{eqn:KKmasses}
m_k=\frac{2\pi\,k}{L_{S^1}},
\eeq
with $L_{S^1}$ the size of the $S^1$. Due to the analogous behaviour between $m_{D0}$ \eqref{eqn:D0masses} and the masses \eqref{eqn:KKmasses}, the D0-branes are interpreted as KK modes.



One can generalise the reduction formula \eqref{eqn:reductionformula}, introducing a constant, $X\in\mathds{Z}$ (preserving periodicity), in the following manner
\begin{align}\label{eqn:reductionformulaGEN}
e^{-\frac{2}{3}\Phi
}ds_{IIA}^2 = ds^2 -  \frac{e^{\frac{4}{3}\Phi
}}{X^2}\Big( d( X\,\psi)+X\,C_1\Big)^2,~~~~~~~~A_3=C_3+\frac{1}{X}B_2\wedge d(X\,\psi),
\end{align}
 where one reduces with respect to $(X\,\psi)$, with $C_1\rightarrow X\, C_1,~e^{\frac{4}{3}\Phi
}\rightarrow \frac{1}{X^2}e^{\frac{4}{3}\Phi
}$ and $B_2\rightarrow \frac{1}{X}B_2$.
\paragraph{Branes}
Recall that M-theory contains M2 branes (charged under $A_3$) and M5 branes (charged under $A_6$). Using the reduction formula \eqref{eqn:reductionformula}, we see that $A_3$ contains within it both $C_3$ and $B_2$. Hence, performing the dimensional reduction on an M2 brane extended along the $S^1$ direction, $\psi$, one derives a one-brane which is charged under $B_2$. The only object in type IIA with this dimension is the fundamental string, $F1$, with $T_{F1}=L_\psi T_{M2}$. In addition, when the M2 brane is extended along a transverse direction to $\psi$, the D2-brane is recovered which is charged under $C_3$, with $T_{D2}=T_{M2}$. In a similar manner, we see the $A_6$ contains a $B_6$ and $C_5$. Hence, an M5-brane extended along the $S^1$ derives a five-brane charged under $B_6$, which must be the NS5, with $T_{NS5}= T_{M5}$. For an M5 extended transversely to $\psi$, one recovers the D4-brane charged under $C_5$, with $T_{D4}=L_\psi T_{M5}$.


\beq\label{eqn:branesunderreduction}
M2\rightarrow 
\begin{cases}
F1\\
D2
\end{cases}~~~~~~~
M5\rightarrow 
\begin{cases}
D4 \\
NS5
\end{cases}.
\eeq
The D6 branes of type IIA are magnetically charged under $C_1$, which from the reduction formula \eqref{eqn:reductionformula}, is pure geometry - forming part of the eleven dimensional metric.


\subsection{T-Duality}
A string theory containing at least one $S^1$ direction (with abelian isometry group $U(1)$) contains a duality called (abelian) T-duality (ATD)- which in the case of type II supergravity, exchanges IIA and IIB theories. Under such a duality, both theories contain an $S^1$ direction but the radii of each is inversely proportional. The natural extension is to consider non-abelian T-duality (NATD), which generalises ATD to non-abelian isometries of the background, however things aren't so nice in these cases. 

\paragraph{Abelian T-Duality (ATD)}
The rules for abelian T-Duality, often called the `Buscher rules' were first presented by Buscher in \cite{Buscher:1987sk}. In this work however, we will utilise the form of the T-Dual rules presented in \cite{Kelekci:2014ima} (see also \cite{Legramandi:2020des}), corresponding to an abelian T-duality along a $U(1)$ direction, $y$, with the following Type IIA decomposition (denoted by $\mathcal{A}$)
   \begin{equation}\label{eqn:TD1}
   \begin{aligned}
  &ds_{10,\mathcal{A}}^2=ds_{9,\mathcal{A}}^2+e^{2C^\mathcal{A}}(dy+A_1^\mathcal{A})^2,~~~~~~~B^\mathcal{A}=B_2^\mathcal{A}+B_1^\mathcal{A} \wedge dy,\\[2mm]
  &F^\mathcal{A}=F_{\perp}^\mathcal{A}+F_{||}^\mathcal{A}\wedge E^y_\mathcal{A} ,~~~~~~~~~E^y_\mathcal{A}  =e^{C^\mathcal{A}}(dy+A_1^\mathcal{A}),
  \end{aligned}
  \end{equation}
with the RR fluxes split into parallel, $||$, and orthogonal, $\perp$, components with respect to the one-form $E^y_\mathcal{A}$. Consequently, the rank of $F_{\perp}^\mathcal{A}$ is one higher than $F_{||}^\mathcal{A}$. Performing a T-duality along $y$ then defines the type IIB solution (denoted by $\mathcal{B}$) in terms of the type IIA components, by the following identifications
  \begin{equation}\label{eqn:TD2}
  \begin{gathered}
  ds_{9,\mathcal{B}}^2=ds_{9,\mathcal{A}}^2,~~~~~\Phi^\mathcal{B}=\Phi^\mathcal{A}-C^\mathcal{A},~~~~~~~C^\mathcal{B}=-C^\mathcal{A},\\
  B_2^\mathcal{B}=B_2^\mathcal{A}+A_1^\mathcal{A} \wedge B_1^\mathcal{A},~~~~~~~~A_1^\mathcal{B}=-B_1^\mathcal{A},~~~~~~~~B_1^\mathcal{B}=-A_1^\mathcal{A},\\
  F_{\perp}^\mathcal{B}=e^{C^\mathcal{A}}F_{||}^\mathcal{A},~~~~~~~F_{||}^\mathcal{B}=e^{C^\mathcal{A}}F_{\perp}^\mathcal{A},
  \end{gathered}
  \end{equation}
with the IIB solution built using the analogue of \eqref{eqn:TD1}, namely
   \begin{equation} 
   \begin{aligned}
  &ds_{10,\mathcal{B}}^2=ds_{9,\mathcal{B}}^2+e^{2C^\mathcal{B}}(dy+A_1^\mathcal{B})^2,~~~~~~~B^\mathcal{B}=B_2^\mathcal{B}+B_1^\mathcal{B} \wedge dy,\\[2mm]
  &F^\mathcal{B}=F_{\perp}^\mathcal{B}+F_{||}^\mathcal{B}\wedge E^y_\mathcal{B} ,~~~~~~~~~E^y_\mathcal{B}  =e^{C^\mathcal{B}}(dy+A_1^\mathcal{B}).
  \end{aligned}
  \end{equation}
Of course, to derive the IIA T-dual of a type IIB theory, one must simply switch $\mathcal{A}\leftrightarrow\mathcal{B}$ in the steps above. One can then show that performing the T-duality for a second time will simply re-derive the original theory, and hence corresponds to a symmetry of string theory - for instance, $F_\perp^\mathcal{A}=e^{C^\mathcal{B}}F_{||}^\mathcal{B}=e^{C^\mathcal{B}}e^{C^\mathcal{A}}F_\perp^\mathcal{A} = e^{-C^\mathcal{A}}e^{C^\mathcal{A}}F_\perp^\mathcal{A} = F_\perp^\mathcal{A}$. One can perform a `TsT transformation' by T-dualising once, followed by an appropriate $SL(2,\mathds{R})$ transformation in the new solution, before T-dualising again. This has the effect of introducing an additional parameter into the original theory.
  
From the identifications \eqref{eqn:TD2}, one can see that the parallel and orthogonal RR flux components are switched under the T-duality, resulting in the rank of the two components, $F_{||}$ and $F_\perp$, increasing and decreasing by one, respectively. This property is associated with the behaviour of D$p$-branes under T-duality, which become D$(p-1)$- (or D$(p+1)$- ) branes when the T-duality is performed on a direction along (or orthogonal to) the D$p$-brane world-volume. In fact, there is a deeper explanation in terms of Dirichlet and Neumann boundary conditions of the open string, which are exchanged under T-duality.

\paragraph{Non-Abelian T-Duality (NATD)}
The obvious extension to ATD is to generalize T-duality to non-abelian isometries of the background (isometries with the structure of a non-abelian group). Historically, NATD was first developed for pure NSNS fields by de la Ossa and Quevedo \cite{delaOssa:1992vci}, before being extended to non-vanishing RR fields with an SU(2) isometry subgroup (and put on a similar footing to the Buscher rules for ATD) by Sfetsos and Thompson \cite{Sfetsos:2010uq}. Extensions beyond $SU(2)$ isometries were then studied in \cite{Kelekci:2014ima,Lozano:2011kb}.
 
Under a NATD, the symmetries of the compact manifold over which the transformations are made are not preserved under the transformation. That is, performing a NATD on a theory with an SU(2) isometry derives a new solution which no longer preserves the original SU(2) isometry. This means, unlike in the ATD case, one cannot transform back to the original theory by performing a second NATD. Hence, NATD should not be interpreted as a symmetry of string theory, but as a solution generating method. Natural questions then arise regarding the dual field theories of NATD backgrounds - see for example \cite{Lozano:2016kum}.

 
 
 
 

\section{Conformal Field Theory (CFT)}
Conformal Field Theories (CFTs) are an important class of quantum field theories, giving rise to exact results which would be difficult to obtain for massive theories. A CFT is a field theory which is invariant under conformal transformations, which is a coordinate change $\sigma^\alpha\rightarrow \tilde{\sigma}^\alpha(\sigma)$ such that the metric remains invariant up to an overall scale  
\beq
g_{\alpha\beta}\rightarrow \Omega^2(\sigma) g_{\alpha\beta}(\sigma).
\eeq
Hence, the physics of a CFT is invariant of length scale, and concerned only with angles (not distances). We only focus on giving a brief review of 4d $\mathcal{N}=2$ 
SCFTs, for further details see for example \cite{DbraneTextbook,StringandMTextbook,Tong:2009np}. 



\subsection{RG flow and the $\beta$-function}
Given that CFTs are independent of length scale, they form a fixed point of the renormalization group. One can then perturb the CFT by adding deformations to the action, which then controls the RG-flow away from the fixed point, 
\begin{equation}
S= S_{CFT} + g \int d^dx~ {\cal O},\label{deformationCFT}
\end{equation}
with $g$ the coupling (of dimension $\Delta_g$) and ${\cal O}$ a scalar primary field (of dimension $\Delta=d-\Delta_g$).
This gives rise to three classes of deformation \cite{Melo:2019jmg,Tong:2009np}:
\begin{itemize}
\item \textbf{Relevant} ($\Delta<d$): deformations relevant at low energies, with the IR physics flowing away from the CFT fixed point - changing the CFT dynamics. This flow is named the `renormalised trajectory', and only stops when it hits a new CFT. Here, 
$\Delta_g>0$. 
\item \textbf{Irrelevant} ($\Delta>d$): deformations irrelevant at low energies, with the IR physics converging to the CFT fixed point -  retaining the CFT dynamics. These deformations form a `critical surface' surrounding the fixed point. Here, 
$\Delta_g<0$. 
\item \textbf{Marginal} ($\Delta=d$): In this case, $g$ is dimensionless, $\Delta_g=0$. This case is more subtle and can be sub-divided into further categories - as once the theory is perturbed under marginal deformations, the dimension of $\Delta_g$ can change under the RG due to quantum corrections (which induces a beta-function for the coupling $g$) \cite{Park:2021nyc}:
\begin{itemize}
\item \textbf{Exactly Marginal}: the coupling $\Delta_g=0$ is unaffected by the RG. These deformations simply define a new CFT, flowing along a more extended fixed point region, such as a fixed line. 
\item \textbf{Marginally relevant}: the coupling $\Delta_g$, along with the perturbation, grows stronger under the flow to IR, $\Delta_g>0$, behaving similarly to relevant operators.
\item \textbf{Marginally irrelevant}: the coupling $\Delta_g$, along with the perturbation, grows weaker under the flow to IR, $\Delta_g<0$, behaving similarly to irrelevant operators.
\end{itemize}
\end{itemize}
Hence, one finds that a perturbation by an exactly marginal operator ${\cal O}$ gives rise to a family of CFTs close to the original fixed point. If more than one of these operators exist, one can have a conformal manifold, which requires the beta function to vanish for all couplings, $g_i$, in \eqref{deformationCFT},
\beq
\beta(g_i)=0.
\eeq
Such solutions are difficult to come by without supersymmetry. In the $d=4$ $\mathcal{N}=1$ case, Leigh and Strassler \cite{Leigh:1995ep} explained how beta functions for gauge and superpotential coupling are related, therefore implying the existence of marginal operators and a conformal manifold. A more powerful approach was then given in \cite{Green:2010da}. Within the context of the AdS/CFT correspondence, conformal manifolds are mapped to AdS supergravity vacua - see for instance \cite{Kol:2002zt,Lunin:2005jy}.

The flow on a renormalized trajectory is highly converging, with distinct physical systems starting from different points in the theory space all flowing towards the same region. As a result, UV physics becomes very challenging from an IR perspective, as there is little hope of extracting any details \cite{Melo:2019jmg}. This is natural as one would expect information loss flowing from the UV to the IR. The $c$ theorem makes these ideas more precise. The $c$ function is a quantity defined at conformal fixed points, and corresponds to the central charge of the CFT \cite{Tong:2009np}. It was proven by Zamolodchikov \cite{Zamolodchikov:1986gt} in two-dimensions, that with reasonable assumptions, $\frac{d c(t)}{dt} \leq 0$ under an RG flow. Here, $c(t)$ is the central charge 
 and $t=-\log\left( \frac{\mu}{\Lambda} \right)$. Hence, the $c(t)$ function is a monotonically decreasing function which goes between the central charges of UV and IR CFTs. This proves that the RG flow is irreversible here, and consequently, that the $c(t)$ function can be used as a measure of the number of degrees of freedom of the CFT at fixed points - see \cite{Komargodski:2011vj}.

\subsection{$\mathcal{N}=2$ SCFTs}\label{sec:SCFTsN2}
In 1994, the work of Seiberg and Witten \cite{Seiberg:1994rs} made an important contribution to the study into $\mathcal{N}=2$ dynamics, in which field theory information was encoded by the `Seiberg-Witten curve'. 
In 2009, using the knowledge that many $\mathcal{N}=2$ CFTs are obtained by compactifications of six dimensional $\mathcal{N}=(0,2)$ theories on a Riemann surface containing punctures, Gaiotto generalised these ideas to the conformal case in \cite{Gaiotto:2009we} - in which an $\mathcal{N}=2$ SCFT can be studied using the moduli space of punctured Riemann surfaces 
\cite{Nunez:2019gbg}.

\begin{figure}[H]
\begin{center}
 \begin{tikzpicture}[square/.style={regular polygon,regular polygon sides=4},scale=1]
        \node at (1,1.75) [square,inner sep=0.4em,draw] (f300) {$F_1$};
        \node at (3,1.75) [square,inner sep=0.4em,draw] (f400) {$F_2$};
          \node at (7,1.75) [square,inner sep=-0.2em,draw] (fp00) {$F_{P-1}$};

        \node at (1,0) [circle,inner sep=0.5em,draw] (c200) {$N_1$};
        \node at (3,0) [circle,inner sep=0.5em,draw] (c300) {$N_2$};
        \node at (7,0) [circle,inner sep=0.1em,draw] (cp00) {$N_{P-1}$};
         

        \draw (c200) -- (f300);
        \draw (c200) -- (c300);
        \draw (c300) -- (f400);
        \draw (cp00) -- (fp00);
       \draw (c300) -- (4.5,0);
        \draw (4.65,0) node[right] {.~.~.};
       \draw (cp00) -- (5.5,0);
    \end{tikzpicture}
    \end{center}
    \caption{ Linear Quiver of $\mathcal{N}=2$ SCFT}
\label{fig:Quiverexample}
\end{figure}
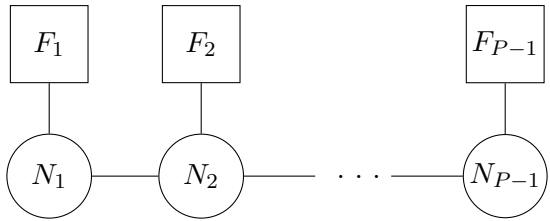

In the case of an $\mathcal{N}=2$ CFT$_4$ with the product gauge group $SU(N_1)\times...\times SU(N_{P-1})$, with $k\in[1,P-1]$, the theory is made up of 
\begin{itemize}
\item $(P-1)$ $\, \mathcal{N}=2$ vector multiplets 
 - with gauge group $SU(N_k)$. 
\item $(P-2)$ hypermultiplets 
- which transform in the fundamental of $SU(N_k)$ and the anti-fundamental of $SU(N_{k+1})$,  known as `bifundamental matter'.
\item  $F_k$ hypermultiplets 
- which transform in the fundamental of each gauge group $SU(N_k)$, with flavour groups represented by $SU(F_k)$ (noting flavour symmetries are global not gauge).
\end{itemize}
One can construct a linear quiver diagram, given in Figure \ref{fig:Quiverexample}, which defines the Lagrangian of the field theory. Here, the lines connecting the nodes represent the bi-fundamental matter, the circular nodes represent the colour symmetry (the gauge nodes) and the square nodes the flavour symmetries (which are global). 

For a linear quiver to be conformal with 8 supercharges, the beta function must vanish for some loop factor, $\lambda$, namely
\beq
\beta=(2N_c-N_f)\lambda=0,
\eeq
which then imposes the condition that the number of colours, $N_c$, must be double the number of flavours, $N_f$, for each gauge node
\beq\label{eqn:balanced}
2N_c=N_f~~~~~~~~~\Rightarrow~~~~~~~2N_k=F_k+N_{k+1}+N_{k-1},~~~~~~k\in[1,P-1],
\eeq
with $F_k$ the number of fundamentals. So the rank of each gauge node, $N_k$, is half the number of fields that transform in the fundamental of $N_k$ (and act as fundamental fields for $N_k$), namely $F_k, N_{k+1}$ and $N_{k-1}$. 

Hence, for a conformal field theory, the quiver diagrams must be finely tuned in order to satisfy \eqref{eqn:balanced}. Quivers satisfying this condition are called `balanced'. In cases where $2N_c>N_f$ and $2N_c<N_f$, the quivers are referred to as `underbalanced' and `overbalanced', respectively (see for instance \cite{Merrikin:2021smb} for some examples of overbalanced quivers). 

Following the arguments outlined in \cite{Cremonesi:2015bld} (see also \cite{Nunez:2019gbg}), one can define forward and backward discrete `lattice derivatives', $\partial_+N_i$ and $\partial_-N_i$ (as in lattice QFT), 
\beq
\partial_+N_k=N_{k+1}-N_k,~~~~\partial_-N_k=N_k-N_{k-1},~~~~~~~
\eeq
which, from the condition above, leads to
\beq
F_k= 2N_k-N_{k+1}-N_{k-1} = \partial_-N_k-\partial_+N_k =-\partial_+\partial_-N_k = -\partial_+s_k~,
\eeq
where the number of flavours, $F_k$, are a kind of discrete double derivative of the number of colours, $N_k$, with the discrete first derivative describing some discrete slope $s_k=\partial_-N_k $. Given that $F_k$ is positive, it follows from basic calculus that a plot of the various values of $N_k$ must be convex, with the slope decreasing with increasing $k$.

One can actually extend this discussion by encoding the information about the colour and flavour nodes into the definition of a `rank function', $\mathcal{R}(\eta)$, which then describes the physics of the CFT. Here $\eta$ is a parametrization of the `theory space', with $\eta\in[0,P]$, and
\begin{itemize}
\item ${\cal R}$ is a continuous, linear by pieces function.
\item The discontinuities of the slope $s=\mathcal{R}'$ occur at integer values of $\eta$.
 \item ${\cal R}$ is a convex function for which the gradient (or slope) between each discontinuity must be integer valued, and decrease in magnitude with increasing $k$.
\item $ {\cal R}(0)={\cal R}(P)=0$.
\end{itemize}
The number of fundamentals is then the double derivative, $F=-\mathcal{R}''$, and one has 
 \begin{figure}[H]
  \hspace{-1.5cm}
\subfigure
{
\centering
  \begin{minipage}{0.55\textwidth}
\begin{equation*}
  \hspace{-0.5cm}
\mathcal{R}(\eta) =
    \begin{cases}
     ~~~~~~~~~~~~~~~ N_1\eta & \eta\in[0,1]\\
      N_k+(N_{k+1}-N_k)(\eta-k) & \eta\in[k,k+1]~~,\\
   ~~~~~~~~~   N_{P-1}(P-\eta)  & \eta\in[P-1,P]
    \end{cases}      ~~~~~
\end{equation*}
\vspace{1.75cm}
  \end{minipage}
     \begin{minipage}{0.5\textwidth}
    \centering
     \begin{tikzpicture}[square/.style={regular polygon,regular polygon sides=4},scale=0.75, every node/.style={scale=0.85}]
 
   \begin{scope}[xshift=0cm,yshift=3.25cm]
 \draw[black,line width=0.3mm] (0,0) -- (1.5,1.2);
\draw[black,line width=0.3mm] (1.5,1.2) -- (3,2);
\draw[black,line width=0.3mm] (3,2) -- (4.5,2.4);
\draw[black,line width=0.3mm] (4.5,2.4)-- (6,2.2);
\draw[black,line width=0.3mm]  (6,2.2)-- (7.5,1.4);
\draw[black,line width=0.3mm] (7.5,1.4)-- (9,0);

\draw (0,0) node[below] {$0$};
\draw[gray,dashed,line width=0.3mm] (1.5,0)-- (1.5,1.2);
\draw (1.5,0) node[below] {$1$};
\draw[gray,dashed,line width=0.3mm] (3,0)-- (3,2);
\draw (3,0) node[below] {$2$};
\draw[gray,dashed,line width=0.3mm] (4.5,0)--  (4.5,2.4);
\draw (4.5,-0.15) node[below] {$.~.~.$};
\draw (6,0) node[below] {$P-2$};
\draw[gray,dashed,line width=0.3mm] (6,0)-- (6,2.2);
\draw (7.5,0) node[below] {$P-1$};
\draw[gray,dashed,line width=0.3mm] (7.5,0)--  (7.5,1.4);
\draw (9,0) node[below] {$P$};

\draw[gray,dashed,line width=0.3mm] (0,1.2)--  (1.5,1.2);
\draw (0,1.2) node[left] {$N_1$};

\draw[gray,dashed,line width=0.3mm] (0,2)-- (3,2);
\draw (0,2) node[left] {$N_2$};

\draw[-stealth, line width=0.53mm] (0,0)--(0,3) node[left ]{$\mathcal{R}(\eta)$};
\draw[-stealth, line width=0.53mm] (0,0) --(9.6,0) node[below ]{$\eta$};
       
         \end{scope}
 
  \begin{scope}[xshift=0cm,yshift=0cm]
        \node at (1.5,1.5) [square,inner sep=0.4em,draw] (f300) {$F_1$};
        \node at (3,1.5) [square,inner sep=0.4em,draw] (f400) {$F_2$};
          \node at (6,1.5) [square,inner sep=-0.2em,draw] (fp00) {$F_{P-2}$};
             \node at (7.5,1.5) [square,inner sep=-0.2em,draw] (fp001) {$F_{P-1}$};
        
        \node at (1.5,0) [circle,inner sep=0.5em,draw] (c200) {$N_1$};
        \node at (3,0) [circle,inner sep=0.5em,draw] (c300) {$N_2$};
        \node at (6,0) [circle,inner sep=0.1em,draw] (cp00) {$N_{P-2}$};
          \node at (7.5,0) [circle,inner sep=0.1em,draw] (cp001) {$N_{P-1}$};
        \draw (c200) -- (f300);
        \draw (c200) -- (c300);
        \draw (c300) -- (f400);
        \draw (cp00) -- (fp00);
        \draw (c300) -- (4,0);
        \draw (4.1,0) node[right] {.~.~.};
        \draw (cp00) -- (5,0);
        \draw (cp00) -- (cp001);
         \draw (cp001) -- (fp001);
         
         \end{scope}
    \end{tikzpicture}
  \end{minipage} 
}

\subfigure
{
\centering
  \begin{minipage}{\textwidth}
  \vspace{-1.75cm}
  \begin{equation*}
  \hspace{-1.85cm}
    \mathcal{R}'(\eta) =
    \begin{cases}
     ~~~~~~ N_1 & \eta\in[0,1]\\
      (N_{k+1}-N_k)& \eta\in[k,k+1]\\
   ~~  - N_{P-1}  & \eta\in[P-1,P]
    \end{cases} , ~~~~~
    \begin{aligned}
    &\\[8mm]
    &  \mathcal{R}''(\eta) =-  (2N_1-N_2)\delta(\eta-1)-(2N_2-N_1-N_3)\delta(\eta-2)...   \\
     \end{aligned}
\end{equation*}
\beq\label{eqn:Rkwithderivs}
~
\eeq
    \end{minipage} 
}

\subfigure
{
  \hspace{-1.25cm}
  \begin{minipage}{0.475\textwidth}
      \vspace{-0.75cm}
    \hspace{-0.83cm}
 \begin{tikzpicture}[scale=0.8, every node/.style={scale=0.95}]

 \draw[black,line width=0.35mm] (1,1.2)--  (2,1.2);
 \draw[black,line width=0.35mm] (2,0.8) -- (3,0.8);
 \draw[black,line width=0.35mm] (3,0.4) -- (4,0.4);
 \draw[black,line width=0.35mm] (4,-0.2) -- (5,-0.2);
 \draw[black,line width=0.35mm] (5,-0.8) -- (6,-0.8);
 \draw[black,line width=0.35mm] (6,-1.4) -- (7,-1.4);

\draw (1,0) node[left] {$0$};
\draw (2,-0.1) node[below] {$1$};
\draw (3,-0.1) node[below] {$2$};
\draw (6,-0.1) node[below] {$P-1$};
\draw (7,-0.1) node[below] {$P$};

\draw[black,line width=0.2mm] (2,0.1)-- (2,-0.1);
\draw[black,line width=0.2mm] (3,0.1)-- (3,-0.1);
\draw[black,line width=0.2mm] (4,0.1)-- (4,-0.1);
\draw[black,line width=0.2mm] (5,0.1)-- (5,-0.1);
\draw[black,line width=0.2mm] (6,0.1)-- (6,-0.1);
\draw[black,line width=0.2mm] (7,0.1)-- (7,-0.1);

\draw (1,1.2) node[left] {$N_1$};
\draw (1,0.8) node[left] {$N_2-N_1$};
\draw (1,-0.8) node[left] {$N_{P-1}-N_{P-2}$};
\draw (1,-1.4) node[left] {$-N_{P-1}$};
\draw (0.5,0.1) node[left] {$\vdots$};

\draw[-stealth, line width=0.53mm] (1,-2)--(1,2) node[left ]{$\mathcal{R}'(\eta)$};
\draw[-stealth, line width=0.53mm] (1,0) --(7.6,0) node[below ]{$\eta$};

\end{tikzpicture}
  \end{minipage}
      \begin{minipage} {0.18\textwidth}
    \centering
    \hspace{-1.5cm}
      \vspace{0.1cm}
    \begin{adjustbox}{minipage=\linewidth,scale=0.4}
\begin{ytableau}
  \none & & \none& \none & \none & \none& \none \\
    \none & & \none& \none & \none & \none& \none \\
      \none&&  & \none & \none & \none& \none \\
   \none & & & \none & \none & \none& \none \\
   \none  & &  &  & \none & \none& \none \\
       \none&  &   &  & \none & \none& \none \\
  \none  & \none& \none & \none &  &  & \\
    \none& \none & \none  &\none  & \none &  & \\
        \none& \none & \none   &\none  & \none &  & \\
  \none & \none& \none   &\none  & \none &  & \\
    \none  & \none& \none  &\none  & \none &\none  & \\
      \none& \none& \none   &\none  & \none &\none  & \\
  \none& \none & \none   &\none  & \none &\none  & \\
  \none&  \none& \none & \none & \none& \none& \none
\end{ytableau}
\end{adjustbox} 
  \end{minipage} 
     \begin{minipage}{0.45\textwidth}
         \vspace{-0.75cm}
                 \hspace{-1.5cm}
    \centering
\begin{tikzpicture}[scale=0.8, every node/.style={scale=0.9}]


\draw (1,0) node[below] {$0$};
\draw[-stealth, line width=0.3mm] (2,0)-- (2,1.2);
\draw (2,0) node[below] {$1$};
\draw[-stealth, line width=0.3mm] (3,0)-- (3,1.2);
\draw (3,0) node[below] {$2$};
\draw[-stealth, line width=0.3mm] (4,0)-- (4,1.8);
\draw (4.3,-0.15) node[below] {$.~.~.$};
\draw[-stealth, line width=0.3mm] (5,0)-- (5,1.8);
\draw (6,0) node[below] {$P-1$};
\draw[-stealth, line width=0.3mm] (6,0)-- (6,1.8);
\draw (7,0) node[below] {$P$};

\draw[gray,dashed,line width=0.3mm] (1,1.2)--  (2,1.2);

\draw (1,1.2) node[left] {$2N_1-N_2$};


\draw[-stealth, line width=0.53mm] (1,0)--(1,3) node[left ]{$\mathcal{R}''(\eta)$};
\draw[-stealth, line width=0.53mm] (1,0) --(7.6,0) node[below ]{$\eta$};

\end{tikzpicture}
  \end{minipage} 
}
  \caption{The generic continuous, linear by pieces and convex rank function, $\mathcal{R}(\eta)$, and its first and second derivatives - encoding the linear quiver and Young diagram.}
    \label{fig:Rankfull}
\end{figure}
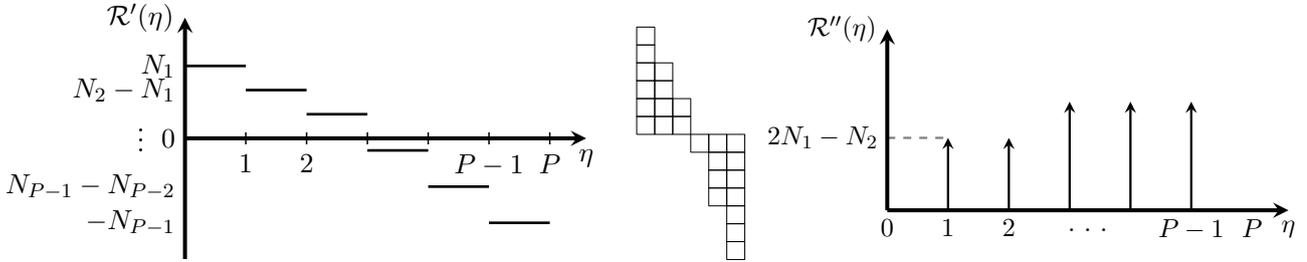
Clearly the rank function, $\mathcal{R}(\eta)$, encodes the rank of each gauge node, $N_k$ (hence the name); the double derivative, $\mathcal{R}''(\eta)=-F$, encodes the number of fundamentals for a balanced quiver, $F_1=2N_1-N_2,~F_2=2N_2-N_1-N_3$ etc; and the slope, $\mathcal{R}'(\eta)$, is a stepwise decreasing function. The position of each flavour in the quiver diagram corresponds to a change in slope of the rank function. Hence, if a flavour node is missing from the quiver diagram, the slope remains the same as the previous value.

The stepwise nature of the slope $\mathcal{R}'(\eta)$ then visually defines two Young diagrams, constructed from the positive slope terms on the left and the negative slope terms on the right. A depiction of this is included in Figure \ref{fig:Rankfull}.

\paragraph
{Central charges}
In the case of 4d supersymmetric field theories, there are two possible c-functions
, labelled $a$ and $c$. In the case of 4d ${\cal N}=2$ SCFTs with long linear quivers, which contain $N_v$ vector multiplets and $N_h$ hypermultiplets, one has (see \cite{Shapere:2008zf} \cite{Akhond:2021xio})
\begin{equation}
a=\frac{5 N_v + N_h}{24\pi},~~~~c=\frac{2 N_v + N_h}{12\pi}.\label{centralsN=2}
\end{equation}
 In the work of Komargodsky and Schwimmer \cite{Komargodski:2011vj}, they demonstrated that (under reasonable assumptions) $a$ is monotonically decreasing towards the IR - namely, $\frac{d a(t)}{dt} \leq 0$. As a result, the quantity $a$ can be used to measure the number of degrees of freedom of the CFT. In the holographic limit, in which the generic gauge rank, $N$, and linear quiver length, $P$, tend to infinity (corresponding to the supergravity approximation), it was shown that 
 \beq
 a=c~~~~~~~~~\text{as}~~~~N,P\rightarrow \infty,
 \eeq
due to the corrections being suppressed by $1/N$ and $1/P$ contributions (see for example \cite{Henningson:1998gx}). In the holographic limit, this quantity has been shown to match a string theory quantity called the `holographic central charge'. This was done using localisation and matrix model techniques in \cite{Nunez:2023loo}. We will return to this discussion in the next section. 

\paragraph
{Generalized quivers} It is worth noting the discussion on generalized quiver field theories provided in \cite{Gaiotto:2009gz}, which look like ordinary quivers but with additional $T_N$ factors (drawn as a triangular node). Here they interpret \eqref{centralsN=2} as either a computation of $(a,c)$ 
as above, or alternatively as the definitions of $N_v$ and $N_h$ for some arbitrary theory. Here, the theory still remains an $\mathcal{N}=2$ SCFT, however the integer nature of $N_h$ and $N_v$ can be broken more generally. These no longer correspond to the number of fields, and are simply viewed as a way to parametrize $a$ and $c$. See \cite{VanGorsel:2019xcw} for continuous quivers.

\paragraph
{Marginal deformations to $\mathcal{N}=1$ SCFTs}\label{sec:SCFTN1}

Finally, we comment on the work of \cite{Nunez:2019gbg} where they investigate the marginal deformation of an $\mathcal{N}=2$ SCFT. It proved useful here to express the theory in terms of $\mathcal{N}=1$ multiplets. 
They use the idea of R-symmetry mixing with flavour symmetries, and find the marginal deformation which breaks $\mathcal{N}=2\rightarrow\mathcal{N}=1$ is due to different interactions between the fields and different global symmetries. They then present the R-charge and superpotential term. 

Note, see the work of \cite{Gauntlett:2005jb} for exactly marginal deformations of the SCFT$_3$ case.
\subsection{Soft-SUSY breaking}\label{sec:softSUSY}
Soft-SUSY breaking involves adding SUSY breaking terms to the effective Lagrangian. If supersymmetry is then broken spontaneously at some high energy scale, the effective Lagrangian below such a scale would  be considered a `softly' broken supersymmetric theory. See for example \cite{Luty:2005sn} for further details.
The soft-SUSY breaking of 4d ${\cal N}=2$ SCFTs was analysed in \cite{Xie:2019aft}, where the representation theory of the superconformal algebra was employed (see \cite{Cordova:2016emh}). The (bosonic) global symmetries of 4d ${\cal N}=2$ SCFTs take the form
 \begin{equation}
 \text{SO}(2,4) \times \text{SU}(2)_R\times \text{U}(1)_R \times \text{G}_F,
 \nonumber
 \end{equation} 
 with SO(2,4) the 4d conformal group, the $\text{SU}(2)_R\times \text{U}(1)_R$ component is the R-symmetry of the theory, and $G_F$ represents other global symmetries, such as the flavour symmetries. A highest weight state is given by $|\Delta, R, r, j_1,j_2\rangle$, with $\Delta$ the scaling dimension, $R$ the $SU(2)_R$ charge, $r$ the $U(1)_R$ charge, and $(j_1,j_2)$ left and right spin for SO(1,3)$\sim$SO(4)$\sim$ SU(2)$\times$ SU(2).
The short representations were classified in \cite{Dolan:2002zh}, \cite{Cordova:2016xhm}.

For our purposes, we are only interested in the Coulomb branch operators, ${\cal E}_{(r,0,0)}$, which have the following component fields: a scalar, $A$; a spinor, $\Psi^i$, in the fundamental of SU(2)$_R$; scalars, $B^{(ij)}$, in the adjoint of SU(2)$_R$; an anti-self-dual two form, $F_{\alpha\beta}$; a spinor, $\Lambda^i$, in the fundamental of $\text{SU}(2)_R$; and a scalar, $C$. Here we are interested in the $[\Delta, \text{SU}(2)_R, \text{U}(1)_R ]$ values of the scalar components, $(A,B^{(ij)},C)$, as they can give rise to deformations of the theory. From \cite{Xie:2019aft}, one has
\begin{table}[H]
\centering
\begin{tabular}{c | c c c c  }
$~$&$A$&$B^{(ij)}$&$C$ \\
\hline
$\Delta$&$r$ &$r+1$&$r+2$ \\
$ \text{SU}(2)_R$&$0$ &$1$&$0$  \\
$\text{U}(1)_R$&$r$ &$r-1$&$r-2$  
\end{tabular}
\end{table}
with deformations in \eqref{deformationCFT} taking the form
 \begin{figure}[H]
 \hspace{-3.5cm}
\centering
\subfigure
{
  \begin{minipage}{0.4\textwidth}
\begin{equation*}
\delta S= g_i \int d^4 x ~{\cal O}_i.
\end{equation*}
  \end{minipage}
     \begin{minipage}{0.5\textwidth}
\begin{tabular}{c c | c c c c  }
$~$&$~$&SUSY&Global Symmetry&$\Delta$\\
\hline
${\cal O}_1:$&$B^{(12)}+cc$&$\mathcal{N}=0$ &$SU(2)_R$&$r+1$ \\
${\cal O}_2:$&$ B^{(11)}+cc$&$\mathcal{N}=1$ &$\frac{2}{r}U(1)_R+(2-\frac{2}{r})I_3$&$r+1$  \\
${\cal O}_3:$&$ B^{(22)}+cc$&$\mathcal{N}=1$ &$\frac{2}{r}U(1)_R-(2-\frac{2}{r})I_3$&$r+1$  
\end{tabular}
  \end{minipage} 
}
  \label{fig:softsusy}
\end{figure}
In the case of marginal operators, one should pick $r=3$ such that $\Delta=d=4$. Deformations with ${\cal O}_1$ breaks the supersymmetry completely to ${\cal N}=0$, but inherits a preserved SU(2) global symmetry from the R-symmetry. Deformations with ${\cal O}_2$ and ${\cal O}_3$ preserve ${\cal N}=1$ SUSY, and in the case of marginal deformations (with $r=3$), inherits $\text{U}(1)= \frac{2}{3} \left( \text{U}(1)_R \pm 2 I_3\right)$ R-symmetry.

\section{The AdS/CFT correspondence}

The original example of the AdS/CFT correspondence was constructed in 1997 by Maldacena \cite{Maldacena:1997re}, stating an equivalence between type IIB string theory on AdS$_5\times S^5$ and 4d $\mathcal{N}=4$ super Yang-Mills. Here the same physics is described using the two different languages, with the dynamics and Hilbert spaces agreeing on both sides. Notably, AdS$_{d+1}$ has an isometry group which is isomorphic to the conformal group of flat Minkowski space, Mink$_{1,d-1}$. 
See for example \cite{Hubeny:2014bla,Ramallo:2013bua,NatasteTextbook} for more detailed discussions on the topic. 
\subsection{Duality Toolbox}\label{sec:dualtoolbox}
We now list some key methods used within Part \ref{Part:SUSYdefs} to gain insight into the dual CFTs.
\subsubsection{Hanany-Witten \& Linear Quivers}
\begin{itemize}
\item On the \textbf{CFT side}: the physics is encoded by \textbf{linear quiver diagrams} 
which are encapsulated by the rank function, with $\mathcal{R}(\eta)$ defining the rank of each colour symmetry (the circular nodes) and $\mathcal{R}''(\eta)$ defining the rank of each flavour symmetry (the square nodes).
\item On the \textbf{Supergravity side}: One can build Hanany-Witten diagrams from the brane set-up of the solution, with the number of each brane encoded in the rank function of the dual CFT. Here, $\mathcal{R}(\eta)$ defines the number of colour D-branes and $\mathcal{R}''(\eta)$ defines the number of each stack of flavour branes.
\end{itemize}

\subsubsection{The Holographic Central charge}
\begin{itemize}
\item On the \textbf{CFT side}: it is a key characteristic quantity - the Free Energy of the CFT on S$^4$, counting the number of degrees of freedom.
\item On the \textbf{Supergravity side}: it measures a weighted effective volume of the internal manifold - see \cite{Bea:2015fja,Macpherson:2014eza,Henningson:1998gx,Klebanov:2007ws,Nunez:2018ags} for further details.
\end{itemize}
Given a supergravity solution, to calculate this internal volume, one can use the methodology outlined in \cite{Macpherson:2014eza}. Given a metric of the form
\beq\label{eqn:HCCmetricbreakup}
ds^2=\alpha(\rho,\vv{\theta})\Big(dx_{1,\hat{d}}^2+\beta(\rho)d\rho^2\Big)+g_{ij}(\rho,\vv{\theta})d\theta^id\theta^j,   
\eeq
one defines
\begin{equation}
\begin{gathered}
V_{int}=\int d\vv{\theta}\sqrt{\text{det}[g_{ij}]e^{-4\Phi}\alpha^{\hat{d}}},~~~~~~~~~~~~~~~~~~~~~~H=V_{int}^2,
\end{gathered}
\end{equation}
for which the corresponding holographic central charge, $c_{hol}$, is given by 
\beq\label{eqn:HCC}
c_{hol}=\frac{\hat{d}^{\hat{d}}}{G_N}\beta^{\hat{d}/2}\frac{H^{\frac{2\hat{d}+1}{2}}}{(H')^{\hat{d}}},
\eeq
where $G_N=8\pi^6 \alpha'^4g_s^2=8\pi^6$ (in the units $\alpha'=g_s=1$). This supergravity calculation will then correspond to the central charge of the dual CFT in the holographic limit (in which, from the rank function, $(N,P)\rightarrow\infty$). In the case of the $\mathcal{N}=2$ SCFTs, there are two central charges \eqref{centralsN=2}, $a$ and $c$, which tend to the same value in this limit, and is reproduced by the above calculation in the dual supergravity. 

Interestingly, it was demonstrated in \cite{Macpherson:2014eza} that the volume form of the holographic central charge remains invariant under an ATD. Using the B$\ddot{\text{u}}$scher rules \cite{Buscher:1987sk}, they show
\beq\label{eqn:HCCtransform}
\tilde{g}_{xx}=\frac{1}{g_{xx}},~~~~\tilde{\Phi}=\Phi-\frac{1}{2}\text{ln}(g_{xx}),~~~~\Rightarrow~~~e^{-2\tilde{\Phi}}\sqrt{\tilde{g}_{xx}}=\frac{e^{-2\Phi+\text{ln}(g_{xx})}}{\sqrt{g_{xx}}} = e^{-2\Phi}\sqrt{g_{xx}}.
\eeq
This result which becomes clear in this work, where we find the holographic central charge remains intact in ATDs from type IIA to type IIB. Note that the NATD case was also considered, see \cite{Macpherson:2014eza} for further details.

In the literature, various results for the holographic central charge have been calculated for diverse Mink$_D$ dimensions (with $D=\hat{d}+1$). These results are written in terms of $\mathcal{R}_n$ which is related to the rank function of the dual quiver, $\mathcal{R}(\eta)$ - see for example \eqref{eqn:potential}. In these backgrounds, the solutions depend on a partial differential equation (along with boundary conditions), which when solved using separation of variables, leads to solutions which depend on the rank function. We will review this explicitly for the Gaiotto-Maldacena class of solutions in the next section. Summarising some literature, we find
\begin{itemize}
          \item \underline{Mink$_6$:} In \cite{Nunez:2018ags}, they found
            \begin{equation}
                        \begin{aligned}
          &  c_{hol_{6D}} =-\frac{2}{3^8\pi^6}\int_0^P\alpha(\eta)\alpha''(\eta)d\eta,~~~~\text{with}  ~~~~~ \frac{\alpha''(\eta)}{81\pi^2}=\mathcal{R}(\eta)=\sum_{n=1}^\infty\mathcal{R}_n\sin\bigg(\frac{k\pi}{P}\eta\bigg),\\
          &~~~~~~~~~~~~~~~~ \Rightarrow ~~~~~~  c_{hol_{6D}}=\frac{P^3}{\pi^4}\sum_{n=1}^\infty \frac{\mathcal{R}_n^2}{n^2},
                       \end{aligned}
            \end{equation}
where $\alpha'''(\eta)=-162\pi^3F_0$. Integrating twice leads to the final result for $ c_{hol_{6D}}$.
            
            \item\underline{Mink$_5$:} In \cite{Legramandi:2021uds}, they found
            \begin{equation}
            \begin{aligned}
           & c_{hol_{5D}}=\frac{1}{2\pi^4}\sum_{n=1}^\infty n\, a_n^2 ,~~~~~~~~\text{with}~~~~~~ a_n= \frac{P}{2\pi n}\mathcal{R}_n,\\
          &~~~~~~~~~ \Rightarrow ~~~~~~ c_{hol_{5D}}= \frac{P^2}{8\pi^6}\sum_{n=1}^\infty  \frac{\mathcal{R}_n^2}{n}.
           \end{aligned}
            \end{equation}
        \item\underline{Mink$_4$:} In \cite{Nunez:2019gbg}, combining equations appropriately leads to
               \begin{equation}
            c_{hol_{4D}}= \frac{P\pi^3}{2^7}\sum_{n=1}^\infty \mathcal{R}_n^2.
            \end{equation}
        \item\underline{Mink$_3$:} In \cite{Akhond:2021ffz}, combining equations in a totally analogous way to the Mink$_5$ case above, one gets
                           \begin{equation}
            c_{hol_{3D}}= \frac{\pi}{2^7}\sum_{n=1}^\infty n\,\mathcal{R}_n^2.
            \end{equation}
\end{itemize}
These results then inspire a proposed general expression for the holographic central charge in Mink$_D$-dimensions, which we derive in Appendix \ref{sec:HCCdiversedim}, namely (with $N_0=N_P=0$)
          \begin{equation}\label{eqn:HCCnew}
\begin{aligned}
c_{hol_D}&=\text{coeff}(D)\sum_{n=1}^\infty \frac{P^{D-3}}{n^{D-4}}\mathcal{R}_n^2\\
&=\text{coeff}(D)\frac{2P^{D-1}}{\pi^4}\sum_{j,l=1}^{P}b_jb_l\,\,\text{Re}\bigg[\text{Li}_D\Big(e^{\frac{i\pi}{P}(j-l)}\Big) - \text{Li}_D\Big(e^{\frac{i\pi}{P}(j+l)}\Big)\bigg],
\end{aligned}
\end{equation}
with the \textit{Polylogarithm Function}, $\text{Li}_s(z)$, and the \textit{Riemann Zeta Function}, $\zeta(s)$, defined as follows
\beq
\text{Li}_s(z)= \sum_{n=1}^\infty \frac{z^n}{n^s},~~~~~\text{with}~~~~~\text{Li}_s(1)= \zeta(s)=\sum_{n=1}^\infty \frac{1}{n^s},~~~~~~\text{Li}_s(-1)=\sum_{n=1}^\infty \frac{(-1)^n}{n^s}=\zeta(s)\Big(2^{1-s}-1\Big).
\eeq
For the full expression including off-sets (namley, including $N_0,N_P\neq0$), see \eqref{eqn:HCCwithOffsets}. This expression proves useful in calculations, which we will see explicitly in the next section within the context of the Giatto-Maldacena class of solutions.

  \subsubsection{Spin 2 fluctuations}  
    \begin{itemize}
\item On the \textbf{Supergravity side}: One can study particular (and consistent) excitations of the supergravity metric, along the directions of the AdS.
\item On the \textbf{CFT side}: This simple fluctuation is associated with states of spin two in the dual CFT.
\end{itemize}
These excitations have been investigated for SCFTs of various dimension, with \cite{Csaki:2000fc} a precursor to this work. 

In the case of an AdS$_5$ geometry, following the work of \cite{Bachas:2011xa,Chen:2019ydk,Itsios:2019yzp}, written in {\it Einstein frame}, 
\begin{equation}
ds_E^2= e^{2A_E}ds^2(\text{AdS}_5) + ds^2(\mathcal{M}_5),
\end{equation}
with spin-two fluctuations only along the AdS$_5$ component of the metric, where the $d=10$ coordinates are labelled $X^M =(x^\mu,y^a)$,
\begin{equation}
\begin{aligned}
ds_E^2 &= e^{2A_E}\bigg[\Big(\tilde{g}_{\mu\nu}(x) +h_{\mu\nu}(x,y)\Big)dx^\mu dx^\nu + \tilde{g}_{ab}(y)dy^a dy^b\bigg],\label{genericglue}
\end{aligned}
\end{equation}
with the condition that the fluctuation $h_{\mu\nu}$ is written in terms of a tensor that is both transverse and traceless, namely
\begin{equation}
h_{\mu\nu}(x,y) = h_{\mu\nu}^{[tt]}(x)\mathcal{F}(y),~~~~~~~~~~~~\tilde{\nabla}^\mu  h_{\mu\nu}^{[tt]}=0,~~~~~~~~~~~~~\tilde{g}^{\mu\nu}h_{\mu\nu}^{[tt]}=0.
\end{equation}
As is discussed in \cite{Bachas:2011xa,Itsios:2019yzp}, the fluctuation of the Maxwell and Dilaton equations are satisfied trivially. However, the Einstein equations lead to the following condition,
\begin{equation}\label{eqn:Spin2-1}
\begin{aligned}
0 &= \tilde{\nabla}^\sigma \tilde{\nabla}_\sigma h_{\mu\nu} +2 h_{\mu\nu} + \tilde{\nabla}^a \tilde{\nabla}_a h_{\mu\nu} + 8 \tilde{\nabla}^a A \tilde{\nabla}_a h_{\mu\nu}\\
&= \tilde{\nabla}^\sigma \tilde{\nabla}_\sigma h_{\mu\nu} +2 h_{\mu\nu} + e^{-8A_E} \tilde{\nabla}^a \Big[e^{8A_E} \tilde{\nabla}_a h_{\mu\nu}\Big] \\
&:= \tilde{\nabla}^\sigma \tilde{\nabla}_\sigma h_{\mu\nu} +2 h_{\mu\nu} + \mathcal{L}(h_{\mu\nu}),
\end{aligned}
\end{equation}
with $h_{\mu\nu}$ acting like a scalar for $\tilde{\nabla}_a$. This is then the equation of motion for a graviton (with mass $M$) propagating on AdS$_5$, given by the `Pauli-Fierz' equation
\begin{equation}
 \tilde{\nabla}^\sigma \tilde{\nabla}_\sigma h_{\mu\nu} =(M^2-2)h_{\mu\nu},
\end{equation}
meaning 
\begin{equation}
\begin{aligned}
 \mathcal{L}(h_{\mu\nu})=- M^2 h_{\mu\nu}. \label{gravitonmass}
\end{aligned}
\end{equation}
For some scalar fluctuation $\mathcal{F}$, using (\ref{eqn:Spin2-1})-(\ref{gravitonmass}), we have
\begin{equation}\label{eqn:Leq}
\mathcal{L}(\mathcal{F}) =\frac{e^{-8A_E}}{\sqrt{\tilde{g}_{\mathcal{M}_5}}}\partial_a \Big(e^{8A_E}\sqrt{\tilde{g}_{\mathcal{M}_5}} \tilde{g}^{ab} \partial_b \mathcal{F}\Big)  =\frac{1}{\sqrt{\tilde{g}_{\mathcal{M}_5}}}\partial_a \Big(\sqrt{\tilde{g}_{\mathcal{M}_5}} \tilde{g}^{ab} \partial_b \mathcal{F}\Big) + 8 \tilde{g}^{ab}  \partial_a A \,\partial_b \mathcal{F}.
\end{equation}
  We will investigate explicit AdS$_5$ examples in Section \ref{sec:commentsCFT}, where we rely on these results - given in \cite{Bachas:2011xa,Chen:2019ydk,Itsios:2019yzp,Lima:2023ggy}.  
\section{The Gaiotto-Maldacena (GM) Class}
In order to construct a holographic dual description to the 4d $\mathcal{N}=2$ SCFTs reviewed in Section \ref{sec:SCFTsN2}, the supergravity solution must contain an AdS$_5$ factor with an $SU(2)_R\times U(1)_R$ R-symmetry and eight Poincar\'e supercharges. Hence, the solution should contain an $S^2$ and $S^1$, giving rise to an $SO(2,4)\times SU(2)_R\times U(1)_R$ bosonic isometry group. The most generic eleven-dimensional candidate was found by Lin-Lunin-Maldacena (LLM) in \cite{Lin:2004nb}, which we now briefly review.

\subsection{Lin-Lunin-Maldacena (LLM)}\label{sec:LLM}
The class of ${\cal N}=2$ AdS$_5$ solutions found by Lin-Lunin-Maldacena (LLM) in \cite{Lin:2004nb}, read
    \begin{align}\label{eqn:LLMgeneral}
    \frac{ds_{11}^2}{\kappa^{\frac{2}{3}}}&=e^{2\lambda}\bigg[4\,ds^2(\text{AdS}_5)+y^2 e^{-6\lambda}ds^2(\text{S}^2)+\frac{4}{1-y \partial_y D}(d\tilde{\chi}+A_a d\hat{x}^a)^2-\frac{\partial_yD}{y}\bigg(dy^2+e^{D}(d\hat{x}_1^2+d\hat{x}_2^2)\bigg)\bigg],\nn\\[2mm]
		A_a&=\epsilon_{ab}\partial_{\hat{x}_b} D,~~~~ e^{-6\lambda}=-\frac{\partial_yD}{y(1- y\partial_yD)},
    \end{align}
    and has a purely magnetic four-form  $G_4$
		\beq
		G_4= 2\kappa \bigg[(d\tilde{\chi}+ A_a d\hat{x}^a)\wedge d(y^3 e^{-6\lambda})+ y(1- y^2 e^{-6\lambda}) d A_a\wedge d\hat{x}^a-\frac{1}{2} \partial y\, e^{D}d\hat{x}^1\wedge d\hat{x}^2\bigg]\wedge \text{vol}(\text{S}^2),
		\eeq
 noting that the AdS$_5$ and S$^2$ metrics have a unit radius, $\tilde{\chi} \in [0,2\pi]$, and the quantity $\kappa$ indicates the size of the space. The bosonic isometry group of this solution is $SO(2,4)\times$\text{SU}(2)$_R\times$\text{U}(1)$_R$, with the latter two factors realising the ${\cal N}=2$ R-symmetry of the solution. The LLM class of backgrounds then depend on a single function and its derivatives, $D=D(y,\hat{x}_1,\hat{x}_2)$, which satisfies the Toda equation
    \beq
    \nabla_{(\hat{x}_1,\hat{x}_2)}^2D+\partial_y^2e^D=0.
    \eeq
    This equation is supplemented with boundary conditions - specified at $y=0$ and $y=y_c$, where the $S^2$ and the circle $(d\tilde{\chi}+A_a d\hat{x}^a)$ shrink smoothly, respectively.
    Using the two sub-manifolds, $\Sigma_4=(y,\hat{x}^1,\hat{x}^2,\tilde{\chi})$ and $\hat{\Sigma}_4=(S^2,\hat{x}^1,\hat{x}^2)$, the number of `colour' and `flavour' M5 branes are defined from the flux $G_4$, see \cite{Nunez:2019gbg}. 
    
    \subsubsection{An additional $U(1)$ isometry}
 One can transform $(\hat{x}_1,\hat{x}_2)$ to a new pair of coordinates $(r,\beta)$ via the following definitions
    \beq\label{eqn:x1x2transformation}
    \hat{x}_1=r\,\cos\tilde{\beta},~~~~~~~\hat{x}_2=r\,\sin\tilde{\beta},
    \eeq
   and imposing that $\tilde{\beta}$ is a $U(1)$ isometry of the background, with $\tilde{\beta} \in [0,2\pi]$. The LLM class then takes the form
   \begin{adjustwidth}{-0.6cm}{}
       \vspace{-0.7cm}
    \begin{align} 
\frac{ds^2}{\kappa^{\frac{2}{3}}}&= e^{2\lambda}\bigg[4\,ds^2(\text{AdS}_5)+y^2 e^{-6\lambda} ds^2(\text{S}^2)+\frac{4}{1-y\partial_yD}\Big(d\tilde{\chi}-\frac{r}{2}\partial_rD d\tilde{\beta}\Big)^2-\frac{\partial_yD}{y}\Big(dy^2+e^D (dr^2+r^2d\tilde{\beta}^2)\Big)\bigg],\nn\\[2mm]
G_4&= \kappa \bigg[-d(2y^3 e^{-6\lambda}) \wedge d\tilde{\chi} +  \bigg(d\bigg(e^{-6\lambda}\frac{y^2}{\partial_yD}r\partial_rD\bigg) - \partial_y(e^D)r\,dr + r\partial_rD\,dy\bigg) \wedge d\tilde{\beta} \bigg]\wedge \text{vol}(\text{S}^2),\label{eqn:LLM}
\end{align}
\end{adjustwidth}
with $D=D(r,y)$ now satisfying
\begin{equation}\label{eqn:LLMPDE}
\frac{1}{r}\partial_r(r \partial_r D) +\partial_y^2e^D=0,~~~~e^{-6\lambda} = \frac{-\partial_yD}{y(1-y\partial_yD)}.
\end{equation}
The new isometry is generated by a  special distribution of punctures (leading to D6 branes in Type IIA), due to a smearing of M5 branes \cite{Nunez:2019gbg}, and moves the cohomogeneity-three backgrounds into backgrounds of cohomogeneity-two.  
 
\subsection{Gaiotto-Maldacena (GM)}\label{sec:GM}
      From the AdS$_5$ $\mathcal{N}=2$ solutions of Lin-Lunin-Maldacena (LLM) \cite{Lin:2004nb}, one can derive the Gaiotto-Maldacena (GM)\cite{Gaiotto:2009gz} electrostatic form of the solution  from \eqref{eqn:LLM}, with bosonic isometry group $SO(2,4)\times SU(2)_R\times U(1)_R$ 
      and magnetic four-form, $G_4$,  via what is called the `B$\ddot{\text{a}}$cklund transformation'
      \begin{equation}\label{eqn:Backlund}
r^2e^D=\sigma^2,~~~~y=\dot{V},~~~~\text{log}(r)=V',
\end{equation}
which replaces $(r,y)$ with $(\sigma,\eta)$. Hence, the resulting class of electrostatic GM solutions which we will now review is only a special case of the more general LLM class given in \eqref{eqn:LLMgeneral}. The transformation from \eqref{eqn:LLM} is given explicitly in Appendix \ref{sec:Transformation}, where we show that the $U(1)$ directions $(\chi,\beta)$ of the GM class are not exactly equivalent to the $U(1)$ directions $(\bar{\chi},\bar{\beta})$ of the LLM solutions. They are in fact related in the following manner
 \beq
 (\chi,\beta)\rightarrow (\tilde{\chi}+ \tilde{\beta},~ -\tilde{\beta}),
 \eeq   
which is a subtlety first pointed out in \cite{Bah:2019jts,Bah:2022yjf,Couzens:2022yjl}. 

\begin{figure}[H] 
\begin{center}
\begin{tikzpicture}
\draw[-stealth] (2.45,1.2) -- (2.45,1.7) node[above] {$\sigma$};
\draw[-stealth] (2.45,1.2) -- (2.95,1.2)   node[right] {$\eta$};


\draw (1,0) -- (1,2);
\draw (1.01,0) -- (1.01,2);
\draw (1.02,0) -- (1.02,2);
\draw (1,0) node[below] {$0$};
\draw (0.65,2.2) node[below] {$\infty$};
\draw (4,0) node[below] {$P$};
\draw (4,0) -- (4,2);
\draw (3.99,0) -- (3.99,2);
\draw (3.98,0) -- (3.98,2);

\draw (1,0) -- (4,0);
\draw (1,0.01) -- (4,0.01);
\draw (1,-0.01) -- (4,-0.01);
\draw (1,0.02) -- (4,0.02);
\draw (1,-0.02) -- (4,-0.02);
\draw (1,0.03) -- (4,0.03);
\draw (1,-0.03) -- (4,-0.03);

\draw (2.5,-0.75) node[above] {$\mathcal{R}(\eta)$};

   \begin{scope}[yshift=2cm]
\draw (0,-1.9) -- (0.6,-1.9);
\draw (0.1,-1.95) -- (0.5,-1.95);
\draw (0.2,-2) -- (0.4,-2);
\draw (0.3,-1.9).. controls (0.3,-1.25)  .. (1,-1.25);

\draw (4.5,-1.9) -- (5.1,-1.9);
\draw (4.6,-1.95) -- (5,-1.95);
\draw (4.7,-2) -- (4.9,-2);
\draw (4.8,-1.9).. controls (4.8,-1.25)  .. (4,-1.25);
\end{scope}

\end{tikzpicture}
\end{center}
\caption{ Electrostatic problem with a line of charge positioned at $\sigma=0$ and extended along $\eta\in [0,P]$, with two parallel conducting plates at $\eta=0,P$. The charge density is defined by the function $\mathcal{R}(\eta)$.}
\label{fig:Electrostatics11}
\end{figure}
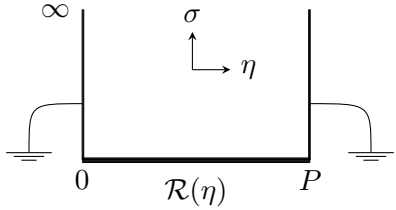

Following this transformation, the resulting $\mathcal{N}=2$ AdS$_5$ metric and potential $A_3$ (with magnetic four form $G_4=dA_3$) are then defined as follows
 \begin{equation}\label{eqn:GM}
\begin{gathered}
ds_{11}^2=f_1\Bigg[4ds^2(\text{AdS}_5) +f_2ds^2(\text{S}^2)+f_3d\chi^2+f_4\big(d\sigma^2 +d\eta^2\big)+f_5\Big(d\beta +f_6d\chi\Big)^2\Bigg],\\
A_3=\Big(f_7d\chi +f_8 d\beta \Big)\,\wedge\text{vol}(\text{S}^2),\\
\end{gathered}
\end{equation}
and an $SU(2)_R\times U(1)_R$ R-symmetry, realised by the $U(1)$ isometries and the presence of the $S^2$ (with $SU(2)$ isometry), where we use 
\beq\label{eqn:S2parametrization}
ds^2(\text{S}^2)= d\theta^2+\sin^2\theta d\phi^2,~~~~\text{vol}(\text{S}^2)=\sin\theta \,d\theta \wedge d\phi,
\eeq
with $  \theta\in [0,\pi],~\phi \in [0,2\pi]$. 
With this parametrization, it is easy to see this class of solutions have three $U(1)$ directions,  $(\partial_\beta,~\partial_\chi,~\partial_\phi)$. This will become important in our analysis given in Part \ref{Part:SUSYdefs}. The GM metric has an AdS$_5$ component, realising the $SO(2,4)$ isometries of the background and suggesting a 4d $\mathcal{N}=2$ SCFT dual. The warp factors $f_i=f_i(\eta,\sigma)$ are all functions of $(\eta,\sigma)$ alone, allowing the isometries of the background to be respected. The range of these coordinates are $ \eta\in [0,P]$ (with $P$ finite) and $\sigma\in [0,\infty)$. As it turns out, by imposing the preservation of eight Poincar\'e supersymmetries \cite{Lin:2004nb}, these eight warp factors can all be written in terms of a single potential $V(\eta,\sigma)$ and its derivatives (along with a constant $\kappa$), as follows
 \begin{equation}\label{eqn:fs}
\begin{gathered}
f_1=\kappa^{\frac{2}{3}}\bigg(\frac{\dot{V}\tilde{\Delta}}{2V''}\bigg)^{\frac{1}{3}},~~~~f_2 = \frac{2V''\dot{V}}{\tilde{\Delta}},~~~~~~~~f_3=\frac{4\sigma^2}{\Lambda},~~~~~~~~f_4 = \frac{2V''}{\dot{V}},~~~~~~~~f_5=\frac{2\Lambda V''}{\dot{V}\tilde{\Delta}},
      \\f_6=\frac{2\dot{V}\dot{V}'}{V''\Lambda},~~~~f_7=-\frac{4\kappa \dot{V}^2V''}{\tilde{\Delta}},~~~~~~~f_8=2\kappa\bigg(\frac{\dot{V}\dot{V}'}{\tilde{\Delta}}-\eta\bigg),
        \\ \tilde{\Delta} =\Lambda(V'')^2+(\dot{V}')^2,~~~~~~~~~\Lambda=\frac{2\dot{V}-\ddot{V}}{V''},
\end{gathered}
\end{equation}
where
        \begin{equation}
\dot{V}\equiv \sigma \partial_\sigma V,~~~~~~~~ V' \equiv \partial_\eta V.
    \end{equation}
This was a very powerful realisation, as it allows the complicated BPS system of eight non-linear coupled PDEs (corresponding to the eight warp factors) to be simplified down to a single linear PDE, namely the following cylindrically symmetric $d=3$ Laplace equation for $V(\eta,\sigma)$ 
    \begin{equation}\label{eqn:laplace}
\frac{1}{\sigma}\partial_\sigma (\sigma \partial_\sigma V)+\partial_\eta^2~\equiv~\ddot{V}+\sigma^2V''=0.
\end{equation}
This plays the analogous role to the Toda equation in the LLM class. In order for the metric to remain regular as the $S^2$ shrinks (up to some $\mathds{Z}_k$ orbifold singularities), the appropriate boundary conditions were found in \cite{Gaiotto:2009gz} (see also \cite{Nunez:2019gbg})
\begin{equation}\label{eqn:BCs}
\dot{V}\Big|_{\eta=0,P} =0,~~~~~~~~~~~\dot{V}\Big|_{\sigma=0} =\mathcal{R}(\eta).
\end{equation}
Due to this Laplace equation, one now has an Electrostatic description of the background \eqref{eqn:GM}, where a line of charge, with charge density, $\mathcal{R}(\eta)$, is positioned at $\sigma=0$ and extended along $\eta$ (with two parallel plates at $\eta=0,P$). See Figure \ref{fig:Electrostatics11} for a depiction.

The charge density, $\mathcal{R}(\eta)$, is then highly constrained by the boundary conditions and the quantisation of flux. To this end, one finds that ${\cal R}$ must be a continuous, linear by pieces and convex function, with $ {\cal R}(0)={\cal R}(P)=0$. In addition, the discontinuities of $\mathcal{R}'$ must occur at integer values of $\eta$, and the gradient between each discontinuity must be integer valued. These conditions then match the description of the rank function given above \eqref{eqn:Rkwithderivs}, with
 \begin{figure}[H]
   \hspace{-1.0cm}
\subfigure
{
\centering
  \begin{minipage}{0.6\textwidth}
\begin{equation*}
\mathcal{R}(\eta) =
    \begin{cases}
     ~~~~~~~~~~~~~~~ N_1\eta & \eta\in[0,1]\\
      N_k+(N_{k+1}-N_k)(\eta-k) & \eta\in[k,k+1]\\
   ~~~~~~~~~   N_{P-1}(P-\eta)  & \eta\in[P-1,P]
    \end{cases}      , 
\end{equation*}
  \end{minipage}
     \begin{minipage}{0.5\textwidth}
    \centering
\begin{tikzpicture}[scale=0.85, every node/.style={scale=0.95}]
\draw[black,line width=0.3mm] (1,0) -- (2,1.2);
\draw[black,line width=0.3mm] (2,1.2) -- (3,2);
\draw[black,line width=0.3mm] (3,2) -- (4,2.4);
\draw[black,line width=0.3mm] (4,2.4)-- (5,2.2);
\draw[black,line width=0.3mm]  (5,2.2)-- (6,1.4);
\draw[black,line width=0.3mm] (6,1.4)-- (7,0);

\draw (1,0) node[below] {$0$};
\draw[gray,dashed,line width=0.3mm] (2,0)-- (2,1.2);
\draw (2,0) node[below] {$1$};
\draw[gray,dashed,line width=0.3mm] (3,0)-- (3,2);
\draw (3,0) node[below] {$2$};
\draw[gray,dashed,line width=0.3mm] (4,0)--  (4,2.4);
\draw (4.3,-0.15) node[below] {$.~.~.$};
\draw[gray,dashed,line width=0.3mm] (5,0)-- (5,2.2);
\draw (6,0) node[below] {$P-1$};
\draw[gray,dashed,line width=0.3mm] (6,0)--  (6,1.4);
\draw (7,0) node[below] {$P$};

\draw[gray,dashed,line width=0.3mm] (1,1.2)--  (2,1.2);
\draw (1,1.2) node[left] {$N_1$};

\draw[gray,dashed,line width=0.3mm] (1,2)-- (3,2);
\draw (1,2) node[left] {$N_2$};

\draw[-stealth, line width=0.53mm] (1,0)--(1,3) node[left ]{$\mathcal{R}(\eta)$};
\draw[-stealth, line width=0.53mm] (1,0) --(7.6,0) node[below ]{$\eta$};
\end{tikzpicture}
  \end{minipage} 
}
  \label{fig:Rank}
\end{figure}
where the $\eta$ axis has been divided into $P$ unit cells, with $k=0,..,P-1$.
 Hence, the charge density of the supergravity description is related to the rank function of the dual quiver, $\mathcal{R}(\eta)$ (which encodes the physics of the dual CFT). This is a non-trivial result, as it relates the supergravity coordinate, $\eta$, with the `field theory space'. We will discuss this relationship a little further for the case of the type IIA reduction.
 
Approaching $\sigma=0$ for some generic $\eta$, and making use of the laplace equation \eqref{eqn:laplace}, one finds $\ddot{V}=0$ to leading order. Consequently, one finds that the sub manifold $(\sigma, \chi)$ will vanish as $\mathbb{R}^2$ in polar coordinates. Using the boundary conditions \eqref{eqn:BCs}, the metric component then becomes
   \beq 
   f_4 d\sigma^2+f_3  d\chi^2  \rightarrow   \frac{2V''}{\mathcal{R}(\eta)}\bigg( d\sigma^2 +\sigma^2d\chi^2 \bigg),~~~~~~~\text{with}~~~\dot{V}\Big|_{\sigma=0} =\mathcal{R}(\eta). 
     \eeq
Approaching the loci of the ${\cal R}'$ discontinuity, with $\eta=k$ and $\sigma=0$, the sub-manifold $(\sigma,\eta,\chi,\beta)$ tends to a $\mathbb{R}^4/\mathbb{Z}_{b_k}$ orbifold singularity, where
\beq
b_k=\mathcal{R}'(k-1)-\mathcal{R}'(k)=2N_k-N_{k+1}-N_{k-1}, 
\eeq
which is the difference in gradient between either side of the discontinuity. We will discuss orbifolds more in Section \ref{sec:Spindles}. In order to investigate the remaining boundaries of the space, one needs to define a specific solution of the Laplace equation. We will return to this analysis more thoroughly in  Part \ref{Part:SUSYdefs} for type IIA and type IIB daughter solutions.


\subsubsection{Particular solutions}



Solutions to the Laplace equation \eqref{eqn:laplace} include a recursive solution presented in \cite{Nunez:2019gbg} (
an extension to the approach of \cite{Itsios:2017nou}\cite{Nunez:2018qcj}) as well as three interesting potentials presented in \cite{Couzens:2022yjl}. However, in this work we adopt the the solution discussed in \cite{Reid-Edwards:2010vpm,Aharony:2012tz}, which defines a potential over the whole range of $\sigma$ by using the separation of variables 
\begin{equation}\label{eqn:potential}
\begin{aligned}
&V(\sigma,\eta) =-\sum_{n=1}^\infty \mathcal{R}_n \sin\bigg(\frac{n\pi}{P}\eta\bigg) K_0\bigg(\frac{n\pi}{P}\sigma\bigg),\\[2mm]
&\mathcal{R}_n = \frac{2}{P}\int_{0}^P \mathcal{R}(\eta)\sin\bigg(\frac{n\pi}{P}\eta\bigg)d\eta = \frac{2P}{(n\pi)^2}\sum_{k=1}^{P}b_k\,\sin\Big(\frac{n\pi\,k}{P}\Big),~~~~~~~~b_k=2N_k-N_{k+1}-N_{k-1},
\end{aligned}
\end{equation}
 where $K_0$ is a modified Bessel function of the second kind. This solution indeed satisfies the necessary boundary conditions \eqref{eqn:BCs}. In general, the rank function can be written as a Fourier series, as follows 
\begin{equation}\label{eqn:rank}
\mathcal{R}(\eta) = \sum_{n=1}^\infty \mathcal{R}_n \sin\bigg(\frac{n\pi}{P}\eta\bigg).
\end{equation}
Using the integral representation of the Bessel function, one can rewrite the potential \eqref{eqn:potential} in the following way
\beq
K_0\bigg(\frac{n\pi}{P}\sigma\bigg)=\int_0^\infty \frac{\cos(\frac{n\pi}{P}\sigma\,t)}{\sqrt{t^2+1}}dt~~~~~~\Rightarrow ~~~~~~~~V(\sigma,\eta)=-\int_0^\infty \frac{\mathcal{R}(u)}{\sqrt{(u-\eta)^2+\sigma^2}}du,
\eeq
using \eqref{eqn:rank} and $u=\eta+\sigma\,t$, with full details given in Appendix A of \cite{Nunez:2019gbg}. This is now the form of an electric potential for an odd-extended density of charge, $\mathcal{R}$, at a point $(\sigma,\eta)$ along the $\eta$ axis. This demonstrates the interpretation as an electrostatic problem, with $\mathcal{R}$ a density of charge.

 It will prove useful to us later, when investigating the boundary of daughter Type IIA solutions, to introduce an alternative form for $\dot{V}$, shown to be equivalent in \cite{Reid-Edwards:2010vpm}
\begin{align}
\dot{V}&=\frac{\pi}{P}\sum_{n=1}^{\infty}n{\cal R}_n\sigma\sin\bigg(\frac{n\pi}{P}\eta\bigg) K_1\bigg(\frac{n\pi}{P}\sigma\bigg),\label{eq:alternative}\\[2mm]
&=\frac{1}{2}\sum_{m=-\infty}^{\infty}\sum_{k=1}^Pb_k\left(\sqrt{\sigma^2+(\eta-2m P+k)^2}-\sqrt{\sigma^2+(\eta-2m P-k)^2}\right),\nn
\end{align}
where $b_k=2N_k-N_{k+1}-N_{k-1}$. Along the $\sigma=0$ boundary, the $m=0$ contribution gives rise to the odd extension of $\cal{R}$, defined in the interval $\eta\in [-P,P]$. Due to the remaining values of $m$, it then becomes $2P$ periodic for $\eta\in \mathbb{R}$.




\subsection{$\mathcal{N}=2$ preserving reduction to Type IIA}\label{sec:GMIIA}
Using the reduction formula \eqref{eqn:reductionformula}, one can dimensionally reduce the $d=11$ GM solution \eqref{eqn:GM} along one of the three $U(1)$ directions, $(\beta,\chi,\phi)$. As it turns out, one can only preserve the full $\mathcal{N}=2$ supersymmetry under a $\beta$ reduction. This will be made more clear in Chapter \ref{chap:setup}, where we also consider new parametric deformations of the following class of solutions, which break some or all of the supersymmetry. For now, we review the $\beta$ reduction case, which derives the following $\mathcal{N}=2$ solution
\beq\label{eq:N=2} 
\begin{aligned}
ds^2&= f_1^{\frac{3}{2}} f_5^{\frac{1}{2}}\bigg[4ds^2(\text{AdS}_5)+f_2ds^2(\text{S}^2)+f_4(d\sigma^2+d\eta^2)+f_3 d\chi^2\bigg],\\[2mm]
e^{\frac{4}{3}\Phi}&=  f_1 f_5,~~~~  H = df_8\wedge \text{vol}(\text{S}^2),~~~~C_1=  f_6d\chi,~~~~ C_3=f_7 d\chi\wedge\text{vol}(\text{S}^2),
\end{aligned}
\eeq
with $\Phi$ the dilaton and the gauge invariant (Maxwell) fluxes $F_2=dC_1,~F_4=dC_3-H\wedge C_1$ and $H=dB_2$. The equations of motion and  Bianchi identities \eqref{eqn:EOMD}-\eqref{eqn:Bianchi} (with $F_0=0$) are implied by their $d=11$ equivalents. The warp factors $f_i$ are defined in \eqref{eqn:fs}, and will remain consistent with this definition throughout this work (unless specified otherwise). For a positive metric, one has the requirement $\frac{\dot{V}}{V''}>0$ (except at the boundaries) - see for example \cite{Reid-Edwards:2010vpm}\cite{Aharony:2012tz}. It was proven in \cite{Macpherson:2016xwk} that \eqref{eq:N=2} is the most general type IIA AdS$_5$ solution to admit an $SU(2)$ R-symmetry, associated with the $S^2$. 

From the flux content and the discussions around \eqref{eqn:branesunderreduction}, the solution contains D6-, D4- and NS5- branes, magnetically charged under $C_1,~C_3$ and $B_2$, respectively. Recall from Section \ref{sec:typeII}, to calculate the quantized charges for these branes, one requires the Page flux and Page charge, given in \eqref{eqn:PageFlux} and \eqref{eqn:Pagecharge} respectively. Re-written here for clarity, they read
 \begin{align}
&\hat{F}^{\text{Page}}=F \wedge e^{-B} = d(C\wedge e^{-B} ),~~~~
Q_{Dp }=\frac{1}{(2\pi  \,l_s)^{7-p}}\int_{\Sigma_{8-p}}\hat{F}_{8-p}^{\text{Page}}
,~~~~Q_{NS5} = \frac{1}{(2\pi  \,l_s)^2}\int_{\Sigma_{3}}dB.\nn
\end{align}
One can define $B_2$ in terms of an integration constant, $k$,
\beq\label{eq:neq2B2def}
B_2=(f_8+2\kappa \,k)\text{vol}(S^2)=2\kappa \Big(\frac{1}{4}\dot{V}f_5f_6-(\eta-k)\Big)\text{vol}(S^2),
\eeq
where $k$ can shift according to large gauge transformations of $B_2$, so need not be fixed globally. In other words, $k$ is able to change as one traverses the internal space. This will become important when investigating the charge of D6 branes on the $\sigma=0$ boundary. The Page charges then read
\beq\label{eqn:PagesforGM}
\hat{F}_2^{\text{Page}} = d(f_6)\wedge d\chi,~~~~~~~~\hat{F}_4^{\text{Page}} =2\kappa\,d\Big(f_6(\eta-k)-\frac{2}{V''}\frac{f_2}{f_5}\Big)\wedge d\chi \wedge \text{vol}(S^2),
\eeq
which are clearly closed locally. To derive the higher form Page fluxes (such as $\hat{F}_6$ etc), it proves easier to use the method of G-structures. We will present them clearly in Section \ref{sec:IIAGstructureN=2}, following the review of G-structures given in Section \ref{sec:Gstructures}.

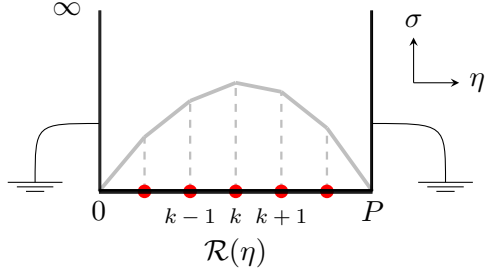
\begin{figure}[H]
\centering  
\begin{tikzpicture}[scale=1.2, every node/.style={scale=1}]


\draw[lightgray,line width=0.5mm] (1,0) -- (1.5,0.6);
\draw[lightgray,line width=0.5mm] (1.5,0.6) -- (2,1);
\draw[lightgray,line width=0.5mm] (2,1) -- (2.5,1.2);
\draw[lightgray,line width=0.5mm] (2.5,1.2)-- (3,1.1);
\draw[lightgray,line width=0.5mm]  (3,1.1)-- (3.5,0.7);
\draw[lightgray,line width=0.5mm] (3.5,0.7)-- (4,0);

\draw[lightgray,dashed,line width=0.3mm] (1.5,0)-- (1.5,0.6);
\draw[lightgray,dashed,line width=0.3mm] (2,0)-- (2,1);
\draw[lightgray,dashed,line width=0.3mm] (2.5,0)--  (2.5,1.2);
\draw[lightgray,dashed,line width=0.3mm] (3,0)-- (3,1.1);
\draw[lightgray,dashed,line width=0.3mm] (3.5,0)--  (3.5,0.7);

\draw[-stealth] (4.45,1.2) -- (4.45,1.7) node[above] {$\sigma$};
\draw[-stealth]  (4.45,1.2) -- (4.95,1.2)  node[right] {$\eta$};

\fill[red] (1.5,0) circle(.075);
\fill[red] (2,0) circle(.075);
\fill[red] (2.5,0) circle(.075);
\fill[red] (3,0) circle(.075);
\fill[red] (3.5,0) circle(.075);

\draw (1,0) -- (1,2);
\draw (1.01,0) -- (1.01,2);
\draw (1.02,0) -- (1.02,2);
\draw (1,0) node[below] {$0$};
\draw (2,-0.075) node[below] {\scriptsize{$k-1$}};
\draw (2.5,-0.075) node[below] {\scriptsize{$k$}};
\draw (3,-0.075) node[below] {\scriptsize{$k+1$}};
\draw (0.65,2.2) node[below] {$\infty$};
\draw (4,0) node[below] {$P$};
\draw (4,0) -- (4,2);
\draw (3.99,0) -- (3.99,2);
\draw (3.98,0) -- (3.98,2);

\draw (1,0) -- (4,0);
\draw (1,0.01) -- (4,0.01);
\draw (1,-0.01) -- (4,-0.01);
\draw (1,0.02) -- (4,0.02);
\draw (1,-0.02) -- (4,-0.02);
%

\draw (2.5,-0.95) node[above] {$\mathcal{R}(\eta)$};
   \begin{scope}[yshift=2cm]
\draw (0,-1.9) -- (0.6,-1.9);
\draw (0.1,-1.95) -- (0.5,-1.95);
\draw (0.2,-2) -- (0.4,-2);
\draw (0.3,-1.9).. controls (0.3,-1.25)  .. (1,-1.25);

\draw (4.5,-1.9) -- (5.1,-1.9);
\draw (4.6,-1.95) -- (5,-1.95);
\draw (4.7,-2) -- (4.9,-2);
\draw (4.8,-1.9).. controls (4.8,-1.25)  .. (4,-1.25);
\end{scope}
\end{tikzpicture}
\caption{An electrostatic problem with a line of charge positioned at $\sigma=0$ and extended along $\eta\in [0,P]$, with two parallel conducting plates at $\eta=0,P$. The charge density is defined by the rank function $\mathcal{R}(\eta)$, which has now been schematically overlaid in grey. The red dots correspond to the position of the D6 branes, which lay along the $\sigma=0$ at the positions of the kinks of the rank function, $k \in \mathds{Z}$.}
  \label{fig:RankonElectrostatic}
\end{figure}

Recall that the $d=11$ solution contained $\mathds{R}^4/\mathds{Z}_{b_k}$ orbifold singularities, positioned at the loci of the discontinuities of $\mathcal{R}'$. Following the reduction to IIA, in which the M-theory circle spanned by $\beta$ vanishes, one finds stacks of $b_k$ D6 branes at these loci. In other words, the D6- branes live at special (integer) points along the $\eta$ direction, corresponding to the different kinks of the rank function. Calculating the charge of the D6 branes leads to the requirement that the integration constant takes integer values, $k\in \mathds{Z}$, and hence corresponds to the position of the kinks. A schematic diagram is given in Figure \ref{fig:RankonElectrostatic}, with the location of the D6-branes represented by red dots.
This can be seen from the Bianchi identity for the $F_2$ given in \eqref{eq:N=2}, 
\beq\label{eqn:F2Bianchi}
d_HF_2=(\partial_\sigma\partial_\eta-\partial_\eta\partial_\sigma)f_6\, d\sigma\wedge d\eta\wedge d\chi,
\eeq
which is of course zero in general, with $\partial_\sigma\partial_\eta f_6=\partial_\eta\partial_\sigma f_6$. However, one needs to be a little more careful here. Approaching the $\sigma= 0$ boundary, and using the laplace equation \eqref{eqn:laplace}, $\ddot{V}\rightarrow 0$ to leading order. Then, using the boundary condition $\mathcal{R}(\eta) = \dot{V}\Big|_{\sigma=0}$ and the warp factors \eqref{eqn:fs}, one finds in this limit
\beq
C_1\Big|_{\sigma\rightarrow 0} = f_6\Big|_{\sigma\rightarrow 0} d\chi=\dot{V}'\Big|_{\sigma\rightarrow 0} d\chi = \mathcal{R}'(\eta)  d\chi,
\eeq
which we recall is a discontinuous function along $\eta$. Taking the derivative carefully, and noting the form of $\mathcal{R}''(\eta)$ given in \eqref{eqn:Rkwithderivs}, one can calculate $F_2$ along this boundary 
\beq
 F_2\Big|_{\sigma\rightarrow 0 } = \sum_{k=1}^{P-1}  \Big(\mathcal{R}'(k)-\mathcal{R}'(k-1)\Big)  d\eta\,\wedge \, d\chi=- \sum_{k=1}^{P-1} b_k\,\delta(\eta-k) \delta(\sigma) d\eta\,\wedge \, d\chi, 
\eeq
which one could integrate appropriately to derive the D6 charge, with $b_k= (2N_k-N_{k+1}-N_{k-1})$. Hence, it is the discontinuities in $\mathcal{R}'(\eta)$ which lead to source terms for D6 branes in the $F_2$ Bianchi identity, with $\partial_\sigma\partial_\eta f_6\neq\partial_\eta\partial_\sigma f_6$ at these loci.
In contrast, one finds $C_3$ and $B_2$ are independent of $f_6$, and consequently the rank function, so the Bianchi identities for $F_4$ and $H$ contain no source terms,  
\begin{align}
dH&=(\partial_\sigma\partial_\eta-\partial_\eta\partial_\sigma)f_8\, d\sigma\wedge d\eta\wedge \text{vol}(S^2)=0,\nn\\[2mm]
d_HF_4&=dF_4-H\wedge F_2=d(C_3-H\wedge C_1)-H\wedge F_2=d(dC_3)\nn\\[2mm]
&=(\partial_\sigma\partial_\eta-\partial_\eta\partial_\sigma)f_7\, d\sigma\wedge d\eta\wedge d\chi\wedge\text{vol}(S^2)=0,
\end{align}
where we note that in the latter case, the contribution from $F_2$ is eliminated using the Leibniz identity \eqref{eqn:Leibniz} (and using $dH=0$). Hence, the only dynamical (physical) objects in the background are D6 branes, with the NS5 and D4 branes considered pure flux.

Calculating the page charges of these branes carefully, leads to
\begin{align}
&Q_{NS5}=-\frac{1}{(2\pi)^2}\int_{S^3} H=P,~~~~~~~~~Q_{D4}^k=-\frac{1}{(2\pi)^3}\int_{\text{S}^2\times \tilde{\text{S}}^2}\hat{F}_4^{\text{Page}}=N_k-N_{k-1},\nn\\[2mm]
&Q_{D6}^k=-\frac{1}{2\pi}\int_{\tilde{S}^2}\hat{F}_2^{\text{Page}}=b_k=2N_k-N_{k+1}-N_{k-1},
\end{align}
which match the results given in \cite{Nunez:2019gbg}. We will derive these results in a more careful manner in Section \ref{sec:BoundaryNeqtwo}, where we investigate the $(\sigma,\eta)$ boundary of the solution and provide the exact definitions of these cycles. However, it is worth noting here that the stack of NS5 charges live at the $\sigma\rightarrow \infty$ boundary, whereas the D6 and D4 branes lie along $\sigma=0$. 

The total charge of the D6 and D4 branes then obey
\beq
Q_{D6}=\sum_{k=1}^{P-1}Q^k_{D6}=N_{P-1}+N_1,~~~~
Q_{D4}=\sum_{k=1}^{P-1}Q^k_{D4}=N_{P-1}.
\eeq
In the case of the D4 branes however, this total charge includes both the `true' colour D4-branes present in the background, and the charge of the D4-branes induced on the D6 and NS5 branes via the Myers effect. There is in fact only  $N_k$ `true' D4 charge in the $[k,k+1]$ interval, and the total charge of D4-branes which avoids this over-counting, proven in \cite{Nunez:2019gbg}, then read
\beq
Q_{D4}^{\text{True},k} = N_k,~~~~~~~~~~~~~~Q_{D4}^{\text{Total}}=\int_0^P R(\eta) d\eta.
\eeq
\subsubsection{The Holographic description}
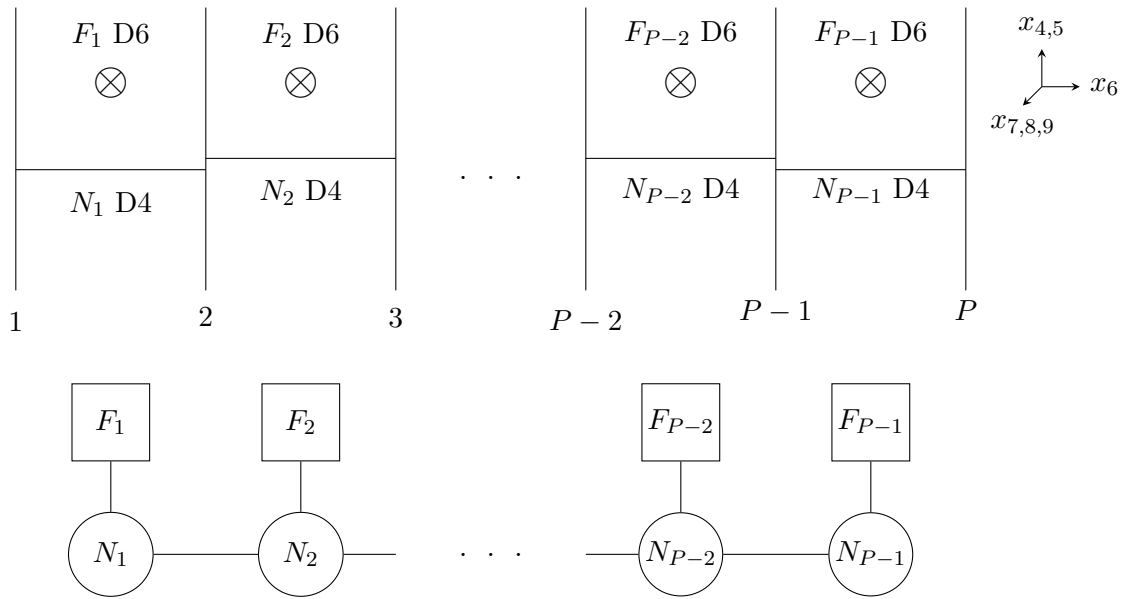
\begin{figure}[H]
\begin{center}
 \begin{tikzpicture}[square/.style={regular polygon,regular polygon sides=4},scale=1]
 




         
   \begin{scope}[xshift=0cm,yshift=5cm]



\node[label=above:$F_1$ D6] at (1.25,1.25){\LARGE $\otimes$};
\node[label=above:$F_2$ D6] at (3.75,1.25){\LARGE $\otimes$};
%
\node[label=above:$F_{P-2}$ D6] at (8.75,1.25){\LARGE $\otimes$};
\node[label=above:$F_{P-1}$ D6] at (11.25,1.25){\LARGE $\otimes$};


\draw (0,-1.5) -- (0,2.25);

\draw (0,-1.66) node[below] {$1$};
\draw(2.5,-1.5) -- (2.5,2.25);
\draw (2.5,-1.59) node[below] {$2$};
\draw (5,-1.5) -- (5,2.25);

\draw (5,-1.6) node[below] {$3$};
\draw (7.5,-1.5) -- (7.5,2.25);

\draw (7.5,-1.63) node[below] {$P-2$};
\draw (10,-1.5) -- (10,2.25);

\draw (10,-1.5) node[below] {$P-1$};
\draw (12.5,-1.5) -- (12.5,2.25);
\draw (12.5,-1.5) node[below] {$P$};

\draw[-stealth] (13.5,1.2) -- (13.5,1.7) node[above] {$x_{4,5}$};
\draw[-stealth]  (13.5,1.2) -- (14,1.2)  node[right] {$x_6$};
\draw[-stealth]  (13.5,1.2) -- (13.25,0.95)  node[below] {$x_{7,8,9}$};


\draw (0,0.1) -- (2.5,0.1);

\draw (2.5,0.25) -- (5,0.25);
\draw (5.7,0) node[right] {.~~.~~.};
\draw (7.5,0.25) -- (10,0.25);

\draw (10,0.1) -- (12.5,0.1);
\draw (1.25,-0.65) node[above] {$N_1$ D4};
\draw (3.75,-.5) node[above] {$N_2$ D4};
\draw (8.75,-0.5) node[above] {$N_{P-2}$ D4};
\draw (11.25,-0.5) node[above] {$N_{P-1}$ D4};

\end{scope}
 
  \begin{scope}[xshift=0cm,yshift=0cm]
        \node at (1.25,1.75) [square,inner sep=0.4em,draw] (f300) {$F_1$};
        \node at (3.75,1.75) [square,inner sep=0.4em,draw] (f400) {$F_2$};
          \node at (8.75,1.75) [square,inner sep=-0.2em,draw] (fp00) {$F_{P-2}$};
             \node at (11.25,1.75) [square,inner sep=-0.2em,draw] (fp001) {$F_{P-1}$};
        
        \node at (1.25,0) [circle,inner sep=0.5em,draw] (c200) {$N_1$};
        \node at (3.75,0) [circle,inner sep=0.5em,draw] (c300) {$N_2$};
        \node at (8.75,0) [circle,inner sep=0.1em,draw] (cp00) {$N_{P-2}$};
          \node at (11.25,0) [circle,inner sep=0.1em,draw] (cp001) {$N_{P-1}$};
        \draw (c200) -- (f300);
        \draw (c200) -- (c300);
        \draw (c300) -- (f400);
        \draw (cp00) -- (fp00);
        \draw (c300) -- (5,0);
        \draw (5.7,0) node[right] {.~~.~~.};
        \draw (cp00) -- (7.5,0);
        \draw (cp00) -- (cp001);
         \draw (cp001) -- (fp001);
         
         \end{scope}
    \end{tikzpicture}
    \end{center}
    \caption{The Hanany-Witten brane set-up of the supergravity description and the linear quiver diagram of the dual 4d $\mathcal{N}=2$ SCFT. Here the vertical lines in the Hanany-Witten diagram denote individual NS5 branes (of which there are $P$ in total), the horizontal lines denote D4 branes extended between NS5 branes and the crossed circles denote D6 branes going into the page, with $F_k=b_k=2N_k-N_{k+1}-N_{k-1}$. All branes share the Mink$_4$ directions on which the dual field theory lives.}
\label{fig:QuiverHW}
\end{figure}

Important field theory results can be obtained by constructing Hanany-Witten diagrams for the brane content. The solutions at hand have NS5, D4 and D6 branes, which all extend along the Mink$_4$ directions. The NS5 branes span two additional directions, which following the conventions of \cite{Nunez:2019gbg}, we will generically call $(x_4,x_5)$. These directions then realise $SO(2)\sim U(1)_R$ rotations. When conformality is broken, these NS5 branes bend in this plane, breaking the $U(1)_R$ component. The NS5 branes are placed along an $x_6$ direction (corresponding to the $\eta$ coordinate) at fixed integer locations, and are connected by extended D4 branes along this direction. At low energies, this leads to an effective field theory in four dimensions. In addition, there are D6 branes which extend along three additional directions, $(x_7,x_8,x_9)$, which then realise the $SO(3)\sim SU(2)_R$ invariance. The $SU(2)_R\times U(1)_R$ R-symmetry of the dual CFT is thus realised. A summary of this discussion is given in the following table, with the dual CFT spanning the Mink$_4$ components (and separated with a bar).
\begin{center}
\begin{tabular}{c c c c c |c c c c c c c c}
&$x_0$&$x_1$&$x_2$&$x_3$&$x_4$&$x_5$&$x_6$&$x_7$&$x_8$&$x_9$\\
D4:&$\times$&$\times$&$\times$&$\times$&$-$&$-$&$\times$&$-$&$-$&$-$ \\
D6:&$\times$&$\times$&$\times$&$\times$&$-$&$-$&$-$&$\times$&$\times$&$\times$ \\
NS5:&$\times$&$\times$&$\times$&$\times$&$\times$&$\times$ &$-$&$-$&$-$&$-$\\
&\multicolumn{4}{c}{\upbracefill} & \multicolumn{2}{c}{\upbracefill}& & \multicolumn{3}{c}{\upbracefill} \\
&\multicolumn{4}{c}{ $SO(1,3)$}& \multicolumn{2}{c}{$SO(2)$} & & \multicolumn{3}{c}{$SO(3)$}
\end{tabular}
\end{center}
Constructing the corresponding Hanany-Witten diagram for this supergravity brane description allows for direct comparison with the dual CFT linear quiver description. This is shown in Figure \ref{fig:QuiverHW}. Here, the integer quantization of charge clearly relates to the rank of the colour and flavour nodes of the linear quiver. It is then clear that D6 branes should be interpreted as `flavour branes' with NS5- and D4- branes considered `colour branes'. Hence, in these examples, the integer quantization of charge is related to the Lagrangian nature of the dual field theory. The quiver structure correspond to strings connecting the various D-branes - see for example \cite{VanGorsel:2019xcw}.
Hence, for a given conformal quiver field theory, one can construct the dual supergravity solution, with potential \eqref{eqn:potential} and warp factors \eqref{eqn:fs}, directly from the rank function of the dual quiver. This will satisfy the laplace equation and associated boundary conditions, which will automatically solve the equations of motion and Bianchi identities.


In fact, the holographic description is not trustable close to the D6 and NS5 branes, which are singular points along the manifold defined by $V(\sigma,\eta)$, with the curvature becoming very large close to these points. The idea of \cite{Aharony:2012tz} was to take $P$ (the range of $\eta$) very large, to force these regions to only very small patches of the manifold, with the holographic description well defined otherwise. This corresponds to dealing with long linear quiver diagrams in the dual CFT. In addition, the rank function can be scaled such that the number of D4 and D6 branes increase independent of the number of NS5 branes. 
One may wonder if large $N$ CFTs are still conformal following $\frac{1}{N}$ corrections, in other words, if going beyond supergravity the isometries of AdS$_5$ (or S$^2\times$S$^1$) are preserved. The SO(2,4) symmetry was shown to survive $\frac{1}{N}$ corrections in \cite{Bashmakov:2017rko}, using a bottom-up perspective.

To calculate the holographic central charge, one should follow the methodology outlined in \eqref{eqn:HCC}, with $d=3$ for the case at hand. Following through the calculation, which is given in more detail in Appendix \ref{sec:HCCcalc}, one finds
\beq\label{eqn:HCCGM}
c_{hol} =\frac{2\kappa^3}{\pi^4}\int_0^P\mathcal{R}(\eta)^2 d\eta = \frac{ \kappa^3}{\pi^4}\sum_{n=1}^\infty P \, \mathcal{R}_n^2,
\eeq
which then matches the results of \cite{Nunez:2019gbg} (up to an appropriate change of variables). The central charge is clearly proportional to the area under $\mathcal{R}(\eta)^2$. Notice also that this expression fits into the formalism of \eqref{eqn:HCCnew}, with $D=4$ (for a Mink$_4$ solution) and
\beq
\text{coeff}(4)= \frac{ \kappa^3}{\pi^4}.
\eeq
On the CFT side, one had two central charges, $a$ and $c$, with expressions given in \eqref{centralsN=2}. These central charges are defined in terms of the number of vector multiplets and hypermultiplets in the dual quiver, $N_v$ and $N_h$. As was just discussed, the holographic comparison is only trustable for $(N,P)\rightarrow \infty$, which also gives $a=c$. See the two example quivers given in Appendix \ref{sec:exampleRanks}.    

    \subsection{The $\gamma$-deformations of NRSZ}\label{sec:gammadefs}

In the work of N\'u\~nez, Roychowdhury, Speziali and Zacar\'\i{}as (NRSZ) \cite{Nunez:2019gbg}, marginally deformed backgrounds to the GM class of solutions were constructed in M-Theory, type IIA and type IIB, containing a parameter, $\gamma$, which recovers the original $\mathcal{N}=2$ solutions (in the first two cases) when fixed to zero. These solutions were then proposed as holographic duals to the $\mathcal{N}=1$ SCFTs mentioned  in Section \ref{sec:SCFTN1}. The supersymmetry of these solutions were not explicitly checked in that work, however some light will be shed on this matter in Chapter \ref{chap:gamma}, using the method of G-structures. We postpone the presentation of these solutions until then. Further analysis of these solutions was also conducted in  \cite{Pal:2023bjz,Roychowdhury:2023lxk,Roychowdhury:2024pvu}. 
     \section{Spindles \& Weighted Projective Spaces}\label{sec:Spindles}
 %
Following the discussion given in Appendix B of \cite{Legramandi:2020txf} (which the reader is directed towards for more details), an odd dimensional sphere, $S^{2n+1}$ (for $n>0$) can always be parametrized as a $U(1)$ bundle over the complex projective space, $\mathds{C}\mathds{P}^n$, where
     \begin{align}
ds^2(S^{2n+1})&=ds^2(\mathds{C}\mathds{P}^n)+(d\psi_n+\eta_n)^2,~~~~~~     ds^2(S^d)=d\alpha^2 +\cos^2\alpha \,d\psi^2 +\sin^2\alpha \,ds^2(S^{d-2}),\nn\\[2mm]
ds^2(\mathds{C}\mathds{P}^n)&=d\theta_n^2+\sin^2\theta_n ds^2(\mathds{C}\mathds{P}^{n-1})+\sin^2\theta_n \cos^2\theta_n (d\psi_{n-1}+\eta_{n-1})^2,\nn\\[2mm]
\text{with}&~~~~~~~   ds^2(\mathds{C}\mathds{P}^1) =\frac{1}{4} ds^2(S^2)=\frac{1}{4}(d\theta_1^2+\sin^2\theta_1 d\phi_1^2),~~~~~~\eta_1=\cos\theta_1 d\phi_1,
     \end{align}
for $\psi_n$ a $U(1)$ isometry 
and $\eta_n=\sin^2\theta_n (d\psi_{n-1}+\eta_{n-1})$ (spanning only the base). Note,
     \begin{align}\label{eqn:CP2}
ds^2(\mathds{C}\mathds{P}^2) &= d\theta_2^2 +\sin^2\theta_2 ds^2(\mathds{C}\mathds{P}^1) +\sin^2\theta_2\cos^2\theta_2 (d\psi_1+\cos\theta_1 \,d\phi_1)^2\\[2mm]\nn
&= d\theta_2^2 +\frac{1}{4}\sin^2\theta_2  (d\theta_1^2+\sin^2\theta_1 d\phi_1^2) +\sin^2\theta_2\cos^2\theta_2 (d\psi_1+\cos\theta_1 \,d\phi_1)^2.
\end{align}
\paragraph{Orbifolds} These are manifolds which have been obtained from flat space, following some identification of points under a discrete group of symmetries. Generally, this leads to singularities in the manifold, but remarkably the resulting string theory remains well behaved. The simplest example, called $\mathbb{R}^1/\mathbb{Z}_2$, involves identifying points on the real line, $\mathbb{R}^1$, under a reflection $X\sim-X$, with the half line $X\geq 0$ the fundamental region and $X=0$ the boundary (which is a fixed point). This orbifold has the property that applying the transformation $X\rightarrow-X$ twice, returns the original coordinate, labelled using $\mathbb{Z}_2$. Alternatively, one can obtain a two-dimensional cone as orbifolds in a similar fashion, labelled $\mathbb{C}/\mathbb{Z}_N$, by identifying points in the 2d (complex) plane, $\mathbb{C}$, by images obtained by a rotation of $2\pi/N$- leading to fundamental domain which describes the surface of a cone, with the fixed point (the origin) forming the tip of the cone. See for example \cite{FirstCourseTextbook,PolchinskiVol2Textbook,PolchinskiVol1Textbook} for more details.

\begin{figure}[H]
  \vspace{-0.3cm}
 \hspace{-0.5cm}
\centering  
\subfigure
{
\centering
  \begin{minipage}{0.35\textwidth}
\beq
\varphi_\pm=2\pi \Big(1-\frac{1}{n_\pm}\Big).
\eeq
  \end{minipage}   
    \hspace{0.4cm}
     \begin{minipage}{.6\textwidth}
    \centering
\begin{tikzpicture}[scale=1.6, every node/.style={scale=1}]
  \draw[black,line width=0.27mm] (-2,-0.25) arc (-60:240:1.25 and 0.4);
    \fill[fill=black] (-2.625,0.1) circle (0.3pt);
       \draw[red,line width=0.35mm] (-2.625,0.1) -- (-2,-0.25) ;
         \draw[red,line width=0.35mm] (-2.625,0.1) -- (-3.25,-0.25) ;
         \draw (-2.625,0.1) node[above] {$o$};
          \draw (-2.625,-0.04) node[below] {$\varphi_\pm$};
        \draw[black] (-2.45,0) arc (-60:-120:0.36 and 0.4);
         \draw[->,>=stealth] (-2.925,-0.35).. controls (-2.625,0-0.39)  ..(-2.325,-0.35) ; 
          \draw [decoration={markings,mark=at position 1 with
    {\arrow[scale=1.5,>=stealth]{>}}},postaction={decorate}] (-1,0.1)--(-0.5,0.1);
         \fill[
  left color=gray!40!black,
  right color=gray!50!black,
  middle color=gray!50,
  shading=axis,
  opacity=0.1,rotate=0
  ] 
 (-3.25,-0.25)  --  (-2.625,0.1)-- (-2,-0.25)arc (-60:240:1.25 and 0.4);
  \draw[black,line width=0.27mm] (0,0.035) arc (180:360:1 and 0.33);
  \draw[gray,dashed] (2,0) arc (0:180:1 and 0.28);
      \draw (1,1) node[above] {$o$};
    \draw[red,line width=0.35mm] (1.5,-0.25) -- (1,1) ;
  \fill[fill=black] (1,0) circle (0.3pt);
     \draw[black,line width=0.27mm](1.999,0.036) -- (1,1) ;
       \draw[black,line width=0.27mm](0.001,0.036) -- (1,1) ;
\fill[
  left color=gray!40!black,
  right color=gray!50!black,
  middle color=gray!50,
  shading=axis,
  opacity=0.1,rotate=0
  ] 
 (1.999,0.036)  -- (1,1) -- (0.001,0.036) -- (0.012,-0.035)arc (188:365:1cm and 0.31cm);
\end{tikzpicture}
  \end{minipage} 
}
\caption{The conical deficit angle, $\varphi_\pm \in[0,2\pi)$, which can be written $\varphi_\pm=2\pi \Big(1-\frac{1}{n_\pm}\Big)$ for $n_\pm\in [1,\infty)$. With $\varphi_\pm=0$ (and $n_\pm=1$), there is no conical (orbifold) singularity.}
    \label{fig:conicaldeficit}
\end{figure}
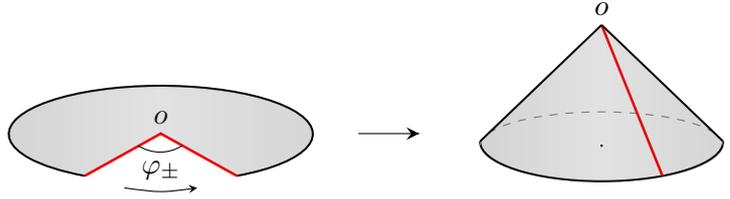

\paragraph{Spindles} Another example of an orbifold is the `spindle', $\mathbb{WCP}^1_{[n_-,n_+]}$, which describes a `weighted complex projective space' - with weights $(n_-,n_+)$. \\[-0.5cm]
\begin{wrapfigure}[6]{r}{0.3\textwidth}
    \centering
    \vspace{-1cm}
            \hspace{0.35cm}
\begin{tikzpicture}[scale=1.5, every node/.style={scale=0.95}]
\node[label=above:] at (0,0.9){ $~$};
\node[label=above:] at (0.6,0.9){ $n_+$};
\node[label=above:] at (0.6,-1.1){ $n_-$};
\node[label=above:] at (2.2,0.75){ $\mathbb{WCP}^1_{[n_-,n_+ ]}$};
  \draw[gray] (0,-0.25) arc (180:360:1 and 0.3);
  \draw[gray,dashed] (2,-0.25) arc (0:180:1 and 0.3);
  \fill[fill=black] (1,-0.25) circle (0.75pt);
  \draw  plot [smooth, tension=0.7] coordinates {(1,-1)
   (2,-0.25)
    (1,1) };
      \draw  plot [smooth, tension=0.7] coordinates {(1,-1)
   (2,-0.25)
    (1,1) };
   \draw plot [smooth, tension=0.7] coordinates {(1,-1) 
   (0,-0.25) 
   (1,1) };
      \draw plot [smooth, tension=0.7] coordinates {(1,-1) 
   (0,-0.25) 
   (1,1) };
\fill[
  left color=gray!40!black,
  right color=gray!50!black,
  middle color=gray!50,
  shading=axis,
  opacity=0.1,rotate=0
  ] 
  (2,-0.32) -- (1.85,-0.53) -- (1.6,-0.73) -- (1,-1) --(0.4,-0.73)--(0.15,-0.53) -- (0,-0.32)arc (188:365:1cm and -0.4cm);
  
  \fill[
  left color=gray!40!black,
  right color=gray!50!black,
  middle color=gray!50,
  shading=axis,
  opacity=0.1,rotate=0
  ] 
  (2,-0.32) --(1.975,-0.09)  --(1.95,-0.05) -- (1.82,0.13) -- (1.5,0.54) -- (1.1,0.9) --  (1,1) -- (0.9,0.9) -- (0.5,0.54) -- (0.16,0.1 ) --(0.05,-0.05)--(0.025,-0.09)  -- (0,-0.32)arc (188:365:1cm and -0.4cm);
\end{tikzpicture}
\end{wrapfigure}
These manifolds have the topology of a 2-sphere with additional $\mathbb{R}^2/\mathbb{Z}_{n_{\mp}}$ orbifold singularities at the south/north poles, with\footnote{$n_-\neq n_+$ describes a `bad' orbifold (not covered by a manifold) with $\chi_E>0$ \cite{Lange1,Caramello}. Examples requiring $n_->n_+$ are discussed in \cite{Boido:2021szx,Ferrero:2021wvk} - describing a `Besse' spindle, see \cite{Lange1}. } $n_-\neq n_+$ 
and $(n_-,n_+)$ relatively prime (or `coprime') positive integers - with\footnote{Here `gcd' stands for `greatest common divisor'. Note some literature uses `hcf' for `highest common factor' - these are interchangeable terms. This condition leads to $n_-\neq n_+$ for the spindle.}
gcd$(n_-,n_+)=1$ \cite{Ferrero:2020twa}. These integers define the `conical deficit angles', $\varphi_\mp$, at each pole of the spindle, with the sphere recovered for $\varphi_\mp=0$ (and $n_\mp=1$)\footnote{If only $n_-=1$, the spindle becomes the $n_+$-`teardrop', with only one orbifold singularity - see \cite{Caramello,Davies}}.\\
The conical deficit angle, $\varphi_\pm$, describes the size of a sector removed from a 2d disc, with the two leftover edges identified to form a cone. The bigger the value of $\varphi_\pm$, the more narrow the cone becomes -with the conical singularity vanishing for $\varphi_\pm=0$, see Figure \ref{fig:conicaldeficit}
In recent years, spindles have received a lot of attention within the context of the near horizon limit of D branes wrapping them and their dual CFTs. Examples include
\cite{Ferrero:2020twa,Ferrero:2020laf,Couzens:2021cpk,Hosseini:2021fge,Boido:2021szx,Ferrero:2021wvk,Bah:2021hei,Ferrero:2021ovq,Couzens:2021rlk,Faedo:2021nub,Ferrero:2021etw,Couzens:2021tnv,Couzens:2022yjl,Arav:2022lzo,Couzens:2022yiv,Suh:2022pkg,Amariti:2023gcx,Inglese:2023tyc,Faedo:2024upq,Bomans:2024mrf}. As a consequence of the spindle within these solutions, they find rational quantization of D-brane charge! 
This is a consequence of the rational nature of the Euler characteristic, $\chi_E$, of the spindle, calculated using the Chern-Gauss-Bonnet theorem, for $\Sigma=\mathbb{WCP}^1_{[n_-,n_+]}$
\beq
Q=\frac{1}{2\pi}\int_\Sigma F=\frac{\lambda}{n_-n_+},~~~\lambda\in\mathds{Z},~~~~~~~~~\chi_E=\frac{1}{4\pi}\int_{\Sigma}R_\Sigma \,\text{vol}(\Sigma) = \frac{n_-+n_+}{n_-n_+}, 
\eeq
where $R_\Sigma$ is the Ricci scalar of the spindle, with a general formula given in \cite{Sakai:1971}. Note $\chi_E=2$ for $n_-=n_+=1$, recovering the $\mathbb{CP}^1$. The discussions in \cite{Ferrero:2021etw} and appendix A of \cite{Ferrero:2020twa} are particularly interesting. Two new mechanisms for the preservation of supersymmetry were classified in \cite{Ferrero:2021etw,Couzens:2021cpk}, called `twist' (a topological topological twist) and `anti-twist'. Mathematics literature, such as \cite{Guillemin,Davies,Caramello,Lange1,Lange} (and references within), proves very useful.

Higher dimensional analogues exist, with the four dimensional version, $\mathbb{WCP}^2_{[1,1,2]}$, also appearing in the literature - see \cite{Gauntlett:2004zh,Bianchi:2021uhn}. Both $\mathbb{WCP}^1$ and $\mathbb{WCP}^2$ play a role in Part \ref{Part:SUSYdefs}.

 



\chapter{The Method Of G-Structures}\label{sec:Gstructures}



Supersymmetric solutions of M-Theory or Type II supergravity must satisfy the necessary supersymmetry conditions, usually given in terms of a spinor and metric, see for example \eqref{eqn:SUSYcondsSp}, as well as the Bianchi identities in each case. The technique of G-structures allows one to recast these supersymmetry conditions in terms of non-spinorial and geometrical objects - forms. 
%
In the G-structure analysis, we are interested in the internal space, so one decomposes the metric into a `warp product' of the external and internal components
\begin{align}
ds^2_d&=e^{2\tilde{A}}ds^2(\text{Mink}_D)+ds^2(\text{M}_{\tilde{d}}),~~~~~~~~\tilde{d}=d-D,
\end{align}
in which $\tilde{A}$ is a function of the internal space, $\text{M}_{\tilde{d}}$. We will follow the convention of using $\hat{A}\equiv\tilde{A}$ for $d=11$, and $ A\equiv\tilde{A} $ for $d=10$.
This internal manifold, $\text{M}_{\tilde{d}}$, is covered by several coordinate systems (or `frames'), related to one another by coordinate changes called `Transition functions' - generally in the group $GL(\tilde{d},\mathds{R})$ of invertible matrices. One can then define a subset of these transition functions, which now form a subgroup $G\subset GL(\tilde{d},\mathds{R})$, as a `G-structure' on $\text{M}_{\tilde{d}}$. Hence, we are picking out a subset of frames, which are now related to one another by the subgroup $G$. A G-structure is typically equivalent to a set of invariant forms $(J,\Omega)$, but is also naturally defined by the stabilizer group of the spinor, $\eta$, namely $G=\text{Stab}(\eta)$ - the group of rotations that keep $\eta$ invariant. This allows the supersymmetry equations involving $\eta$ to be recast in terms of these invariant forms \cite{TomTextbook}.

As we have reviewed in the previous chapter, many supergravity solutions are known with various AdS$_{D+1}$ factors (of $D$ spatial dimensions). One can utilise the Poincar\'e patch to express these solutions within this warped product formalism, as follows
\begin{equation}\label{eqn:PPatch}
ds^2(AdS_{D+1})=e^{2\rho}ds^2(\text{Mink}_D) +d\rho^2.
\end{equation}
In the case of type II supergravity, solutions with a general Mink$_D$ factor can be decomposed into a `warped product' geometry, which following the conventions of \cite{Macpherson:2024qfi}, takes the form
\newpage
\begin{align}\label{eqn:generalDform}
&ds_{10}^2=e^{2A}ds^2(\text{Mink}_D)+ds^2(\text{M}_{\tilde{d}}),\\[2mm]
&F =g^{(\tilde{d})}+ e^{4A}\text{vol}(\text{Mink}_{D})\wedge \star_{\tilde{d}}\lambda (g^{(\tilde{d})}),~~~~~~~
H=e^{D A}H_{3-D}\wedge \text{vol}(\text{Mink}_{D})+H_3,\nn
\end{align}
with $\tilde{d}=10-D$. 
In order for the isometries of $\text{Mink}_D$ to be respected, the functions $(e^{2A}, H_3,g^{(\tilde{d})},H_{3-D})$ and dilaton, $\Phi$, must only depend on the internal space $\text{M}_{\tilde{d}}$ coordinates, $y^m$ (with $\mu=0,...,D-1$ and $m=D,...,\tilde{d}$). The only non trivial $H_{3-D}$ is $D=1,2,3$. 

We will focus our attention on the supersymmetry conditions of solutions with a Mink$_4$ factor. 
This presentation will be primarily aimed towards setting up the G-structure analysis for the $d=11$ AdS$_5$ GM solution \eqref{eqn:GM}, derived in Part \ref{Part:SUSYdefs}. This class of solutions describe a Mink$_4\times M_7$ background, with an $SU(3)$ structure on $M_7$, and dimensionally reduces to Mink$_4\times M_6$ backgrounds, with an $SU(2)$ structure on $M_6$. We will review this G-structure analysis more generally, before discussing the G-structure description of the LLM solution explicitly. These are the results which derive the GM G-structure analysis presented throughout Part \ref{Part:SUSYdefs}. 
The interested reader is directed to \cite{Grana:2005sn,Grana:2005jc,Gauntlett:2004zh,Kaste:2003zd,Gauntlett:2003cy,TomTextbook,Grana:2006kf,TomsNotes} for more detailed and insightful discussions on the topic.

\section{Mink$_4$}\label{sec:Mink4}

In an M-Theory with Mink$_4\times M_7$, one has the following warp product
\beq\label{eqn:MTheorymetric}
ds_{11}^2=e^{2\hat{A}}ds^2(\text{Mink}_4)+ds^2(M_7),
\eeq
where $\hat{A}$ is a function of the internal space, $M_7$, coordinates. We decompose the spinors on the 
external/internal parts of the space, as $\epsilon = \psi_+\otimes \theta_++\psi_-\otimes \theta_-$ (following the notation of \cite{Kaste:2003zd})
where $\theta$ are 7 dimensional spinors, with $\theta_+=\theta_-^*$.  

In the case of Mink$_4\times M_6$ supergravity, one would have the warp product
\beq\label{eqn:M6metric}
ds_{10}^2=e^{2A}ds^2(\text{Mink}_4)+ds^2(M_6),
\eeq
with the function, $A$, and the field strengths depending only on the internal space coordinates, $M_6$, where 
\begin{align}\label{eqn:relstuff}
&~~~~~~~~~F=g+e^{4A}\text{Vol}(\text{Mink}_4)\wedge *_6\lambda(g),~~~~~~~~~~~~H=H_3\nn\\[2mm]
&\text{Type IIA: }
~~~~~~g=F_0+F_2+F_4+F_6,~~~~~~~\lambda(g)=F_0-F_2+F_4-F_6,\\[2mm]
&\text{Type IIB: }
~~~~~~g=F_1+F_3+F_5,~~~~~~~~~~~~~~\lambda(g)=F_1-F_3+F_5.\nn
\end{align}
The ten dimensional (real) gamma matrices, $\Gamma_M$, can now be decomposed in the following manner
\beq
\Gamma_\mu=e^A\gamma_\mu \otimes 1,~~~~\Gamma_m=\gamma_5\otimes \gamma_m,~~~~~~~~\gamma_5=i\,\gamma^{0123},
\eeq
such that $\gamma_\mu$  are four dimensional and real, and $\gamma_m$ are six dimensional and imaginary (giving four Lorentzian and six Euclidean dimensions).

One can now decompose the spinors, $(\varepsilon^1,\varepsilon^2)$, on the internal and external components of the warped product geometry. We recall from \eqref{eqn:Chiralities} that these spinors have $(+,-)$ and $(+,+)$ chirality in IIA and IIB, respectively. The decomposition will take the general form $\zeta \otimes\eta$, with $\zeta$ living on Mink$_4$  and $\eta$ living on $M_6$. Here we are interested in $\mathcal{N}=1$ solutions in four dimensions, requiring four independent Killing spinors, leading to 
\begin{align}
&\text{Type IIA: }~~~~~~~~~\epsilon^1(y)=\zeta_+ \otimes \eta_+^1(y)+\zeta_-\otimes \eta_-^1(y),~~~~~~~~
\epsilon^2(y)=\zeta_+ \otimes \eta_-^2(y)+\zeta_-\otimes \eta_+^2(y),\nn\\[2mm]
&\text{Type IIB: }~~~~~~~~~\epsilon^{(a)}(y)=\zeta_+ \otimes \eta_+^{(a)}(y)+\zeta_-\otimes \eta_-^{(a)}(y),
\end{align}
for some generic constant (and automatically Weyl), $\zeta_+$, with the Majorana condition $\varepsilon=\varepsilon^*$ satisfied for $\zeta_-=(\zeta_+)^*,~\eta_-^{(a)}=(\eta_+^{(a)})^*$. Hence, the $\mathcal{N}=1$ solution is associated to two internal spinors $(\eta^1_+,\eta^2_+)$. These will in general describe an $SU(3)\times SU(3)$ structure.

When plugging these decompositions back into the SUSY conditions given in \eqref{eqn:SUSYcondsSp}, one can factor out $\zeta$ from the equations. This then leads to six fairly ugly equations in terms of $\eta^{1,2}$, given in \cite{Martucci:2005ht}. Hence, the original supersymmetry conditions then split into two parts: an algebraic part and a differential part (see for example \cite{Grana:2005sn} for a nice discussion)
\begin{itemize}
\item \textbf{Algebraic part: } This condition corresponds to the existence of a non-vanishing and globally well defined Spinor, $\eta$.
\item \textbf{Differential part: }This corresponds to differential conditions on the spinor. In the absence of flux, the six differential equations reduce to
\beq\label{eqn:LeviCivita1}
\nabla_M\eta_\pm=0,
\eeq
with $\nabla_M$ the Levi-Civita connection. This is the Killing Spinor equation, and determines the amount of supersymmetry a solution has by the number of spinors which satisfy the equation. The more supersymmetry, the more structure is required on the manifold. 
The Killing spinors which satisfy the Killing spinor equation are then rotated into each other by the R-symmetry of the theory.
\end{itemize}
One can now re-write these conditions in a more elegant manner, in the language of generalized complex geometry and differential forms. 
The algebraic and differential parts of the supersymmetry conditions become
\newpage
\begin{itemize}
\item \textbf{Algebraic part}: 
This is a topological requirement on the manifold\footnote{That $T\otimes T^*$ must have an $SU(3)\times SU(3)$ structure.}, and now implies the existence of two nowhere vanishing, globally defined Clifford$(6,6)$ pure spinors, $\Psi_+,\Psi_-$
\beq\label{eqn:AlgCond}
\Psi_+\equiv \eta_+^1 \otimes \eta_+^{2\dagger},~~~~~~~~~~~~~~~\Psi_-\equiv \eta_+^1 \otimes \eta_-^{2\dagger}.
\eeq
Using the Clifford map \eqref{eqn:Clifford}, these are equivalent to the sum of forms
\beq
\Psi_\pm=\sum_k \Psi_{\pm_k},
\eeq
with $k$ even and odd for $\Psi_+$ and $\Psi_-$, respectively.
\item \textbf{Differential part}: 
The preservation of supersymmetry then imposes differential conditions on the metric, derived from the Killing conditions via the Clifford map in \cite{Grana:2005sn}, for which the two pure spinors should satisfy 
\beq\label{eqn:diffcons}
e^{-2A+\Phi}d_H(e^{2A-\Phi}\Psi_2)=0,~~~~~~~~~~~~~~~~~~e^{-2A+\Phi}d_H(e^{2A-\Phi}\Psi_1)=dA\wedge \bar{\Psi}_1+F,
\eeq
with an additional normalization requirement
\beq
d|a|^2=|b|^2dA,~~~~~d|b|^2=|a|^2dA,~~~~~\text{where}~~~~~~|a|^2=||\eta^1||^2,~~~~~|b|^2=||\eta^2||^2.
\eeq
These equations are proven to be equivalent to the supersymmetric Killing conditions, see \cite{Grana:2005sn,Martucci:2005ht} for details, meaning they are considered necessary and sufficient conditions for a supersymmetric solution. These conditions then contain the same information as the supersymmetry variations given in \eqref{eqn:SUSYcondsSp} (and \eqref{eqn:LeviCivita1} in the absence of flux)
, and must be supplemented with the Bianchi identities and equations of motion for the fluxes. The more supersymmetry a solution has, the more structure is required on the manifold.\\
 Writing this more schematically as in \cite{Barranco:2013fza}, we have
 
 \beq\label{eqn:schematic}
 \begin{gathered}
 d_{H}\Psi_2=0,~~~~~~~~~~d_{H}\Psi_1=F_{RR},
 \end{gathered}
 \eeq
\\[-1cm]
 \beq\label{eqn:IIAIIBPsis}
 \text{Type IIA:}~~~~~~\Psi_1=\Psi_-,~\Psi_2=\Psi_+,~~~~~~~~~~~~~  \text{Type IIB:}~~~~~~\Psi_1=\Psi_+,~\Psi_2=\Psi_- ,
 \eeq
with the conditions identical in form for type IIA and type IIB, under the exchange
\beq\label{eqn:PsiFrelations}
\Psi_\pm\leftrightarrow \Psi_\mp,~~~~~~F_A\leftrightarrow F_B,
\eeq
which follows from $F_{RR}$ having an even form in IIA and an odd form in IIB. 
Clearly the pure spinors must transform in essentially the same manner under T-Duality as the Ramond fields \eqref{eqn:TD1}, this will become important to us later. 
\end{itemize}

\section{Example Structures}

We now summarise some key (and relevant) examples. See \cite{TomTextbook,Grana:2005sn,Grana:2006kf} for further details.\\\\
Having two spinors on $M_6$, $\eta^{1,2}$, describes an $SU(3)\times SU(3)$ structure. When $\eta^1$ and $\eta^2$ are parallel, they describe an $SU(3)$ structure - now defined by a single nowhere vanishing spinor $\eta$ (which is covariantly constant in the case of a Calabi-Yau 3-fold). When they are nowhere parallel, they define what is called a static $SU(2)$ structure.
\subsubsection{$SU(3)$ Structure}
An $SU(3)$ structure on an $M_6$ can be defined in three ways: via a metric and spinor $(g,\eta)$; a metric and a complex, decomposable and non-degenerate three-form $(g,\Omega)$; or a real two-form and the complex three-form $(J,\Omega)$. Hence there is a bijection between the descriptions, notably $(g,\eta)\leftrightarrow (J,\Omega)$. 

In terms of the metric and the spinor, the differential condition \eqref{eqn:LeviCivita1} now read in general (for non-zero flux)
\beq\label{eqn:etaCov}
\nabla_m \eta=i\,q_m\gamma_7\,\eta+i\,q_{mn} \gamma^n\,\eta,~~~~~~~~~~~~~\gamma_7=-\frac{i}{6!}\epsilon_{mnpqrs}\gamma^{mnpqrs},
\eeq
defining some real valued $q_m$ and $q_{mn}$ - see for example \cite{Grana:2005sn,Fidanza:2003zi,Grana:2004bg}.

Using the `Fierz identities', one gets the mapping
\begin{equation*}
\Omega_{mnp}=-\eta_-^\dagger \gamma_{mnp}\eta_+,~~~~~~~\eta_-=(\eta_+)^c,~~~~~~~~~~~~J_{mn}=-i\,\eta_+^\dagger \gamma_{mn}\eta_+.
\end{equation*}
The following conditions are necessary
\beq\label{eqn:JOmegacondition}
J\wedge \Omega=0,~~~~~~~~~~~~J\wedge J\wedge J=\frac{3}{4}i\,\Omega \wedge \bar{\Omega},
\eeq
with the positive definite metric defined as $g=-J I_\Omega$, and $\text{vol}(M_6)=-\frac{1}{3!}J^3=-\frac{i}{8}\Omega\wedge \bar{\Omega}$.\\
  \begin{adjustwidth}{1cm}{}
Here $I_\Omega$ is an `Almost Complex Structure' - that is, a tensor $I^m_n$ with $I^2=-1$. This allows one to embed the $GL(d/2,\mathds{C})\in GL(d,\mathds{R})$. The stabilizer group is given as follows
\begin{equation*}
\text{Stab}_{GL(d,\mathds{R})}(I)=GL(d/2,\mathds{C}),~~~~~~~~~~~~\text{Stab}(I)=O(d)\,\cap \,GL(d/2,\mathds{C}) \cong U(d/2),
\end{equation*}
so adding a metric on $M_6$ means the stabilizer group of $I$ is $U(3)$. One then defines $J\equiv gI$, or alternatively, $g=-JI$. To move to an $SU(3)$ one requires a nowhere vanishing holomorphic three-form $\Omega$, with $\Omega\rightarrow \text{det}(U)\Omega$ and det$(U)=1$. See \cite{TomTextbook} for further details.\\
\end{adjustwidth}
One can then build the normalized pure spinors, as follows
\beq\label{eqn:SU(3)spinors}
\Psi_+=\eta_+\otimes \eta_+^\dagger =\frac{1}{8}e^{-iJ},~~~~~~~~~~~~~~~~~~~~\Psi_-=\eta_+\otimes \eta_-^\dagger = -\frac{i}{8}\Omega.
\eeq
The differential conditions then take the form
\beq\label{eqn:JOmegaEqsgen}
\begin{aligned}
dJ&=\frac{3}{2}\text{Im}(\bar{W}_1\Omega)+W_4\wedge J+W_3   ,\\[2mm]
d\Omega&= W_1J^2+W_2\wedge J +\bar{W}_5\wedge \Omega    ,
\end{aligned}
\eeq
where
\beq
(q_m,q_{mn})\leftrightarrow W_i,
\eeq
that is, the exterior derivatives of $(J,\Omega)$ given in \eqref{eqn:JOmegaEqsgen} contains the same information as the covariant derivative of the spinor, $\eta$, given in \eqref{eqn:etaCov}. Hence, in the $SU(3)$ on M$_6$ case, the existence of the real two-form and complex three-form, $(J,\Omega)$, corresponds to the algebraic condition \eqref{eqn:AlgCond}, with the exterior derivative \eqref{eqn:JOmegaEqsgen} corresponding to the differential condition \eqref{eqn:diffcons}. 

In \eqref{eqn:JOmegaEqsgen}, $W_i$ correspond to five `torsion classes', which classify the geometry of the manifold based on the `intrinsic torsion' - see Table 3.1 of \cite{Grana:2005jc}. The intrinsic torsion is then intimately linked with the flux. In the zero torsion case, for which all torsion classes (and flux) are fixed to zero, one defines a Calabi-Yau 3-fold.

Finally, to generalize this discussion to an $SU(3)$ structure on an $M_7$, one must add a real one-form, $K$, where
\beq\label{eqn:ds7defn}
ds^2(M_7)=ds^2(M_6)+K^2,~~~~~~~~\text{vol}(M_7)=\text{vol}(M_6)\wedge K,
\eeq
so an $SU(3)$ structure on an $M_7$ is described by $(J,\Omega,K)$. As we will soon discuss, this is the case for both the $d=11$ LLM and GM solutions.


\subsubsection{$SU(d)$ in $d=$ even Structure}
More generally, this is extended to $SU(d/2)$ in $d=$ even dimensions. Here the structure is specified by a real 2-form $J$ and a complex $d/2$ form, $\Omega$ 
\beq\label{eqn:JOmegadefs}
\Omega=E^1\wedge ...\wedge E^{\frac{d}{2}},~~~~~~~~~~~~~~~~J=\frac{i}{2}\sum_{a=1}^{d/2}E^a \wedge \bar{E}^a,
\eeq
where $E^a$ is a holomorphic vielbein $E^a=e^a+i\,e^{a+d/2}$, and the metric is defined by
\beq\label{eqn:gdefn}
g=\sum_{a=1}^{d/2}E^a\bar{E}^a.
\eeq
Hence, for an $SU(3)$ structure on $M_6$, one has
\beq\label{eqn:OmegaJ}
ds^2(M_6) = \sum_{a=1}^{3}E^a\bar{E}^a,~~~~~~~~~~~\Omega= E^1\wedge E^2\wedge E^3  ,~~~~~~~~~~~ J=\frac{i}{2}(E^1\wedge \overline{E}^1+E^2\wedge \overline{E}^2+E^3\wedge \overline{E}^3),
\eeq
which then satisfies \eqref{eqn:JOmegacondition}.

Another noteworthy example is an $SU(2)$ structure on $M_4$, which by \eqref{eqn:JOmegadefs}, is defined by a real two-form, $j$, and complex two-form, $\omega$. The following necessary conditions are then met
\beq\label{eqn:jomegaconditions}
j\wedge \omega=\omega\wedge \omega=0,~~~~~~~~~~~~\omega\wedge \bar{\omega}=2j \wedge j.
\eeq

To generalize to an $SU(2)$ structure on an $M_6$, one must gain two additional dimensions. Extending the approach of \eqref{eqn:ds7defn} by adding now a complex one-form, $z$, one has
\beq
ds^2(M_6)=ds^2(M_4)+z\bar{z},~~~~~~~~~~~~~~~~~~~~~z\equiv u+i\,v,
\eeq
with $u$ and $v$ real one-forms. 

The pure spinors are then constructed as follows
\beq \label{eqn:Psi}
\Psi_+=\frac{1}{8} e^{\frac{1}{2}z\wedge \overline{z}}\wedge \omega,~~~~~\Psi_-=\frac{i}{8}  e^{-i j}\wedge z.
\eeq

\subsubsection{$SU(3)\times SU(3)$ Structure}
In general, having two spinors, $\eta^{1,2}$, describing an $SU(3)\times SU(3)$ structure, we have
\beq\label{eqn:SU(3)SU(3)}
\Psi_+=\eta_+^1 \otimes \eta_+^{2\dagger} = \frac{1}{8}  e^{\frac{1}{2}z\wedge \bar{z}} \wedge (\bar{c}e^{-ij}-i\omega) ,~~~~~~~~~~~~~~~~~\Psi_-=\eta_+^1\otimes \eta_-^{2\dagger}=-\frac{1}{8}(e^{-ij}+ic\,\omega)\wedge z.
\eeq

The generalisations to \eqref{eqn:etaCov} and \eqref{eqn:JOmegaEqsgen} for an $SU(3)\times SU(3)$ structure can be found in \cite{Grana:2005sn}. 
The differential conditions given in \eqref{eqn:schematic}, written in the form outlined in \cite{Tomasiello:2007zq}, now become
\begin{align}\label{eqn:origSUSYcondsSU3SU3}
&d_H\Psi_2=0,~~~~~~~d_H(e^{-A}\text{Re}\Psi_1)=0,~~~~~~~~~d_H(e^A\text{Im}\Psi_1)=\frac{e^{4A}}{8}\star_6 \lambda(g),\nn\\[2mm]
&||\Psi_1||=||\Psi_2||=\frac{e^{3A-\Phi}}{\sqrt{8}},~~~~~~~~F=g+e^{4A}\text{Vol}(\text{Mink}_4)\wedge *_6\lambda(g),
\end{align}
which, in a more schematic form, read 
\beq
\begin{aligned}
\mathcal{N}=1 ~(\text{RR}\neq0):&~~~ d_H\Psi_2=0,~~~~d_H\Psi_1=F_{RR},\\[2mm]
\mathcal{N}=2 ~(\text{RR}=0):&~~~d_H\Psi_2=0,~~~~d_H\Psi_1=0,
\end{aligned}
\eeq
with $\mathcal{N}=2$ vacua given when $F_{RR}=0$. See the discussion given in \cite{Grana:2005sn}. 

In the case of type II backgrounds with an $SU(3)\times SU(3)$ structure, the \eqref{eqn:origSUSYcondsSU3SU3} conditions with the relation \eqref{eqn:PsiFrelations} can be seen as a `generalised mirror symmetry'. This was first discussed in \cite{Martucci:2005ht}.

In \cite{Tomasiello:2007zq}, the \eqref{eqn:origSUSYcondsSU3SU3} conditions were reformulated in the language of generalized complex geometry, improving the formalism by eliminating the dependence on the Hodge star (and hence the metric). For the purposes of the analysis conducted in this work however, \eqref{eqn:origSUSYcondsSU3SU3} will be sufficient.

To then specialise the pure spinors \eqref{eqn:SU(3)SU(3)} to the $SU(3)$ \eqref{eqn:SU(3)spinors} and static $SU(2)$ \eqref{eqn:Psi} cases, one requires
\beq
J=j+\frac{i}{2}z\wedge \bar{z},~~~~~~~~~~\Omega=\omega\wedge z,
\eeq
in the $SU(3)$ case and $c=0$ (with $\Psi\rightarrow i\,\Psi$) in the static $SU(2)$ case. Using the $SU(3)$ pure spinors \eqref{eqn:SU(3)spinors} reduces the pure spinor conditions \eqref{eqn:origSUSYcondsSU3SU3} to differential conditions of $(J,\Omega)$ more directly, as in \eqref{eqn:JOmegaEqsgen}.


\subsection{An $SU(3)$ structure on M$_7$
}\label{sec:LLMGstructures}
 We will now discuss the case of a Mink$_4$ solution describing an $SU(3)$ structure on an $M_7$, for which
 \begin{align} 
&ds^2 =e^{2\hat{A}}ds^2(\text{Mink}_4) +ds^2(M_7),~~~~~~~~~~~~~~~
ds^2(M_7)  = \sum_{a=1}^{3}E^a \bar{E}^{\bar{a}}+K^2.\nn
\end{align}
The necessary $\mathcal{N}=1$ G-Structure conditions were derived in \cite{Kaste:2003zd} (see also \cite{Gauntlett:2004zh}), defined by a real two-form, $J$, a holomorphic three-form, $\Omega$, (giving an SU(3) structure on $d=6$), with an additional (orthogonal and unit normed) real one-form, $K$ (to move to $d=7$), as follows
\beq\label{eqn:Geqns}
\begin{aligned}
&d(e^{2\hat{A}} K)=0,~~~~~~~~~~~~~~~~~~
d(e^{4\hat{A}} J)=e^{4\hat{A}} *_7G_4,\\
&d(e^{3\hat{A}} \Omega)=0,~~~~~~~~~~~~~~~~~~
d(e^{2\hat{A} }J\wedge J )=-2 e^{2\hat{A} }G_4\wedge K,
\end{aligned}
\eeq
  where $e^{\hat{A}}$ and $G_4$ live on $M_7$, with \eqref{eqn:JOmegacondition}. 
  In a case without flux, they would clearly reduce to three conditions, one for each of the three forms $(J,\Omega,K)$.
In addition to \eqref{eqn:Geqns}, one must impose the Bianchi identities separately. The conditions \eqref{eqn:JOmegacondition} allow $(J,\Omega)$ to be written in the manner outlined in \eqref{eqn:OmegaJ}, namely
     \begin{equation*}
\Omega= E^1\wedge E^2\wedge E^3  ,~~~~~~~~~~~~~~~ J=\frac{i}{2}(E^1\wedge \overline{E}^1+E^2\wedge \overline{E}^2+E^3\wedge \overline{E}^3),
\end{equation*}
where the complex vielbein, $E^a$, is orthogonal to $K$ (with $a=1,2,3$). 

One can then utilise the Poincar\'e patch \eqref{eqn:PPatch} to express an AdS$_5$ solution within this formalism. Alternatively, one could use the AdS$_5$ G-structure conditions given in \cite{Gauntlett:2004zh}. However, the Mink$_4$ conditions will be more convenient for the cases discussed in this work, allowing for supersymmetric embeddings for objects extended along Mink$_4$ and orthogonal to $\rho$. 

\subsubsection{Lin-Lunin-Maldacena (LLM)}
A nice example of an $SU(3)$ structure on an $M_7$ is the LLM background \eqref{eqn:LLMgeneral}, for which we will now discuss the G-Structure formalism explicitly.
 
Using the Poincar\'e patch \eqref{eqn:PPatch} along with \eqref{eqn:ds7defn} and \eqref{eqn:OmegaJ}, one can re-write the LLM metric in the appropriate form 
\begin{align}\label{eqn:LLMmetriccondition}
&ds^2 =e^{2\hat{A}}ds^2(\text{Mink}_4) +ds^2(M_7),~~~~~~~~~~~~~~~e^{2\hat{A}}= 4\,\kappa^{\frac{2}{3}}e^{2(\rho+\lambda)},\nn\\[2mm]
&ds^2(M_7)  = \sum_{a=1}^{3}E^a \bar{E}^{a}+K^2\\[2mm]
&~~~~~~~~~~~= \kappa^{\frac{2}{3}}e^{2\lambda}\bigg[4\,d\rho^2+y^2 e^{-6\lambda}ds^2(\text{S}^2)+\frac{4}{1-y \partial_y D}(d\tilde{\chi}+A_a d\hat{x}^a)^2-\frac{\partial_yD}{y}\bigg(dy^2+e^{D}(d\hat{x}_1^2+d\hat{x}_2^2)\bigg)\bigg]\nn.
\end{align}
The complex vielbeins for this background were then presented in \cite{Macpherson:2016xwk} (and typo corrected in \cite{Macpherson:2024frt}), and read
\begin{align}\label{eq:LLMniceframe}
K&= \kappa^{\frac{1}{3}}e^{-2(\lambda+\rho)}d\left(e^{2\rho}y y_3\right),~~~~E_1= \kappa^{\frac{1}{3}}\sqrt{\frac{-\partial_yD}{y}}e^{\lambda+\frac{1}{2}D}\bigg(d\hat{x}_1+ i \,d \hat{x}_2\bigg),\nn\\[2mm]
E_2&= \kappa^{\frac{1}{3}}e^{-2(\lambda+\rho)}d\left(e^{2\rho}y(y_1+i y_2)\right),\nn\\[2mm]
E_3&= -\kappa^{\frac{1}{3}} e^{i \tilde{\chi}}\frac{2}{\sqrt{1- y \partial_yD}}e^{\lambda}\bigg(d\rho+\frac{1}{2}\partial_y Ddy + i (d\tilde{\chi}+ A_a d\hat{x}^a)\bigg),
\end{align}
where $y_i$ are a set of embedding coordinates for a 2-sphere of unit radius, such that
\begin{align}
&ds^2(S^2)=dy_1^2+dy_2^2+dy_3^2,~~~~~~~~~~y_1^2+y_2^2+y_3^2=1,   \nn\\[2mm]
&\text{where}~~~~~ y_1\equiv \cos\phi\sin\theta,~~~~~~ y_2\equiv \sin\phi\sin\theta,~~~~~~y_3\equiv \cos\theta.
\end{align}
Hence, the coordinates on M$_7$ are $(\rho, y, \hat{x}_1, \hat{x}_2,\tilde{\chi})$ along with either $(y_1,y_2,y_3)$ or $ (\theta,\phi)$ (which define the $S^2$). 

The complex vielbeins given in \eqref{eq:LLMniceframe} now define the G-structure forms $(J,\Omega,K)$, using \eqref{eqn:OmegaJ}, which are clearly charged under an $SU(2)_R\times U(1)_R$ R-symmetry. The $\Omega$ three-form has an overall phase of $e^{i\tilde{\chi}}$ coming from the form of $E_3$, and charged under $\tilde{\chi}$ which spans the $U(1)_R$ component. 
One can then check the supersymmetry conditions given in \eqref{eqn:Geqns} are satisfied.


A Mink$_4$ solution has minimal supersymmetry with four real supercharges, which gets doubled by the $SU(2)_R$ R-symmetry, and doubled again by the $U(1)_R$ R-symmetry. This then realises the 16 real supercharges of the $\mathcal{N}=2$ AdS$_5$ solution of LLM. 

Making the coordinate transformations \eqref{eqn:x1x2transformation}, such that an additional $U(1)$ isometry is imposed, the following modifications to \eqref{eq:LLMniceframe} are required
\beq\label{eq:LLMniceframeModified}
V=-\frac{r}{2}\partial_r\,D d\tilde{\beta},~~~~~E_1= \kappa^{\frac{1}{3}} e^{i\tilde{\beta}}\sqrt{\frac{-\partial_yD}{y}}e^{\lambda+\frac{1}{2}D}\Big(dr+i\,r\,d\tilde{\beta}\Big),
\eeq
with $d\hat{x}_1\wedge d\hat{x}_2=r\,dr\wedge d\tilde{\beta}$ and \eqref{eqn:LLMPDE}. The metric \eqref{eqn:LLM} is then built in the analogous manner to \eqref{eqn:LLMmetriccondition} following this modification. One can now observe, from the results of $E_1$ and $E_3$, the $\Omega$ three-form defined in \eqref{eqn:OmegaJ} now depends on the overall phase $e^{i(\tilde{\chi}+\tilde{\beta})}$. One can then conclude that it is in fact $ \tilde{\chi}+\tilde{\beta}\equiv \tilde{\psi}$ which corresponds to the $U(1)_R$ component of the $U(1)_R\times SU(2)_R$ R-symmetry. Hence, in order to dimensionally reduce along $\tilde{\beta}$ in a supersymmetry preserving manner, one must first transform $\tilde{\chi}\rightarrow \tilde{\chi}=\tilde{\psi}-\tilde{\beta}$. The solution now contains the two $U(1)$ components, $\tilde{\psi}$ and $\tilde{\beta}$, with $\tilde{\psi}$ corresponding to the $U(1)$ component of the R-symmetry - remaining unaffected by a $\tilde{\beta}$ reduction. This line of analysis will be extended to the GM solution in Part \ref{Part:SUSYdefs}, following an $SL(3,\mathds{R})$ transformation, and will play a crucial role in keeping track of supersymmetry under dimensional reduction. See \cite{Macpherson:2016xwk} for further details.

\subsection{An $SU(2)$ structure on M$_6$ - via dimensional reduction
}\label{sec:SU(2)Gs}
When a solution with an $SU(3)$ structure on $M_7$ is dimensionally reduced along a $U(1)$ living along $K$ (such that $K\rightarrow 0$ under the reduction), the resulting theory will simply describe an $SU(3)$ structure on an $M_6$.\\
However, when the dimensional reduction is performed along a $U(1)$ which lies strictly outside $K$, the $M_6$ of the resulting theory will now have an $SU(2)$ structure, where
  \begin{align}\label{eqn:SU(2)metric}
&ds_{10}^2=e^{2A}ds^2(\text{Mink}_4)+ds^2(M_6),\nn\\[2mm]
&ds^2(M_6) = \sum_{a=1}^{2}\hat{E}^a \bar{\hat{E}}^{a}+z\bar{z},~~~~~~~~~~~~~~~~~~~~~z\equiv u+i\,v.
\end{align}
The G-structure analysis is then defined by a real two-form, $j$, and a holomorphic two-form, $\omega$ (defining an SU(2) structure on $d=4$), 
      \beq\label{eqn:IIAomegaj1}
\omega= \hat{E}^1\wedge \hat{E}^2 ,~~~~~~~~~~~~~~~ j=\frac{i}{2}(\hat{E}^1\wedge \overline{\hat{E}}^1+\hat{E}^2\wedge \overline{\hat{E}}^2),
\eeq
satisfying \eqref{eqn:jomegaconditions}, with an additional complex one-form, $z$ (to move to $d=6$). 
 

For a IIA solution, these forms can then be derived directly from the $d=11$ G-structures by the analogue of the reduction formula \eqref{eqn:reductionformula}, given in \cite{Macpherson:2015tka}, where
\begin{equation}\label{eqn:11Dto10Dforms}
\begin{aligned}
&J = j\,e^{-\frac{2}{3}\Phi} +e^{\frac{1}{3}\Phi}u \wedge (d\psi+C_1),~~~~~~~~~~K= v\,  e^{-\frac{1}{3}\Phi},\\[2mm]
&\Omega  = \omega \wedge \Big(e^{-\Phi}u +i(d\psi+C_1)\Big),
~~~~~~~~~~~~~~~~z=u+i\,v,
\end{aligned}
\end{equation}
with $e^A=e^{\hat{A}+\frac{1}{3}\Phi}$ and $\Phi$ the dilaton. This reduction formula is very useful, allowing one to build the $SU(2)$ pure spinors given in \eqref{eqn:Psi}. However, it is worthwhile deriving the G-structure description at the level of the complex vielbeins, which then allows one to build both the $d=10$ metric \eqref{eqn:SU(2)metric} and the forms \eqref{eqn:IIAomegaj1}. To do this, we can re-write \eqref{eqn:11Dto10Dforms} using the analysis given in \cite{Macpherson:2015tka}, in which the $d=10$ metric takes the form
\beq
ds_{10}^2=e^{2A}ds^2(\text{Mink}_4)+ \sum_{i=1}^3(\hat{h}^i)^2+\sum_{j=1}^2(\hat{e}^j)^2,
\eeq
with the definitions
\beq
j=\hat{h}^1\wedge \hat{h}^2+\hat{e}^1\wedge \hat{e}^2,~~~~~~~~~~~\omega =(\hat{h}^1+i \hat{h}^2)\wedge (\hat{e}^1+i \hat{e}^2),~~~~~z=\hat{h}^3+i\hat{e}^3,
\eeq
leading to
\beq
\begin{aligned}\label{eqn:SU(3)forms2}
&\Omega = e^{i\hat{\theta}_+}(\hat{h}^1+i \hat{h}^2)\wedge (\hat{e}^1+i \hat{e}^2) \wedge \Big(e^{-\Phi}\hat{h}^3+i(d\psi +C_1)\Big),~~~~~~~~K=\hat{e}^3e^{-\frac{1}{3}\Phi},\\
&J=e^{-\frac{2}{3}\Phi}(\hat{h}^1\wedge \hat{h}^2+\hat{e}^1\wedge \hat{e}^2)+e^{\frac{1}{3}\Phi}\hat{h}^3 \wedge (d\psi+C_1).
\end{aligned}
\eeq
One can now directly compare \eqref{eqn:OmegaJ} and \eqref{eqn:SU(3)forms2}. In the following result, we pick $E_2$ to include the overall phase (including contributions from $E_1$ and $E_3$), along with $e^{i\hat{\theta}_+}$ and any overall sign of $E_3$. We then find
\beq
\begin{aligned}
&E^1 = e^{-\frac{1}{3}\Phi}(\hat{h}^1+i \hat{h}^2) = e^{-\frac{1}{3}\Phi} \hat{E}^1 ,~~~~~~~E^2 =e^{-\frac{1}{3}\Phi}e^{i\hat{\theta}_+}(\hat{e}^1+i \hat{e}^2) = e^{-\frac{1}{3}\Phi} e^{i\hat{\theta}_+}\hat{E}^2,\\
&E^3 = e^{\frac{2}{3}\Phi}\Big(e^{-\Phi}\hat{h}^3+i (d\psi+C_1)\Big).
\end{aligned}
\eeq
Finally, we can now derive the IIA complex vielbeins directly from their $d=11$ equivalent
\beq\label{eqn:IIAviels}
\hat{E}^1 = e^{\frac{1}{3}\Phi} E^1,~~~~~~\hat{E}^2 =e^{\frac{1}{3}\Phi} e^{-i\hat{\theta}_+} E^2,~~~~~~z=e^{\Phi}\Big( e^{-\frac{2}{3}\Phi}E^3-i (d\psi+C_1)\Big)+i\,e^{\frac{1}{3}\Phi}K,
\eeq
from which one can build the metric and G-structure forms. Of course, in practice, the reduction formula \eqref{eqn:11Dto10Dforms} is the easiest and most direct approach to derive $(j,\omega,z)$.

Once the G-structure forms are calculated, for a solution with an $SU(2)$ structure on $M_6$, one is free to construct the $SU(2)$ pure spinors 
from \eqref{eqn:Psi}, namely
  \begin{equation*}
\Psi_+=\frac{1}{8} e^{\frac{1}{2}z\wedge \overline{z}}\wedge \omega,~~~~~\Psi_-=\frac{i}{8}  e^{-i j}\wedge z,
\end{equation*}
which when expanded, take the more practically convenient form
    \begin{equation}\label{eqn:PsiConvenient}
    \begin{aligned}
    \Psi_{+}&=\frac{1}{8}\Big(1+\frac{1}{2}z\wedge \bar{z}\Big)\wedge \omega,\\
 \Psi_{-}&= \frac{i}{8}z \wedge \Big(1-i\,j-\frac{1}{2}j\wedge j\Big)=\frac{1}{8}\bigg[u\wedge j-v\wedge\Big(1-\frac{1}{2}j\wedge j\Big)\bigg]+\frac{i}{8}\,\bigg[u\wedge \Big(1-\frac{1}{2}j\wedge j\Big)+v\wedge j\bigg],
    \end{aligned}
    \end{equation}
    with the real and imaginary parts of $ \Psi_{-}$ made clear. Now, for the preservation of $\mathcal{N}=1$ SUSY, the differential conditions on the metric \eqref{eqn:origSUSYcondsSU3SU3} must be satisfied, which can be re-written as follows
 \begin{subequations}\label{eqn:IIAGconditions}
\begin{align}
d_{H}(e^{3A-\Phi}\Psi_2)&=0,\label{eqn:IIAGconditions1}\\
d_{H}(e^{2A-\Phi}\text{Re}\Psi_1)&=0,\label{eqn:IIAGconditions2}\\
d_{H}(e^{4A-\Phi}\text{Im}\Psi_1)&=\frac{e^{4A}}{8}*_6\lambda(g),\label{eqn:Calibrationform}
\end{align}
 \end{subequations}
where $(A,\Phi,H,g$) are all functions of the $M_6$ coordinates exclusively, with $g$ the total internal RR flux. 

In the case of a type IIA solution, $\Psi_1=\Psi_-,~\Psi_2=\Psi_+$ and $\lambda(g)=F_0-F_2+F_4-F_6$. From \eqref{eqn:relstuff}, one has $e^{4A}\text{vol}_4 \wedge *_6\lambda(g)= F_6+F_8+F_{10}$, allowing \eqref{eqn:Calibrationform} to be rewritten in the following way
\begin{align}
F_6+F_8+F_{10}&= 8\,\text{vol}(\text{Mink}_4)\wedge d_{H_3}(e^{4A-\Phi}\text{Im}\Psi_-),\nn\\[2mm]
\Rightarrow C_5+C_7+C_9&=8\,e^{4A-\Phi}\text{vol}(\text{Mink}_4)\wedge \text{Im}\Psi_-,\nn\\[2mm]
\Rightarrow C_m&=8\,e^{4A-\Phi}\text{vol}(\text{Mink}_4)\wedge \text{Im}\Psi_{-_{m-4}},\label{eq:usefulpotentials}
\end{align}
which then derives the potential for the higher form fluxes directly from the G-structure analysis, where $\Psi_{-_{m-4}}$ is the $(m-4)$-form part of $\Psi_-$. We will see in Section \ref{sec:calibration} that this result also becomes very useful in investigating the supersymmetry of D-brane sources.

In Appendix \ref{sec:ATDGstructureConds}, we demonstrate that under an ATD to type IIB, the differential conditions \eqref{eqn:IIAGconditions} remain intact. In the IIB case however, recalling the arguments given in \eqref{eqn:IIAIIBPsis} and \eqref{eqn:PsiFrelations}, one now requires $\Psi_1=\Psi_+,~\Psi_2=\Psi_-$ to account for the opposite dimensionality of the total RR flux. 

We now write the higher form fluxes in general
\beq\label{eqn:Cmgen}
C_m=8\,e^{4A-\Phi_{\mathcal{A}/\mathcal{B}}}\text{vol}(\text{Mink}_4)\wedge \text{Im}\Psi_{1_{m-4}},
\eeq
where $\Psi_{1_{m-4}}$ is the $(m-4)$-form part of $\Psi_1$, with $\Psi_1=\Psi_-$ in IIA (with $\Phi_{\mathcal{A}}$ the dilaton) and $\Psi_1=\Psi_+$ in IIB (with $\Phi_{\mathcal{B}}$ the dilaton).

\section{Calibrations}\label{sec:calibration}
In addition to determining the level of supersymmetry preservation of a solution, G structures provide tools to establish whether there is a supersymmetric embedding of the background sources. The sources of interest are D-branes, which have the action, $S_{D_p}$, given in \eqref{eq:branactions}. It was shown in \cite{Martucci:2005ht} that the $\mathcal{N}=1$ supersymmetry conditions given in \eqref{eqn:IIAGconditions} can be interpreted as a calibration condition for a D-brane, which then provides important information into both the stability and geometry of the brane. Such an interpretation then allows for stability investigations into both SUSY and non-SUSY solutions. 

\subsection{Supersymmetric D-branes}\label{sec:susyDbranes}
In this subsection, we will review supersymmetric D-brane configurations living on $\mathcal{N}=1$ backgrounds of Mink$_4\times M_6$ warped product geometry with general NSNS and RR fields.

The supersymmetry conditions for a D-brane are obtained from $\kappa$-symmetry constraints, and were re-written in terms of generalised calibrations in \cite{Gutowski:1999iu,Gutowski:1999tu}, which were in turn elegantly expressed in terms of the pure spinors $\Psi_\pm$ in \cite{Martucci:2005ht}. See also \cite{Gauntlett:2001ur,Gauntlett:2002sc,Gauntlett:2003cy,Martelli:2003ki}.

The notion of a `generalised calibration' is that it should minimise the energy of the D$p$-brane, which may not necessarily match the volume which the brane wraps. A generalised calibration is defined as a sum of forms of varying degree, satisfying
\beq
\hat{\omega}=\sum_k\hat{\omega}_k,
\eeq
such that, for some D$p$-brane with energy density $\mathcal{E}(\Sigma,\mathcal{F})$, characterised by the cycle over which it wraps, $\Sigma$, and the worldvolume field strength, $\mathcal{F}$, defined in \eqref{eq:branactions}, one has\footnote{noting that some referenced literature use $\hat{\omega}|_\Sigma\wedge e^\mathcal{F}$ due to their alternative definition of the twisted exterior derivative $d_H=d+H\wedge$.}
\beq
\hat{\omega}|_\Sigma\wedge e^\mathcal{-F} \leq \mathcal{E}(\Sigma,\mathcal{F}).
\eeq
The D$p$-brane is then considered `calibrated' by $\hat{\omega}$ if 
\beq\label{eqn:calcond}
\hat{\omega}|_\Sigma\wedge e^\mathcal{-F} = \mathcal{E}(\Sigma,\mathcal{F}),
\eeq
which is a minimum energy bound, and under continuous deformations, can be viewed as a stability condition. The energy density then corresponds to the contribution from the DBI action \cite{Martucci:2005ht},
\beq
\mathcal{E}(\Sigma,\mathcal{F}) = e^{qA-\Phi}\sqrt{\text{det}(g+\mathcal{F})}|_{\Sigma}.
\eeq
One can consider a static supersymmetric D$p$-branes which span four, three or two of the external Mink$_4$ directions - corresponding to a space-filling, domain wall or string configuration in the external space. The D-branes then must wrap some $\Sigma_{p+1-q}$ cycle in the internal manifold, $M_6$, with $q=4,3,2$, in the three cases respectively. This cycle must then be a calibrated generalised sub-manifold for the D-brane to be supersymmetric (BPS). The generalised calibration in each case takes the form
 \begin{subequations}\label{eqn:GenCal}
\begin{align}
\hat{\omega}^{DW}& = 8e^{3A-\Phi}\Psi_2,\\
\hat{\omega}^{String}&=8e^{2A-\Phi}\text{Re}\Psi_1,\\
\hat{\omega}^{sf}&= \pm 8e^{4A-\Phi}\text{Im}\Psi_1, 
\end{align}
 \end{subequations}
with the following differential conditions
 \begin{subequations}\label{eqn:GenCalDiff}
\begin{align}
d_H(\hat{\omega}^{DW})&=0,~~~~~~~~~~\text{domain-wall BPSness}\\
d_H(\hat{\omega}^{String})&=0,~~~~~~~~~~\text{D-string BPSness}\\
d_H(\hat{\omega}^{sf})&=e^{4A}\tilde{F}, ~~~~~\text{gauge BPSness}
\end{align}
 \end{subequations}
 where $\pm$ differentiates between branes and anti-branes. See \cite{Martucci:2005ht} for further details.
These results are in fact equivalent to the supersymmetry conditions \eqref{eqn:IIAGconditions}, which should now be physically interpreted as generalised calibration conditions for each of the allowed configurations of a supersymmetric D-brane. Hence, following \cite{Lust:2008zd}, the three supersymmetry conditions correspond to domain-wall, D-string and gauge BPSness.

In the case of a D$p$-brane extended in Mink$_4$, such that it wraps some $\Sigma_{p-3}$ cycle, the calibration condition \eqref{eqn:calcond} reads 
 \beq
\Psi^{(\text{cal})}_{\text{Dp}}\bigg\lvert_{\Sigma_{p-3}} = e^{4A-\Phi}\sqrt{\det(g+{\cal F})}\bigg\lvert_{\Sigma_{p-3}}d^{p-3}w,~~~~~~~~~\Psi^{(\text{cal})}_{\text{Dp}}\equiv \pm 8e^{4A-\Phi}\text{Im}\Psi_1\wedge e^{-{\cal F}},\label{eq:Dpsusy}
\eeq
which recalling the result for the higher form fluxes \eqref{eq:usefulpotentials}, one has the relation 
\beq\label{eq:Dpsusy2}
\text{vol}(\text{Mink}_4)\wedge\Psi^{(\text{cal})}_{\text{Dp}}\Big|_{\Sigma_{p-3}} =\pm 8\,e^{4A-\Phi}\text{vol}(\text{Mink}_4)\wedge \text{Im}\Psi_{1}\wedge e^{-{\cal F}}|_{\Sigma_{p-3}} =  \pm C|_{D_p}\wedge e^{-{\cal F}},
\eeq
which is equivalent to the integrand of the (negative) Wess-Zumino action given in \eqref{eq:branactions}. In addition, from the calibration condition \eqref{eq:Dpsusy}, one has
\beq
\hspace{-1.25cm}
\text{vol}(\text{Mink}_4)\wedge\Psi^{(\text{cal})}_{\text{Dp}}\Big|_{\Sigma_{p-3}} =e^{4A-\Phi}\sqrt{\det(g+{\cal F})}\bigg\lvert_{\Sigma_{p-3}}\text{vol}(\text{Mink}_4)\,\wedge\, d^{p-3}w =e^{-\Phi}\sqrt{\det(g \lvert_{D_p}+{\cal F})} \,d^{p+1}w,
\eeq
which in turn is equivalent to the integrand of the DBI action. Hence, from the form of D-brane action \eqref{eq:branactions}, one can see that the calibration condition leads to  
\beq
e^{-\Phi}\sqrt{\det(g \lvert_{D_p}+{\cal F})} \,d^{p+1}w = \pm C|_{D_p}\wedge e^{-{\cal F}},~~~~~\Rightarrow~~~~~S_{\text{Dp}}= S_{\text{DBI}}+S_{\text{WZ}} = 0,
\eeq
which then corresponds to a minimum energy D-brane. Hence, a supersymmetric embedding of a D-brane implies that it has minimum energy - but the converse is not true.


\subsection{Supersymmetry Breaking}\label{sec:SUSYbreakingGs}
The $\mathcal{N}=1$ G-structure approach can be extended to consider more general $\mathcal{N}=0$ Type II supergravity solutions, for which the supersymmetry is broken in a controllable manner by switching on some 
parameter (which can lead to `no-scale' supersymmetry breaking - see \cite{Camara:2007cz}). In practice, this approach proves very useful as finding non-supersymmetric solutions directly is a very difficult task.
The $\mathcal{N}=1$ supersymmetry conditions \eqref{eqn:IIAGconditions} should then be modified by additional terms which include this SUSY-breaking parameter as an overall factor - hence recovering the original $\mathcal{N}=1$ conditions when the parameter is switched off. As we have just reviewed however, these $\mathcal{N}=1$ conditions have the physical interpretation as calibration conditions for a supersymmetric D-brane - with a domain-wall, string or space-filling configuration in the external space, respectively. Hence, the additional terms will then be referred to as domain-wall (DWSB), string-like (SSB) or space-filling (SFSB) supersymmetry breaking, depending on which of the three conditions \eqref{eqn:GenCalDiff} they effect. This will only be a very brief summary of some key results, the reader is pointed towards \cite{Koerber:2010bx,Lust:2008zd,Legramandi:2019ulq,Menet:2023rnt,Menet:2023rml,Menet:2024afb,Held:2010az} for more detailed and rigorous analysis.  

\subsubsection{DWSB}
In DWSB, the differential conditions \eqref{eqn:GenCalDiff} and \eqref{eqn:IIAGconditions} will now have the following modification
\begin{align}
d_H(\hat{\omega}^{DW})=d_H( 8e^{3A-\Phi}\Psi_2)=\{\text{SUSY-breaking terms}\}.~~~~~~~~~~\text{domain-wall (non)-BPSness}\nn
\end{align}
An example of a DWSB solution with a single supersymmetry-breaking parameter was studied in \cite{Lust:2008zd}, along with a more general deformation which they include in an appendix. They present the SUSY-breaking modification to the Killing spinor equation, along with corresponding DWSB modification to the pure spinor conditions. After some manipulation, the supersymmetry conditions \eqref{eqn:IIAGconditions} have the following modification\footnote{with $\Psi\rightarrow \frac{i}{8}\Psi$ and $H\rightarrow -H$ to match their convention},
 \begin{subequations}\label{eqn:DWSB1}
\begin{align}
d_{H}(e^{3A-\Phi}\Psi_2)&=\frac{i\,r}{8}\,j,\label{eqn:DWSB1a}\\
d_{H}(e^{2A-\Phi}\text{Re}\Psi_1)&=0,\label{eqn:DWSB1b}\\
d_{H}(e^{4A-\Phi}\text{Im}\Psi_1)&=\frac{e^{4A}}{8}*_6\lambda(g), \label{eqn:DWSB1c}
\end{align}
\end{subequations}
where $r$ is a supersymmetry-breaking parameter, and 
\beq
j=4(-1)^{|\Psi_2|}e^{3A-\Phi}\frac{\sqrt{\text{det}\,g|_\Sigma}}{\sqrt{\text{det}\,(g|_\Sigma+\mathcal{F})}}e^{\mathcal{F}}\wedge \lambda(d\text{vol}_\perp),
\eeq
where 
$\text{vol}_6=\text{vol}_\Sigma \wedge \text{vol}_\perp$, with $\text{vol}_\perp$ the volume form of the space orthogonal to the $\Sigma$ cycle. Here $j$ is interpreted geometrically as a (smeared) generalised current associated with the sources of the background, and $r$ has been chosen such that the right hand side is $d_H$ closed. 


\subsubsection{SSB}
In the SSB case, the modification will now take the form
\begin{align}
d_H(\hat{\omega}^{string})=d_H(8e^{2A-\Phi}\text{Re}\Psi_1)=\{\text{SUSY-breaking terms}\}.~~~~~~~~~~\text{D-string (non)-BPSness}\nn
\end{align}
An example of a SSB solution was given in \cite{Legramandi:2019ulq} for type IIA, where they claim that imposing the Bianchi identities and a particular consistency condition on $H$, only \eqref{eqn:IIAGconditions1} and \eqref{eqn:Calibrationform} are required to satisfy the equations of motion. As an example, \eqref{eqn:IIAGconditions} can be modified in the following way
 \begin{subequations}\label{eqn:SSB1}
\begin{align}
d_{H}(e^{3A-\Phi}\Psi_2)&=0,\label{eqn:SSB1a}\\
d_{H}(e^{2A-\Phi}\text{Re}\Psi_1)&=\frac{c}{8}e^{6A-2\Phi}\text{vol}(M_6),\label{eqn:SSB1b}\\
d_{H}(e^{4A-\Phi}\text{Im}\Psi_1)&=\frac{e^{4A}}{8}*_6\lambda(g), \label{eqn:SSB1c}
\end{align}
\end{subequations}
which includes a supersymmetry-breaking parameter, $c$, in \eqref{eqn:SSB1b}. This example can stem from a non-supersymmetric modification to the Imamura class \cite{Imamura:2001cr} (describing a D6-D8-NS5 system) via two T-dualities, and involves adding new supersymmetry-breaking terms to $H$. The analogue of \eqref{eqn:SSB1} is also given for the non-supersymmetric Imamura class in \cite{Legramandi:2019ulq}, for which $e^{6A-2\Phi}\text{vol}(M_6)\rightarrow e^{8A-2\Phi}\text{vol}(M_4)$.

In \cite{Menet:2023rnt}, a similar approach was taken as the DWSB analysis of \cite{Lust:2008zd}, which then leads to the following modification to \eqref{eqn:SSB1}
\begin{align}
d_{H}(e^{2A-\Phi}\text{Re}\Psi_1)&= \frac{\alpha_m}{8}[\gamma^m\,j+(-1)^{|j|}j\,\gamma^m] = \frac{\alpha_m}{4} dy^m\wedge \,j ,\label{eqn:SSB2b}
\end{align}
where $\alpha_m$ are real supersymmetry-breaking parameters.

The case in which both the DWSB and SSB conditions are broken was studied in \cite{Menet:2023rnt} for which \eqref{eqn:DWSB1} is combined with \eqref{eqn:SSB2b}.

\subsubsection{SFSB}
In Part \ref{Part:SUSYdefs} of this work, we will derive multiple-parameter families of $\mathcal{N}=0$ type IIA and type IIB solutions, which enhance to $\mathcal{N}=2$ and $\mathcal{N}=1$ via appropriate choice of SUSY-breaking parameters. These new solutions will in fact break all three supersymmetry conditions, and provide the first explicit example of solutions which have SFSB controlled by a parameter.


\part{SUSY breaking deformations of Gaiotto-Maldacena}\label{Part:SUSYdefs}
\section*{Overview}
Our attention will now turn to deriving new multi-parameter families of $\mathcal{N}=0$ type IIA and type IIB AdS$_5$ solutions, which enhance to one-parameter ${\cal N}=1$ backgrounds (with four Poincar\'e SUSYs) in some special cases. In the type IIA backgrounds, there is an additional ${\cal N}=2$ solution (with eight Poincar\'e SUSYs), which corresponds to the Gaiotto-Maldacena class \cite{Gaiotto:2009gz} (reviewed in Section \ref{sec:GMIIA}). The full (two-parameter) type IIA solution then represents SUSY breaking deformations of GM, reducing to the background when both parameters are fixed to zero. The derivation of these solution stem from a $\beta$- dimensional reduction of the $d=11$ GM solutions (reviewed in Section \ref{sec:GM}), following an $SL(3,\mathds{R})$ transformation amongst the three $U(1)$ directions of the background. Performing dimensional reductions along the two alternative $U(1)$ directions then lead to additional two-parameter $\mathcal{N}=0$ solutions, which at best enhance to $\mathcal{N}=1$. A subsequent ATD to type IIB can then be performed in all cases, deriving new three-parameter solutions, for which the $\mathcal{N}=2$ supersymmetry is necessarily broken under the ATD. 

Following careful investigations at the boundary, we find that the spindle (along with its higher dimensional analogue) plays a central role in the type IIA analysis - giving rise to rational quantization of D-brane charge. In type IIB, some solutions also contain orbifold singularities, but with a less clear physical interpretation. In fact, as a consequence of T-dualising within the spindle-like orbifold, the quantization of charge simply becomes a remnant of the IIA solution, and is now disconnected from the orbifold nature of the metric (which can be broken under T-duality). Consequently, for certain type IIB solutions, one finds integer (and rational) quantization with (and without) an orbifold structure.

Central to our supersymmetry analysis is the method of G-structures, with the $\mathcal{N}=1$ analysis presented explicitly in both IIA and IIB. We also include some discussion on the breaking of the G-structure conditions, and include the explicit G-structure forms for both the two-parameter type IIA and three-parameter type IIB solutions.

Calculations of the holographic central charge, in both the type IIA and type IIB, then demonstrate that all SUSY-breaking parameters drop out neatly - 
reproducing \eqref{eqn:HCCGM}. One can then consider these multi-parameter families as dual descriptions of marginal deformations of the `parent' $\mathcal{N}=2$ CFT (holographically dual to the original GM class). A field theory discussion of the dual CFTs is then included. 

These backgrounds have similarities with the $\gamma$- deformed solutions of N\'u\~nez, Roychowdhury, Speziali and Zacar\'\i{}as (NRSZ) \cite{Nunez:2019gbg}, re-deriving their IIB solution as a sub-class of our results. This thesis then closes by presenting G-structure analysis for these solutions - uncovering that the IIB solution preserves $\mathcal{N}=1$ supersymmetry when $\gamma=-1$, whilst in M-Theory and type IIA, the supersymmetry is broken completely for all values of $\gamma$.

\chapter{Set-up}\label{chap:setup}
In this chapter, we return to the ${\cal N}=2$ Gaiotto-Maldacena system, presenting the G-structure and calibration forms as new results (in both M-Theory and type IIA). In addition, SUSY preserving probes are studied, along with careful investigation of the Page charges and the behaviour at the boundaries. This sets up the analysis presented in the remaining chapters - dedicated to the type IIA, type IIB and M-theory deformations.  Throughout our investigations, the definitions of the eight warp factors $f_i$, given in \eqref{eqn:fs}, will remain consistent.

\section{G-Structure description of GM in $d=11$}\label{sec:G-Structure description of GM}
In order to keep track of SUSY under the various deformations uncovered in the upcoming chapters, it proves fruitful to derive the G-structure description of the original GM class.

We begin by recalling the form of the GM class \eqref{eqn:GM}, namely
\begin{align} 
ds_{11}^2&=f_1\Bigg[4ds^2(\text{AdS}_5) +f_2ds^2(\text{S}^2)+f_3d\chi^2+f_4\big(d\sigma^2 +d\eta^2\big)+f_5\Big(d\beta +f_6d\chi\Big)^2\Bigg],~~~~~f_i=f_i(\eta,\sigma),\nn\\[2mm]
A_3&=\Big(f_7d\chi +f_8d\beta \Big)\,\wedge\text{vol}(\text{S}^2),~~~~~~ds^2(\text{S}^2)= d\theta^2+\sin^2\theta d\phi^2,~~~~\text{vol}(\text{S}^2)=\sin\theta d\theta \wedge d\phi.\nn
\end{align} 
The three $U(1)$ directions,  $(\beta,~\chi,~\phi)$
, will give rise to the SUSY breaking deformations, following an $SL(3,\mathds{R})$ transformation. We will return to this in the next section.
 
As in the LLM class of solutions, reviewed in sections \ref{sec:LLM} and \ref{sec:LLMGstructures}, the GM backgrounds describe an $SU(3)$ structure on an $M_7$. Hence, the description will be analogous to the LLM case, in which the $\mathcal{N}=1$ G-Structure conditions are defined in terms of a real two-form, $J$, a holomorphic three-form, $\Omega$ (giving an $SU(3)$ structure on $d=6$), with an additional (orthogonal and unit normed) real one-form, $K$ (to move to $d=7$). We recall from Section \ref{sec:LLMGstructures}, the G-Structure conditions then read
\beq 
\begin{aligned}
&d(e^{2\hat{A}} K)=0,~~~~~d(e^{3\hat{A}} \Omega)=0,~~~~~
d(e^{4\hat{A}} J)=e^{4\hat{A}} *_7G_4,~~~~~
d(e^{2\hat{A} }J\wedge J )=-2 e^{2\hat{A} }G_4\wedge K,\nn\\[2mm]
&~~~~~\text{where}~~~~~~~~~~~~~\Omega= E^1\wedge E^2\wedge E^3  ,~~~~~~~~~~~~ J=\frac{i}{2}(E^1\wedge \overline{E}^1+E^2\wedge \overline{E}^2+E^3\wedge \overline{E}^3),\nn
\end{aligned}
\eeq
 with the complex vielbein, $E^a$, orthogonal to $K$ (with $a=1,2,3$), and recalling $e^{\hat{A}}$ and $G_4$ span $M_7$. The Bianchi identities are then imposed separately. 

In order to express the GM background in this formalism, we utilise the Poincar\'e patch \eqref{eqn:PPatch}, leading to
\begin{align}\label{eqn:metriccondition}
ds^2 &=e^{2\hat{A}}ds^2(\text{Mink}_4) +ds^2(M_7),~~~~~~~~~~~~~~~e^{2\hat{A}}=4f_1e^{2\rho},\\
ds^2(M_7)&= \sum_{a=1}^{3}E^a \bar{E}^{\bar{a}}+K^2= f_1\bigg[4d\rho^2 +f_2 ds^2(S^2) +f_3 d\chi^2 + f_4(d\sigma^2 + d\eta^2) +f_5\big(d\beta + f_6 d\chi\big)^2\bigg],\nn
\end{align}
where we now need to derive the appropriate $K$ and complex vielbein $E^a$ which satisfies these conditions. To do this, we simply utilise the B$\ddot{\text{a}}$cklund transformation to derive the GM G-Structure forms from the LLM results reviewed in \eqref{eq:LLMniceframe} and \eqref{eq:LLMniceframeModified}, giving 
   \begin{align}\label{eqn:InitialGstructureForms}
   &~~~~~~~~~~~~~~~~~~~~~~~~~~~~~~~   \textbf{$\chi$ reduction frame}\nn\\
&K= \frac{\kappa\, e^{-2\rho}}{f_1}d(\cos\theta e^{2\rho}\dot{V}),~~~~~~~~~~~~~
E_1 =  \sqrt{\frac{f_1f_3}{f_6^2 +\frac{f_3}{f_5}}}  \bigg( d(V') -i\,d\beta\bigg),  \nn\\[2mm]
&E_2=\frac{e^{i\phi}}{f_1} \bigg[ \kappa \,e^{-2\rho}d(\sin \theta e^{2\rho}\dot{V})+i\, f_1^{\frac{3}{2}}f_2^{\frac{1}{2}}\sin \theta d\phi\bigg]=\frac{\kappa \, e^{-2\rho}}{f_1}d\Big(e^{2\rho}\dot{V} e^{i\phi} \sin\theta\Big),\\[2mm]
&E_3 =-e^{i\chi} \sqrt{f_1 f_5\Big(f_6^2+\frac{f_3}{f_5}\Big)} \,\Bigg[ d\rho - \frac{V''}{(\dot{V}')^2-\ddot{V}V''} d(\dot{V})+ i\,\bigg(d\chi +  \frac{f_6}{f_6^2 +\frac{f_3}{f_5}}  d\beta\bigg)\Bigg], \nn
\end{align}
which are in a frame corresponding to a dimensional reduction along $\chi$ - which can be seen directly from $E_3$ using the reduction formula \eqref{eqn:reductionformula}, 
where $\psi=\chi$,  $C_1= f_6\big(f_6^2 +\frac{f_3}{f_5}\big)^{-1}  d\beta$ and $e^{\frac{4}{3}\Phi_{IIA}}=f_1 f_5\big(f_6^2+\frac{f_3}{f_5}\big)$. 

As in the LLM solutions, we can now read off the $U(1)_R$ component of the $SU(2)_R \times U(1)_R$ R-symmetry directly from the overall phase of $\Omega$, namely 
\beq\label{eqn:origU(1)}
U(1)_R=\chi+\phi.
\eeq
This is a significant advantage of deriving the G-Structure description, and will allow us to keep track of supersymmetry under dimensional reduction and T-duality.  

Given that the background has three $U(1)$ directions $(\beta,\chi,\phi)$ one could dimensionally reduce along each direction in turn. However, as it stands, a dimensional reduction along either $\chi$ or $\phi$ would break the $U(1)_R$ component of the R-symmetry, and consequently break the supersymmetry. Hence, the only KK reduction which will preserve any supersymmetry is a dimensional reduction along $\beta$, which derives the $\mathcal{N}=2$ GM solution of Section \ref{sec:GMIIA}.

As it stands, the complex vielbeins given in \eqref{eqn:InitialGstructureForms} correspond to a $\chi$ reduction frame. In order to derive the G-Structure description of the $\mathcal{N}=2$ Type IIA solution, we must first perform an appropriate frame rotation to a new set of vielbeins which corresponds to a $\beta$ reduction frame. We discuss this in a more general fashion following the next section.

As we will now discuss, further possibilities emerge by performing an $SL(3,\mathds{R})$ transformation prior to reduction, including additional $\mathcal{N}=1$ Type IIA and Type IIB solutions (however no other $\mathcal{N}=2$ solutions can be derived by this approach). 

\section{$SL(3,\mathds{R})$ Transformation}\label{sec:SL(3,R)}
The GM class of solutions can be dimensionally reduced to Type IIA in a more general manner to reduction of Section \ref{sec:GMIIA}, by first performing the following $SL(3,\mathds{R})$ transformation amongst the three $U(1)$ directions of the background $(\beta,~\chi,~\phi)$
, 
   \begin{equation}\label{eqn:S2breakingdefns}
 \begin{gathered}
d\beta \rightarrow a\, d\chi+b\, d\beta +c\, d\phi,~~~~~~~~~~~~~d\chi \rightarrow p\, d\chi +q\, d\beta + m \,d\phi,~~~~~~~~~~~~~d\phi \rightarrow s\, d\chi +v\,d\beta +u\,d\phi,\\
\begin{vmatrix}
~p&q&m~\\
~a&b&c~\\
~s&v&u~\\
\end{vmatrix}
=p(bu-vc)-q(au-sc)+m(av-sb)=1.
\end{gathered}
\end{equation}
We restrict the $GL(3,\mathds{R})$ transformation to $SL(3,\mathds{R})$ (with determinant equal to one) in order to preserve the scale of the background. The nine transformation parameters can then be reduced, without loss of generality, to three free parameters (corresponding to the three $U(1)$ directions being mixed). The choice of these remaining parameters provide options when reducing to Type IIA. However, in order to preserve the periodicity of $(\beta,~\chi,~\phi)$, these parameters must be integer valued - hence describing an $SL(3,\mathds{Z})$ transformation. For our purposes, we fix $(p,b,u)=1$ by trivially absorbing them into the definitions of $(\chi,\beta,\phi)$, respectively. This avoids re-defining the three $U(1)$ directions amongst themselves, and immediately eliminates three of the nine parameters. There can be some utility in choosing different values for these parameters however, which is briefly discussed in sections \ref{sec:naive} and \ref{sec:followingbeta} - hence, it can be beneficial to perform the analysis in full generality first.

The $U(1)$ component of the $SU(2)_R \times U(1)_R$ R-symmetry now generalises to 
\beq\label{eqn:U(1)}
\begin{aligned}
U(1)_R&=\chi+\phi \rightarrow (p+s)\chi +(q+v)\beta + (m+u)\phi,
\end{aligned}
\eeq
which, after fixing $(p,b,u)=1$, becomes
\beq\label{eqn:U(1)lessgen}
U(1)_R= (1+s)\chi +(q+v)\beta + (m+1)\phi,
\eeq
opening up additional possibilities to preserve supersymmetry under reduction. 

In the case of the $\beta$ reduction, one can now preserve the $U(1)_R$ component by fixing $v=-q$. It is clear however, from \eqref{eqn:S2parametrization} and \eqref{eqn:S2breakingdefns}, that the $S^2$ (and hence the $SU(2)_R$ component of the R-symmetry) will be broken under reduction for $v\neq0$. Hence, when $v=-q=0$, one re-derives the $\mathcal{N}=2$ IIA solution with a $U(1)_R\times SU(2)_R$ R-symmetry, when $v=-q\neq0$, one derives a family of $\mathcal{N}=1$ IIA solutions with a $U(1)_R$ R-symmetry, and for all other values of $(v,q)$, one derives $\mathcal{N}=0$ IIA solutions, with broken supersymmetry.

In the case of the $\chi$ and $\phi$ reductions, in order to preserve the $U(1)_R$ component given in \eqref{eqn:U(1)lessgen}, things are a little less general. This is a consequence of the $U(1)_R$ being $\chi+\phi$ prior to transformation. Nevertheless, the $SL(3,\mathds{R})$ transformation has opened up the possibility to preserve the $U(1)_R$ under $\chi$ and $\phi$ reductions, by fixing $s=-1$ and $m=-1$, respectively. Once again, from \eqref{eqn:S2parametrization} and \eqref{eqn:S2breakingdefns}, it is clear that these two conditions in fact break the $S^2$ (and the $SU(2)_R$ component of the R-symmetry) in each case. That is, in the case of a $\chi$ and $\phi$ reduction, the condition to preserve the $U(1)_R$ component of the R-symmetry necessarily breaks the $SU(2)_R$ component, so the most supersymmetry one can preserve under these reductions is $\mathcal{N}=1$. 

The $U(1)_R$ in \eqref{eqn:U(1)lessgen} also provides vital insight into the preservation of supersymmetry under a subsequent abelian T-duality to Type IIB. For example, in the $\beta$ reduction case, fixing $v=-q$ will preserve the $U(1)_R$ component under dimensional reduction - with either $U(1)_R =  (1+s)\chi + \phi$ or $U(1)_R =  \chi + (m+1)\phi$ (depending on the third free parameter we choose). For instance, in this first example, we can preserve the $U(1)_R$ R-symmetry under an ATD along $\chi$ by fixing $s=-1$. As in the $\chi$ reduction case, fixing $s=-1$ necessarily breaks the $S^2$ (and the $SU(2)_R$ component), so we would derive an $\mathcal{N}=1$ IIB solution with a $U(1)_R=\phi$ R-symmetry. Analogous arguments indeed hold for the second example, along with the $\chi$ and $\phi$ reductions. This means that the most supersymmetry we can preserve for a IIB solution is $\mathcal{N}=1$, as the $SU(2)_R$ component of the R-symmetry is necessarily broken either by the dimensional reduction to IIA or by the abelian T-duality to IIB.

When deriving the these IIA and IIB daughter solutions, we will require the new form for $A_3$ \eqref{eqn:GM}, which when expanded in full generality, reads
 \begin{align}
A_3 &=\sin\theta \,\Big((up-sm)f_7+(ua-sc)f_8 \Big)d\chi  \wedge d\theta \wedge d\phi\\
 &~~~+\sin\theta \,\bigg(\Big((v p -sq )f_7+(v a -sb)f_8 \Big)d\chi  +\Big((v m - uq)f_7 +(v c-ub)f_8 \Big)d\phi \bigg) \wedge d\theta \wedge d\beta.\nn
 \end{align}

We will investigate all these backgrounds in further detail in the next chapter, but in order to derive the IIA G-structure descriptions for each of the SUSY preserved (and SUSY broken) backgrounds, we will first need to rotate our complex vielbeins in \eqref{eqn:InitialGstructureForms} to the appropriate reduction frames in each case. This is the focus of the next section.


\section{Frame rotation of G-Structure forms}\label{sec:framerotation}
In Section \ref{sec:G-Structure description of GM}, the G-Structure description for the GM class of solutions was presented, in a form corresponding to the $\chi$ reduction frame. We now need to rotate these forms to a reduction frame which accounts for the $SL(3,\mathds{R})$ transformations being performed. It proves useful to do this in the most general fashion, in terms of all nine transformation parameters, allowing one to easily restrict to the particular reduction frames of interest by simply fixing these parameters appropriately.

We begin by performing the $SL(3,\mathds{R})$ transformations \eqref{eqn:S2breakingdefns} on the GM metric \eqref{eqn:GM} before re-writing it in the following manner (for a dimensional reduction along $\phi_3$)
   \begin{align}\label{eqn:genGforms}
        ds_{11}^2&=f_1\bigg[4ds^2(\text{AdS}_5)+f_2d\theta^2+f_4(d\sigma^2+d\eta^2)\bigg]+f_1^2e^{-\frac{4}{3}\Phi_{IIA}}ds^2_2+ e^{\frac{4}{3}\Phi_{IIA}}(d\phi_3+ C_{1,\phi_1}d\phi_1+C_{1,\phi_2}d\phi_2)^2,\nn\\[2mm]
    ds^2_2 &=  h_{\phi_1}d\phi_1^2 + h_{\phi_2}d\phi_2^2 + h_{\phi_1\phi_2}d\phi_1 d\phi_2 = h_{\phi_1} \bigg(d\phi_1 + \frac{1}{2}\frac{h_{\phi_1\phi_2} }{h_{\phi_1} }d\phi_2\bigg)^2 + \bigg( h_{\phi_2} - \frac{1}{4} \frac{h_{\phi_1\phi_2}^2}{h_{\phi_1} }\bigg)d\phi_2^2,\nn\\[2mm]
    &C_{1,\phi_i}= C_{1,\phi_i}(\eta,\sigma,\theta),~~~~h_{\phi_i}= h_{\phi_i}(\eta,\sigma,\theta),~~~~h_{\phi_1\phi_2}= h_{\phi_1\phi_2}(\eta,\sigma,\theta),
    \end{align}
   with $(\phi_1,\phi_2,\phi_3)$ representing any arrangement of $(\beta,\chi,\phi)$ as required. This then derives the corresponding $(h_{\phi_1},h_{\phi_2},h_{\phi_1\phi_2},C_{1,\phi_1},C_{1,\phi_2})$ in each case. The G-Structure forms for the GM background then read in general
  \begin{align}\label{eqn:Gstructuregenforms}
K&= \frac{\kappa\, e^{-2\rho}}{f_1}d(\cos\theta e^{2\rho}\dot{V}),\nn\\[2mm]
E_1 &= f_1 e^{-\frac{2}{3}\Phi}\sqrt{h_1}  \bigg[b_1d\sigma +b_2d\eta + b_3d\rho +b_4d\theta -i  \bigg(d\phi_1 + \frac{1}{2}\frac{h_{\phi_1\phi_2}}{h_{\phi_1}}d\phi_2\bigg)\bigg],\nn\\[2mm]
E_2&= e^{i\bar{\phi}} \bigg[b_5d\sigma +b_6d\eta + b_7d\rho +b_8d\theta +i\,  f_1 e^{-\frac{2}{3}\Phi}\sqrt{ h_{\phi_2}- \frac{1}{4} \frac{h_{\phi_1\phi_2}^2}{h_{\phi_1}}}  d\phi_2\bigg],\nn\\[2mm]
E_3 &=-\sqrt{f_1}e^{i\bar{\chi}} \,\Bigg[b_9d\sigma +b_{10}d\eta + b_{11}d\rho + b_{12}d\theta  + i \frac{e^{\frac{2}{3}\Phi}}{\sqrt{f_1}}\big(d\phi_3 + C_{1,\phi_1}d\phi_1+C_{1,\phi_2}d\phi_2\big)\Bigg],\nn\\[2mm]
 &b_i=b_i(\eta,\sigma,\theta),~~~~~~~~~\bar{\phi}=  s\,\chi +v\,\beta +u\,\phi,~~~~~~~~~~~~~~\bar{\chi}= p\, \chi +q\, \beta + m \,\phi,
\end{align}
 where $(b_1,..,~b_{12}) $ are functions which must be derived.
 
 We will now outline a method to derive these twelve functions for each reduction frame. This approach is motivated by observing in \eqref{eqn:genGforms} that $K$ remains untouched under the rotation of frames, as it is independent of $(\beta,\chi,\phi)$. Hence, in order to satisfy all four G-structure conditions in \eqref{eqn:Geqns}, there is a requirement on $J$ to remain intact under a frame rotation. By performing the $SL(3,\mathds{R})$ transformation \eqref{eqn:S2breakingdefns} on the original forms for the $\chi$ reduction frame \eqref{eqn:InitialGstructureForms}, and enforcing that this $J$ matches the one defined by  \eqref{eqn:Gstructuregenforms}, expressions for $(b_1,..,~b_{12}) $ are easily derived using Mathematica. This procedure must be repeated for each arrangement of $(\phi_1,\phi_2,\phi_3)$, giving a separate set of functions for each. Hence, there are two sets of equivalent vielbeins for each $U(1)$ being reduced - which should both define the same $(J,\Omega)$ and metric.
 These sets of functions, in their most general form, then depend on the nine $SL(3,\mathds{R})$ parameters and the derivatives of $V(\eta,\sigma)$. 
 
This framework is general enough to easily derive the G-Structure forms for all reduction frames of interest, including those leading to $\mathcal{N}=1$ Type IIA and Type IIB solutions, by simply picking the transformation parameters appropriately. We can then verify the preservation of supersymmetry in the daughter solutions by deriving the IIA G-Structure results directly from the $d=11$ description, as outlined in Section \ref{sec:SU(2)Gs}. For the IIB solutions, the G-Structure description will be derived in turn from the IIA description, following an appropriate T-duality of the results. If the corresponding G-Structure conditions are satisfied in each case, supersymmetry is of course preserved. In the cases where the conditions are broken, interesting insight can be gained into the breaking of supersymmetry, as we outlined in Section \ref{sec:SUSYbreakingGs}. Hence, the G structure analysis can be of interest even when supersymmetry is broken! 
 
  As we go through the remaining chapters, we will require multiple sets of the complex vielbeins, corresponding to the different solutions being investigated. These can be quite cumbersome, including some non-zero transformation parameters, so we will refrain from presenting them until they are required. We will however move onto to discuss the non-deformed reduction frames, 
with no  $SL(3,\mathds{R})$ transformations taking place. In these cases, only the $\beta$ reduction will preserve any SUSY, deriving the GM solution of Section \ref{sec:GMIIA}. 



\subsection{The non-deformed reduction frames}\label{sec:naive}


To derive the G-structure forms corresponding to the $\beta$ reduction frame, we could choose $(\phi_1,\phi_2,\phi_3)=(\chi,\phi,\beta)$. One must then specify the relevant values of the nine $SL(3,\mathds{R})$ parameters. In the case of a simple frame rotation, with no $SL(3,\mathds{R})$ transformation, we simply fix $(p,b,u)=1$ with all other parameters set to zero. The resulting set of complex vielbeins then correspond to the $\beta$ reduction frame, deriving the IIA $\mathcal{N}=2$ solution, 
   \begin{align}\label{eqn:GstructureForms}
&~~~~~~~~~~~~~~~~~~~~~~~~~~~~~~~   \textbf{$\beta$ reduction frame}\nn\\
&K= \frac{\kappa\, e^{-2\rho}}{f_1}d(\cos\theta e^{2\rho}\dot{V}),~~~~~~~~~~~~
E_1 =  -\sqrt{f_1f_3}\bigg(\frac{1}{\sigma} d\sigma+ d\rho   +i d\chi\bigg),\nn\\[2mm]
&E_2= \frac{e^{i\phi}}{f_1} \bigg[ \kappa \,e^{-2\rho}d(\sin \theta e^{2\rho}\dot{V})+i\, f_1^{\frac{3}{2}}f_2^{\frac{1}{2}}\sin \theta d\phi\bigg]=\frac{\kappa \, e^{-2\rho}}{f_1}d\Big(e^{2\rho}\dot{V} e^{i\phi} \sin\theta\Big),\\[2mm]
&E_3 =-e^{i\chi}\sqrt{f_1f_5} \,\Bigg[-\frac{1}{4}f_3 \frac{\dot{V}'}{\sigma} d\sigma -V''d\eta +f_6 d\rho  + i \Big(d\beta +f_6d\chi\Big)\Bigg],\nn
\end{align}
where, using this new form of $E_3$ along with \eqref{eqn:reductionformula}, we now have $\psi=\beta$,  $C_1=f_6 d\chi$ and $e^{\frac{4}{3}\Phi_{IIA}}=f_1f_5$ (which matches the background of Section \ref{sec:GMIIA}). Notice that $K$ and $E_2$ are the same for both \eqref{eqn:InitialGstructureForms} and \eqref{eqn:GstructureForms}. This is because they are independent of $(\beta,\chi)$ and so remain untouched by the frame rotation. In addition, from the conditions given in \eqref{eqn:OmegaJ} and \eqref{eqn:metriccondition}, it is easy to verify that the following necessary relations indeed hold
\beq\label{eqn:E1E3conditions}
\begin{aligned}
E_1\wedge \bar{E_1}+E_3\wedge \bar{E_3}&=\tilde{E}_1\wedge \bar{\tilde{E}}_1+\tilde{E}_3\wedge \bar{\tilde{E}}_3,\\
E_1\wedge E_3&=\tilde{E}_1\wedge \tilde{E}_3,\\
E_1\bar{E_1}+E_3\bar{E_3}&=\tilde{E}_1\bar{\tilde{E}}_1+\tilde{E}_3\bar{\tilde{E}}_3,
\end{aligned}
\eeq
where $E_i$ and $\tilde{E}_i$ are the vielbeins given in \eqref{eqn:InitialGstructureForms} and \eqref{eqn:GstructureForms}, respectively.

For a rotation to the $\phi$ reduction frame, $K$ is of course independent of $\phi$, but the $E_2$ (in \eqref{eqn:InitialGstructureForms} and \eqref{eqn:GstructureForms}) does indeed depend on $\phi$. However, as it turns out, this $E_2$ has the appropriate form for the $E_3$ of the $\phi$ reduction (with $C_1=0$). Hence, by simply relabelling the vielbeins, our $E_2$ now becomes the $E_3$ of the $\phi$ reduction frame. The $(E_1,E_2)$ are then simply the $(E_1,E_3)$ from either \eqref{eqn:InitialGstructureForms} or \eqref{eqn:GstructureForms} (satisfying the conditions in \eqref{eqn:E1E3conditions}). In this case, we then find $\psi=\phi$,  $C_1=0$ and $e^{\frac{4}{3}\Phi_{IIA}}=f_1f_2 \sin^2\theta$. Alternatively, one can re-derive these vielbeins directly from \eqref{eqn:Gstructuregenforms} (with $(p,b,u)=1$ and all other parameters set to zero). Picking $(\phi_1,\phi_2,\phi_3)=(\chi,\beta,\phi)$ and $(\phi_1,\phi_2,\phi_3)=(\beta,\chi,\phi)$ re-derives \eqref{eqn:InitialGstructureForms} and \eqref{eqn:GstructureForms}, respectively (up to re-labelling). From \eqref{eqn:E1E3conditions}, both sets of vielbeins are equivalent - defining the same metric and forms, $(J,\Omega)$ .

In the case of the $\chi$ reduction frame, to  re-derive \eqref{eqn:InitialGstructureForms}, we simply pick $(\phi_1,\phi_2,\phi_3)=(\beta,\phi,\chi)$ with $(p,b,u)=1$ and all other parameters zero. Alternatively, one can in fact derive the $\chi$ reduction forms (up to $E_2\rightarrow -E_2$) from the general $\beta$ reduction results by utilising the $SL(3,\mathds{R})$ transformation appropriately, fixing $(a,q)=1,u=-1$ with all others set to zero. From the determinant in \eqref{eqn:S2breakingdefns}, this now requires $qau=-1$. Hence,  in this case, we have made the transformations $\beta\rightarrow\chi,~\chi\rightarrow\beta,~\phi\rightarrow-\phi$. This is an additional advantage of performing the frame rotations in full generality, in terms of all nine parameters.  

Before moving onto study the numerous IIA and IIB solutions derived following an $SL(3,\mathds{R})$ transformation, it will prove instructive to first investigate the G-structure description for the daughter IIA solutions using the set of non-deformed reduction vielbeins just presented. As we have already discussed, in the $\beta$ reduction case of \eqref{eqn:GstructureForms}, one derives the $\mathcal{N}=2$ Type IIA GM solution. In the case of the $\chi$ and $\phi$ reduction frames, we instead derive $\mathcal{N}=0$ solutions - we will refrain from presenting these result here, but the calculation follows in an identical manner to the following sub-section. 

\section{Returning to the Type IIA $\mathcal{N}=2$ solution}\label{sec:N=2IIAagain}
Armed with the knowledge of the previous section, where the G-structure description for the GM class in $d=11$ was derived, we can perform some preliminary investigations into the non-deformed dimensional reductions (prior to an $SL(3,\mathds{R})$ transformation). 
Recall the GM solution from Section \ref{sec:GMIIA}
\beq
\begin{aligned}
ds^2_{10}&= f_1^{\frac{3}{2}} f_5^{\frac{1}{2}}\bigg[4ds^2(\text{AdS}_5)+f_2ds^2(\text{S}^2)+f_4(d\sigma^2+d\eta^2)+f_3 d\chi^2\bigg],\\[2mm]
e^{\frac{4}{3}\Phi}&=  f_1 f_5,~~~~  H_3 = df_8\wedge \text{vol}(\text{S}^2),~~~~C_1=  f_6d\chi,~~~~ C_3=f_7 d\chi\wedge\text{vol}(\text{S}^2).\nn 
\end{aligned}
\eeq
We begin by deriving the G-structure description of this IIA solution, before performing investigations into the boundary of this background. This analysis will lay the foundations for later discussions into the more general multi-parameter families of solutions, following an $SL(3,\mathds{R})$ transformation. 
\subsection{G-Structure description}\label{sec:IIAGstructureN=2}
Using the $d=11$ G-structure forms defined by \eqref{eqn:OmegaJ} and \eqref{eqn:GstructureForms}, along with the reduction formula \eqref{eqn:11Dto10Dforms}, it is straightforward to derive the following $SU(2)$ structure forms for this background
    \begin{align}\label{eqn:uandv}
&v=  \kappa e^{-2\rho}f_5^{\frac{1}{4}}f_1^{-\frac{3}{4}} d\left(e^{2\rho} \dot{V}\cos\theta\right),~~~~~~u=  (f_1f_5)^{\frac{3}{4}}\left(\frac{f_3 \dot{V}'}{4 \sigma}d\sigma +V''d\eta-f_6d\rho \right),\nn\\[2mm]
&\omega=  -2\kappa^2  f_1^{-\frac{3}{2}}e^{-3\rho} d\Big(e^{2\rho}\dot{V} e^{i\phi}\sin\theta\, d(e^{i\chi}e^{\rho}\sigma)\Big) ,\nn\\[2mm]
&j= \frac{f_1^{\frac{3}{2}}f_5^{\frac{1}{2}} f_3}{\sigma} e^{-\rho}d\left(e^{\rho}\sigma\right)\wedge d\chi +\kappa\frac{\sigma f_2 f_3^{-\frac{1}{2}}e^{-4\rho}}{ \dot{V}^2}d(e^{4\rho}\sin^2\theta \dot{V}^2)\wedge d\phi  ,
\end{align}
which are charged under $SU(2)_R\times U(1)_R$. In addition, we can construct the complex vielbeins from \eqref{eqn:IIAviels}, \eqref{eqn:GstructureForms} and $e^{\frac{4}{3}\Phi}=  f_1 f_5$, leading to
\begin{align}
\hat{E}^1 &= e^{\frac{1}{3}\Phi} E^1 = -   f_1^{\frac{3}{4}} f_5^{\frac{1}{4}} f_3^{\frac{1}{2}}\bigg(\frac{1}{\sigma} d\sigma+ d\rho   +i d\chi\bigg) ,\nn\\[2mm]
\hat{E}^2 &=e^{\frac{1}{3}\Phi} e^{-i\hat{\theta}_+} E^2 =- \kappa \, f_1^{-\frac{3}{4}} f_5^{\frac{1}{4}} e^{-2\rho}e^{i\chi}d\Big(e^{2\rho}\dot{V} e^{i\phi} \sin\theta\Big),
\end{align}
which re-derive the above $(j,\omega)$ via \eqref{eqn:IIAomegaj1}, and define the metric in the following manner
  \begin{align} 
&ds_{10}^2=e^{2A}ds^2(\text{Mink}_4)+
\hat{E}^1\bar{\hat{E}}^{1}+ \hat{E}^2 \bar{\hat{E}}^{2}+u^2+v^2,~~~~~~~~
e^{2A}=e^{2\hat{A}+\frac{2}{3}\Phi}=4  f_1^{\frac{3}{2}} f_5^{\frac{1}{2}} e^{2\rho} .\nn
\end{align}
By constructing the pure spinors \eqref{eqn:Psi}, one can now verify the SUSY conditions \eqref{eqn:IIAGconditions} are indeed satisfied. 
In addition, using \eqref{eqn:PsiConvenient}, \eqref{eq:usefulpotentials}  and \eqref{eqn:uandv}, one can easily derive the higher form fluxes directly from the G-structure description:
\beq 
\begin{aligned}
C_5=
e^{4A-\Phi}\text{vol}(\text{Mink}_4) \wedge u,~~~~
C_7=
e^{4A-\Phi}\text{vol}(\text{Mink}_4) \wedge v\wedge j,~~~
C_9=
-\frac{1}{2}e^{4A-\Phi}\text{vol}(\text{Mink}_4) \wedge u\wedge j\wedge j,
\end{aligned}
\eeq
which we write explicitly
\begin{align}\label{eqn:C5andC7}
C_5
&=-2^5\kappa^2 \frac{\dot{V}\dot{V}'}{V''}\text{vol}(\text{AdS}_5)+2^4\kappa^2 e^{4\rho} \text{vol}(\text{Mink}_4)\wedge\bigg(\sigma \dot{V}'d\sigma +(2\dot{V}-\ddot{V})d\eta\bigg) ,\nn\\[2mm]
C_7
&= 2^6\kappa^3 \sigma \bigg( \text{vol}(\text{AdS}_5)\wedge    \Big(3\dot{V} \cos\theta\, d\sigma - d(\sigma \dot{V} \cos\theta) \Big) + e^{4\rho} \text{vol}(\text{Mink}_4)\wedge d(\dot{V}\cos\theta) \wedge d\sigma  \bigg) \wedge d\chi \nn\\[2mm]
&~~~~~ -2^4 f_1^3f_5f_7 \bigg( \text{vol}(\text{AdS}_5) +\frac{e^{4\rho}}{2\dot{V}}d(\dot{V})\wedge \text{vol}(\text{Mink}_4)\bigg)\wedge \text{vol}(S^2) ,\nn\\[2mm]
C_9&=-2^7\kappa^4   \sigma (2\dot{V}-\ddot{V}) \bigg(\text{vol}(\text{AdS}_5)\wedge \Big( \sin^2\theta\, d\eta \wedge d\sigma\wedge d\phi -\frac{\sigma \dot{V}}{\tilde{\Delta}}  \cos\theta \,d(V')\wedge \text{vol}(S^2)\Big)\nn\\[2mm]
&~~~~~~~+e^{4\rho}\frac{\dot{V}V''}{\tilde{\Delta}} \cos\theta \,\text{vol}(\text{Mink}_4)\wedge \text{vol}(S^2)\wedge d\eta\wedge d\sigma \bigg)\wedge d\chi.
\end{align}

\subsubsection{Calibrations}
The type IIA GM solution has stacks of D6 branes along the $\sigma=0$ boundary, located at the various kinks of the rank function (where $\dot{V}$ is discontinuous). These branes are charged under $C_1$, so descend from pure geometry in $d=11$. Hence, one expects them to be supersymmetric - which we will now confirm.

For branes extended in $\text{Mink}_4$ with zero world-volume flux ($\tilde{f_2}=0$), we have from \eqref{eq:Dpsusy} and \eqref{eq:Dpsusy2} the following calibration forms
\begin{align}\label{eqn:cals}
   \text{vol(Mink}_4\text{)}\wedge \Psi^{(\text{cal})}_{\text{D4}}&=  8 e^{4A-\Phi}\text{vol(Mink}_4\text{)}\wedge  \text{Im}\Psi_{1}=C_5 ,\nn\\[2mm]
 \text{vol(Mink}_4\text{)}\wedge \Psi^{(\text{cal})}_{\text{D6}} &= 8 e^{4A-\Phi}\text{vol(Mink}_4\text{)}\wedge ( \text{Im}\Psi_{3}-B_2^k\wedge \text{Im}\Psi_{1} ) = C_7-B_2^k\wedge C_5  ,\nn\\[2mm]
  \text{vol(Mink}_4\text{)}\wedge \Psi^{(\text{cal})}_{\text{D8}}&= 8 e^{4A-\Phi}\text{vol(Mink}_4\text{)}\wedge ( \text{Im}\Psi_{5}-B_2^k\wedge \text{Im}\Psi_{3} ) =  C_9-B_2^k\wedge C_7 ,
\end{align}
where
\begin{align}
\Psi^{(\text{cal})}_{\text{D4}}&=e^{4A-\Phi}u,~~~~~\Psi^{(\text{cal})}_{\text{D6}}&=e^{4A-\Phi}(v\wedge j-B_2^k\wedge u),~~~~~\Psi^{(\text{cal})}_{\text{D8}}&=e^{4A-\Phi}\Big(-\frac{1}{2}u\wedge j\wedge j-B_2^k\wedge v\wedge j \Big),\nn
\end{align}
with the SU(2)-structure forms defined in \eqref{eqn:uandv} and $B_2$ given in \eqref{eq:neq2B2def}, with $B_2\wedge B_2=0$. 
In the case of the D6 branes, in order for them to preserve AdS$_5\times$S$^2$ symmetry, we should take $\Sigma_{p-3}=(\rho,~\text{S}^2)$ at $\sigma=0$. We note from above
\begin{align}
\left(C_7-B^k_2\wedge C_5\right)
&=\frac{1}{2}(4\kappa)^3\bigg[\frac{2\dot{V}^2}{V''}\text{vol}(\text{AdS}_5)\wedge \text{vol}(\text{S}^2)- e^{4\rho}\dot{V}\text{vol}(\text{Mink}_4)\wedge \text{vol}(\text{S}^2)\wedge (\sigma d\sigma)\nn\\[2mm]
+2 \sigma^4&\text{vol}(\text{AdS}_5)\wedge d\chi\wedge d(\sigma^{-2}\dot{V}\cos\theta)-2 \sigma e^{4\rho}\text{vol}(\text{Mink}_4)\wedge d\chi\wedge d\sigma\wedge d(\dot{V}\cos\theta)\nn\\[2mm]
-(\eta-k)&\bigg(\frac{4\sigma^2 f_6}{f_3}\text{vol}(\text{AdS}_5)-e^{4\rho}\text{vol}(\text{Mink}_4)\wedge \left(\dot{V}'d\sigma-\sigma^2\partial_{\sigma}(\sigma^2\dot{V})d\eta\right)\bigg)\wedge \text{vol}(\text{S}^2) \bigg],\nn
\end{align}
with only the first term relevant for the D6 branes, leading to
\begin{adjustwidth}{-1.25cm}{}
\vspace{-0.6cm}
\begin{align}\label{eqn:caltake1}
\Psi^{(\text{cal})}_{\text{D6}}\bigg\lvert_{(\rho,\text{S}^2)}&=(4\kappa)^3 e^{4\rho}\frac{\dot{V}^2}{V''}d\rho\wedge \text{vol}(\text{S}^2),\\[2mm]
\hspace{-3mm}e^{4A-\Phi}\sqrt{\det(g+B_2)}d^{p-3}w\bigg\lvert_{(\rho,\text{S}^2)}&=(4\kappa)^3 e^{4\rho}\frac{\dot{V}^2}{V''}\sqrt{1+\frac{\sigma^2 V''}{2 \dot{V}}}\sqrt{1-2 (\eta-k)\frac{\dot{V}'}{\dot{V}}-4 \kappa(\eta-k)^2\frac{V''}{f_7}}d\rho\wedge \text{vol}(\text{S}^2),\nn
\end{align}
\end{adjustwidth}
which satisfies \eqref{eq:Dpsusy} at $(\sigma=0,\eta=k)$. Recall that $k$ originally appeared as an integration constant in $B_2$, and can change due to a large gauge transformation (with $\frac{1}{(2\pi)^2}\int_{\text{S}^2} B_2$ shifting by an integer). Hence, we find that the D6 branes are restricted to lie along $\sigma=0$, and at integer values of $\eta$, by the unbroken supersymmetry - with $k$ integer for consistency.

Let us consider the supersymmetry of other probe branes which extend along the AdS$_5$. Performing the computation tells us that this is not possible for the D4 branes.
 For the D8 brane, every term in $\Psi^{(\text{cal})}_{\text{D8}}$ 
  is charged under SU(2)$_R$, so D8s cannot be added without breaking supersymmetry. We do find however, from the first term in $C_9$ \eqref{eqn:C5andC7} and $\Sigma_{5}=(\rho,\phi,\sigma,\eta,\chi)$, that 
\begin{align}\label{eqn:cals1}
e^{4A-\Phi}\sqrt{\det(g+B_2)}\bigg\lvert_{\Sigma_{5}}d^5\xi&=\frac{1}{2}(4\kappa)^4 e^{4\rho}\sigma(2 \dot{V}-\ddot{V})\sin\theta d\rho\wedge d\phi\wedge d\sigma \wedge d\eta \wedge d\chi,\\[2mm]
\Psi^{(\text{cal})}_{\text{D8}}\bigg\lvert_{\Sigma_{5}}&=\frac{1}{2}(4\kappa)^4 e^{4\rho}\sigma(2 \dot{V}-\ddot{V})\sin^2\theta d\rho\wedge d\phi\wedge d\sigma \wedge d\eta \wedge d\chi,\nn
\end{align}
allowing half BPS D8 branes to be placed at $\sin\theta=1$.

\subsection{Investigations at the boundary}\label{sec:BoundaryNeqtwo}

The laplace equation given in \eqref{eqn:laplace} is an elliptical PDE, for which the extrema of the solutions lie only at the boundaries. Hence, we know the solutions are regular in the interior of the Riemann surface, which as we recall, is bounded by $\sigma\in [0,\infty),\eta\in [0,P]$. We therefore need to conduct investigations into the behaviour along these boundaries. The behaviour of the warp factors \eqref{eqn:fs} along each boundary are presented in Appendix \ref{sec:fs}, which are very useful in the following analysis.

  \subsubsection{The $\sigma\rightarrow \infty$ boundary}
  Let us first consider the boundary at $\sigma\rightarrow \infty$ (depicted in red in the following diagram). In this limit, as $x\to \infty$, the Bessel function becomes $K_0(x)\to\sqrt{\pi} (2x)^{-\frac{1}{2}}e^{-x}$. Hence, the leading term in \eqref{eqn:potential} is simply the $n=1$ contribution,  
 \begin{figure}[H]
\centering  
\subfigure
{
\centering
  \begin{minipage}{0.6\textwidth}
\beq\label{eqn:Vsigmainfty}
V=-{\cal R}_1 e^{-\frac{\pi}{P}\sigma}\sqrt{\frac{P}{2\sigma}}\sin\left( \frac{2\pi}{P} \eta\right) +...~~~.
\eeq
  \end{minipage}
     \begin{minipage}{.5\textwidth}
    \centering
\begin{tikzpicture}[scale=1, every node/.style={scale=1}]

\draw[lightgray,line width=0.5mm] (1,0) -- (1.5,0.6);
\draw[lightgray,line width=0.5mm] (1.5,0.6) -- (2,1);
\draw[lightgray,line width=0.5mm] (2,1) -- (2.5,1.2);
\draw[lightgray,line width=0.5mm] (2.5,1.2)-- (3,1.1);
\draw[lightgray,line width=0.5mm]  (3,1.1)-- (3.5,0.7);
\draw[lightgray,line width=0.5mm] (3.5,0.7)-- (4,0);

\draw[lightgray,dashed,line width=0.3mm] (1.5,0)-- (1.5,0.6);
\draw[lightgray,dashed,line width=0.3mm] (2,0)-- (2,1);
\draw[lightgray,dashed,line width=0.3mm] (2.5,0)--  (2.5,1.2);
\draw[lightgray,dashed,line width=0.3mm] (3,0)-- (3,1.1);
\draw[lightgray,dashed,line width=0.3mm] (3.5,0)--  (3.5,0.7);

\draw[-stealth] (4.45,1.2) -- (4.45,1.7) node[above] {$\sigma$};
\draw[-stealth]  (4.45,1.2) -- (4.95,1.2)  node[right] {$\eta$};


\draw[red,line width=1mm] (1,2.05)--(4,2.05);


\draw (1,0) -- (1,2);
\draw (1.01,0) -- (1.01,2);
\draw (1.02,0) -- (1.02,2);
\draw (1,0) node[below] {$0$};
\draw (0.65,2.2) node[below] {$\infty$};
\draw (4,0) node[below] {$P$};
\draw (4,0) -- (4,2);
\draw (3.99,0) -- (3.99,2);
\draw (3.98,0) -- (3.98,2);

\draw (1,0) -- (4,0);
\draw (1,0.01) -- (4,0.01);
\draw (1,-0.01) -- (4,-0.01);
\draw (1,0.02) -- (4,0.02);
\draw (1,-0.02) -- (4,-0.02);
%

\draw (2.5,-0.75) node[above] {$\mathcal{R}(\eta)$};

   \begin{scope}[yshift=2cm]
\draw (0,-1.9) -- (0.6,-1.9);
\draw (0.1,-1.95) -- (0.5,-1.95);
\draw (0.2,-2) -- (0.4,-2);
\draw (0.3,-1.9).. controls (0.3,-1.25)  .. (1,-1.25);

\draw (4.5,-1.9) -- (5.1,-1.9);
\draw (4.6,-1.95) -- (5,-1.95);
\draw (4.7,-2) -- (4.9,-2);
\draw (4.8,-1.9).. controls (4.8,-1.25)  .. (4,-1.25);
\end{scope}
\end{tikzpicture}
  \end{minipage} 
}
  \label{fig:sigmainfinityboundary}
\end{figure}
$~$\\[-1.6cm]
We then find, to leading order, the solution reduces to
\begin{align}\label{eqn:sigmainftymetric}
ds^2&=\kappa \bigg[4\sigma\bigg(ds^2(\text{AdS}_5)+d\chi^2\bigg)+\frac{2P}{\pi}\bigg(d\left(\frac{\pi}{P}\sigma\right)^2+ d\left(\frac{\pi}{P}\eta\right)^2+ \sin^2\left(\frac{\pi}{P}\eta\right)ds^2(\text{S}^2)\bigg)\bigg],\nn\\[2mm]
e^{-\Phi}&=\frac{{\cal R}_1\pi^2}{2 P^{\frac{3}{2}}\sqrt{\kappa}}e^{-\frac{\pi}{P}\sigma}\left(\frac{\pi}{P}\sigma\right)^{-\frac{1}{2}},~~~~H_3=-\frac{4\kappa P}{\pi}\sin^2\left(\frac{\pi}{P}\eta\right) d\left(\frac{\pi}{P}\eta\right)\wedge \text{vol}(\text{S}^2),
\end{align}
where we can observe $(\frac{\pi}{P}\eta,S^2)$ span a round (unit radius) 3-sphere, and we note the RR fluxes have vanished at leading order.

This is very similar to the metric found in section  3.9.2 of \cite{DHoker:2017mds}, where they parametrise $ds^2(\text{AdS}_5)= e^{2\rho}\eta_{\mu\nu}dx^{\mu}dx^{\nu}+d\rho^2$ then redefine $(x^{\mu},\rho,\chi)=(4\kappa\sigma)^{-\frac{1}{2}}(\tilde{x}^{\mu},\tilde{\rho},\tilde{\chi})$. This then yields $4\kappa\sigma \left(ds^2(\text{AdS}_5)+d\chi^2\right)= \eta_{\mu\nu}d\tilde x^{\mu}d\tilde x^{\nu}+d\tilde \rho^2+d\tilde{\chi}^2$ to leading order in $\sigma$, which is Mink$_6$. Hence, $4\kappa\sigma \left(ds^2(\text{AdS}_5)+d\chi^2\right)\to ds^2(\text{Mink}_6)$ up to sub-leading terms in $\sigma$.
Introducing the new coordinate $ \tilde{r}= e^{-\frac{\pi}{P}\sigma}(\frac{\pi}{P}\sigma)^{-\frac{1}{2}}$, we then find to leading order
\beq\label{eqn:sigmainftymetric2}
ds^2=ds^2(\text{Mink}_6)+\frac{2P \kappa}{\pi \tilde{r}^2}\bigg(d\tilde{r}^2+\tilde{r}^2ds^2(\text{S}^3)\bigg),~~~H_3=-\frac{4\kappa P}{\pi}\text{vol}(\text{S}^3),~~~e^{-\Phi}=\frac{{\cal R}_1\pi^2}{2 P^{\frac{3}{2}}\sqrt{\kappa}}\tilde{r},
\eeq
corresponding to the near horizon limit of a spherically symmetric stack of NS5 branes in flat space, see \eqref{eqn:NS5metrics}.
Fixing $2\kappa=\pi$, gives an appropriate quantization for the charge for these NS5 branes, namely
\beq\label{eqn:NS5charge}
Q_{\text{NS5}}=-\frac{1}{(2\pi)^2}\int_{S^3} H_3=P.
\eeq
Hence, in this limit, we find a stack of $P$ NS5 branes!

  \subsubsection{The $\eta=0$ boundary, with $\sigma\neq 0$}
At $\eta=0$ for $\sigma\neq0$ (in red), we find to leading order
$~$\\[-1cm]
 \begin{figure}[H]
\centering  
\subfigure
{
\centering
  \begin{minipage}{0.6\textwidth}
   \begin{align}
&\dot V'= f,~~~\dot{V}= f \eta,~~~V''=-\eta\sigma^{-2}\dot f,~~~\ddot{V}=\eta \dot f,\nn\\[2mm]
&\text{where}~~~f(\sigma)=\frac{\pi^2}{P^2}\sum_{n=1}^{\infty}{\cal R}_n \sigma n^2 K_1\left(\frac{n\pi}{P}\sigma\right).\label{eq:fdef}
\end{align}
  \end{minipage}
     \begin{minipage}{.5\textwidth}
    \centering
\begin{tikzpicture}[scale=1, every node/.style={scale=1}]

\draw[lightgray,line width=0.5mm] (1,0) -- (1.5,0.6);
\draw[lightgray,line width=0.5mm] (1.5,0.6) -- (2,1);
\draw[lightgray,line width=0.5mm] (2,1) -- (2.5,1.2);
\draw[lightgray,line width=0.5mm] (2.5,1.2)-- (3,1.1);
\draw[lightgray,line width=0.5mm]  (3,1.1)-- (3.5,0.7);
\draw[lightgray,line width=0.5mm] (3.5,0.7)-- (4,0);

\draw[lightgray,dashed,line width=0.3mm] (1.5,0)-- (1.5,0.6);
\draw[lightgray,dashed,line width=0.3mm] (2,0)-- (2,1);
\draw[lightgray,dashed,line width=0.3mm] (2.5,0)--  (2.5,1.2);
\draw[lightgray,dashed,line width=0.3mm] (3,0)-- (3,1.1);
\draw[lightgray,dashed,line width=0.3mm] (3.5,0)--  (3.5,0.7);

\draw[-stealth] (4.45,1.2) -- (4.45,1.7) node[above] {$\sigma$};
\draw[-stealth]  (4.45,1.2) -- (4.95,1.2)  node[right] {$\eta$};

\draw[red,line width=1mm] (1,0.05)--(1,1.95);

\draw (1,0) -- (1,2);
\draw (1.01,0) -- (1.01,2);
\draw (1.02,0) -- (1.02,2);
\draw (1,0) node[below] {$0$};
\draw (0.65,2.2) node[below] {$\infty$};
\draw (4,0) node[below] {$P$};
\draw (4,0) -- (4,2);
\draw (3.99,0) -- (3.99,2);
\draw (3.98,0) -- (3.98,2);

\draw (1,0) -- (4,0);
\draw (1,0.01) -- (4,0.01);
\draw (1,-0.01) -- (4,-0.01);
\draw (1,0.02) -- (4,0.02);
\draw (1,-0.02) -- (4,-0.02);
%

\draw (2.5,-0.75) node[above] {$\mathcal{R}(\eta)$};

   \begin{scope}[yshift=2cm]
\draw (0,-1.9) -- (0.6,-1.9);
\draw (0.1,-1.95) -- (0.5,-1.95);
\draw (0.2,-2) -- (0.4,-2);
\draw (0.3,-1.9).. controls (0.3,-1.25)  .. (1,-1.25);

\draw (4.5,-1.9) -- (5.1,-1.9);
\draw (4.6,-1.95) -- (5,-1.95);
\draw (4.7,-2) -- (4.9,-2);
\draw (4.8,-1.9).. controls (4.8,-1.25)  .. (4,-1.25);
\end{scope}

\end{tikzpicture}
  \end{minipage} 
}
  \label{fig:eta0boundary}
\end{figure}

Here $f$ is a positive monotonically decreasing function which has a finite maximum at $\sigma=0$. This then means that $|\dot{f}|=-\dot{f}$, and the metric tends to
\beq\label{eqn:eta0metric}
ds^2= \kappa\sqrt{\frac{ 2 f+|\dot f|}{|\dot{f}|}}\bigg[4 \sigma\bigg( ds^2(\text{AdS}_5)+ \frac{|\dot{f}|}{2 f+|\dot{f}|}d\chi^2\bigg)+\frac{2 |\dot{f}|}{\sigma f}\bigg(d\sigma^2+d\eta^2+ \eta^2 ds^2(\text{S}^2)\bigg)\bigg].
\eeq
It is clear to see that $(\eta,S^2)$ now vanishes as $\mathds{R}^3$ in polar coordinates, so we know the solution is regular away from the bounds of $\sigma$. In a similar fashion, we now look at the $\eta=P$ boundary for $\sigma\neq0$. 
$~$\\[-1cm]
   \begin{figure}[H]
\centering  
\subfigure
{
\centering
\hspace{-0.25cm}
  \begin{minipage}{0.61\textwidth}
 \paragraph{The $\eta=P$ boundary, with $\sigma\neq 0$} 
 This boundary is qualitatively equivalent to the $\eta=0$ boundary above.
  \end{minipage}
     \begin{minipage}{.5\textwidth}
    \centering
\begin{tikzpicture}[scale=1, every node/.style={scale=1}]

\draw[lightgray,line width=0.5mm] (1,0) -- (1.5,0.6);
\draw[lightgray,line width=0.5mm] (1.5,0.6) -- (2,1);
\draw[lightgray,line width=0.5mm] (2,1) -- (2.5,1.2);
\draw[lightgray,line width=0.5mm] (2.5,1.2)-- (3,1.1);
\draw[lightgray,line width=0.5mm]  (3,1.1)-- (3.5,0.7);
\draw[lightgray,line width=0.5mm] (3.5,0.7)-- (4,0);

\draw[lightgray,dashed,line width=0.3mm] (1.5,0)-- (1.5,0.6);
\draw[lightgray,dashed,line width=0.3mm] (2,0)-- (2,1);
\draw[lightgray,dashed,line width=0.3mm] (2.5,0)--  (2.5,1.2);
\draw[lightgray,dashed,line width=0.3mm] (3,0)-- (3,1.1);
\draw[lightgray,dashed,line width=0.3mm] (3.5,0)--  (3.5,0.7);

\draw[-stealth] (4.45,1.2) -- (4.45,1.7) node[above] {$\sigma$};
\draw[-stealth]  (4.45,1.2) -- (4.95,1.2)  node[right] {$\eta$};

\draw[red,line width=1mm] (4,0.05)--(4,1.95);

\draw (1,0) -- (1,2);
\draw (1.01,0) -- (1.01,2);
\draw (1.02,0) -- (1.02,2);
\draw (1,0) node[below] {$0$};
\draw (0.65,2.2) node[below] {$\infty$};
\draw (4,0) node[below] {$P$};
\draw (4,0) -- (4,2);
\draw (3.99,0) -- (3.99,2);
\draw (3.98,0) -- (3.98,2);

\draw (1,0) -- (4,0);
\draw (1,0.01) -- (4,0.01);
\draw (1,-0.01) -- (4,-0.01);
\draw (1,0.02) -- (4,0.02);
\draw (1,-0.02) -- (4,-0.02);
%

\draw (2.5,-0.75) node[above] {$\mathcal{R}(\eta)$};

   \begin{scope}[yshift=2cm]
\draw (0,-1.9) -- (0.6,-1.9);
\draw (0.1,-1.95) -- (0.5,-1.95);
\draw (0.2,-2) -- (0.4,-2);
\draw (0.3,-1.9).. controls (0.3,-1.25)  .. (1,-1.25);

\draw (4.5,-1.9) -- (5.1,-1.9);
\draw (4.6,-1.95) -- (5,-1.95);
\draw (4.7,-2) -- (4.9,-2);
\draw (4.8,-1.9).. controls (4.8,-1.25)  .. (4,-1.25);
\end{scope}

\end{tikzpicture}
  \end{minipage} 
}
  \label{fig:etaPboundary}
\end{figure}
$~$\\[-2.8cm]
  \subsubsection{The $\sigma=0$ boundary, with $\eta\in (k,k+1)$}
 Along the entire $\sigma=0$ boundary, one can show $\ddot V=0$ to leading order (using \eqref{eqn:VderivVals} and the Laplace equation \eqref{eqn:laplace}). Hence, along this boundary, the solution tends to
\begin{align}\label{eqn:sigma0metric}
\frac{ds^2}{\kappa}&=\sqrt{\frac{2 \dot{V}}{V''}}\bigg[4ds^2(\text{AdS}_5)+ \frac{2\dot{V}V''}{2\dot{V}V''+(\dot{V}')^2}ds^2(\text{S}^2)+ 2\frac{V''}{\dot{V}}\bigg(d\eta^2+d\sigma^2+ \sigma^2 d\chi^2\bigg)\bigg],\nn\\[2mm]
e^{-4\Phi}&=\frac{V''(2\dot{V}V''+ (\dot{V}')^2)^2}{2^5\kappa^2\dot{V}},~~~~H_3=2\kappa d\left(-\eta+\frac{\dot{V} \dot{V}'}{2 \dot{V} V''+ (\dot{V}')^2}\right)\wedge \text{vol}(\text{S}^2),\nn\\[2mm]
C_1&= \dot{V}'d\chi,~~~~C_3=-4\kappa\left(\frac{\dot{V}^2V''}{2 \dot{V}V''+(\dot{V}')^2}\right) d\chi\wedge \text{vol}(\text{S}^2).
\end{align}
At $\sigma=0$ and non-integer values of $0<\eta<P$, from the boundary conditions \eqref{eqn:BCs}, we have $\dot{V}'$ is a constant and $\dot{V}$ is non vanishing by definition. Provided $V''$ neither blows up or vanishes as $\sigma\rightarrow0$, we would find the $(\sigma,\chi)$ sub-manifold vanishes as $\mathds{R}^2$. Let us now investigate the $V''$. Choosing $\eta\in (k,k+1)$ (depicted in red) and using the double series parametrisation given in \eqref{eq:alternative}, one finds
$~$\\[-1cm]
 \begin{figure}[H]
\centering  
\subfigure
{
\centering
  \begin{minipage}{0.45\textwidth}
   \beq
V''= P_k \quad\quad\quad\quad \label{eqPdef}
\eeq
  \end{minipage}
     \begin{minipage}{.8\textwidth}
    \centering
\begin{tikzpicture}[scale=1, every node/.style={scale=1}]

\draw[lightgray,line width=0.5mm] (1,0) -- (1.5,0.6);
\draw[lightgray,line width=0.5mm] (1.5,0.6) -- (2,1);
\draw[lightgray,line width=0.5mm] (2,1) -- (2.5,1.2);
\draw[lightgray,line width=0.5mm] (2.5,1.2)-- (3,1.1);
\draw[lightgray,line width=0.5mm]  (3,1.1)-- (3.5,0.7);
\draw[lightgray,line width=0.5mm] (3.5,0.7)-- (4,0);

\draw[lightgray,dashed,line width=0.3mm] (1.5,0)-- (1.5,0.6);
\draw[lightgray,dashed,line width=0.3mm] (2,0)-- (2,1);
\draw[lightgray,dashed,line width=0.3mm] (2.5,0)--  (2.5,1.2);
\draw[lightgray,dashed,line width=0.3mm] (3,0)-- (3,1.1);
\draw[lightgray,dashed,line width=0.3mm] (3.5,0)--  (3.5,0.7);

\draw[-stealth] (4.45,1.2) -- (4.45,1.7) node[above] {$\sigma$};
\draw[-stealth]  (4.45,1.2) -- (4.95,1.2)  node[right] {$\eta$};

\draw[red,line width=1.5mm] (1.1,0)--(1.45,0);
\draw[red,line width=1.5mm] (1.55,0)--(1.95,0);
\draw[red,line width=1.5mm] (2.05,0)--(2.45,0);
\draw[red,line width=1.5mm] (2.55,0)--(2.95,0);
\draw[red,line width=1.5mm] (3.05,0)--(3.45,0);
\draw[red,line width=1.5mm] (3.55,0)--(3.9,0);

\draw (1,0) -- (1,2);
\draw (1.01,0) -- (1.01,2);
\draw (1.02,0) -- (1.02,2);
\draw (1,0) node[below] {$0$};
\draw (0.65,2.2) node[below] {$\infty$};
\draw (4,0) node[below] {$P$};
\draw (4,0) -- (4,2);
\draw (3.99,0) -- (3.99,2);
\draw (3.98,0) -- (3.98,2);

\draw (1,0) -- (4,0);
\draw (1,0.01) -- (4,0.01);
\draw (1,-0.01) -- (4,-0.01);
\draw (1,0.02) -- (4,0.02);
\draw (1,-0.02) -- (4,-0.02);
%

\draw (2.5,-0.75) node[above] {$\mathcal{R}(\eta)$};

   \begin{scope}[yshift=2cm]
\draw (0,-1.9) -- (0.6,-1.9);
\draw (0.1,-1.95) -- (0.5,-1.95);
\draw (0.2,-2) -- (0.4,-2);
\draw (0.3,-1.9).. controls (0.3,-1.25)  .. (1,-1.25);

\draw (4.5,-1.9) -- (5.1,-1.9);
\draw (4.6,-1.95) -- (5,-1.95);
\draw (4.7,-2) -- (4.9,-2);
\draw (4.8,-1.9).. controls (4.8,-1.25)  .. (4,-1.25);
\end{scope}
\end{tikzpicture}
  \end{minipage} 
}
\subfigure
{
\centering
  \begin{minipage}{\textwidth}
     \vspace{-0.4cm}
    \centering
\beq
\hspace{-1cm}
P_k= \sum_{j=k+1}^P\frac{b_j}{j-\eta}+\frac{1}{2P}\sum_{j=1}^Pb_j\left(\psi\left(\frac{\eta+j}{2P}\right)-\psi\left(\frac{\eta-j}{2P}\right)+\frac{\pi}{2}\left(\cot\left(\frac{\pi(\eta+j)}{2P}\right)-\cot\left(\frac{\pi(\eta-j)}{2P}\right)\right)\right),\nn
\eeq
  \end{minipage}
}
  \label{fig:eta0boundary}
\end{figure}
where $\psi$ is the digamma function. Indeed, within the bounds of $\eta$, $V''$ neither blows up or vanishes. Hence, along $\sigma=0,\eta\notin \mathds{Z}$, the solution is regular.

      \subsubsection{The $\sigma=0,~\eta=0$ boundary}
      In order to approach the boundary at $\sigma=\eta=0$, we make the coordinate change $(\eta=r\,\cos\alpha,~\sigma=r\,\sin\alpha)$ for small $r$. We then find, to leading order
 \begin{figure}[H]
\centering  
\subfigure
{
\centering
  \begin{minipage}{0.65\textwidth}

      \begin{align}
&\dot{V}=N_1 r \cos\alpha,~~~~~\dot{V}'= N_1,~~~V''= r\,Q\,\cos\alpha,\nn\\[2mm]
&Q=\frac{1}{4P^2}\sum_{j=1}^Pb_k\left(2 \psi^1\left(\frac{j}{2P}\right)-\pi^2\csc^2\left(\frac{j\pi}{2P}\right)\right),\label{eqQdef2}
\end{align}
  \end{minipage}
     \begin{minipage}{.46\textwidth}
    \centering
\begin{tikzpicture}[scale=1, every node/.style={scale=1}]

\draw[lightgray,line width=0.5mm] (1,0) -- (1.5,0.6);
\draw[lightgray,line width=0.5mm] (1.5,0.6) -- (2,1);
\draw[lightgray,line width=0.5mm] (2,1) -- (2.5,1.2);
\draw[lightgray,line width=0.5mm] (2.5,1.2)-- (3,1.1);
\draw[lightgray,line width=0.5mm]  (3,1.1)-- (3.5,0.7);
\draw[lightgray,line width=0.5mm] (3.5,0.7)-- (4,0);

\draw[lightgray,dashed,line width=0.3mm] (1.5,0)-- (1.5,0.6);
\draw[lightgray,dashed,line width=0.3mm] (2,0)-- (2,1);
\draw[lightgray,dashed,line width=0.3mm] (2.5,0)--  (2.5,1.2);
\draw[lightgray,dashed,line width=0.3mm] (3,0)-- (3,1.1);
\draw[lightgray,dashed,line width=0.3mm] (3.5,0)--  (3.5,0.7);

\draw[-stealth] (4.45,1.2) -- (4.45,1.7) node[above] {$\sigma$};
\draw[-stealth]  (4.45,1.2) -- (4.95,1.2)  node[right] {$\eta$};

\fill[red] (1,0) circle(.15);

\draw (1,0) -- (1,2);
\draw (1.01,0) -- (1.01,2);
\draw (1.02,0) -- (1.02,2);
\draw (1,-0.1) node[below] {$0$};
\draw (0.65,2.2) node[below] {$\infty$};
\draw (4,-0.1) node[below] {$P$};
\draw (4,0) -- (4,2);
\draw (3.99,0) -- (3.99,2);
\draw (3.98,0) -- (3.98,2);

\draw (1,0) -- (4,0);
\draw (1,0.01) -- (4,0.01);
\draw (1,-0.01) -- (4,-0.01);
\draw (1,0.02) -- (4,0.02);
\draw (1,-0.02) -- (4,-0.02);
%

\draw (2.5,-0.75) node[above] {$\mathcal{R}(\eta)$};

   \begin{scope}[yshift=2cm]
\draw (0,-1.9) -- (0.6,-1.9);
\draw (0.1,-1.95) -- (0.5,-1.95);
\draw (0.2,-2) -- (0.4,-2);
\draw (0.3,-1.9).. controls (0.3,-1.25)  .. (1,-1.25);

\draw (4.5,-1.9) -- (5.1,-1.9);
\draw (4.6,-1.95) -- (5,-1.95);
\draw (4.7,-2) -- (4.9,-2);
\draw (4.8,-1.9).. controls (4.8,-1.25)  .. (4,-1.25);
\end{scope}

\end{tikzpicture}
  \end{minipage} 
}
  \label{fig:sigma0eta0boundary}
\end{figure}
$~$\\[-1.3cm]
with $\psi^1$ the trigamma function. The metric and dilaton then become
  \begin{align}\label{eqn:sigma0eta0metric}
&\frac{ds^2}{\kappa}=\sqrt{\frac{2N_1}{Q}}\bigg[ 4ds^2(\text{AdS}_5)+ \frac{2Q}{N_1}\bigg(dr^2+r^2d\alpha^2+r^2\cos^2\alpha \,ds^2(\text{S}^2) + r^2\sin^2\alpha d\chi^2\bigg)\bigg], ~~~~e^{-4\Phi}=\frac{N_1^3\,Q}{2^5\kappa^2},
  \end{align}
where we see that the internal metric vanishes as $\mathds{R}^5$ in polar coordinates. The dilaton and AdS$_5$ warp factor remain constant in the limit of small $r$, so again, the solution remains regular here.
  $~$\\[-1cm]
          \begin{figure}[H]
\centering  
\subfigure
{
\centering
\hspace{-0.25cm}
  \begin{minipage}{0.6\textwidth}
 \paragraph{The $\sigma=0,~\eta=P$ boundary} 
 This boundary is qualitatively the same as the $(\sigma,\eta)=0$ above.
  \end{minipage}
     \begin{minipage}{.57\textwidth}
    \centering
\begin{tikzpicture}[scale=1, every node/.style={scale=1}]

\draw[lightgray,line width=0.5mm] (1,0) -- (1.5,0.6);
\draw[lightgray,line width=0.5mm] (1.5,0.6) -- (2,1);
\draw[lightgray,line width=0.5mm] (2,1) -- (2.5,1.2);
\draw[lightgray,line width=0.5mm] (2.5,1.2)-- (3,1.1);
\draw[lightgray,line width=0.5mm]  (3,1.1)-- (3.5,0.7);
\draw[lightgray,line width=0.5mm] (3.5,0.7)-- (4,0);

\draw[lightgray,dashed,line width=0.3mm] (1.5,0)-- (1.5,0.6);
\draw[lightgray,dashed,line width=0.3mm] (2,0)-- (2,1);
\draw[lightgray,dashed,line width=0.3mm] (2.5,0)--  (2.5,1.2);
\draw[lightgray,dashed,line width=0.3mm] (3,0)-- (3,1.1);
\draw[lightgray,dashed,line width=0.3mm] (3.5,0)--  (3.5,0.7);

\draw[-stealth] (4.45,1.2) -- (4.45,1.7) node[above] {$\sigma$};
\draw[-stealth]  (4.45,1.2) -- (4.95,1.2)  node[right] {$\eta$};

\fill[red] (4,0) circle(.15);

\draw (1,0) -- (1,2);
\draw (1.01,0) -- (1.01,2);
\draw (1.02,0) -- (1.02,2);
\draw (1,-0.1) node[below] {$0$};
\draw (0.65,2.2) node[below] {$\infty$};
\draw (4,-0.1) node[below] {$P$};
\draw (4,0) -- (4,2);
\draw (3.99,0) -- (3.99,2);
\draw (3.98,0) -- (3.98,2);

\draw (1,0) -- (4,0);
\draw (1,0.01) -- (4,0.01);
\draw (1,-0.01) -- (4,-0.01);
\draw (1,0.02) -- (4,0.02);
\draw (1,-0.02) -- (4,-0.02);
%

\draw (2.5,-0.75) node[above] {$\mathcal{R}(\eta)$};

   \begin{scope}[yshift=2cm]
\draw (0,-1.9) -- (0.6,-1.9);
\draw (0.1,-1.95) -- (0.5,-1.95);
\draw (0.2,-2) -- (0.4,-2);
\draw (0.3,-1.9).. controls (0.3,-1.25)  .. (1,-1.25);

\draw (4.5,-1.9) -- (5.1,-1.9);
\draw (4.6,-1.95) -- (5,-1.95);
\draw (4.7,-2) -- (4.9,-2);
\draw (4.8,-1.9).. controls (4.8,-1.25)  .. (4,-1.25);
\end{scope}

\end{tikzpicture}
  \end{minipage} 
}
  \label{fig:sigma0etaPboundary}
\end{figure}
  $~$\\[-2.5cm]
      \subsubsection{The $\sigma=0$ boundary, with $\eta=k$}
Finally, at $\sigma=0,\eta=k$, for $0<k<P$ (depicted as red dots), we now use the coordinate change $(\eta=k- r\cos\alpha,\sigma=r\sin\alpha)$ for small $r$, with
  $~$\\[-1cm]
 \begin{figure}[H]
\centering  
\subfigure
{
\centering
  \begin{minipage}{0.7\textwidth}
\beq
\dot{V}=N_k,~~~V''=\frac{b_k}{2r },~~~~\dot{V}'=\frac{b_k}{2}(1+\cos\alpha)+N_{k+1}-N_{k},\label{eq:VsattheD6s}
\eeq
  \end{minipage}
     \begin{minipage}{.35\textwidth}
    \centering
\begin{tikzpicture}[scale=1, every node/.style={scale=1}]

\draw[lightgray,line width=0.5mm] (1,0) -- (1.5,0.6);
\draw[lightgray,line width=0.5mm] (1.5,0.6) -- (2,1);
\draw[lightgray,line width=0.5mm] (2,1) -- (2.5,1.2);
\draw[lightgray,line width=0.5mm] (2.5,1.2)-- (3,1.1);
\draw[lightgray,line width=0.5mm]  (3,1.1)-- (3.5,0.7);
\draw[lightgray,line width=0.5mm] (3.5,0.7)-- (4,0);

\draw[lightgray,dashed,line width=0.3mm] (1.5,0)-- (1.5,0.6);
\draw[lightgray,dashed,line width=0.3mm] (2,0)-- (2,1);
\draw[lightgray,dashed,line width=0.3mm] (2.5,0)--  (2.5,1.2);
\draw[lightgray,dashed,line width=0.3mm] (3,0)-- (3,1.1);
\draw[lightgray,dashed,line width=0.3mm] (3.5,0)--  (3.5,0.7);

\draw[-stealth] (4.45,1.2) -- (4.45,1.7) node[above] {$\sigma$};
\draw[-stealth]  (4.45,1.2) -- (4.95,1.2)  node[right] {$\eta$};
\fill[red] (1.5,0) circle(.15);
\fill[red] (2,0) circle(.15);
\fill[red] (2.5,0) circle(.15);
\fill[red] (3,0) circle(.15);
\fill[red] (3.5,0) circle(.15);

\draw (1,0) -- (1,2);
\draw (1.01,0) -- (1.01,2);
\draw (1.02,0) -- (1.02,2);
\draw (1,0) node[below] {$0$};
\draw (0.65,2.2) node[below] {$\infty$};
\draw (4,0) node[below] {$P$};
\draw (4,0) -- (4,2);
\draw (3.99,0) -- (3.99,2);
\draw (3.98,0) -- (3.98,2);

\draw (1,0) -- (4,0);
\draw (1,0.01) -- (4,0.01);
\draw (1,-0.01) -- (4,-0.01);
\draw (1,0.02) -- (4,0.02);
\draw (1,-0.02) -- (4,-0.02);
%

\draw (2.5,-0.75) node[above] {$\mathcal{R}(\eta)$};

   \begin{scope}[yshift=2cm]
\draw (0,-1.9) -- (0.6,-1.9);
\draw (0.1,-1.95) -- (0.5,-1.95);
\draw (0.2,-2) -- (0.4,-2);
\draw (0.3,-1.9).. controls (0.3,-1.25)  .. (1,-1.25);

\draw (4.5,-1.9) -- (5.1,-1.9);
\draw (4.6,-1.95) -- (5,-1.95);
\draw (4.7,-2) -- (4.9,-2);
\draw (4.8,-1.9).. controls (4.8,-1.25)  .. (4,-1.25);
\end{scope}

\end{tikzpicture}
  \end{minipage} 
}
  \label{fig:eta0etakboundary}
\end{figure}
  $~$\\[-1.7cm]
at leading order, the solution then tends to
\begin{align}\label{eqn:etakorigmetric}
&\frac{ds^2}{2\kappa\sqrt{N_k}}= \frac{1}{\sqrt{\frac{b_k}{r}}}\bigg(4ds^2(\text{AdS}_5)+ ds^2(\text{S}^2)\bigg)+ \frac{\sqrt{\frac{b_k}{r}}}{N_k}\bigg(dr^2+ r^2 ds^2(\tilde{\text{S}}^2)\bigg),~~~e^{-\Phi}=\left(\frac{N_kb_k^3}{2^6\kappa^2r^3 }\right)^{\frac{1}{4}},\nn\\[2mm]
&B_2=0,~~~C_1=\left(\frac{b_k}{2}(1+\cos\alpha)+N_{k+1}-N_k\right)d\chi,~~~C_3=-2\kappa N_k d\chi\wedge \text{vol}(\text{S}^2),
\end{align}
using \eqref{eqn:Dbranemetrics}, this is the near horizon limit of a D6 brane stack wrapping AdS$_5\times$S$^2$, with $\tilde{\text{S}}^2$ spanned by $(\alpha,\chi)$. In addition, the D6 branes have appropriate quantization of charge
\beq
F_2=-\frac{1}{2}b_k \text{vol}(\tilde{\text{S}}^2)~~~\Rightarrow~~~Q^k_{D6}=-\frac{1}{2\pi}\int_{\tilde{S}^2}F_2= b_k=2N_k-N_{k-1}-N_{k+1},
\eeq
recovering the results of Section \ref{sec:GMIIA}, with D6s located at the kinks of the rank function. 

\begin{figure}[H]
 \centering
\begin{tikzpicture}
\draw[-stealth] (6.45,1) -- (6.45,1.5)  node[above] {$\sigma$};
\draw[-stealth]  (6.45,1) -- (6.95,1)  node[right] {$\eta$};
 
 \draw[-stealth] (4,0) -- (3.5,0.56)  node[above] {$r$};
\fill[red] (4,0) circle(.15);
\fill[red] (2,0) circle(.15);
\fill[red] (6,0) circle(.15);

\fill[black] (1.1,0) circle(.04);
\fill[black] (1,0) circle(.04);
\fill[black] (0.9,0) circle(.04);

\fill[black] (7.1,0) circle(.04);
\fill[black] (7,0) circle(.04);
\fill[black] (6.9,0) circle(.04);

 \draw (2 ,-0.4) node[below] {$k-1$};
 \draw (4,-0.4) node[below] {$k$};
  \draw (6,-0.4) node[below] {$k+1$};
  \draw (3.2,0) node[below] {$\alpha=0$};
   \draw (4.8,-0.05) node[below] {$\alpha=\pi$};
 
 \begin{scope}
    \clip (3.25,0) rectangle (4.75,1);
    \draw (4,0) circle(0.75);
\end{scope}
 \begin{scope}
    \clip (3.65,0) rectangle (4,0.25);
    \draw (4,0) circle(0.35);
\end{scope}

 \draw (2 ,-0.2) -- (2 ,0.2);
  \draw (4,-0.2) -- (4,0.2);
    \draw (6,-0.2) -- (6,0.2);

\draw (0.3,-0.2) -- (0.3,2);
\draw (0.31,-0.2) -- (0.31,2);
\draw (0.32,-0.2) -- (0.32,2);
\draw (0.3,-0.4) node[below] {$0$};
\draw (7.7,-0.4) node[below] {$P$};
\draw (7.7,-0.2) -- (7.7,2);
\draw (7.69,-0.2) -- (7.69,2);
\draw (7.68,-0.2) -- (7.68,2);

\draw (1.25,0) -- (6.75,0);
\draw (1.25,0.01) -- (6.75,0.01);
\draw (1.25,-0.01) -- (6.75,-0.01);
\draw (1.25,0.02) -- (6.75,0.02);
\draw (1.25,-0.02) -- (6.75,-0.02);

\draw (0.3,0) -- (0.75,0);
\draw (0.3,0.01) -- (0.75,0.01);
\draw (0.3,-0.01) -- (0.75,-0.01);
\draw (0.3,0.02)-- (0.75,0.02);
\draw (0.3,-0.02) -- (0.75,-0.02);
%
\draw (7.25,0) -- (7.7,0);
\draw (7.25,0.01) -- (7.7,0.01);
\draw (7.25,-0.01) -- (7.7,-0.01);
\draw (7.25,0.02) -- (7.7,0.02);
\draw (7.25,-0.02) -- (7.7,-0.02);

\end{tikzpicture}
  \caption{Semi-circular contour 
  with $0\leq \alpha \leq \pi$ and infinitesimal radius, $r$.}
  \label{fig:semicircularcontour}
\end{figure}
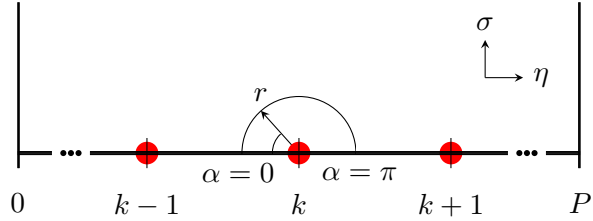

A Page Charge for D4 branes can also be defined, using \eqref{eqn:PagesforGM}. For $k<\eta<k+1$,
\beq\label{eqn:pageF4betweenkandkplusone}
\hat F_4= 2\kappa{\cal R}'' (\eta-k)d\eta\wedge d\chi\wedge \text{vol}(\text{S}^2),
\eeq
which is zero in this range due to $\mathcal{R}''(\eta)$, see \eqref{eqn:Rkwithderivs}, which consists of delta functions at the loci of the kinks, namely at $\eta=k$. Likewise, for the same reason, we cannot define a non-trivial Page flux at $\eta=0$ or $\eta=P$. Things work better at the position of the kinks, close to $\eta=k$ with \eqref{eq:VsattheD6s}. Here, we can integrate on $(\chi,\text{S}^2)$ and the semi-circular contour defined by $(\eta=k- r\cos\alpha,\sigma=r\sin\alpha)$, with $0\leq \alpha \leq \pi$ and infinitesimal $r$. It is important to note here that $\hat{F}_4$ given in \eqref{eqn:PagesforGM} is the Page flux for the $k$'th unit cell, however the semi-circular contour we are following actually starts in the $(k-1)$'th cell (when $\alpha=0$) and crosses into the $k$'th unit cell at $\alpha=\pi/2$ - see Figure \ref{fig:semicircularcontour}.

The calculation then requires a bit of care when performing the integration, leading to
\beq
Q^k_{D4}=-\frac{1}{(2\pi)^3}\int_{\text{S}^2\times \tilde{\text{S}}^2}\hat F_4=N_k-N_{k-1},\label{eq:neq2F4pagecharge}
\eeq
which we recall are interpreted as colour branes - not flavour branes. 

We can now repeat the calibration calculation conducted in \eqref{eqn:caltake1}, using \eqref{eqn:C5andC7}, \eqref{eqn:cals}, \eqref{eq:VsattheD6s}, \eqref{eqn:etakorigmetric} and \eqref{eqn:fsfork}, finding
\begin{align}\label{eqn:callimit1}
\Psi^{(\text{cal})}_{\text{D6}}\bigg\lvert_{(\rho,\text{S}^2)}&= -2^4 f_1^3f_5f_7  e^{4\rho}d\rho\wedge \text{vol}(\text{S}^2) =  \frac{2^7\kappa^3N_k^2}{b_k}\,e^{4\rho} \,r \,d\rho\wedge \text{vol}(\text{S}^2),\nn\\[2mm]
\hspace{-5mm}e^{4A-\Phi}\sqrt{\det(g+B_2)}d^{p-3}w\bigg\lvert_{(\rho,\text{S}^2)}&=\Big(4  f_1^{\frac{3}{2}} f_5^{\frac{1}{2}} e^{2\rho}\Big)^2\left(\frac{N_kb_k^3}{2^6\kappa^2r^3 }\right)^{\frac{1}{4}}
\sqrt{4(2\kappa)^3 \Big(\frac{N_k\,r}{b_k}\Big)^{\frac{3}{2}}}\sin\theta d\rho\wedge d\theta\wedge d\phi \nn\\[2mm]
&=\frac{2^7\kappa^3N_k^2}{b_k}\,e^{4\rho} \,r \,d\rho\wedge \text{vol}(\text{S}^2), 
\end{align}
which clearly satisfies \eqref{eq:Dpsusy}.

\chapter{Type IIA - via Dimensional Reduction}\label{chap:typeIIA}
In this chapter we will investigate dimensional reductions along each of the three $U(1)$ directions of the $d=11$ GM class in turn, with the full solutions (including all nine $GL(3,\mathds{R})$ transformation parameters) provided in Appendix \ref{sec:GL(3R)}.
These results are of course too general, as for an $SL(3,\mathds{R})$ transformation, one can reduce these nine parameters to just three free parameters in each case, without loss of generality. They will however prove very useful to us, allowing one to simply plug in the desired values of the parameters in each unique example, with all of the calculation already in place. These solutions are derived using the generalised reduction formula 
 \begin{equation*} 
e^{-\frac{2}{3}\Phi
}ds_{IIA}^2 = ds^2 -  \frac{e^{\frac{4}{3}\Phi
}}{X^2}\Big( d( X\,\psi)+ X\,C_1\Big)^2,~~~~~~~~A_3= C_3+\frac{1}{X}B_2\wedge d(X\,\psi),
\end{equation*}
which will provide options when considering the quantization of charge.

The solutions presented in this chapter, which in general depend on two parameters, break the $\mathcal{N}=2$ supersymmetry either partially or completely. Our main focus is the solution generated following a $\beta$ reduction, as this background is a direct supersymmetry-breaking deformation of the GM class. In this case, investigations at the boundary were performed, demonstrating that NS5 and D6 branes are present in all backgrounds - with D4 branes appearing only for a preserved $S^2$. In addition, we find that the D6 brane sources of the ${\cal N}=2$ solution are mapped to sources backreated onto spindles - in fact, there is a different spindle for each stack of D6 branes (located at each kink of the rank function). The conical deficit angles of each spindle is then governed by the slope of the rank function at each stack, along with one of the SUSY-breaking parameters (which recovers the $S^2$ when fixed to zero). This leads to a peculiar quantization condition for these sources, giving rise to rational charge (with the conical deficit angles of the spindle as a denominator). In the case of the D4 brane, the effect of the spindle can be eliminated by introducing a new term to the large gauge transformation of $B_2$, restoring integer quantization. This is not possible for the D6 branes, however interestingly they take the same rational form for all solutions. Stability of some D6 probes is then studied, with a discussion of some aspects of the corresponding dual CFTs provided. Given that the holographic central charge (identified with $a$-central charge) remains untouched by the parameters, we propose the solutions are dual to marginal deformations of the `parent' ${\cal N}=2$ CFTs. A proposal for the operators which deform this parent theory is then given, within the context of soft-SUSY breaking. Recall that in the holographic limit (with long linear quivers of large ranks), both central charges of the CFT become equal, $a=c$. Comments on a mirror-like symmetry relation between two different quivers is then included. Lastly, equations describing spin-two fluctuations in the CFT are presented, with simple universal solutions included, and a bound on the dimension (mass) for these operators provided.

We then move on to investigate the backgrounds derived via dimensional reductions along the remaining two $U(1)$ directions (namely, $\chi$ and $\phi$). These reductions lead to further new and unique solutions, including a one-parameter family of $\mathcal{N}=1$ backgrounds. This solution can be mapped to the $\beta$ reduction solution, via transforming one of the parameters into its reciprocal (which is not trivial given that they are integer valued), but does contain a new and unique zero-parameter $\mathcal{N}=1$ sub-class (when fixing this parameter to zero). The G-Structure description of this background is also given.

From here on in, we will begin to relabel the three free parameters as $(\xi,\zeta,\gamma)$. In most cases, $\zeta$ will keep track of the SUSY under dimensional reduction to Type IIA, $\gamma$ will keep track of the SUSY under an abelian T-duality to Type IIB, and the remaining parameter, $\xi$, will be left over as a free parameter in the resulting backgrounds. In fact, $\gamma$ plays only a trivial role in the type IIA solutions.
   
 
We will now present the $\beta$ reduction case, before turning to the $\chi$ and $\phi$ reductions in turn, where the analysis will be largely analogous. 
\newpage
\section{$\beta$ Reduction}\label{sec:betaRed}
The full solution, with all $GL(3,\mathds{R})$ transformation parameters intact, is given in \eqref{eqn:beta-Gen}.
\subsection{Two-Parameter Family}
 Motivated by the form of the $U(1)_R$ given in \eqref{eqn:U(1)lessgen}, we first investigate keeping $(q,v)$ free. 
 
 \subsubsection{Fixing $(q\equiv \xi,v\equiv \zeta)$}
 We will now relabel $q\equiv \xi$ and $v\equiv \zeta$, and use $\gamma$ for the third free parameter. From the determinant in \eqref{eqn:S2breakingdefns}, fixing $(q\equiv \xi,v\equiv \zeta)$ as free parameters requires $(a,c)=0$. The determinant then becomes 
\beq\label{eqn:pums}
pu-ms=1,
\eeq
which reduces to $ms=0$ after enforcing $(p,b,u)=1$. This condition then gives the third free parameter as either $m\equiv \gamma$ (with $s=0$) or $s\equiv \gamma$ (with $m=0$). In the solution which we will now present, $\gamma$ is in fact carried through the calculation trivially, and can be fixed to zero without loss of generality.
More specifically, when $m\equiv\gamma$ (or $s\equiv \gamma$), one simply derives \eqref{eqn:generalresult1} but with the redefinition $\chi\rightarrow \chi+\gamma\phi$ (or $\phi\rightarrow\phi+\gamma \chi$). Later, we will see that this $\gamma$ will in fact play a vital role when performing an abelian T-duality to Type IIB, allowing for the preservation of $\mathcal{N}=1$ supersymmetry! We will return to this in the next chapter, but for now we fix $\gamma=0$. Hence, in $d=11$, we have performed the following transformations prior to a $\beta$ reduction 
\beq\label{eqn:exacttransformation}
d\beta \rightarrow   d\beta  ,~~~~~~~~~~~~~~~~~d\chi \rightarrow d\chi +\xi\, d\beta  ,~~~~~~~~~~~~~~~~~~d\phi \rightarrow d\phi + \zeta\,d\beta,
\eeq
which when using \eqref{eqn:beta-Gen}, derives the following two-parameter family of solutions 
           \begin{align}\label{eqn:generalresult1}
     &   ds_{10,st}^2=\frac{1}{X}f_1^{\frac{3}{2}}f_5^{\frac{1}{2}}\sqrt{\Xi}\bigg[4ds^2(\text{AdS}_5)+f_4(d\sigma^2+d\eta^2)+ ds^2(M_3)\bigg],\nn\\[2mm]
   &     ds^2(M_3)=f_2 \bigg(d\theta^2+ \frac{\Delta}{\Xi}\sin^2\theta D\phi^2\bigg) +\frac{f_3}{\Delta}d\chi^2 =f_2 \bigg(d\theta^2+ \frac{1}{\Pi}\sin^2\theta d\phi^2\bigg) + \frac{\Pi}{\Xi} f_3 D\chi^2,\nn\\[2mm]
    &    B_2  =\frac{1}{X} \sin\theta \,\bigg[\zeta  f_7 d\chi  - (f_8+\xi f_7 ) d\phi \bigg] \wedge d\theta,~~~~~~~~~~~~~~~~        e^{\frac{4}{3}\Phi}= \frac{1}{X^2}f_1 f_5 \Xi , \\[2mm]
  &     C_1=\frac{X}{\Xi}\bigg(\Big(f_6(1 + \xi f_6) +\,\xi \frac{f_3}{f_5} \Big)d\chi +  \zeta  \frac{f_2}{f_5}  \sin^2\theta d\phi \bigg),~~~~~~~~~~~~~~~ C_3= f_7  \,d\chi \wedge \text{vol}(S^2),\nn\\[2mm]
&  \Xi=\Delta+\zeta^2\frac{f_2}{f_5}\sin^2\theta,~~~\Delta=(1+ \xi f_6)^2+\xi^2\frac{f_3}{f_5},~~~\Pi=1+\zeta^2f_2\frac{f_5 f_6^2+f_3}{f_3 f_5}\sin^2\theta,\nn\\[2mm]
&D\phi =d\phi -\frac{\zeta}{\Delta}\bigg(f_6(1+\xi f_6)+\xi \frac{f_3}{f_5}\bigg) d\chi,~~~~~~~D\chi=d\chi -\frac{\zeta}{\Pi}\frac{f_2}{f_3}\bigg(f_6(1+\xi f_6) +\xi \frac{f_3}{f_5}\bigg)\sin^2\theta\,d\phi,\nn
    \end{align} 
    with
    \beq
    H=dB_2,~~~~~~F_2=dC_1,~~~~~~F_4=dC_3-H\wedge C_1,
    \eeq
where we have rewritten the solution in a form which will become useful when investigating the boundary. This solution is clearly a parametric deformation of \eqref{eq:N=2}, reducing to it exactly when $\xi=\zeta=0$ (and $X=1$).

Following the $\mathcal{N}=2$ analysis, we consider approaching the $\sigma= 0$ boundary and use the laplace equation \eqref{eqn:laplace}, where we find $\ddot{V}\rightarrow 0$ to leading order. Then, using the boundary condition $\mathcal{R}(\eta) = \dot{V}\Big|_{\sigma=0}$ and the warp factors \eqref{eqn:fs}, we now find the relation
\beq\label{eqn:C1sourcebeta}
C_1\Big|_{\sigma\rightarrow 0} =X \frac{ \mathcal{R}' (1+\xi \mathcal{R}') d\chi+\frac{1}{2}\zeta V'' \mathcal{R} \sin^2\theta d\phi }{(1+\xi \mathcal{R}')^2+\frac{1}{2}\zeta^2 V'' \mathcal{R} \sin^2\theta },~~\Rightarrow~~~~~~C_1\Big|_{\sigma\rightarrow 0}^{\zeta/\sin\theta=0}=X \frac{\mathcal{R}' (\eta)}{1+\xi \mathcal{R}'(\eta)} d\chi ,
\eeq
which reduces for $\zeta=0$ or $\sin\theta=0$ (corresponding to a pole of the deformed $S^2$). 
Recalling from \eqref{eqn:Rkwithderivs} that $\mathcal{R}'(\eta)$ is discontinuous in $\eta$, which leads to source terms for the D6 branes in the Bianchi identity. Calculating the derivative carefully, we find  
\begin{adjustwidth}{-0.8cm}{}
\vspace{-0.6cm}
\begin{align}\label{eqn:betaF2val}
&F_2\Big|_{\sigma\rightarrow 0,\eta=k}^{\zeta/\sin\theta=0}=X  \bigg(\frac{\mathcal{R}' (k)}{1+\xi \mathcal{R}'(k)}-\frac{\mathcal{R}' (k-1)}{1+\xi \mathcal{R}'(k-1)}\bigg)   d\eta\wedge  d\chi = \frac{ X\big(\mathcal{R}' (k)-\mathcal{R}' (k-1)\big)}{\big(1+\xi \mathcal{R}'(k)\big)\big(1+\xi \mathcal{R}'(k-1)\big)}  d\eta\wedge  d\chi\nn \\[2mm]
&~~~~~~~\Rightarrow ~~~~ F_2\Big|_{\sigma\rightarrow 0 }^{\zeta/\sin\theta=0}= - X\sum_{k=1}^{P-1} \frac{2N_k-N_{k+1}-N_{k-1}}{\big(1+\xi  (N_{k+1}-N_k)\big)\big(1+\xi  (N_{k}-N_{k-1})\big)} \delta(\eta-k)\delta(\sigma)d\eta\wedge  d\chi,
\end{align}
\end{adjustwidth}
which would now lead to rational D6 charge under integration, due to the additional parameter $\xi$ in the denominator. To investigate this more thoroughly, we will need to perform careful analysis at the boundary - which we return to in the next sub-section. Note that the Bianchi identities for the $F_4$ and $H$ fluxes still read,
\beq
dH=0,~~~~~d_HF_4=0.
\eeq
Hence we find that this more general solution still contains D6 branes as sources, with NS5 and D4 branes considered pure flux.

   It is worth noting, one can satisfy \eqref{eqn:pums} by instead fixing $pu=0$ and $ms=-1$. We will still derive the solution given in \eqref{eqn:generalresult1}, but now we have trivially re-defined the $U(1)$ directions amongst themselves. This is clear from the $SL(3,\mathds{R})$ transformation. Specifically, when $(p=0,u\equiv\gamma,s=-1,m=1)$ we require $(\chi\rightarrow\phi,~\phi\rightarrow -\chi+\gamma \phi)$ to map to \eqref{eqn:generalresult1}, and when $(p\equiv\gamma, u=0,s=-1,m=1)$ we require the mapping $(\chi\rightarrow\phi +\gamma \chi,~\phi\rightarrow \chi)$.

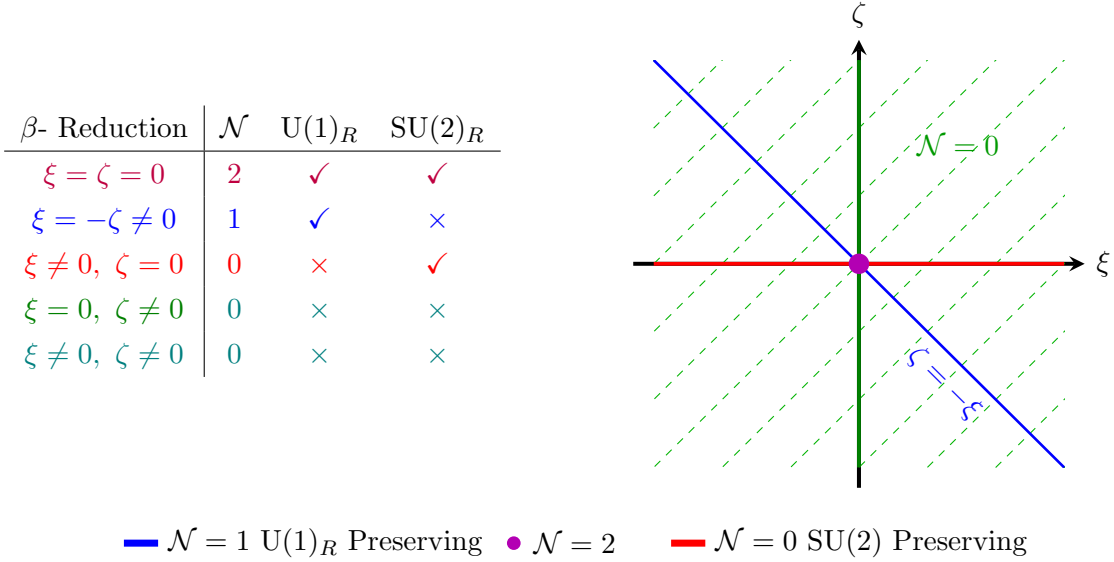
\begin{figure}[H]
\centering  
\subfigure
{
\centering
  \begin{minipage}{0.5\textwidth}
\begin{tabular}{c | c c c c  }
$\beta$- Reduction&$\mathcal{N}$&U(1)$_R$&SU(2)$_R$  \\
\hline
$\textcolor{purple}{\xi= \zeta=0}$&$ \textcolor{purple}{2}$ &$\textcolor{purple}{\checkmark}$&$\textcolor{purple}{\checkmark}$ \\
$\textcolor{blue}{\xi=-\zeta\neq 0}$&$ \textcolor{blue}{1}$ &$\textcolor{blue}{\checkmark}$&$\textcolor{blue}{\times}$  \\
$\textcolor{red}{\xi\neq 0,~\zeta=0}$&$ \textcolor{red}{0}$ &$\textcolor{red}{\times}$&$\textcolor{red}{\checkmark}$  \\
$\textcolor{green!50!black}{\xi=0,~\zeta\neq0}$&$ \textcolor{teal}{0}$ &$\textcolor{teal}{\times}$ &$\textcolor{teal}{\times}$  \\
$\textcolor{teal}{\xi\neq0,~\zeta\neq 0}$&$ \textcolor{teal}{0}$ &$\textcolor{teal}{\times}$&$\textcolor{teal}{\times}$ 
\end{tabular}
  \end{minipage}
     \begin{minipage}{.5\textwidth}
    \centering
\begin{tikzpicture}[scale=0.9]
\draw[-stealth, line width=0.53mm] (-3.3,0)--(3.3,0) node[right ]{$\xi$};
\draw[-stealth, line width=0.53mm] (0,-3.3)--(0,3.3) node[above ]{$\zeta$};
\clip (-3,-3) rectangle (3,3);
\begin{scope}[cm={0.5,-0.5,  50,50,  (0,0)}]  
\draw[green!70!black,
dashed] (-6,-6) grid (6,6);
\end{scope}
\draw[blue,line width=0.35mm](-3,3)--(3,-3);
\draw[line width=0.5mm,red](-3,0)--(3,0);
\draw[line width=0.5mm,green!50!black ](0,-3)--(0,3);
\fill[violet!60!magenta] (0,0) circle(.15);
\draw[green!60!black ] (1.45,1.45) node[above] {$\mathcal{N}=0$};
 \node[blue ,rotate=-45] at (1.25,-1.75) {$\zeta=-\xi$};
\end{tikzpicture}
  \end{minipage} 
}
\subfigure
{
\centering
  \begin{minipage}{\textwidth}
    \centering
\begin{tikzpicture}[scale=0.9]
\draw[ blue,line width=0.93mm] (-6,-4.5)--(-5.5,-4.5);
\draw (-5.5,-4.5) node[right]{$ \mathcal{N}=1$ U(1)$_R$ Preserving};
\draw[ red,line width=0.93mm] (2,-4.5)--(2.5,-4.5);
\draw  (2.5,-4.5) node[right]{$ \mathcal{N}=0$ SU(2) Preserving};
\fill[violet!60!magenta] (-0.3,-4.5) circle(.1);
\draw(-0.3,-4.5) node[right]{$~\mathcal{N}=2$};
\end{tikzpicture}
  \end{minipage}
}
\caption{In the general case, for arbitrary $(\xi,\zeta)$ (in green dashed lines), the background breaks all SUSY. Along the $\zeta=-\xi$ line (in blue), the $U(1)_R$-symmetry is preserved, leading to $\mathcal{N}=1$ solutions. Along the $\zeta=0$ line (in red), the background preserves $SU(2)$ isometry (descending from the original R-symmetry) with the SUSY completely broken in general. At the point where the red and blue lines cross, with $(\zeta,\xi)=(0,0)$ (in purple), the background preserves $SU(2)_R\times U(1)_R$ R-symmetry - leading to an $\mathcal{N}=2$ solution.  
}
    \label{fig:IIAtableplot}
\end{figure}

From the transformation \eqref{eqn:exacttransformation} and parametrization \eqref{eqn:S2parametrization}, it is clear that $\zeta\neq 0$ breaks the S$^2$ of the IIA background under $\beta$ reduction (and hence the SU(2)$_R$ component of the R-Symmetry). We can see this directly from \eqref{eqn:generalresult1} using the forms of $ds^2(M_3)$, $\Pi$ and $D\chi$, which recover the $S^2$ for $\zeta=0$ (when $\Pi=1$ and $D\chi=d\chi$). Consequently, from the $U(1)_R$ component in \eqref{eqn:U(1)}, choosing $\zeta=-\xi \neq 0$ defines a background which preserves $\mathcal{N}=1$ supersymmetry (with a U(1)$_R$ R-symmetry). When both parameters are zero, we preserve the full $SU(2)_R\times U(1)_R$ R-Symmetry, recovering the $\mathcal{N}=2$ solution \eqref{eq:N=2}. In all other cases, the supersymmetry is completely broken. A summary of this discussion is given in Figure \ref{fig:IIAtableplot}.

\subsubsection{Alternative free parameters}
We now make a brief aside and investigate what solutions we find if we relax the condition that both $q$ and $v$ are free. We will in fact demonstrate that all roads lead back to the two-parameter family given in \eqref{eqn:generalresult1}, as follows:
\begin{itemize}
\item \textbf{Keeping $(q,c)$ free}\\
With $(p,b,u)=1$, to keep $q$ as a free parameter, one must fix $(a,s)=0$. This condition comes from the determinant \eqref{eqn:S2breakingdefns}, which now becomes $vc=0$. To avoid the previous case, we fix $v=0$ with $(q\equiv\xi,m\equiv\gamma)$. The solution derived can be mapped to the $\zeta=0$ solution of \eqref{eqn:generalresult1}, via
\beq
C_1\rightarrow C_1+ c\, d\phi,~~~~\chi\rightarrow \chi+(\gamma-c\,\xi)\phi.
\eeq

\item \textbf{Keeping $(v,s)$ free}\\
The other possibility to keep $v$ a free parameter is to fix $(c,m)=0$, leaving the determinant $qa=0$. We of course should fix $q=0$ with $(v\equiv \zeta,~s\equiv \gamma)$. This now derives the $\xi=0$ solution of \eqref{eqn:generalresult1}, following
\beq
C_1\rightarrow C_1+ a\, d\chi,~~~~~~C_3\rightarrow C_3 + a\,d\chi \wedge B_2,~~~~~~\phi\rightarrow \phi+(\gamma-a\,\zeta)\phi.
\eeq

\item \textbf{Keeping $(a,c)$ free}\\
The final possibility is to enforce that neither $q$ or $v$ are free parameters by fixing both to zero. This then makes $(a,c)$ free, with the determinant becoming $ms=0$. This solution re-derives the $\mathcal{N}=2$ background (i.e. \eqref{eqn:generalresult1} with $(\xi,\zeta)=0$), with the gauge transformations
\beq
C_1\rightarrow C_1+(a d\chi+c d\phi),~~~~~~~~~~~~C_3\rightarrow C_3+(a d\chi+c d\phi)\wedge B_2,
\eeq
and $\phi\rightarrow \phi+s\chi$ or $\chi\rightarrow\chi+m\phi$ (depending on which choice of parameter is made).
\end{itemize}
Let us now return to our two parameter family of solutions and perform investigations at the boundary.
\subsection{Investigations at the boundary}\label{sec:boundary}

We now investigate the two-parameter solution \eqref{eqn:generalresult1} along the $(\sigma,\eta)$ boundary, using the same approach as Section \ref{sec:BoundaryNeqtwo} for the $\mathcal{N}=2$ solution, with the limits of the warp factors given in Appendix \ref{sec:fs}.

Before we begin, we see for general values of $(\eta,\sigma,\theta)$, that $(\Xi,\Pi)$ are non-zero and finite. The deformed $S^2$ given by $(\theta,\phi)$ has $\Pi\rightarrow 1$ at the poles, which given the expression for $M_3$ in \eqref{eqn:generalresult1}, means it still behaves topologically as an $S^2$. To demonstrate this, consider the parametrisation $\sin\theta=x$, for $x$ small at the poles. This leads to $D\chi\to d\chi+ q(\eta,\sigma)x^2d\phi$, with $q$ easily determined. To leading order in $x$, $D\chi\to d\chi$ through $\chi\to \chi-\frac{1}{3}q x^3$, demonstrating that the fibration is topologically trivial at the poles.

 We now turn to the behaviour at the boundaries.
 

  \subsubsection{The $\sigma\rightarrow \infty$ boundary}
   At the $\sigma\rightarrow \infty$ boundary, using \eqref{eqn:Vsigmainfty} to leading order, we find all parameters neatly drop out of the calculation. Hence, the behaviour remains the same as the $\mathcal{N}=2$ solution, re-deriving \eqref{eqn:sigmainftymetric} but now with $X$ switched on, namely
   \begin{align} 
ds^2&=\frac{\kappa  }{X} \bigg[4\sigma\bigg(ds^2(\text{AdS}_5)+d\chi^2\bigg)+\frac{2P}{\pi}\bigg(d\left(\frac{\pi}{P}\sigma\right)^2+ d\left(\frac{\pi}{P}\eta\right)^2+ \sin^2\left(\frac{\pi}{P}\eta\right)ds^2(\text{S}^2)\bigg)\bigg],\nn\\[2mm]
e^{-\Phi}&=X^{\frac{3}{2}}\frac{{\cal R}_1\pi^2}{2 P^{\frac{3}{2}}\sqrt{\kappa}}e^{-\frac{\pi}{P}\sigma}\left(\frac{\pi}{P}\sigma\right)^{-\frac{1}{2}},~~~~H_3=-\frac{1}{X}\frac{4\kappa P}{\pi}\sin^2\left(\frac{\pi}{P}\eta\right) d\left(\frac{\pi}{P}\eta\right)\wedge \text{vol}(\text{S}^2),\nn
\end{align}
with the new coordinate $ \tilde{r}= e^{-\frac{\pi}{P}\sigma}(\frac{\pi}{P}\sigma)^{-\frac{1}{2}}$, we then find to leading order
\begin{equation*}
ds^2=ds^2(\text{Mink}_6)+\frac{2P \kappa}{X \pi\, \tilde{r}^2}\bigg(d\tilde{r}^2+\tilde{r}^2ds^2(\text{S}^3)\bigg),~~~H_3=-\frac{4\kappa P}{X \pi}\text{vol}(\text{S}^3),~~~e^{-\Phi}=X^{\frac{3}{2}}\frac{{\cal R}_1\pi^2}{2 P^{\frac{3}{2}}\sqrt{\kappa}}\tilde{r}.
\end{equation*}
In order to recover the NS5 charge, we simply fix $2\kappa=\pi \widehat{\kappa} X $ (with $\widehat{\kappa}\in\mathds{Z}$). Repeating the procedure in \eqref{eqn:sigmainftymetric2}, leads to
\beq 
Q_{\text{NS5}}=-\frac{1}{(2\pi)^2}\int_{S^3} H_3= \widehat{\kappa}\, P,
\eeq
with a stack of $\widehat{\kappa}P$ NS5 branes at the $\sigma=\infty$ boundary (we will return to the quantization, but for now we keep things general).
\subsubsection{The $\eta=0$ boundary, with $\sigma\neq 0$}
At $\eta=0$ with $\sigma\neq 0$, using \eqref{eq:fdef} and \eqref{eqn:feq}, one finds using the second form for $M_3$ in \eqref{eqn:generalresult1} that $(\eta,\theta,\phi)$ vanish as $\mathds{R}^3$ in polar coordinates, namely
\beq
f_4 d\eta^2+f_2 \bigg(d\theta^2+ \frac{1}{\Pi}\sin^2\theta d\phi^2\bigg)  = \frac{2|\dot{f}|}{\sigma^2\,f}\Big( d\eta^2+\eta^2 ds^2(S^2) \Big),
\eeq
which once again matches the $\mathcal{N}=2$ case, given in \eqref{eqn:eta0metric}.
\paragraph{The $\eta=P$ boundary, with $\sigma\neq 0$} This boundary is qualitatively equivalent to the $\eta=0$ boundary.

\subsubsection{The $\sigma=0$ boundary, with $\eta\in (k,k+1)$}
At $\sigma=0,~\eta\in (k,k+1)$, we recall that along the $\sigma=0$ boundary, $\ddot{V}=0$ to leading order. Following the $\mathcal{N}=2$ procedure, and using \eqref{eqn:fsat0}, we now find
\begin{align}
  &  \Xi=l_k^2+ \frac{1}{2}\zeta^2 \mathcal{R} V'' \sin^2\theta,~~~\Delta\rightarrow l_k^2,~~~D\phi =d\phi -\frac{\zeta (N_{k+1}-N_k)}{l_k} d\chi,~~~~l_k=1+\xi  (N_{k+1}-N_k),\nn
    \end{align} 
   where $\mathcal{R}=\dot{V}$ from the boundary conditions \eqref{eqn:BCs}, and hence $\dot{V}'=\mathcal{R}'=N_{k+1}-N_k$. Note also that $\Xi$ is a nowhere vanishing and finite function of $(\eta,\theta)$, with the connection of $D\phi$ topologically trivial. In addition, from the form of $M_3$ in \eqref{eqn:generalresult1}, we find an $\mathds{R}^2/\mathds{Z}_{l_k}$ orbifold singularity in $(\sigma,\chi)$, as follows
   \beq\label{eqn:orbifold1}
   f_4 d\sigma^2+\frac{f_3 }{\Delta}  d\chi^2  \rightarrow   \frac{2V''}{\mathcal{R}}\bigg( d\sigma^2 +\frac{\sigma^2}{l_k^2}d\chi^2 \bigg).
     \eeq
 Fixing $\xi=0$ recovers the $\mathcal{N}=2$ case and the discussion given below \eqref{eqn:sigma0metric}- as it should. Hence, this deformation parameter, which was introduced by the $SL(3,\mathds{R})$ transformation in $d=11$, has introduced an orbifold singularity to the $\sigma=0$ boundary!
 
     
     \subsubsection{The $\sigma=0,~\eta=0$ boundary}
   As in the $\mathcal{N}=2$ case, to approach $\sigma=0,~\eta=0$, we make the coordinate change \\$(\eta=r \cos\alpha,~\sigma=r \sin\alpha)$, expanding about $r=0$. Using \eqref{eqQdef2} and \eqref{eqQdef}, we now find
  \begin{align}
 & \Xi\rightarrow \Delta \rightarrow l_0^2 ,~~~~~~l_0=1+\xi N_1,~~~~~~~~~D\phi=d\phi-\frac{\zeta N_1}{l_0}d\chi,\\[2mm]
 & f_4(d\sigma^2+d\eta^2) +ds^2(M_3) \rightarrow \frac{2Q}{N_1}\bigg(dr^2+r^2d\alpha^2+r^2\cos^2\alpha \bigg(d\theta^2 +\sin^2\theta \,D\phi^2 \bigg) + r^2\sin^2\alpha \frac{d\chi^2}{l_0^2}\bigg),\nn
  \end{align}
  where we say that the internal space vanishes as $\mathds{R}^5/\mathds{Z}_{l_0}$, with the external space finite. Again, $\xi$ has introduced an orbifold singularity compared to the $\mathcal{N}=2$ case in \eqref{eqn:sigma0eta0metric}.
 
      \paragraph{The $\sigma=0,~\eta=P$ boundary} This boundary is qualitatively equivalent to the $(\sigma,\eta)=0$ boundary.  
\subsubsection{The $\sigma=0$ boundary, with $\eta=k$}
At the $\sigma=0$ boundary with $\eta=k$, we will again follow the $\mathcal{N}=2$ approach by making the coordinate change $(\eta=k-r \cos\alpha,~\sigma=r \sin\alpha)$ for $r\sim0$, using \eqref{eq:VsattheD6s} and \eqref{eqn:fsfork}. We now find for the forms of $\Xi$ and $C_1$ given in \eqref{eqn:generalresult1} that the $f_2/f_5$ terms will dominate, unless we either fix $\zeta=0$, or we are at a pole of the deformed $S^2$ (where $\sin\theta=0$). Hence, we first investigate the $S^2$ preserved case, with $\zeta=0$. We then switch on $\zeta$ and investigate this boundary both away from the pole and approaching it.

\begin{itemize}
\item \underline{$\zeta=0$}: Let us first investigate the $S^2$ preserved ($\zeta=0$) case, which corresponds to the SU$(2)\times$U$(1)$ preserving $\mathcal{N}=0$ deformations (which will be presented explicitly in the next section). We postpone performing a gauge transformation on the $B_2$ for the moment, leading to
\beq
\hspace{-0.5cm}
\frac{ds^2}{2\kappa\sqrt{N_k}}=  \frac{1}{X}\sqrt{\Delta_k}\bigg[\frac{1}{\sqrt{\frac{b_k}{r}}}\bigg(4ds^2(\text{AdS}_5)+ ds^2(\text{S}^2)\bigg)+ \frac{\sqrt{\frac{b_k}{r}}}{N_k}\bigg(dr^2+ r^2 \left(d\alpha^2+\frac{\sin^2\alpha}{\Delta_k} d\chi^2\right)\bigg)\bigg],\nn
\eeq
\begin{align}
&e^{\frac{4}{3}\Phi}=\frac{4\,\kappa^{\frac{2}{3}}\,r}{X^{2}N_k^{\frac{1}{3}}b_k}\Delta_k ,
~~~~~
C_1= X \frac{\frac{1}{4}\xi k^2b_k^2\sin^2\alpha+ g(\alpha)(1+\xi g(\alpha))}{\Delta_k}d\chi,\label{eq:neq0D6s}\nn\\[2mm]
&C_3=-2\kappa  N_k d\chi\wedge \text{Vol}(\text{S}^2),~~~~~~~~~B_2=-\frac{2\kappa }{X} (k+\xi N_k) \text{Vol}(S^2),
\end{align}
with
\beq
\Delta_k= \frac{1}{4}\xi^2b_k^2  \sin^2\alpha+\left(1+\xi g(\alpha)\right)^2,~~~g(\alpha)=\cos^2\left(\frac{\alpha}{2}\right)(N_k-N_{k-1})+\sin^2\left(\frac{\alpha}{2}\right)(N_{k+1}-N_{k})\label{eq:usefulfunction}.
\eeq

At the poles of the deformed 2-sphere spanned by $(\alpha,\chi)$, we have
\beq
\Delta_k(\alpha=0)=l^2_{k-1},~~~~\Delta_k(\alpha=\pi)=l^2_{k},~~~~~~~~l_k=
1+\xi (N_{k+1}-N_k)
,\label{eq:usefulfunctionatpoles}
\eeq
with $\Delta_k$ finite and non zero between these bounds, and observing $l_k=1+\xi \,\mathcal{R}'_{[k,k+1]}$. When $l_k=l_{k-1}=1$, the deformed 2-sphere becomes a round one. However, from the definition of $l_k$ above, we see
\beq
 l_{k-1}-l_{k} = \xi \,b_k,~~~~~~~~b_k=2N_k-N_{k+1}-N_{k-1},
\eeq
with $b_k$ the $\mathcal{N}=2$ D6 brane charge at $\eta=k$ (and $\xi= 0$). Given that $k$ is the position of a kink, we have $b_k\neq0$ here. Hence, we necessarily have $l_k\neq  l_{k-1}$ for $\xi\neq0$, which leads to $\mathbb{R}^2/\mathbb{Z}_{l_{k}}$ and $\mathbb{R}^2/\mathbb{Z}_{l_{k-1}}$ conical singularities at the north and south poles, respectively. This is then the behaviour of the ``spindle'', reviewed in Section \ref{sec:Spindles}, which in the case at hand, is the weighted projective space $\mathbb{WCP}^1_{[l_{k-1},l_{k}]}$. With the behaviour of $\Delta_k$ at the poles, which only depends on $(l_{k-1}^2,l_k^2)$, the metric of the spindle reads
\\[-1cm]
 \begin{figure}[H]
\centering  
\subfigure
{
\centering
  \begin{minipage}{0.6\textwidth}
\beq\label{eqn:Spindlefig}
ds^2\Big(\mathbb{WCP}^1_{[l_{k-1},l_{k}]}\Big)= d\alpha^2+\frac{\sin^2\alpha}{\Delta_k} d\chi^2,
\eeq
  \end{minipage}
     \begin{minipage}{.5\textwidth}
    \centering
\begin{tikzpicture}[scale=1.5, every node/.style={scale=0.9}]
\node[label=above:] at (0.6,0.9){ $|l_k|$};
\node[label=above:] at (0.52,-1.1){ $|l_{k-1}|$};
\node[label=above:] at (2.2,0.75){ $\mathbb{WCP}^1_{[ l_{k-1},l_k ]}$};
  \draw[lightgray] (0,-0.25) arc (180:360:1 and 0.3);
  \draw[lightgray,dashed] (2,-0.25) arc (0:180:1 and 0.3);
  \fill[fill=black] (1,-0.25) circle (0.75pt);
  \draw  plot [smooth, tension=0.7] coordinates {(1,-1)
   (2,-0.25)
    (1,1) };
   \draw plot [smooth, tension=0.7] coordinates {(1,-1) 
   (0,-0.25) 
   (1,1) };
\fill[
  left color=gray!40!black,
  right color=gray!50!black,
  middle color=gray!50,
  shading=axis,
  opacity=0.1,rotate=0
  ] 
  (2,-0.32) -- (1.85,-0.53) -- (1.6,-0.73) -- (1,-1) --(0.4,-0.73)--(0.15,-0.53) -- (0,-0.32)arc (188:365:1cm and -0.4cm);
  
  \fill[
  left color=gray!40!black,
  right color=gray!50!black,
  middle color=gray!50,
  shading=axis,
  opacity=0.1,rotate=0
  ] 
  (2,-0.32) --(1.975,-0.09)  --(1.95,-0.05) -- (1.82,0.13) -- (1.5,0.54) -- (1.1,0.9) --  (1,1) -- (0.9,0.9) -- (0.5,0.54) -- (0.16,0.1 ) --(0.05,-0.05)--(0.025,-0.09)  -- (0,-0.32)arc (188:365:1cm and -0.4cm);
\end{tikzpicture}
  \end{minipage} 
}
    \label{fig:spindle}
\end{figure}
with $\big(|l_{k-1}|,|l_k|\big)$ defining the conical deficit angles, $(\varphi_{k-1}, \varphi_{k})\in [0,2\pi)$, at the south and north poles, respectively. The definitions read (see Figure \ref{fig:conicaldeficit})
\begin{align}
&\varphi_{k-1}=2\pi \Big(1-\frac{1}{|l_{k-1}|}\Big),~~~~~~~\varphi_k=2\pi \Big(1-\frac{1}{|l_k|}\Big),\\[2mm]
&~~~~~\text{with}~~~~~~(| l_{k-1}|,|l_k|)\in [1,\infty),\nn
\end{align}
recalling that $\big(|l_{k-1}|,|l_k|\big)$ are considered coprime (or `relatively prime') positive integers with $\text{gcd}\big(|l_{k-1}|,|l_k|\big)=1$ and $|l_{k-1}|\neq |l_k|$. 

These conditions on the conical deficit angles are not trivial, and provide fairly strong constraints on the possible rank functions which give rise to a nice spindle description. Given the values of $\Delta_k$ at the poles depend only on the square of $l_k$ (and $l_{k-1}$), and the requirement that conical deficit angles depend on positive integers (greater or equal to one), we conclude the conical deficit angles depend only on the absolute value of $l_k$ (not $l_k$ itself). Recall from the form of $\mathcal{R}'$ in Figure \ref{fig:Rankfull} that $(N_{k+1}-N_k)$ is a step-wise decreasing function, but the value of $|l_k|=|1+\xi (N_{k+1}-N_k)|$ is more subtle. For instance, we can now have situations where e.g. $|l_{k-1}|=|a|=|-a|=|l_{k}|$ for some $a\in\mathds{Z}$. These values of $(l_{k-1},l_k)$ are perfectly allowed from the perspective of the rank function, but would not satisfy the spindle constraints - firstly because $|a|=1$ would recover the $S^2$, and secondly because one requires $|l_{k-1}|\neq|l_k|$. 
The $\text{gcd}\big(|l_{k-1}|,|l_k|\big)=1$ condition further constrains the values of $N_k$ for a given $\xi$. In addition, we can find examples of the `tear-drop' within these solutions (with either $|l_{k-1}|$ or $|l_{k}|$ equal to one). It would be nice to consider further whether these tear-drop orbifolds are also `allowed' within our solutions. We will take a closer look at the Triangular rank function example in the next section.

Now, using the form of the metric and dilaton given in \eqref{eq:neq0D6s}, along with the result for a generic Dp-brane \eqref{eqn:Dbranemetrics}, 
we can easily see the metric describes D6 branes extended in (AdS$_5$,~S$^2$) and back-reacted on a cone whose base is $\mathbb{WCP}^1_{[l_{k-1},l_{k}]}$ - see Figure \ref{fig:D6orthogonaltospindle1}. Note that this is different to the cases discussed in Section \ref{sec:Spindles}, where the D branes actually wrap the spindle.
\begin{figure}[H]
\begin{center}
\begin{tikzpicture}[scale=1.2, every node/.style={scale=0.9}]
\draw[draw=black,line width=0.25mm] (-3,-3) rectangle ++(2,3);
  \shade[ color = gray!40, opacity = 0.5]  (-3,-3) rectangle ++(2,3);
  
\draw(-1,0)--(1,1);
\draw(-1,0)--(0.52,-0.78);
\fill[
  left color=gray!50!black,
  right color=gray!50!black,
  middle color=gray!50,
  shading=axis,
  opacity=0.1,rotate=90
  ] 
 (0.2,-0.21)-- (0.4,-0.39)
  -- (1,-1) -- (0,1) -- (0,-0.1) arc (184:360:0cm and 0cm);
  \fill[
  left color=gray!50!black,
  right color=gray!50!black,
  middle color=gray!50,
  shading=axis,
  opacity=0.1,rotate=90
  ] 
  (-0.2,0)-- (-0.3,0.005) -- (-0.5,-0.12) -- (-0.7,-0.4) -- (0,1) -- (0,-0.1) arc (184:360:0cm and 0cm);
  

\node[label=above:] at (1.55,0.9){ $|l_k|$};
\node[label=above:] at (1.73,-1.02){ $|l_{k-1}|$};
\node[label=above:] at (1.5,-1.6){ $\mathbb{WCP}^1_{[ l_{k-1},l_k ]}$};
\node[label=above:] at (-1.8,-0.25){AdS$_5\times S^2$};
\node[label=above:] at (-2.6,-2.8){\textbf{D6}$^k$};
  \draw[lightgray] (0,-0.25) arc (180:360:1 and 0.3);
  \draw[lightgray,dashed] (2,-0.25) arc (0:180:1 and 0.3);
  \fill[fill=black] (1,-0.25) circle (0.75pt);
  \draw  plot [smooth, tension=0.7] coordinates {(1,-1)
   (2,-0.25)
    (1,1) };
   \draw plot [smooth, tension=0.7] coordinates {(1,-1) 
   (0,-0.25) 
   (1,1) };
\fill[
  left color=gray!40!black,
  right color=gray!50!black,
  middle color=gray!50,
  shading=axis,
  opacity=0.1,rotate=0
  ] 
  (2,-0.32) -- (1.85,-0.53) -- (1.6,-0.73) -- (1,-1) --(0.4,-0.73)--(0.15,-0.53) -- (0,-0.32)arc (188:365:1cm and -0.4cm);
  
  \fill[
  left color=gray!40!black,
  right color=gray!50!black,
  middle color=gray!50,
  shading=axis,
  opacity=0.1,rotate=0
  ] 
  (2,-0.32) --(1.975,-0.09)  --(1.95,-0.05) -- (1.82,0.13) -- (1.5,0.54) -- (1.1,0.9) --  (1,1) -- (0.9,0.9) -- (0.5,0.54) -- (0.16,0.1 ) --(0.05,-0.05)--(0.025,-0.09)  -- (0,-0.32)arc (188:365:1cm and -0.4cm);
\end{tikzpicture}
\end{center}
\caption{At each kink of the rank function, $\eta=k$, one finds a stack of D6 branes extended in (AdS$_5$,~S$^2$) and orthogonal to a cone whose base is the $\mathbb{WCP}^1_{[l_{k-1},l_{k}]}$ spindle - with conical deficit angles defined by the slope of the rank function at that kink. }
\label{fig:D6orthogonaltospindle1}
\end{figure}
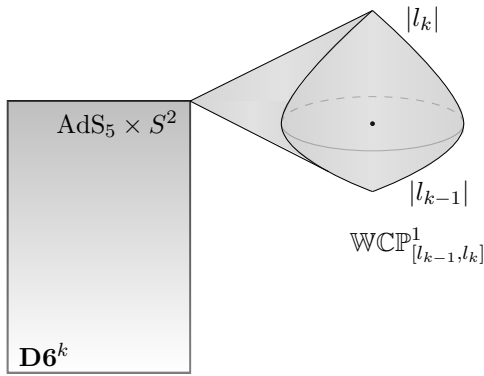
Calculating the charge of each stack of D6 branes, we find
\beq
Q^k_6=-\frac{1}{2\pi}\int_{\mathbb{WCP}^1_{[l_{k-1},l_{k}]}}F_2=-\frac{1}{2\pi}\int_{\chi=0}^{\chi=2\pi} C_1\bigg\lvert_{\alpha=0}^{\alpha=\pi}= \frac{X }{l_{k}l_{k-1}}(2N_k-N_{k+1}-N_{k-1}),\label{quantcond1}
\eeq
yielding precisely the rational quantisation condition one should get when integrating over the spindle. 
Using analogous arguments to \eqref{eqn:gluing} (see \cite{Ferrero:2021etw}), this follows because the Euler characteristic on the spindle is itself rational, namely
\beq\label{eqn:Eulerspindle}
\chi_E=\frac{1}{2\pi}\int_{\mathbb{WCP}^1_{[l_{k-1},l_{k}]}}R\, \text{vol}_2= \frac{|l_{k-1}|+|l_k|}{|l_{k-1}l_k|}=2-\left(1-\frac{1}{|l_k|}\right)-\left(1-\frac{1}{|l_{k-1}|}\right),
\eeq
where $\text{vol}_2$ is the volume form on $\mathbb{WCP}^1_{[l_{k-1},l_{k}]}$. For a $\mathbb{CP}^1$ the Euler characteristic is $\chi_E=2$, which is clearly recovered when $l_k=l_{k-1}=1$ (and $\xi=0$), as it should be. The total charge of the D6 branes then read
\beq
Q_{D6}=\sum_{k=1}^{P-1}Q^k_{D6}=\frac{X}{l_0 l_P}(N_{P-1}+N_1),~~~~~~l_0=1+\xi N_1,~~~~~l_P=1-\xi N_{P-1}.
\eeq
It is important to observe from \eqref{quantcond1} that each D6 brane stack, located at each of the different kinks of the rank function, has a different multiplicate factor for its rational charge - given by $1/(l_kl_{k-1})$. This is because each stack of D6 branes is orthogonal to a different spindle - \textit{i.e} the D6 brane stack at $\eta=k-1$ is orthogonal to $\mathbb{WCP}^1_{[l_{k-2},l_{k-1}]}$, whereas the stack at $\eta=k$ is orthogonal to $\mathbb{WCP}^1_{[l_{k-1},l_{k}]}$ (with different conical deficit angles). This is once again different to the cases reviewed in Section \ref{sec:Spindles}. 
Instinctively, the conical deficit angle at the north pole of the $\eta=k-1$ spindle matches the south pole conical deficit angle of the $\eta=k$ spindle (defined by $|l_{k-1}|$), and so on.  See Figure \ref{fig:D6orthogonaltospindless2}. Note, imposing the condition $|l_{k-1}|>|l_k|$ (as in \cite{Boido:2021szx,Ferrero:2021wvk}) for a `Besse' spindle (see \cite{Lange1}), would imply further restrictions onto the allowed rank functions. Consequently, the $k=1$ spindle (on the far left of  Figure \ref{fig:D6orthogonaltospindless2}) would require a narrow peak at the south pole, becoming ever more broad for increasing $\eta$. However, such restrictions could be loosened by considering flipping the orientation of neighbouring spindles, such that the conical deficit angle at the north pole of the $\eta=k$ spindle matches the north pole conical deficit angle of the (flipped) $\eta=k+1$ spindle (defined by $l_k$). Hence, we simply leave things more general here, considering cases for which $|l_{k-1}|\neq |l_k|$.  

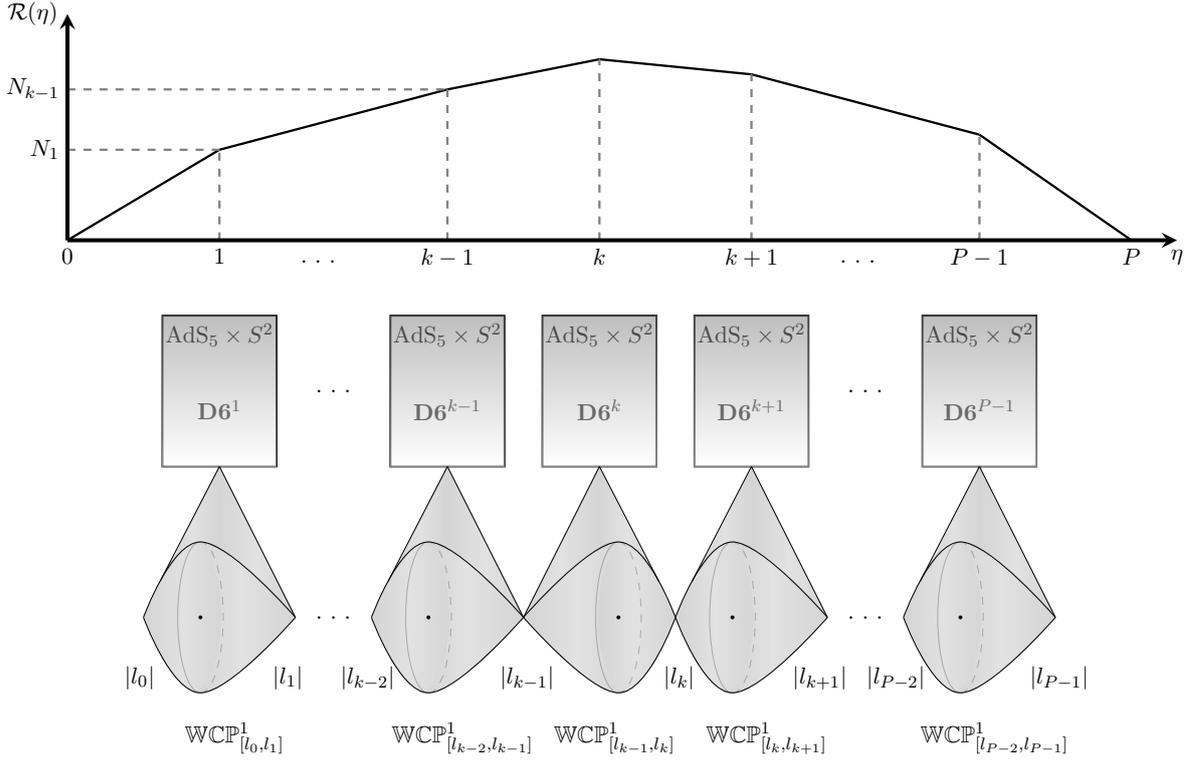
\begin{figure}[H]
\begin{center}
\begin{tikzpicture}[scale=1, every node/.style={scale=0.8}]

  \begin{scope}[xshift=-8cm,yshift=4cm]
\draw[black,line width=0.3mm] (1,0) -- (3,1.2);
\draw[black,line width=0.3mm] (3,1.2) -- (6,2);
\draw[black,line width=0.3mm] (6,2) -- (8,2.4);
\draw[black,line width=0.3mm] (8,2.4)-- (10,2.2);
\draw[black,line width=0.3mm]  (10,2.2)-- (13,1.4);
\draw[black,line width=0.3mm] (13,1.4)-- (15,0);

\draw (1,0) node[below] {$0$};
\draw[gray,dashed,line width=0.3mm] (3,0)-- (3,1.2);
\draw (3,0) node[below] {$1$};
\draw[gray,dashed,line width=0.3mm] (6,0)-- (6,2);
\draw (6,0) node[below] {$k-1$};
\draw[gray,dashed,line width=0.3mm] (8,0)--  (8,2.4);
\draw (8,0) node[below] {$k$};
\draw (4.3,-0.15) node[below] {$.~.~.$};
\draw[gray,dashed,line width=0.3mm] (10,0)-- (10,2.2);
\draw (10,0) node[below] {$k+1$};
\draw (13,0) node[below] {$P-1$};
\draw[gray,dashed,line width=0.3mm] (13,0)--  (13,1.4);
\draw (15,0) node[below] {$P$};
\draw (11.4,-0.15) node[below] {$.~.~.$};

\draw[gray,dashed,line width=0.3mm] (1,1.2)--  (3,1.2);
\draw (1,1.2) node[left] {$N_1$};
\draw[gray,dashed,line width=0.3mm] (1,2)-- (6,2);
\draw (1,2) node[left] {$N_{k-1}$};

\draw[-stealth, line width=0.53mm] (1,0)--(1,3) node[left ]{$\mathcal{R}(\eta)$};
\draw[-stealth, line width=0.53mm] (1,0) --(15.6,0) node[below ]{$\eta$};
\end{scope}

   \begin{scope}[xshift=-3cm]
\node[label=above:] at (-1.1,-1.8){ $| l_1 |$};
\node[label=above:] at (-3.05,-1.8){ $| l_{0}|$};
\node[label=above:] at (-1.8,-2.6){ $\mathbb{WCP}^1_{[l_{0},l_1]}$};
\node[label=above:] at (-2,2.75){AdS$_5\times S^2$};
\node[label=above:] at (-2,1.75){\textbf{D6}$^1$};
\node[label=above:] at (-0.5,2){$.~.~.$};
 \node[label=above:] at (-0.5,-1){$.~.~.$};


\draw[draw=black,line width=0.25mm] (-2.75,1) rectangle ++(1.5,2);
  \shade[ color = gray!40, opacity = 0.5]  (-2.75,1) rectangle ++(1.5,2);
  
   \begin{scope}[xshift=-2cm,rotate=270]
  \draw[lightgray] (0,-0.25) arc (180:360:1 and 0.3);
  \draw[lightgray,dashed] (2,-0.25) arc (0:180:1 and 0.3);
  \fill[fill=black] (1,-0.25) circle (0.75pt);
  
\draw(-1,0)--(1,1);
\draw(-1,0)--(0.52,-0.78);

  \fill[
  left color=gray!40!black,
  right color=gray!50!black,
  middle color=gray!50,
  shading=axis,
  opacity=0.1,rotate=0
  ] 
  (2,-0.32) -- (1.85,-0.53) -- (1.6,-0.73) -- (1,-1) --(0.4,-0.73)--(0.15,-0.53) -- (0,-0.32)arc (188:365:1cm and -0.4cm);
  
  \fill[
  left color=gray!40!black,
  right color=gray!50!black,
  middle color=gray!50,
  shading=axis,
  opacity=0.1,rotate=0
  ] 
  (2,-0.32) --(1.975,-0.09)  --(1.95,-0.05) -- (1.82,0.13) -- (1.5,0.54) -- (1.1,0.9) --  (1,1) -- (0.9,0.9) -- (0.5,0.54) -- (0.16,0.1 ) --(0.05,-0.05)--(0.025,-0.09)  -- (0,-0.32)arc (188:365:1cm and -0.4cm);
  
  \draw  plot [smooth, tension=0.7] coordinates {(1,-1)
   (2,-0.25)
    (1,1) };
   \draw plot [smooth, tension=0.7] coordinates {(1,-1) 
   (0,-0.25) 
   (1,1) };
   
\fill[
  left color=gray!50!black,
  right color=gray!50!black,
  middle color=gray!50,
  shading=axis,
  opacity=0.1,rotate=90
  ] 
 (0.2,-0.21)-- (0.4,-0.39)
  -- (1,-1) -- (0,1) -- (0,-0.1) arc (184:360:0cm and 0cm);
  \fill[
  left color=gray!50!black,
  right color=gray!50!black,
  middle color=gray!50,
  shading=axis,
  opacity=0.1,rotate=90
  ] 
  (-0.2,0)-- (-0.3,0.005) -- (-0.5,-0.12) -- (-0.7,-0.4) -- (0,1) -- (0,-0.1) arc (184:360:0cm and 0cm);
  \end{scope}
    \end{scope}

   \begin{scope}[xshift=0]
\node[label=above:] at (-0.95,-1.8){ $|l_{k-1}|$};
\node[label=above:] at (-3.05,-1.8){ $|l_{k-2}|$};
\node[label=above:] at (-1.8,-2.6){ $\mathbb{WCP}^1_{[l_{k-2},l_{k-1}]}$};
\node[label=above:] at (-2,2.75){AdS$_5\times S^2$};
\node[label=above:] at (-2,1.75){\textbf{D6}$^{k-1}$};


\draw[draw=black,line width=0.25mm] (-2.75,1) rectangle ++(1.5,2);
  \shade[ color = gray!40, opacity = 0.5]  (-2.75,1) rectangle ++(1.5,2);
  
    \begin{scope}[xshift=-2cm,rotate=270]
  \draw[lightgray] (0,-0.25) arc (180:360:1 and 0.3);
  \draw[lightgray,dashed] (2,-0.25) arc (0:180:1 and 0.3);
  \fill[fill=black] (1,-0.25) circle (0.75pt);
  
\draw(-1,0)--(1,1);
\draw(-1,0)--(0.52,-0.78);

  \fill[
  left color=gray!40!black,
  right color=gray!50!black,
  middle color=gray!50,
  shading=axis,
  opacity=0.1,rotate=0
  ] 
  (2,-0.32) -- (1.85,-0.53) -- (1.6,-0.73) -- (1,-1) --(0.4,-0.73)--(0.15,-0.53) -- (0,-0.32)arc (188:365:1cm and -0.4cm);
  
  \fill[
  left color=gray!40!black,
  right color=gray!50!black,
  middle color=gray!50,
  shading=axis,
  opacity=0.1,rotate=0
  ] 
  (2,-0.32) --(1.975,-0.09)  --(1.95,-0.05) -- (1.82,0.13) -- (1.5,0.54) -- (1.1,0.9) --  (1,1) -- (0.9,0.9) -- (0.5,0.54) -- (0.16,0.1 ) --(0.05,-0.05)--(0.025,-0.09)  -- (0,-0.32)arc (188:365:1cm and -0.4cm);
  
  \draw  plot [smooth, tension=0.7] coordinates {(1,-1)
   (2,-0.25)
    (1,1) };
   \draw plot [smooth, tension=0.7] coordinates {(1,-1) 
   (0,-0.25) 
   (1,1) };
   
\fill[
  left color=gray!50!black,
  right color=gray!50!black,
  middle color=gray!50,
  shading=axis,
  opacity=0.1,rotate=90
  ] 
 (0.2,-0.21)-- (0.4,-0.39)
  -- (1,-1) -- (0,1) -- (0,-0.1) arc (184:360:0cm and 0cm);
  \fill[
  left color=gray!50!black,
  right color=gray!50!black,
  middle color=gray!50,
  shading=axis,
  opacity=0.1,rotate=90
  ] 
  (-0.2,0)-- (-0.3,0.005) -- (-0.5,-0.12) -- (-0.7,-0.4) -- (0,1) -- (0,-0.1) arc (184:360:0cm and 0cm);
  \end{scope}
    \end{scope}

      \begin{scope}[xshift=2cm]
\node[label=above:] at (-0.95,-1.8){ $|l_k|$};
\node[label=above:] at (-1.8,-2.6){ $\mathbb{WCP}^1_{[l_{k-1},l_k]}$};
\node[label=above:] at (-2,2.75){AdS$_5\times S^2$};
\node[label=above:] at (-2,1.75){\textbf{D6}$^k$};


\draw[draw=black,line width=0.25mm] (-2.75,1) rectangle ++(1.5,2);
  \shade[ color = gray!40, opacity = 0.5]  (-2.75,1) rectangle ++(1.5,2);
  
   \begin{scope}[xshift=-2cm,rotate=90,yscale=1,xscale=-1]
    \draw[lightgray,dashed] (0,-0.25) arc (180:360:1 and 0.3);
  \draw[lightgray] (2,-0.25) arc (0:180:1 and 0.3);
  \fill[fill=black] (1,-0.25) circle (0.75pt);
\draw(-1,0)--(1,1);
\draw(-1,0)--(0.52,-0.78);
  \fill[
  left color=gray!40!black,
  right color=gray!50!black,
  middle color=gray!50,
  shading=axis,
  opacity=0.1,rotate=0
  ] 
  (2,-0.32) -- (1.85,-0.53) -- (1.6,-0.73) -- (1,-1) --(0.4,-0.73)--(0.15,-0.53) -- (0,-0.32)arc (188:365:1cm and -0.4cm);
  
  \fill[
  left color=gray!40!black,
  right color=gray!50!black,
  middle color=gray!50,
  shading=axis,
  opacity=0.1,rotate=0
  ] 
  (2,-0.32) --(1.975,-0.09)  --(1.95,-0.05) -- (1.82,0.13) -- (1.5,0.54) -- (1.1,0.9) --  (1,1) -- (0.9,0.9) -- (0.5,0.54) -- (0.16,0.1 ) --(0.05,-0.05)--(0.025,-0.09)  -- (0,-0.32)arc (188:365:1cm and -0.4cm);
  
  \draw  plot [smooth, tension=0.7] coordinates {(1,-1)
   (2,-0.25)
    (1,1) };
   \draw plot [smooth, tension=0.7] coordinates {(1,-1) 
   (0,-0.25) 
   (1,1) };
\fill[
  left color=gray!50!black,
  right color=gray!50!black,
  middle color=gray!50,
  shading=axis,
  opacity=0.1,rotate=90
  ] 
 (0.2,-0.21)-- (0.4,-0.39)
  -- (1,-1) -- (0,1) -- (0,-0.1) arc (184:360:0cm and 0cm);
  \fill[
  left color=gray!50!black,
  right color=gray!50!black,
  middle color=gray!50,
  shading=axis,
  opacity=0.1,rotate=90
  ] 
  (-0.2,0)-- (-0.3,0.005) -- (-0.5,-0.12) -- (-0.7,-0.4) -- (0,1) -- (0,-0.1) arc (184:360:0cm and 0cm);
  \end{scope}
    \end{scope}

      \begin{scope}[xshift=4cm]
\node[label=above:] at (-1.1,-1.8){ $|l_{k+1}|$};
\node[label=above:] at (-1.8,-2.6){ $\mathbb{WCP}^1_{[l_{k},l_{k+1}]}$};
\node[label=above:] at (-2,2.75){AdS$_5\times S^2$};
\node[label=above:] at (-2,1.75){\textbf{D6}$^{k+1}$};


\draw[draw=black,line width=0.25mm] (-2.75,1) rectangle ++(1.5,2);
  \shade[ color = gray!40, opacity = 0.5]  (-2.75,1) rectangle ++(1.5,2);
  
    \begin{scope}[xshift=-2cm,rotate=270]
  \draw[lightgray] (0,-0.25) arc (180:360:1 and 0.3);
  \draw[lightgray,dashed] (2,-0.25) arc (0:180:1 and 0.3);
  \fill[fill=black] (1,-0.25) circle (0.75pt);
  
\draw(-1,0)--(1,1);
\draw(-1,0)--(0.52,-0.78);

  \fill[
  left color=gray!40!black,
  right color=gray!50!black,
  middle color=gray!50,
  shading=axis,
  opacity=0.1,rotate=0
  ] 
  (2,-0.32) -- (1.85,-0.53) -- (1.6,-0.73) -- (1,-1) --(0.4,-0.73)--(0.15,-0.53) -- (0,-0.32)arc (188:365:1cm and -0.4cm);
  
  \fill[
  left color=gray!40!black,
  right color=gray!50!black,
  middle color=gray!50,
  shading=axis,
  opacity=0.1,rotate=0
  ] 
  (2,-0.32) --(1.975,-0.09)  --(1.95,-0.05) -- (1.82,0.13) -- (1.5,0.54) -- (1.1,0.9) --  (1,1) -- (0.9,0.9) -- (0.5,0.54) -- (0.16,0.1 ) --(0.05,-0.05)--(0.025,-0.09)  -- (0,-0.32)arc (188:365:1cm and -0.4cm);
  
  \draw  plot [smooth, tension=0.7] coordinates {(1,-1)
   (2,-0.25)
    (1,1) };
   \draw plot [smooth, tension=0.7] coordinates {(1,-1) 
   (0,-0.25) 
   (1,1) };
   
\fill[
  left color=gray!50!black,
  right color=gray!50!black,
  middle color=gray!50,
  shading=axis,
  opacity=0.1,rotate=90
  ] 
 (0.2,-0.21)-- (0.4,-0.39)
  -- (1,-1) -- (0,1) -- (0,-0.1) arc (184:360:0cm and 0cm);
  \fill[
  left color=gray!50!black,
  right color=gray!50!black,
  middle color=gray!50,
  shading=axis,
  opacity=0.1,rotate=90
  ] 
  (-0.2,0)-- (-0.3,0.005) -- (-0.5,-0.12) -- (-0.7,-0.4) -- (0,1) -- (0,-0.1) arc (184:360:0cm and 0cm);
  \end{scope}
    \end{scope}

   \begin{scope}[xshift=7cm]
\node[label=above:] at (-0.95,-1.8){ $|l_{P-1}|$};
\node[label=above:] at (-3.1,-1.8){ $|l_{P-2}|$};
\node[label=above:] at (-1.8,-2.6){ $\mathbb{WCP}^1_{[l_{P-2},l_{P-1}]}$};
\node[label=above:] at (-2,2.75){AdS$_5\times S^2$};
\node[label=above:] at (-2,1.75){\textbf{D6}$^{P-1}$};
 \node[label=above:] at (-3.5,2){$.~.~.$};
 \node[label=above:] at (-3.5,-1){$.~.~.$};
 

\draw[draw=black,line width=0.25mm] (-2.75,1) rectangle ++(1.5,2);
  \shade[ color = gray!40, opacity = 0.5]  (-2.75,1) rectangle ++(1.5,2);
  
    \begin{scope}[xshift=-2cm,rotate=270]
  \draw[lightgray] (0,-0.25) arc (180:360:1 and 0.3);
  \draw[lightgray,dashed] (2,-0.25) arc (0:180:1 and 0.3);
  \fill[fill=black] (1,-0.25) circle (0.75pt);
  
\draw(-1,0)--(1,1);
\draw(-1,0)--(0.52,-0.78);

  \fill[
  left color=gray!40!black,
  right color=gray!50!black,
  middle color=gray!50,
  shading=axis,
  opacity=0.1,rotate=0
  ] 
  (2,-0.32) -- (1.85,-0.53) -- (1.6,-0.73) -- (1,-1) --(0.4,-0.73)--(0.15,-0.53) -- (0,-0.32)arc (188:365:1cm and -0.4cm);
  
  \fill[
  left color=gray!40!black,
  right color=gray!50!black,
  middle color=gray!50,
  shading=axis,
  opacity=0.1,rotate=0
  ] 
  (2,-0.32) --(1.975,-0.09)  --(1.95,-0.05) -- (1.82,0.13) -- (1.5,0.54) -- (1.1,0.9) --  (1,1) -- (0.9,0.9) -- (0.5,0.54) -- (0.16,0.1 ) --(0.05,-0.05)--(0.025,-0.09)  -- (0,-0.32)arc (188:365:1cm and -0.4cm);
  
  \draw  plot [smooth, tension=0.7] coordinates {(1,-1)
   (2,-0.25)
    (1,1) };
   \draw plot [smooth, tension=0.7] coordinates {(1,-1) 
   (0,-0.25) 
   (1,1) };
   
\fill[
  left color=gray!50!black,
  right color=gray!50!black,
  middle color=gray!50,
  shading=axis,
  opacity=0.1,rotate=90
  ] 
 (0.2,-0.21)-- (0.4,-0.39)
  -- (1,-1) -- (0,1) -- (0,-0.1) arc (184:360:0cm and 0cm);
  \fill[
  left color=gray!50!black,
  right color=gray!50!black,
  middle color=gray!50,
  shading=axis,
  opacity=0.1,rotate=90
  ] 
  (-0.2,0)-- (-0.3,0.005) -- (-0.5,-0.12) -- (-0.7,-0.4) -- (0,1) -- (0,-0.1) arc (184:360:0cm and 0cm);
  \end{scope}
    \end{scope}

\end{tikzpicture}
\end{center}
\caption{The stack of D6 branes located at each kink of the rank function are orthogonal to a different spindle, with conical deficit angles defined by that kink. The conical deficit angle at the north pole of the $k-1$ spindle, $|l_{k-1}|$, matches that of the south pole of the $k$ spindle. One is then led to the depicted set up, with $\text{gcd}\big(|l_{k-1}|,|l_k|\big)=1$ and $|l_{k-1}| \neq |l_{k}|$. }
\label{fig:D6orthogonaltospindless2}
\end{figure}

Given that our solutions stem from a $d=11$ origin, it should not be surprising that we find the behaviour of D6 branes back-reacted on a spindle. Indeed, beginning with the following embedding of the Taub-Nut metric into $d=11$ 
\beq
ds^2=ds^2(\mathbb{R}^{1,6})+  h\left(dr^2+ r^2(d\alpha^2+\sin^2\alpha d\chi^2)\right)+ \frac{1}{h}(d\beta+N \cos\alpha d\chi)^2,~~~~h=1+\frac{M}{r}\nn,\label{eq:tndef}
\eeq
and transforming $\beta\to \beta+\frac{1}{2}(N_{k+1}-N_{k-1})\chi$ followed by $\chi\to\chi+\xi \beta$, before reducing to IIA along $\beta$, the resulting metric and 2-form are very similar to \eqref{eq:neq0D6s} (but for D6  branes extended on $\mathbb{R}^{1,6}$). The singular behaviour of \eqref{eq:neq0D6s} close to $r=0$ is reproduced, but with AdS$_5\to \text{Mink}_5$ and $M=b_k$. This solution also breaks supersymmetry for $\xi\neq0$.

We now investigate the Page charge for D4 branes at $\eta=k$ (and the loci of the D6 branes). Naively adopting the large gauge transformation used in the $\mathcal{N}=2$ analysis for the $B_2$ given in \eqref{eqn:generalresult1}, we find for the $k$'th cell 
\begin{align}\label{eqn:Gauge1}
&  B_2^k\rightarrow B_2^k+\frac{2\kappa  }{X}k \text{vol}(S^2),~~~~~~~~B^{k}_2=\frac{1}{X}\left( f_8+\xi f_7+2\kappa k\right) \text{vol}(\text{S}^2), 
\nn\\[2mm]
&\hat F_4=  d\left(f_7-\frac{ \Big(f_6(1 + \xi f_6) +\,\xi \frac{f_3}{f_5} \Big) (f_8+\xi f_7+2\kappa k)}{\Delta}\right)\wedge d\chi\wedge\text{vol}(\text{S}^2).
\end{align}
 Using \eqref{eqn:fsfork} for this boundary, we find $B_2^k=-\frac{2\kappa}{X} \xi N_k\text{Vol}(\text{S}^2)$. Calculating the limit of $\hat F_4$ and integrating carefully, recalling the semi-circular contour discussion given below \eqref{eqn:pageF4betweenkandkplusone}, we find (using $2\kappa=\pi \widehat{\kappa} X $)
\begin{align}\label{eqn:D4orig}
&Q_{D4}^k=-\frac{1}{(2\pi)^3}\int_{S^2\times\mathbb{WCP}^1_{[l_{k-1},l_k]}}\hat{F}_4=\widehat{\kappa} X  \bigg( \frac{N_k}{l_k}-\frac{N_{k-1}}{l_{k-1}}\bigg),
\end{align}
with total charge
\beq\label{eqn:D4origT}
Q_{D4}=\sum_{k=1}^{P-1}Q^k_{D4}= \frac{ \widehat{\kappa} X }{l_P}N_{P-1} .
\eeq
These colour branes now have a rational charge due to the spindle. However, if we add an additional term to the gauge transformation, noting $(\xi,N_k)\in \mathds{Z}$,
\begin{align}
&   B_2^k\rightarrow B_2^k+\frac{2\kappa}{X} (k+ \xi N_k)\text{Vol}(S^2),~~~~~~~~B^{k}_2=\frac{1}{X}\Big( f_8+\xi f_7+2\kappa  (k+ \xi N_k)\Big) \text{vol}(\text{S}^2), 
\nn\\[2mm]
&\hat F_4=  d\left(f_7-\frac{ \Big(f_6(1 + \xi f_6) +\,\xi \frac{f_3}{f_5} \Big)\Big( f_8+\xi f_7+2\kappa  (k+ \xi N_k)\Big) }{\Delta}\right)\wedge d\chi\wedge\text{vol}(\text{S}^2),
\end{align}
we instead find $B_2^k=0$ at this boundary. Taking the limit of this new $\hat F_4$, and integrating carefully, we derive
\begin{align}\label{eqn:D4orig2}
&Q_{D4}^k=-\frac{1}{(2\pi)^3}\int_{S^2\times \mathbb{WCP}^1_{[l_{k-1},l_k]}}\hat{F}_4= \widehat{\kappa} X (N_k - N_{k-1} ),
\end{align}
eliminating the rational charge and recovering the $\mathcal{N}=2$ quantization. Hence, we have eliminated the rational D4 charge by an appropriate gauge transformation of $B_2^k$. The total D4 charge now becomes
\beq
Q_{D4}=\sum_{k=1}^{P-1}Q^k_{D4}=  \widehat{\kappa} X   N_{P-1}.
\eeq
This is potentially allowed due to the non-dynamical (pure flux) nature of the D4s.

\item\underline{$\zeta\neq0$}:
switching on $\zeta$, we find that in general, the $f_2/f_5$ term dominates in the $\Xi$ and $C_1$ given in \eqref{eqn:generalresult1} - unless we are at a pole of the deformed $S^2$ (where $\sin\theta=0$). Let us first assume that we are indeed away from one of the poles.
\paragraph{Away from a pole:} in this case, we once again expand $(\eta=k-r \cos\alpha,~\sigma=r \sin\alpha)$ in small $r$, finding to leading order
\begin{align}
&\Xi\rightarrow \frac{b_k\zeta^2 N_k\sin^2\theta}{4r},~~~~~~~~~\Delta\rightarrow  \Delta_k,~~~~~~~~~~~   
\Delta_k\equiv  \frac{1}{4}\xi^2 b_k^2 \sin^2\alpha +\Big(1+ \xi g(\alpha)\Big)^2,\nn\\[2mm]
&\Delta_k(\alpha=0)=l_{k-1}^2,~~~~~~\Delta_k(\alpha=\pi)=l_{k}^2,
\end{align}
with $\Delta_k$ the smooth and nowhere zero function given in \eqref{eq:usefulfunction}. 
Using $r=z^2$, 
\begin{align}\label{eqn:B3eq}
& ds^2= \frac{|\zeta| \kappa }{X} \sin\theta \Big[N_k\Big(4ds^2(AdS_5) +d\theta^2\Big) +4b_k\Big(dz^2 +z^2 ds^2(\mathbb{B}_3)\Big)\Big],\nn\\[2mm]
&ds^2(\mathbb{B}_3)= \frac{1}{4}\Big(d\alpha^2 + \frac{\sin^2\alpha}{\Delta_k}d\chi^2\Big)+\frac{\Delta_k}{\zeta^2 b_k^2}(d\phi+\mathcal{A}_k)^2,~~~~~~e^{4\Phi}= \frac{\kappa^2 \zeta^6}{X^6} N_k^2\sin^6\theta, \nn\\[2mm]
&\mathcal{A}_k=-\frac{\zeta}{\Delta_k}\Big[g(\alpha)\Big(1+\xi g(\alpha)\Big)+\frac{\xi}{4}b_k^2\sin^2\alpha\Big]d\chi, ~~~~~~~C_1=\frac{ X}{\zeta}d\phi, \\[2mm]
&B_2= -\frac{2\kappa}{X} N_k\,\sin\theta \Big(\zeta  d\chi - \xi  d\phi\Big)\wedge d\theta,~~~~~C_3=-2\kappa N_k\sin\theta  d\theta\wedge d\phi\wedge d\chi.\nn
\end{align}
 We see here that the $(z,\mathbb{B}_3)$ sub-manifold describes a cone of base $\mathbb{B}_3$, and the rest of the space has a constant warping (ignoring the overall $\sin\theta$). In addition, the sub-manifold $\mathbb{B}_3$ is itself clearly a $U(1)$ fibration over $\mathbb{WCP}^1_{[l_{k-1},l_k]}$, with
\begin{align}\label{eqn:B3stuff}
&ds^2(\mathbb{B}_3)\bigg|_{\alpha\sim 0} =  \frac{1}{4}\Big(d\alpha^2 + \frac{\alpha^2}{l_{k-1}^2}d\chi^2\Big)+\frac{l_{k-1}^2}{\zeta^2 b_k^2}\bigg(d\phi-\frac{\zeta (N_k-N_{k-1})}{l_{k-1}}  d\chi\bigg)^2,\nn\\[2mm]
&ds^2(\mathbb{B}_3)\bigg|_{\alpha\sim \pi} =  \frac{1}{4}\Big(d\alpha^2 + \frac{(\pi-\alpha)^2}{l_{k}^2}d\chi^2\Big)+\frac{l_{k}^2}{\zeta^2 b_k^2}\bigg(d\phi-\frac{\zeta (N_{k+1}-N_{k})}{l_{k}}  d\chi\bigg)^2,\nn\\[2mm]
&\frac{1}{2\pi}\int_{\mathbb{WCP}_{[l_{k-1},l_k]}}d\mathcal{A}=\frac{\zeta b_k}{l_{k-1}l_k},~~~~~~~~~~~~~~~~~b_k=2N_k-N_{k+1}-N_{k-1},
\end{align}
which is consistent with this claim (see for example \cite{Ferrero:2021etw}). Hence, the sub-manifold $(z,\mathbb{B}_3)$ describes a cone whose base is a $U(1)$ fibration over $\mathbb{WCP}^1_{[l_{k-1},l_k]}$.
In the recent work of \cite{Inglese:2023tyc}, similar three- dimensional orbifolds were considered within the context of supersymmetric localisation. Note that the deformed Taub-Nut space discussed above \eqref{eq:tndef} reproduces the cone over $\mathbb{B}_3$ precisely, meaning it is the the orbifold singularity associated with this generalised space. 

We note for $\zeta\neq0$, at this boundary, the limit of $C_3-C_1\wedge B_2=0$. Consequently, there are no D4 branes present (for $\zeta\neq0$), as $\hat{F}_4=d(C_3\wedge e^{-B_2})$. This is in contrast to the $S^2$ preserved case presented above (with $\zeta=0$), where D4 branes are indeed recovered. It is worth noting that one could recover a D4 charge for the $\zeta\neq0$ case by adding an additional $2\kappa \zeta N_k \sin\theta d\chi\wedge d\theta$ term to the $B_2$ gauge transformation. However, although this make sense locally, it may not be well defined globally - it is not clear that this additional term is a 2-cycle.

We now study this limit when approaching the pole of the deformed $S^2$.

\paragraph{Approaching the pole:} here we study the behaviour when approaching $(\sigma=0,$ $\eta=k,\sin\theta=0)$, which motivates the following coordinate change
\beq
\eta=k-\bar{r} \cos\alpha \sin^2\mu,~~~~~\sigma=\bar{r} \sin\alpha \sin^2\mu,~~~~~~\sin\theta = 2\sqrt{\frac{b_k\bar{r}}{N_k}}\cos\mu,
\eeq
with $r= \bar{r} \sin^2\mu$ (for small $\bar{r}$), such that
\beq
f_4(d\sigma^2+d\eta^2)+f_2 (d\theta^2+\sin^2\theta d\phi^2)= f_4\bigg(d\bar{r}^2+\bar{r}^2 \Big(d\mu^2+\cos^2\mu \,d\phi^2 + \frac{1}{4}\sin^2\mu\, d\alpha^2\Big)\bigg),
\eeq
which then closes nicely as an $\mathds{R}^4$  (noting that $f_4= b_k/(\bar{r}N_k)$ and $f_2=1$ in this limit). 
This then gives rise to the function $\tilde{\Xi}_k$, which is nowhere zero and smooth, given by
 \beq
 \sin^2\mu \,\Xi \rightarrow \tilde{\Xi}_k =\Delta_k \sin^2\mu + \zeta^2 b_k^2 \cos^2\mu,~~~~~~~~~~~ B_2\rightarrow B_2+\frac{4\kappa \,k\,b_k \cos^2\mu}{X N_k}  d\bar{r}\wedge d\phi,
 \eeq
for the gauge transformation in \eqref{eqn:Gauge1}, leading to
 \begin{align}\label{eqn:B4eqmetric}
 & \frac{ds^2}{ 2\kappa\sqrt{N_k} } =\frac{1}{X}\sqrt{\tilde{\Xi}_k} \Bigg[\frac{4}{\sqrt{\frac{b_k}{\bar{r}}}}ds^2(AdS_5)+ \frac{\sqrt{\frac{b_k}{\bar{r}}}}{N_k}\bigg(d\bar{r}^2+4\bar{r}^2 ds^2(\mathbb{B}_4)\bigg)\Bigg],~~~~~~e^{-\Phi}=\bigg(\frac{X^6b_k^3N_k}{2^6\kappa^2\tilde{\Xi}_k^3\bar{r}^3}\bigg)^{\frac{1}{4}},\nn\\[2mm]
 &B_2=\frac{4\kappa }{X} b_k \cos^2\mu \Big(\xi d\phi -\zeta d\chi\Big)\wedge d\bar{r},~~~~~~~~~~~~C_3=-4\kappa \, b_k \cos^2\mu \,d\bar{r}\wedge d\phi\wedge d\chi,\nn\\[2mm]
 &C_1=\frac{ X}{\tilde{\Xi}_k}\bigg[b_k^2 \zeta \cos^2\mu\, d\phi + \bigg(g(\alpha)\Big(1+\xi g(\alpha)\Big)+\frac{1}{4}\xi b_k^2 \sin^2\alpha\bigg) \sin^2\mu\,d\chi\bigg],
 \end{align}
with the 4 manifold defined as 
\beq\label{eqn:B4eq}
ds^2(\mathbb{B}_4) = d\mu^2 +\frac{1}{4}\sin^2\mu \bigg(d\alpha^2 + \frac{\sin^2\alpha}{\Delta_k}d\chi^2\bigg) +\frac{\sin^2\mu \cos^2\mu\,\Delta_k}{\tilde{\Xi}_k}(d\phi+\mathcal{A}_k)^2.
\eeq
This $\mathbb{B}_4$ is topologically a $\mathbb{CP}^2$, see \eqref{eqn:CP2}, but with additional orbifold singularities arising from the $(\alpha,\chi)$ spindle - namely, at $\mathbb{R}^4/\mathbb{Z}_{l_{k-1}}$ and $\mathbb{R}^4/\mathbb{Z}_{l_{k}}$, along with a further orbifold singularity as $\sin\mu \to 0$ (where $\mathbb{B}_4$  approaches a cone over the $\mathbb{B}_3$ defined in \eqref{eqn:B3eq}), arising from the $\tilde{\Xi}_k$ contribution.

Calculating the Euler Characteristic, using
\beq\label{eqn:Chern-Gauss}
\chi_E=\frac{1}{32\pi^2}\int_{\mathbb{B}_4}(R_{abcd}R^{abcd}-4R_{ab}R^{ab}+R^2)\text{Vol}(\mathbb{B}_4),
\eeq
from the Chern-Gauss-Bonnet theorem, leads to
\beq\label{eqn:WCP2beta}
\chi_E
=3-\bigg(1-\frac{1}{|l_{k-1}|}\bigg)-\bigg(1- \frac{1}{|\frac{\zeta}{\xi}(l_{k-1}-l_k)|}\bigg)-\bigg(1-\frac{1}{|l_{k|}}\bigg),
\eeq
where $|\zeta (l_{k-1}-l_k)/\xi|=|\zeta \,b_k|\in \mathds{Z}$, which is the contribution from the $\tilde{\Xi}_k$. For a round $\mathbb{CP}^2$, $\chi_E=3$ - hence, $\mathbb{B}_4$ is the weighted projective space $\mathbb{WCP}^2_{[l_{k-1},l_k,\frac{\zeta}{\xi}(l_{k-1}-l_k)]}$. This is then a four dimensional analogue of the spindle!  

By analogous arguments to the spindles in the $\zeta=0$ example, because the orbifold singularities depend on the values of both $\mathcal{R}'$ and $\mathcal{R}''$ at each kink, we appear to have multiple neighbouring $\mathbb{WCP}^2$ manifolds along the $\eta$ direction- one for each $k$. We can then imagine a similar set up to Figure \ref{fig:D6orthogonaltospindless2} for these higher dimensional analogues. 

Distinct examples of restricted forms of this orbifold appear in the literature \cite{Gauntlett:2004zh,Bianchi:2021uhn}, specifically $\mathbb{WCP}^2_{[1,1,2]}$, however we have here a generalization. To our knowledge, this is the first time it has appeared in a solution of supergravity! While a $\mathbb{WCP}^2_{[k_1,k_2,k_3]}$ embedding into supergravity probably does exists (for truly independent $k_i$), it is hard to see how it would be consistent with supersymmetry -  it isn't for the case of $\mathbb{WCP}^2_{[1,1,1]}=\mathbb{CP}^2$ (see \cite{Wulff:2017zbl}), which such a solution would accommodate. 

Note in the $\xi=0$ case, the only orbifold singularity arises from the $\zeta\, b_k$ term in $\tilde{\Xi}_k$ (as $\Delta_k=l_k=1$)
, we now find
\begin{align}\label{eqn:B4eqZetaZero}
ds^2(\mathbb{B}_4) &= d\mu^2 +\frac{1}{4}\sin^2\mu \bigg(d\alpha^2 +  \sin^2\alpha\, d\chi^2\bigg) +\frac{\sin^2\mu \cos^2\mu }{  \sin^2\mu + \zeta^2 b_k^2 \cos^2\mu}\Big(d\phi - \zeta g(\alpha)d\chi\Big)^2, \nn\\[2mm]
& \text{with}~~~~~~~\chi_E= 3-\bigg(1- \frac{1}{|\zeta \,b_k|}\bigg),~~~~~~~~~(\xi=0,\zeta\neq0).
\end{align}
 We have already considered the $\zeta=0$ case above.
 
Now calculating the charge at $\mu=\frac{\pi}{2}$, we notice that all $\zeta$ dependence drops out of the calculation, re-deriving the D6 charge given in \eqref{quantcond1}
\beq
Q_{D6}^k=-\frac{1}{2\pi}\int_{\mathbb{WCP}^1_{[l_{k-1},l_k]}}F_2 =-\frac{1}{2\pi}\int_{\chi=0}^{\chi=2\pi}C_1\Big|_{\alpha=0}^{\alpha=\pi} =\frac{X}{l_kl_{k-1}}(2N_k-N_{k+1}-N_{k-1}),
\eeq
where the rational charge is a consequence of the spindle, see for example \cite{Ferrero:2021etw}. Note the orbifold singularity due to $\zeta b_k$ does not effect the charge. The boundary analysis then agrees with the discussion around \eqref{eqn:C1sourcebeta}, where we had nice D6 sources for $\zeta=0$ or $\sin\theta=0$.

It is known that reducing $\mathbb{R}^{1,5}\times \text{TN}_{M}$ on the Hopf fiber of the Taub-Nut space derives a stack of $M$ D6 branes in flat space, described by the $d=7$ KK monopole geometry. In \eqref{eqn:B4eqmetric}, we appear to have the singularity associated to a $d=5$ KK monopole extended in AdS$_5$, descending (via dimensional reduction) from the embedding of a conical Calabi-Yau 3-fold (with orbifold singularities) into $d=11$. We refer the reader to appendix C of \cite{Macpherson:2024frt}, where the flat space analogue of the above singular behaviour is shown to be realised by the dimensional reduction of $d=11$ supergravity on an $\mathbb{R}^{1,4}\times \mathbb{R}^{6}$ orbifold. As for the D6 branes in the previous deformation, the charge of this  KK monopole is found by integrating $F_2$ at $\mu= \frac{\pi}{2}$, namely
\beq
-\frac{1}{2\pi}\int_{\mathbb{WCP}^1_{[l_{k-1},l_k]}}F_2= \frac{1}{l_{k}l_{k-1}}(2N_k-N_{k+1}-N_{k-1}),\label{quantcond2}
\eeq
 in this case however, supersymmetry is not broken.

\end{itemize}
\subsubsection{Summary}
For all values of $(\xi,\zeta)$, we have a stack of colour NS5 branes at the $\sigma\rightarrow \infty$ boundary. In addition, we find stacks of D6 branes along the $\sigma=0$ boundary, and located at each kink of the rank function.
 In the $\zeta=0$ case (along the red line in Figure \ref{fig:IIAtableplot}), with a preserved $S^2$, the D6 branes are extended in $(AdS_5,S^2)$ and orthogonal to a different spindle at each kink, with conical deficit angles defined by that kink - giving rise to a different rational charge for each D6 stack. 
 This is the only solution with D4 (colour) branes. In that case, the rational charge of the D4 branes, derived from integrating over the spindle, can be eliminated by an additional term in the gauge transformation of $B_2$. In the $\zeta\neq0$ case, the spindle at each kink is now replaced by its higher dimensional analogue, giving rise to the same rational quantization of charge. When $\xi=0$ (along the dark green line in Figure \ref{fig:IIAtableplot}), integer quantization is recovered. The D6 branes are the only physical objects in the background. 
Hence, in summary
 \begin{equation}\label{eqn:ChargesIIA}
 \begin{aligned}
&Q_{NS5}=\widehat{\kappa} \,P,~~~~~~~Q_{D6}^k =\frac{X}{l_kl_{k-1}}b_k,~~~~~~ Q_{D4}^k=\widehat{\kappa} X  \begin{cases}
     ~~~~~~~~~0 & (\zeta\neq0)\\
     ( N_k - N_{k-1})  & (\zeta=0)
    \end{cases}  ,   \\[2mm]
    &Q_{D6}=\sum_{k=1}^{P-1}Q^k_{D6}=\frac{X}{l_0 l_P}(N_{P-1}+N_1),~~~~~Q_{D4}=\sum_{k=1}^{P-1}Q^k_{D4}=\widehat{\kappa} X  \begin{cases}
     ~~~~0 & (\zeta\neq0)\\
     N_{P-1} & (\zeta=0)
    \end{cases},\\[2mm]
    &l_k=1+\xi (N_{k+1}-N_k),~~~~~~~~l_0=1+\xi N_1,~~~~~~l_P=1-\xi N_{P-1},~~~~~~~b_k=(2N_k-N_{k+1}-N_{k-1}).
    \end{aligned} 
\end{equation}
Calculating the holographic central charge, and using $2\kappa=\pi \widehat{\kappa} X $, one finds
\begin{equation*}
c_{hol} 
= \frac{ \widehat{\kappa}^3X^2}{8\pi}\sum_{n=1}^\infty P \, \mathcal{R}_n^2.
\end{equation*}
Given that the D6 branes are physical, it is worthwhile using the parameters $(\widehat{\kappa},X)$ to investigate quantizing their charge. Following this discussion, we will return to the $S^2$ preserved $\mathcal{N}=0$ solution, before investigating further the one-parameter $\mathcal{N}=1$ solutions defined by $\zeta=-\xi$.

\subsection{Quantization \& the Holographic Central Charge}\label{sec:quantconds}
Let us now turn to quantization, keeping in mind any effects on the holographic central charge. 
We have a few options:
\begin{itemize}
\item In the usual case, with $\widehat{\kappa}=X=1$, we simply get
\begin{equation}\label{eqn:X1vals} 
 \begin{aligned}
&Q_{D6}^k =\frac{2N_k-N_{k+1}-N_{k-1}}{l_kl_{k-1}},~~~~~~~Q_{NS5}= P, ~~~~~~~Q_{D4}^{k,(\zeta=0)}=  ( N_k - N_{k-1}), 
   \\[2mm]
       &Q_{D6}=\sum_{k=1}^{P-1}Q^k_{D6}=\frac{(N_{P-1}+N_1)}{l_0 l_P},~~~~~Q_{D4}^{(\zeta=0)}=\sum_{k=1}^{P-1}Q_{D4}^{k,(\zeta=0)}=N_{P-1} ,
   \\[2mm]
&~~~\text{with}~~~~~c_{hol}=  \frac{1}{8\pi}\sum_{n=1}^\infty P \, \mathcal{R}_n^2,~~~~~~~~~l_k=1+\xi (N_{k+1}-N_k),
    \end{aligned} 
\end{equation}
which gives nice results for the holographic central charge and the NS5- and D4- charges, however the D6 brane charge has rational quantization due to the spindle. This simply recovers the GM charges when $\xi=0$ - where the spindle orbifold singularities vanish (returning to an $S^2$ for $\zeta=0$), recovering the integer quantization. In the $\xi=0$ solution, the remaining $\zeta\neq0$ parameter breaks the supersymmetry from $\mathcal{N}=2$ to $\mathcal{N}=0$ (along with the $S^2$), corresponding to the vertical axis in Figure \ref{fig:IIAtableplot}. Of course, there may exist example rank functions (such as certain triangular rank functions) which give rise to integer charges (as well as satisfying all the requirements imposed by the spindle). After cycling through hundreds of triangular rank function examples, when $(|l_{S-1}|,|l_S|)>1$, no such case has yet been found (see Figure \ref{fig:Mathematica2}). 
We note, in the limit $N\rightarrow\infty$, we find $Q_{D6}\sim N$ (for $\xi=0$) and  $Q_{D6}\sim 1/N$ (for $\xi\neq 0$).  
\item Alternatively, the additional parameter, $X$, allows us to impose the quantization of each and every D6 brane stack, where we require
\beq\label{eqn:Xval}
X\equiv \prod_{j=0}^{P}l_j,~~~~~~~~l_j=1+\xi (N_{j+1}-N_j),
\eeq
leading to
\beq
Q_{D6}^k = \prod_{\substack{j=0 \\ j\neq (k, k-1)}}^{P}\hspace{-3mm}l_j (2N_k-N_{k+1}-N_{k-1}), ~~~~~~~~Q_{D6}=\sum_{k=1}^{P-1}Q^k_{D6}=(N_{P-1}+N_1) \prod_{j=1}^{P-1} l_j,
\eeq
with $Q_{D6}$ and $Q_{D6}^k$ integer, for all $k$. Alternatively, one could focus on quantizing only the total D6 charge, in which case this condition could be lessened to just $X\equiv l_0l_P$ - with the charge of each D6 stack (which is now rational in general) summing to an integer overall. In practice however, either choice lead to similar consequences on the remaining charges - which we now investigate. From here there are two notable options 
\begin{itemize}
\item Imposing $\widehat{\kappa}=X^{-\frac{2}{3}}$, we can recover the GM holographic central charge,
\beq
c_{hol}=  \frac{1}{8\pi}\sum_{n=1}^\infty P \, \mathcal{R}_n^2,
\eeq
however, as a consequence, the NS5 and D4 charges now read
 \begin{equation}\label{eqn:Xtwothirdsval}
 \begin{aligned}
&  Q_{NS5}= P \prod_{j=0}^{P}(l_j)^{-\frac{2}{3}},  ~~~~~~~ Q_{D4}^{k,(\zeta=0)}= ( N_k - N_{k-1}) \prod_{j=0}^{P}(l_j)^{\frac{1}{3}},  ~~~~~~Q_{D4}^{(\zeta=0)}= N_{P-1}\prod_{j=0}^{P}(l_j)^{\frac{1}{3}}.
    \end{aligned} 
\end{equation}
Of course, in order to quantize the D4 here, we could simply redefine $X\rightarrow X^3$ (which would still give integer D6 quantization). However, the quantization of NS5 is necessarily broken in general. Hence, we have quantized the D6 charge and kept the $\mathcal{N}=2$ holographic central charge, but as a consequence, broken the quantization of the NS5 branes. Again, many Triangular rank functions have been tested. When $(|l_{S-1}|,|l_S|)>1$, no integer NS5 examples have been found as yet. However, loosening this condition, leads to special teardrop examples with integer quantization (see Figure \ref{fig:Mathematica4}). Perhaps rational NS5 charge is physically superior to rational D6 charge, given that the D6 branes are in fact dynamical (unlike the NS5s). 
\item Finally, let us impose the quantization of all charges by fixing $\widehat{\kappa}=1$, hence
 \begin{equation} 
 \begin{aligned}
&Q_{NS5}= P, ~~~~~~~~~  Q_{D4}^{k,(\zeta=0)}= ( N_k - N_{k-1}) \prod_{j=0}^{P}l_j ,~~~~~~~  Q_{D4}^{(\zeta=0)}=N_{P-1} \prod_{j=0}^{P}l_j.
        \end{aligned} 
\end{equation}
Note here that this approach would still quantize the D4 branes even without the additional gauge transformation, namely for \eqref{eqn:D4orig} and \eqref{eqn:D4origT}. \\
However, as a consequence of quantizing all three charges, we find 
\beq
 c_{hol}=  \frac{1}{8\pi} \prod_{j=0}^{P}(l_j)^2\,\sum_{n=1}^\infty P \, \mathcal{R}_n^2, ~~~~~~~~l_j=1+\xi (N_{j+1}-N_j),
\eeq
where it is clear that $\xi$ has now been introduced into the holographic central charge. Hence, in this case, the parameter $\xi$ would change the number of degrees of freedom of the dual field theory. This could suggest that in order to quantize all charges, the parameter $\xi$ must not correspond to a marginal deformation. However, given that the isometries of the AdS$_5$ are untouched by $\xi$, the conformality of the dual field theory should also remain unaffected by this parameter. Hence, it is most likely that $\xi$ does correspond to a marginal operator, but perhaps requiring the quantization of all charges is simply too much to ask here. The most likely interpretation is that $\xi$ corresponds to a marginal, but non-Lagrangian, deformation of the field theory. Perhaps the generalized quiver theories discussed in \cite{Gaiotto:2009gz} may be of particular interest here.
\end{itemize}
\end{itemize}
Moving forward we will leave the parameters $(X,\kappa)$ totally general, but we will consider $\xi$ as a marginal non-Lagrangian deformation - intuitively fixing $X=1$ and $\kappa=\pi/2$. Given $\zeta$ does not effect quantization, we can consider it as a marginal Lagrangian deformation. We will return to this discussion in Section \ref{sec:commentsCFT}, where we make some comments on the dual field theories to our backgrounds.



\subsection{SU$(2)\times$U$(1)$ preserving $\mathcal{N}=0$ deformations}\label{sec:N=0}
In this section we study in more detail the $\zeta=0$ solution of the more general two parameter family given in \eqref{eqn:generalresult1}, which is the only $S^2$ preserving deformation of the $\mathcal{N}=2$ solution reviewed in Section \ref{sec:GM}. This solution then retains the SU(2)$\times$U(1) isometry, but preserves none of the supersymmetry - this solution corresponds to the horizontal red line in Figure \ref{fig:IIAtableplot}. Whilst the SU$(2)$ isometry descends from the SU$(2)_R$ part of the ${\cal N}=2$ R-symmetry in $d=11$, the U$(1)$ isometry does not - the $U(1)_R$ R symmetry component is broken under the reduction.

Fixing $\zeta=0$ in \eqref{eqn:generalresult1}, we find
 \begin{align}
ds^2&= \frac{1}{X}f_1^{\frac{3}{2}} f_5^{\frac{1}{2}}\sqrt{\Delta}\bigg[4ds^2(\text{AdS}_5)+f_2ds^2(\text{S}^2)+f_4(d\sigma^2+d\eta^2)+\frac{f_3}{\Delta} d\chi^2\bigg],~~~~e^{\frac{4}{3}\Phi}=  \frac{1}{X^2} f_1 f_5\Delta, \label{N=0metric}\nn\\[2mm]
\Delta&=(1+\xi f_6)^2+\xi^2  \frac{f_3}{f_5},~~~~  H_3 =  \frac{1}{X} d(f_8+\xi f_7)\wedge \text{vol}(\text{S}^2),\\[2mm]
C_1&=   \frac{X}{\Delta}\Big(f_6(1 + \xi f_6) +\,\xi \frac{f_3}{f_5} \Big)d\chi~~~~ C_3=f_7 d\chi\wedge\text{vol}(\text{S}^2).\nn 
\end{align}
As we uncovered in the previous section, this solution has some very interesting properties. As depicted in Figure \ref{fig:D6orthogonaltospindless2}, the background contains stacks of D6 branes at each kink of the rank function, each wrapping AdS$_5\times$S$^2$ and orthogonal to a spindle whose conical deficit angle is defined by the kink (specifically, the slope of the rank function at the kink). 
In addition to D6 branes, which are the only physical objects, the solution has a stack of NS5 branes at $\sigma\rightarrow\infty$ as well as colour D4 branes. These D4 branes can have integer charge, recovered via an appropriate gauge transformation in $B_2$. The charges from the previous section \eqref{eqn:ChargesIIA} are summarised below,
\begin{equation*}
Q_{NS5}=\widehat{\kappa} \,P,~~~~~~~~~Q_{D6}^k = \frac{X}{l_kl_{k-1}}(2N_k-N_{k+1}-N_{k-1}),~~~~~Q_{D4}^k = \widehat{\kappa} X (N_k - N_{k-1}) .
\end{equation*}

We postpone any non-SUSY G-structure discussion until section \eqref{sec:nonSUSYfor2parameter}, where we investigate the full two-parameter solution. We will however turn to checking the stability of probe D6 branes for this background. We will close the subsection by investigating an example rank function - demonstrating the spindle characteristics and quantization conditions explicitly.
 
\subsubsection{Probe Stability}
Now that supersymmetry is broken, it is no longer guaranteed that the D6 branes are stable. A stable D brane configuration should have minimal energy \cite{Martucci:2005ht,Koerber:2005qi}, for an action
defined in \eqref{eq:branactions}, with the WZ action requiring the pull back of $C_7-B_2^k\wedge C_5$ onto AdS$_5\times$S$^2$ (the world volume of the D6 branes). Hence, to investigate the solution at hand, we need to derive the higher form fluxes $C_5$ and $C_7$. In the $\xi=0$ case, we can use G-structures to derive these forms, which are given in \eqref{eqn:C5andC7}. As supersymmetry is now broken by $\xi$, this method is no longer an option for us. However, after a bit of work, using the higher form fluxes for $\xi=0$ \eqref{eqn:C5andC7}, we can derive the appropriate parametric deformations via direct integration. For the purposes of D6 branes, we are particularly interested in $(C_5,C_7)$, which read in full
\beq\label{eqn:C5andC7S2broken}
\begin{aligned}
C_5 &= -\frac{2^4}{X}  f_1^3( f_5f_6+4\xi) \text{vol}(\text{AdS}_5)+\frac{2^4\kappa^2}{X} e^{4\rho} \text{vol}(\text{Mink}_4)\wedge\bigg(\sigma \dot{V}'d\sigma +(2\dot{V}-\ddot{V})d\eta\bigg), \\[2mm]
C_7&= 
\frac{2^6\kappa^3}{X^2} \sigma \bigg( \Big(d(\sigma \dot{V} \cos\theta) -3\dot{V} \cos\theta\, d\sigma\Big)\wedge  \text{vol}(\text{AdS}_5)  + e^{4\rho} d(\dot{V}\cos\theta) \wedge d\sigma \wedge \text{vol}(\text{Mink}_4) \bigg) \wedge d\chi\\[2mm]
& -\frac{2^4}{X^2} f_1^3f_5f_7 \Bigg[ \text{vol}(\text{AdS}_5) +e^{4\rho}\frac{1}{2\dot{V}}\bigg(1 -\xi^2 \frac{2\dot{V}\tilde{\Delta}}{\Lambda V''} \bigg) d(\dot{V})\wedge \text{vol}(\text{Mink}_4)  \\[2mm]
&+\xi e^{4\rho} \bigg[  \frac{1}{4}f_6 \bigg( f_2\Big(1+\frac{\sigma^2}{2}f_4\Big) \frac{1}{V''}d(V'')+ \Big( \frac{2(\dot{V}')^2}{\tilde{\Delta}}-1\Big)\frac{1}{\dot{V}'} d(\dot{V}')  \bigg)    \\[2mm]
&  -  \bigg(\frac{(\dot{V}')^2(3(\dot{V}')^2-5\ddot{V}V'')+2(V'')^2(\ddot{V})^2}{2V'' \Lambda \tilde{\Delta}} +\frac{4\dot{V}}{\Lambda}\Big(1-\frac{1}{2}f_2\Big)  \bigg)d\eta + \frac{\sigma \dot{V}'}{2\Lambda}  d\sigma 
\bigg]\wedge \text{vol}(\text{Mink}_4)    \Bigg]\wedge \text{vol}(S^2),
\end{aligned}
\eeq
recovering \eqref{eqn:C5andC7} for $\xi=0$ (noting $ f_1^3f_5f_6 = 2\kappa^2 \frac{\dot{V}\dot{V}'}{V''}$). 
From \eqref{eqn:C5andC7S2broken} and the form of $B_2$ given in \eqref{eqn:generalresult1}, the $(\text{AdS}_5,S^2)$ components read
\begin{align}
C_7-B^k_2\wedge C_5&=-\frac{2^4}{X^2}f_1^3\bigg(f_5f_7 -(f_5f_6+4\xi)  (f_8+\xi f_7 +2\kappa k) \bigg)\text{vol}(\text{AdS}_5)\wedge  \text{vol}(S^2)+\frac{1}{2}\frac{(2\kappa)^3}{X^2} \bar{Z}_7 \nn\\[2mm]
&=\left(C_7-B^k_2\wedge C_5\right)\bigg\lvert_{\xi=0}+\frac{1}{2}\frac{(2\kappa)^3}{X^2}\left( \bar{Z}_7+\xi \bar{X}_7+\xi^2 \bar{Y}_7\right),
\end{align}
where
\begin{align}
\left(C_7-B^k_2\wedge C_5\right)\bigg\lvert_{\xi=0} &= -\frac{2^4}{X^2}f_1^3f_5\Big(f_7- f_6(f_8+2\kappa k)\Big) \text{vol}(\text{AdS}_5)\wedge  \text{vol}(S^2)\nn\\[2mm]
&=\frac{(4\kappa)^3}{X^2}\frac{(\dot{V}^2-(\eta-k)\dot{V}\dot{V}')}{V''}  \text{vol}(\text{AdS}_5)\wedge  \text{vol}(S^2),
\end{align}
with $\bar{Z}_7$ some gauge transformation containing terms of any order in $\xi$, and decomposed in terms of an arbitrary function $\bar{p}=\bar{p}(\eta,\sigma)$ as
\beq
\bar{Z}_7=\left( \bar{p} \,\text{vol}(\text{AdS}_5)\wedge \text{vol}(\text{S}^2)+\frac{1}{4}e^{4\rho} \text{vol}(\text{Mink}_4)\wedge \text{vol}(\text{S}^2)\wedge d\bar{p}\right),
\eeq
such that it is manifestly closed, with $d\bar{Z}_7=0$, and contains only forms on the external space whose exterior derivatives respect the isometries of Mink$_4$. 

We can absorb all additional $(\text{AdS}_5,S^2)$ components into a single function $p(\eta,\sigma)$ in the following manner
\begin{align}
\bar{Z}_7+\xi \bar{X}_7+\xi^2 \bar{Y}_7&= (\bar{p}+\xi \,p_X+\xi^2\,p_Y) \,\text{vol}(\text{AdS}_5)\wedge \text{vol}(\text{S}^2)+\frac{1}{4}e^{4\rho} \text{vol}(\text{Mink}_4)\wedge \text{vol}(\text{S}^2)\wedge d\bar{p}\nn\\[2mm]
&=p\,\text{vol}(\text{AdS}_5)\wedge \text{vol}(\text{S}^2)+\frac{1}{4}e^{4\rho} \text{vol}(\text{Mink}_4)\wedge \text{vol}(\text{S}^2)\wedge d (p-\xi \,p_X-\xi^2\,p_Y)\nn\\[2mm]
&\equiv Z_7+\xi X_7+\xi^2 Y_7 ,\nn
\end{align}
\begin{equation*}
p\equiv\bar{p}+\xi \,p_X+\xi^2\,p_Y,~~~~p_X \equiv \frac{2^2}{\kappa^3} f_1^3 \Big(f_5f_6f_7+4(f_8+2\kappa k)\Big),~~~~~p_Y\equiv \frac{2^4}{\kappa^3} f_1^3f_7 = - 2^5\dot{V}^3,
\end{equation*}
with
\begin{align}
Y_7&=  e^{4\rho} \text{vol}(\text{Mink}_4)\wedge \text{vol}(\text{S}^2)\wedge d(\dot{V}^3),\nn\\[2mm]
X_7&=\frac{1}{2}e^{4\rho} \text{vol}(\text{Mink}_4)\wedge \text{vol}(\text{S}^2)\wedge\Bigg[d\left(\frac{\dot{V}^2 \dot{V}'}{V''}\right)-\frac{2\dot{V}^2}{f_2}d\eta\nn\\[2mm]
&~~~~~~~~~~~~~~~~~~~~~~~~~~~~~~~~~~~~~~~~~~~~+(\eta-k)\left(d\left(\frac{\dot{V}\dot{V}'}{V''}-\ddot{V} \dot{V}\right)+ 4 \dot{V}(\dot{V}'d\eta- \sigma V''d\sigma)\right)\Bigg].\nn
\end{align}
The WZ action for a D6 brane of world volume (AdS$_5$,S$^2$) then reads
\beq\label{eqn:WSp}
S_{\text{WZ}}= \frac{(4\kappa)^3}{X^2}T_6\int \left(\frac{\dot{V}^2}{V''}+\frac{p}{2} \right)\text{vol}(\text{AdS}_5)\wedge \text{vol}(\text{S}^2).
\eeq
We find to leading order about $(\sigma=0,\eta=k)$ that
\beq
e^{-\Phi}\sqrt{\det(g+ B_2)}\bigg\lvert_{(\text{AdS}_5,~ \text{S}^2)}=\frac{(4\kappa)^3}{X^2}\sin\theta\frac{\dot{V}^2}{V''}\xi\sqrt{2\dot{V}V''\Delta_k}.
\eeq
Hence, we can fix $p$ such that $S= S_{\text{DBI}}+S_{\text{WZ}}=0$ for a D6 brane at $(\sigma=0,\eta=k)$,
\beq
p=2\xi \frac{\dot{V}^2}{V''}\sqrt{2\dot{V}V''}(1+\xi \dot{V}')=4\xi\frac{\dot{V}^2}{V''}\sqrt{\frac{\Delta}{f_5}},
\eeq
which achieves the desired goal (noting that $\dot{V}V''$ dominates in this limit), but so does the sum of $p$ and any function tending to zero at the loci of the D6 branes.
 \subsubsection{The Triangle Rank Function}
Let us now investigate a simple example, using the Triangular rank function given in \eqref{eqn:triangle}. These solutions contain a single stack of D6 branes orthogonal to a spindle, positioned at $\eta=k=S$ , with
\beq
l_{S-1}=1+\xi N,~~~~~~~~~l_S=1-\xi \frac{NS}{P-S},~~~~~~~Q_{D6}  =X\frac{NP}{(1+\xi N)\big(1-\xi \frac{NS}{P-S}\big)(P-S)}.
\eeq
Assuming $X=\widehat{\kappa}=1$, we should simply update the D6 charge in the Hanany-Witten set-up of Figure \ref{fig:Triangular} to include $\xi$. In this case, the holographic central charge equals the $\mathcal{N}=2$ solution, meaning the holographic limit discussed in Appendix \ref{sec:exampleRanks} still holds. Of course, the linear quiver diagram of the dual CFT, given in Figure \ref{fig:Triangular}, will no longer be valid (as the rank of the square flavour node should be integer). Possibly the linear quiver should be replaced by a generalized quiver diagram, for which the value of the central charge \eqref{centralsN=2} remains the same, but the requirement that $(N_h,N_v)$ are integer is loosened. 

\begin{figure}[H]
\centering  
\subfigure[$\xi>0$]
{
\centering
      \vspace{-1cm}
  \begin{minipage}{0.475\textwidth}
      \centering
\begin{tikzpicture}[scale=0.75, every node/.style={scale=0.95}]
\draw (1,0) node[left] {$0$};
\draw (5,0) node[below] {$S$};
\draw (7,0) node[below] {$P$};
\draw[black,line width=0.3mm] (1,1)-- (5,1);
\draw[black,line width=0.3mm] (5,-1)-- (7,-1);
\draw (1,1) node[left] {$|\xi| N$};
\draw (1,-1) node[left] {$- |\xi|\frac{NS}{P-S}$};
\draw[-stealth, line width=0.53mm] (1,-1.5)--(1,2) node[left ]{$|\xi|\mathcal{R}' $};
\draw[-stealth, line width=0.53mm] (1,0) --(7.6,0) node[below ]{$\eta$};
\end{tikzpicture}
  \end{minipage}
      \begin{minipage} {0.1\textwidth}
$~~\Rightarrow$
  \end{minipage} 
     \begin{minipage}{0.4\textwidth}
    \centering
\begin{tikzpicture}[scale=0.75, every node/.style={scale=0.95}]

\draw (1,0) node[below] {$0$};
\draw[gray,dashed,line width=0.3mm] (5,0)-- (5,2);
\draw (5,0) node[below] {$S$};
\draw (7,0) node[below] {$P$};

\draw[black,line width=0.3mm] (1,2)-- (5,2);
\draw (1,2) node[left] {$1+|\xi| N$};

\draw[black,line width=0.3mm] (5,1)-- (7,1);
\draw[gray,dashed,line width=0.3mm] (7,0)-- (7,1);
\draw (1,1) node[left] {$\Big|1- |\xi|\frac{NS}{P-S}\Big|$};

\draw[-stealth, line width=0.53mm] (1,0)--(1,3) node[left ]{$\Big|1+|\xi|\mathcal{R}'\Big|$};
\draw[-stealth, line width=0.53mm] (1,0) --(7.6,0) node[below ]{$\eta$};
\end{tikzpicture}
  \end{minipage} 
}

\subfigure[$\xi<0$]
{
\centering
 \hspace{0.1cm}
      \vspace{-1cm}
  \begin{minipage}{0.475\textwidth}
      \centering
\begin{tikzpicture}[scale=0.75, every node/.style={scale=0.95}]
\draw (1,0) node[left] {$0$};
\draw (5,0) node[below] {$S$};
\draw (7,0) node[below] {$P$};
\draw[black,line width=0.3mm] (1,-1)-- (5,-1);
\draw[black,line width=0.3mm] (5,1)-- (7,1);
\draw (1,-1) node[left] {$-|\xi| N$};
\draw (1,1) node[left] {$ |\xi|\frac{NS}{P-S}$};
\draw[-stealth, line width=0.53mm] (1,-1.5)--(1,2) node[left ]{$-|\xi|\mathcal{R}' $};
\draw[-stealth, line width=0.53mm] (1,0) --(7.6,0) node[below ]{$\eta$};
\end{tikzpicture}
  \end{minipage}
      \begin{minipage} {0.1\textwidth}
$~\Rightarrow$
  \end{minipage} 
     \begin{minipage}{0.45\textwidth}
    \centering
\begin{tikzpicture}[scale=0.75, every node/.style={scale=0.95}]

\draw (1,0) node[below] {$0$};
\draw[gray,dashed,line width=0.3mm] (5,0)-- (5,1.5);
\draw (5,0) node[below] {$S$};
\draw (7,0) node[below] {$P$};

\draw[black,line width=0.3mm] (1,1.5)-- (5,1.5);
\draw (1,1.5) node[left] {$\Big|1-|\xi| N\Big|$};

\draw[black,line width=0.3mm] (5,0.5)-- (7,0.5);
\draw[gray,dashed,line width=0.3mm] (7,0)-- (7,0.5);
\draw (1,0.5) node[left] {$1+ |\xi|\frac{NS}{P-S}$};

\draw[-stealth, line width=0.53mm] (1,0)--(1,3) node[left ]{$\Big|1-|\xi|\mathcal{R}'\Big|$};
\draw[-stealth, line width=0.53mm] (1,0) --(7.6,0) node[below ]{$\eta$};
\end{tikzpicture}
  \end{minipage} 
}
  \caption{Plots of conical deficit angles, $(|l_{S-1}|,|l_S|)$, for a generic Triangular rank function - for both $\xi>0$ and $\xi<0$. Here we assume $|l_{S-1}|\neq|l_S|\geq1$ - which of course is only satisfied by a subset of $(N,S,P,\xi)$ values, further restricted by $\text{gcd}\big(|l_{S-1}|,|l_S|\big)=1$.}
    \label{fig:conicaldefTriang}
\end{figure}
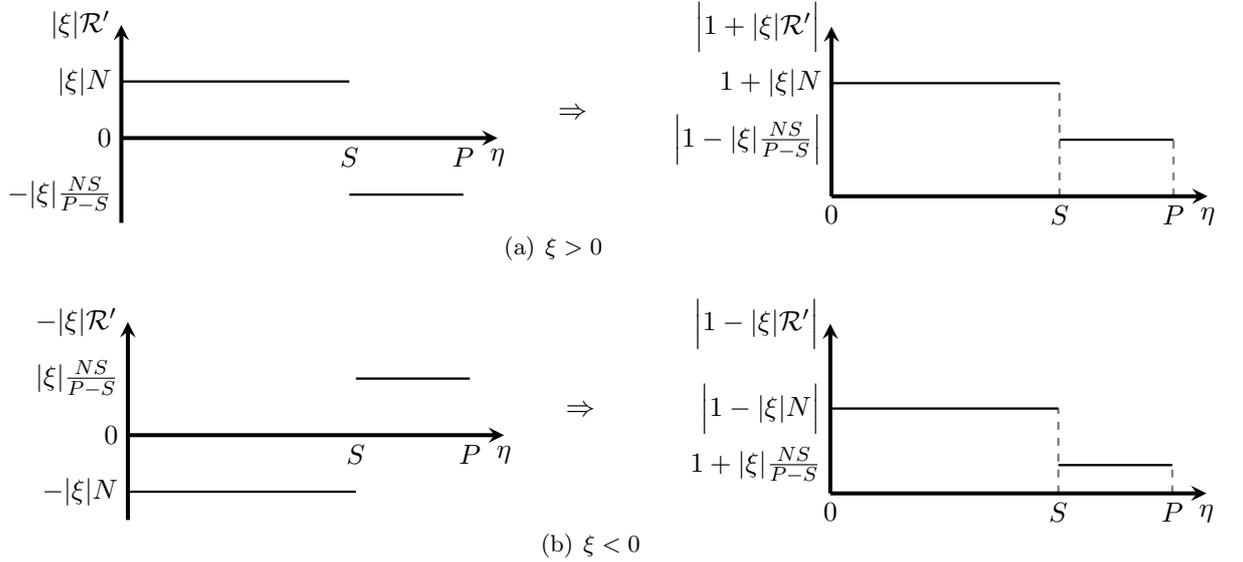

Along the $\sigma=0$ boundary, we find from \eqref{eqn:orbifold1} an $\mathds{R}^2/\mathds{Z}_{|l_{S-1}|}$ and $\mathds{R}^2/\mathds{Z}_{|l_{S}|}$ orbifold singularity, for $\eta<S$ and $\eta>S$, respectively. Recall the conditions on the conical deficit angles on the spindle, require $|l_{S-1}|\neq|l_S|$ 
and $\text{gcd}\big(|l_{S-1}|,|l_S|\big)=1$. We know both $N$ and $\frac{NS}{P-S}$ are positive valued, but $\xi$ can be either positive or negative. Replacing $\xi $ with $\pm|\xi|$, leads to the two plots in Figure \ref{fig:conicaldefTriang}. 

\begin{figure}[H]
 \hspace*{-1.5cm}
 \includegraphics[width=1.125 \textwidth]{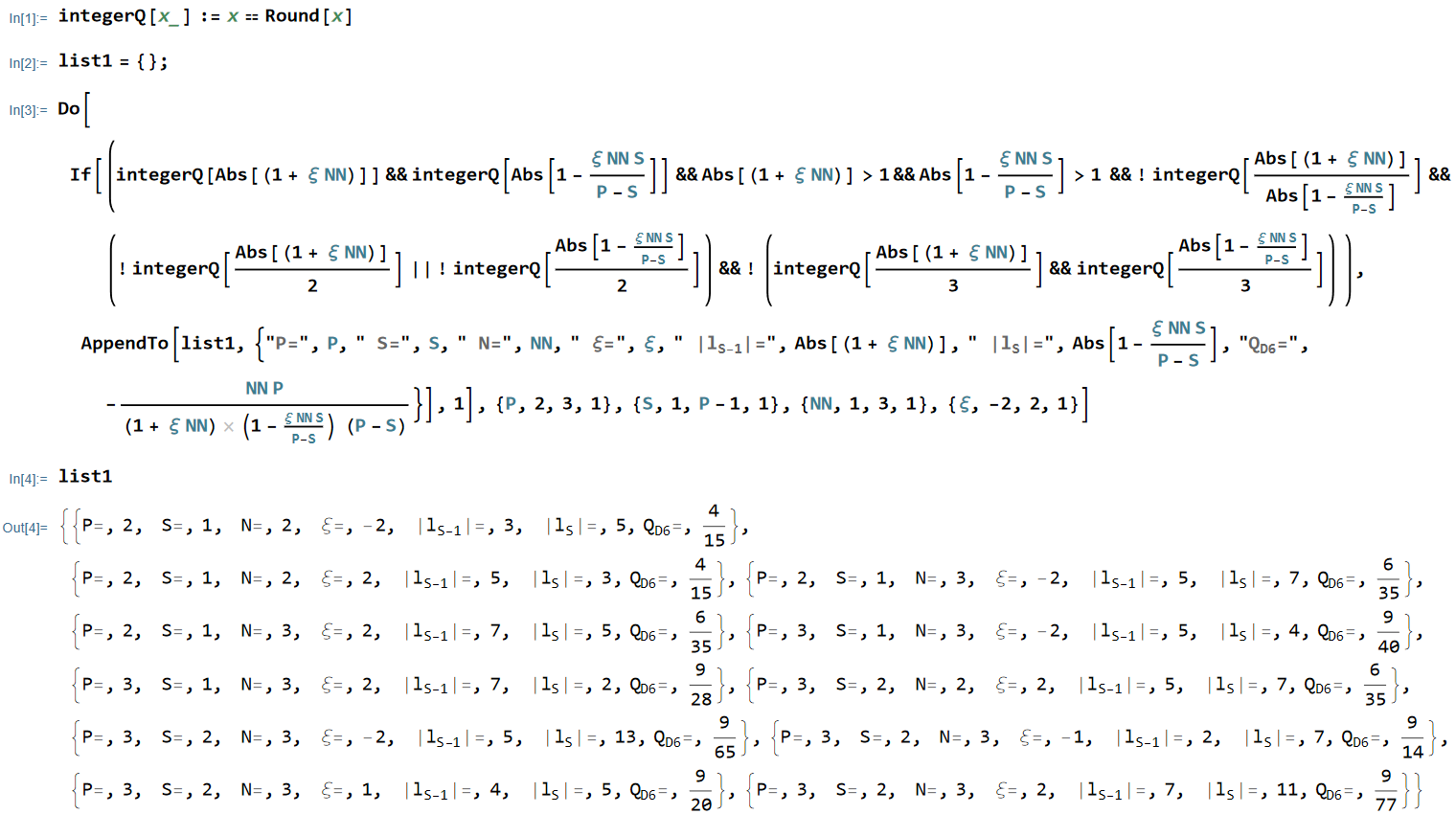}
  \caption{Some simple Triangular rank function examples which satisfy the spindle constraints (assuming $X=1$, $l_{S-1}\neq l_S>1$). One can find more examples by increasing the range of each variable - note that one would need to include the final condition (which checks whether both numbers are divisible by 3) for each odd number. For instance, checking whether both numbers are divisible by 5 would be required to eliminate examples such as 
  gcd$(15,10)\neq1$. The sign of $Q_{D6}$ is chosen to be positive.}
\label{fig:Mathematica}
\end{figure}

\begin{figure}[H]
 \hspace*{-1.5cm}
   \includegraphics[width=1.125 \textwidth]{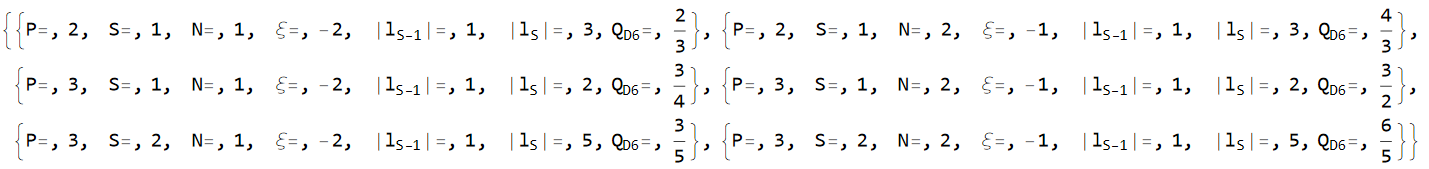}
   \caption{Some examples of `teardrop' orbifold solutions.}
   \label{fig:Mathematica4}
\end{figure}

\begin{figure}[H]
\includegraphics[width=1.125 \textwidth]{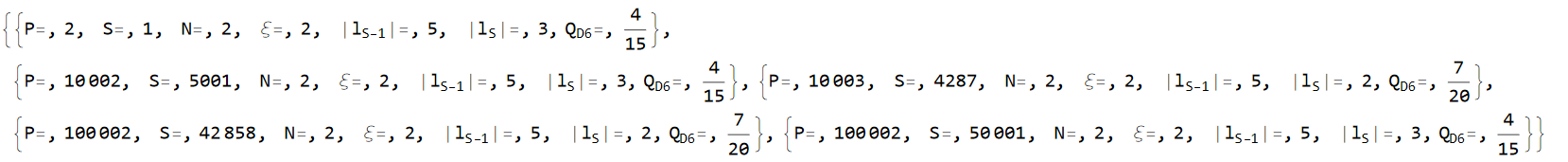}
   \caption{Testing large values of $P$ and $S$ for small $N$, for which the values $(|l_{S-1}|,|l_S|)$ and $Q_{D6}$ remain consistent. Note, for large $N$, $Q_{D6}\rightarrow 0$.}
\label{fig:Mathematica3}
\end{figure}

\begin{figure}[H]
\includegraphics[width=1.125 \textwidth]{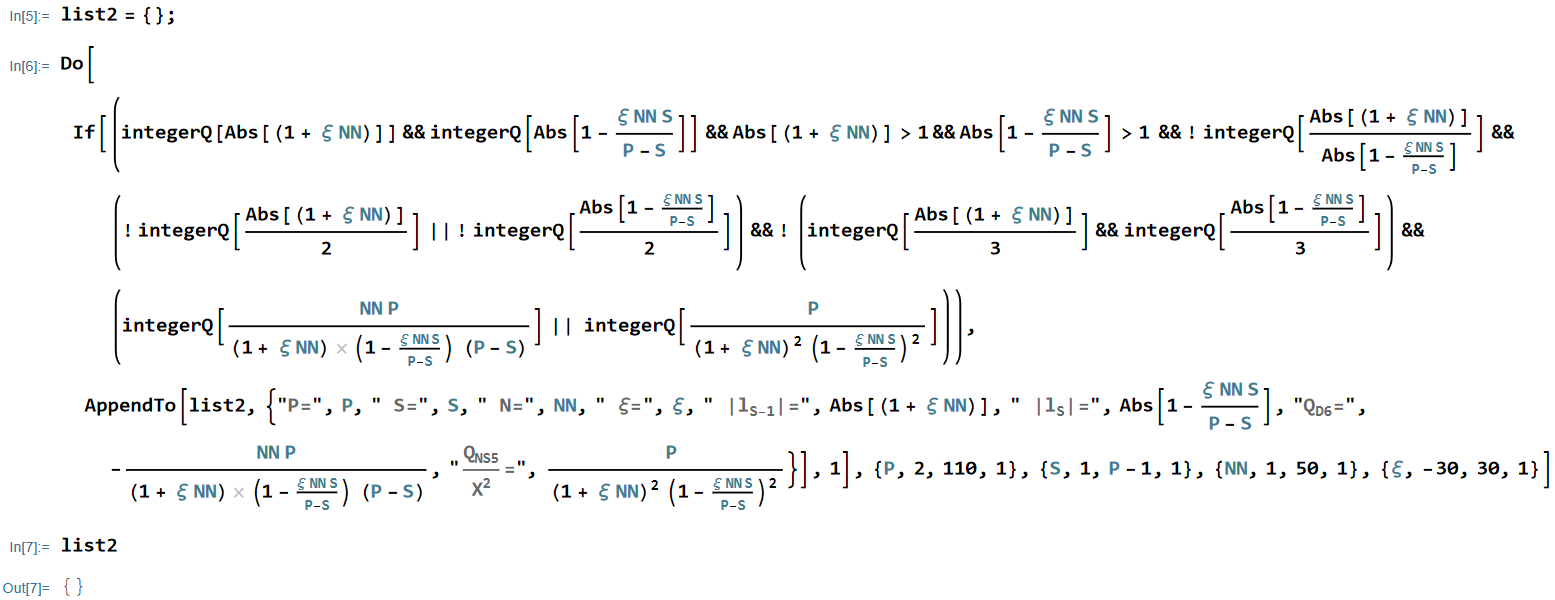}
   \caption{Checking examples for integer quantization of charge, with $(|l_{S-1}|,|l_S|)>1$ (for either $Q_{D6}$ or $Q_{NS5}$ quantization), finding no examples within these ranges.}
\label{fig:Mathematica2}
\end{figure}

\begin{figure}[H]
\includegraphics[width=1.125\textwidth]{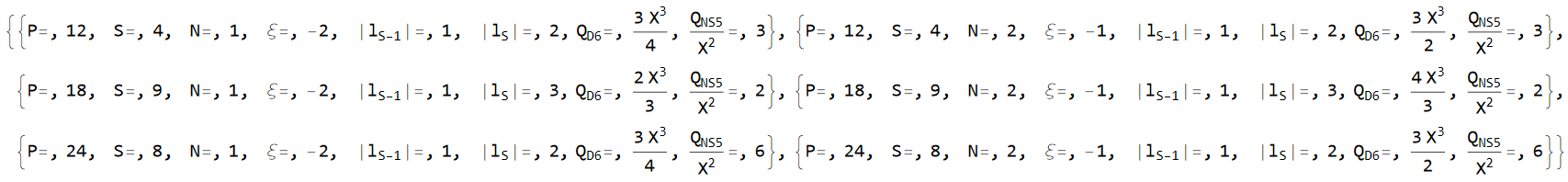}
   \caption{Checking integer quantization of charge for teardrop solutions.}
\label{fig:Mathematica5}
\end{figure}

One can use Mathematica to cycle through various examples. For instance, the code given in Figure \ref{fig:Mathematica} seems sufficient for small values. We can then easily restrict further to $|l_{S-1}|>|l_S|$ cases if required. Here we ignored the teardrop solutions by assuming $(l_{S-1},l_S)>1$, but some examples are given separately in Figure \ref{fig:Mathematica4}. 
As we noted below \eqref{eqn:X1vals}, taking $N\rightarrow\infty$ leads to $Q_{D6}\rightarrow 0$ (for $\xi\neq0$). Recall that the holographic limit is only trustable for $P\rightarrow \infty$, hence we test some examples for high values of $P$ in Figure \ref{fig:Mathematica3}. We find the conical deficit angles and D6 charge remain unaffected in this limit.
We can extend this approach to search for examples which either have integer D6 charge (from the \eqref{eqn:X1vals}  quantization) or integer NS5 charge (from the \eqref{eqn:Xtwothirdsval} quantization for $X\rightarrow X^3$). Using the code in Figure \ref{fig:Mathematica2} (and extending the ranges of each parameter), we find no such examples. Of course, this is not an exhaustive check. If we include teardrop solutions as a possibility, we do find examples - see Figure \ref{fig:Mathematica5}. In these examples, the holographic central charge can be preserved along with the integer quantization - using \eqref{eqn:Xtwothirdsval} with $X\rightarrow X^3$. In the first example of Figure \ref{fig:Mathematica5}, we have
\begin{align}
 &Q_{NS5}=   (l_{S-1}l_S)^{-2}P,  ~~~~~~~ Q_{D4}=  (l_{S-1}l_S) N ,~~~~~~~~~Q_{D6}= \frac{ (l_{S-1}l_S)^3NP}{(1+\xi N)\big(1-\xi \frac{NS}{P-S}\big)(P-S)} \nn\\[2mm]
&(P=12, S=4,N=1, \xi=-2):~~~~~~Q_{NS5}=3,~~~~~~~-Q_{D4}=2,~~~~~~~~~Q_{D6}=6.\nn
\end{align}

\subsection{U$(1)\times$U$(1)$ preserving $\mathcal{N}=1$ deformations}\label{sec:N1IIA1}
We now investigate a little further the $\zeta=-\xi$ case of \eqref{eqn:generalresult1}, which is the only ${\cal N}=1$ preserving deformation of the $\mathcal{N}=2$ solution reviewed in Section \ref{sec:GM},  and retains a U(1)$\times$U(1) isometry. This solution now takes the following form
\begin{align}\label{eq:neq1sol}
&ds^2 = \frac{1}{X}f_1^{\frac{3}{2}} f_5^{\frac{1}{2}}\sqrt{\Xi}\bigg[4ds^2(\text{AdS}_5)+f_4(d\sigma^2+d\eta^2)+ds^2(\text{M}_3)\bigg],~~~~e^{\frac{4}{3}\Phi}=  \frac{1}{X^2} f_1 f_5\Xi\nn\\[2mm]
&ds^2(\text{M}_3)=f_2\left(d\theta^2+\frac{\Delta}{\Xi}\sin^2\theta D\phi^2\right)+\frac{f_3}{\Delta} d\chi^2=f_2\left(d\theta^2+\frac{1}{\Pi}\sin^2\theta d\phi^2\right)+\frac{\Pi}{\Xi}f_3D\chi^2,\nn\\[2mm]
&H_3 = \frac{1}{X}\Big( df_8\wedge \text{vol}(\text{S}^2)+ \xi \sin\theta df_7\wedge d\theta\wedge (d\phi+d\chi)\Big), \\[2mm]
&C_1=\frac{X}{\Xi}\left(f_6+\xi\left(f_6^2+\frac{f_3}{f_5}\right)d\chi-\xi \frac{f_2}{f_5}\sin^2\theta d\phi\right),~~~~ C_3=f_7 d\chi\wedge\text{vol}(\text{S}^2),\nn 
\end{align}
where for this solution
\beq
\begin{aligned}
\Xi&=\Delta+\xi^2\frac{f_2}{f_5}\sin^2\theta,~~~\Delta=(1+ \xi f_6)^2+\xi^2\frac{f_3}{f_5},~~~\Pi=1+\xi^2f_2\frac{f_3+f_5 f_6^2}{f_3 f_5}\sin^2\theta,\\[2mm]
D\phi&= d\phi+\frac{\Delta-1-\xi f_6}{\Delta}d\chi,~~~D\chi=d\chi+\frac{\Pi-1+\frac{f_2 f_6}{f_3}\sin^2\theta}{\Pi}d\phi. 
\end{aligned}
\eeq
This is a one-parameter family of $\mathcal{N}=1$ solutions, recovering the $\mathcal{N}=2$ solution for $\xi=0$. This solution corresponds to the diagonal blue line in Figure \ref{fig:IIAtableplot}.
Investigations at the boundary of this solution are given in full in Section \ref{sec:boundary} (fixing $\zeta=-\xi$). There we found the background again has $P$ NS5 branes at the $\sigma=\infty$ boundary. The most interesting modifications to the $\mathcal{N}=2$ solution, of Sections \ref{sec:GM} and \ref{sec:N=2IIAagain}, occur along the $\sigma=0$ boundary - where we find D6 brane stacks positioned at each kink of the rank function. 
Approaching each kink, away from the pole of the deformed $S^2$, we find a sub manifold $(z,\mathbb{B}_3)$ which describes a cone of base $\mathbb{B}_3$ - where this $\mathbb{B}_3$ is a $U(1)$ fibration over a spindle. See the \eqref{eqn:B3eq} and the discussion given there (for $\zeta=-\xi$). Approaching each kink at one of the poles, we find a 4-manifold $\mathbb{B}_4$ which described a $\mathbb{WCP}^2_{[l_{k-1},l_k,l_{k-1}-l_k]}$ - a higher dimensional analogue of the spindle. See \eqref{eqn:B4eq} and the discussion there. This background does not have $D4$ colour charges, as the limit of $\hat{F}_4$ vanishes for $\zeta\neq0$. Summarising the charges \eqref{eqn:ChargesIIA}, we have
\begin{equation*}
Q_{NS5}=\widehat{\kappa} \,P,~~~~~~~~~Q_{D6}^k = \frac{X}{l_kl_{k-1}}(2N_k-N_{k+1}-N_{k-1}).
\end{equation*}
Let us now investigate the G-structure description of this background.
\newpage
\subsubsection{G-structure description}
We now present the G-structure that this ${\cal N}=1$ preserving deformation preserves. We include in these results the third parameter, $\gamma$, purely for the purposes of the next chapter, where it plays an important non-trivial role in deriving the G structure description for a IIB $\mathcal{N}=1$ solution (following a T-duality along $\chi$ with $\gamma=-1$). For the purposes of this $\mathcal{N}=1$ IIA solution \eqref{eq:neq1sol}, we should simply fix $\gamma=0$. Using the approach outlined in sections \ref{sec:G-Structure description of GM} - \ref{sec:IIAGstructureN=2}, the $d=11$ G-structure forms corresponding to the ${\cal N}=1$ preserving reduction frame can be derived. 
In $d=11$, the vielbein frame of \eqref{eqn:InitialGstructureForms} gets modified to
 \begin{align}\label{eqn:N=1Gs11d}
 &K= -\frac{\kappa e^{-2\rho}}{f_1}d(e^{2\rho}\dot{V}\cos\theta),\nn\\[2mm]
& E_1=-\sqrt{\frac{f_1f_3}{\Xi\,\Sigma_1}} \bigg[\frac{1}{\sigma}d\sigma +d\rho+i  \Big( \Sigma_1 \,d\chi +\Sigma_2\,\sin^2\theta d\phi \Big)  +\xi \,d(V')-\frac{\gamma\xi}{2\sigma^2}\dot{V}\sin^2\theta\, d\eta\nn\\[2mm]
 &~~~~~ +\frac{e^{-2\rho}}{2\dot{V}}\bigg( \Sigma_2\, d(e^{2\rho}\dot{V}\sin^2\theta) +\bigg(\Sigma_2 +\xi \frac{f_2}{f_3} \Big(\gamma f_6 -4\xi (1+\gamma)\frac{f_2}{f_5}\Big)\bigg)\sin^2\theta d(e^{2\rho}\dot{V}) \nn\\[2mm]
 &~~~~~+ \frac{4\xi^2(1+\gamma)f_2}{f_3f_5} \sin^2\theta \, \dot{V}d(e^{2\rho})\bigg)\bigg],\nn\\[2mm]
& E_2=\frac{e^{i(\phi+\gamma\chi)}}{\sqrt{\Sigma_1}} \Bigg[\frac{\kappa}{f_1} \bigg[e^{-2\rho} d(e^{2\rho}\dot{V}\sin\theta) - \dot{V}\sin\theta \bigg(\xi (\gamma+1) d(V') + \gamma \Big(\frac{1}{\sigma}d\sigma+d\rho\Big)\bigg)\bigg]+i \sqrt{f_1f_2}\sin\theta d\phi \Bigg] ,\nn\\[2mm]
& E_3= -\frac{1}{X}e^{i\chi} \,\Xi^{\frac{1}{2}}\sqrt{f_1f_5}\bigg[ \frac{X}{\Xi}\left(-\frac{f_3 \dot{V}'}{4 \sigma}d\sigma -V''d\eta+f_6d\rho + \frac{\xi}{f_5}\Big(\frac{\kappa^2e^{-4\rho}}{2f_1^3}d(e^{4\rho}\dot{V}^2(\cos^2\theta-3))+4 d\rho\Big)\right)\nn\\[2mm]
 &~~~~~~~~~+i \Big(d(X\beta) + C_1\Big)\bigg],  
 \end{align}	
where we define
\begin{align}
 &\Sigma_1=1+\frac{f_2}{f_3}\bigg(\big(\gamma+\xi(1+\gamma)f_6\big)^2+\xi^2(1+\gamma)^2\frac{f_3}{f_5}\bigg)\sin^2\theta\equiv \frac{h_\chi}{f_3f_5},\nn\\[2mm]
  &\Sigma_2=\frac{f_2}{f_3}\bigg((1+\xi f_6)\Big(\gamma+\xi (1+\gamma)f_6\Big)+\xi^2(1+\gamma)\frac{f_3}{f_5}\bigg)\equiv \frac{h_{\chi\phi}}{2f_3f_5 \sin^2\theta },\nn\\[2mm]
   & \Xi=(1+\xi f_6)^2 +\xi^2 \frac{f_3}{f_5}+\xi^2 \frac{f_2}{f_5}\sin^2\theta,~~~~C_1= \frac{X}{\Xi}\bigg[\bigg(f_6+ \xi \Big(f_6^2+ \frac{f_3}{f_5}-\gamma  \frac{f_2}{f_5}\sin^2\theta\Big)\bigg)d\chi -\xi \frac{f_2}{f_5}\sin^2\theta d\phi\bigg].
        \end{align}
It is worth making clear that these results still describe the $\mathcal{N}=2$ solution in $d=11$, because at this stage, the $SL(3,\mathds{R})$ transformation and rotation of frames play a somewhat trivial role - it is only under reduction that the $\mathcal{N}=2$ is broken by these parameters.
        
In Type IIA, the SU(2) structure forms for \eqref{eq:neq1sol} can be extracted from these results following the reduction formulae in \eqref{eqn:11Dto10Dforms}, we then find
\begin{adjustwidth}{-1cm}{}
\vspace{-0.6cm}
     \begin{align}\label{eqn:uandv1}
&v=  \frac{\kappa}{\sqrt{X}} e^{-2\rho}f_5^{\frac{1}{4}}f_1^{-\frac{3}{4}} \Xi^{\frac{1}{4}}d\left(e^{2\rho} \dot{V}\cos\theta\right),\nn\\[2mm]
&u=  \frac{1}{\sqrt{X}}(f_1f_5)^{\frac{3}{4}}\Xi^{-\frac{1}{4}}\left(\frac{f_3 \dot{V}'}{4 \sigma}d\sigma +V''d\eta-f_6d\rho - \frac{\xi}{f_5}\Big(\frac{\kappa^2e^{-4\rho}}{2f_1^3}d(e^{4\rho}\dot{V}^2(\cos^2\theta-3))+4 d\rho\Big)\right),\nn\\[2mm]
&\omega=  -\frac{2\kappa^2}{X}  f_1^{-\frac{3}{2}}e^{-3\rho} d\Big(e^{2\rho}\dot{V} e^{-\xi V'}e^{i(\phi+\gamma\chi)}\sin\theta\, d(e^{i\chi}e^{\rho}e^{\xi V'}\sigma)\Big) ,\\[2mm]
&j=\frac{1}{X\sqrt{\Xi}}\Bigg[ \frac{f_1^{\frac{3}{2}}f_5^{\frac{1}{2}} f_3}{\sigma} e^{-\rho}e^{-\xi V'}d\left(e^{\rho}\sigma e^{\xi V'}\right)\wedge d\chi+ \kappa f_2 f_3^{\frac{1}{2}}\hat{X}_1 \wedge  \Big((\gamma+1) d\chi+ d\phi\Big) +\hat{X}_2\wedge (d\phi+\gamma\, d\chi)\Bigg]\nn,
\end{align}
\end{adjustwidth}
where we have defined
\begin{adjustwidth}{-1cm}{}
\vspace{-0.6cm}
     \begin{align}\label{eqn:XhatsN1}
&\hat{X}_1=\xi \frac{\dot{V}' e^{-4\rho}}{2 \dot{V} V''\sigma}d(e^{4\rho}\sin^2\theta\dot{V}^2)+\xi^2\frac{(-V'' \ddot{V}+(\dot{V}')^2 )e^{-6\rho}}{2\sigma \dot{V} V''}d(e^{6\rho}\sin^2\theta\dot{V}^2)+\xi^2 \dot{V}\sin^2\theta\Big(V'' d\sigma-\frac{\dot{V}'}{\sigma}d\eta\Big)\nn,\\[2mm]
&\hat{X}_2=\kappa\Bigg(\frac{\sigma f_2 e^{-4\rho}}{f_3^{\frac{1}{2}} \dot{V}^2}(1+\xi f_6)d(e^{4\rho}\sin^2\theta \dot{V}^2)-\frac{\xi}{2} f_2f_3^{\frac{1}{2}}\sin^2\theta\left(\dot{V}'\left(d\sigma-\frac{2 \dot{V}}{\sigma V''}d\rho\right)+\frac{d\eta}{\sigma}\left(2 \dot{V}- \ddot{V}\right)\right)\Bigg),\nn\\[2mm]
\end{align}
\end{adjustwidth}
 with $z=u+iv$. The Pure Spinors are then constructed using \eqref{eqn:Psi}, and one can show that the supersymmetry conditions in \eqref{eqn:IIAGconditions} are indeed satisfied. 
 
 It is worth observing that we have simply defined $\phi\rightarrow\phi+\gamma \chi$ in these results, which will correspond to a SUSY preserving T-duality along $\chi$. Hence, because $u$ and $v$ are independent of $\phi$, they are consequently independent of $\gamma$. For a SUSY preserving T-duality along $\phi$, one should instead define $\chi\rightarrow\chi+\gamma \phi$.  We will return to this in the next chapter, where we will show that the T-duality along both $U(1)$ directions will in fact derive mathematically equivalent IIB $\mathcal{N}=1$ solutions. 

We now fix $\gamma=0$, and write the complex vielbeins explicitly 
\begin{adjustwidth}{-1cm}{}
\vspace{-0.6cm}
 \begin{align}
\hat{E}^1 &= e^{\frac{1}{3}\Phi} E^1 =  - \frac{1}{\sqrt{X}} f_1^{\frac{3}{4}} f_5^{\frac{1}{4}}f_3^{\frac{1}{2}} \Xi^{-\frac{1}{4}}\Sigma_1^{-\frac{1}{2}} \bigg[\frac{1}{\sigma}d\sigma +d\rho+i  \Big( \Sigma_1 \,d\chi +\Sigma_2\,\sin^2\theta d\phi \Big)  +\xi \,d(V') \nn\\[2mm]
 &~~~~~ +\frac{e^{-2\rho}}{2\dot{V}}\bigg( \Sigma_2\, d(e^{2\rho}\dot{V}\sin^2\theta) +\bigg(\Sigma_2 -4\xi^2  \frac{f_2^2}{f_3f_5} \bigg)\sin^2\theta d(e^{2\rho}\dot{V})  + \frac{4\xi^2 f_2}{f_3f_5} \sin^2\theta \, \dot{V}d(e^{2\rho})\bigg)\bigg] ,\nn\\[2mm]
\hat{E}^2 &=e^{\frac{1}{3}\Phi} e^{-i\hat{\theta}_+} E^2 =-\frac{e^{i(\phi+\chi)}}{\sqrt{X}} \frac{ (f_1 f_5 \Xi  )^{\frac{1}{4}}}{\sqrt{\Sigma_1}} \Bigg[\frac{\kappa}{f_1} \Big(e^{-2\rho} d(e^{2\rho}\dot{V}\sin\theta) -\xi \dot{V}\sin\theta  d(V')  \Big)+i \sqrt{f_1f_2}\sin\theta d\phi \Bigg],
\end{align}
\end{adjustwidth} 
which re-derive the above $(j,\omega)$ via \eqref{eqn:IIAomegaj1}, and define the metric in the following manner
  \begin{align} 
&ds_{10}^2=e^{2A}ds^2(\text{Mink}_4)+
\hat{E}^1\bar{\hat{E}}^{1}+ \hat{E}^2 \bar{\hat{E}}^{2}+u^2+v^2,~~~~~~~~
e^{2A}=e^{2\hat{A}+\frac{2}{3}\Phi}=\frac{4}{X}  f_1^{\frac{3}{2}} f_5^{\frac{1}{2}}\Xi^{\frac{1}{2}} e^{2\rho} .\nn
\end{align}
In addition, we construct the higher form fluxes $(C_5,C_7)$, as follows
\begin{align}\label{eqn:C5C7N1}
C_5&=e^{4A-\Phi}\text{vol(Mink}_4\text{)}\wedge  u\nn \\[2mm]
&=\frac{16\kappa^2}{X}\Lambda\, e^{4\rho} \, \text{vol}(\text{Mink}_4) \wedge \bigg[\frac{\dot{V}'f_3}{4\sigma} d\sigma +V'' d\eta -f_6 d\rho -\frac{\xi}{f_5} \bigg(\frac{ \kappa^2 e^{-4\rho}}{2f_1^3}d\Big(e^{4\rho}\dot{V}^2 (\cos^2\theta-3)\Big)+4  d\rho\bigg)  \bigg],\nn \\[2mm]
C_7 &=  e^{4A-\Phi}\text{vol(Mink}_4\text{)}\wedge v\wedge  j \nn\\[2mm]
&=\frac{2^4\kappa}{X^2} \,e^{2\rho}f_1^{\frac{3}{2}}f_5^{\frac{1}{2}} \text{vol(Mink}_4\text{)}\wedge d(e^{2\rho} \dot{V} \cos\theta) \wedge\bigg( \frac{f_1^{\frac{3}{2}}f_5^{\frac{1}{2}} f_3}{\sigma} e^{-\rho}e^{-\xi V'}d\left(e^{\rho}\sigma e^{\xi V'}\right)\wedge d\chi \nn\\[2mm]
&~~~~+ \kappa f_2 f_3^{\frac{1}{2}}\hat{X}_1 \wedge  \Big(  d\chi+ d\phi\Big) +\hat{X}_2\wedge d\phi  \bigg).
\end{align}
 In the next section, we will generalise the G-structure discussion to the full two-parameter non-SUSY solution. 
\subsection{Supersymmetry breaking}\label{sec:nonSUSYfor2parameter}
We now generalise the G-structure description of the $\mathcal{N}=1$ solution discussed in the previous section, and investigate what insight can be gained into the breaking of supersymmetry for the general two-parameter $\mathcal{N}=0$ solution. As in the previous section, we first present the complex vielbeins in $d=11$, including all three parameters - as $\gamma$ becomes important when moving to type IIB. We then derive the IIA G-structure description from there. We follow the approach outlined in sections \ref{sec:G-Structure description of GM} - \ref{sec:IIAGstructureN=2}.

Following the review of supersymmetry breaking given in Section \ref{sec:SUSYbreakingGs}, by deriving the IIA G-structure forms for the general two-parameter solution, we will gain insight into how supersymmetry is broken from the three G-structure conditions given in \eqref{eqn:IIAGconditions}. In the general solution, we know that supersymmetry is preserved when $\zeta=-\xi$. Hence, we expect the right hand side of these conditions to contain supersymmetry breaking terms with an overall $(\xi+\zeta)$ multiplicative factor out the front. It will be interesting to see which of the three conditions are broken. 

The original motivation for extending the G-structure analysis to the full two-parameter family of $\mathcal{N}=0$ solutions actually occurred when investigating the stability of probe D6 branes for this general background, where one requires calculating the appropriate parametric deformation for the higher form fluxes $C_5$ and $C_7$ - extending the stability discussion of the $\mathcal{N}=0$ solution given in Section \ref{sec:N=0}. However, assuming the third supersymmetry condition \eqref{eqn:Calibrationform} is broken, we could no longer derive these higher form fluxes directly from the G-structure description. 
This raises an interesting question into possible parametric deformations of the Pure spinors, for which this third condition would be automatically satisfied by construction, and consequently, the higher form fluxes derived directly from these new Pure Spinors. If this approach proves fruitful, one could potentially extend this method to derive new sets of Pure spinors for each of the three SUSY conditions - satisfying them by construction. We will return to this discussion later in this section.
\subsubsection{$d=11$ forms}
We now present the appropriate modification to the vielbein frame of \eqref{eqn:InitialGstructureForms}, deriving \eqref{eqn:generalresult1} following a dimensional reduction along $\beta$
\begin{adjustwidth}{-1.5cm}{}
\vspace{-0.5cm}
\begin{align}\label{eqn:Gen11dGs}
K& = -\frac{ \kappa e^{-2\rho}}{f_1}   d(e^{2\rho} \dot{V} \cos\theta) ,\nn\\[2mm]
E_1&=-\sqrt{\frac{f_1f_3}{\Xi \Sigma_1}}\Bigg[\frac{1}{\sigma} d\sigma +d\rho +i\Big(\Sigma_1 d\chi +\Sigma_2 \sin^2\theta d\phi\Big) -\zeta d(V') +\frac{\gamma\zeta}{2\sigma^2}\dot{V}\sin^2\theta d\eta +(\zeta+\xi)d(V')\nn\\[2mm]
&~~~~~+\frac{e^{-2\rho}}{2\dot{V}} \Bigg(\Sigma_2 d(e^{2\rho}\dot{V}\sin^2\theta) +\bigg(\Sigma_2 -\zeta \frac{f_2}{f_3} \bigg(\gamma f_6 -4 (\gamma\xi-\zeta)\frac{f_2}{f_5}\bigg)\bigg)\sin^2\theta d(e^{2\rho}\dot{V})\nn\\[2mm]
&~~~~~ -\zeta \frac{4(\gamma\xi-\zeta)f_2}{f_3f_5}\sin^2\theta \dot{V}d(e^{2\rho})\Bigg) \Bigg], \nn\\[2mm]
E_2&=\frac{e^{i(\phi+\zeta \beta +\gamma\chi)}}{\sqrt{\Sigma_1}} \Bigg[\frac{\kappa}{f_1} \Bigg(e^{-2\rho} d(e^{2\rho} \dot{V}\sin\theta)- \dot{V}\sin\theta \bigg((\gamma\xi-\zeta) d(V') +\gamma \Big(\frac{1}{\sigma}d\sigma +d\rho\Big)\bigg)\Bigg)+ i \sqrt{f_1f_2} \sin\theta d\phi\Bigg],\nn\\[2mm]
E_3& = -\frac{e^{i(\chi+\xi \beta)}}{X}\Xi^{\frac{1}{2}}\sqrt{f_1f_5}\Bigg[\frac{X}{\Xi} \Bigg(- \frac{\dot{V}'f_3}{4\sigma} d\sigma -V'' d\eta +f_6 d\rho +\frac{1}{f_5} \bigg(\frac{-\zeta \kappa^2 e^{-4\rho}}{2f_1^3}d\Big(e^{4\rho}\dot{V}^2 (\cos^2\theta-3)\Big)+4\xi d\rho\bigg) \nn\\[2mm]
&~~~~~ -(\zeta+\xi)\frac{e^{-4\rho}}{\Lambda}d (\dot{V}^2e^{4\rho} )\Bigg)+i \big(d(X\beta) +C_1\big) \Bigg], 
\end{align}
\end{adjustwidth}
where
\begin{adjustwidth}{-1.25cm}{}
\vspace{-0.5cm}
\begin{align}
&\Sigma_1 = 1+\frac{f_2}{f_3} \bigg(\Big(\gamma +(\gamma \xi-\zeta) f_6\Big)^2 +(\gamma \xi-\zeta)^2 \frac{f_3}{f_5}\bigg)\sin^2\theta \equiv \frac{h_\chi}{f_3f_5},\nn\\[2mm]
&\Sigma_2= \frac{f_2}{f_3}\bigg((1+\xi f_6) \Big(\gamma+(\gamma\xi-\zeta)f_6\Big)  +\xi(\gamma \xi-\zeta) \frac{f_3}{f_5}\bigg)\equiv \frac{h_{\chi\phi}}{2f_3f_5\sin^2\theta}, \\[2mm]
&C_1=\frac{X}{\Xi}\bigg[\bigg(f_6(1+\xi f_6) +\xi \frac{f_3}{f_5} +\gamma\zeta \frac{f_2}{f_5}\sin^2\theta \bigg)d\chi   +\zeta \frac{f_2}{f_5}\sin^2\theta d\phi  \bigg],~~~~\Xi = (1+\xi f_6)^2+\xi^2\frac{f_3}{f_5}+ \zeta^2 \frac{f_2}{f_5}\sin^2\theta.\nn
\end{align}
\end{adjustwidth}
One then re-derives the $\mathcal{N}=1$ case given in \eqref{eqn:N=1Gs11d} by fixing $\zeta=-\xi$, noting that the $(\zeta+\xi)$ terms in \eqref{eqn:Gen11dGs} totally vanish in that case. Hence, these terms should contribute to the breaking of supersymmetry under the reduction!
\subsubsection{IIA initial forms}
Using the reduction formulae in \eqref{eqn:11Dto10Dforms} and recalling $ e^{\frac{4}{3}\Phi}=\frac{1}{X^2} f_1 f_5 \Xi$, we find
\begin{align}\label{eqn:GenIIAbetauandv}
v&= \hat{e}^3=e^{\frac{1}{3}\Phi}K =  \frac{\kappa}{\sqrt{X}} e^{-2\rho}f_5^{\frac{1}{4}}f_1^{-\frac{3}{4}}\Xi^{\frac{1}{4}} d(e^{2\rho} \dot{V} \cos\theta),\\[2mm]
u&=\hat{h}^3 =\frac{1}{\sqrt{X}}  (f_1f_5)^{\frac{3}{4}}\Xi^{-\frac{1}{4}}
  \bigg[\frac{\dot{V}'f_3}{4\sigma} d\sigma +V'' d\eta -f_6 d\rho -\frac{1}{f_5} \bigg(\frac{-\zeta \kappa^2 e^{-4\rho}}{2f_1^3}d\Big(e^{4\rho}\dot{V}^2 (\cos^2\theta-3)\Big)+4\xi d\rho\bigg)\nn\\[2mm]
  &~~~~~~ ~~~~~  +(\zeta+\xi)\frac{e^{-4\rho}}{\Lambda}d (\dot{V}^2e^{4\rho} )\bigg],\nn
\end{align}
with $E^1_{IIA} =e^{\frac{1}{3}\Phi}E^1$ and $E^2_{IIA} =-e^{i(\chi+\xi \beta)}
e^{\frac{1}{3}\Phi}E^2$ given in terms of the $d=11$ vielbeins \eqref{eqn:Gen11dGs} (where the multiplicative factor of $E^3$ has been included in the definition of $E_{IIA}^2$ with $\beta\rightarrow 0$). Using the Poincar\'e patch \eqref{eqn:PPatch}, one can now check that indeed we re-derive the \eqref{eqn:generalresult1} metric (with $\phi\rightarrow \phi+\gamma\,\chi$) in the following manner
 \begin{align}
     &   ds_{10,st}^2=  e^{2A} \,ds^2(\text{Mink}_4) +ds_6^2,~~~~~~~~~ e^{2A} = \frac{4}{X} e^{2\rho} f_1^{\frac{3}{2}}f_5^{\frac{1}{2}}\sqrt{\Xi}, \\[2mm]
     & ds_6^2=E^1_{IIA} \overline{E}^1_{IIA}+E^2_{IIA} \overline{E}^2_{IIA} +u^2+v^2= \frac{1}{X}f_1^{\frac{3}{2}}f_5^{\frac{1}{2}}\sqrt{\Xi}\bigg[4d\rho^2+f_4(d\sigma^2+d\eta^2)+ ds^2(M_3)\bigg],\nn
    \end{align} 
noting $z\bar{z}=u^2+v^2$, with $z=u+i\,v$.
We can now derive the corresponding $j$ and $\omega$, using \eqref{eqn:11Dto10Dforms}, or directly from the vielbeins
     \beq\label{eqn:IIAomegaj}
\omega= E^1_{IIA}\wedge E^2_{IIA} ,~~~~~~~~~~~~~~~ j=\frac{i}{2}(E^1_{IIA}\wedge \overline{E}^1_{IIA}+E^2_{IIA}\wedge \overline{E}^2_{IIA}),
\eeq
where we can check the following relations still hold
\beq\label{eqn:jomegarelations}
j\wedge \omega=\omega\wedge\omega =0,~~~~~~~\omega\wedge \bar{\omega}=2\, j\wedge j,
\eeq
which they do. The results for $\omega$ and $j$ then read
\begin{adjustwidth}{0cm}{}
\vspace{-0.5cm}
\begin{align}\label{eqn:GenIIAbetajandomega}
&\omega=  -\frac{2\kappa^2}{X} 
f_1^{-\frac{3}{2}}\bigg[e^{-3\rho} d\Big(e^{2\rho}\dot{V} e^{\zeta V'}e^{i(\phi+\gamma\,\chi)}\sin\theta\, d(e^{i\chi}e^{\rho}e^{-\zeta V'}\sigma)\Big)\nn\\[2mm]
&~~~~~~~~~~+(\zeta+\xi) e^{i\chi}e^{-2\rho} \sigma \,d\Big(e^{2\rho}\dot{V} e^{i(\phi+\gamma \,\chi)}\sin\theta \,d(V')\Big)   \bigg] ,\nn\\[2mm]
&j = \frac{1}{X\sqrt{\Xi}}\bigg[\frac{f_1^{\frac{3}{2}}f_5^{\frac{1}{2}}f_3}{\sigma} e^{-\rho}e^{\zeta V'} d\big(e^\rho \sigma e^{-\zeta V'}\big)\wedge d\chi +\kappa f_2 f_3^{\frac{1}{2}}\Big((X_1-\zeta \hat{X}_1)\wedge d\chi +(X_1+\xi  \hat{X}_1)\wedge (d\phi+\gamma\,d\chi) \Big)\nn\\[2mm]
&~~~~~~~~~~~~+ X_2 \wedge (d\phi+\gamma\,d\chi) -(\zeta +\xi) \frac{4\kappa \dot{V}}{\tilde{\Delta} \sqrt{\Lambda}}\Big( X_3\wedge  (d\phi+\gamma\,d\chi) +\hat{X}_3 \wedge d\chi\Big) \bigg],\nn\\[2mm]
&X_1= -\zeta \bigg(\frac{\dot{V}'e^{-4\rho}}{2\dot{V}V''\sigma} d\big(e^{4\rho}\sin^2\theta \dot{V}^2\big) +\xi \frac{(-V''\ddot{V} +(\dot{V}')^2)e^{-6\rho}}{2\sigma \dot{V}V''} d\big(e^{6\rho}\sin^2\theta \dot{V}^2\big)\bigg),\nn
\end{align}
\end{adjustwidth}
\begin{adjustwidth}{-1.5cm}{}
\vspace{-0.5cm}
\begin{align}
&\hat{X}_1=-\zeta \dot{V}\sin^2\theta  \Big(V''d\sigma -\frac{\dot{V}'}{\sigma}d\eta \Big),\\[2mm]
&X_2 = \kappa\,\bigg(\frac{\sigma f_2 e^{-4\rho}}{f_3^{\frac{1}{2}}\dot{V}^2}(1-\zeta f_6) d(e^{4\rho}\sin^2\theta \dot{V}^2) +\frac{\zeta}{2} f_2f_3^{\frac{1}{2}}\sin^2\theta \,\bigg(\dot{V}' \Big(d\sigma - \frac{2\dot{V}}{\sigma V''} d\rho\Big) +\frac{d\eta}{\sigma}(2\dot{V}-\ddot{V})\bigg)\bigg),\nn\\[2mm]
&X_3= e^{-2\rho}\sin\theta \bigg(2\dot{V}' +\xi \Big((\dot{V}')^2 -V''\ddot{V}\Big)\bigg)d(e^{2\rho}\sin\theta \dot{V}),~~~\hat{X}_3 = \frac{\sigma^2\tilde{\Delta}}{\dot{V}}d(V')+\zeta \sin^2\theta \,\dot{V} \Big((\dot{V}')^2 -V''\ddot{V}\Big)d\rho.\nn
\end{align}
\end{adjustwidth}
One can see by observation that the $\mathcal{N}=1$ result \eqref{eqn:uandv} is recovered when $\zeta=-\xi$. 

Notice, by naively following the derivation for the IIA forms, we would get an overall $e^{i(\zeta+\xi)\beta}$ multiplicative factor in the result for $\omega$ - which is the remnant of the broken $U(1)_R$ component in the general $\mathcal{N}=0$ case, vanishing in the SUSY preserved backgrounds. Of course, following the reduction, $\beta$ is no longer a coordinate of the background - so we must send $\beta\rightarrow 0$ and eliminate this overall factor.

We are now free to build the Pure spinors for the two-parameter background, using their definition \eqref{eqn:Psi}, and check the corresponding supersymmetry conditions \eqref{eqn:IIAGconditions}. For the remainder of this section, we fix $\gamma=0$. We can switch it on again by simply sending $\phi\rightarrow \phi+\gamma\,\chi$ in the results.
After some effort, we find
 \begin{subequations}\label{eqn:SUSYbrokenIIA}
\begin{align}
d_{H_3}(e^{3A-\Phi}\Psi_+)&=(\zeta+\xi) (Y_3+Y_5),\\[2mm]
d_{H_3}(e^{2A-\Phi}\text{Re}\Psi_-)&=(\zeta+\xi) \tilde{Y}_1
,\\
d_{H_3}(e^{4A-\Phi}\text{Im}\Psi_-)-\frac{e^{4A}}{8}*_6\lambda(g)&= (\zeta+\xi)\tilde{Y}_2,
\label{eqn:susybrokencond}
\end{align}
 \end{subequations}
where it is clear that the right hand side vanishes for $\zeta=-\xi$, with
\begin{align}
& Y_3= -\frac{2\kappa^2}{X}e^{i\chi}e^\rho \sigma \,d\bigg(\sigma e^\rho e^{i\chi} d\Big(e^\rho e^{i(\phi-\chi)}\sin\theta \frac{\dot{V}}{\sigma}d(V')\Big)\bigg),\\
&Y_5 = -\frac{4\kappa^3}{X^2}e^{i(\phi+\chi)}e^\rho\dot{V}\Bigg[\zeta e^{2\rho}\dot{V}V''\sin^2\theta d\eta\wedge d\sigma\wedge d\rho\wedge d\theta \wedge d\phi\nn \\
&+\bigg[\Big(\dot{V}'+\xi \big((\dot{V}')^2-\ddot{V}V''\big)\Big)\bigg(2e^{2\rho} d\eta \wedge d\sigma \wedge d\rho\wedge d\theta+i\,\sin\theta \Big(\frac{\dot{V}}{\sigma\tilde{\Delta}}\sin\theta \,d\big(e^{2\rho}\sigma^2 d(V')\big)\wedge d\theta \nn\\
&+2e^{2\rho}\cos\theta \,d\eta \wedge d\sigma\wedge d\rho\Big)\wedge d\phi\bigg)+i\,\zeta \frac{\sigma \dot{V}V''}{\tilde{\Delta}}\sin^2\theta \,d\Big(e^{2\rho} \dot{V} d(V')\Big) \wedge d\theta \wedge d\phi\bigg]\wedge d\chi \Bigg],\nn
\end{align}
and $(\tilde{Y}_1,\tilde{Y}_2)$ cumbersome poly-forms of even dimension.


\subsubsection{Deriving the higher form fluxes}
Following the discussion given in Section \ref{sec:N=0}, in order to investigate the stability of probe D branes, we will need to derive the higher form fluxes for the general two-parameter family. In the SUSY preserved solutions, one can use the third G-structure condition to derive the higher form fluxes directly from \eqref{eq:usefulpotentials}. Of course, in the general non-SUSY case, this final condition is broken by $(\zeta+\xi)$ terms on the right hand side of \eqref{eqn:susybrokencond}. This would suggest that, in general, the higher form fluxes one would derive using \eqref{eq:usefulpotentials} (which we will label $\tilde{C}_5$ and $\tilde{C}_7$) should be parametric deformations of the true forms ($C_5$ and $C_7$) - recovering the SUSY preserved case for $(\zeta+\xi)=0$. That is, 
 \beq\label{eqn:higherformDef1}
 C_5=\tilde{C}_5 +(\zeta+\xi)\tilde{C}'_5,~~~~~ C_7=\tilde{C}_7 +(\zeta+\xi)\tilde{C}'_7,
 \eeq
 with $\tilde{C}'_5$ and $\tilde{C}'_7$ the appropriate deformations, which can be derived via the Maxwell equations (where $H_3\wedge C_3=0$)
 \begin{align}
 F_6=-\star F_4 = dC_5,~~~~~
 F_8=\star F_2=dC_7-H_3\wedge C_5,
 \end{align}
 up to some gauge transformation of $C_7$.
We should then be able to split the higher form fluxes into a SUSY preserved part plus some SUSY-breaking parametric deformation
  \beq\label{eqn:higherformDef2}
 C_5=C_5\Big|_{(\zeta+\xi)=0}+(\zeta+\xi)C'_5,~~~~~ C_7=C_7\Big|_{(\zeta+\xi)=0}+(\zeta+\xi)C'_7.
 \eeq
 Following through the lengthy calculation, and writing each result in the manner given in \eqref{eqn:higherformDef1} and \eqref{eqn:higherformDef2} in turn, we find
    \begin{adjustwidth}{-0.5cm}{}
       \vspace{-0.7cm}
\begin{align}\label{eqn:C5gen}
C_5&=e^{4A-\Phi}\text{Vol(Mink}_4\text{)}\wedge \bigg(u+  \frac{(\zeta+\xi)}{\sqrt{X}}  (f_1f_5)^{\frac{3}{4}}\Xi^{-\frac{1}{4}} \Lambda^{-1}e^{-4\rho} d(\dot{V}^2e^{4\rho} )\bigg)\nn\\[2mm]
&=\frac{16\kappa^2}{X} \text{Vol}(\text{Mink}_4) \wedge \bigg[e^{4\rho}\Lambda \bigg(\frac{\dot{V}'f_3}{4\sigma} d\sigma +V'' d\eta -f_6 d\rho -\frac{1}{f_5} \bigg(\frac{-\zeta \kappa^2 e^{-4\rho}}{2f_1^3}d\Big(e^{4\rho}\dot{V}^2 (\cos^2\theta-3)\Big)+4\xi d\rho\bigg) \bigg) \nn\\
&~~~~~~~~~~~+2(\zeta+\xi)d (\dot{V}^2e^{4\rho} )\bigg], 
\end{align}
   \end{adjustwidth}
where we can see that the additional term vanishes in $F_6$, because $ d\big(d (\dot{V}^2e^{4\rho} )\big) = 0$, and 
\begin{align}\label{eqn:C7gen}
C_7 &=  e^{4A-\Phi}\text{Vol(Mink}_4\text{)}\wedge v\wedge\bigg[ j - \frac{ (\zeta+\xi)}{X\sqrt{\Xi}}\frac{4\kappa \dot{V}}{\tilde{\Delta} \sqrt{\Lambda}}\bigg(  \Big(X_3-e^{-2\rho}  \sin\theta \dot{V}' d(e^{2\rho}\sin\theta \dot{V})\Big) \wedge d\phi\nn\\
& ~~~~~-\,\bigg(\frac{\sigma^2\tilde{\Delta}}{\dot{V}}d(V') +\zeta \dot{V}\Big((\dot{V}')^2 -V''\ddot{V}\Big) \text{tan}\theta \,d\theta \bigg)\wedge d\chi  \bigg)\bigg]\nn\\
&=\frac{2^4\kappa}{X^2} \,e^{2\rho}f_1^{\frac{3}{2}}f_5^{\frac{1}{2}}\Xi^{\frac{1}{2}}\text{Vol(Mink}_4\text{)}\wedge d(e^{2\rho} \dot{V} \cos\theta) \wedge \Bigg[ \frac{1}{\sqrt{\Xi}}\bigg(\frac{f_1^{\frac{3}{2}}f_5^{\frac{1}{2}}f_3}{\sigma} e^{-\rho}e^{\zeta V'} d\big(e^\rho \sigma e^{-\zeta V'}\big)\wedge d\chi \nn \\
&~~~~ +\kappa f_2 f_3^{\frac{1}{2}}\Big((X_1-\zeta \hat{X}_1)\wedge d\chi +(X_1+\xi  \hat{X}_1)\wedge d\phi \Big)+ X_2 \wedge d\phi \bigg) \nn \\
&~~~~- \frac{ (\zeta+\xi)}{\sqrt{\Xi}}\frac{4\kappa \dot{V}}{\tilde{\Delta} \sqrt{\Lambda}}\bigg( \Big(2X_3-e^{-2\rho}  \sin\theta \dot{V}' d(e^{2\rho}\sin\theta \dot{V})\Big) \wedge d\phi \nn \\
&~~~~~~~~~~~~~~~~~~+\,\bigg(\hat{X}_3 -\frac{\sigma^2\tilde{\Delta}}{\dot{V}}d(V') -\zeta \dot{V}\Big((\dot{V}')^2 -V''\ddot{V}\Big) \text{tan}\theta \,d\theta \bigg)\wedge d\chi  \bigg) \Bigg],
\end{align}
 using the definitions given in \eqref{eqn:GenIIAbetauandv} and \eqref{eqn:GenIIAbetajandomega}, observing that $v$ remains undeformed.
 
 Now that we have the higher form fluxes, we could investigate the probe stability of the general two-parameter family, generalising the analysis given in Section \ref{sec:N=0} for the $S^2$ preserved solution. However, this is left for future study.

Let us now comment on an alternative set of pure spinors, inspired by the above analysis.
\subsubsection{Imposing the calibration condition}
  From the results of the higher form fluxes given in \eqref{eqn:C5gen} and \eqref{eqn:C7gen}, we have
  \beq
  C_5=e^{4A-\Phi}\text{Vol(Mink}_4\text{)}\wedge \hat{u},~~~~~~~C_7 =  e^{4A-\Phi}\text{Vol(Mink}_4\text{)}\wedge \hat{v}\wedge \hat{j},
  \eeq
  where
  \beq\label{eqn:jhatuhat}
  \begin{aligned}
 & \hat{u}=u+(\zeta+\xi)u',~~~~~~~\hat{v}=v+(\zeta+\xi)v',~~~~~~~\hat{j}=j+(\zeta+\xi)j',\\[2mm]
 &u' =   \frac{1}{\sqrt{X}}(f_1f_5)^{\frac{3}{4}}\Xi^{-\frac{1}{4}} \Lambda^{-1}e^{-4\rho} d(\dot{V}^2e^{4\rho} ),~~~~~~~~~v'=0,\\[2mm]
 & j' = - \frac{ 1}{X\sqrt{\Xi}}\frac{4\kappa \dot{V}}{\tilde{\Delta} \sqrt{\Lambda}}\bigg(  \Big(X_3-e^{-2\rho}  \sin\theta \dot{V}' d(e^{2\rho}\sin\theta \dot{V})\Big) \wedge d\phi \\[2mm]
 &~~~~~~-\,\bigg(\frac{\sigma^2\tilde{\Delta}}{\dot{V}}d(V') +\zeta \dot{V}\Big((\dot{V}')^2 -V''\ddot{V}\Big) \text{tan}\theta \,d\theta \bigg)\wedge d\chi  \bigg),
  \end{aligned}
  \eeq
  which then motivates the possibility of building a new set of parametrically deformed pure spinors from these new hatted forms, as follows
  \beq
\hat{\Psi}_+=\frac{1}{8} e^{\frac{1}{2}\hat{z}\wedge \overline{\hat{z}}}\wedge \hat{\omega},~~~~~\hat{\Psi}_{-} = \frac{i}{8}e^{-i\,\hat{j}}\wedge \hat{z},~~~~~~~~~~~\hat{z}=\hat{u}+i\,\hat{v},
\eeq
with some $ \hat{\omega}=\omega+(\zeta+\xi)\omega'$. The final G-structure condition \eqref{eqn:Calibrationform} would then be automatically satisfied by this new pure spinor (by construction), namely
\beq
d_H (e^{4A-\Phi}\text{Im}\hat{\Psi}_-)=\frac{e^{4A}}{8}\star_6 \lambda (g).
\eeq
In other words, using this more general set of (parametrically deformed) pure spinors, one could derive the higher form fluxes directly from the G-structure conditions for all backgrounds (no matter the SUSY). Obvious questions then arise regarding the meaning of such spinors, and whether they would make any sense physically. 

In order to preserve the $SU(2)$ description, and the conditions given in \eqref{eqn:jomegarelations}, we are motivated to generalise the definitions \eqref{eqn:IIAomegaj} by defining a new set of deformed complex vielbeins $(\hat{E}_{IIA}^1,\hat{E}_{IIA}^2)$ given by
\beq\label{eqn:newVielbeins}
\hat{E}^1_{IIA} = E^1_{IIA}+(\zeta+\xi)E'^1_{IIA},~~~~~~~\hat{E}^2_{IIA} = E^2_{IIA}+(\zeta+\xi)E'^2_{IIA},
\eeq
such that
     \beq\label{eqn:IIAomegajnew}
\hat{\omega}= \hat{E}^1_{IIA}\wedge \hat{E}^2_{IIA} ,~~~~~~~~~ \hat{j}=\frac{i}{2}(\hat{E}^1_{IIA}\wedge \overline{\hat{E}}^1_{IIA}+\hat{E}^2_{IIA}\wedge \overline{\hat{E}}^2_{IIA}),~~~~~~~~~g=\hat{E}^1_{IIA} \overline{\hat{E}}^1_{IIA}+\hat{E}^2_{IIA} \overline{\hat{E}}^2_{IIA} +\hat{u}^2+\hat{v}^2 .
\eeq
Hence, the new set of vielbeins should re-derive the $\hat{j}$ given in \eqref{eqn:jhatuhat}, 
whilst still defining the original metric, where
\beq\label{eqn:metricCondition}
\hat{E}^1_{IIA} \overline{\hat{E}}^1_{IIA}+\hat{E}^2_{IIA} \overline{\hat{E}}^2_{IIA} +\hat{u}^2\equiv E^1_{IIA} \overline{E}^1_{IIA}+E^2_{IIA} \overline{E}^2_{IIA} +u^2,
\eeq
with $\hat{v}\equiv v$ dropping out. These vielbeins would then define the corresponding $\hat{\omega}$. 

Using \eqref{eqn:IIAomegaj}, \eqref{eqn:newVielbeins}, \eqref{eqn:IIAomegajnew} and \eqref{eqn:jhatuhat}, we find from the conditions on $\hat{j}$, along with the metric condition \eqref{eqn:metricCondition}, the following requirements
\begin{align}
&E'^1_{IIA}   \bar{E}^1_{IIA}+E^1_{IIA}  \bar{E}'^1_{IIA}+E'^2_{IIA}  \bar{E}^2_{IIA}+E^2_{IIA}  \bar{E}'^2_{IIA}+2uu' +(\zeta+\xi) \Big( E'^1_{IIA}  \bar{E}'^1_{IIA}+E'^2_{IIA}  \bar{E}'^2_{IIA}+u'^2\Big) = 0,\nn\\[2mm]
&j'=\frac{i}{2} \bigg(E'^1_{IIA} \wedge \bar{E}^1_{IIA}+E^1_{IIA} \wedge \bar{E}'^1_{IIA}+E'^2_{IIA} \wedge \bar{E}^2_{IIA}+E^2_{IIA} \wedge \bar{E}'^2_{IIA}\\[2mm]
&~~~~~~+(\zeta+\xi)\Big(E'^1_{IIA} \wedge \bar{E}'^1_{IIA}+E'^2_{IIA} \wedge \bar{E}'^2_{IIA}\Big)\bigg),\nn
\end{align}
where the second condition is a consequence of $j'$, given in \eqref{eqn:jhatuhat}, having no $(\zeta+\xi)$ terms. It is now a case of cranking the Mathematica handle and seeing whether such a set of vielbeins exist. Again, this is left for future analysis.

\subsection{Comments on dual CFTs}\label{sec:commentsCFT}
In this sub-section, we propose that the CFT deformations we encountered in the dual supergravity description \eqref{eqn:generalresult1} represents marginal deformations. In Section \ref{sec:quantconds}, we analysed the holographic central charge within the context of the quantization of charge. There we concluded that forcing integer quantization onto the system is simply too much to ask, giving rise to the same holographic central charge for the whole family of solutions. Here we comment on a mirror-like relation that our CFTs satisfy, and study spin-two fluctuations for both the $S^2$ preserved ${\cal N}=0$ background as well as the $\mathcal{N}=1$ solutions.

The CFT dual to our two-parameter family of solutions should admit a large $N$ expansion, and all single trace operators with spin greater than two must have a very large dimension. CFTs with these characteristics were studied in \cite{Bashmakov:2017rko}.

Considering the reduction to five dimensional gravity of the $S^2$ preserved ($\zeta=0$) case of \eqref{eqn:generalresult1}, we find the scalars in the AdS-bulk (which correspond to marginal operators) have mass $m^2=\Delta(\Delta-4)=0$. The non-normalisable mode of these scalars is then dual to the $g$ coupling of the perturbed CFT \eqref{deformationCFT}, with the conformal manifold associated with the moduli space of AdS$_5$ vacua in the reduced theory.

 \paragraph{Holographic Central Charge}  
Following the calculation of Appendix \ref{sec:HCCcalc} and the discussion given in Section \ref{sec:quantconds}, 
one arrives at the result
 \begin{equation}\label{eqn:HCC}
\begin{aligned}
c_{hol}=\frac{1}{4\pi}\int_0^P \mathcal{R}(\eta)^2d\eta =\frac{1}{8\pi}\sum_{k=1}^\infty P \, \mathcal{R}_k^2,
\end{aligned}
\end{equation}
where all dependence on the dilaton drops out neatly, matching the $\mathcal{N}=2$ result of \cite{Nunez:2019gbg} (up to appropriate conversion of notation). It would be interesting to calculate \eqref{eqn:HCC} for more rank function examples, comparing with the results of \eqref{centralsN=2}. Various examples along these lines are given in  \cite{Nunez:2019gbg,Lozano:2016kum,Nunez:2023loo}.

\paragraph{The dual to our backgrounds}
Let us now go into more detail about the marginal deformations represented in \eqref{eqn:generalresult1}, focusing our attention on the $U(1)_R$ preserving ${\cal N}=1$ case (for $\xi=-\zeta$) and in the $\text{SU}(2)$ preserving ${\cal N}=0$ case (with $\zeta=0$). We recall from the soft-SUSY breaking discussion given in Section \ref{sec:softSUSY}, there are ${\cal N}=0$ deformations with operator ${\cal O}_1$, with a preserved SU(2) global symmetry inherited from the R-symmetry. We then associate this deformation with the CFTs described by the parameter $\xi$ (with $\zeta=0$), corresponding to the red line in Figure \ref{fig:IIAtableplot}. Similarly, the operators ${\cal O}_2$ and ${\cal O}_3$ preserve ${\cal N}=1$, with R-symmetry $\text{U}(1)= \frac{2}{3} \left( \text{U}(1)_R \pm 2 I_3\right)$. We then associate these deformations with the CFTs described by $\xi=-\zeta$ in \eqref{eqn:generalresult1}, corresponding to the blue line in Figure \ref{fig:IIAtableplot}. The discussion into the quantization of D6 branes given in Section \ref{sec:quantconds}, which is inherited from the presence of spindles and the parameter $\xi$, suggest that these deformations are non-Lagrangian. When $\xi=0$, the integer D6 quantization is recovered (with an $S^2$ recovered for $\zeta=0$). This suggests that the parameters $\zeta$ and $\xi$ give rise to Lagrangian and non-Lagrangian deformations, respectively. The arguments presented above are not air tight, and a more careful analysis of these exactly marginal deformations is required.

\begin{figure}[H]
 \includegraphics[width=1.125 \textwidth]{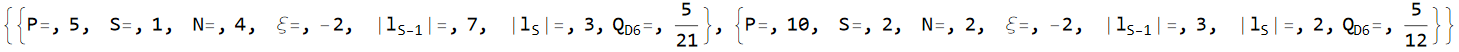}
   \caption{Mirror relation between two (spindle appropriate) Triangular rank functions}
\label{fig:Mathematica9}
\end{figure}

 \paragraph{A mirror-like relation} It is worth noting a mirror symmetry relation between the linear quiver $\mathcal{R}(\eta)$ (of length $P-1$) \eqref{eqn:Rkwithderivs} and $\hat{\mathcal{R}}(\hat{\eta})$ (of length $F-1$), 
 \begin{equation*} 
    \hat{\mathcal{R} }(\hat{\eta}) =
    \begin{cases}
     ~~~~~~~~~~~~~~~  \hat{N}_1\hat{\eta} & \hat{\eta} \in [0,\frac{F}{P}]\\
      \hat{N}_k+(\hat{N}_{k+1}-\hat{N}_k) ( \hat{\eta}-k) & \hat{\eta} \in  [k \frac{F}{P},(k+1)\frac{F}{P} ] \\
   ~~~~~~~~~   \hat{N}_{F-1}(F-\hat{\eta})  & \hat{\eta}\in [F(1-\frac{1}{P}),F],
    \end{cases}     
\end{equation*}
 with 
 $(\frac{F}{P},\hat{N}_k=\frac{P}{F}N_k)\in\mathds{Z}$, identical Fourier coefficients, $\mathcal{R}_n=\hat{\mathcal{R}}_n$, and equivalent `holographic central charge per unit length', $\frac{c_{hol}}{P}=\frac{\hat{c}_{hol}}{F}$. See \cite{Macpherson:2024frt} for further details. It would be interesting to test whether the preservation of the spindle conditions survive under such a mirror symmetry - \textit{i.e.} for a given rank function, $\mathcal{R}(\eta)$, satisfying all conditions, would its mirror, $\hat{\mathcal{R}}(\hat{\eta})$, still satisfy these conditions? As an example, the triangular rank function defined by $P=5,S=1,N=4,\xi=-2$ leads to a mirror rank function defined by $\hat{P}=F=10,\hat{S}=\frac{SF}{P}=2,\hat{N}=\frac{PN}{F}=2,\xi=-2$. Indeed, both rank functions satisfy the spindle conditions, see Figure \ref{fig:Mathematica9}, but a more thorough investigation could be conducted for a wider range of examples.


  \paragraph{Spin 2 fluctuations}  
For our two example solutions \eqref{N=0metric} and \eqref{eq:neq1sol}, we now study particular excitations of the metric (along the AdS$_5$ directions), using the results of \cite{Bachas:2011xa,Chen:2019ydk,Itsios:2019yzp,Lima:2023ggy}. This simple fluctuation is consistent and can be associated with states of spin two in the CFT. See Section \ref{sec:dualtoolbox} for a brief review. 
\begin{itemize}
\item \textbf{$\mathcal{N}=0$ Reduction:}
Using the form of the  SU$(2)$ preserving $\mathcal{N}=0$ metrics presented in \eqref{N=0metric}, after moving to Einstein Frame, one finds
\begin{equation}
ds_{E}^2=4(f_1^9 f_5 \Delta)^{\frac{1}{8}}\bigg[ds^2(\text{AdS}_5)+\frac{1}{4}\bigg(f_2 ds^2(\text{S}^2)+f_4(d\sigma^2+d\eta^2)+\frac{f_3 }{\Delta} d\chi^2\bigg)\bigg].
\end{equation} 
    Using \eqref{genericglue}, we have
    \begin{equation}
    e^{2A_E} = 4 (f_1^9 f_5 \Delta)^{\frac{1}{8}} ,~~~~~~~~\tilde{g}_{\mathcal{M}_5}=\frac{1}{4^5\Delta}f_2^2 f_4^2 f_3 \sin^2\theta,~~~~~~~~e^{8A_E}\sqrt{\tilde{g}_{\mathcal{M}_5}} = {2^3}f_1^{\frac{9}{2}}f_3^{\frac{1}{2}}f_5^{\frac{1}{2}}f_2f_4 \sin\theta.
    \end{equation}
Then using \eqref{eqn:Leq} along with the definitions of $f_i$ \eqref{eqn:fs}, we find
\begin{align}
&\frac{2\left((2\dot{V}-\ddot{V})V''+(\dot{V}')^2\right)}{V''\dot{V}}\nabla_{\text{S}^2}^2 \mathcal{F}  +\frac{ 2\dot{V}- \ddot{V}}{\sigma^2 V''}\Delta\partial_\chi^2  \mathcal{F}\nn\\[2mm]
&+ \frac{2}{V''\dot{V}\sigma }\bigg[\partial_\eta\Big( \sigma \dot{V}^2   \partial_\eta\Big)  +\partial_\sigma\Big( \sigma \dot{V}^2  \partial_\sigma\Big) \bigg] \mathcal{F} +M^2 \mathcal{F}  =0,
\end{align}
noting the Laplacian on the two sphere reads 
\beq
\nabla_{\text{S}^2}^2\equiv \frac{1}{\sin\theta}\partial_\theta (\sin\theta \partial_\theta)+\frac{1}{\sin^2\theta}\partial_\phi^2.
\eeq
This result only differs from the ${\cal N}=2$ GM result by $\Delta$, which goes to 1 for $\xi=0$. In that case, universal spin-two modes were already considered in \cite{Chen:2019ydk}. We therefore follow the same procedure and expand the mass eigenfunction as
\begin{equation}
\mathcal{F} = \sum_{lmn}\phi_{lmn}Y_{lm}e^{in\chi},~~~~\nabla_{\text{S}^2}^2Y_{lm} = -l(l+1) Y_{lm},\label{fredef}
\end{equation}
where $l,m,n\in\mathbb{Z}$ and $l\geq1$. Note here that $\mathcal{F}$ and $\phi_{lmn}$ are complex in general. This leads to 
\begin{align}
&-l(l+1)\frac{2\left((2\dot{V}-\ddot{V})V''+(\dot{V}')^2\right)}{V''\dot{V}} \mathcal{F}  -n^2\frac{ 2\dot{V}- \ddot{V}}{\sigma^2 V''}\Delta   \mathcal{F}\nn\\[2mm]
&+ \frac{2}{V''\dot{V}\sigma }\bigg[\partial_\eta\Big( \sigma \dot{V}^2   \partial_\eta\Big)  +\partial_\sigma\Big( \sigma \dot{V}^2  \partial_\sigma\Big) \bigg] \mathcal{F} +M^2 \mathcal{F}  =0,\label{eq:lnformneq0}
\end{align}
which is in fact a version of the Sturm-Liouville problem, fitting into the `universal form'  given in \cite{Lima:2023ggy} by the following identification
\begin{eqnarray}
& &\partial_a(\bar{p}\partial^a\psi)+\bar{q}\psi =-\bar{m}^2 \bar{w} \psi, ~~~~\text{with} ~~\psi=\phi_{\hat{l}\hat{m}\hat{n}} ~~~\text{and}  \label{eqn:UF}\\
& & \bar{p}=\sigma \dot{V}^2,~~~~~\bar{w}=\frac{1}{2}\sigma V'' \dot{V}, ~~~~~\bar{m}^2=M^2,~~~ \bar{q}= - \sigma \dot{V}V'' \left( \frac{\tilde{\Delta} \,\hat{l}(\hat{l}+1)}{\dot{V} V''} +\frac{\hat{n}^2 \Lambda \Delta}{2\sigma^2} \right).\nonumber
\end{eqnarray}
By redefining $\phi_{lmn}$, the ${\cal N}=2$ analogue of \eqref{eq:lnformneq0} was mapped to a more useful form in \cite{Chen:2019ydk}. We follow the same procedure, defining
\beq
\phi_{lmn}= e^{n \xi V'}\sigma^n (\dot{V})^l\tilde{\phi}_{lmn},~~~~ M^2=\mu^2+(2l+n)(2l+n+4),
\eeq
upon which \eqref{eq:lnformneq0} becomes
\beq
\sigma^n \dot{V}^le^{n\xi V'}\left(\frac{2 e^{-2n \xi V'}}{\sigma^{2n-1}\dot{V}^{2l+1}\ddot{V}}\partial_a\left(\sigma^{2n+1}\dot{V}^{2+2l}e^{2n\xi V'}\partial_a\tilde{\phi}_{lmn}\right)-\mu^2\right)=0,\label{eq:spin2finalneq0}
\eeq
where $a\in(\eta,\sigma)$. This can then be solved by $\tilde{\phi}_{lmn}=$ constant and $\mu=0$, for all $n$, which then leads to the universal solution
\beq
\phi^r_{lmn}= e^{n \xi V'}\sigma^n (\dot{V})^l\phi_0,~~~~ M^2=-4+(2+2l+n)^2,
\eeq
where $\phi_0=$ constant. 

For the $V$ defined in \eqref{eqn:potential}, this solution remains finite at all points on the Riemann surface -  just as they were shown to be for the ${\cal N}=2$ case in \cite{Chen:2019ydk}.

In order to define the norm of fluctuations, we can derive an integration measure by computing the quadratic fluctuation of  $S= \frac{1}{2\kappa_{10}^2}\int\sqrt{-\det g}\, R\, dx^{10}$. In general, we find
\beq
\delta^{(2)} S= \frac{1}{2\kappa_{10}^2}\int dx^5 dy^5 e^{3A}\sqrt{-\det g_{\text{AdS}_5}}\sqrt{\det {\cal M}_5}
h_{\mu\nu}\left(\nabla^2_{\text{AdS}_5}+2-M^2\right)h^{\mu\nu}.
\eeq
Hence, when integrating over $y$, we find the appropriate integration measure is $e^{3A}\sqrt{\det {\cal M}_5}$. We can now use this to derive a bound on $M^2$. 

Returning to \eqref{eq:spin2finalneq0}, contracting with $\overline{\phi}_{lmn}$ and integrating with respect to the above measure, we find
\beq
-\int d\eta d\sigma\bar{\tilde{\phi}}_{lmn}\left(\partial_a\left(\sigma^{2n+1}\dot{V}^{2+2l}e^{2n\xi V'}\partial_a\tilde{\phi}_{lmn}\right)+\frac{1}{2}\mu^2 e^{2n\xi V'}\sigma^{2n+1}\dot{V}^{2l+1}V'' \tilde{\phi}_{lmn}\right)=0.
\eeq
Integrating by parts, and assuming no boundary contributions (which is the case for regular fluctuations -  finite everywhere on the Riemann surface), leads to
\beq
\int d\eta d\sigma\left(\sigma^{2n+1}\dot{V}^{2+2l}e^{2n\xi V'}|\partial_a\tilde{\phi}_{lmn}|^2-\frac{1}{2}\mu^2 e^{2n\xi V'}\sigma^{2n+1}\dot{V}^{2l+1}V'' |\tilde{\phi}_{lmn}|^2\right)=0,\nn
\eeq
where the first term is positive definite and the second term is negative definite. We then see that the minimal value $\mu^2$ can take is $\mu^2=0$, for which $\tilde{\phi}_{lmn}$ is a constant. We then conclude that
\beq
M^2\geq (2l+n)(2l+n+4),
\eeq
just as the ${\cal N}=2$ case of \cite{Chen:2019ydk}\cite{Itsios:2019yzp}. As a bound on the scaling dimension of operators, this then becomes $\Delta \geq 4+2l+n$.
\item \textbf{$\mathcal{N}=1$ Reduction}:
Using the $\mathcal{N}=1$ metric presented in \eqref{eq:neq1sol}, after moving to Einstein Frame,
    \begin{align}
        ds_{E}^2=4(f_1^9f_5\Xi)^{\frac{1}{8}}\bigg[ds^2(\text{AdS}_5)+h_{\mu\nu} dx^\mu dx^\nu+\frac{1}{4}\bigg(f_4(d\sigma^2+d\eta^2)+ds^2(M_3) \bigg)\bigg],\nn
    \end{align}
and using \eqref{genericglue}, gives
    \begin{equation}
    e^{2A_E} =4(f_1^9f_5\Xi)^{\frac{1}{8}},~~~~~~~~\tilde{g}_{\mathcal{M}_5}=\frac{1}{4^5\Xi}f_2^2 f_4^2  f_3 \sin^2\theta ,~~~~e^{8A_E}\sqrt{\tilde{g}_{\mathcal{M}_5}} = 2^3 f_1^{\frac{9}{2}}f_3^{\frac{1}{2}}f_5^{\frac{1}{2}}f_2f_4\sin\theta,
    \end{equation}
leading to
\begin{align}
&\frac{2\left((2\dot{V}-\ddot{V})V''+(\dot{V}')^2\right)}{V''\dot{V}}\nabla_{\text{S}^2}^2 \mathcal{F} 
+\left(4\xi^2\left(\frac{1}{f_5}+\frac{f_6^2}{f_3}\right)\partial_{\phi}^2+8 \xi \frac{\xi f_3 +f_5 f_6(1+\xi f_6)}{f_3f_5}\partial_{\chi}\partial_{\phi}\right){\cal F}\nn\\[2mm]
& +\frac{ 2\dot{V}- \ddot{V}}{\sigma^2 V''}\Delta\,\partial_\chi^2  \mathcal{F}+ \frac{2}{V''\dot{V}\sigma }\bigg[\partial_\eta\Big( \sigma \dot{V}^2   \partial_\eta\Big)  +\partial_\sigma\Big( \sigma \dot{V}^2  \partial_\sigma\Big) \bigg] \mathcal{F} +M^2 \mathcal{F}  =0,
\end{align}
which actually behaves remarkably similar to the equation defining the mass of spin 2 fluctuations in the ${\cal N}=0$ deformation.  Once again, we redefine ${\cal F}$ in terms of  \eqref{fredef}, and now further redefine 
\beq
\phi_{lmn}= e^{(n-m) \xi V'}\sigma^n (\dot{V})^l\tilde{\phi}_{lmn},~~~~ M^2=\mu^2+(2l+n)(2l+n+4).
\eeq
We again make use of $\nabla_{\text{S}^2}^2Y_{lm} = -l(l+1) Y_{lm}$ and $\partial_{\phi}Y_{lm}=im Y_{lm}$, where we find the resulting PDE for $\tilde{\phi}_{lmn}$ takes an almost identical form to \eqref{eq:spin2finalneq0}, but with $e^{n \xi V'}$ replaced with $e^{(n-m) \xi V'}$ everywhere it appears. The remaining arguments of the previous $\mathcal{N}=0$ example then follow through the same. That is, a universal regular fluctuation, valid when $\mu^2=0$, is given by
\beq
\phi^r_{lmn}= e^{(n-m) \xi V'}\sigma^n (\dot{V})^l\phi_0,~~~~ M^2=-4+(2+2l+n)^2,
\eeq
with $\phi_0$ a constant. Again, this solution saturates the bound on the mass on can derive, which is again
\beq
M^2\geq (2l+n)(2l+n+4),
\eeq
which once more matches the ${\cal N}=2$ result.
\end{itemize}
\newpage
\section{$\chi$ Reduction}\label{sec:chiRed}
The full solution, with all $GL(3,\mathds{R})$ transformation parameters intact, is given in \eqref{eqn:chi-Gen}.
\subsection{Two-Parameter Family}
Given that we are now performing a dimensional reduction along $\chi$, in order to preserve the $U(1)_R$ component \eqref{eqn:U(1)lessgen} under reduction, we must now fix $s=-1$. Recall from Section \ref{sec:SL(3,R)} that the maximum supersymmetry which can be achieved under this reduction is $\mathcal{N}=1$, because the very condition required to preserve the $U(1)_R$ component is what breaks the $SU(2)_R$ component - a consequence of the $U(1)_R$ component being $\chi+\phi$ prior to the $SL(3,\mathds{R})$ transformation, given in \eqref{eqn:origU(1)}.

We will first investigate keeping the parameter $s$ free, allowing us to turn the $\mathcal{N}=1$ supersymmetry on and off at will.
\subsubsection{Fixing $s\equiv\zeta$}
With $s\equiv \zeta$ and $(p,b,u)=1$, from the determinant given in \eqref{eqn:S2breakingdefns}, we must fix $m=0$ - with either $q=0$ (where $a\equiv \xi$) or $c=0$ (where $v\equiv \xi$). The remaining free parameter will be labelled $\gamma$ in each case. We will now investigate both options in turn.
\begin{itemize}
\item \textbf{Fixing $(a\equiv \xi, v\equiv\gamma)$}\\
With $(m,q)=0$, the determinant \eqref{eqn:S2breakingdefns} reduces to $vc=0$. Hence, the third free parameter will be $v\equiv \gamma$ (with $c=0$) or $c\equiv \gamma$ (with $v=0$). In the solution we will now present, as in the $\beta$ reduction case, the parameter $\gamma$ will in fact play a trivial role and can be fixed to zero without loss of generality. More specifically, when $c\equiv \gamma$ (or $v\equiv \gamma$), one derives \eqref{eqn:chireduction1} but with the redefinition $\beta=\beta+\gamma\,\phi$ (or $\phi=\phi+\gamma\,\beta$). Notice from the $U(1)_R$ component \eqref{eqn:U(1)lessgen}, $v\equiv\gamma$ plays an important role when performing an ATD to IIB, we will return to this discussion in the next chapter. 

Hence, in $d=11$ , we have the following transformations being performed
\beq\label{eqn:chi1}
\begin{gathered}
d\beta \rightarrow d\beta + \xi \, d\chi,~~~~~~~~~~~~~~~~~d\chi \rightarrow d\chi  ,~~~~~~~~~~~~~~~~~~d\phi \rightarrow d\phi+ \zeta \, d\chi .
\end{gathered}
\eeq
Utilising \eqref{eqn:chi-Gen}, we now derive the following two-parameter family of solutions
  \begin{align}\label{eqn:chireduction1}
  &      ds_{10,st}^2=  \frac{1}{X}f_1^{\frac{3}{2}}\big(f_5f_6^2+ f_3  \big)^{\frac{1}{2}}\sqrt{\Xi_2}\bigg[4ds^2(\text{AdS}_5)+f_4(d\sigma^2+d\eta^2)+     ds^2(M_3)  \bigg],\nn\\[2mm]
   &          ds^2(M_3) =  f_2\bigg(d\theta^2 + \frac{\Delta_2}{\Xi_2}\sin^2\theta \,D\phi^2\bigg) +\frac{1}{\Delta_2}\frac{f_3f_5}{(f_5f_6^2+f_3)}d\beta^2 \nn\\[2mm]
   &~~~~~~~~~~~ =f_2\bigg(d\theta^2+\frac{1}{\Pi_2}\sin^2\theta \,d\phi^2\bigg) +\frac{\Pi_2}{\Xi_2}\frac{f_3f_5}{(f_5f_6^2+f_3)}D\beta^2 ,\nn\\[2mm]
    &   B_2 =  \frac{1}{X}\sin\theta \bigg[  \zeta f_8 d\beta -\Big( f_7+ \xi f_8\Big)d\phi\bigg] \wedge \, d\theta,~~~~~~~~  e^{\frac{4}{3}\Phi}= \frac{1}{X^2} f_1\big(f_5f_6^2+ f_3  \big)\Xi_2 ,\nn\\[2mm]
   &   C_1=  \frac{X}{ \big(f_5f_6^2+ f_3  \big)\Xi_2} \bigg(f_5 (f_6+\xi) d\beta  +\zeta f_2\sin^2\theta \, d\phi\bigg)  ,~~~~~~~~   C_3=  f_8    \,d\beta  \wedge \text{vol}(S^2),\nn\\[2mm]
    &\Xi_2= \Delta_2+\zeta^2 \frac{f_2}{\big(f_5f_6^2+ f_3  \big)} \sin^2\theta,~~~~~~~~\Delta_2=1+\xi \frac{f_5(2f_6+\xi)}{\big(f_5f_6^2+ f_3 \big)},~~~~~~~~~~\Pi_2=1+\zeta^2 \frac{f_2}{f_3}\sin^2\theta,\nn\\[2mm]
    &D\phi =d\phi -\frac{\zeta}{\Delta_2} \frac{f_5(f_6+\xi)}{(f_5f_6^2+f_3)}d\beta ,~~~~~~~D\beta=d\beta -\frac{\zeta}{\Pi_2}\frac{f_2}{f_3}(f_6+\xi)\sin^2\theta\,d\phi,
    \end{align}
        with
    \beq\label{eqn:fluxeschi}
    H=dB_2,~~~~~~F_2=dC_1,~~~~~~F_4=dC_3-H\wedge C_1,~~~~~~dH=0,~~~~~d_HF_4=0,
    \eeq
where we have rewritten the solution in a manner which becomes useful when performing analysis at the boundaries. Notice the roles of $(f_7,f_8)$ in the definitions of $(B_2,C_3)$ have swapped with respect to the $\beta$ reduction case \eqref{eqn:generalresult1}.

Following previous analysis, we consider approaching the $\sigma= 0$ boundary where $\ddot{V}\rightarrow 0$ to leading order. With the boundary condition $\mathcal{R}(\eta) = \dot{V}\Big|_{\sigma=0}$ and warp factors \eqref{eqn:fs}, we find
\beq\label{eqn:C1sourcechi}
C_1\Big|_{\sigma\rightarrow 0} =X \frac{  (\xi+ \mathcal{R}') d\beta+\frac{1}{2}\zeta V'' \mathcal{R} \sin^2\theta d\phi }{(\xi+ \mathcal{R}')^2+\frac{1}{2}\zeta^2 V'' \mathcal{R} \sin^2\theta },~~\Rightarrow~~~~~~C_1\Big|_{\sigma\rightarrow 0}^{\zeta/\sin\theta=0}= \frac{ X }{ \xi+ \mathcal{R}'  }d\beta ,
\eeq
which reduces for $\zeta=0$ or $\sin\theta=0$, as shown. We recall from \eqref{eqn:Rkwithderivs} that $\mathcal{R}'(\eta)$ is discontinuous in $\eta$. Hence, following a reduction along $\chi$, we still have a source term for D6 branes in the $F_2$ Bianchi identity. Taking the derivative carefully leads to
\begin{align}
&F_2\Big|_{\sigma\rightarrow 0,\eta=k}^{\zeta/\sin\theta=0}=X  \bigg(\frac{1}{\xi+ \mathcal{R}'(k)}-\frac{1}{\xi+ \mathcal{R}'(k-1)}\bigg)   d\eta\wedge  d\beta = \frac{ X\big(\mathcal{R}' (k-1)-\mathcal{R}' (k)\big)}{\big(\xi+ \mathcal{R}'(k)\big)\big(\xi+ \mathcal{R}'(k-1)\big)}  d\eta\wedge  d\beta, \nn\\[2mm]
&\Rightarrow ~~~~ F_2\Big|_{\sigma\rightarrow 0 }^{\zeta/\sin\theta=0}= X\sum_{k=1}^{P-1} \frac{2N_k-N_{k+1}-N_{k-1}}{\big(\xi+ (N_{k+1}-N_k)\big)\big(\xi+  (N_{k}-N_{k-1})\big)} \delta(\eta-k)\delta(\sigma)d\eta\wedge  d\beta,
\end{align}
where we now have a different denominator compared to the $\beta$ reduction case \eqref{eqn:betaF2val}. We can see immediately, following an integrating to find the D6 charge, fixing $\xi=0$ no longer lead to integer quantization. The forms of $B_2$ and $C_3$ are still independent of $f_6$, leading to source free Bianchi identities for $H$ and $F_4$ - given in \eqref{eqn:fluxeschi}. We will return to this discussion more thoroughly in the next sub-section, where we investigate the boundary.

As in the $\beta$ reduction case, we once again find that $\zeta\neq0$ breaks the $S^2$ of this background. We can see this from \eqref{eqn:chireduction1} using the forms of $ds^2(M_3), \Pi_2$ and $D\beta$. In addition, $\zeta$ is now the only parameter which determines the supersymmetry of the solution, which is in general $\mathcal{N}=0$ and enhances to $\mathcal{N}=1$ when $\zeta=-1$. This condition then preserves the $U(1)_R$ component of the R-symmetry, but necessarily breaks the $SU(2)_R$ component - this can be seen graphically in Figure \ref{fig:IIAtableplotchi}, where the $U(1)_R$ preserving $\mathcal{N}=1$ line (given in blue) never crosses the $SU(2)$ preserving line (given in red). This is in contrast to Figure \ref{fig:IIAtableplot}, where they cross at the origin, giving rise to an $\mathcal{N}=2$ solution there.

  \begin{figure}[H]
\centering  
\subfigure
{
\centering
  \begin{minipage}{0.5\textwidth}
\begin{tabular}{c | c c c c  }
$\chi$- Reduction&$\mathcal{N}$&U(1)$_R$&SU(2)$_R$  \\
\hline
$\textcolor{blue}{\zeta=-1}$&$ \textcolor{blue}{1}$ &$\textcolor{blue}{\checkmark}$&$\textcolor{blue}{\times}$  \\
$\textcolor{red}{\zeta=0}$&$ \textcolor{red}{0}$ &$\textcolor{red}{\times}$&$\textcolor{red}{\checkmark}$  \\
$\textcolor{teal}{\zeta \in \mathds{Z}/\{0,-1\}}$&$ \textcolor{teal}{0}$ &$\textcolor{teal}{\times}$ &$\textcolor{teal}{\times}$
\end{tabular}
  \end{minipage}
     \begin{minipage}{.5\textwidth}
    \centering
\begin{tikzpicture}[scale=0.9]
\draw[-stealth, line width=0.53mm] (-3.3,0)--(3.3,0) node[right ]{$\xi$};
\draw[-stealth, line width=0.53mm] (0,-3.3)--(0,3.3) node[above ]{$\zeta$};
\clip (-3,-3) rectangle (3,3);
\begin{scope}[cm={0.5,-0.5,  50,50,  (0,0)}]  
\draw[green!70!black,
dashed] (-6,-6) grid (6,6);
\end{scope}
\draw[white,line width=0.35mm,dashed](-3,3)--(3,-3);
\draw[blue,line width=0.5mm](-3,-1)--(3,-1);
\draw[line width=0.5mm,red](-3,0)--(3,0);
\draw[green!60!black ] (1.45,1.45) node[above] {$\mathcal{N}=0$};
 \node[blue] at (1.25,-1.5) {$\zeta=-1$};
\end{tikzpicture}
  \end{minipage} 
}
\subfigure
{
\centering
  \begin{minipage}{\textwidth}
    \centering
\begin{tikzpicture}[scale=0.9]
\draw[ blue,line width=0.93mm] (-6,-4.5)--(-5.5,-4.5);
\draw (-5.5,-4.5) node[right]{$ \mathcal{N}=1$ U(1)$_R$ Preserving};
\draw[ red,line width=0.93mm] (2,-4.5)--(2.5,-4.5);
\draw  (2.5,-4.5) node[right]{$ \mathcal{N}=0$ SU(2) Preserving};
\end{tikzpicture}
  \end{minipage}
}
\caption{In the general case, for arbitrary $(\xi,\zeta)$ (in green dashed lines), the background breaks all SUSY. Along the $\zeta=-1$ line (in blue), the $U(1)_R$-symmetry is preserved, leading to $\mathcal{N}=1$ solutions. Along the $\zeta=0$ line (in red), the background preserves $SU(2)$ isometry (descending from the original R-symmetry) with the SUSY completely broken in general. Given the red and blue lines do not intersect (as they are now parallel), there are no $\mathcal{N}=2$ solutions here - as no background preserves the necessary $SU(2)_R\times U(1)_R$ R-symmetry. }
    \label{fig:IIAtableplotchi}
\end{figure}
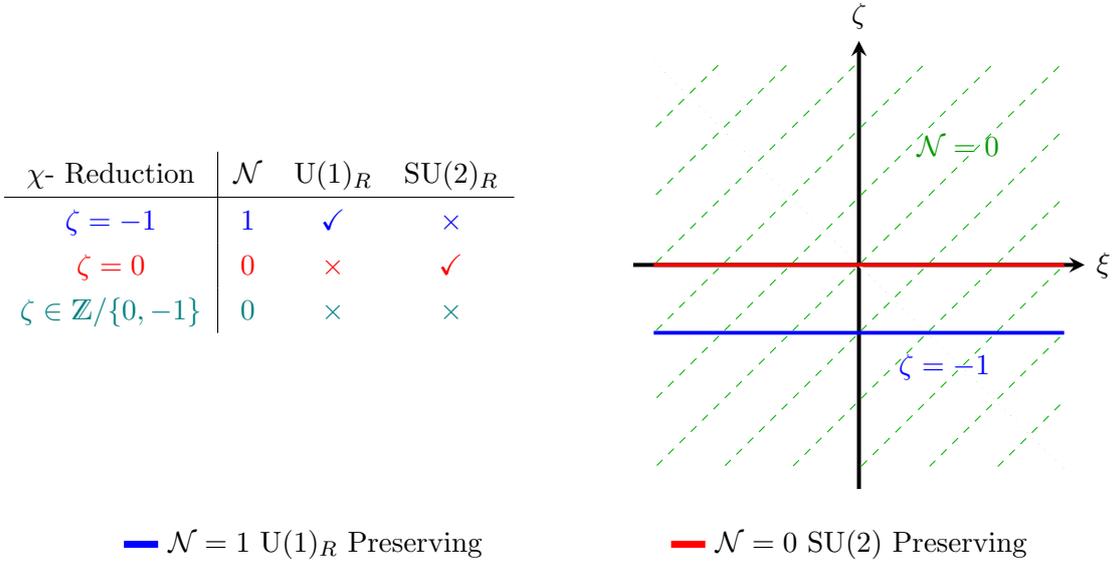

We can actually map this solution to \eqref{eqn:generalresult1}, using the following transformations
    \begin{align}\label{eqn:Transformation}
    1:&~~F_n \rightarrow k_1^{\frac{n}{2}} F_n,~~~~~~~~e^{-\Phi} \rightarrow k_1^{\frac{1}{2}} e^{-\Phi},~~~~~~~~~  g_{MN} \rightarrow k_1 g_{MN},~~~~~~~~~~~~~H_3 \rightarrow k_1 H_3,\nn\\[2mm]
    2:&~~F_n \rightarrow k_2 F_n,~~~~~~~~~e^{-\Phi} \rightarrow k_2 \,e^{-\Phi},
    \end{align}
    with $k_1=\xi^{-1},~k_2=\xi^{2}$, giving
    \beq\label{eqn:mapping1}
        g_{MN} \rightarrow \frac{1}{\xi} g_{MN},~~~~~~~~~~~~~B_2 \rightarrow \frac{1}{\xi} B_2,~~~~~~~~~C_1\rightarrow \xi\, C_1,~~~~~~~~~~~C_3\rightarrow C_3,~~~~~~~~~e^{\frac{4}{3}\Phi} \rightarrow \frac{1}{\xi^2} e^{\frac{4}{3}\Phi},
    \eeq
followed (in order from left to right) by
\begin{align}\label{eqn:mapping1-1}
&C_1\rightarrow C_1 -\,d\beta,~~~~~~~~~~~~C_3\rightarrow C_3- \, d\beta \wedge B_2, ~~~~~~~~~~~   \zeta \rightarrow \xi\, \zeta,\nn\\[2mm]
&\phi\rightarrow \phi-\zeta\,\xi\,\chi,~~~~~~~~~~~~~~\beta \rightarrow  - \xi \,\chi,~~~~~~~~~~~~~~~~~~~~~~~~
\xi\rightarrow\frac{1}{\xi}.\nn
\end{align}
Notice that we require $\xi \rightarrow 1/\xi$, despite the requirement that $\xi\in \mathds{Z}$ (which stems from the preservation of $U(1)$ peridocity in the $d=11$ solution - see the discussion below \eqref{eqn:S2breakingdefns}). This transformation is then a non-trivial one. Mathematically, the two solutions can be mapped to one another (for arbitrary $\xi\in\mathds{R}$); however, given the physical (integer) requirements on $\xi$, this is no longer true. Hence, the two solutions should describe different physics. Of course, fixing $\xi=0$ will derive a unique solution which can not be mathematically mapped to the $\beta$ reduction case.

\item \textbf{Fixing $(q\equiv \xi, v\equiv\gamma)$}\\
The determinant \eqref{eqn:S2breakingdefns} now reduces to $qa=0$. To avoid repeating the previous case, we must fix $a=0$ with $q\equiv \xi$. After re-labelling $(s\equiv \zeta,q\equiv \xi, v\equiv \gamma)$, we derive a solution which maps to the $\xi=0$ solution of \eqref{eqn:chireduction1}, with
\beq
C_1\rightarrow C_1 +\xi d\beta,~~~~~~~~~ C_3\rightarrow C_3+\xi d\beta \wedge B_2,~~~~~~~~~~~\phi\rightarrow \phi-\zeta\xi \beta.
\eeq
\end{itemize}
 We now investigate the case where $s$ is not a free parameter, breaking supersymmetry in all cases.
 \subsubsection{Fixing $s=0$ (with $m$ free)}
 The other possibility is to ensure $s$ is not a free parameter by fixing it to zero. In these cases, $m$ is now the free parameter.
\begin{itemize}
\item  \textbf{$(s,a)=0$ with $(m,q)$ free parameters}\\
Here the determinant reduces to $vc=0$. We now look at each case in turn. These cases can be derived from the $(\xi,\zeta)=0$ solution of \eqref{eqn:chireduction1}, by the following transformations.
\begin{itemize}
\item \textbf{$c=0$ (with $v$ free)}: Re-defining $\phi\rightarrow \phi+ v\beta$ followed by
\beq
C_1\rightarrow C_1+q\, d\beta+m\, d\phi,~~~~~~~~~~~~~C_3\rightarrow C_3+(q \,d\beta+m\, d\phi)\wedge B_2.
\eeq
\item \textbf{$v=0$ (with $c$ free)}: Re-defining $\beta\rightarrow \beta+ c\phi$ followed by 
\beq
C_1\rightarrow C_1+q\, d\beta+m\, d\phi,~~~~~~~~~~~~~C_3\rightarrow C_3+(q \,d\beta+m\, d\phi)\wedge B_2.
\eeq
\end{itemize}
\item  \textbf{$(s,v,q)=0$ with $(m,c,a)$ free parameters}\\
The remaining case can be derived from the $\zeta=0$ solution of \eqref{eqn:chireduction1} by re-defining $\xi\equiv a$ then $\beta\rightarrow \beta+(c-a\,m)\phi$, followed by
\beq
C_1 \rightarrow C_1+m\,d\phi,~~~~~~~~~~~~~C_3 \rightarrow C_3 +m\,d\phi \wedge B_2.
\eeq
\end{itemize} 

\subsection{Investigations at the boundary}

We begin by observing that $(\Xi_2,\Pi_2)$ are non-zero and finite for general values of $(\eta,\sigma,\theta)$. As in the $\beta$ reduction case, the deformed $S^2$ given by $(\theta,\phi)$ has $\Pi_2\rightarrow 1$ at the poles. Hence, given the expression for $M_3$ in \eqref{eqn:chireduction1}, we conclude that the deformed sphere still behaves as an $S^2$ topologically.  

\subsubsection{The $\sigma\rightarrow \infty$ boundary}
   At the $\sigma\rightarrow \infty$ boundary, where we use \eqref{eqn:Vsigmainfty} to leading order, we find $\Pi_2=1$ and
   \beq
   \Xi_2=\frac{\xi^2 P^2 e^{\frac{2\pi \sigma}{P}}}{\pi^3 \mathcal{R}_1^2}~~~(\xi\neq0),~~~~~~~~~~\Xi_2=1~~~(\xi=0),
   \eeq
   hence, we need to investigate this boundary for both cases independently. 
   \begin{itemize}
   \item \underline{$\xi\neq0$}: We begin with the $\xi\neq0$ case, where
   \begin{adjustwidth}{-1.5cm}{}
   \begin{align} 
ds^2&=\frac{\xi\,\kappa  }{X} \bigg[4\sigma\bigg(ds^2(\text{AdS}_5)+\frac{1}{\xi^2}d\beta^2\bigg)+\frac{2P}{\pi}\bigg(d\left(\frac{\pi}{P}\sigma\right)^2+ d\left(\frac{\pi}{P}\eta\right)^2+ \sin^2\left(\frac{\pi}{P}\eta\right)ds^2(\text{S}^2)\bigg)\bigg],\nn\\[2mm]
e^{-\Phi}&=X^{\frac{3}{2}}\frac{{\cal R}_1\pi^2}{2 (\xi\,P)^{\frac{3}{2}}\sqrt{\kappa}}e^{-\frac{\pi}{P}\sigma}\left(\frac{\pi}{P}\sigma\right)^{-\frac{1}{2}},~~~~H_3=-\frac{4\xi \kappa P}{X\pi}\sin\theta \sin^2\left(\frac{\pi}{P}\eta\right) d\left(\frac{\pi}{P}\eta\right)\wedge  d\theta \wedge \Big(d\phi- \frac{\zeta}{\xi }d\beta \Big),\nn
\end{align}
\end{adjustwidth}
noting that $C_1=\frac{X}{\xi}d\beta$ and $C_3-C_1\wedge B_2=0$ (up to a gauge transformation in $B_2$). 
Introducing the new coordinate $ \tilde{r}= e^{-\frac{\pi}{P}\sigma}(\frac{\pi}{P}\sigma)^{-\frac{1}{2}}$, we then find to leading order
\beq 
\hspace{-0.5cm}
ds^2=ds^2(\text{Mink}_6)+\frac{2\xi \kappa  P}{X \pi \,\tilde{r}^2}\bigg(d\tilde{r}^2+\tilde{r}^2ds^2(\text{S}^3)\bigg),~~~H_3=-\frac{4\xi \kappa P}{X\pi}\text{vol}(\text{S}^3),~~~e^{-\Phi}=X^{\frac{3}{2}}\frac{{\cal R}_1\pi^2}{2 (\xi P)^{\frac{3}{2}}\sqrt{\kappa}}\tilde{r},\nn
\eeq
following analogous arguments to those presented below \eqref{eqn:Vsigmainfty}, with $d\hat{\beta} =(\frac{4\xi\,\kappa  }{X} \sigma)^{-\frac{1}{2}}d\tilde{\hat{\beta}}$ to leading order in $\sigma$ (and $\hat{\beta}\equiv \frac{1}{\xi}\beta$). As in the $\beta$ reductions, this clearly describes a stack of NS5 branes, with the results matching the form given in \eqref{eqn:NS5metrics}.

To calculate the NS5 charge, we simply fix $2\kappa=\pi \widehat{\kappa} X $ as before (with $\widehat{\kappa}\in\mathds{Z}$), giving
\beq 
Q_{\text{NS5}}=-\frac{1}{(2\pi)^2}\int_{S^3} H_3= \xi\, \widehat{\kappa}\, P,
\eeq
with a the stack of $\xi \,\widehat{\kappa}P$ NS5 branes at the $\sigma=\infty$ boundary.

  \item \underline{$\xi=0$}: When $\xi=0$, things become a little less clean
  \begin{adjustwidth}{-1.5cm}{}
   \begin{align} 
ds^2&=\frac{ \kappa\,  \pi^{\frac{3}{2}}\mathcal{R}_1 e^{-\frac{\pi \sigma}{P}}}{P X} \bigg[4\sigma\bigg(ds^2(\text{AdS}_5)+\frac{P^2 e^{\frac{2\pi \sigma}{P}}}{\pi^3 \mathcal{R}_1^2 }d\beta^2\bigg)+\frac{2P}{\pi}\bigg(d\left(\frac{\pi}{P}\sigma\right)^2+ d\left(\frac{\pi}{P}\eta\right)^2+ \sin^2\left(\frac{\pi}{P}\eta\right)ds^2(\text{S}^2)\bigg)\bigg],\nn\\[2mm]
e^{-\Phi}&=\frac{X^{\frac{3}{2}}e^{\frac{\pi}{2P}\sigma} }{2\pi^{\frac{1}{4}}(\kappa \mathcal{R}_1)^{\frac{1}{2}}}\left(\frac{\pi}{P}\sigma\right)^{-\frac{1}{2}}, ~~~~~~B_2=
\begin{cases}
\frac{\zeta \kappa  P}{X \pi}\Big(\sin\Big(\frac{2\pi \eta}{P}\Big)-2\Big(\frac{\pi \eta}{P}\Big) \Big) \sin\theta \,d\beta \wedge d\theta &(\zeta\neq0)  \\
-\frac{2^{\frac{3}{2}}\sqrt{\pi}\,\kappa \mathcal{R}_1}{X}e^{-\frac{\pi}{P}\sigma}\left(\frac{\pi}{P}\sigma\right)^{-\frac{1}{2}} \sin^3\left(\frac{\pi}{P}\eta\right) \text{vol}(S^2) &(\zeta=0) 
\end{cases},\nn\\[2mm]
C_1&=\frac{\sqrt{2}\,PX}{\pi^{\frac{3}{2}}\mathcal{R}_1}e^{\frac{\pi}{P}\sigma}\left(\frac{\pi}{P}\sigma\right)^{-\frac{1}{2}} \cos\Big(\frac{\pi \eta}{P}\Big)d\beta ,~~~~~~C_3=\frac{P\kappa}{\pi}\bigg(\sin\Big(\frac{2\pi \eta}{P}\Big)-2\Big(\frac{\pi \eta}{P}\Big) \bigg) d\beta \wedge \text{vol}(S^2),\nn\\[2mm]
\hat{F}_2&=-\frac{\sqrt{2}\,P X}{\pi^{\frac{3}{2}}\mathcal{R}_1} e^{\frac{\pi}{P}\sigma}\left(\frac{\pi}{P}\sigma\right)^{-\frac{1}{2}} \bigg( \sin\left(\frac{\pi}{P}\eta\right) d\left(\frac{\pi}{P}\eta\right) - \cos\left(\frac{\pi}{P}\eta\right) d\left(\frac{\pi}{P}\sigma\right)  \bigg)\wedge d\beta.
\end{align}
  \end{adjustwidth}
When $(\zeta\neq0) $
\beq
\begin{aligned}
 H_3&=\frac{4  \zeta \kappa P}{X\pi}\sin\theta \sin^2\left(\frac{\pi}{P}\eta\right) d\left(\frac{\pi}{P}\eta\right)\wedge  d\theta \wedge  d\beta,\nn\\[2mm]
\hat{F}_4&=d(C_3-B_2\wedge C_1)=- \frac{4 \kappa P}{\pi}  \sin^2\left(\frac{\pi}{P}\eta\right) d\left(\frac{\pi}{P}\eta\right)\wedge d\beta \wedge \text{vol}(S^2),
\end{aligned}
\eeq
however, $(\eta,\theta,\beta)$ no longer appears to be a valid cycle - so we restrict to D4 charge, 
\begin{align}
&
Q_{D4}=-\frac{1}{(2\pi)^3}\int_{S^3\times \beta}\hat{F}_4=\frac{2\kappa}{\pi}P=X\,\widehat{\kappa}\,P.
\end{align}
For ($\zeta=0$), with $B_2\rightarrow B_2+\lambda\, \text{vol}(S^2)$
   \begin{adjustwidth}{-0.5cm}{}
       \vspace{-0.7cm}
 \begin{align}
H_3&=-\frac{2^{\frac{3}{2}}\kappa^2 P \mathcal{R}_1}{\pi^{\frac{1}{2}} X^2}e^{-\frac{\pi \sigma}{P}} \left(\frac{\pi}{P}\sigma\right)^{-\frac{1}{2}} \sin^2\left(\frac{\pi}{P}\eta\right) \bigg(3 \cos\left(\frac{\pi}{P}\eta\right)d\left(\frac{\pi}{P}\eta\right)-\sin\left(\frac{\pi}{P}\eta\right)  d\left(\frac{\pi}{P}\sigma\right)\bigg)\wedge \text{vol}(S^2) ,\nn\\[2mm]
\hat{F}_4&= \begin{cases}
- \frac{4 \kappa P}{\pi}  \sin^2\left(\frac{\pi}{P}\eta\right) d\left(\frac{\pi}{P}\eta\right)\wedge d\beta \wedge \text{vol}(S^2)  &  (\lambda=0)\\
\lambda \frac{\sqrt{2}\,P X}{\pi^{\frac{3}{2}}\mathcal{R}_1} e^{\frac{\pi}{P}\sigma}\left(\frac{\pi}{P}\sigma\right)^{-\frac{1}{2}} \bigg( \sin\left(\frac{\pi}{P}\eta\right) d\left(\frac{\pi}{P}\eta\right) - \cos\left(\frac{\pi}{P}\eta\right) d\left(\frac{\pi}{P}\sigma\right)  \bigg)\wedge d\beta \wedge \text{vol}(S^2)  & (\lambda\neq 0)
\end{cases},\nn
 \end{align}
    \end{adjustwidth} 
 giving
 \begin{align}
&Q_{D4}=-\frac{1}{(2\pi)^3}\int_{S^3\times \beta}\hat{F}_4=\frac{2\kappa}{\pi}P=X\,\widehat{\kappa}\,P,
\end{align}
hence, we find D4 charge in this limit for $\xi=0$ .
    \end{itemize}

 \subsubsection{The $\eta=0$ boundary, with $\sigma\neq 0$}

At $\eta=0$ with $\sigma\neq 0$, using \eqref{eq:fdef} and \eqref{eqn:feq}, one finds using the second form for $M_3$ in \eqref{eqn:chireduction1} that $(\eta,\theta,\phi)$ vanish as $\mathds{R}^3$ in polar coordinates, namely
\beq
f_4 d\eta^2+f_2 \bigg(d\theta^2+ \frac{1}{\Pi_2}\sin^2\theta d\phi^2\bigg)  = \frac{2|\dot{f}|}{\sigma^2\,f}\Big( d\eta^2+\eta^2 ds^2(S^2) \Big),
\eeq
which matches the $\beta$ reduction solutions.
\paragraph{The $\eta=P$ boundary, with $\sigma\neq 0$} This boundary is qualitatively equivalent to the $\eta=0$ boundary.

\subsubsection{The $\sigma=0$ boundary, with $\eta\in (k,k+1)$}
In the case of $\sigma=0,~\eta\in (k,k+1)$, we recall that along the $\sigma=0$ boundary, $\ddot{V}=0$ to leading order. Following the $\beta$ reduction procedure, and using \eqref{eqn:fsat0}, we find (with $\dot{V}=\mathcal{R}$ and $\dot{V}'=\mathcal{R}'=N_{k+1}-N_k$ at this boundary)
\beq
      \Xi_2= \frac{l_k^2+ \frac{1}{2}\zeta^2 \mathcal{R} V'' \sin^2\theta}{(N_{k+1}-N_k)^2},~~~\Delta_2\rightarrow \frac{l_k^2 }{(N_{k+1}-N_k)^2},~~~D\phi =d\phi -\frac{\zeta}{l_k} d\beta,~~~~l_k=\xi+ (N_{k+1}-N_k),
    \eeq
which, assuming $N_{k+1}-N_k\neq 0$ (for $\zeta\neq0$), makes $\Xi_2$ a nowhere vanishing and finite function of $(\eta,\theta)$. 
From the form of $M_3$ in \eqref{eqn:chireduction1}, we find an $\mathds{R}^2/\mathds{Z}_{l_k}$ orbifold singularity in $(\sigma,\beta)$, as follows
   \beq\label{eqn:orbifold1}
      f_4 d\sigma^2+\frac{1}{\Delta_2}\frac{f_3f_5}{(f_5f_6^2+f_3)}d\beta^2  \rightarrow   \frac{2V''}{\mathcal{R}}\bigg( d\sigma^2 +\frac{\sigma^2}{l_k^2}d\beta^2 \bigg). 
     \eeq 

     \subsubsection{The $\sigma=0,~\eta=0$ boundary}
To approach $\sigma=0,~\eta=0$, we make the coordinate change $(\eta=r \cos\alpha,~\sigma=r \sin\alpha)$, expanding about $r=0$. Using \eqref{eqQdef2} and \eqref{eqQdef}, we now find
  \begin{align}
  & \Xi_2\rightarrow \Delta_2 \rightarrow \frac{l_0^2}{N_1^2} ,~~~~~~l_0=\xi+N_1,~~~~~~~~~D\phi=d\phi-\frac{\zeta}{l_0 }d\chi,\\[2mm]
 & f_4(d\sigma^2+d\eta^2) +ds^2(M_3) \rightarrow \frac{2Q}{N_1}\bigg(dr^2+r^2d\alpha^2+r^2\cos^2\alpha \bigg(d\theta^2 +\sin^2\theta \,D\phi^2 \bigg) + r^2\sin^2\alpha \frac{d\beta^2}{l_0^2}\bigg),\nn
  \end{align}
  where we say that the internal space vanishes as $\mathds{R}^5/\mathds{Z}_{l_0}$, with the external space finite. 
   
    \paragraph{The $\sigma=0,~\eta=P$ boundary} This boundary is qualitatively equivalent to the $(\sigma,\eta)=0$ boundary, with $\mathds{R}^5/\mathds{Z}_{l_{P-1}}$.  

\subsubsection{The $\sigma=0$ boundary, with $\eta=k$}
At the $\sigma=0$ boundary with $\eta=k$, we will again follow the $\beta$ reduction approach by making the coordinate change $(\eta=k-r \cos\alpha,~\sigma=r \sin\alpha)$ for $r\sim0$, using \eqref{eq:VsattheD6s} and \eqref{eqn:fsfork}. 
Here we find for the forms of $\Xi_2$ and $\Pi_2$ given in \eqref{eqn:chireduction1} that the $f_2/f_3$ and $f_2/(f_5f_6^2+f_3)$ terms will dominate, unless we either fix $\zeta=0$, or we are at a pole of the deformed $S^2$ (where $\sin\theta=0$). Hence, in a similar manner to the $\beta$ reduction solutions, we first investigate the $S^2$ preserved case, with $\zeta=0$. We then switch on $\zeta$ and investigate this boundary both away from the pole and approaching it.
\begin{itemize}
\item \underline{$\zeta=0$:} We first investigate the $S^2$ preserved case. Postponing the $B_2$ gauge transformation, we find
\begin{align}
&\frac{ds^2}{2\kappa\sqrt{N_k}}=  \frac{1}{X}\sqrt{\Delta_k}\bigg[\frac{1}{\sqrt{\frac{b_k}{r}}}\bigg(4ds^2(\text{AdS}_5)+ ds^2(\text{S}^2)\bigg)+ \frac{\sqrt{\frac{b_k}{r}}}{N_k}\bigg(dr^2+ r^2 \left(d\alpha^2+\frac{\sin^2\alpha}{\Delta_k} d\beta^2\right)\bigg)\bigg],\nn\\[2mm]
&e^{\frac{4}{3}\Phi}=\frac{4\,\kappa^{\frac{2}{3}}\,r}{X^{2}N_k^{\frac{1}{3}}b_k}\Delta_k , 
~~~~~
C_1=  \frac{ X\big(\xi+ g(\alpha)\big)}{\Delta_k}d\beta, ~~~~
C_3=-2\kappa\, k \,d\beta \wedge \text{vol}(\text{S}^2),\nn\\[2mm]
&B_2=-\frac{2\kappa }{X} (\xi\,k+ N_k) \text{vol}(\text{S}^2),
\end{align}
with
\begin{equation*}
\Delta_k=\Big(\xi+g(\alpha)\Big)^2+\frac{1}{4}b_k^2\sin^2\alpha,~~~~~~g(\alpha)=\cos^2\left(\frac{\alpha}{2}\right)(N_k-N_{k-1})+\sin^2\left(\frac{\alpha}{2}\right)(N_{k+1}-N_{k}),
\end{equation*}
where we observe that the metric and dilaton take the identical form to the $\beta$ reduction case \eqref{eq:neq0D6s} (up to the definition of $\Delta_k$ and sending $\beta\leftrightarrow\chi$). The potentials are different however, observing that the roles of $k$ and $N_k$ have swapped in $B_2$ and $C_3$ compared to the $\beta$ reduction case. It is worth reiterating that this is due to the mapping between the two solutions, which mathematically involves sending $\xi\rightarrow 1/\xi$. Because we require $\xi \in \mathds{Z}$ however, we are dealing with a different physical system. 

At the poles of the deformed 2-sphere spanned by $(\alpha,\beta)$, we have
\beq
\Delta_k(\alpha=0)=l^2_{k-1},~~~~\Delta_k(\alpha=\pi)=l^2_{k},~~~~~~~~l_k=
\xi+ (N_{k+1}-N_k)
,\label{eq:usefulfunctionatpoles2}
\eeq
with $\Delta_k$ finite and non zero between these bounds. So we once again find the presence of spindles, but with different conical deficit angles compared to the $\beta$ reduction case - where we now observe that $l_k=\xi+ \,\mathcal{R}'_{[k,k+1]}$. Recall the pictorial representation given next to \eqref{eqn:Spindlefig}
. When $l_k=l_{k-1}=1$, the deformed 2-sphere becomes a round one. In the $\beta$ reduction case, this simply required fixing $\xi=0$ (recovering the $\mathcal{N}=2$ solution). In the present case however, there is no finite value of $\xi$ which will recover a round 2-sphere for all values of $k$ (other than in a Sfetsos-Thompson potential where the slope of the rank function is constant - with no D6 brane sources). Taking the $\xi\rightarrow\infty$ limit, one would find $l_k\sim\xi$. By absorbing $\xi$ into the definition of $\beta$, one could recover a round $S^2$ - however, this would just correspond to the $\xi=0$ case of the $\beta$ reduction (a consequence of $\xi\rightarrow 1/\xi$ in the mapping). Hence, the fact that there is no finite value of $\xi$ which gives a round 2-sphere (for a generic rank function) is then a reflection of the breaking of $\mathcal{N}=2$ supersymmetry for all $\xi$ (and vice versa). 

Given the form of the metric and dilaton, and using \eqref{eqn:Dbranemetrics}, we once again find D6 branes extended in $(\text{AdS}_5,S^2)$ and orthogonal to a cone whose base is $\mathbb{WCP}^1_{[l_{k-1},l_{k}]}$ - see Figure \ref{fig:D6orthogonaltospindle1}. Given that the conical deficit angles of the spindle once again depend on the slope of the rank function at each kink, we again find that each stack of D6 branes (located at $k\in\mathds{Z}$ intervals along $\eta$) are orthogonal to a different spindle. See Figure \ref{fig:D6orthogonaltospindless2} for a pictorial representation.

Calculating the charge of each stack of D6 branes, we find
\begin{align}
Q^k_6&=\frac{1}{2\pi}\int_{\mathbb{WCP}^1_{[l_{k-1},l_{k}]}}F_2=\frac{1}{2\pi}\int_{\beta=0}^{\beta=2\pi} C_1\bigg\lvert_{\alpha=0}^{\alpha=\pi}= \frac{X }{l_{k}l_{k-1}}(2N_k-N_{k+1}-N_{k-1}), \nn\\[2mm]
Q_{D6}&=\sum_{k=1}^{P-1}Q^k_{D6}=\frac{X}{l_0 l_P}(N_{P-1}+N_1),~~~~~~~~l_0=\xi+N_1,~~~~l_P=\xi-N_{P-1},
\end{align}
yielding the rational quantisation condition one should get when integrating over the spindle - following from the rational nature of the Euler characteristic on the spindle, given in \eqref{eqn:Eulerspindle}- with the new values of $l_k$ defined in \eqref{eq:usefulfunctionatpoles2}. Notice that in this case, for a general rank function, fixing $\xi=0$ no longer eliminates the orbifold singularity (as $l_k=N_{k+1}-N_k\neq 1$ in general). Hence the $\mathbb{CP}^1$ (with $\chi_E=2$) is not recovered for $\xi=0$.

We now turn to the D4 branes, noting
\begin{align}
&  B_2^k\rightarrow B_2^k+\frac{2  \kappa  }{X}\lambda \text{vol}(S^2),~~~~~~~~B^{k}_2=\frac{1}{X}\Big( f_7+\xi f_8+2\kappa \lambda\Big) \text{vol}(\text{S}^2), \nn\\[2mm]
&\hat F_4=  d\left(f_8-\frac{ f_5(  f_6+\xi ) \Big(f_7+\xi f_8+ 2\kappa \lambda\Big)}{(f_5f_6^2+f_3)\Delta}\right)\wedge d\beta\wedge\text{vol}(\text{S}^2),
\end{align}
with $\lambda=N_k+\xi\,k$ and integrating carefully, recalling the discussion in Figure \ref{fig:semicircularcontour}, we find
\begin{align}\label{eqn:D4chired}
&Q_{D4}^k=-\frac{1}{(2\pi)^3}\int_{S^2\times \mathbb{WCP}^1_{[l_{k-1},l_k]}}\hat{F}_4= \frac{2\kappa}{\pi}=\widehat{\kappa}X,
\end{align}
which is a constant in each interval.
\item $\zeta \neq 0$: switching on $\zeta$, we find that in general, the $f_2/f_3$ and $f_2/(f_5f_6^2+f_3)$ terms dominates in the $\Xi_2$ and $\Pi_2$ - unless we are at a pole of the deformed $S^2$ (where $\sin\theta=0$). Let us first assume that we are indeed away from one of the poles.
\paragraph{Away from a pole:} in this case, we once again expand $(\eta=k-r \cos\alpha,~\sigma=r \sin\alpha)$ in small $r$, finding to leading order
\begin{equation*}
\Xi_2\rightarrow \frac{\zeta^2 b_k N_k\sin^2\theta}{ 4r \big(g(\alpha)^2+ \frac{1}{4} b_k^2\sin^2\alpha\big)},~~~~~~\Delta_2\rightarrow \frac{\Delta_k}{g(\alpha)^2+ \frac{1}{4} b_k^2\sin^2\alpha},~~~~~~~~\Pi_2\rightarrow \frac{\zeta^2 N_k\sin^2\theta}{r\,b_k\,\sin^2\alpha}.
\end{equation*}
\newpage
Using $r=z^2$, we find 
\begin{align}\label{eqn:B3eqchi}
& ds^2= \frac{|\zeta| \kappa }{X} \sin\theta \Big[N_k\Big(4ds^2(AdS_5) +d\theta^2\Big) +4b_k\Big(dz^2 +z^2 ds^2(\mathbb{B}_3)\Big)\Big],~~~~~~e^{4\Phi}= \frac{\kappa^2 \zeta^6}{X^6} N_k^2\sin^6\theta, \nn\\[2mm]
&ds^2(\mathbb{B}_3)= \frac{1}{4}\Big(d\alpha^2 + \frac{\sin^2\alpha}{\Delta_k}d\beta^2\Big)+\frac{\Delta_k}{\zeta^2 b_k^2}(d\phi+\mathcal{A}_k)^2,~~~~~~~~~\mathcal{A}_k=-\frac{\zeta}{\Delta_k}\Big(\xi+ g(\alpha)\Big) d\beta, \nn\\[2mm]
&B_2= -\frac{2\kappa}{X} \,\sin\theta \Big(\zeta \,k\, d\beta - (N_k+\xi\,k)  d\phi\Big)\wedge d\theta,~~~~C_1=\frac{ X}{\zeta}d\phi,~~~~C_3=-2\kappa \,k\,\sin\theta  d\theta\wedge d\phi\wedge d\beta.
\end{align}
 We see once again the form of the metric and dilaton remain the same as the $\beta$ reduction - up to a change in $\mathcal{A}_k$, $\Delta_k$ and switching $\beta$ for $\chi$. Hence, we again find that the $(z,\mathbb{B}_3)$ sub-manifold describes a cone of base $\mathbb{B}_3$, and the rest of the space has a constant warping (ignoring the overall $\sin\theta$). In addition, the sub-manifold $\mathbb{B}_3$ is itself clearly a $U(1)$ fibration over $\mathbb{WCP}^1_{[l_{k-1},l_k]}$, with
\begin{align}
&ds^2(\mathbb{B}_3)\bigg|_{\alpha\sim 0} =  \frac{1}{4}\Big(d\alpha^2 + \frac{\alpha^2}{l_{k-1}^2}d\beta^2\Big)+\frac{l_{k-1}^2}{\zeta^2 b_k^2}\bigg(d\phi-\frac{\zeta}{l_{k-1}}  d\beta\bigg)^2,\nn\\[2mm]
&ds^2(\mathbb{B}_3)\bigg|_{\alpha\sim \pi} =  \frac{1}{4}\Big(d\alpha^2 + \frac{(\pi-\alpha)^2}{l_{k}^2}d\beta^2\Big)+\frac{l_{k}^2}{\zeta^2 b_k^2}\bigg(d\phi-\frac{\zeta}{l_{k}}  d\beta\bigg)^2,\nn\\[2mm]
&-\frac{1}{2\pi}\int_{\mathbb{WCP}_{[l_{k-1},l_k]}}d\mathcal{A}=\frac{\zeta b_k}{l_{k-1}l_k},~~~~~~~~~~~~~~~~~b_k=2N_k-N_{k+1}-N_{k-1},
\end{align}
which is consistent with this claim (see for example \cite{Ferrero:2021etw}). 
We note for $\zeta\neq0$, at this boundary, the limit of $(C_3-C_1\wedge B_2)\sim z^2\rightarrow0$, so we say there are no D4 branes here - given that $\hat{F}_4=d(C_3-C_1\wedge B_2)$.

\paragraph{Approaching the pole:} we now study the behaviour when approaching $(\sigma=0,$ $\eta=k,\sin\theta=0)$, making use of the coordinate change
\beq
\eta=k-\bar{r} \cos\alpha \sin^2\mu,~~~~~\sigma=\bar{r} \sin\alpha \sin^2\mu,~~~~~~\sin\theta = 2\sqrt{\frac{b_k\bar{r}}{N_k}}\cos\mu,
\eeq
with $r= \bar{r} \sin^2\mu$, and expanding about $\bar{r}=0$. We now find
\beq
(g(\alpha)^2+ \frac{1}{4} b_k^2\sin^2\alpha)\sin^2\mu\, \Xi \rightarrow \tilde{\Xi}_k= \sin^2\mu \,\Delta_k+\zeta^2 b_k^2 \cos^2\mu,~~~
\eeq
leading to
 \begin{align}\label{eqn:B4eqmetricchi}
 & \frac{ds^2}{ 2\kappa\sqrt{N_k} } =\frac{1}{X}\sqrt{\tilde{\Xi}_k} \Bigg[\frac{4}{\sqrt{\frac{b_k}{\bar{r}}}}ds^2(AdS_5)+ \frac{\sqrt{\frac{b_k}{\bar{r}}}}{N_k}\bigg(d\bar{r}^2+4\bar{r}^2 ds^2(\mathbb{B}_4)\bigg)\Bigg],~~~~~~~~e^{-\Phi}=\bigg(\frac{X^6b_k^3N_k}{2^6\kappa^2\tilde{\Xi}_k^3\bar{r}^3}\bigg)^{\frac{1}{4}},\nn\\[2mm]
& ds^2(\mathbb{B}_4) = d\mu^2 +\frac{1}{4}\sin^2\mu \bigg(d\alpha^2 + \frac{\sin^2\alpha}{\Delta_k}d\beta^2\bigg) +\frac{\sin^2\mu \cos^2\mu\,\Delta_k}{\tilde{\Xi}_k}(d\phi+\mathcal{A}_k)^2,\nn\\[2mm]
& B_2=\frac{4\kappa  b_k }{X N_k}\cos^2\mu \Big((N_k+\xi \,k) d\phi -\zeta \,k\,d\beta \Big)\wedge d\bar{r},~~~~C_3=-\frac{4\kappa \,k\, b_k}{N_k} \cos^2\mu \,d\bar{r}\wedge d\phi\wedge d\beta,\nn\\[2mm]
 &C_1=\frac{ X}{\tilde{\Xi}_k}\bigg[\zeta \,b_k^2  \cos^2\mu\, d\phi +  \Big(\xi+ g(\alpha)\Big)  \sin^2\mu\,d\beta\bigg]. 
 \end{align}
Calculating the Euler Characteristic of the the four-manifold, $\mathbb{B}_4$, using the Chern-Gauss-Bonnet theorem \eqref{eqn:Chern-Gauss} leads to
\beq
\chi_E
=3-\bigg(1-\frac{1}{|l_{k-1}|}\bigg)-\bigg(1- \frac{1}{|\zeta(l_{k-1}-l_k)|}\bigg)-\bigg(1-\frac{1}{|l_{k}|}\bigg),
\eeq 
with $|\zeta(l_{k-1}-l_k)|=|\zeta b_k|\in \mathds{Z}$. Hence, $\mathbb{B}_4$ is the weighted projective space \\$\mathbb{WCP}^2_{[l_{k-1},l_k,\zeta(l_{k-1}-l_k)]}$. This is similar to the $\beta$ reduction case \eqref{eqn:WCP2beta}, but varies in the forms of $1/|l_{k-1}|$ and $1/|l_{k}|$ (which no longer equal one for $\xi=0$). Hence, fixing $\xi=0$ does not simplify things much, and the above form of $\chi_E$ remains (with $l_k= N_{k+1}-N_k$). As in the $\beta$ reduction case, we again find neighbouring $\mathbb{WCP}^2$ manifolds, one at each $\eta=k$, and when $\sin\mu\rightarrow 0$, we find $\mathbb{B}_4$ approaches a cone over the $\mathbb{B}_3$ given in \eqref{eqn:B3eqchi}. 

We now calculate the D6 charge at $\mu=\frac{\pi}{2}$, noticing that all $\zeta$ dependence drops out of the calculation,
\beq
Q_{D6}^k=-\frac{1}{2\pi}\int_{\mathbb{WCP}^1_{[l_{k-1},l_k]}}F_2 =-\frac{1}{2\pi}\int_{\beta=0}^{\beta=2\pi}C_1\Big|_{\alpha=0}^{\alpha=\pi} =\frac{X}{l_kl_{k-1}}(2N_k-N_{k+1}-N_{k-1}),
\eeq
with the rational charge a consequence of the spindle. This charge takes the same form as the $\beta$ reduction result (up to the definitions of $l_k$). The boundary analysis then agrees with the discussion around \eqref{eqn:C1sourcechi}, where we have nice D6 sources for $\zeta=0$ or $\sin\theta=0$.
\end{itemize}

\subsubsection{Summary}
For $\xi\neq 0$, we have a stack of colour NS5 branes at the $\sigma\rightarrow \infty$ boundary, analogous to the $\beta$ reduction case. When $\xi=0$, things now differ, as we instead find D4 charge at this limit
. Along the $\sigma=0$ boundary, we still find stacks of D6 branes located at each kink of the rank function. In the $\zeta=0$ case (along the red line in Figure \ref{fig:IIAtableplotchi}), with a preserved $S^2$, the D6 branes are extended in $(AdS_5,S^2)$ and orthogonal to a different spindle at each kink, with conical deficit angles defined by that kink - giving rise to a different rational charge for each D6 stack. The conical deficit angles now take a different form compared to the $\beta$ reduction case. Again, this is the only solution with D4 (colour) branes. 
 In the $\zeta\neq0$ case, the spindle at each kink is now replaced by its higher dimensional analogue, giving rise to the same rational quantization of charge. Due to the new form of $l_k$, when $\xi=0$, integer quantization is no longer recovered. This then differs from the $\beta$ reduction case. The D6 branes are the only physical objects in the background. 
Hence, in summary
 \begin{equation}\label{eqn:ChargesIIAchi}
 \begin{aligned}
&Q_{D6}^k =\frac{X}{l_kl_{k-1}}b_k,~~~~~~Q_{D6}=\sum_{k=1}^{P-1}Q^k_{D6}=\frac{X}{l_0 l_P}(N_{P-1}+N_1),~~~~~~~Q_{NS5}=  \begin{cases}
     \xi\,\widehat{\kappa} P& (\xi\neq0)\\
    ~~0  & (\xi=
    0)
    \end{cases} ,\nn\\[2mm]
    &Q_{D4}^{\sigma\rightarrow\infty}=  \begin{cases}
     0& (\xi\neq0)\\
       X \,\widehat{\kappa} P& (\xi=0 ) 
    \end{cases},~~~~~ Q_{D4}^{\sigma=0,\eta=k}=  \begin{cases}
     ~0 & (\zeta\neq0)\\
    \widehat{\kappa} X  & (\zeta=0)
    \end{cases},  ~~~~~Q_{D4}^{\sigma=0} = (P-1) Q_{D4}^{\sigma=0,\eta=k},\\[2mm]
    &l_k=\xi+ (N_{k+1}-N_k),~~~~~~l_0=\xi+ N_1,~~~~~~l_P=\xi- N_{P-1},~~~~~b_k=(2N_k-N_{k+1}-N_{k-1}).
    \end{aligned} 
\end{equation}
Calculating the holographic central charge, and using $2\kappa=\pi \widehat{\kappa} X $, one finds
\begin{equation*}
c_{hol} 
= \frac{ \widehat{\kappa}^3X^2}{8\pi}\sum_{n=1}^\infty P \, \mathcal{R}_n^2.
\end{equation*}

\paragraph{$\mathcal{N}=0$ deformations:} One could now repeat the analysis of Section \ref{sec:N=0}, however this is left to future study. We instead move straight on to the $\mathcal{N}=1$ solutions.

\subsection{U$(1)\times$U$(1)$ preserving $\mathcal{N}=1$ deformations}\label{sec:chiN1}
Let us look closer at the one parameter family of $\mathcal{N}=1$ solutions, derived following a $\chi$ reduction with $\zeta=-1$. The solution reads
  \begin{align}\label{eqn:chireductionNeq1}
  &      ds_{10,st}^2=  \frac{1}{X}f_1^{\frac{3}{2}}\big(f_5f_6^2+ f_3  \big)^{\frac{1}{2}}\sqrt{\Xi_2}\bigg[4ds^2(\text{AdS}_5)+f_4(d\sigma^2+d\eta^2)+     ds^2(M_3)  \bigg],\nn\\[2mm]
    &   B_2 = - \frac{1}{X}\sin\theta \bigg[ f_8 d\beta +\Big( f_7+ \xi f_8\Big)d\phi\bigg] \wedge \, d\theta,~~~~~~~~  e^{\frac{4}{3}\Phi}= \frac{1}{X^2} f_1\big(f_5f_6^2+ f_3  \big)\Xi_2 ,\nn\\[2mm]
   &   C_1=  \frac{X}{ \big(f_5f_6^2+ f_3  \big)\Xi_2} \bigg(f_5 (f_6+\xi) d\beta  - f_2\sin^2\theta \, d\phi\bigg)  ,~~~~~~~~   C_3=  f_8    \,d\beta  \wedge \text{vol}(S^2),\nn\\[2mm]
    &\Xi_2= \Delta_2+  \frac{f_2}{\big(f_5f_6^2+ f_3  \big)} \sin^2\theta,~~~~~~~~\Delta_2=1+\xi \frac{f_5(2f_6+\xi)}{\big(f_5f_6^2+ f_3 \big)},~~~~~~~~~~\Pi_2=1+  \frac{f_2}{f_3}\sin^2\theta,\nn\\[2mm]
    &D\phi =d\phi +\frac{1}{\Delta_2} \frac{f_5(f_6+\xi)}{(f_5f_6^2+f_3)}d\beta ,~~~~~~~D\beta=d\beta +\frac{1}{\Pi_2}\frac{f_2}{f_3}(f_6+\xi)\sin^2\theta\,d\phi,
    \end{align}
with the form of $ds^2(M_3)$ given in \eqref{eqn:chireduction1}. One can map this solution to the $\mathcal{N}=1$ background given in \eqref{eq:neq1sol} via the transformations outlined in \eqref{eqn:mapping1} followed by
\begin{equation*}
\begin{aligned}
&C_1\rightarrow C_1 -\,d\beta,~~~~~~~~~C_3\rightarrow C_3- \, d\beta \wedge B_2,\\
&\phi\rightarrow \phi +\,\chi,~~~~~~~~~~~~~~\beta \rightarrow  - \xi \,\chi,~~~~~~~~~~~~~~~~\xi\rightarrow\frac{1}{\xi}.
\end{aligned}
\end{equation*}
Hence, we will focus on the G-structure description for the (unique) zero parameter solution, derived by fixing $\xi=0$ (and more specifically $(p,b,u)=1,(m,q,c)=0,v\equiv \gamma=0,a\equiv \xi=0,s\equiv\zeta=-1$). 
   \subsubsection{G-structure description}
   We now derive the $\zeta=-1,\xi=0$ G-structure description for the zero parameter $\mathcal{N}=1$ solution. We begin with the appropriate $d=11$ vielbeins, which read
   \begin{adjustwidth}{-1.1cm}{}
   \vspace{-0.7cm}
    \begin{align}\label{eqn:chiredd11viels}
 K&= -\frac{\kappa e^{-2\rho}}{f_1}d(e^{2\rho}\dot{V}\cos\theta)\nn\\[2mm]
 E_1&=\frac{2\kappa^2 f_1^{-\frac{5}{2}}f_5^{-1}}{ \sqrt{\Sigma}\sqrt{f_3+f_2\sin^2\theta}}\bigg[\dot{V}f_2\sin^2\theta d\eta +2\sigma^2\Big(1+\frac{1}{4}f_2\sin^2\theta\Big)d(V')\nn\\[2mm]
 &~~~~-\frac{\dot{V}}{4}f_5f_6 \Big(e^{-6\rho}d(e^{6\rho}\dot{V}\sin^2\theta)+\sin^2\theta d(\dot{V})\Big) 
 -\frac{i}{2\kappa^2}f_1^3f_5(f_3+f_2\sin^2\theta)\Big(d\beta +\frac{f_2f_6\sin^2\theta}{f_3+f_2\sin^2\theta}d\phi\Big)\bigg],\nn\\[2mm]
 E_2&= e^{i\phi}\Bigg[  \frac{\kappa}{f_1\sqrt{\frac{1}{4}\big(1+\frac{4}{ \sigma^2f_4}\big)f_2 \sin^2\theta +1}}\Big(e^{-3\rho}d(e^{3\rho}\dot{V}\sin\theta) +\frac{\dot{V}}{\sigma}\sin\theta \,d\sigma\Big)   + i\,\sqrt{\frac{f_1f_2f_3}{f_3+f_2\sin^2\theta}}\sin\theta d\phi \Bigg],\nn\\[2mm]
 E_3&=   - \frac{1}{X}  \bigg[   \frac{X\sqrt{8f_1} (f_5\Sigma)^{-\frac{1}{4}}}{\big(1+\frac{f_2}{4} (\sin^2\theta-4)\big)^{\frac{1}{4}}}\Big[d\rho + \frac{1}{8}f_2 \Big(d(\sin^2\theta) -\frac{2}{\dot{V}}e^{-2\rho}d\big(e^{2\rho}\dot{V}(\sin^2\theta+2)\big)\Big)\Big]\nn\\[2mm]
 &~~~~~~~~~+i\sqrt{f_1f_5\Sigma}\,\Big(d(X\chi)+C_1\Big)\bigg],\nn\\
 &\Sigma=f_6^2 +\frac{f_3}{f_5}+\frac{f_2}{f_5}\sin^2\theta,~~~~~~~~~~~~~~~~  C_1=\frac{ X}{ \Sigma}\bigg( f_6 d\beta  - \frac{f_2}{f_5}  \sin^2\theta d\phi\bigg).
 \end{align}
 \end{adjustwidth} 
 As in the $\beta$ reduction case, these results still describe the $\mathcal{N}=2$ solution in $d=11$, because at this stage the supersymmetry has not been broken by the reduction to type IIA.
 
 The G-structure forms for the SU(2) structure in IIA is then derived from the usual formula \eqref{eqn:11Dto10Dforms}, and read
    \begin{adjustwidth}{-1.25cm}{}
   \vspace{-0.7cm}
     \begin{align}\label{eqn:UnqIIAGs}
 u^\mathcal{A}&=\frac{1}{\sqrt{X}}\frac{2\sqrt{2} f_1^{\frac{3}{4}}}{\Big(1+\frac{f_2}{4} (\sin^2\theta-4)\Big)^{\frac{1}{4}}}\bigg[d\rho + \frac{1}{8}f_2 \bigg(d(\sin^2\theta) -\frac{2}{\dot{V}}e^{-2\rho}d\Big(e^{2\rho}\dot{V}(\sin^2\theta+2)\Big)\bigg)\bigg],\nn\\[2mm]
 v^\mathcal{A}&=-\frac{\kappa}{\sqrt{X}}\sqrt{2}f_1^{-\frac{3}{4}}\Big(1+\frac{f_2}{4} (\sin^2\theta-4)\Big)^{\frac{1}{4}} e^{-2\rho}d(e^{2\rho}\dot{V}\cos\theta),~~~~~~z^\mathcal{A}=u^\mathcal{A}+i\,v^\mathcal{A},\nn\\[2mm]
 j^\mathcal{A}&= \frac{\kappa f_2^{\frac{1}{2}}}{2X\Big(1+\frac{f_2}{4} (\sin^2\theta-4)\Big)^{\frac{1}{2}}}\Bigg[\bigg( (1-f_2) \Big(d(2\dot{V}\sin^2\theta)+e^{-6\rho}  \sin^2\theta d(2e^{6\rho} \dot{V})\Big) -2f_2 \sin^2\theta d(\dot{V})\bigg)\wedge d\phi\nn\\[2mm]
 &~~~~~~~~~~~~ + \bigg[f_5f_6\Big(e^{-3\rho}d(\sin^2\theta e^{3\rho}\dot{V}) -\frac{1}{2}d(\sin^2\theta \dot{V}) +\sin^2\theta d(\dot{V})\Big)- \frac{\Lambda f_3}{\dot{V}} d(V')-2\sin^2\theta d\eta \bigg]\wedge d\beta \Bigg],\nn\\[2mm]
 \omega^\mathcal{A}&=-\frac{ 2\kappa}{X}   \frac{\sqrt{f_2}}{\dot{V}} \bigg( \sigma \dot{V} e^{-3\rho}  d(V')\wedge d(e^{3\rho}\sin\theta e^{i\phi})+   \tilde{\Delta} \Big(1-\frac{3 }{2}f_2\Big) e^{i\phi} \sin\theta d\sigma\wedge d\eta  \nn\\
 &~~~~~~~~~~~~~~~+i e^{-3\rho} \, d(\sigma \dot{V}e^{3\rho}e^{i\phi}\sin\theta )\wedge d\beta \bigg),
\end{align}
   \end{adjustwidth} 
where we can verify the preservation of $\mathcal{N}=1$ supersymmetry by building the pure spinors \eqref{eqn:Psi} and testing the conditions \eqref{eqn:IIAGconditions}. In addition, one can derive $E^1_{IIA} =e^{\frac{1}{3}\Phi}E^1$ and $E^2_{IIA} =-
e^{\frac{1}{3}\Phi}E^2$ from the $d=11$ vielbeins \eqref{eqn:chiredd11viels}.

\paragraph{A comment on Supersymmetry breaking:} We leave the G-structure analysis of the full solution to future work, however repeating the procedures outlined in the $\beta$ reduction case, we expect analogous equations to \eqref{eqn:SUSYbrokenIIA} - with the $(\zeta+\xi)$ terms replaced by $(\zeta+1)$ analogues. 

\section{$\phi$ Reduction}\label{sec:phiRed}
We finally consider a dimensional reduction along $\phi$. The full solution, with all $GL(3,\mathds{R})$ transformation parameters intact, is given in \eqref{eqn:phi-Gen}.
\subsection{Two-Parameter Family}
Now that we are performing a dimensional reduction along $\phi$, in order to preserve the $U(1)_R$ component \eqref{eqn:U(1)lessgen} under reduction, we must fix $m=-1$. As in the $\chi$ reduction case, the maximum supersymmetry which can be achieved under this reduction is $\mathcal{N}=1$ - this is a consequence of the $U(1)_R$ component being $\chi+\phi$ prior to the $SL(3,\mathds{R})$ transformation, given in \eqref{eqn:origU(1)}.

We will first investigate keeping the parameter $m$ free, allowing us to turn the $\mathcal{N}=1$ supersymmetry on and off at will.
\subsubsection{Fixing $m\equiv \zeta$}
With $m\equiv \zeta$ and $(p,b,u)=1$, from the determinant given in \eqref{eqn:S2breakingdefns}, we must fix $s=0$ - with either $a=0$ or $v=0$. We will investigate both options in turn.

\begin{itemize}
\item \textbf{Fixing $(c\equiv \xi, q\equiv \gamma)$}\\
With $(s,a)=0$, the determinant in \eqref{eqn:S2breakingdefns} reduces to $vc=0$, meaning we can either fix $v=0$ or $c=0$. We first look at the $v=0$ case, with the following 11D transformations 
    \beq
\begin{gathered}
d\beta = d\beta +\xi\, d\phi,~~~~~~~~~~~~~~~~~d\chi =  d\chi +\gamma\, d\beta + \zeta \,d\phi,~~~~~~~~~~~~~~~~~~d\phi =d\phi ,\\
U(1)_R = \chi +\gamma \beta + (\zeta+1)\phi,
\end{gathered}
\eeq
where we would need to fix $\gamma=0$ in order to T-Dualise in a SUSY preserving manner (along $\beta$). Following the dimensional reduction to type IIA, this $\gamma$ only plays a trivial role, and derives the following background with $\chi\rightarrow \chi+\gamma\,\beta$. We are then free to fix $\gamma=0$ without loss of generality, deriving the following two-parameter solution
   \begin{adjustwidth}{-0.75cm}{}
   \vspace{-0.7cm}
      \begin{align}\label{eqn:phieqmatch}
    &    ds_{10,st}^2= \frac{1}{X}f_1^{\frac{3}{2}}f_2^{\frac{1}{2}}\sin\theta \,\sqrt{\Xi_3}\bigg[4ds^2(\text{AdS}_5)+f_4(d\sigma^2+d\eta^2)+ds^2(M_3) \bigg],\nn\\[2mm]
        &ds^2(M_3)=f_2d\theta^2+ \frac{\Delta_3}{\Xi_3}f_5\,D\beta^2+\frac{f_3}{\Delta_3} d\chi^2 =f_2d\theta^2+  \frac{\widehat{\Delta}_3}{\Xi_3}(f_5f_6^2+f_3)D\chi^2+\frac{1}{\widehat{\Delta}_3}\frac{f_3f_5}{(f_5f_6^2+f_3)} d\beta^2,  \nn\\[2mm]
        &  B_2= \frac{1}{X} \sin\theta\Big(   f_7 d\chi+  f_8 d\beta\Big)\wedge d\theta, ~~~~~~~~~~~~     e^{\frac{4}{3}\Phi}=  \frac{1}{X^2} f_1  f_2 \sin^2\theta \,\Xi_3,\nn\\[2mm]
    &   C_1=  \frac{X}{  f_2 \sin^2\theta \,\Xi_3} \Big[  \Big(\xi  f_5f_6+\zeta ( f_5f_6^2+  f_3 ) \Big)  d\chi+ f_5 (\xi+\zeta f_6) d\beta\Big],   ~~~~~~~~~~~   C_3 =0, \nn\\[2mm]
    & \Xi_3= 1+\frac{f_5(\xi  +\zeta f_6)^2 + \zeta^2 f_3}{ f_2 \sin^2\theta}  \equiv 1+\frac{(\Delta_3+ \widehat{\Delta}_3)}{f_3}\bigg[f_5f_6^2 \Big[1-\frac{2}{\Delta_3+ \widehat{\Delta}_3}\Pi_3\Big]+f_3\Big[1-\frac{2 }{\Delta_3+ \widehat{\Delta}_3}\Big]  \bigg] ,\nn\\[2mm]
    &\Delta_3=1+\zeta^2 \frac{f_3}{f_2\,\sin^2\theta},~~~~~~\widehat{\Delta}_3= 1+\xi^2 \frac{f_3f_5}{(f_5f_6^2+f_3)f_2\sin^2\theta},~~~~~~\Pi_3 = 1-\xi\zeta \frac{f_3}{f_6f_2\sin^2\theta},\nn\\[2mm]
  &  D\beta = d\beta + \frac{\Pi_3}{\Delta_3}f_6\, d\chi,~~~~~~~~D\chi = d\chi+ \frac{\Pi_3}{\widehat{\Delta}_3}\frac{f_5f_6}{(f_5f_6^2+f_3)}d\beta,
    \end{align}
       \end{adjustwidth}
        with
    \beq\label{eqn:fluxesphi}
    H=dB_2,~~~~~~F_2=dC_1,~~~~~~F_4=dC_3-H\wedge C_1,~~~~~~dH=0,~~~~~d_HF_4=0,
    \eeq
   where we have written the solution in an analogous manner to the $\beta$ and $\chi$ reduction cases - in terms of functions which reduce to one when $(\xi,\zeta)=0$. Given that we are reducing along $\phi$, we clearly no longer have a preserved $S^2$ in any case. As in the $\chi$ reduction, the only parameter which determines the supersymmetry of this $\mathcal{N}=0$ background is $\zeta$, enhancing the solution to $\mathcal{N}=1$ when $\zeta=-1$. See Figure \ref{fig:IIAtableplotphi} for a summary.
  
     Following previous analysis, we consider approaching the $\sigma= 0$ boundary where $\ddot{V}\rightarrow 0$ to leading order. With the boundary condition $\mathcal{R}(\eta) = \dot{V}\Big|_{\sigma=0}$ and warp factors \eqref{eqn:fs}, we find
\beq\label{eqn:C1sourcephi}
C_1\Big|_{\sigma\rightarrow 0} =X \frac{  (\xi+ \zeta \,\mathcal{R}') (d\beta+\mathcal{R}'  d\chi )}{(\xi+ \zeta\,\mathcal{R}')^2+\frac{1}{2} V'' \mathcal{R} \sin^2\theta },~~\Rightarrow~~~~~~C_1\Big|_{\sigma\rightarrow 0}^{\sin\theta=0}=X \frac{ d\beta+\mathcal{R}'  d\chi  }{ \xi+ \zeta\,\mathcal{R}'  }, 
\eeq
which reduces for $\sin\theta=0$ (which is no longer a pole of the two-sphere given that the $S^2$ has been broken under the reduction). Notice also, in contrast to the $\beta$ and $\chi$ reduction cases, fixing $\zeta=0$ no longer reduces the $C_1$. In addition, the $C_1$ now includes both $d\beta$ and $d\chi$ contributions, but still depends on the discontinuous $\mathcal{R}'(\eta)$. 
     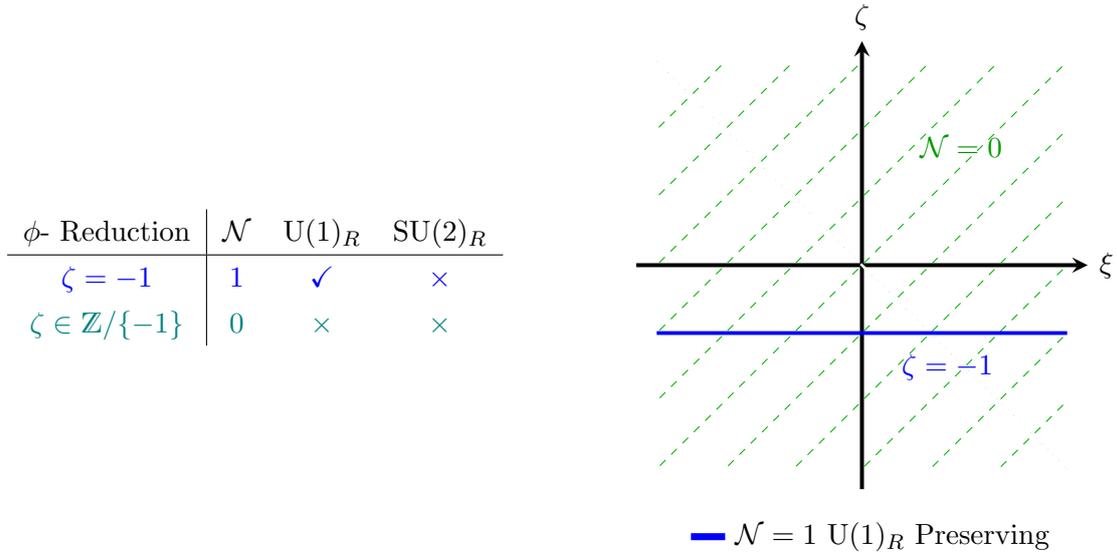
\begin{figure}[H]
\centering  
\subfigure
{
\centering
  \begin{minipage}{0.5\textwidth}
\begin{tabular}{c | c c c c  }
$\phi$- Reduction&$\mathcal{N}$&U(1)$_R$&SU(2)$_R$  \\
\hline
$\textcolor{blue}{\zeta=-1}$&$ \textcolor{blue}{1}$ &$\textcolor{blue}{\checkmark}$&$\textcolor{blue}{\times}$  \\
$\textcolor{teal}{\zeta \in \mathds{Z}/\{-1\}}$&$ \textcolor{teal}{0}$ &$\textcolor{teal}{\times}$ &$\textcolor{teal}{\times}$
\end{tabular}
  \end{minipage}
     \begin{minipage}{.5\textwidth}
    \centering
\begin{tikzpicture}[scale=0.9]
\draw[-stealth, line width=0.53mm] (-3.3,0)--(3.3,0) node[right ]{$\xi$};
\draw[-stealth, line width=0.53mm] (0,-3.3)--(0,3.3) node[above ]{$\zeta$};
 \draw[ blue,line width=0.93mm] (-2.5,-4)--(-2,-4);
\draw (-2,-4) node[right]{$ \mathcal{N}=1$ U(1)$_R$ Preserving};
\clip (-3,-3) rectangle (3,3);
\begin{scope}[cm={0.5,-0.5,  50,50,  (0,0)}]  
\draw[green!70!black,
dashed] (-6,-6) grid (6,6);
\end{scope}
\draw[white,line width=0.35mm,dashed](-3,3)--(3,-3);
\draw[blue,line width=0.5mm](-3,-1)--(3,-1);
\draw[green!60!black ] (1.45,1.45) node[above] {$\mathcal{N}=0$};
 \node[blue] at (1.25,-1.5) {$\zeta=-1$};

\end{tikzpicture}
  \end{minipage} 
}
\caption{In the general case, for arbitrary $(\xi,\zeta)$ (in green dashed lines), the background breaks all SUSY. Along the $\zeta=-1$ line (in blue), the $U(1)_R$-symmetry is preserved, leading to $\mathcal{N}=1$ solutions. No backgrounds preserve an $SU(2)$ isometry, hence there are no $\mathcal{N}=2$ solutions here - as no backgrounds preserve $SU(2)_R\times U(1)_R$ R-symmetry.}
    \label{fig:IIAtableplotphi}
\end{figure}
Hence, following a reduction along $\phi$, we now find two source terms for D6 branes in the Bianchi identity (picking up an additional contribution compared to the $\beta$ and $\chi$ reduction cases)!  Taking the derivative carefully (assuming $(\xi,\zeta)\neq0$), leads to
\begin{align}
&F_2\Big|_{\sigma\rightarrow 0,\eta=k}^{ \sin\theta=0}=-X  \bigg(\frac{ d\beta+\mathcal{R}'(k)  d\chi  }{ \xi+ \zeta\,\mathcal{R}'(k)  }  -\frac{ d\beta+\mathcal{R}'(k-1)  d\chi  }{ \xi+ \zeta\,\mathcal{R}'(k-1)  } \bigg) \wedge  d\eta  \nn\\[2mm]
&~~~~~~~~~~~~~~~= \frac{ X\big(\mathcal{R}' (k)-\mathcal{R}' (k-1)\big)}{\big(\xi+ \zeta\,\mathcal{R}'(k)  \big)\big(\xi+ \zeta\,\mathcal{R}'(k-1)  \big)} \Big(\zeta\,d\beta-\xi\,d\chi\Big)  \wedge  \,d\eta  \\[2mm]
&\Rightarrow ~ F_2\Big|_{\sigma\rightarrow 0 }^{ \sin\theta=0}= X\sum_{k=1}^{P-1} \frac{2N_k-N_{k+1}-N_{k-1}}{\big(\xi+\zeta (N_{k+1}-N_k)\big)\big(\xi+ \zeta (N_{k}-N_{k-1})\big)} \delta(\eta-k)\delta(\sigma)\,d\eta\,\wedge \Big(\zeta\,d\beta-\xi\,d\chi\Big),\nn
\end{align}
where again, the denominator has changed compared to the $\beta$ and $\chi$ reduction cases. The forms of $B_2$ and $C_3$ are still independent of $f_6$, leading to source free Bianchi identities for $H$ and $F_4$ - given in \eqref{eqn:fluxesphi}. We will return to this discussion more thoroughly in the next sub-section, where we investigate the boundary.

To map to \eqref{eqn:generalresult1}, one requires the transformations of \eqref{eqn:Transformation}
    with $k_1=\xi^{-1},~k_2=\xi^{2}$, giving
    \beq\label{eqn:mappingphi1}
        g_{MN} \rightarrow \frac{1}{\xi} g_{MN},~~~~~~~~~~~~~B_2 \rightarrow \frac{1}{\xi} B_2,~~~~~~~~~C_1\rightarrow \xi\, C_1,~~~~~~~~~~~C_3\rightarrow C_3,~~~~~~~~~e^{\frac{4}{3}\Phi} \rightarrow \frac{1}{\xi^2} e^{\frac{4}{3}\Phi},
    \eeq
followed by (in order)
\beq\label{eqn:mappingphi2}
\begin{aligned}
&C_1\rightarrow C_1 +\xi \,d\phi,~~~~~~C_3\rightarrow C_3+\xi \, d\phi \wedge B_2,~~~~~~~~\zeta \rightarrow \frac{\tilde{\zeta}}{\zeta},~~~~~~~~~~\xi\rightarrow \frac{1}{\zeta},\\[2mm]
&\tilde{\zeta}\rightarrow \xi,~~~~~~~~~~~~~~~~~~~\chi\rightarrow \chi-\frac{\xi}{\zeta}\phi,~~~~~~~~~~~~~~~~~~~\beta \rightarrow -\frac{1}{\zeta}\phi.
\end{aligned}
\eeq
To map to \eqref{eqn:chireduction1}, one requires  
\beq
\begin{aligned}
&g_{MN}\rightarrow g_{MN},~~~~~~~~~~~B_2\rightarrow B_2,~~~~~~~C_1\rightarrow \zeta^2 C_1,~~~~~~~~e^{\frac{4}{3}\Phi} \rightarrow \frac{1}{\zeta^2} e^{\frac{4}{3}\Phi},\\[2mm]
&C_1\rightarrow C_1-\zeta d\chi,~~~~~~~~C_3\rightarrow C_3-\zeta d\chi \wedge B_2,\\[2mm]
&\zeta\rightarrow \frac{1}{\zeta},~~~~~~~~~~~~~~~~~~~~\xi\rightarrow \frac{\xi}{\zeta},~~~~~~~~~~\chi\rightarrow -\phi,~~~~~~~~~~~~\beta\rightarrow \zeta \beta-\xi \phi,
\end{aligned}
\eeq
which does not appear to fit into the conditions of \eqref{eqn:Transformation}. 
Fixing $\zeta=-1$ in \eqref{eqn:phieqmatch} defines the $\mathcal{N}=1$ background, which re-derives \eqref{eqn:chireductionNeq1} via the following set of gauge and coordinate transformations, 
\beq\label{eqn:N1trans}
\begin{aligned}
&C_1\rightarrow -(C_1+d\chi),~~~~~~~~B_2\rightarrow -B_2,~~~~~~~~~~~~~~~C_3\rightarrow C_3 -d\chi\wedge B_2,\\[2mm]
&\beta\rightarrow \beta-\xi \phi,~~~~~~~~~~~~~~~~\chi\rightarrow \phi,~~~~~~~~~~~~~~~~~~~~~\xi\rightarrow -\xi.
\end{aligned}
\eeq
Notice that the mapping to \eqref{eqn:generalresult1} requires both $1/\xi$ and $1/\zeta$, which again is non-trivial as both parameters are integers from the $SL(3,\mathds{R})$ transformation. Nevertheless, fixing $\zeta=0$ and $\xi=0$ derives two new and unique solutions. The $\xi=0$ solution is then a one-parameter family of $\mathcal{N}=0$ solutions which enhances to the zero-parameter $\mathcal{N}=1$ background discussed in Section \ref{sec:chiN1} when $\zeta=-1$, and following the above gauge transformations.

 \item \textbf{Fixing $(v\equiv \xi, q\equiv \gamma)$}\\
 In this case, one re-derives the $\xi=0$ solution of \eqref{eqn:phieqmatch} with
 \beq
C_1\rightarrow C_1+\xi \,d\beta,~~~~~~~~~~~~~~~~C_3\rightarrow C_3+\xi \,d\beta \wedge B_2,~~~~~~~~~~~\chi\rightarrow \chi-\zeta\xi \,\beta.
\eeq

 \item \textbf{Fixing $c\equiv \xi $}\\
 The determinant in \eqref{eqn:S2breakingdefns} now becomes $qa=0$. Of course, fixing $a=0$ corresponds to the case just studied, and taking $q=0$ with $a$ free re-derives \eqref{eqn:phieqmatch} as well (with $\beta=\beta+a\chi$, allowing one to set $a=0$ without loss of generality).
\end{itemize}

\subsubsection{Fixing $m=0$ (with $s$ free)}

The alternative possibility is to ensure $m$ is not a free parameter by fixing it to zero. In these cases, $s$ is now the free parameter.

\begin{itemize}
\item  \textbf{$(m,c)=0$ with $(s,v)$ free parameters}\\
Here the determinant reduces to $qa=0$. These cases can be derived from the $(\xi,\zeta)=0$ solution of \eqref{eqn:phieqmatch} by the following transformations
\begin{itemize}
\item \textbf{$a=0$ (with $q$ free)}: Re-defining $\chi\rightarrow \chi+ q\,\beta$ followed by 
\beq
C_1\rightarrow C_1 +v\,d\beta+s\,d\chi,~~~~~~~~~~~C_3\rightarrow C_3+(v\,d\beta+s\,d\chi)\wedge B_2.
\eeq
\item  \textbf{$q=0$ (with $a$ free)}: Re-defining $\beta\rightarrow \beta+a\,\chi$ followed by
\beq
C_1\rightarrow C_1 +v\,d\beta+s\,d\chi,~~~~~~~~~~~C_3\rightarrow C_3+(v\,d\beta+s\,d\chi)\wedge B_2.
\eeq
\end{itemize}
\item  \textbf{$(m,q,v)=0$ with $(s,a,c)$ free parameters}\\
The remaining case can be derived from the $\zeta=0$ solution of \eqref{eqn:phieqmatch} after re-defining $\xi\equiv c$ and $\beta \rightarrow \beta+(a-s\,c)\chi$, followed by
\beq
C_1\rightarrow C_1+s\,d\chi,~~~~~~~~~~~~C_3\rightarrow C_3 +s\,d\chi\wedge B_2.
\eeq
\end{itemize} 

\subsection{Investigations at the boundary}
We begin by observing that for general values of $(\eta,\sigma,\theta)$, the components of the background are non-zero and finite. In this case, we have no $S^2$ as it was broken under reduction.

\subsubsection{The $\sigma\rightarrow \infty$ boundary}
   At the $\sigma\rightarrow \infty$ boundary, where we use \eqref{eqn:Vsigmainfty} to leading order, we find  
\beq
\begin{aligned}
&\Xi_3^{(\xi\neq0)}=\widehat{\Delta}_3^{(\xi\neq0)}=\frac{2\xi^2 P \,\sigma e^{\frac{2\pi \sigma}{P}}}{\pi^2 \mathcal{R}_1^2\sin^2(\frac{\pi \eta}{P})\sin^2\theta},~~~~~~~~~~\Xi_3^{(\xi=0)}=\Delta_3^{(\zeta\neq0)}=\frac{2\pi \zeta^2 \sigma}{P\sin^2(\frac{\pi \eta}{P})\sin^2\theta},\nn\\[2mm]
&\Pi_3^{(\xi,\zeta\neq0)}=\frac{\zeta\,\xi \sqrt{2} \sigma^{\frac{3}{2}}e^{\frac{\pi \sigma}{P}}}{\sqrt{P}\mathcal{R}_1 \sin^2(\frac{\pi \eta}{P})\cos(\frac{\pi \eta}{P})\sin^2\theta},~~~~~~~~\Big(\Xi_3^{(\xi,\zeta=0)},\Delta_3^{(\zeta=0)},\widehat{\Delta}_3^{(\xi=0)},\Pi_3^{(\zeta=0)},\Pi_3^{(\xi=0)}\Big)=1 ,
\end{aligned}
\eeq
hence we need to investigate this boundary for multiple cases independently. Given that $B_2$ is parameter independent, we find in all cases (using $2\kappa =\pi \widehat{\kappa}X$)
\beq
H_3=-\frac{4  \kappa P}{X\pi}\sin\theta \sin^2\left(\frac{\pi}{P}\eta\right) d\left(\frac{\pi}{P}\eta\right)\wedge d\beta\wedge  d\theta ,
\eeq
and we note $C_3=0$.

When $\xi\neq0$, we find in all cases
\beq
e^{-\Phi}=X^{\frac{3}{2}}\frac{{\cal R}_1\pi^2}{2 (\xi\,P)^{\frac{3}{2}}\sqrt{\kappa}}e^{-\frac{\pi}{P}\sigma}\left(\frac{\pi}{P}\sigma\right)^{-\frac{1}{2}}, ~~~~~~~~~~~~~C_1=\frac{X}{\xi}d\beta,\nn
\eeq
when $\zeta\neq0$, the metric takes the form
\beq
ds^2=\frac{\xi\,\kappa  }{X} \bigg[4\sigma\bigg(ds^2(\text{AdS}_5)+\frac{1}{\xi^2}\Big(\zeta d\beta-\xi d\chi\Big)^2\bigg)+\frac{2P}{\pi}\bigg(d\left(\frac{\pi}{P}\sigma\right)^2+ d\left(\frac{\pi}{P}\eta\right)^2+ \sin^2\left(\frac{\pi}{P}\eta\right) d\theta^2 
\bigg)\bigg],
\eeq
where naively $(\eta,\theta)$ appear to close as an $S^2$, however $H_3$ doesn't take the appropriate form. Hence, $(\eta,\beta,\theta)$ doesn't seem like a valid cycle to integrate over, so we conclude there are no NS5 branes in this case. 
However, when $\zeta=0$, the metric now becomes
\beq
\hspace{-0.5cm}
ds^2=\frac{\xi\,\kappa  }{X} \bigg[4\sigma\bigg(ds^2(\text{AdS}_5)+ d\chi^2\bigg)+\frac{2P}{\pi}\bigg(d\left(\frac{\pi}{P}\sigma\right)^2+ d\left(\frac{\pi}{P}\eta\right)^2+ \sin^2\left(\frac{\pi}{P}\eta\right) \Big(d\theta^2 +\frac{1}{\xi^2} \sin^2\theta\, d\beta^2\Big)\bigg)\bigg],\nn
\eeq
which has a very similar form to the $\beta$ reduction metric in this limit (up to an additional $\xi$ factor out front and $d\beta^2/\xi^2$ replacing $d\phi^2$ in the $S^2$), hence describing a stack of NS5 branes by analogous arguments to those leading to \eqref{eqn:sigmainftymetric2}. Thus
\beq
Q_{NS5}=-\frac{1}{(2\pi)^2}\int_{S^3}H_3=\widehat{\kappa}P.
\eeq
When $\xi=0$, we find for $\zeta\neq0$
   \begin{align} 
ds^2&=\frac{\zeta\, \kappa\,  \pi^{\frac{3}{2}}\mathcal{R}_1 e^{-\frac{\pi \sigma}{P}}}{P X} \bigg[4\sigma\bigg(ds^2(\text{AdS}_5)+\frac{P^2 e^{\frac{2\pi \sigma}{P}}}{\pi^3 \mathcal{R}_1^2 }d\beta^2\bigg)+\frac{2P}{\pi}\bigg(d\left(\frac{\pi}{P}\sigma\right)^2+ d\left(\frac{\pi}{P}\eta\right)^2\nn\\[2mm]
&~~~~~~~~~~~+ \sin^2\left(\frac{\pi}{P}\eta\right)\Big(d\theta^2+\frac{1}{\zeta^2}\sin^2\theta d\chi^2\Big)\bigg)\bigg],\nn\\[2mm]
&e^{-\Phi}=\frac{X^{\frac{3}{2}}e^{\frac{\pi}{2P}\sigma} }{2\pi^{\frac{1}{4}}\zeta^{\frac{3}{2}}(\kappa \mathcal{R}_1)^{\frac{1}{2}}}\left(\frac{\pi}{P}\sigma\right)^{-\frac{1}{2}},  ~~~~~~~C_1=\frac{\xi \,XP^2}{\zeta^2\pi^3 \mathcal{R}_1^2}e^{\frac{2\pi \sigma}{P}}d\beta,
\end{align}
in this case, $(\eta,\theta,\chi)$ close as an $S^3$, and would form a nice cycle - however, $H_3$ does not span this cycle. For $\zeta=0$
   \begin{align} 
ds^2&= \frac{\sqrt{2\pi}\kappa\, \mathcal{R}_1 }{ X} \sin\left(\frac{\pi}{P}\eta\right)\sin\theta\left(\frac{\pi}{P}\sigma\right)^{-\frac{1}{2}} e^{-\frac{\pi \sigma}{P}}\bigg[2\left(\frac{\pi}{P}\sigma\right) \bigg(ds^2(\text{AdS}_5)+d\chi^2+\frac{P^2 e^{\frac{2\pi \sigma}{P}}}{\pi^3 \mathcal{R}_1^2 }d\beta^2\bigg)\nn\\[2mm]
&~~~~~~~~~~~+ d\left(\frac{\pi}{P}\sigma\right)^2+ d\left(\frac{\pi}{P}\eta\right)^2+ \sin^2\left(\frac{\pi}{P}\eta\right) d\theta^2  \bigg],\nn\\[2mm]
&e^{-4\Phi}=\frac{X^6e^{\frac{2\pi}{P}\sigma}\left(\frac{\pi}{P}\sigma\right)  }{2\pi  \kappa^2 \mathcal{R}_1^2 \sin^6(\frac{\pi}{P}\eta)\sin^6\theta }  ,  ~~~~~~~C_1=\frac{2\xi \,XP^2   e^{\frac{2\pi \sigma}{P}}\left(\frac{\pi}{P}\sigma\right) }{ \pi^3 \mathcal{R}_1^2 \sin^2(\frac{\pi}{P}\eta)\sin^2\theta  }  d\beta,
\end{align}
so again the cycle doesn't seem appropriate. Hence, we conclude there exists a stack of NS5 branes in this limit for $\xi\neq0,\zeta=0$ only.

 \subsubsection{The $\eta=0$ boundary, with $\sigma\neq 0$}
At $\eta=0$ with $\sigma\neq 0$, using \eqref{eq:fdef}, \eqref{eqn:feq} and the second form for $M_3$ in \eqref{eqn:phieqmatch}, we find
\beq
f_4 (d\eta^2+d\sigma^2)+ds^2(M_3)  = \frac{2|\dot{f}|}{\sigma^2\,f}\Big( d\eta^2+\eta^2 d\theta^2+d\sigma^2+\frac{\sigma^2}{f^2}d\beta^2\Big)+4\Big(d\chi^2-\frac{\zeta}{\xi}d\beta\Big)^2,
\eeq
where $(\eta,\theta)$ vanish as $\mathds{R}^2$ in polar coordinates.

\paragraph{The $\eta=P$ boundary, with $\sigma\neq 0$} This boundary is qualitatively equivalent to the $\eta=0$ boundary.
 
\subsubsection{The $\sigma=0$ boundary, with $\eta\in (k,k+1)$}
In the case of $\sigma=0,~\eta\in (k,k+1)$, we recall that along the $\sigma=0$ boundary, $\ddot{V}=0$ to leading order. Following the usual procedure, and \eqref{eqn:fsat0}, we find (with $\dot{V}=\mathcal{R}$ and $\dot{V}'=\mathcal{R}'=N_{k+1}-N_k$ at this boundary)
 \beq
 \begin{aligned}
 &\Xi_3=1+\frac{ l_k^2}{\frac{1}{2}\mathcal{R} V''\sin^2\theta},~~~~~~\Delta_3=\widehat{\Delta}_3=\Pi_3=1,~~~~~~l_k=\xi+\zeta (N_{k+1}-N_k),\\
& D\beta = d\beta + (N_{k+1}-N_k)\, d\chi,~~~~~~~~D\chi = d\chi+ \frac{1}{N_{k+1}-N_k} d\beta,
 \end{aligned}
\eeq
leading to
\beq
\hspace{-0.7cm}
\begin{aligned}
&f_4(d\sigma^2+d\eta^2)+ds^2(M_3) \\[2mm]
&\rightarrow \frac{2V''}{\mathcal{R}}\bigg[d\eta^2 +\Big(d\sigma^2+\sigma^2 d\chi^2\Big) +\frac{\mathcal{R}^2}{(N_{k+1}-N_k)^2+2V''\mathcal{R}}\bigg(d\theta^2+\frac{\sin^2\theta}{l_k^2+\frac{1}{2}V''\mathcal{R} \sin^2\theta}D\beta^2\bigg) \bigg]\nn\\[2mm]
&=\frac{2V''}{\mathcal{R}}\bigg[d\eta^2 +\Big(d\sigma^2+\frac{\sigma^2}{(N_{k+1}-N_k)^2} d\beta^2 \Big)+\frac{\mathcal{R}^2}{(N_{k+1}-N_k)^2+2V''\mathcal{R}}\bigg( d\theta^2+\sin^2\theta \frac{(N_{k+1}-N_k)^2}{l_k^2+\frac{1}{2}V''\mathcal{R} \sin^2\theta}D\chi^2\bigg) \bigg],\nn\\[2mm]
\end{aligned}
\eeq 
where we find two alternative forms, with $(\sigma,\chi)$ closing as $\mathds{R}^2$, or $(\sigma,\beta)$ closing as an $\mathds{R}^2/\mathds{Z}_{(N_{k+1}-N_k)}$ orbifold.

     \subsubsection{The $\sigma=0,~\eta=0$ boundary}
To approach $\sigma=0,~\eta=0$, we make the coordinate change $(\eta=r \cos\alpha,~\sigma=r \sin\alpha)$, expanding about $r=0$, using \eqref{eqQdef2} and \eqref{eqQdef}.

In the general case, we can write two equivalent forms for the metric component
\begin{align}
&f_4(d\sigma^2+d\eta^2)+ds^2(M_3) \\[2mm]
&\rightarrow \frac{2Q}{N_1}\bigg[dr^2+r^2 \bigg(d\alpha^2 +\cos^2\alpha\Big(d\theta^2+\frac{1}{l_0^2}\big(N_1^2 \sin^2\theta +\xi^2 \text{tan}^2\alpha\big)D\chi^2\Big) +\frac{\sin^2\alpha}{N_1^2+\xi^2\text{tan}^2\alpha\, \text{cosec}^2\theta } d\beta^2\bigg)\bigg]\nn\\[2mm]
&= \frac{2Q}{N_1}\bigg[dr^2+r^2 \bigg( d\alpha^2+\cos^2\alpha \Big(d\theta^2+\frac{\sin^2\theta}{l_0^2}D\beta^2\Big)+\sin^2\alpha \bigg(\frac{d\chi^2}{1+\zeta^2 \text{tan}^2\alpha\, \text{cosec}^2\theta}+\frac{\zeta^2}{l_0^2}D\beta^2\Big)\bigg)\bigg],\nn\\[2mm]
&l_0=\xi+\zeta N_1,~~~~~~~~D\beta=d\beta +\frac{N_1-\zeta\,\xi\,\text{tan}^2\alpha\, \text{cosec}^2\theta}{1+\zeta^2 \text{tan}^2\alpha\, \text{cosec}^2\theta}d\chi,~~~~~~D\chi=d\chi+\frac{N_1-\xi\,\zeta\,\text{tan}^2\alpha\, \text{cosec}^2\theta}{N_1^2+\xi^2 \text{tan}^2\alpha\, \text{cosec}^2\theta}d\beta ,\nn
\end{align}
where we can observe that the space closes nicely in $r$. When we fix either $\xi=0$ or $\zeta=0$, using the top and bottom lines receptively (with $l_0=\zeta N_1$ and $l_0=\xi$), the internal space closes as an $\mathds{R}^5
$ with additional orbifold singularities.
Due to the form of $l_0$, it is clear that we need to consider the special case of $\xi=\zeta=0$ separately, where we find
\beq
f_4(d\sigma^2+d\eta^2)+ds^2(M_3)\rightarrow \frac{2Q}{N_1}\bigg(dr^2+r^2 \Big(d\alpha^2 +\cos^2\alpha \,d\theta^2 +\sin^2\alpha\, d\chi^2\Big)\bigg)+\frac{4}{N_1}(d\beta+N_1 d\chi)^2,\nn
\eeq
with the $r$ components closing nicely as an $\mathds{R}^4$.

    \paragraph{The $\sigma=0,~\eta=P$ boundary} This is qualitatively equivalent to the $(\sigma,\eta)=0$ boundary.  

\subsubsection{The $\sigma=0$ boundary, with $\eta=k$}
At the $\sigma=0$ boundary with $\eta=k$, we will follow the usual approach by making the coordinate change $(\eta=k-r \cos\alpha,~\sigma=r \sin\alpha)$ for $r\sim0$, using \eqref{eq:VsattheD6s} and \eqref{eqn:fsfork}, we find to leading order (with $r=z^2$)
\begin{align}\label{eqn:B3eqphi}
& ds^2= \frac{ \kappa }{X} \sin\theta \Big[N_k\Big(4ds^2(AdS_5) +d\theta^2\Big) +4b_k\Big(dz^2 +z^2 ds^2(\mathbb{B}_3)\Big)\Big],~~~~~~e^{ 4 \Phi}=  \frac{ \kappa^2 }{X^6}N_k^2   \sin^6\theta,  
\nn\\[2mm]
&ds^2(\mathbb{B}_3)= \frac{1}{4}\Big(d\alpha^2 +  \sin^2\alpha\, d\chi^2\Big)+\frac{1}{  b_k^2}\Big(d\beta+g(\alpha)d\chi \Big)^2, ~~~B_2= - \frac{2\kappa}{X} \sin\theta\Big( N_k \,d\chi+  k \,d\beta\Big)\wedge d\theta, 
\end{align}
with $\mathbb{B}_3$ now a $U(1)$ fibration over an $S^2$.

Inspired by the discussion around \eqref{eqn:C1sourcephi}, where we had two nice D6 brane sources at $\sin\theta=0$, let us now approach this point using the same coordinate change as before - namely, 
\beq
\eta=k-\bar{r} \cos\alpha \sin^2\mu,~~~~~\sigma=\bar{r} \sin\alpha \sin^2\mu,~~~~~~\sin\theta = 2\sqrt{\frac{b_k\bar{r}}{N_k}}\cos\mu,
\eeq
with $r= \bar{r} \sin^2\mu$, and expanding about $\bar{r}=0$. We find
\beq
\hspace{-1cm}
\begin{aligned}
&b_k^2\text{cot}^2\mu \,\Xi_3\rightarrow\tilde{\Xi}_k=\big(\xi+\zeta g(\alpha)\big)^2+b_k^2 \Big(\text{cot}^2\mu+\frac{1}{4}\zeta^2 \sin^2\alpha\Big),~~~~~\Delta_3\rightarrow \Delta_k=1+\frac{\zeta^2}{4}\text{tan}^2\mu \,\sin^2\alpha,\\[2mm]
&(g(\alpha)^2+\frac{1}{4}b_k^2 \sin^2\alpha)\widehat{\Delta}_3 \rightarrow \widehat{\Delta}_k =  g(\alpha)^2+\frac{1}{4} \sin^2\alpha \Big(b_k^2+\xi^2 \text{tan}^2\mu\Big),~~~\Pi_3\rightarrow \Pi_k= 1-\zeta\,\xi  \frac{\text{tan}^2\mu \sin^2\alpha}{4\,g(\alpha)},\\[2mm]
&\Delta_k(\alpha=0)=\Delta_k(\alpha=\pi)=1,~~~~\widehat{\Delta}_k(\alpha=0)=(N_k-N_{k-1})^2,~~~~~\widehat{\Delta}_k(\alpha=\pi)=(N_{k+1}-N_{k})^2,
\end{aligned}
\eeq
leading to
  \begin{align}\label{eqn:B4eqmetricphi}
 & \frac{ds^2}{ 2\kappa\sqrt{N_k} } =\frac{ 1}{X} \sin\mu \sqrt{\tilde{\Xi}_k}\Bigg[\frac{4}{\sqrt{\frac{b_k}{\bar{r}}}}ds^2(AdS_5)+ \frac{\sqrt{\frac{b_k}{\bar{r}}}}{N_k}\bigg(d\bar{r}^2+4\bar{r}^2  ds^2(\mathbb{B}_4 ) \bigg)\Bigg],\nn\\[2mm]
&  ds^2(\mathbb{B}_4 )= d\mu^2 +\frac{1}{4}\sin^2\mu \bigg(d\alpha^2 + \frac{\sin^2\alpha}{\Delta_k}d\chi^2\bigg) +  \cos^2\mu \frac{ \Delta_k}{\tilde{\Xi}_k}\Big(d\beta+\frac{\Pi_k}{\Delta_k}g(\alpha)d\chi\Big)^2 \nn\\[2mm]
  &~~~~~~~~~~= d\mu^2 +\frac{1}{4}\sin^2\mu \bigg(d\alpha^2 + \frac{\sin^2\alpha}{ \widehat{\Delta}_k }d\beta^2\bigg) +  \cos^2\mu \frac{ \widehat{\Delta}_k}{\tilde{\Xi}_k}\Big(d\chi +\frac{\Pi_k}{\widehat{\Delta}_k}g(\alpha)d\beta\Big)^2, \nn\\[2mm]
&B_2 =-\frac{4\kappa  b_k }{X N_k}\cos^2\mu \Big(k\, d\beta +N_k\,d\chi \Big)\wedge d\bar{r},~~~~~~~~~~~~C_3=0,~~~~~~~~~~e^{ 4 \Phi}=  \frac{2^6 \kappa^2  \tilde{\Xi}_k^3   }{ X^6 b_k^3N_k}\,\bar{r}^3 \sin^6\mu, \nn\\[2mm]
& C_1 = \frac{ X }{\tilde{\Xi}_k}\bigg(\big(\xi+\zeta\, g(\alpha)\big) \Big(d\beta + g(\alpha) d\chi\Big)+\frac{1}{4}\zeta \,b_k^2\,\sin^2\alpha \,d\chi\bigg),
 \end{align}
where the four-manifold, $\mathbb{B}_4$, has an orbifold singularity coming from the $\tilde{\Xi}_k$ contribution - unique to each position $\eta=k$. Calculating the Euler Characteristic using the Chern-Gauss-Bonnet theorem \eqref{eqn:Chern-Gauss} gives
\beq
\chi_E
=3-\bigg(1-\frac{1}{|l_{k-1}|}\bigg)-\bigg(1- \frac{1}{|\frac{1}{\zeta}(l_{k-1}-l_k)|}\bigg)-\bigg(1-\frac{1}{|l_{k}|}\bigg),
\eeq 
with $|\frac{1}{\zeta}(l_{k-1}-l_k)|=| b_k|\in \mathds{Z}$. When $\xi=0$, we find $l_k=\zeta (N_{k+1}-N_k)$, and for $\zeta=0$, we have $l_k=l_{k-1}=\xi$. In both cases, we still have the $b_k$ contribution arising from the $\tilde{\Xi}_k$.

We now calculate the D6 charge at $\mu=\frac{\pi}{2}$. Integrating over $(\alpha,\chi)$ gives
\beq
Q_{D6}^k=-\frac{1}{2\pi}\int_{(\alpha,\chi)}F_2 =-\frac{1}{2\pi}\int_{\chi=0}^{\chi=2\pi}C_1\Big|_{\alpha=0}^{\alpha=\pi} =\frac{\xi \,X }{l_kl_{k-1}}(2N_k-N_{k+1}-N_{k-1}).
\eeq
Alternatively, integrating over $(\alpha,\beta)$ leads to
\beq
Q_{D6}^k=\frac{1}{2\pi}\int_{(\alpha,\beta)}F_2 =\frac{1}{2\pi}\int_{\beta=0}^{\beta=2\pi}C_1\Big|_{\alpha=0}^{\alpha=\pi} =\frac{\zeta \,X }{l_kl_{k-1}}(2N_k-N_{k+1}-N_{k-1}).
\eeq

We clearly need to consider the $\xi=\zeta=0$ case separately (due to the form of $l_k$), where
\beq
\begin{aligned}
& \tilde{\Xi}_k= b_k^2  \text{cot}^2\mu ,~~~~~  \Delta_k=\Pi_k= 1 ,~~~~~~ \widehat{\Delta}_k =  g(\alpha)^2+\frac{b_k^2}{4} \sin^2\alpha   , 
\end{aligned}
\eeq
with
  \begin{align}\label{eqn:B4eqmetricphizero}
 & \frac{ds^2}{ 2\kappa\sqrt{N_k} } =\frac{ b_k}{X} \cos\mu  \Bigg[\frac{4}{\sqrt{\frac{b_k}{\bar{r}}}}ds^2(AdS_5)+ \frac{\sqrt{\frac{b_k}{\bar{r}}}}{N_k}\bigg(d\bar{r}^2+4\bar{r}^2  ds^2(\mathbb{B}_4 ) \bigg)\Bigg],~~~~~~ ds^2(\mathbb{B}_4 )= d\mu^2 +\sin^2\mu \,ds^2(\mathbb{B}_3 )\nn\\[2mm]
&B_2 =-\frac{4\kappa  b_k }{X N_k}\cos^2\mu \Big(k\, d\beta +N_k\,d\chi \Big)\wedge d\bar{r},~~~~~~~~~~~~C_1=C_3=0,~~~~~~~~~~e^{ 4 \Phi}=  \frac{2^6 \kappa^2  b_k^3   }{ X^6  N_k}\,\bar{r}^3 \cos^6\mu,
 \end{align}
 where there are no D6 branes. In a similar manner to both the $\beta$ and $\chi$ reduction cases, when $\sin\mu\rightarrow 0$, we find $\mathbb{B}_4$ approaches a cone over the $\mathbb{B}_3$ given in \eqref{eqn:B3eqphi}. The Euler characteristic of $\mathbb{B}_4 $ now becomes
 \beq
\chi_E = \frac{1}{|b_k|} = 1-\bigg(1- \frac{1}{|b_k|}\bigg).
\eeq 
 



\subsubsection{Summary}
At the $\sigma\rightarrow\infty$ limit, the form of $H_3$ is parameter independent, but the integration cycle only appears valid in the $\xi\neq0,\zeta=0$ case - for which the form of the metric describes NS5 branes. Hence, we conclude a stack of NS5 branes only exist at $\sigma\rightarrow\infty$ for $\xi\neq0,\zeta=0$. Along the $\sigma=0$ boundary (at $\sin\theta=0$) for $(\zeta,\xi)\neq0$, we find two stacks of D6 branes at each kink of the rank function (orthogonal to either $\chi$ or $\beta$). These branes have rational quantization due to orbifold singularities. However, when $\zeta=\xi=0$, we find no D6 branes. Note, when $\zeta=0$, an $S^2$ is no longer recovered. No backgrounds contain D4 branes.
Hence, in summary
 \begin{equation}\label{eqn:ChargesIIA}
 \hspace{-0cm}
 \begin{aligned}
&Q_{NS5}=\begin{cases}
     \widehat{\kappa} \,P & \xi\neq0,\zeta=0\\
     ~0  & \text{otherwise}
    \end{cases}  ,~~~~~l_k=\xi+\zeta (N_{k+1}-N_k),~~~~~b_k=(2N_k-N_{k+1}-N_{k-1}),  
    \nn\\[2mm]
    &Q_{D6}^{1,k} =\frac{X}{l_kl_{k-1}} \begin{cases}
      0 & \zeta=\xi=0\\
   \xi \, b_k   &\text{otherwise}
    \end{cases}  ,~~~~~~Q_{D6}^{2,k} =\frac{X}{l_kl_{k-1}} \begin{cases}
      0 & \zeta=\xi=0\\
   \zeta \, b_k   &\text{otherwise}
    \end{cases}  ,  \\[2mm]
    &Q_{D6}^1
    =\frac{X}{l_0 l_P} \begin{cases}
   ~~~~~~~   0 & \zeta=\xi=0\\
     \xi(N_{P-1}+N_1) &\text{otherwise}
    \end{cases},~~~~~~~~Q_{D6}^2  =\frac{X}{l_0 l_P} \begin{cases}
   ~~~~~~~   0 & \zeta=\xi=0\\
     \zeta (N_{P-1}+N_1) &\text{otherwise}
    \end{cases}.
    \end{aligned} 
\end{equation}
Calculating the holographic central charge, and using $2\kappa=\pi \widehat{\kappa} X $, one finds
\begin{equation*}
c_{hol} 
= \frac{ \widehat{\kappa}^3X^2}{8\pi}\sum_{n=1}^\infty P \, \mathcal{R}_n^2.
\end{equation*}

\paragraph{$\mathcal{N}=1$ Solution}
Fixing $\zeta=-1$ in \eqref{eqn:phieqmatch} defines a one-parameter family of $\mathcal{N}=1$ backgrounds, which simply re-derives \eqref{eqn:chireductionNeq1} via the set of gauge and coordinate transformations given in \eqref{eqn:N1trans}.

\paragraph{A comment on Supersymmetry breaking:} We leave this analysis for future work, however repeating the procedures outlined in the $\beta$ reduction case, we expect analogous equations to \eqref{eqn:SUSYbrokenIIA} - with the $(\zeta+\xi)$ terms replaced by $(\zeta+1)$ analogues.     
\newpage
\section{Summary so far}
\begin{itemize}
\item Following an SL$(3,\mathds{R})$ transformation in $d=11$, we constructed new two-parameter families of type IIA solutions, each labelled by a potential $V(\sigma,\eta)$. 
We studied supersymmetry preservation using the method of G-structures, and studied the quantization of charges - leading to the discovery of D6 branes orthogonal to spindles. Most solutions contain colour NS5 branes, with D4 branes present for a preserved $S^2$.
\item Analysis for the dual SCFTs was provided for the $\beta$ reduction, proposing the operators which give rise to marginal deformations of ${\cal N}=2$ SCFTs. A mirror-like relation between quivers was discussed, along with spin-two fluctuations in the CFT. 
\item Reductions along $\chi$ and $\phi$ lead to additional families, including an $\mathcal{N}=1$ background. 
\item In the $\beta$ reduction case:
\begin{itemize}
\item when  $\xi=0$: the $\mathcal{N}=2$ is broken by the $\zeta$ parameter (corresponding to a marginal Lagrangian deformation) which physically breaks the $S^2$ and introduces a $\zeta b_k$ orbifold singularity into the four manifold $\mathbb{B}_4$.
\item when  $\zeta=0$: the $\mathcal{N}=2$ is broken by the $\xi$ parameter (corresponding to a marginal non-Lagrangian deformation) which physically introduces orbifold singularities to the poles of the $S^2$ - defining spindles.
\item when $\zeta\neq0,~\xi\neq0$: the $\mathcal{N}=2$ is broken by both parameters in general, with the broken $S^2$ gaining orbifold singularities, leading to a higher dimensional analogue of the spindle.
\end{itemize}
\item The $\chi$ reduction case behaves analogously, but now the $\mathcal{N}=2$ is broken in all cases - containing orbifold singularities for all finite values of $\xi$ (for a generic rank function).
\item The $\phi$ reduction case necessarily breaks the $S^2$, but still includes orbifold singularities with rational quantization. Two alternative cycles for D6 charge can now be used. The solutions contain no D4 branes, and at the $\sigma\rightarrow\infty$, the NS5 charge is only valid for $\xi\neq0,\zeta=0$.
\end{itemize}

\chapter{Type IIB - via Abelian T-Duality (ATD)}\label{chap:typeIIB}

In this chapter we will investigate the abelian T-Duality (ATD) of our various two-parameter families of Type IIA solutions, deriving new three-parameter families of Type IIB solutions - with the additional non-trivial parameter picked up in the T-Duality. These solutions are in general $\mathcal{N}=0$, but under two separate conditions on the parameters, can enhance to new one-parameter families of $\mathcal{N}=1$ backgrounds. Performing boundary analysis then leads to the discovery of orbifold singularities in some cases, however the physical interpretation is more complicated than the spindles in IIA. In fact, as an artefact of T-dualising within this spindle-like orbifold, we find solutions for which the orbifold structure has been completely broken, but the rational charge still inherited. In all backgrounds investigated, we find NS5 and D7 branes, with D5s present for a preserved $S^2$. 

Following the methodology of the previous two chapters, the supersymmetry is kept track of using the method of G-structures. An ATD of the G-structure forms and conditions are performed, deriving new expressions for the pure spinors in IIB, written in terms of their type IIA ancestors. Explicit results are then presented for multiple infinite families of $\mathcal{N}=1$ solutions. Notably, these $\mathcal{N}=1$ solutions have a zero five-form flux, which is an interesting result given that very few supersymmetric solutions are known with this property. The first two examples of such backgrounds were found in \cite{Macpherson:2014eza}, evading the work of \cite{Gauntlett:2005ww}, and leading to \cite{Couzens:2016iot} - where the classification was completed.

The one-parameter background presented in \cite{Nunez:2019gbg} is re-derived as a sub-class of our more general solution. 
The original holographic central charge of the GM class is still preserved for the IIB solutions, suggesting that these backgrounds are once again dual to marginal deformations of the `parent' $\mathcal{N}=2$ SCFT. However, 
switching these parameters off in the supergravity does not recover an $\mathcal{N}=2$ background.

\section{General ATD Transformation}
We now turn to the abelian T-Duality (ATD) calculations performed in this work. We utilise the T-Dual rules presented in \cite{Kelekci:2014ima}, and given in \eqref{eqn:TD1} and \eqref{eqn:TD2}, corresponding to an abelian T-duality along a $U(1)$ direction, $y$. For convenience, we re-write the rules here. With a Type IIA decomposition 
   \begin{equation*}
   \begin{aligned}
  &ds_{10}^2=ds_{9,\mathcal{A}}^2+e^{2C}(dy+A_1)^2,~~~~~~~B=B_2+B_1 \wedge dy,\\[2mm]
  &F=F_{\perp}+F_{||}\wedge E^y,~~~~~~~~~E^y =e^{C}(dy+A_1),
  \end{aligned}
  \end{equation*}
the type IIB T-dual solution is then defined by  
  \begin{equation*} 
  \begin{gathered}
  ds_{9,\mathcal{B}}^2=ds_{9,\mathcal{A}}^2,~~~~~\Phi^\mathcal{B}=\Phi^\mathcal{A}-C^\mathcal{A},~~~~~~~C^\mathcal{B}=-C^\mathcal{A},\\
  B_2^\mathcal{B}=B_2^\mathcal{A}+A_1^\mathcal{A} \wedge B_1^\mathcal{A},~~~~~~~~A_1^\mathcal{B}=-B_1^\mathcal{A},~~~~~~~~B_1^\mathcal{B}=-A_1^\mathcal{A},\\
  F_{\perp}^\mathcal{B}=e^{C^\mathcal{A}}F_{||}^\mathcal{A},~~~~~~~F_{||}^\mathcal{B}=e^{C^\mathcal{A}}F_{\perp}^\mathcal{A}.
  \end{gathered}
  \end{equation*}
  
  We begin by utilising these rules to derive general type IIB formulae, specific to the IIA backgrounds studied in the previous section, making life easy when considering the various multi-parameter families in question.
  \subsection{General IIB Backgrounds}
Let us begin with the following general form for the IIA backgrounds in question (with $(\phi_1,\phi_2,\phi_3)$ representing three $U(1)$ directions)
      \begin{equation}\label{eqn:GenGenIIA}
    \begin{aligned}
 & ~~~~~~~~~~~~~~~~~~~~~~~~~~~~~~~~~~~~~~~~~~~~~~~~~~~~~~~  \textbf{\underline{IIA}}\\[2mm]
   &     ds_{10,st}^2= ds_7^2 + \,ds^2_3,~~~~~~~~~~~~~~~~    \Phi_A,\\[2mm]
  &         ds^2_3 = h_{\phi_1}(\eta,\sigma,\theta)d\phi_1^2 + h_{\phi_2}(\eta,\sigma,\theta)d\phi_2^2 + h_{\theta}(\eta,\sigma,\theta)d\theta^2 + h_{\phi_1\phi_2}(\eta,\sigma,\theta)d\phi_1 d\phi_2 \\[2mm]
  &~~~~~~~~~~~+ h_{\phi_1\theta}(\eta,\sigma,\theta)d\phi_1 d\theta+ h_{\phi_2\theta}(\eta,\sigma,\theta)d\phi_2 d\theta,\\[2mm]
   &   B_2 = B_{2,\phi_1\theta}d\phi_1 \wedge d\theta  +B_{2,\phi_2\theta}d\phi_2 \wedge d\theta +B_{\phi_1,\phi_2}d\phi_1 \wedge d\phi_2  ,\\[2mm]
    &    C_1= C_{1,\phi_1} d\phi_1  + C_{1,\phi_2} d\phi_2+ C_{1,\theta} d\theta    ,~~~~~~~~~~~~~~~~~
       C_3=C_{3,\phi_1\theta\phi_2} d\phi_1 \wedge d\theta \wedge d\phi_2.
    \end{aligned}
    \end{equation} 
In order to T-dualise such a solution, the first step is to re-write the metric in the following manner (for an ATD along $\phi_1$)
    \beq
\begin{gathered}
     ds^2_3 = h_{\phi_1}\Big(d\phi_1+\frac{1}{2h_{\phi_1}}(h_{\phi_1\phi_2}d\phi_2+h_{\phi_1\theta}d\theta)\Big)^2+\frac{4h_\theta h_{\phi_1}-h_{\phi_1\theta}^2}{4h_{\phi_1}}d\theta^2~~~~~~~~~~~\\
  +\frac{4h_{\phi_1} h_{\phi_2}-h_{\phi_1\phi_2}^2}{4h_{\phi_1}}d\phi_2^2 +\frac{2h_{\phi_1}h_{\phi_2\theta}-h_{\phi_1\theta}h_{\phi_1\phi_2}}{2h_{\phi_1}}d\theta d\phi_2,
     \end{gathered}
    \eeq
 noting
  \begin{equation}
\begin{aligned}
F_2&=dC_1 =dC_{1,\phi_1}\wedge  d\phi_1  + dC_{1,\phi_2} \wedge d\phi_2+ dC_{1,\theta} \wedge d\theta,\\
F_4&=dC_3-H_3\wedge C_1\\
&= \Big( dC_{3,\phi_1\theta\phi_2} + C_{1,\phi_1}dB_{2,\phi_2\theta}  + C_{1,\theta}dB_{2,\phi_1\phi_2}- C_{1,\phi_2} dB_{2,\phi_1\theta}\Big)\wedge  d\phi_1 \wedge d\theta \wedge d\phi_2 .
\end{aligned}
  \end{equation}
  We are now free to use the rules in \eqref{eqn:TD1} and \eqref{eqn:TD2} (presented again at the start of this section) to derive the following IIB result
        \begin{align}\label{eqn:betaredchitdualgeneral}
 &  ~~~~~~~~~~~~~~~~~~~~~~~~~~~~~~~~~~~~~~~~~~~~~   \textbf{\underline{IIB (ATD along $\phi_1$)}}\nn\\[2mm]
   &    ds_{10,B}^2=ds_9^2   +     \,  \frac{1}{ h_{\phi_1}}\Big(d\phi_1+B_{2,\phi_1\theta}d\theta + B_{2,\phi_1\phi_2}d\phi_2\Big)^2,\nn\\[2mm]
  &     ds_9^2=ds_7^2+\bigg(\frac{4h_\theta h_{\phi_1}-h_{\phi_1\theta}^2}{4h_{\phi_1}}\bigg)d\theta^2+\bigg(\frac{4h_{\phi_1} h_{\phi_2}-h_{\phi_1\phi_2}^2}{4h_{\phi_1}}\bigg)d\phi_2^2 +\bigg(\frac{2h_{\phi_1}h_{\phi_2\theta}-h_{\phi_1\theta}h_{\phi_1\phi_2}}{2h_{\phi_1}}\bigg)d\theta d\phi_2 ,\nn\\[2mm]
& e^{\frac{4}{3}\Phi_B}=\Big(\frac{1}{h_{\phi_1}}\Big)^{\frac{2}{3}}e^{\frac{4}{3}\Phi_A},
~~~~~~~~~~~~~~~~~~~~~~~C_0=C_{1,\phi_1},\nn\\[2mm]
& B_2=\bigg(B_{2,\phi_2\theta}+ \frac{1}{2h_{\phi_1}}  (h_{\phi_1\theta}  B_{2,\phi_1\phi_2}-h_{\phi_1\phi_2}  B_{2,\phi_1\theta}) \bigg) d\phi_2 \wedge d\theta -\frac{1}{2h_{\phi_1}} (h_{\phi_1\phi_2}d\phi_2+h_{\phi_1\theta}d\theta)\wedge d\phi_1 ,\nn\\[2mm]
 &  C_2=\bigg[ C_{3,\phi_1\theta\phi_2}+B_{2,\phi_1\phi_2} \Big(C_{1,\theta} - \frac{1}{2}\frac{h_{\phi_1\theta}}{h_{\phi_1}} C_{1,\phi_1} \Big) - B_{2,\phi_1\theta} \Big(C_{1,\phi_2} - \frac{1}{2}\frac{h_{\phi_1\phi_2}}{h_{\phi_1}} C_{1,\phi_1} \Big) \bigg] d\theta \wedge d\phi_2\nn \\[2mm]
&~~~~~~~~~~  + \bigg(C_{1,\phi_2} -\frac{1}{2}\frac{h_{\phi_1\phi_2}}{h_{\phi_1}}C_{1,\phi_1}\bigg) d\phi_2\wedge d\phi_1  + \bigg(C_{1,\theta} -\frac{1}{2}\frac{h_{\phi_1\theta}}{h_{\phi_1}}C_{1,\phi_1}\bigg) d\theta\wedge d\phi_1. 
  \end{align}  
One is now free to plug in the functions corresponding to the specific IIA example in question, deriving the IIB ATD solution without the need to perform the calculation each time. 

We can extend this approach to a TsT transformation,  performing the coordinate change $\phi_2 \rightarrow \phi_2+\gamma_1 \phi_1$ in \eqref{eqn:betaredchitdualgeneral}, before performing the ATD along $\phi_1$ for a second time (to return to a IIA theory)
\begin{adjustwidth}{-1cm}{}
     \vspace{-0.7cm}
       \begin{align}
  &    ~~~~~~~~~~~~~~~~~~~~~~~~~~~~~~~~~~~~~~~~~~~~~~~~~~~~~~~~~  \textbf{\underline{TsT }}\nn\\[2mm]
  &     ds_{10,A}^2=ds_9^2+\frac{h_{\phi_1}(1-\gamma_1 \alpha_1)}{1+\gamma_1 B_{2,\phi_1\phi_2}}\bigg[d\phi_1 +\frac{h_{\phi_1\phi_2}}{2h_{\phi_1}}d\phi_2+\frac{h_{\phi_1\theta}}{2h_{\phi_1}}d\theta\nn\\[2mm]
  &~~~~+\gamma_1 \Big[B_{2,\phi_2\theta}+\frac{1}{2h_{\phi_1}}(h_{\phi_1\theta}B_{2\phi_1\phi_2}-h_{\phi_1\phi_2}B_{2,\phi_1\theta})\Big]d\theta\bigg]^2, ~~~~~~~   e^{\frac{4}{3}\Phi_A^{TsT}}=\bigg(\frac{ 1-\gamma_1 \alpha_1}{1+\gamma_1 B_{2,\phi_1\phi_2}}\bigg)^{\frac{2}{3}} e^{\frac{4}{3}\Phi_A}, \nn \\[2mm]
  &     ds_9^2=ds_7^2 + \frac{1}{h_{\phi_1}}\bigg(B_{2,\phi_1\theta}^2 +h_\theta h_{\phi_1} -\frac{1}{4}h_{\phi_1\theta}^2-\alpha_3\alpha_2^2\bigg)d\theta^2 + \bigg(\frac{h_{\phi_1}h_{\phi_2}-\frac{1}{4}h_{\phi_1\phi_2}^2}{h_{\phi_1}\alpha_3}\bigg) d\phi_2^2~~~~~~~~~~~~~~~~~~~~~~~~~~~~~~~~ \nn\\[2mm]
 &    ~~~~~~~  + \frac{1}{h_{\phi_1}\alpha_3}\bigg[(1+\gamma_1 B_{2,\phi_1\phi_2}) \bigg(h_{\phi_1}h_{\phi_2\theta}-\frac{1}{2}h_{\phi_1\theta}h_{\phi_1\phi_2}\bigg) - 2\gamma_1 B_{2,\phi_1\theta}  \bigg(h_{\phi_1}h_{\phi_2}-\frac{1}{4} h_{\phi_1\phi_2}^2\bigg) \bigg] d\theta d\phi_2 ,\nn
     \end{align}
       \end{adjustwidth} 
     \begin{adjustwidth}{0cm}{}
     \vspace{-0.7cm}
            \begin{align}\label{eqn:TSTgenlgeneral}
   &    B_2= \bigg[B_{2,\phi_2\theta} +\frac{(1-\gamma_1 \alpha_1)}{2h_{\phi_1}} \Big(h_{\phi_1\theta} B_{2,\phi_1\phi_2} -h_{\phi_1\phi_2} B_{2,\phi_1\theta}\Big) +\alpha_2 \frac{h_{\phi_1\phi_2}}{2h_{\phi_1}}-\alpha_1 \bigg(\frac{h_{\phi_1\theta}}{2h_{\phi_1}}+\gamma_1 B_{2,\phi_2\theta}\bigg)\bigg]d\phi_2\wedge d\theta \nn\\[2mm]
  &  ~~~~~~~~~~   -(\alpha_1 d\phi_2 + \alpha_2d\theta)\wedge d\phi_1,\nn\\[2mm]
  &     C_1 = C_{1,\phi_2}d\phi_2 +C_{1,\phi_1}d\phi_1 +C_{1,\theta}d\theta+\gamma_1 \Big(C_{3,\phi_1\theta\phi_2} + C_{1,\theta}B_{2,\phi_1\phi_2}+ C_{1,\phi_1}B_{2,\phi_2\theta} -C_{1,\phi_2}B_{2,\phi_1\theta}\Big)d\theta,\nn\\[2mm]
  &     C_3 = \bigg[(1-\gamma_1 \alpha_1) \bigg(C_{3,\phi_1\theta\phi_2} +C_{1,\theta} B_{2,\phi_1\phi_2} -C_{1,\phi_2}B_{2,\phi_1\theta} +\frac{C_{1,\phi_1}}{2h_{\phi_1}}\Big( h_{\phi_1\phi_2}B_{2,\phi_1\theta} -h_{\phi_1\theta}B_{2,\phi_1\phi_2}\Big)\bigg) \nn\\[2mm]
    &~~~~~~~~~~~~~   + \alpha_1\bigg(C_{1,\phi_1} \frac{h_{\phi_1\theta}}{2h_{\phi_1}} - C_{1,\theta}\bigg) -\alpha_2 \bigg(C_{1,\phi_1} \frac{h_{\phi_1\phi_2}}{2h_{\phi_1}} - C_{1,\phi_2}\bigg)\bigg] d\theta \wedge d\phi_2 \wedge d\phi_1,  
    \end{align}
  \end{adjustwidth} 
  with
      \begin{align}
  &     \alpha_1= \frac{1}{\alpha_3}\Big[B_{2,\phi_1\phi_2}(1+\gamma_1 B_{2,\phi_1\phi_2}) +\gamma_1 \Big(h_{\phi_1}h_{\phi_2} -\frac{1}{4}h_{\phi_1\phi_2}^2\Big)\Big],\nn\\[2mm]
 &      \alpha_2=\frac{1}{\alpha_3}\Big[B_{2,\phi_1\theta}(1+\gamma_1 B_{2,\phi_1\phi_2})  +\frac{\gamma_1}{2} \Big(h_{\phi_1}h_{\phi_2\theta} -\frac{1}{2}h_{\phi_1\theta}h_{\phi_1\phi_2}\Big) \Big],\nn\\[2mm]
    &   \alpha_3=(1+\gamma_1 B_{2,\phi_1\phi_2})^2 +\gamma_1^2 \Big(h_{\phi_1}h_{\phi_2} -\frac{1}{4}h_{\phi_1\phi_2}^2\Big).
  \end{align}
     In doing such a calculation, we have picked up the transformation parameter $\gamma_1$. One can see by observation that setting $\gamma_1=0$ in \eqref{eqn:TSTgenlgeneral} reproduces \eqref{eqn:GenGenIIA}, as required.
     
 Following this TsT transformation along $\phi_1$, and given that there are multiple $U(1)$ directions in these backgrounds, we could in fact perform a second TsT transformation along $\phi_2$ (with transformation parameter $\gamma_2$). However, this turns out to be simply a trivial re-definition of the TsT parameter, for instance $\gamma_1\rightarrow \gamma_1+ \gamma_2$. 
  
  The formulae we have just presented are in fact a little too general for the GM backgrounds themselves, where $h_{\phi_1\theta}=h_{\phi_2\theta}=0$. These results are then unnecessarily cumbersome for our purposes. 
  \subsubsection{Less general forms}
The Type IIA backgrounds presented in this chapter all fit into the following form, where $(\phi_1,\phi_2)$ are the two $U(1)$ directions
      \begin{equation}\label{eqn:GenIIA}
    \begin{aligned}
 &~~~~~~~~~~~~~~~~~~~~~~~~~~~~~~~~~~~~~~~~~~~~~~~~~~~~~~~   \textbf{\underline{IIA}}\\[2mm]
&        ds_{10,st}^2= ds_8^2+\Gamma \,ds^2_2,~~~~~~~\Phi_A,~~~~~  ds^2_2 = h_{\phi_1}(\eta,\sigma,\theta)d\phi_1^2 + h_{\phi_2}(\eta,\sigma,\theta)d\phi_2^2 + h_{\phi_1\phi_2}(\eta,\sigma,\theta)d\phi_1 d\phi_2, \\[2mm]
 &     B_2 =  \Big(B_{2,\phi_1}d\phi_1  +B_{2,\phi_2}d\phi_2 \Big) \wedge d\theta,~~~~~~~
        C_1= C_{1,\phi_1} d\phi_1  + C_{1,\phi_2} d\phi_2  ,~~~~~~~ C_3=C_{3,\phi_1\phi_2} d\phi_1 \wedge d\theta \wedge d\phi_2,
    \end{aligned}
    \end{equation} 
      where specifically, we have 
 \begin{equation} 
  \begin{aligned}
 &ds_8^2=  ds_7^2 +f_\theta d\theta^2,~~~~~~~~~~~~~~~~~  ds_7^2=e^{\frac{2}{3}\Phi_A}f_1\bigg[4ds^2(\text{AdS}_5)+f_4(d\sigma^2+d\eta^2)\bigg],   \\[2mm]
 &  \Gamma=f_1^2e^{-\frac{2}{3}\Phi_A},~~~~~~~~~~~~f_\theta=e^{\frac{2}{3}\Phi_A}f_1 f_2 ,
  \end{aligned}
  \end{equation}
  noting that in this case,
  \begin{equation}
\begin{aligned}
F_2&=dC_1 =dC_{1,\phi_1}\wedge  d\phi_1  + dC_{1,\phi_2} \wedge d\phi_2,\\[2mm]
F_4&=dC_3-H_3\wedge C_1
= \Big( dC_{3,\phi_1\phi_2} + C_{1,\phi_1}dB_{2,\phi_2} - C_{1,\phi_2} dB_{2,\phi_1}\Big)\wedge  d\phi_1 \wedge d\theta \wedge d\phi_2.
\end{aligned}
  \end{equation}
  We play the same game as before, re-writing the metric as follows
    \beq\label{eqn:A1Aval}
  \begin{aligned}
    ds^2_2 &= h_{\phi_1} d\phi_1^2 + h_{\phi_2} d\phi_2^2 + h_{\phi_1\phi_2} d\phi_1 d\phi_2 = h_{\phi_1}\bigg(d\phi_1 +\frac{1}{2}\frac{h_{\phi_1\phi_2}}{h_{\phi_1}}d\phi_2\bigg)^2+\bigg(h_{\phi_2}-\frac{1}{4}\frac{h_{\phi_1\phi_2}^2}{h_{\phi_1}}\bigg)d\phi_2^2,
    \end{aligned}
  \eeq
before using the ATD rules given in \eqref{eqn:TD1} and \eqref{eqn:TD2}. Of course, one could instead use the more general forms above. 
Now, performing a T-Duality along $\phi_1$ on \eqref{eqn:GenIIA}, one gets
    \begin{align}\label{eqn:betaredchitdual}
 &   ~~~~~~~~~~~~~~~~~~~~~~~~~~~~~~~~~~~~~~~~~~  \textbf{\underline{IIB (ATD along $\phi_1$)}}\nn\\[2mm]
 &      ds_{10,B}^2=ds_8^2+\delta_1\, d\phi_2^2+\delta_2\,\bigg(d\phi_1 +B_{2,\phi_1} d\theta\bigg)^2,~~~~~~~  \delta_1=\Gamma  \bigg(h_{\phi_2}-\frac{1}{4}\frac{h_{\phi_1\phi_2}^2}{h_{\phi_1}}\bigg),
       ~~~~~\delta_2=\frac{1}{h_{\phi_1} \Gamma },\nn\\[2mm]
&e^{\frac{4}{3}\Phi_B}=\delta_2^{\frac{2}{3}}e^{\frac{4}{3}\Phi_A},
~~~~B_2=\bigg(B_{2,\phi_2}- \frac{1}{2}\frac{h_{\phi_1\phi_2}}{h_{\phi_1}}  B_{2,\phi_1} \bigg) d\phi_2 \wedge d\theta -\frac{1}{2}\frac{h_{\phi_1\phi_2}}{h_{\phi_1}}d\phi_2\wedge d\phi_1 ,~~~~  C_0=C_{1,\phi_1},\nn\\[2mm]
&  C_2=\bigg( C_{3,\phi_1\phi_2}- B_{2,\phi_1} \Big(C_{1,\phi_2} - \frac{1}{2}\frac{h_{\phi_1\phi_2}}{h_{\phi_1}} C_{1,\phi_1} \Big) \bigg) d\theta \wedge d\phi_2 + \bigg(C_{1,\phi_2} -\frac{1}{2}\frac{h_{\phi_1\phi_2}}{h_{\phi_1}}C_{1,\phi_1}\bigg) d\phi_2\wedge d\phi_1.
    \end{align}
Once again, we now write the TsT solution by making the coordinate transformation $\phi_2 \rightarrow \phi_2+\gamma_1 \phi_1$ in \eqref{eqn:betaredchitdual}, before performing the ATD along $\phi_1$ for a second time (to return to a IIA theory)
\begin{adjustwidth}{-1.65cm}{}
\vspace{-0.7cm}
     \begin{align}\label{eqn:TSTgenl}
 &    ~~~~~~~~~~~~~~~~~~~~~~~~~~~~~~~~~~~~~~~~~~~~~~~~~~~~~~~~~~~~~~~~~   \textbf{\underline{TsT }}\nn\\[2mm]
&       ds_{10,A}^2=ds_8^2 +\frac{1}{\delta_2 + \gamma_1^2 \delta_1}\bigg[\delta_1 \delta_2(d\phi_2 -\gamma_1 B_{2,\phi_1}d\theta)^2+ \bigg(d\phi_1 +\frac{1}{2}\frac{h_{\phi_1\phi_2}}{h_{\phi_1}}d\phi_2 +\gamma_1 \Big(B_{2,\phi_2}-\frac{1}{2}\frac{h_{\phi_1\phi_2}}{h_{\phi_1}}B_{2,\phi_1}\Big)d\theta\bigg)^2\,\bigg],\nn\\[2mm]
 &      e^{\frac{4}{3}\Phi_A^{TsT}}= \bigg(\frac{\delta_2}{\delta_2 +\gamma_1^2 \delta_1}\bigg)^{\frac{2}{3}}e^{\frac{4}{3}\Phi_A},~~~~~~~B_2= \frac{\delta_2}{\delta_2+\gamma_1^2\delta_1}(B_{2,\phi_2}d\phi_2 +B_{2,\phi_1}d\phi_1)\wedge d\theta -\frac{\gamma_1 \delta_1}{\delta_2+\gamma_1^2\delta_1}d\phi_2 \wedge d\phi_1,\nn\\[2mm]
&       C_1 = C_{1,\phi_2}d\phi_2 +C_{1,\phi_1}d\phi_1 +\gamma_1 (C_{3,\phi_1\phi_2} + C_{1,\phi_1}B_{2,\phi_2} -C_{1,\phi_2}B_{2,\phi_1})d\theta,\\[2mm]
 &      C_3=\frac{\delta_2}{\delta_2+\gamma_1^2 \delta_1}C_{3,\phi_1\phi_2}d\theta \wedge d\phi_2 \wedge d\phi_1,~~~~~~~~~~~
       \delta_1=\Gamma  \bigg(h_{\phi_2}-\frac{1}{4}\frac{h_{\phi_1\phi_2}^2}{h_{\phi_1}}\bigg),
       ~~~~~~~~~~\delta_2=\frac{1}{h_{\phi_1} \Gamma }.\nn
  \end{align}
  \end{adjustwidth} 
 One can see by observation that fixing $\gamma_1=0$ in \eqref{eqn:TSTgenl} reproduces \eqref{eqn:GenIIA}. 
 
We will return to the TsT discussion in the next section, where we will utilise \eqref{eqn:TSTgenl} to re-derive the NRSZ supersymmetry breaking deformation of the $d=11$ GM background - via a dimensional reduction to IIA, a subsequent TsT transformation and finally uplifting. 
   
Later in this section, we will utilise \eqref{eqn:betaredchitdual} to derive Type IIB theories, keeping track of the supersymmetry under the transformation. We will see that one can indeed get $\mathcal{N}=1$ IIB theories using this method, but one must be careful in order to preserve the $U(1)_R$ R-Symmetry component. To investigate the supersymmetry properly, we must investigate the G-structure description of these IIB solutions, following the ATD. This is where we now turn. 
 
\subsection{G-Structure ATD Transformation}\label{sec:ATDGssumarry}
 We will now discuss the abelian T-Duality of the type IIA G-Structure description, with the detailed derivations given in Appendix \ref{sec:ATDGstructure}. 
  
Motivated by the G-Structure condition given in \eqref{eq:usefulpotentials}, quoted again here for convenience
\begin{equation}\label{eqn:Gconditionrepeated}
  \text{Vol}_4\wedge d_{H^\mathcal{A}_3}(e^{4A-\Phi_\mathcal{A}}\text{Im}\Psi^\mathcal{A}_-)=    \frac{1}{8}( F_6+F_8+F_{10}),
\end{equation}
along with the more schematic form given in \eqref{eqn:schematic}, one can see that under T-duality, the pure spinors will transform in the same manner as the Ramond fields. We then use the T-dual rules \eqref{eqn:TD2} to make the following decomposition of the pure spinors defined by \eqref{eqn:Psi} and \eqref{eqn:PsiConvenient}
 \beq
\begin{gathered}
 \Psi_\pm^\mathcal{A}=\Psi_{\pm_{\perp}}^\mathcal{A} + \Psi_{\pm_{||}}^\mathcal{A}\wedge E^y_\mathcal{A},~~~~~~~~~~E^y_\mathcal{A}=  e^{C^\mathcal{A}}(dy+A_1^\mathcal{A}),\\
\omega^\mathcal{A}=\omega^\mathcal{A}_{\perp}+\omega^\mathcal{A}_{||}\wedge E^y_\mathcal{A},~~~~~~~~~~~~~~~~~
j^\mathcal{A}=j^\mathcal{A}_{\perp}+j^\mathcal{A}_{||}\wedge E^y_\mathcal{A},~~~~~~~~~~~~~
z^\mathcal{A}=z^\mathcal{A}_{\perp}+z^\mathcal{A}_{||}\wedge E^y_\mathcal{A},\\
z^\mathcal{A}_{\perp}=u^\mathcal{A}_{\perp}+i\,v^\mathcal{A}_{\perp},~~~~~~~~~~~~~~~~~~z^\mathcal{A}_{||}=u^\mathcal{A}_{||}+i\,v^\mathcal{A}_{||},
\end{gathered}
\eeq  
 which then leads to the following IIB pure spinors
\begin{equation}\label{eqn:IIBPureSpinors}
e^{-\Phi_\mathcal{B}}  \Psi_\mp^\mathcal{B}=e^{-\Phi_\mathcal{A}}\Big[e^{C^\mathcal{A}}\Psi_{\pm_{||}}^\mathcal{A} + \Psi_{\pm_{\perp}}^\mathcal{A}\wedge (dy-B_1^\mathcal{A})\Big].
\end{equation}
 The factors of $\Phi_\mathcal{A}$ and $\Phi_\mathcal{B}$ were introduced to derive the  IIB G-Structure conditions written solely in terms of IIB quantities, namely
 \begin{subequations}\label{eqn:IIBGconditions}
\begin{align}
d_{H^\mathcal{B}_3}(e^{3A-\Phi_\mathcal{B}}\Psi^\mathcal{B}_-)&=0,\label{eqn:CalibrationformIIB1}\\
d_{H^\mathcal{B}_3}(e^{2A-\Phi_\mathcal{B}}\text{Re}\Psi^\mathcal{B}_+)&=0,\\
d_{H^\mathcal{B}_3}(e^{4A-\Phi_\mathcal{B}}\text{Im}\Psi^\mathcal{B}_+)&=\frac{e^{4A}}{8}*_6\lambda(g). \label{eqn:CalibrationformIIB}
\end{align}
 \end{subequations}
Notice that the roles of $\Psi_{\pm}$ have switched when moving from type IIA to type IIB. This is because the Ramond fields are even in IIA and odd in IIB, so the roles of the pure spinors must swap in the G-structure conditions to account for this - allowing \eqref{eqn:CalibrationformIIB} to have matching dimensionality on both sides of the equation. This third condition of course leads to the IIB analogue of \eqref{eqn:Gconditionrepeated}
\begin{equation}\label{eqn:GconditionrepeatedIIB}
  \text{Vol}_4\wedge d_{H^\mathcal{B}_3}(e^{4A-\Phi_\mathcal{B}}\text{Im}\Psi^\mathcal{B}_+)=    \frac{1}{8}( F_5+F_7+F_{9}).
\end{equation}
We can now derive the IIB pure spinors written in terms of the IIA G-structure forms
\begin{align}
  \Psi_-^\mathcal{B} &=\frac{1}{8}e^{\Phi_\mathcal{B}-\Phi_\mathcal{A}}\bigg(e^{\frac{1}{2}z^\mathcal{A}_{\perp}\wedge \overline{z}^\mathcal{A}_{\perp}}\wedge\Big( e^{C^\mathcal{A}}\omega^\mathcal{A}_{||}+   \omega^\mathcal{A}_{\perp}\wedge (dy-B_1^\mathcal{A})\Big)+\frac{1}{2}e^{C^\mathcal{A}}( z^\mathcal{A}_{\perp}\wedge \overline{z}^\mathcal{A}_{||} -z^\mathcal{A}_{||}\wedge \overline{z}^\mathcal{A}_{\perp})\wedge \omega^\mathcal{A}_{\perp}\bigg),\nn \\
   \Psi_+^\mathcal{B}&=\frac{i }{8} e^{\Phi_\mathcal{B}-\Phi_\mathcal{A}}e^{-ij^\mathcal{A}_{\perp}}\wedge \bigg( e^{C^\mathcal{A}}z^\mathcal{A}_{||}+ z^\mathcal{A}_{\perp}\wedge (dy-B_1^\mathcal{A})+i  e^{C^\mathcal{A}} j^\mathcal{A}_{||}\wedge z^\mathcal{A}_{\perp}\bigg),
\end{align}
with
\beq
e^{\frac{1}{2}z^\mathcal{A}_{\perp}\wedge \overline{z}^\mathcal{A}_{\perp}} = 1+\frac{1}{2}z^\mathcal{A}_\perp\wedge \bar{z}^\mathcal{A}_\perp=1- i\, u^\mathcal{A}_\perp \wedge v^\mathcal{A}_\perp ,~~~~~~~~~~e^{-ij^\mathcal{A}_{\perp}} = 1-i j_\perp^\mathcal{A}-\frac{1}{2}j_\perp^\mathcal{A}\wedge j_\perp^\mathcal{A},
\eeq
giving explicitly
   \beq\label{eqn:IIBfullspinors}
     \begin{aligned}
     \text{Re}\Psi_+^\mathcal{B}&=\frac{1}{8} e^{\Phi_\mathcal{B}-\Phi_\mathcal{A}} \bigg[j^\mathcal{A}_\perp\wedge \Big(e^{C^\mathcal{A}}u^\mathcal{A}_{||}+u^\mathcal{A}_\perp \wedge  (dy-B_1^\mathcal{A}) -e^{C^\mathcal{A}}j^\mathcal{A}_{||}\wedge v^\mathcal{A}_\perp\Big) -\Big(e^{C^\mathcal{A}}v^\mathcal{A}_{||}+e^{C^\mathcal{A}}j^\mathcal{A}_{||}\wedge u^\mathcal{A}_\perp\\[2mm]
  &~~~~~~~~~~~ +v^\mathcal{A}_\perp \wedge (dy-B_1^\mathcal{A}) \Big)+\frac{1}{2}j^\mathcal{A}_\perp\wedge j^\mathcal{A}_\perp\wedge \Big(e^{C^\mathcal{A}} v^\mathcal{A}_{||}+v^\mathcal{A}_\perp\wedge (dy-B_1^\mathcal{A}) + e^{C^\mathcal{A}}j^\mathcal{A}_{||}\wedge u^\mathcal{A}_\perp \Big) \bigg],\\[2mm]
  \text{Im}\Psi_+^\mathcal{B}&=\frac{1}{8}  e^{\Phi_\mathcal{B}-\Phi_\mathcal{A}}\bigg[j^\mathcal{A}_\perp\wedge \Big(e^{C^\mathcal{A}}v^\mathcal{A}_{||}+v^\mathcal{A}_\perp \wedge  (dy-B_1^\mathcal{A}) +e^{C^\mathcal{A}}j^\mathcal{A}_{||}\wedge u^\mathcal{A}_\perp\Big) +\Big(e^{C^\mathcal{A}}u^\mathcal{A}_{||}-e^{C^\mathcal{A}}j^\mathcal{A}_{||}\wedge v^\mathcal{A}_\perp \\[2mm]
  &~~~~~~~~~~~ +u^\mathcal{A}_\perp \wedge (dy-B_1^\mathcal{A}) \Big)-\frac{1}{2}j^\mathcal{A}_\perp\wedge j^\mathcal{A}_\perp\wedge \Big(e^{C^\mathcal{A}} u^\mathcal{A}_{||}+u^\mathcal{A}_\perp\wedge (dy-B_1^\mathcal{A}) - e^{C^\mathcal{A}}j^\mathcal{A}_{||}\wedge v^\mathcal{A}_\perp \Big) \bigg],
     \end{aligned}
     \eeq
where $(j_{\perp}^\mathcal{A},\omega_{\perp}^\mathcal{A})$ are two-forms, $(j_{||}^\mathcal{A},\omega_{||}^\mathcal{A},u_{\perp}^\mathcal{A},v_{\perp}^\mathcal{A})$ are one-forms and $(u_{||}^\mathcal{A},v_{||}^\mathcal{A})$ are zero-forms.

For the IIB solutions under consideration in this chapter, we find $u^\mathcal{A}_{||}=v^\mathcal{A}_{||}=0$ (with  $z^\mathcal{A}_{||}=\bar{z}^\mathcal{A}_{||}=0$) and $u^\mathcal{A}_\perp=u^\mathcal{A},~v^\mathcal{A}_\perp=v^\mathcal{A}$ (with $u^\mathcal{A}$ and $v^\mathcal{A}$ the IIA one forms).  This reduces the pure spinors to
\beq\label{eqn:IIBspinorsSimp}
\begin{aligned}
  \Psi_-^\mathcal{B} &=\frac{1}{8}e^{\Phi_\mathcal{B}-\Phi_\mathcal{A}} e^{\frac{1}{2}z^\mathcal{A}_{\perp}\wedge \overline{z}^\mathcal{A}_{\perp}}\wedge\Big( e^{C^\mathcal{A}}\omega^\mathcal{A}_{||}+   \omega^\mathcal{A}_{\perp}\wedge (dy-B_1^\mathcal{A})\Big),  \\
   \Psi_+^\mathcal{B}&=\frac{i }{8} e^{\Phi_\mathcal{B}-\Phi_\mathcal{A}}e^{-ij^\mathcal{A}_{\perp}}\wedge    z^\mathcal{A}_{\perp}\wedge \Big( (dy-B_1^\mathcal{A})-i    e^{C^\mathcal{A}} j^\mathcal{A}_{||}\Big).
\end{aligned}
\eeq
Comparing these results with \eqref{eqn:Psi} (and \eqref{eqn:SU(3)SU(3)} more generally), from the form of $\Psi_+^\mathcal{B}$, this would suggest the following relations for an $SU(2)$ structure (noting $\frac{1}{2}z^\mathcal{B}\wedge \bar{z}^\mathcal{B}=iv^\mathcal{B}\wedge u^\mathcal{B}$) 
\beq
v^\mathcal{B}\wedge u^\mathcal{B} = -j_\perp^\mathcal{A},~~~~~~~\omega^\mathcal{B}=i\,  e^{\Phi_\mathcal{B}-\Phi_\mathcal{A}}\, z^\mathcal{A}_{\perp}\wedge \Big( (dy-B_1^\mathcal{A})-i    e^{C^\mathcal{A}} j^\mathcal{A}_{||}\Big).
\eeq
The $\Psi_-^\mathcal{B}$ comparison is less clear but the dimensions of the Polyform match as they should.

\subsubsection{Higher form fluxes \& Calibrations}
From the fully expanded pure spinor expressions given in \eqref{eqn:IIBfullspinors}, 
we find in general
 \begin{align}
  8 \,\text{Im}\Psi_{+_{0}}^\mathcal{B}&= e^{\Phi_\mathcal{B}-\Phi_\mathcal{A}}e^{C^\mathcal{A}}u^\mathcal{A}_{||}  ,\nn\\[2mm]
 8 \,\text{Im}\Psi_{+_{2}}^\mathcal{B}&=e^{\Phi_\mathcal{B}-\Phi_\mathcal{A}}  \Big(  e^{C^\mathcal{A}} v^\mathcal{A}_{||} j^\mathcal{A}_\perp -e^{C^\mathcal{A}}j^\mathcal{A}_{||}\wedge v^\mathcal{A}_\perp +u^\mathcal{A}_\perp \wedge (dy-B_1^\mathcal{A}) \Big),\nn\\[2mm]
  8\, \text{Im}\Psi_{+_{4}}^\mathcal{B}&=e^{\Phi_\mathcal{B}-\Phi_\mathcal{A}}   j^\mathcal{A}_\perp\wedge \Big( v^\mathcal{A}_\perp \wedge  (dy-B_1^\mathcal{A}) +e^{C^\mathcal{A}}j^\mathcal{A}_{||}\wedge u^\mathcal{A}_\perp -\frac{1}{2}e^{C^\mathcal{A}} u^\mathcal{A}_{||}  j^\mathcal{A}_\perp   \Big),\nn\\[2mm]
  8  \, \text{Im}\Psi_{+_{6}}^\mathcal{B}&=- \frac{1}{2}e^{\Phi_\mathcal{B}-\Phi_\mathcal{A}}  j^\mathcal{A}_\perp\wedge j^\mathcal{A}_\perp\wedge \Big( u^\mathcal{A}_\perp\wedge (dy-B_1^\mathcal{A}) - e^{C^\mathcal{A}}j^\mathcal{A}_{||}\wedge v^\mathcal{A}_\perp \Big),
 \end{align}
 where we recall from \eqref{eqn:Cmgen} that we can now build the higher form fluxes
 \beq
C_m=8\,e^{4A-\Phi_{\mathcal{B}}}\text{vol}(\text{Mink}_4)\wedge \text{Im}\Psi_{+_{m-4}},
\eeq
namely $(C_4,C_6,C_8,C_{10})$. For our purposes however, with $u^\mathcal{A}_{||}=v^\mathcal{A}_{||}=0$, we find $\Psi_{+_{0}}^\mathcal{B}=0$ (and hence $C_4=0$). 
 
Recalling from the discussions in sections \ref{sec:susyDbranes}, \ref{sec:IIAGstructureN=2} and \ref{sec:N1IIA1}, the calibrations will take the following form 
\begin{align}
 \text{vol(Mink}_4\text{)}\wedge \Psi^{(\text{cal})}_{\text{Dp}} &= 8 e^{4A-\Phi_B}\text{vol(Mink}_4\text{)}\wedge ( \text{Im}\Psi_{+_{p-3}}-B_{\mathcal{B}}^k\wedge \text{Im}\Psi_{+_{p-5}} ) = C_{p+1}-B_{\mathcal{B}}^k\wedge C_{p-1}  ,\nn
\end{align}
with $B_{\mathcal{B}}=B_2^{\mathcal{B}}+B_1^{\mathcal{B}}\wedge dy$ using \eqref{eqn:TD2} (or the $B_2$ defined in \eqref{eqn:betaredchitdualgeneral}). In the case of a D7 brane (with $u^\mathcal{A}_{||}=v^\mathcal{A}_{||}=0$), we find
\begin{align}\label{eqn:calsIIB}
\Psi^{(\text{cal})}_{\text{D7}}&=8e^{4A-\Phi_\mathcal{B}}( \text{Im}\Psi_{+_{4}}-B_{\mathcal{B}}^k\wedge \text{Im}\Psi_{+_{2}} ),\nn\\[2mm]
&=e^{4A-\Phi_\mathcal{A}}  \bigg[  j^\mathcal{A}_\perp\wedge \Big( v^\mathcal{A}_\perp \wedge  (dy-B_1^\mathcal{A}) +e^{C^\mathcal{A}}j^\mathcal{A}_{||}\wedge u^\mathcal{A}_\perp\Big)-B_{\mathcal{B}}^k\wedge  \Big( -e^{C^\mathcal{A}}j^\mathcal{A}_{||}\wedge v^\mathcal{A}_\perp +u^\mathcal{A}_\perp \wedge (dy-B_1^\mathcal{A}) \Big) \bigg].
\end{align}

We have now built the necessary framework to derive multi-parameter type IIB solutions and their corresponding G-structure description via an abelian T-duality of the IIA analysis presented in the previous section. We will now be able to verify the preservation of supersymmetry using the IIB G-structure conditions, in an analogous manner to the IIA solutions. This is the focus of the next few section.
\newpage
\section{Following the IIA $\beta$ Reduction}\label{sec:followingbeta}
One could now use \eqref{eqn:betaredchitdual} to calculate the T-dual solution to the nine-parameter background given in \eqref{eqn:beta-Gen}, but here we restrict to the two-parameter solution of \eqref{eqn:generalresult1}.

 Recall that the most general 11D G-structure forms must preserve the $U(1)_R$ R-Symmetry component given in \eqref{eqn:U(1)} under dimensional reduction, in order to preserve supersymmetry. In the case of the $\beta$ reduction, this corresponds to the condition $\zeta=-\xi$ (with $c=a=0$ and $q\equiv \xi,~v\equiv \zeta$). This derives $\mathcal{N}=1$ solutions for $\xi\neq 0$, promoting to $\mathcal{N}=2$ for $\xi=0$ (where the $SU(2)_R$ component is recovered). The $U(1)_R$ component now becomes in general $(p+s)\chi+(m+u)\phi$, which in turn must be preserved under T-Duality to preserve the supersymmetry in Type IIB. Thus, for a T-duality along $\chi$, we must fix $p+s=0$; and for a T-duality along $\phi$, we must fix $m+u=0$. Recalling that the determinant given in \eqref{eqn:S2breakingdefns} becomes $pu-ms=1$, we must either have $(pu=1,~ms=0)$ or $(pu=0,~ms=-1)$.\\
 We look first at the case where $pu=1$ and $ms=0$ (where $m$ or $s$ is a free parameter and can be set to zero in the IIA case without loss of generality). Following our conventions, we will use $\gamma$ for this free parameter. Hence, we now have the following possibilities (using the $SL(3,\mathds{R})$ transformation given in \eqref{eqn:S2breakingdefns})
 \begin{enumerate}
\item $(p=1,~s=\gamma,~u=1,~m=0)$ with a T-duality along $\chi$
  \beq\label{eqn:1}
d\chi = d\chi+ \xi\, d\beta ,~~~~~~~~~~~~~~~~~~d\phi =d\phi+ \gamma\, d\chi +\zeta \,d\beta ,\\
  \eeq
  with $\gamma=-1,~\zeta=-\xi$ for $\mathcal{N}=1$ (noting $p=-1,~\gamma=1$ corresponds to the same background after $\chi\rightarrow-\chi$ and $\phi\rightarrow -\phi$).
  \item $(u=1,~m=\gamma,~p=1,~s=0)$ with a T-duality along $\phi$
    \beq\label{eqn:2}
d\chi = d\chi+ \gamma d\phi +\xi\, d\beta ,~~~~~~~~~~~~~~~~~~d\phi=d\phi+\zeta \,d\beta ,\\
  \eeq
   with $\gamma=-1,~\zeta=-\xi$ for $\mathcal{N}=1$ (noting as above, $u=-1,~\gamma=1$ corresponds to the same background after $\chi\rightarrow-\chi$ and $\phi\rightarrow -\phi$).
\end{enumerate}
 We now instead fix $pu=0$ and $ms=-1$ (where in this case $\gamma=1$ corresponds to supersymmetry preservation), giving
 \begin{enumerate}[resume]
\item  $(s=1,m=-1,p=0,u=\gamma)$ with a T-duality along $\phi$
\beq \label{eqn:3}
d\chi =   -d\phi +\xi\, d\beta ,~~~~~~~~~~~~~~~~~~d\phi =  d\chi + \gamma d\phi+\zeta d\beta .\\
\eeq
  \item $(m=1,s=-1,u=0,p=\gamma)$ with a T-duality along $\chi$
  \beq\label{eqn:4}
  d\chi = d\phi +\gamma d\chi +\xi\, d\beta ,~~~~~~~~~~~~~~~~~~d\phi =- d\chi +\zeta\,d\beta.
  \eeq
\end{enumerate}

We now see that \eqref{eqn:3} maps to \eqref{eqn:1} by $\gamma\rightarrow-\gamma,~\phi \rightarrow -\chi,~\chi\rightarrow \phi$ and \eqref{eqn:4} maps to \eqref{eqn:2} by $\gamma\rightarrow-\gamma,~\chi \rightarrow -\phi,~\phi\rightarrow \chi$. The consequence of this is that the coordinate transformations of \eqref{eqn:2} followed by a T-duality along $\phi$ is equivalent to the coordinate transformations of \eqref{eqn:4} followed by a T-duality along $\chi$. Thus, calculating the most general form for a T-duality along $\chi$ will automatically contain within it the T-duality along $\phi$, and vice versa. Analogous arguments should of course hold for the $\chi$ and $\phi$ reductions, with ATDs along $(\beta,\phi)$ and $(\beta,\chi)$, respectively.

\subsection{Three-Parameter Families}
We will now perform an ATD on the two-parameter family of solutions given in \eqref{eqn:generalresult1}, where we have already fixed $(p,b,u)=1$. For this discussion we will need to switch on the SUSY parameter $\gamma$ (which plays a trivial role in the Type IIA background and set to zero) as it will become vital in the following analysis. Of course, there are many $\mathcal{N}=0$ solutions contained within the mathematics, but we will focus here on deriving an $\mathcal{N}=1$ background. The type IIA background in question now have the following $U(1)_R$ component
\beq
U(1)_R = (1+s)\chi+(1+m)\phi,~~~~~~~~~~~~~~~\text{with}~~~ms=0.
\eeq
Hence, to preserve the R-Symmetry under ATD, one must either fix $(s\equiv \gamma=-1,m=0)$ and T-dualise along $\chi$, or fix $(m\equiv \gamma=-1,s=0)$ and T-Dualise along $\phi$. We will see in fact that both approaches lead to the same $\mathcal{N}=1$ IIB background, following appropriate transformations.

\subsubsection{Fixing $s\equiv \gamma$ }
We begin with fixing $s\equiv \gamma$ (with $m=0$). From the $U(1)_R$ component given in \eqref{eqn:U(1)}, fixing $(p,b,u)=1,~(a,c,m)=0,~q\equiv \xi,v\equiv \zeta,s\equiv \gamma$, we have
\beq\label{eqn:newU(1)orig}
\begin{aligned}
U(1)_R&= (1+\gamma)\chi +(\xi+\zeta)\beta + \phi.
\end{aligned}
\eeq
We then perform a dimensional reduction along $\beta$, followed by an ATD to IIB. 
We will start with the ATD which will give rise to an $\mathcal{N}=1$ background, which in this case is along $\chi$. We then T-dualise along the other $U(1)$ direction, which is along $\phi$ in this case.  
\begin{itemize}
\item \textbf{Performing an ATD along $\chi$}\\
Using the T-dual formula \eqref{eqn:betaredchitdual}, we derive the following three-parameter family of type IIB solutions
\newpage
\begin{adjustwidth}{-1.25cm}{}
  \begin{align}\label{eqn:ATD1}
 &      ds_{10,B}^2= \frac{1}{X}f_1^{\frac{3}{2}}f_5^{\frac{1}{2}}\sqrt{\Xi}\Bigg[4ds^2(\text{AdS}_5)+f_4(d\sigma^2+d\eta^2) +ds^2(M_3)\Bigg],~~~~~~~e^{2\Phi_\mathcal{B}}=\frac{1}{X^2}\frac{f_5}{f_3}\frac{\Xi^2}{\Pi } ,\nn\\[2mm]
&   ds^2(M_3)=f_2\bigg(d\theta^2  + \frac{1}{\Pi} \sin^2\theta\, d\phi^2\bigg)+\frac{X^2}{f_1^3f_3f_5\Pi }\,D\chi^2,~~~~~~~~~~~~~~\Xi=\Delta+\zeta^2 \frac{f_2}{f_5}\sin^2\theta,
  \nn\\[2mm]
&  C_0=
  \frac{X}{\Xi}\bigg(f_6(1+\xi f_6) +\xi \frac{f_3}{f_5}+\gamma \zeta \frac{f_2}{f_5}\sin^2\theta\bigg),~~~~~~~~~~~~D\chi=d\chi -\frac{1}{X}\big((\gamma \xi -\zeta )f_7 +\gamma f_8 \big)\sin\theta  d\theta,\nn \\[2mm]
&B_2=-\frac{1}{X\Pi} \bigg( f_8+\xi f_7 -\zeta \frac{f_2}{f_3} \Big((\gamma\xi-\zeta)\frac{f_3}{f_5}f_8 -(f_7-f_6f_8)\big(\gamma +(\gamma\xi-\zeta)f_6\big)\Big)\sin^2\theta \bigg) \sin\theta\,d\phi \wedge d\theta \nn\\[2mm]
&~~~~~~~~-  \frac{1}{\Pi} \frac{f_2}{f_3}\Big(\xi (\gamma\xi-\zeta)\frac{f_3}{f_5} +(1+\xi f_6)\big(\gamma+(\gamma\xi-\zeta)f_6\big)\Big)\sin^2\theta  d\phi\wedge d\chi ,\nn\\[2mm]
&  C_2=\frac{1}{\Pi}\bigg( f_7-\gamma \frac{f_2}{f_3}\Big((\gamma\xi-\zeta)\frac{f_3}{f_5}f_8+(f_6f_8-f_7)\big(\gamma+(\gamma\xi-\zeta)f_6\big)\Big)\sin^2\theta \bigg)\sin\theta  d\theta \wedge d\phi\nn\\[2mm]
&~~~~~~~  - \frac{X}{\Pi}  \frac{f_2}{f_3}\Big( (\gamma\xi-\zeta)\frac{f_3}{f_5} +f_6\big(\gamma+(\gamma\xi-\zeta)f_6\big)\Big)\sin^2\theta  d\phi\wedge d\chi, \\[2mm]
& \Delta = (1+\xi f_6)^2 +\xi^2\frac{f_3}{f_5},  ~~~~~~~\Pi= 1+\frac{f_2}{f_3}\bigg(\frac{f_3}{f_5}(\gamma\xi-\zeta)^2 +\big(\gamma +(\gamma\xi-\zeta)f_6\big)^2\bigg)\sin^2\theta,\nn
  \end{align}
  \end{adjustwidth}
  with $F_5=dC_4- H\wedge C_2=0$, and
  \beq
  H=dB_2,~~~~~~~~~F_1=dC_0,~~~~~~~~~F_3=dC_2- C_0 H
  .
  \eeq
  Following the IIA analysis, we consider approaching the $\sigma= 0$ boundary where $\ddot{V}\rightarrow 0$ to leading order. With the boundary condition $\mathcal{R}(\eta) = \dot{V}\Big|_{\sigma=0}$ and warp factors \eqref{eqn:fs}, we find
\beq\label{eqn:C1sourcebetaIIB}
C_0\Big|_{\sigma\rightarrow 0} =X \frac{ \mathcal{R}' (1+\xi \mathcal{R}')  +\frac{1}{2}\gamma\,\zeta \,V'' \mathcal{R} \sin^2\theta  }{(1+\xi \mathcal{R}')^2+\frac{1}{2}\zeta^2 V'' \mathcal{R} \sin^2\theta },~~\Rightarrow~~~~~~C_0\Big|_{\sigma\rightarrow 0}^{\zeta/\sin\theta=0}=X \frac{ \mathcal{R}'    }{ 1+\xi \mathcal{R}' },
\eeq
which reduces for $\zeta=0$ or $\sin\theta=0$, and depends the discontinuous $\mathcal{R}'(\eta)$.  Hence, we find a source term for D7 branes in the $F_1$ Bianchi identity. 
 Taking the derivative, we then find
\begin{align}\label{eqn:betaF2val}
&F_1\Big|_{\sigma\rightarrow 0,\eta=k}^{\zeta/\sin\theta=0}=X  \bigg(\frac{\mathcal{R}' (k)}{1+\xi \mathcal{R}'(k)}-\frac{\mathcal{R}' (k-1)}{1+\xi \mathcal{R}'(k-1)}\bigg)   d\eta  = \frac{ X\big(\mathcal{R}' (k)-\mathcal{R}' (k-1)\big)}{\big(1+\xi \mathcal{R}'(k)\big)\big(1+\xi \mathcal{R}'(k-1)\big)}  d\eta \nn \\[2mm]
&\Rightarrow ~~~~ F_1\Big|_{\sigma\rightarrow 0 }^{\zeta/\sin\theta=0}= - X\sum_{k=1}^{P-1} \frac{2N_k-N_{k+1}-N_{k-1}}{\big(1+\xi  (N_{k+1}-N_k)\big)\big(1+\xi  (N_{k}-N_{k-1})\big)} \delta(\eta-k)\delta(\sigma)d\eta .
\end{align}
In addition, when $(\zeta,\gamma)\neq0$, we find
\begin{align}
B_2\Big|_{\sigma\rightarrow 0} &=- \frac{2 \kappa \,\zeta}{X} \frac{\mathcal{R}-\eta\, \mathcal{R}'}{(\gamma+(\gamma\xi-\zeta)\mathcal{R}')}\sin\theta d\theta\wedge d\phi -  \frac{1+\xi \mathcal{R}'  }{ \gamma+(\gamma\xi-\zeta)\mathcal{R}' }\sin^2\theta\, d\phi\wedge d\chi,\nn\\[2mm]
C_2\Big|_{\sigma\rightarrow 0} &= -2\kappa\,\gamma\frac{  \mathcal{R}-\eta \,\mathcal{R}' }{\gamma+(\gamma\xi-\zeta)\mathcal{R}'}\sin\theta d\theta\wedge d\phi -X \frac{\mathcal{R}' }{(\gamma+(\gamma\xi-\zeta)\mathcal{R}')}\sin^2\theta\, d\phi\wedge d\chi,
\end{align}
which appear to lead to source terms for both D4 and NS5 branes. However, when $(\zeta,\gamma)\neq0$, the $(\theta,\phi)$ sphere is broken, so it is unclear whether these terms will lead to valid integration cycles for the corresponding charges.
We will return to this discussion more thoroughly in the next sub-section, where we investigate the boundary.

  We can preserve the $U(1)_R$ component given in \eqref{eqn:newU(1)orig} by first fixing $\zeta=-\xi$ (as in the IIA case), followed by fixing $\gamma=-1$ under the T-Duality. These conditions together mean $\gamma\xi-\zeta=0$. Due to the transformation given in \eqref{eqn:S2breakingdefns}, such conditions break the $SU(2)_R$ component of the R-Symmetry - leading to an $\mathcal{N}=1$ background. All other solutions are of course $\mathcal{N}=0$ backgrounds.
One can recover the $S^2$ by enforcing $\gamma=\zeta=0$, giving a one-parameter family of $S^2$ preserved solutions (corresponding to the ATD of \eqref{N=0metric} - the $S^2$ preserved $\mathcal{N}=0$ solution studied in the previous chapter). Unlike in the IIA case however, this is $\mathcal{N}=0$ for all values of $\xi$ (including $\xi=0$) - see Figure \ref{fig:IIBtableplot}. 
We present the $\mathcal{N}=1$ solution explicitly in a later subsection. 

  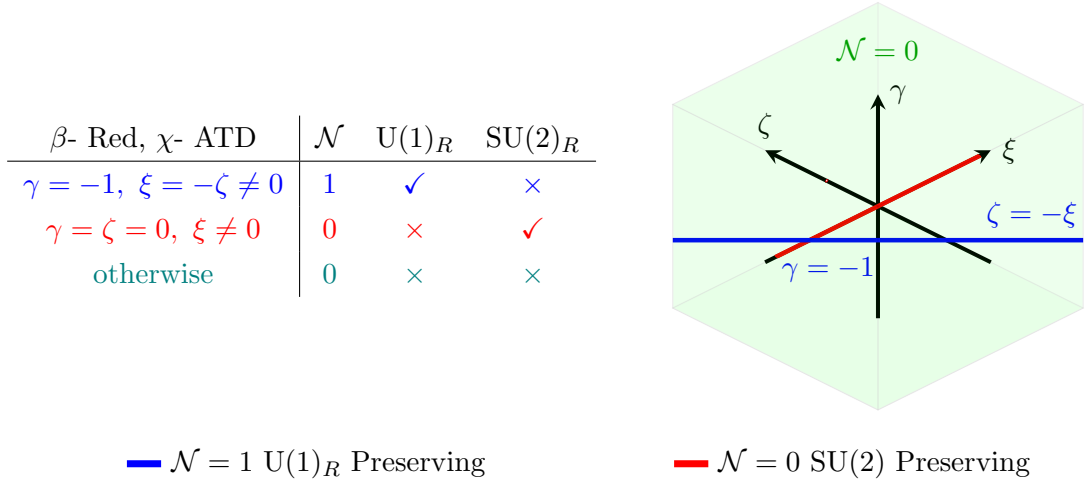
\begin{figure}[H]
\centering  
\subfigure
{
\centering
  \begin{minipage}{0.5\textwidth}
\begin{tabular}{c | c c c c  }
$\beta$- Red, $\chi$- ATD&$\mathcal{N}$&U(1)$_R$&SU(2)$_R$  \\
\hline
$\textcolor{blue}{ \gamma=-1,~\xi=-\zeta\neq 0}$&$ \textcolor{blue}{1}$ &$\textcolor{blue}{\checkmark}$&$\textcolor{blue}{\times}$  \\
$\textcolor{red}{\gamma=\zeta=0,~\xi\neq 0}$&$ \textcolor{red}{0}$ &$\textcolor{red}{\times}$&$\textcolor{red}{\checkmark}$  \\
\textcolor{teal}{otherwise}&$ \textcolor{teal}{0}$ &$\textcolor{teal}{\times}$ &$\textcolor{teal}{\times}$  
\end{tabular}
  \end{minipage}
     \begin{minipage}{.5\textwidth}
    \centering
\begin{tikzpicture}[scale=0.45,y={(1cm,0.5cm)},x={(-1cm,0.5cm)}, z={(0cm,1cm)}]
\draw[-stealth, line width=0.53mm] (0,0,-3.3)--(0,0,3.3) node[right ]{$\mathbf{\gamma}$};
\draw[green!60!black ] (2,2,2) node[above] {$\mathcal{N}=0$};
\begin{scope}[canvas is yx plane at z=0]
\draw[-stealth, line width=0.53mm] (-3.3,0)--(3.3,0) node[right ]{$\xi$};
\draw[-stealth, line width=0.53mm] (0,-3.3)--(0,3.3) node[above ]{$\zeta$};
\clip (-3,-3) rectangle (3,3);
\begin{scope}[cm={0.5,-0.5,  50,50,  (0,0)}]  
\end{scope}
\draw[line width=0.6mm,red](-3,0)--(3,0);
\end{scope}
\begin{scope}[canvas is yx plane at z=-1]
\draw[blue,line width=0.6mm](-3,3)--(3,-3);
 \node[blue ,rotate=0] at (3,-1.5) {$\mathbf{\zeta=-\xi}$};
  \node[blue ,rotate=0] at (-1.6,-0.15) {$\gamma=-1$};
\end{scope}


\draw[fill=green,opacity=0.025] (-3,3,-3) -- (3,3,-3) -- (3,3,3) -- (-3,3,3) -- cycle;
\draw[fill=green,opacity=0.035] (3,-3,-3) -- (3,3,-3) -- (3,3,3) -- (3,-3,3) -- cycle;
\draw[fill=green,opacity=0.05] (3,-3,-3) -- (3,3,-3) -- (-3,3,-3) -- (-3,-3,-3) -- cycle;

\draw[fill=green,opacity=0.05] (-3,-3,-3) -- (3,-3,-3) -- (3,-3,3) -- (-3,-3,3) -- cycle;
\draw[fill=green,opacity=0.05] (-3,-3,-3) -- (-3,-3,3) -- (-3,3,3) -- (-3,3,-3) -- cycle;
\draw[fill=green,opacity=0.05] (-3,-3,3) -- (-3,3,3) -- (3,3,3) -- (3,-3,3) -- cycle;
\begin{scope}[canvas is yz plane at x=1.5]
\coordinate (Origin) at (0,0);
\shade[ball color=red] (Origin) circle (0.05cm);
\end{scope}
\begin{scope}[canvas is yz plane at x=0]
\end{scope}
\end{tikzpicture}
  \end{minipage} 
}
\subfigure
{
\centering
  \begin{minipage}{\textwidth}
    \centering
\begin{tikzpicture}[scale=0.9]
\draw[ blue,line width=0.93mm] (-6,-4.5)--(-5.5,-4.5);
\draw (-5.5,-4.5) node[right]{$ \mathcal{N}=1$ U(1)$_R$ Preserving};
\draw[ red,line width=0.93mm] (2,-4.5)--(2.5,-4.5);
\draw  (2.5,-4.5) node[right]{$ \mathcal{N}=0$ SU(2) Preserving};
\end{tikzpicture}
  \end{minipage}
}
\caption{In the general case, for arbitrary $(\xi,\zeta)$ (in green), the background breaks all SUSY. Along the $(\zeta=-\xi,\gamma=-1)$ line (in blue), the $U(1)_R$-symmetry is preserved, leading to $\mathcal{N}=1$ solutions. Along the $\gamma=\zeta=0$ line (in red), the background preserves $SU(2)$ isometry (descending from the original R-symmetry) with the SUSY completely broken in general. Given the red and blue lines do not intersect (as they are now separated in the $\gamma$ axis), there are no $\mathcal{N}=2$ solutions here - as no background preserves the necessary $SU(2)_R\times U(1)_R$ R-symmetry.}
    \label{fig:IIBtableplot}
\end{figure}

One could split \eqref{eqn:ATD1} into two categories of 2-parameter families, each promoting to the $\mathcal{N}=1$ solution in a different manner
  \begin{itemize}
  \item Case 1: with $\zeta=-\xi$ (or $\zeta=\gamma\xi$) and $\gamma\in\mathds{Z}$ (enhancing to $\mathcal{N}=1$ for $\gamma=-1$)
   \item Case 2: with $\gamma=-1$ and $(\zeta,\xi)\in \mathds{Z}$ (enhancing to $\mathcal{N}=1$ for $\zeta=-\xi$).
  \end{itemize}
  We will however focus on a more general discussion of the full three parameter family, investigating the boundary by the same procedure as the IIA solutions. We will return to this in the next section.
  
It may prove useful to relabel $\gamma\equiv \hat{\gamma}-1$ such that the $\mathcal{N}=1$ solution is recovered for $\hat{\gamma}=0$. However, given that $\gamma=0$ can lead to a recovered $S^2$, we leave our notation as it is. 

\item \textbf{Performing an ATD along $\phi$}\\
Performing the ATD along $\phi$ leads to a two parameter $\mathcal{N}=0$ family of solutions.
 This background will be re-derived as the $\gamma=0$ solution of the upcoming solution given in \eqref{eqn:ATD2} (up to appropriate gauge transformation). 

\end{itemize}

\subsubsection{Fixing $m\equiv \gamma$}
We now investigate the solutions with $m\equiv \gamma$ (and $s=0$). From the $U(1)_R$ component given in \eqref{eqn:U(1)}, fixing $(p,b,u)=1,~(a,c,s)=0,~q\equiv\xi,v\equiv\zeta,m\equiv\gamma$, we have
\beq\label{eqn:newU(1)orig}
\begin{aligned}
U(1)_R&= \chi +(\xi+\zeta)\beta + (1+\gamma)\phi.
\end{aligned}
\eeq
We then perform a dimensional reduction along $\beta$, followed by an ATD to IIB. Once again, we begin with the case which will derive a SUSY background, which is now an ATD along $\phi$. We then once again T-Dualise along the other $U(1)$ direction in question, now $\chi$.  
\begin{itemize}
\item \textbf{Performing an ATD along $\phi$}\\
Using the T-dual formula \eqref{eqn:betaredchitdual}, we now derive the following three-parameter family of type IIB solutions (noting that $C_{3,\phi\chi}=-C_{3,\chi\phi}$) 
\begin{adjustwidth}{-2cm}{}
\vspace{-0.7cm}
    \begin{align}
      & ds_{10,B}^2=\frac{1}{X}f_1^{\frac{3}{2}}f_5^{\frac{1}{2}}\sqrt{\Xi}\Bigg[4ds^2(\text{AdS}_5)+f_4(d\sigma^2+d\eta^2)+ds^2(M_3) \Bigg],~~~~~~~~~~e^{2\Phi_\mathcal{B}}=\frac{1}{X^2}\frac{f_5}{f_2 \sin^2\theta} \frac{\Xi^2}{\Pi}, \nn\\[2mm]
       &ds^2(M_3)= f_2d\theta^2+\frac{1}{\Pi }\bigg( f_3\, d\chi^2+\frac{X^2}{f_1^3 f_2f_5 \sin^2\theta }\,D\phi^2\bigg), ~~~~~~~~~ \Xi=\Delta+ \zeta^2 \frac{f_2}{f_5} \sin^2\theta , \nn\\[2mm]
&C_0=\frac{ X}{ \Xi}\bigg(\gamma  \bigg(f_6 (1+\xi f_6) + \xi \frac{f_3}{f_5}\bigg) +\zeta \frac{f_2}{f_5} \sin^2\theta\bigg),~~~~~~D\phi= d\phi -\frac{1}{X} \big(f_8+(\xi-\zeta \gamma)f_7\big)\sin\theta d\theta, \nn\\[2mm]
& B_2=\frac{1}{X}\bigg[\zeta f_7- \frac{1}{\Pi}\big(f_8+(\xi-\gamma \zeta)f_7\big)\bigg(\zeta  \Big(f_6\big(1+(\xi-\gamma \zeta)f_6 \big) +(\xi-\gamma \zeta)\frac{f_3}{f_5}\Big)-\gamma \frac{f_3}{ f_2 \sin^2\theta}  \bigg) \bigg] \sin\theta d\chi \wedge d\theta\nn
    \end{align}
\end{adjustwidth}  
   \begin{align}\label{eqn:ATD2}
&~~~~~~+ \frac{1}{\Pi}\bigg(\zeta  \Big(f_6\big(1+(\xi-\gamma \zeta)f_6 \big) +(\xi-\gamma \zeta)\frac{f_3}{f_5}\Big)-\gamma \frac{f_3}{ f_2 \sin^2\theta} \bigg)  d\chi\wedge d\phi ,\nn\\[2mm]
 & C_2=
  -  \frac{1}{\Pi}\bigg( (f_7-f_6f_8)\big(1+(\xi-\gamma\zeta)f_6\big)-(\xi-\gamma\zeta)\frac{f_3f_8}{f_5} +\gamma^2 \frac{f_3 f_7}{ f_2 \sin^2\theta} \bigg)  \sin\theta\, d\theta \wedge d\chi\nn\\[2mm]
&  ~~~~~~~~ + \frac{X}{\Pi}  \bigg(f_6\big(1+(\xi-\gamma \zeta)f_6 \big) +(\xi-\gamma \zeta)\frac{f_3}{f_5}\bigg)    d\chi \wedge d\phi, \\[2mm]
 &\Delta= (1+\xi f_6)^2 +\xi^2\frac{f_3}{f_5},~~~~~~~
 \Pi= \big(1+(\xi-\gamma\zeta)f_6\big)^2+(\xi-\gamma\zeta)^2 \frac{f_3}{f_5} +\gamma^2 \frac{f_3 }{f_2 \sin^2\theta } .\nn
    \end{align}  
By the discussion at the start of this section, one can derive this result by fixing $p=\gamma,u=0,m=1,s=-1,b=1,c=a=0,q=\xi,v=\zeta$, performing an ATD along $\chi$ and then redefining $\gamma=-\gamma$ with $\phi\rightarrow \chi, \chi\rightarrow -\phi$. 

 One can map this solution to \eqref{eqn:ATD1} via the set of transformations given in \eqref{eqn:Transformation}, with $k_1=1,k_2=\frac{1}{\gamma}$
 \beq\label{eqn:IIBtrans1}
 \begin{aligned}
 &e^{2\Phi_B}\rightarrow \gamma^2 e^{2\Phi_B},~~~~~~~~~C_0\rightarrow \frac{1}{\gamma} C_0,~~~~~~~C_2\rightarrow \frac{1}{\gamma}C_2,~~~~~~~~B_2\rightarrow B_2-\frac{1}{\gamma}d\phi\wedge d\chi,\\
& \phi\rightarrow \gamma \chi,~~~~~~~~~~~~~~~~~~~\chi\rightarrow -\gamma \phi,~~~~~~~~~\gamma\rightarrow \frac{1}{\gamma}.
 \end{aligned}
 \eeq
Given that the mapping requires $\gamma\rightarrow 1/\gamma$, with the knowledge from the original $SL(3,\mathds{R})$ transformation that $\gamma\in\mathds{Z}$, these steps are a little non-trivial. Nonetheless, we can now say that the $\gamma=0$ case is a new and unique family of $\mathcal{N}=0$ solutions. 

 As before, when $\gamma=-1$ and $\zeta=-\xi$ (where $\xi -\gamma\zeta=0$), one derives an $\mathcal{N}=1$ background. This solution re-derives the $\mathcal{N}=1$ case of \eqref{eqn:ATD1}, 
 via the following transformations
  \begin{equation*}
  \begin{aligned}
  &C_0\rightarrow -C_0,~~~~~~~~~~C_2\rightarrow -C_2,~~~~~~~~~~~~~~~~B_2\rightarrow B_2+d\phi\wedge d\chi,\\
  &\phi\rightarrow -\chi,~~~~~~~~~~~~~\chi\rightarrow \phi.
  \end{aligned}
  \end{equation*}
Noting that now, for $\gamma=-1$, the transformation $\gamma\rightarrow 1/\gamma$ is of course integer on both sides.
An analogous T-Duality along the $U(1)$ of the $S^2$ can generate singularities in the dual description, which is worth keeping in mind here - see the discussion in \cite{Itsios:2013wd}.

\item \textbf{Performing an ATD along $\chi$}\\
Finally, by T-dualising along $\chi$, one derives a background which corresponds to the $\gamma=0$ solution of \eqref{eqn:ATD1} (up to a gauge transformation in $B_2$).

\end{itemize}

\paragraph{Re-deriving the $\gamma$- deformation of NRSZ} We note here that by fixing $\zeta=-\xi=0$ in \eqref{eqn:ATD1}, we have re-derived the TsT background presented in \cite{Nunez:2019gbg}. We will return to this discussion in Section \ref{sec:gammaIIB}.

\subsection{Investigations at the boundary}
We will now investigate the solution given in \eqref{eqn:ATD1} along the $(\sigma,\eta)$ boundary, using the same approach as the IIA solutions. We leave the \eqref{eqn:ATD2} solution for future study. Recall that we give the limits of the warp factors in Appendix \ref{sec:fs}.
\subsection*{Solution 1}
We now investigate the solution given in \eqref{eqn:ATD1}. Here, for general values of $(\eta,\sigma,\theta)$, $(\Xi,\Pi)$ are non-zero and finite. The deformed $S^2$ given by $(\theta,\phi)$ has $\Pi\rightarrow 1$ at the poles, which given the expression for $M_3$ in \eqref{eqn:ATD1}, means it still behaves topologically as an $S^2$. This follows the argument given in Section \ref{sec:boundary}. We now turn to the behaviour at the boundaries, keeping all three parameters non-zero. 
  
  
  \subsubsection{The $\sigma\rightarrow \infty$ boundary}
  At the $\sigma\rightarrow \infty$ boundary, we use utilise \eqref{eqn:Vlimit} and \eqref{eqn:finf} to find
    \begin{equation}
  \begin{gathered}
       ds_{10,B}^2=\frac{\kappa}{X}  \Bigg[4\sigma ds^2(\text{AdS}_5) +\frac{2P}{\pi}\bigg(d\Big(\frac{\pi}{P}\sigma\Big)^2+d\Big(\frac{\pi}{P}\eta\Big)^2+  \sin^2\bigg(\frac{\pi\eta}{P}\bigg) (d\theta^2+ \sin^2\theta\, d\phi^2)\bigg) \\
     ~~~~~~~~~~~~~~~~~~~~~~~~~~~~+\frac{X^2}{4\kappa^2\sigma } \,\bigg(d\chi +\gamma \frac{\kappa P}{X\pi} \Big(\frac{2\pi}{P}\eta-\sin\Big(\frac{2\pi}{P}\eta\Big) \Big)\sin\theta  d\theta\bigg)^2 \Bigg],
       \\
  e^{2\Phi_\mathcal{B}}=\frac{P^2}{X^2\pi^3\mathcal{R}_1^2 }e^{\frac{2\pi \sigma}{P}} ,
~~~~~~~~~~
H_3= - \frac{4\kappa P}{X\pi} \sin^2\Big(\frac{\pi}{P}\eta\Big)\sin\theta\,  d\Big(\frac{\pi}{P}\eta\Big)\wedge d \theta\wedge d\phi  .
  \end{gathered}
  \end{equation}  
  It is immediately clear that $\gamma$ plays a special role here, as it is the only parameter of the three which remains in this limit - breaking the $S^2$. However, $(\theta,\phi)$ still behaves like an $S^2$ topologically for $\gamma\neq0$. When $\gamma=0$, the $S^2$ is recovered, along with the $S^3$ spanned by $(\eta,S^2)$. This makes sense given that the $\mathcal{N}=1$ solution (with a broken $SU(2)$) is derived when $\gamma=-1$.   The background tends to a stack of NS5 branes. The $S^1$ shrinks in this limit, in contrast to the IIA cases where it grows alongside the AdS$_5$ - this makes intuitive sense following the T-Duality. Again we find $ \widehat{\kappa}\,P$ NS5 branes, as follows (with $2\kappa=\pi \widehat{\kappa} X$)
  \beq
Q_{NS5}=-\frac{1}{(2\pi)^2}\int_{(\eta,\theta,\phi)}H_3= \widehat{\kappa}\,P.
\eeq
When $\gamma=0$ (or $\gamma\neq0$ close to one of the poles of the deformed $S^2$), we can redefine $(x^\mu,\rho)=(\frac{4\kappa}{X}\sigma)^{-\frac{1}{2}}(\tilde{x}^\mu,\tilde{\rho})$ and $\chi = (\frac{4\kappa}{X}\sigma)^{\frac{1}{2}}\tilde{\chi }$, which to leading order in $\sigma$, gives $\frac{4\kappa}{X}\sigma ds^2(AdS_5)+ \frac{X}{4\kappa  }\frac{1}{\sigma}d\chi^2\rightarrow \eta_{\mu\nu}d\tilde{x}^\mu d\tilde{x}^\nu +d\tilde{\rho}^2+d\tilde{\chi}^2=ds^2(\text{Mink}_6)$. Introducing the coordinate $\bar{r}=e^{-\frac{\pi}{P}\sigma}$, we find to leading order
\beq
ds^2=ds^2(\text{Mink}_6) +\frac{2P\kappa}{X \pi \bar{r}^2}\Big(d\bar{r}^2 +\bar{r}^2 ds^2(S^3)\Big),~~~~e^{ \Phi_\mathcal{B}}=\frac{P }{X \pi^\frac{3}{2}\mathcal{R}_1 }\bar{r}^{-1} ,
\eeq
corresponding to the near horizon limit of a spherically symmetric stack of NS5 branes in flat space, see \eqref{eqn:NS5metrics}. When $\gamma\neq 0$, and away from the poles of the deformed $S^2$, the behaviour is not so clean - and the $\frac{1}{\sigma}$ contribution to $d\theta$ now dominates. 
  \subsubsection{The $\eta=0$ boundary, with $\sigma\neq0$}
   At $\eta=0$ with $\sigma\neq 0$, using the warp factor limits in \eqref{eqn:feq}, gives 
%
        \begin{align}
  &   f_4(d\sigma^2+d\eta^2)+ds^2(M_3)=\frac{1}{\sigma^2}\bigg[   \frac{2|\dot{f}|}{ f}\bigg(d\sigma^2+d\eta^2 +\eta^2 (d\theta^2+\sin^2\theta d\phi^2)\bigg)+\frac{X^2}{4\kappa^2}D\chi^2\bigg], \nn\\[2mm]
&D\chi=d\chi + \frac{2\kappa\,\eta^3 |\dot{f}|}{X\sigma^2} \Big[2(\gamma \xi-\zeta) +\frac{\gamma}{f}\bigg(2+\frac{|\dot{f}|}{f}\bigg) \Big]\sin\theta d\theta.
  \end{align}  
    In this limit we notice that the $\mathds{R}^3$ of $(\eta,S^2)$ is recovered only when all three parameters are zero.

  \paragraph{The $\eta=P$ boundary, with $\sigma\neq 0$} This boundary is qualitatively equivalent to the $\eta=0$ boundary.
  
    \subsubsection{The $\sigma=0$ boundary, with $\eta\in (k,k+1)$}
 At $\sigma=0,~\eta\in (k,k+1)$, recall that along the $\sigma=0$ boundary, $\ddot{V}=0$ to leading order. We now see
     \beq\label{eqn:lkandlkhat}
     \begin{gathered}
     \Xi \rightarrow l_k^2 +\frac{1}{2}\zeta^2 \mathcal{R}V'' \sin^2\theta,~~~~~~~~~~~~~\frac{\Pi}{f_2}\rightarrow  \frac{\mathcal{R}}{2\sigma^2 V''}\hat{l}_k^2 \sin^2\theta,\\ 
     l_k\equiv 1+\xi (N_{k+1}-N_k),~~~~~~~~~~~~~~\hat{l}_k \equiv \gamma+(\gamma \xi-\zeta )(N_{k+1}-N_k) =\gamma \, l_k-\zeta (N_{k+1}-N_k),
     \end{gathered}
     \eeq
     where
   \beq
   f_4(d\sigma^2+d\eta^2) +\frac{f_2 }{\Pi}\sin^2\theta\, d\phi^2  \rightarrow   \frac{2V''}{\mathcal{R}}\bigg(d\eta^2+\Big(d\sigma^2 +\frac{\sigma^2}{\hat{l}_k^2}d\phi^2\Big)\bigg).
     \eeq
     So there is a $\mathds{R}^2/\mathds{Z}_{\hat{l}_k}$ orbifold singularity in $(\sigma,\phi)$, with $\hat{l}_k$. Recall, in the IIA case, this orbifold singularity was over $(\sigma,\chi)$, with $l_k$ - see \eqref{eqn:orbifold1}. So in moving from IIA to IIB, $\chi$ has been replaced by $\phi$, and the orbifold singularity made a little more general ($l_k\rightarrow \hat{l}_k$). 
     It is worth noting that the case of $(\gamma,\zeta)=0$ must be treated separately because $\hat{l}_k=0$ (this is the $S^2$ preserving condition). Now we turn to the remaining internal metric component
     \beq
          \begin{gathered}
       f_2 d\theta^2  +\frac{X^2}{f_1^3f_3f_5\Pi}\,D\chi^2 \rightarrow \frac{2\mathcal{R}V''}{2\mathcal{R}V''+(\mathcal{R}')^2}d\theta^2 + \frac{2\mathcal{R}V''+(\mathcal{R}')^2}{4\kappa^2\mathcal{R}^2}\frac{X^2}{ \hat{l}_k^2 \sin^2\theta}D\chi^2,\\
                D\chi\rightarrow d\chi +\frac{2\kappa}{X}\bigg( \gamma\, \eta +\frac{\mathcal{R} }{2\mathcal{R}V''+(\mathcal{R}')^2} \Big(2(\gamma\xi-\zeta)\mathcal{R} V'' -\gamma \mathcal{R}' \Big)\bigg)\sin\theta d\theta    .
               \end{gathered}
     \eeq
    Notice that for the $\mathcal{N}=1$ condition, $\gamma\xi-\zeta=0$, we have $\hat{l}_k^2=\gamma^2=1$. It is worth pointing out that although $(\gamma\xi-\zeta)=0$ is satisfied by the $\mathcal{N}=1$ solution, the reverse isn't always true - for instance, $\gamma=0,\zeta=0$ would satisfy the condition without preserving supersymmetry.

    \subsubsection{The $\sigma=0, \eta=0$ boundary}
    At $\sigma=0,~\eta=0$, we again adopt the coordinate change \eqref{eqQdef}, where
    \beq
    \Xi\rightarrow l_0^2,~~~~~~~~~~ \Pi \rightarrow 1+ \hat{l}_0^2 \,\text{cot}^2\alpha\sin^2\theta,~~~~~~~~~~~~l_0=1+\xi N_1,~~~~~~~~~~~\hat{l}_0 =\gamma\,l_0-\zeta  N_1 . 
    \eeq
 We note in general
  \begin{align}
 & f_4(d\sigma^2+d\eta^2) +ds^2(M_3) \rightarrow \frac{2Q}{N_1}\bigg(dr^2+r^2d\alpha^2+r^2\cos^2\alpha \bigg(d\theta^2 +\frac{\sin^2\theta \,d\phi^2}{1+ \hat{l}_0^2\, \text{cot}^2\alpha\sin^2\theta}\bigg)\bigg) \nn\\[2mm]
& ~~~~~~ +\frac{X^2D\chi^2}{4\kappa^2  r^2 \sin^2\alpha ( 1+\hat{l}_0^2\, \text{cot}^2\alpha \sin^2\theta)},~~~~~~~~~D\chi=d\chi+\frac{4\kappa}{X} (\gamma\xi-\zeta)Qr^3 \cos^3\alpha \sin\theta d\theta.
  \end{align}
So we see, as one would expect following a T-Duality, the internal space no longer vanishes as $\mathds{R}^5/\mathds{Z}_{l_0}$. The orbifold singularity here is also less clean.
Note that for the $\mathcal{N}=1$ case, $D\chi\rightarrow d\chi$ and again, $\hat{l}_0^2=1$.
      \paragraph{The $\sigma=0,~\eta=P$ boundary} This boundary is qualitatively equivalent to the $(\sigma,\eta)=0$ boundary, with $\mathds{R}^5/\mathds{Z}_{l_{P-1}}$.  
 \subsubsection{The $\sigma=0$ boundary, with $\eta=k$}
 At the $\sigma=0$ boundary with $\eta=k$,  we use the coordinate change $(\eta=k-r \cos\alpha,~\sigma=r \sin\alpha)$ for $r\sim 0$ (where $0<k<P$ and $k\in \mathds{Z}$) and \eqref{eqn:fsfork}, as in the IIA case. We recall in this limit that the $f_2/f_5$ term dominates away from the pole of the deformed $S^2$ (in the expressions for $\Xi$ and $C_0$ given in \eqref{eqn:ATD1}), unless we fix $\zeta=0$. Let us first investigate the $\zeta=0$ case, before switching on $\zeta$ and investigating the boundary both away from the pole and approaching it.
 \begin{itemize}
 \item \underline{$\zeta=0$}:  Let us first investigate the $\zeta=0$ case. We first look at the case where $\gamma=0$, which corresponds to the $S^2$ preserved case, before switching $\gamma$ on. We use \eqref{eqn:fsfork} with $r=z^2$. 
  \begin{itemize}
  \item \underline{$\gamma= 0$}: In this case, we have a preserved $S^2$, with
    \begin{align}
  & \Xi\rightarrow \Delta_k,~~~~~~~~\Pi\rightarrow1,~~~~~~~~\Delta_k\equiv \Big(1+ \xi g(\alpha)\Big)^2+ \frac{1}{4}\xi^2 b_k^2 \sin^2\alpha ,
  \end{align}
  which, postponing the gauge transformation in $B_2$ for the moment, leads to
     \begin{adjustwidth}{-0.5cm}{}
       \vspace{-0.7cm}
    \begin{align} 
 &      ds_{10,B}^2= \frac{2\kappa}{X}\sqrt{\frac{N_k}{b_k}}\sqrt{\Delta_k} \,\,z\Bigg[4ds^2(\text{AdS}_5)+ ds^2(S^2)+\frac{b_k}{N_k}\Big(4dz^2+z^2 d\alpha^2\Big) +\frac{X^2}{ 4\kappa^2}\frac{1}{z^4\sin^2\alpha}\,d\chi^2\Bigg],\nn\\[2mm]
&  
  e^{\Phi_\mathcal{B}}=\frac{ 2\Delta_k}{X b_k \sin\alpha }  ,~~~~~~~~  C_0= \frac{X}{\Delta_k}\bigg(g(\alpha)\Big(1+\xi g(\alpha)\Big) +\frac{1}{4}\xi \,b_k^2\sin^2\alpha \bigg), \nn\\[2mm]
& B_2= -\frac{2\kappa}{X } (k+\xi N_k )\text{vol}(S^2) ,~~~~~~~~~ C_2=  -2\kappa N_k \text{vol}(S^2).
  \end{align}
     \end{adjustwidth}
    Calculating the D7 charge, we find
   \beq\label{eqn:D7charge1}
Q_{D7}=- \int_\alpha F_1=C_0\Big|_{\alpha=0}^{\alpha=\pi}=\frac{X}{l_kl_{k-1}}(2N_k-N_{k+1}-N_{k-1}),
 \eeq
 which gives a rational charge analogous to the D6 case in IIA. However, it is worth observing from the form of the above metric that there is no orbifold singularity in this case. The rational quantization comes from the $\Delta_k^{-1}$ factor in $C_0$, which is inherited from the IIA solution - even in the absence of an orbifold. We will return to this observation later in this section. Let us first investigate the D5 charge here, and the effects of the gauge transformations of $B_2$. We first take (using $2\kappa=\pi \widehat{\kappa} X$)
  \begin{align}
  &  B_2\rightarrow B_2+\frac{2\kappa}{X} k\text{Vol}(S^2), ~~~~~~~~~~~~~~~B_2= -\frac{2\kappa}{X }  \xi N_k  \text{vol}(S^2),\nn \\[2mm]
   &Q_{D5}=-\frac{1}{(2\pi)^2} \int_{(\alpha,S^2)} \hat{F}_3 =- \frac{1}{(2\pi)^2} \int_{\phi=0}^{\phi=2\pi}\int_{\theta=0}^{\theta=\pi} \Big[C_2-C_0  B_2\Big]_{\alpha=0}^{\alpha=\pi} =X\bigg(\frac{ N_k}{l_k}-\frac{N_{k-1}}{l_{k-1}}\bigg),
  \end{align}
which again gives a rational charge (analogous to the D4 branes). As in the IIA case, we can include an additional term in the gauge transformation
  \begin{align}
  & B_2\rightarrow B_2+\frac{2\kappa}{X} (k+\xi N_k)\text{Vol}(S^2), ~~~~~~~~~~~~~~~B_2=0,\nn \\[2mm]
& Q_{D5} =-\frac{1}{(2\pi)^2} \int_{(\alpha,S^2)} \hat{F}_3 = \widehat{\kappa} X(N_k-N_{k-1}), 
  \end{align}
which then leads to integer quantization. Given we are using the semi-circular contour given in Figure \ref{fig:semicircularcontour}, we assume $(\alpha,S^2)$ to be a valid closed cycle.
  \item \underline{$\gamma\neq 0$}: Let us switch on $\gamma$, where we now have $\Pi_k\rightarrow \gamma^2\Delta_k$, hence
  \begin{align}
  & \Xi\rightarrow \Delta_k,~~~~~~~~\Pi\rightarrow \gamma^2 \frac{N_k\sin^2\theta}{b_k r\sin^2\alpha}\Delta_k,~~~~~~~~\Delta_k\equiv \Big(1+ \xi g(\alpha)\Big)^2+ \frac{1}{4}\xi^2 b_k^2 \sin^2\alpha ,\nn
  \end{align}
  we then see (again postponing the gauge transformation for $B_2$)
  \begin{align}
  &ds^2= \frac{2\kappa}{X} \sqrt{\frac{N_k}{b_k}}\sqrt{\Delta_k}\, z \Big[4ds^2(AdS_5)+ ds^2(\mathds{B}_5)\Big], ~~~~~~~e^{2\Phi_{\mathcal{B}}} =\frac{4z^2 \Delta_k}{\gamma^2X^2b_k N_k \sin^2\theta}, \nn\\[2mm]
  &ds^2(\mathds{B}_5) = d\theta^2 +\frac{4b_k}{N_k}\bigg(dz^2+\frac{1}{4}z^2 \Big(d\alpha^2 + \frac{\sin^2\alpha}{\gamma^2 \Delta_k}d\phi^2\Big) +\frac{X^2}{ z^2}\frac{(d\chi+\mathcal{A}_k)^2}{16\gamma^2 \kappa^2\Delta_k \sin^2\theta} \bigg),\nn\\[2mm]
  &
 B_2=-\frac{1}{\gamma}d\phi\wedge d\chi , ~~~~~~~C_0=\frac{X}{\Delta_k}\Big(g(\alpha)\Big(1+\xi g(\alpha)\Big) +\frac{1}{4}b_k^2 \xi \sin^2\alpha\Big),\nn\\[2mm]
  &C_2=\frac{1}{\Delta_k} \bigg[2\kappa \bigg(\Big(1+\xi g(\alpha)\Big)\Big(k g(\alpha) -N_k\Big)+\frac{1}{4}b_k^2 \xi k \sin^2\alpha\bigg) \sin\theta d\theta \wedge d\phi \nn\\[2mm]
  &~~~~~~~~~~~~~~~~~~~-\frac{X}{\gamma} \bigg(g(\alpha)\Big(1+\xi g(\alpha)\Big) +\frac{1}{4} b_k^2 \xi \sin^2\alpha \bigg) d\phi\wedge d\chi \bigg],\nn\\[2mm]
 & 
 \mathcal{A}_k =\frac{2\kappa\,\gamma}{X} \big( k+  \xi N_k\big)\sin\theta  d\theta .
  \end{align}
In a similar manner to the IIA case, we find the presence of a spindle here. The form of $C_0$ remains the same as above, so calculating the D7 charge re-derives \eqref{eqn:D7charge1}.
 We could calculate D5 charge by integrating over $(\alpha,\theta,\phi)$, which depending on the gauge transformation of $B_2$, would derive either rational or integer charge (equal to the D4 counterparts in IIA, given in \eqref{eqn:D4orig} and \eqref{eqn:D4orig2}). However, although $(\alpha,\phi)$ describes a spindle, there is no $\sin\theta$ factor out front - hence, this does not appear to be a valid (closed) cycle. We therefore conclude there is no D5 charge here.


\end{itemize}
 
  \item \underline{$\zeta\neq0$}: we now switch on $\zeta$, where the $f_2/f_5$ term dominates unless we are at a pole of the deformed $S^2$ (where $\sin\theta=0$). We first assume then that we are indeed away from the poles.
  \paragraph{Away from a pole: } In this case, we once again expand $(\eta=k-r \cos\alpha,~\sigma=r \sin\alpha)$ in small $r$, finding to leading order 
\begin{align}
&\Xi\rightarrow \frac{b_k\zeta^2 N_k\sin^2\theta}{4r},~~~~\Pi\rightarrow \frac{N_k\sin^2\theta}{b_k r\sin^2\alpha}\Pi_k(\alpha),~~~~~\Pi_k(\alpha=0) = \hat{l}_{k-1}^2,~~~\Pi_k(\alpha=\pi) =\hat{l}_k^2  ,\nn\\[2mm]
& \Pi_k\equiv (\gamma\xi-\zeta)^2 \frac{b_k^2}{4}\sin^2\alpha +\Big(\gamma+(\gamma\xi-\zeta) g(\alpha)\Big)^2.
\end{align}
We make the same coordinate transformation as the IIA case, $r=z^2$, giving    
    \begin{align}\label{eqn:zeq}
 &      ds_{10,B}^2=\frac{|\zeta| \kappa}{X} N_k\sin\theta \Big(4ds^2(\text{AdS}_5)+ds^2(\mathds{B}_5)\Big),\nn\\[2mm]
&       ds^2(\mathds{B}_5) =d\theta^2  + \frac{4b_k}{N_k }\bigg( dz^2+\frac{z^2}{4} \Big(d\alpha^2 +   \frac{\sin^2\alpha  }{\Pi_k}\,  d\phi^2 \Big)+ \frac{X^2}{z^2\sin^2\theta}\frac{ (d\chi +\mathcal{A}_k )^2 }{16 \,\kappa^2 \Pi_k}  \bigg),\nn\\[2mm]
&  e^{2\Phi_\mathcal{B}}=\frac{\zeta^4 N_k b_k\sin^2\theta  }{ 4X^2z^2 \Pi_k},~~~~~~~~~~~~~~\mathcal{A}_k =\frac{2\kappa}{X} \big((\gamma \xi -\zeta )N_k+\gamma k\big)\sin\theta  d\theta ,~~~~~~~C_0=  \frac{ \gamma }{\zeta}X\nn\\[2mm]
&B_2=\frac{1}{\Pi_k} \bigg(\Big(1+\xi g(\alpha)\Big)\Big(\gamma+(\gamma\xi-\zeta)g(\alpha)\Big) +\frac{1}{4}\xi b_k^2 (\gamma \xi-\zeta)\sin^2\alpha\bigg) \nn\\[2mm]
&~~~~~~~~~~~~~~~~~~~~~~~~\times \bigg(\frac{2\kappa}{X} \Big(\gamma\,k +(\gamma \xi-\zeta)N_k\Big)\sin\theta d\theta \wedge d\phi -d\phi\wedge d\chi\bigg), \\[2mm]
&   C_2=\frac{1}{\Pi_k } \bigg(X\Big( (\gamma\xi-\zeta)\frac{1}{4}b_k^2\sin^2\alpha +g(\alpha)\big(\gamma+(\gamma\xi-\zeta)g(\alpha)\big)\Big)  d\chi\wedge d\phi  ~\nn  \\[2mm]
&~~~~~~~~~~~~~~~~~~+2\kappa\gamma \Big((\gamma\xi-\zeta)\, \frac{ k}{4}\,  b_k^2\sin^2\alpha+( k\,g(\alpha)-N_k)\big(\gamma+(\gamma\xi-\zeta)g(\alpha)\big)\Big) \sin\theta  d\theta \wedge d\phi\bigg).\nn
  \end{align} 
We note that $\mathds{B}_5$ corresponds to the ATD of the $(\theta,~z,~\mathds{B}_3)$ components of \eqref{eqn:B3eq} (using the appropriate gauge transformation for $B_2$).  Now, we find
  \begin{align}
 &    ds^2(\mathds{B}_5) \Big|_{\alpha\sim0}= d\theta^2  + \frac{4b_k}{N_k }\bigg( dz^2+\frac{z^2}{4} \Big(d\alpha^2 +   \frac{\alpha^2  }{\hat{l}_{k-1}^2}\,  d\phi^2 \Big)+ \frac{X^2}{z^2\sin^2\theta}\frac{ (d\chi +\mathcal{A}_k )^2 }{16 \,\kappa^2 \hat{l}_{k-1}^2}  \bigg),\\[2mm]
  &    ds^2(\mathds{B}_5) \Big|_{\alpha\sim\pi}= d\theta^2  + \frac{4b_k}{N_k }\bigg( dz^2+\frac{z^2}{4} \Big(d\alpha^2 +   \frac{(\pi-\alpha)^2  }{\hat{l}_{k}^2}\,  d\phi^2 \Big)+ \frac{X^2}{z^2\sin^2\theta}\frac{ (d\chi +\mathcal{A}_k )^2 }{16 \,\kappa^2 \hat{l}_{k}^2}  \bigg).\nn
  \end{align}
  Note the presence of a $\sin\theta$ out the front of the whole metric! We observe that $(\alpha,\phi)$ form a $\mathds{W}\mathds{C}\mathds{P}^1_{[\hat{l}_{k-1},\hat{l}_k]}$, with Euler characteristic
\beq
\chi_E=\frac{1}{2\pi}\int_{\mathds{W}\mathds{C}\mathds{P}^1_{[\hat{l}_{k-1},\hat{l}_k]}}R \,\text{Vol}_2 = 2-\bigg(1- \frac{1}{|\hat{l}_{k-1}|}\bigg)-\bigg(1-\frac{1}{|\hat{l}_{k}|}\bigg).
\eeq

In this limit, $C_2-C_0 B_2=\frac{1}{\zeta}d\phi\wedge d\chi-\frac{\gamma}{\zeta}\tilde{B}\text{Vol}(S^2)$, following the gauge transformation, $B_2\rightarrow B_2+2\kappa \tilde{B} \text{Vol}(S^2)$. It is reasonable to conclude that there are no D5 branes present for $\zeta\neq0$, noting $\hat{F}_3=d(C_2\wedge e^{-B_2})$. Note, due to the semi-circular contour we are using, which begins in the $(k-1)$th cell and crosses into the $k$th cell at $\alpha=\pi/2$, one could rescue the charge by fixing $\tilde{B}=\frac{\zeta}{\gamma}N_k$. However, as in the previous case, the $(\alpha,\theta,\phi)$ cycle does not appear to be a valid one, and we conclude no D5 charge.

\paragraph{Approaching the pole:} As in the IIA case, we approach the pole $(\sigma=0,\eta=k,\sin\theta=0)$ using
\beq
\eta=k-\bar{r} \cos\alpha \sin^2\mu,~~~~~\sigma=\bar{r} \sin\alpha \sin^2\mu,~~~~~~\sin\theta = 2\sqrt{\frac{b_k\bar{r}}{N_k}}\cos\mu,
\eeq
recalling that $r= \bar{r} \sin^2\mu$, and expanding about $\bar{r}=0$. Now
 \begin{align}
&\sin^2\alpha\sin^2\mu\,\Pi\rightarrow \sin^2\alpha \sin^2\mu+4\cos^2\mu\,\Pi_k,~~~~~~~~~~ \sin^2\mu \,\Xi \rightarrow \tilde{\Xi}_k =\Delta_k \sin^2\mu + \zeta^2 b_k^2 \cos^2\mu,\nn\\[2mm]
&\Delta_k\equiv \Big(1+ \xi g(\alpha)\Big)^2+ \frac{1}{4}\xi^2 b_k^2 \sin^2\alpha ,
~~~~~~~~~~~ B_2\rightarrow B_2+\frac{4\kappa b_k \cos^2\mu}{X N_k}(k+\xi N_k)d\bar{r}\wedge d\phi,
 \end{align}
 giving
 \begin{adjustwidth}{-1.75cm}{}
 \begin{align}\label{eqn:B5eq}
 &\frac{ds^2}{2\kappa \sqrt{N_k}}=\frac{\sqrt{\tilde{\Xi}_k}}{X}\Bigg[\frac{4}{\sqrt{\frac{b_k}{\bar{r}}}}ds^2(AdS_5)+\frac{\sqrt{\frac{b_k}{\bar{r}}}}{N_k}ds^2(\mathds{B}_5)\Bigg],~~~e^{\Phi_{\mathcal{B}}} =\frac{2\,\tilde{\Xi}_k}{Xb_k\sin\mu \sqrt{\sin^2\alpha\sin^2\mu +4\cos^2\mu\,\Pi_k}},\nn\\[2mm]
& ds^2(\mathds{B}_5) = d\bar{r}^2 +4\bar{r}^2 \bigg[d\mu^2+\frac{1}{4}\sin^2\mu\bigg( \,d\alpha^2 +\frac{\sin^2\alpha}{\Pi_k+\frac{1}{4}\sin^2\alpha\tan^2\mu} d\phi^2\bigg)\bigg]\nn\\[2mm]
&~~~~~~~~~~~~~~~~~~+\frac{1}{\bar{r}}\frac{X^2 N_k}{4\kappa^2b_k\sin^2\mu} \frac{(d\chi+\mathcal{A}_k)^2}{(\sin^2\alpha\sin^2\mu+4\,\cos^2\mu\,\Pi_k)},\nn\\[2mm]
&\mathcal{A}_k=\frac{4\kappa b_k \cos\mu}{X N_k}\Big(\gamma k+(\gamma \xi-\zeta)N_k\Big)\Big(\cos\mu\,d\bar{r}-2\bar{r} \sin\mu\,d\mu\Big),\nn\\[2mm]
&B_2=\frac{4\cos^2\mu}{\sin^2\alpha\sin^2\mu +4\cos^2\mu\,\Pi_k} \bigg(\Big(1+\xi g(\alpha)\Big)\Big(\gamma+(\gamma\xi-\zeta)g(\alpha)\Big)+\frac{1}{4}\xi b_k^2 (\gamma\xi-\zeta)\sin^2\alpha\bigg)\nn\\[2mm]
&~~~~~~~~~~~~~~~\times \bigg(\frac{4\kappa b_k}{X N_k} \cos^2\mu \Big(\gamma k +(\gamma \xi-\zeta )N_k\Big) d\bar{r}\wedge d\phi -d\phi \wedge d\chi\bigg), \nn\\[2mm]
&C_0=\frac{X}{\tilde{\Xi}_k}\bigg(g(\alpha)\Big(1+\xi g(\alpha)\Big)\sin^2\mu +\frac{1}{4}b_k^2 \Big(4\gamma \zeta\,\cos^2\mu +\xi \sin^2\alpha\sin^2\mu\Big)\bigg),\nn\\[2mm]
&C_2=\frac{4}{\sin^2\alpha\sin^2\mu +4\cos^2\mu\,\Pi_k} \bigg[\frac{b_k\kappa \cos^2\mu}{N_k} \bigg(\gamma b_k^2k (\gamma\xi-\zeta) \cos^2\mu \sin^2\alpha +4\gamma k (\gamma \xi-\zeta)g(\alpha)^2\, \cos^2\mu \nn\\[2mm]
&~~~~~~~ -N_k (\sin^2\alpha\sin^2\mu +4\gamma^2 \cos^2\mu) +4\gamma \,g(\alpha)\,\cos^2\mu \Big(\gamma k -(\gamma\xi-\zeta)N_k\Big)\bigg)d\bar{r}\wedge d\phi  \nn\\[2mm]
&~~~~~~~-X \cos^2\mu \bigg(\frac{1}{4}b_k^2\sin^2\alpha (\gamma\xi-\zeta) +g(\alpha) \Big(\gamma+(\gamma\xi-\zeta)g(\alpha)\Big)\bigg)d\phi\wedge d\chi\bigg].
 \end{align}
 \end{adjustwidth}
 So again, we see that the $S^2$ has inherited orbifold singularities. On first sight, these appear more complicated than the spindle found in IIA (in the general case at least). This is because $\mathds{B}_5$ in \eqref{eqn:B5eq} is the ATD of the $(\rho,~\mathds{B}_4)$ components of \eqref{eqn:B4eq} (using the appropriate gauge transformation in $B_2$). One can then conclude that $z$ and $\rho$, in \eqref{eqn:zeq} and \eqref{eqn:B5eq} respectively, aren't a part of the orbifolds themselves but the cones over them. 
 It would be nice to understand this further topologically in the future.  
 Now we calculate the charge of the D7 branes at $\mu=\frac{\pi}{2}$ on one of the components of the `spindle like' space, $\alpha$, as follows
 \beq\label{eqn:D7charge}
Q_{D7}=- \int_\alpha F_1=C_0\Big|_{\alpha=0}^{\alpha=\pi}=\frac{X}{l_kl_{k-1}}(2N_k-N_{k+1}-N_{k-1}).
 \eeq
 This is the result for all values of $(\gamma,\zeta)$ - including the $\mathcal{N}=1$ background. The boundary analysis then agrees with the discussion around \eqref{eqn:C1sourcebetaIIB}, where we had nice D7 sources for $\zeta=0$ or $\sin\theta=0$. 
 It is important to note that this rational quantization of charge is due to the $\tilde{\Xi}_k$ (and hence $\Delta_k$) in the denominator of $C_0$, and not $\Pi_k$ which is what gives rise to singularities in the internal manifold. In other words, the charge still depends on $l_k$ and not $\hat{l}_k$, see \eqref{eqn:lkandlkhat}. It appears then that this breaking of quantization is simply a remnant of the IIA orbifold singularity. As a result of this, one can find situations where the background has the above rational charge (with $l_k\neq1$) without having orbifold singularities in its metric (i.e. with $|\hat{l}_k|=1$). In other words, by T-dualising within the spindle like orbifold in Type IIA, we can break the orbifold structure completely in Type IIB whilst still inheriting the rational charge. The $\mathcal{N}=1$ and $S^2$ preserved $\mathcal{N}=0$ cases are two example. 
 Conversely, fixing $\xi=0$ recovers the integer quantization of charge (with $l_k=1$) but does not necessarily eliminate orbifold singularities, as $\hat{l}_k = \gamma-\zeta (N_{k+1}-N_k)$.
  \end{itemize}
  \subsubsection{Summary}
  For all values of $(\gamma,\zeta,\xi)$, we find NS5 charge at the $\sigma\rightarrow \infty$ boundary. In these solutions, $(\theta,\phi)$ behaves only topologically as an $S^2$ in general, recovering it when $\gamma=0$ (which is the only parameter to remain in this limit). At this boundary, for either $\gamma=0$ (with a preserved $S^2$) or $\sin\theta\rightarrow 0$ (close to a pole of the deformed $S^2$), the background describes the near horizon limit of a spherically symmetric stack of NS5 branes. Along the $\sigma=0$ boundary, we find stacks of D7 branes at each kink of the rank function. These charges have rational quantization in all $\xi\neq0$ cases, however some backgrounds do not contain orbifold singularities in the metric! In these cases, the rational quantization is a remnant of the IIA orbifold singularities, which can be broken by T-dualising within the orbifold structure. Additionally, one can find solutions which contain orbifold singularities, due to the parameters $(\gamma,\zeta)$, which contain integer quantization of charge (for $\xi=0$). Naively, one can find D5 charge for all cases, however it is not clear that the integration cycles are always closed. We propose the D5 branes are present only for the preserved $S^2$ case (with $\gamma=\zeta=0$), by considering $(\alpha,S^2)$ a valid cycle - with integer quantization recoverable via an appropriate gauge transformation. In summary, we find
   \begin{equation}\label{eqn:ChargesIIA}
 \begin{aligned}
&Q_{NS5}=\widehat{\kappa} \,P,~~~~~~~Q_{D7}^k =\frac{X}{l_kl_{k-1}}b_k,~~~~~~ Q_{D5}^k=\widehat{\kappa} X  \begin{cases}
     ( N_k - N_{k-1}) & (\gamma=\zeta=0)\\
     ~~~~~~~~~0  & \text{otherwise}
    \end{cases},     \\[2mm]
    &Q_{D7}=\sum_{k=1}^{P-1}Q^k_{D7}=\frac{X}{l_0 l_P}(N_{P-1}+N_1),~~~~~Q_{D5}=\sum_{k=1}^{P-1}Q^k_{D5}=\widehat{\kappa} X  \begin{cases}
     N_{P-1} &  (\gamma=\zeta=0)\\
     ~~~~0 &  \text{otherwise}
    \end{cases},\\[2mm]
    &l_k=1+\xi (N_{k+1}-N_k),~~~~~~~~l_0=1+\xi N_1,~~~~~~l_P=1-\xi N_{P-1},~~~~~~~b_k=(2N_k-N_{k+1}-N_{k-1}),
    \end{aligned} 
\end{equation}
with the holographic central charge given by
\begin{equation*}
c_{hol} 
= \frac{ \widehat{\kappa}^3X^2}{8\pi}\sum_{n=1}^\infty P \, \mathcal{R}_n^2.
\end{equation*}
The quantization discussion is then identical to the one given in Section \ref{sec:quantconds}. One could possibly introduce an extra parameter, $\lambda$, by T-dualising along $\lambda \chi$. If so, this could possibly aid quantization, but this would need investigating further. 
  
  


In solutions without orbifold singularities (such as the $\mathcal{N}=1$ and $SU(2)$ preserved $\mathcal{N}=0$ cases), rank functions such as those given in Figure \ref{fig:Mathematica5} seem the most valid (as they will lead to integer quantization). A potential interpretation in terms of rotating branes is also a possibility, and offered in Appendix E of \cite{Merrikin:2024bmv}, but is omitted from this thesis.

\subsection{U(1)$\times$U(1) preserving $\mathcal{N}=1$ solution}\label{sec:IIBN1one}
We now investigate a little further the $\zeta=-\xi,~\gamma=-1$ case of \eqref{eqn:ATD1}, which leads to a one-parameter family of $\mathcal{N}=1$ solutions (corresponding to the blue line in Figure \ref{fig:IIBtableplot}), given by
  \begin{align}\label{eqn:IIBN1}
 &      ds_{10,B}^2= \frac{1}{X}f_1^{\frac{3}{2}}f_5^{\frac{1}{2}}\sqrt{\Xi}\Bigg[4ds^2(\text{AdS}_5)+f_4(d\sigma^2+d\eta^2) +ds^2(M_3)\Bigg],\nn\\[2mm]
&   ds^2(M_3)=f_2\bigg(d\theta^2  + \frac{1}{\Pi} \sin^2\theta\, d\phi^2\bigg)+\frac{X^2}{f_1^3f_3f_5\Pi }\,\Big(d\chi +\frac{1}{X} f_8\sin\theta  d\theta\Big)^2,\nn\\[2mm]
&  C_0=
  \frac{X}{\Xi}\bigg(f_6(1+\xi f_6) +\xi \frac{f_3}{f_5}+\xi \frac{f_2}{f_5}\sin^2\theta\bigg),
~~~~~~~~
  e^{2\Phi_\mathcal{B}}=\frac{1}{X^2}\frac{f_5}{f_3}\frac{\Xi^2}{\Pi } ,  \nn \\[2mm]
&B_2=-\frac{1}{X\Pi} \bigg( f_8+\xi f_7 +\xi \frac{f_2}{f_3} (f_7-f_6f_8) \sin^2\theta \bigg) \sin\theta\,d\phi \wedge d\theta + \frac{1}{\Pi} \frac{f_2}{f_3}(1+\xi f_6)\sin^2\theta  d\phi\wedge d\chi ,\nn\\[2mm]
&  C_2=\frac{1}{\Pi}\bigg( f_7- \frac{f_2}{f_3}(f_6f_8-f_7)\sin^2\theta \bigg)\sin\theta  d\theta \wedge d\phi + \frac{X}{\Pi}  \frac{f_2}{f_3}f_6\sin^2\theta  d\phi\wedge d\chi,\nn\\[2mm]
&  \Xi=\Delta+\xi^2 \frac{f_2}{f_5}\sin^2\theta,~~~~~\Delta = (1+\xi f_6)^2 +\xi^2\frac{f_3}{f_5} ,~~~~~~~\Pi= 1+\frac{f_2}{f_3}\sin^2\theta.
  \end{align}
Investigations at the boundary were conducted in the previous sub-section, with charges given in \eqref{eqn:ChargesIIA} for $\zeta=-\xi,~\gamma=-1$.
\subsubsection{G-Structure description}

We have performed an ATD along $\chi$, meaning $dy\equiv d\chi$, with $(u^\mathcal{A},v^\mathcal{A},\omega^\mathcal{A},j^\mathcal{A})$ given in \eqref{eqn:uandv1}. Here we have $u^\mathcal{A}_{||}=v^\mathcal{A}_{||}=0$ and $u^\mathcal{A}_\perp=u^\mathcal{A},~v^\mathcal{A}_\perp=v^\mathcal{A}$, namely
     \begin{align} 
&z_{\perp}^\mathcal{A} =u_{\perp}^\mathcal{A}+i\,v_{\perp}^\mathcal{A} ,~~~~~~v_{\perp}^\mathcal{A}=  \frac{\kappa}{\sqrt{X}} e^{-2\rho}f_5^{\frac{1}{4}}f_1^{-\frac{3}{4}} \Xi^{\frac{1}{4}}d\left(e^{2\rho} \dot{V}\cos\theta\right),\nn\\[2mm]
&u_{\perp}^\mathcal{A}=  \frac{1}{\sqrt{X}}(f_1f_5)^{\frac{3}{4}}\Xi^{-\frac{1}{4}}\left(\frac{f_3 \dot{V}'}{4 \sigma}d\sigma +V''d\eta-f_6d\rho - \frac{\xi}{f_5}\Big(\frac{\kappa^2e^{-4\rho}}{2f_1^3}d(e^{4\rho}\dot{V}^2(\cos^2\theta-3))+4 d\rho\Big)\right),\nn
\end{align}
and recall the following relations (see \eqref{eqn:A1Aval} with $\phi_1=\chi,\phi_2=\phi$)
\beq\label{eqn:N1A1}
 E^\chi_\mathcal{A}=  e^{C^\mathcal{A}}(d\chi +A_1^\mathcal{A}),~~~~~~~~~~~A_1^\mathcal{A}=\frac{1}{2}\frac{h_{\chi\phi}}{h_{\chi}}d\phi,~~~~~~~~~~~\frac{1}{2}\frac{h_{\chi\phi}}{h_{\chi}} =-\frac{f_2(1+\xi f_6)}{f_3+f_2\sin^2\theta}\sin^2\theta  ,
 \eeq
now using \eqref{eqn:beta-Gen} to calculate $A_1^\mathcal{A}$ (with $(p,b,u)=1,(a,c,m)=0,q=-v=\xi,s=\gamma=-1$). Then to derive the remaining elements we re-write the forms given in \eqref{eqn:uandv1}, with $\gamma=-1$, as follows
\beq\label{eqn:jomegasplit}
\begin{aligned}
&j^\mathcal{A}= j_1^\mathcal{A}\wedge d\chi + j_2^\mathcal{A}\wedge d\phi =\Big(j_2^\mathcal{A} -\frac{1}{2}\frac{h_{\chi\phi}}{h_{\chi}} j_1^\mathcal{A} \Big)\wedge d\phi+ e^{-C^\mathcal{A}}j_1^\mathcal{A}\wedge E^\chi_\mathcal{A} ,\\[2mm]
&\omega^\mathcal{A} =  \omega_1^\mathcal{A}\wedge d\chi+\omega_2^\mathcal{A} =\Big(\omega_2^\mathcal{A}-\frac{1}{2}\frac{h_{\chi\phi}}{h_{\chi}} \omega_1^\mathcal{A} \wedge d\phi\Big)+ e^{-C^\mathcal{A}}\omega_1^\mathcal{A}\wedge E^\chi_\mathcal{A} ,
\end{aligned}
\eeq
where the slight difference in expression between the $j$ and $\omega$ is because $j$ is made of only $d\chi$ and $d\phi$ components, whereas $\omega$ is a little more general. Hence
\beq\label{eqn:jomegadecomps}
\begin{aligned}
&j^\mathcal{A} = j^\mathcal{A}_{\perp}+j^\mathcal{A}_{||}\wedge  E^\chi_\mathcal{A} ,~~~~~~~  j^\mathcal{A}_{\perp} =  \Big(j_2^\mathcal{A} -\frac{1}{2}\frac{h_{\chi\phi}}{h_{\chi}} j_1^\mathcal{A} \Big)\wedge d\phi    ,~~~~~~~~j^\mathcal{A}_{||} = e^{-C^\mathcal{A}}j_1^\mathcal{A} \\[2mm]
&\omega^\mathcal{A} =\omega^\mathcal{A}_{\perp}+\omega^\mathcal{A}_{||}\wedge  E^\chi_\mathcal{A} ,     ~~~~~~~  \omega^\mathcal{A}_{\perp} =    \omega_2^\mathcal{A}-\frac{1}{2}\frac{h_{\chi\phi}}{h_{\chi}} \omega_1^\mathcal{A} \wedge d\phi, ~~~~~~~~\omega^\mathcal{A}_{||} = e^{-C^\mathcal{A}}\omega_1^\mathcal{A},
\end{aligned}
\eeq

leading to
  \beq\label{eqn:IIBfamGs}
\begin{aligned}
j_{||}^\mathcal{A}&=   \frac{ e^{-C^\mathcal{A}}}{X\sqrt{\Xi}} \bigg( \frac{f_1^{\frac{3}{2}}f_5^{\frac{1}{2}} f_3}{\sigma} e^{-\rho}e^{-\xi V'}d\left(e^{\rho}\sigma e^{\xi V'}\right)-\hat{X}_2\bigg) ,\\[2mm]
j_\perp^\mathcal{A} &=\frac{1}{X\sqrt{\Xi}}\Bigg[\Big(\kappa f_2 f_3^{\frac{1}{2}}\hat{X}_1 +\hat{X}_2 \Big) +\frac{f_2(1+\xi f_6)}{f_3+f_2\sin^2\theta} \bigg( \frac{f_1^{\frac{3}{2}}f_5^{\frac{1}{2}} f_3}{\sigma} e^{-\rho}e^{-\xi V'}d\left(e^{\rho}\sigma e^{\xi V'}\right)-\hat{X}_2\bigg)\sin^2\theta    \Bigg] \wedge d\phi \\[2mm]
\omega_{||}^\mathcal{A}&=  -i \,\frac{2\kappa^2}{X}   e^{-C^\mathcal{A}}  f_1^{-\frac{3}{2}}e^{-3\rho}   d(e^{3\rho}\dot{V}e^{i\phi}\sigma \sin\theta )    ,\\[2mm]
 \omega_\perp^\mathcal{A} &= -\frac{2\kappa^2}{X}  f_1^{-\frac{3}{2}}e^{-3\rho} \bigg[d\Big(e^{2\rho}\dot{V}e^{-\xi V'} e^{i\phi} \sin\theta\, d(e^\rho e^{\xi V'}\sigma)\Big)+ i \frac{f_2(1+\xi f_6)\sin^2\theta}{f_3+f_2\sin^2\theta} \,d(e^{3\rho}\dot{V}e^{i\phi}\sigma \sin\theta ) \wedge d\phi \bigg],
\end{aligned}
\eeq
with $(\hat{X}_{1},\hat{X}_{2})$ defined in \eqref{eqn:XhatsN1}. Using \eqref{eqn:IIBspinorsSimp}, and noting
\begin{equation*}
B_1^\mathcal{A} = -  \frac{1}{X} f_8 \sin\theta\,d\theta,~~~~~  e^{2\Phi_\mathcal{B}}=\frac{1}{X^2}\frac{f_5}{f_3}\frac{\Xi^2}{\Pi },~~~~ e^{4\Phi_\mathcal{A}}=\frac{1}{X^6}f_1^3f_5^3\Xi^3,
\end{equation*}
we can now build the IIB pure spinors as follows
\begin{equation*}
\begin{aligned}
  \Psi_-^\mathcal{B} &=\frac{1}{8}e^{\Phi_\mathcal{B}-\Phi_\mathcal{A}} e^{\frac{1}{2}z^\mathcal{A}_{\perp}\wedge \overline{z}^\mathcal{A}_{\perp}}\wedge\Big( e^{C^\mathcal{A}}\omega^\mathcal{A}_{||}+   \omega^\mathcal{A}_{\perp}\wedge \Big(d\chi+ \frac{1}{X} f_8 \sin\theta\,d\theta\Big)\Big),  \\
   \Psi_+^\mathcal{B}&=\frac{i }{8} e^{\Phi_\mathcal{B}-\Phi_\mathcal{A}}e^{-ij^\mathcal{A}_{\perp}}\wedge    z^\mathcal{A}_{\perp}\wedge \Big( \Big(d\chi + \frac{1}{X} f_8 \sin\theta\,d\theta\Big)-i    e^{C^\mathcal{A}} j^\mathcal{A}_{||}\Big),
\end{aligned}
\end{equation*}
with the $ e^{C^\mathcal{A}}$ terms cancelling out of the expressions. These spinors then satisfy the supersymmetry conditions given in \eqref{eqn:IIBGconditions}.

\subsubsection{Higher form fluxes 
}
Now that we have derived the pure spinors for the IIB solution, we can now easily derive the higher form fluxes for the solution. We use the expression given in \eqref{eqn:calsIIB} (noting that $C_{10}=0$)
\begin{align}
C_6&
=e^{4A-\Phi_\mathcal{A}}  \text{vol(Mink}_4\text{)} \wedge \Big( -e^{C^\mathcal{A}}j^\mathcal{A}_{||}\wedge v^\mathcal{A}_\perp +u^\mathcal{A}_\perp \wedge  \Big(d\chi + \frac{1}{X} f_8 \sin\theta\,d\theta\Big) \Big), \nn\\[2mm]
C_8&
= e^{4A-\Phi_\mathcal{A}}  \text{vol(Mink}_4\text{)} \wedge j^\mathcal{A}_\perp\wedge \Big( v^\mathcal{A}_\perp \wedge  \Big(d\chi + \frac{1}{X} f_8 \sin\theta\,d\theta\Big) +e^{C^\mathcal{A}}j^\mathcal{A}_{||}\wedge u^\mathcal{A}_\perp\Big),\nn
\end{align}
with $e^{2A}=\frac{4}{X}  f_1^{\frac{3}{2}} f_5^{\frac{1}{2}}\Xi^{\frac{1}{2}} e^{2\rho}$ from the IIA solution. One can verify these results using \eqref{eqn:origfieldstengths} and \eqref{eqn:fieldrelation} as follows
\beq
F_7=-\star F_3=dC_6,~~~~~~~~F_9=\star F_1=dC_8-H\wedge C_6,
\eeq
with $F_3=dC_2-C_0H,~F_1=dC_0,~H=dB_2$ (and $C_4=0$) given in \eqref{eqn:IIBN1}.
\subsection{Supersymmetry breaking}\label{sec:GstructuresIIB}

We now derive the G-structure description of the three parameter family given in \eqref{eqn:ATD1}. 
Using the approach discussed in Section \ref{sec:ATDGssumarry}, we have $dy\equiv d\chi$, with $u^\mathcal{A},v^\mathcal{A}$ given in \eqref{eqn:GenIIAbetauandv} and $\omega^\mathcal{A},j^\mathcal{A}$ given in \eqref{eqn:GenIIAbetajandomega}. 
We notice, as $u$ and $v$ are independent of $\phi$, they are independent of $\gamma$. In addition, we recall that $u^\mathcal{A}_{||}=v^\mathcal{A}_{||}=0$ and $u^\mathcal{A}_\perp=u^\mathcal{A},~v^\mathcal{A}_\perp=v^\mathcal{A}$, namely
\begin{align} 
z_{\perp}^\mathcal{A} &=u_{\perp}^\mathcal{A}+i\,v_{\perp}^\mathcal{A} ,~~~~~~~~~~v^\mathcal{A}_\perp=  \frac{\kappa}{\sqrt{X}} e^{-2\rho}f_5^{\frac{1}{4}}f_1^{-\frac{3}{4}}\Xi^{\frac{1}{4}} d(e^{2\rho} \dot{V} \cos\theta),\nn\\[2mm]
u^\mathcal{A}_\perp& =\frac{1}{\sqrt{X}}  (f_1f_5)^{\frac{3}{4}}\Xi^{-\frac{1}{4}}
  \bigg[\frac{\dot{V}'f_3}{4\sigma} d\sigma +V'' d\eta -f_6 d\rho -\frac{1}{f_5} \bigg(\frac{-\zeta \kappa^2 e^{-4\rho}}{2f_1^3}d\Big(e^{4\rho}\dot{V}^2 (\cos^2\theta-3)\Big)+4\xi d\rho\bigg)\nn\\[2mm]
  &~~~~~~ ~~~~~  +(\zeta+\xi)\frac{e^{-4\rho}}{\Lambda}d (\dot{V}^2e^{4\rho} )\bigg].\nn
\end{align}
 Using \eqref{eqn:TD1} and \eqref{eqn:A1Aval}, with $\phi_1=\chi,\phi_2=\phi$, we find  
\beq
\hspace{-1.5cm}
 E^\chi_\mathcal{A}=  e^{C^\mathcal{A}}(d\chi +A_1^\mathcal{A}),~~~~~~~~A_1^\mathcal{A}= \frac{1}{2}\frac{h_{\chi\phi}}{h_{\chi}}d\phi,~~~~~~~~  \frac{1}{2}\frac{h_{\chi\phi}}{h_{\chi}}= \frac{1}{\Pi} \frac{f_2}{f_3}\Big(\xi (\gamma\xi-\zeta)\frac{f_3}{f_5} +(1+\xi f_6)\big(\gamma+(\gamma\xi-\zeta)f_6\big)\Big)\sin^2\theta ,
 \eeq
 with $\Pi$ defined in \eqref{eqn:ATD1}. In the $\mathcal{N}=1$ case (with $\zeta=-\xi$ and $\gamma=-1$), this reduces to \eqref{eqn:N1A1}. 
 This result can be derived using \eqref{eqn:beta-Gen} (with $(p,b,u)=1,~(a,c,m)=0,~q\equiv \xi,v\equiv \zeta,s\equiv \gamma$), or alternatively by inspection, using the forms of \eqref{eqn:ATD1} and \eqref{eqn:betaredchitdual}. We can now derive the remaining elements, but to do so we must first split up the forms given in \eqref{eqn:GenIIAbetajandomega} using \eqref{eqn:jomegadecomps} and \eqref{eqn:jomegasplit}, leading to
 \beq
\begin{aligned}
& j_{1}^\mathcal{A} = \frac{1}{X\sqrt{\Xi}}\bigg[\frac{f_1^{\frac{3}{2}}f_5^{\frac{1}{2}}f_3}{\sigma} e^{-\rho}e^{\zeta V'} d\big(e^\rho \sigma e^{-\zeta V'}\big)+\kappa f_2 f_3^{\frac{1}{2}}\Big((\gamma+1) X_1+(\gamma\xi-\zeta) \hat{X}_1   \Big)+\gamma\, X_2\nn\\[2mm]
&~~~~~~ -(\zeta +\xi) \frac{4\kappa \dot{V}}{\tilde{\Delta} \sqrt{\Lambda}}\Big(\hat{X}_3  + \gamma X_3  \Big) \bigg],\\[2mm]
& j_2^\mathcal{A}=   \frac{1}{X\sqrt{\Xi}}\bigg[ \kappa f_2 f_3^{\frac{1}{2}} (X_1+\xi  \hat{X}_1) +X_2  -(\zeta +\xi) \frac{4\kappa \dot{V}}{\tilde{\Delta} \sqrt{\Lambda}}  X_3\bigg], \nn\\[2mm]
&\omega_{1}^\mathcal{A}=  -i\,\frac{2\kappa^2}{X}  
f_1^{-\frac{3}{2}}e^{-2\rho}e^{i(\gamma+1)\chi} \bigg[ e^{-\zeta V'}\sigma \, d\Big(\dot{V}e^{2\rho} e^{\zeta V'}e^{i\phi}\sin\theta \Big)\nn\\[2mm]
&~~~~~~~-\gamma \, e^{i\phi}e^{\rho} \dot{V} \sin\theta\Big(e^{\zeta V'}d(e^{\rho} e^{-\zeta V'}\sigma)+ (\zeta+\xi) e^{\rho} \sigma   \, d(V') \Big)   \bigg]  ,\\[2mm]
& \omega_2^\mathcal{A} =  -\frac{2\kappa^2}{X} 
f_1^{-\frac{3}{2}} e^{-3\rho} e^{i(\gamma+1)\chi}\bigg[  d\Big(e^{2\rho}\dot{V} e^{\zeta V'}e^{i \phi }\sin\theta\, d(e^{\rho}e^{-\zeta V'}\sigma)\Big)+(\zeta+\xi)   e^{\rho} \sigma \,d\Big(e^{2\rho}\dot{V} e^{i\phi}\sin\theta \,d(V')\Big)   \bigg],
\end{aligned}
\eeq
with
\begin{equation*}
\begin{aligned}
 j^\mathcal{A}_{\perp} =  \Big(j_2^\mathcal{A} -\frac{1}{2}\frac{h_{\chi\phi}}{h_{\chi}} j_1^\mathcal{A} \Big)\wedge d\phi    ,~~~~~~j^\mathcal{A}_{||} = e^{-C^\mathcal{A}}j_1^\mathcal{A},~~~~~ 
  \omega^\mathcal{A}_{\perp} =    \omega_2^\mathcal{A}-\frac{1}{2}\frac{h_{\chi\phi}}{h_{\chi}} \omega_1^\mathcal{A} \wedge d\phi, ~~~~~\omega^\mathcal{A}_{||} = e^{-C^\mathcal{A}}\omega_1^\mathcal{A},
\end{aligned}
\end{equation*}
which we can substitute into \eqref{eqn:IIBspinorsSimp} to build the IIB pure spinors, noting the definitions in \eqref{eqn:generalresult1} and \eqref{eqn:ATD1},
we have
\begin{equation*}
\begin{aligned}
  \Psi_-^\mathcal{B} &=\frac{1}{8}e^{\Phi_\mathcal{B}-\Phi_\mathcal{A}} e^{\frac{1}{2}z^\mathcal{A}_{\perp}\wedge \overline{z}^\mathcal{A}_{\perp}}\wedge\bigg[ e^{C^\mathcal{A}}\omega^\mathcal{A}_{||}+   \omega^\mathcal{A}_{\perp}\wedge \bigg(d\chi -\frac{1}{X}\Big((\gamma \xi -\zeta )f_7 +\gamma f_8 \Big)\sin\theta  d\theta\bigg)\bigg],  \\
   \Psi_+^\mathcal{B}&=\frac{i }{8} e^{\Phi_\mathcal{B}-\Phi_\mathcal{A}}e^{-ij^\mathcal{A}_{\perp}}\wedge    z^\mathcal{A}_{\perp}\wedge \bigg[ \bigg(d\chi -\frac{1}{X}\Big((\gamma \xi -\zeta )f_7 +\gamma f_8 \Big)\sin\theta  d\theta\bigg)-i    e^{C^\mathcal{A}} j^\mathcal{A}_{||}\bigg],
\end{aligned}
\end{equation*}
with the $ e^{C^\mathcal{A}}$ terms cancelling out of the expressions. We are now in a position to investigate the supersymmetry conditions given in \eqref{eqn:IIBGconditions}. 

Checking \eqref{eqn:CalibrationformIIB1} leads to $(\xi+\zeta)$ and $(\gamma+1)$ terms on the right hand side - which is no surprise  given that both conditions are separately required for supersymmetry. In addition, there are $(\gamma\xi-\zeta)$ terms included, which also vanish when supersymmetry is preserved. However, these terms can vanish even in the absence of supersymmetry (e.g. for $\gamma=\zeta=0$), so must be supplemented with $(\xi+\zeta)$ and/or $(\gamma+1)$ terms. The remaining two conditions need to be investigated further (as these equations take much computer time and power), but judging by the $\zeta=\xi=0$ case that we will study in Section \ref{sec:gammaIIB}, they are likely to include $(\zeta+\xi)$ terms only (as in the IIA case \eqref{eqn:SUSYbrokenIIA}) with no requirements from $\gamma$. A more thorough investigation is left for future work.


 
\newpage
\section{Following the IIA $\chi$ Reduction}\label{sec:followingchi}
We now investigate performing an abelian T-duality following a dimensional reduction along $\chi$, using the same procedure as the previous section. Once again, in order to preserve supersymmetry, we must leave the $U(1)_R$ component of the R-symmetry intact.
\subsection{Three-Parameter Families}
Here we will perform an ATD on the two-parameter family of solutions given in \eqref{eqn:chireduction1}. As we will discover, only an ATD along $\beta$ (with $\zeta=-1$ and $\gamma=0$) will lead to a SUSY preserved solution. This is a little different to the $\beta$ reduction case, where one could perform an ATD along both $\chi$ and $\phi$ to derive an $\mathcal{N}=1$ background. This is a remnant of the GM $U(1)_R$ component being $\chi+\phi$ prior to the $SL(3,\mathds{R})$ transformation. 
We now investigate the ATD for the $\chi$ reduction with $a\equiv \xi$ and $q \equiv \xi$, in turn.

\subsubsection{Fixing $a\equiv \xi$}
We begin with the case where $(p,b,u)=1,~(q,c,m)=0,~a\equiv \xi,s\equiv \zeta,v\equiv \gamma$, meaning \eqref{eqn:U(1)} becomes
  \begin{equation}\label{eqn:chiredU(1)1}
U(1)_R = (1+\zeta)\chi +\gamma \beta +\phi.
\end{equation}
 We then perform a dimensional reduction along $\chi$, followed by an ATD to IIB. We will investigate an ATD along $\beta$ and $\phi$ in turn, beginning with the case which will give rise to an $\mathcal{N}=1$ solution.
\begin{itemize}
\item \textbf{Performing an ATD long $\beta$}\\
Using the T-dual formula \eqref{eqn:betaredchitdual}, we get the following three-parameter family of solutions
     \begin{align}\label{eqn:ATD3}
   &    ds_{10,st}^2=\frac{1}{X} f_1^{\frac{3}{2}}(f_5f_6^2+f_3)^{\frac{1}{2}} \sqrt{\Xi_2}\bigg[4ds^2(\text{AdS}_5)+f_4(d\sigma^2+d\eta^2)+ ds^2(M_3)\bigg] , \nn\\[2mm]
   & ds^2(M_3) = f_2 \bigg(d\theta^2+\frac{1}{\Pi_2} \sin^2\theta d\phi^2  \bigg)  +\frac{X^2}{f_1^3 f_3f_5\Pi_2 }D\beta^2,\nn\\[2mm]
 &     C_0=\frac{X}{(f_5f_6^2+f_3) \Xi_2}\Big( f_5 (f_6+\xi) +\gamma \zeta\, f_2\sin^2\theta\Big) ,~~~~~e^{2\Phi_\mathcal{B}}=\frac{ (f_5f_6^2+f_3)^2}{X^2 f_3f_5}\frac{\Xi_2^2}{\Pi_2},   \nn\\[2mm]
&B_2=  \frac{\sin\theta}{\Pi_2} \bigg[ \frac{ f_2}{f_3 }\bigg((f_6+\xi)(\gamma f_6+\gamma\xi-\zeta)+\gamma \frac{f_3}{f_5} \bigg)\sin\theta \,d\beta      \nn\\[2mm]
&  ~~~+\frac{1}{X} \bigg((f_7+\xi f_8)+\zeta \frac{ f_2}{f_3 }  \bigg((f_6f_8-f_7)(\gamma f_6+\gamma\xi-\zeta)+\gamma \frac{f_3f_8}{f_5}\bigg)\sin^2\theta\bigg) d\theta\bigg] \wedge d\phi , \nn\\[2mm]
&  C_2= \frac{ \sin\theta }{\Pi_2}\bigg[X\frac{f_2}{f_3 } (\gamma f_6+\gamma\xi-\zeta)\sin\theta\, d\beta    \nn\\[2mm]
& ~~~~~~~~~~ +\bigg( f_8 +\gamma  \frac{f_2}{f_3}\bigg((f_6f_8-f_7)(\gamma f_6+\gamma\xi-\zeta)+\gamma \frac{f_3f_8}{f_5}\bigg)\sin^2\theta\bigg)d\theta\bigg] \wedge d\phi , \nn\\[2mm]
   &\Xi_2= \Delta_2+\zeta^2 \frac{f_2}{\big(f_5f_6^2+ f_3  \big)} \sin^2\theta,~~~D\beta=d\beta -\frac{1}{X}\big((\gamma\xi - \zeta)f_8+\gamma f_7\big)\sin\theta \, d\theta,\nn\\[2mm]
 &
 \Pi_2=1+\frac{f_2}{f_3}\bigg(\big(\gamma f_6+\gamma\xi-\zeta\big)^2+\gamma^2\frac{f_3}{f_5}\bigg)\sin^2\theta~~~~~~~~\Delta_2=1+\xi \frac{f_5(2f_6+\xi)}{\big(f_5f_6^2+ f_3 \big)}.
  \end{align}
We can preserve the $U(1)_R$ component given in \eqref{eqn:chiredU(1)1} by first fixing $\zeta=-1$ (to derive the $\mathcal{N}=1$ IIA solution) followed by fixing $\gamma=0$ under the T-duality. Due to the transformations \eqref{eqn:S2breakingdefns}, we find the $S^2$ is recovered for $(\gamma,\zeta)=0$  - see Figure \ref{fig:IIBtableplotchired}. 

We find once again that one can map this solution to \eqref{eqn:ATD1} via the transformations given in \eqref{eqn:Transformation} (with $k_1=\frac{1}{\xi},k_2=\xi^{\frac{5}{2}}$)
 \beq\label{eqn:IIBtrans2}
 \begin{aligned}
&g_{MN}\rightarrow \frac{1}{\xi} g_{MN},~~~~~ e^{2\Phi_B}\rightarrow \frac{1}{\xi^4} e^{2\Phi_B},~~~~~~C_0\rightarrow \xi^2 C_0,~~~~~~~C_2\rightarrow \xi C_2,~~~~~~~B_2\rightarrow \frac{1}{\xi}  B_2,\\
&B_2\rightarrow- B_2,~~~~~~~~~~C_0\rightarrow -C_0+\xi,~~~~~~~C_2\rightarrow C_2 - \xi B_2,\\
& \phi\rightarrow -\phi ,~~~~~~~~~~~~~~\beta \rightarrow - \chi,~~~~~~~~~~~~~~~\zeta \rightarrow \zeta \,\xi,~~~~~~~~~~~~\gamma\rightarrow \zeta -\frac{\gamma}{\xi},~~~~~~\xi\rightarrow \frac{1}{\xi}.
 \end{aligned}
 \eeq
 However, it does not appear possible to map this to \eqref{eqn:ATD2}. The above mapping requires $1/\xi$ terms, which again is non-trivial given that $\xi$ is an integer. We nonetheless find that the $\xi=0$ solution is new and unique.
 
 We derive the $\mathcal{N}=1$ solution by fixing $(\zeta=-1,\gamma=0)$,
which after using the transformations given in \eqref{eqn:Transformation}, with $k_1=\xi^{-1},~k_2=\xi^{\frac{5}{2}}$ 
        \begin{equation}\label{eqn:N=1transforms}
    g_{MN} \rightarrow \frac{1}{\xi}\, g_{MN},~~~~~~~~~~~~~B_2 \rightarrow  \frac{1}{\xi}\,  B_2~~~~~~~~~C_0 \rightarrow \xi^2\, C_0,~~~~~~~~~~~~C_2 \rightarrow \xi \,C_2~~~~~~~~~e^{2\Phi} \rightarrow \frac{1}{\xi^4}e^{2\Phi},
    \end{equation}
  followed by
  \beq\label{eqn:N=1transforms2}
  C_0\rightarrow -C_0+\xi,~~~~~~~~~~~~~C_2\rightarrow -C_2 + \xi B_2,~~~~~~~~~\xi\rightarrow \frac{1}{\xi},~~~~~~~~~~~~\beta\rightarrow -\chi,
  \eeq
one maps to \eqref{eqn:IIBN1}. Again, we argue that this is non-trivial, with the $\xi=0$ case new and unique. We will return to this $\mathcal{N}=1$ solution later in this section. 

  \begin{figure}[H]
\centering  
\subfigure
{
\centering
  \begin{minipage}{0.5\textwidth}
\begin{tabular}{c | c c c c  }
$\chi$- Red, $\beta$- ATD&$\mathcal{N}$&U(1)$_R$&SU(2)$_R$  \\
\hline
$\textcolor{blue}{ \gamma=0,~\zeta=-1,~\xi\neq 0}$&$ \textcolor{blue}{1}$ &$\textcolor{blue}{\checkmark}$&$\textcolor{blue}{\times}$  \\
$\textcolor{red}{\gamma=0,~\zeta=0,~\xi\neq 0}$&$ \textcolor{red}{0}$ &$\textcolor{red}{\times}$&$\textcolor{red}{\checkmark}$  \\
\textcolor{teal}{otherwise}&$ \textcolor{teal}{0}$ &$\textcolor{teal}{\times}$ &$\textcolor{teal}{\times}$  
\end{tabular}
  \end{minipage}
     \begin{minipage}{.5\textwidth}
    \centering
\begin{tikzpicture}[scale=0.45,y={(1cm,0.5cm)},x={(-1cm,0.5cm)}, z={(0cm,1cm)}]
\draw[-stealth, line width=0.53mm] (0,0,-3.3)--(0,0,3.3) node[right ]{$\mathbf{\gamma}$};
\draw[green!60!black ] (2,2,2) node[above] {$\mathcal{N}=0$};
\begin{scope}[canvas is yx plane at z=0]
\draw[-stealth, line width=0.53mm] (-3.3,0)--(3.3,0) node[right ]{$\xi$};
\draw[-stealth, line width=0.53mm] (0,-3.3)--(0,3.3) node[above ]{$\zeta$};
\clip (-3,-3) rectangle (3,3);
\begin{scope}[cm={0.5,-0.5,  50,50,  (0,0)}]  
\end{scope}
\draw[line width=0.6mm,red](-3,0)--(3,0);
\draw[line width=0.6mm,blue](-3,-1)--(3,-1);
\end{scope}
\begin{scope}[canvas is yx plane at z=-1]
 \node[blue ,rotate=0] at (3,-1.5) {$\mathbf{\zeta=-1}$};
\end{scope}


\draw[fill=green,opacity=0.025] (-3,3,-3) -- (3,3,-3) -- (3,3,3) -- (-3,3,3) -- cycle;
\draw[fill=green,opacity=0.035] (3,-3,-3) -- (3,3,-3) -- (3,3,3) -- (3,-3,3) -- cycle;
\draw[fill=green,opacity=0.05] (3,-3,-3) -- (3,3,-3) -- (-3,3,-3) -- (-3,-3,-3) -- cycle;

\draw[fill=green,opacity=0.05] (-3,-3,-3) -- (3,-3,-3) -- (3,-3,3) -- (-3,-3,3) -- cycle;
\draw[fill=green,opacity=0.05] (-3,-3,-3) -- (-3,-3,3) -- (-3,3,3) -- (-3,3,-3) -- cycle;
\draw[fill=green,opacity=0.05] (-3,-3,3) -- (-3,3,3) -- (3,3,3) -- (3,-3,3) -- cycle;
\begin{scope}[canvas is yz plane at x=1.5]
\coordinate (Origin) at (0,0);
\shade[ball color=red] (Origin) circle (0.05cm);
\end{scope}
\begin{scope}[canvas is yz plane at x=0]
\end{scope}
\end{tikzpicture}
  \end{minipage} 
}
\subfigure
{
\centering
  \begin{minipage}{\textwidth}
    \centering
\begin{tikzpicture}[scale=0.9]
\draw[ blue,line width=0.93mm] (-6,-4.5)--(-5.5,-4.5);
\draw (-5.5,-4.5) node[right]{$ \mathcal{N}=1$ U(1)$_R$ Preserving};
\draw[ red,line width=0.93mm] (2,-4.5)--(2.5,-4.5);
\draw  (2.5,-4.5) node[right]{$ \mathcal{N}=0$ SU(2) Preserving};
\end{tikzpicture}
  \end{minipage}
}
\caption{In the general case, for arbitrary $(\xi,\zeta)$ (in green), the background breaks all SUSY. Along the $(\zeta=-1,\gamma=0)$ line (in blue), the $U(1)_R$-symmetry is preserved, leading to $\mathcal{N}=1$ solutions. Along the $\gamma=\zeta=0$ line (in red), the background preserves $SU(2)$ isometry (descending from the original R-symmetry) with the SUSY completely broken in general. Given the red and blue lines do not intersect (as they are parallel in the $\zeta$ axis), there are no $\mathcal{N}=2$ solutions here - as no background preserves the necessary $SU(2)_R\times U(1)_R$ R-symmetry.}
    \label{fig:IIBtableplotchired}
\end{figure}

\item \textbf{Performing an ATD long $\phi$}\\
Performing an ATD along $\phi$ leads to the following $\mathcal{N}=0$ family of solutions
  \begin{adjustwidth}{-2cm}{}
     \begin{align}\label{eqn:chiredphiatd}
&       ds_{10,st}^2= \frac{1}{X}f_1^{\frac{3}{2}}(f_5 f_6^2+f_3)^{\frac{1}{2}}\sqrt{\Xi_2}\bigg[4ds^2(\text{AdS}_5)+f_4(d\sigma^2+d\eta^2)+ds^2(M_3)\bigg], \nn\\[2mm]
& ds^2(M_3)=  f_2d\theta^2  +\frac{1 }{(f_5f_6^2+f_3)\Delta_2}\bigg( f_3f_5  d\beta^2
       +\frac{X^2}{f_1^3f_2 \sin^2\theta} D\phi^2 \bigg),   ~~~D\phi = d\phi -\frac{1}{X}(f_7+\xi f_8)\sin\theta\,  d\theta, \nn\\[2mm]
& C_2= -\frac{\gamma}{\zeta} X d\phi\wedge d\beta,~~~~~~~~~~~  C_0=   \zeta X \frac{f_2}{(f_5 f_6^2+f_3)\Xi_2} \sin^2\theta -\frac{X}{\zeta} ,~~~~~~~~e^{2\Phi^B}=\frac{(f_5 f_6^2+f_3)}{X^2 f_2 \sin^2\theta}\frac{\Xi_2^2 }{\Delta_2}, 
\nn\\[2mm]
&B_2=-\frac{ \zeta }{   (f_5f_6^2+f_3)\Delta_2 } \bigg( f_5(f_6+\xi)d\phi+\frac{1}{X}\Big(f_3f_8 +f_5(f_6f_8-f_7)(f_6+\xi)\Big)\sin\theta d\theta \bigg) \wedge d\beta + \gamma \,d\phi \wedge d\beta ,\nn\\[2mm]
  &\Xi_2= \Delta_2+\zeta^2 \frac{f_2}{\big(f_5f_6^2+ f_3  \big)} \sin^2\theta,~~~~~~~~\Delta_2=1+\xi \frac{f_5(2f_6+\xi)}{\big(f_5f_6^2+ f_3 \big)},~~~~~~~~~~
  \end{align}
  \end{adjustwidth}
where we see that $\gamma$ only emerges as a large gauge transformation of $B_2$. Here we have introduced a gauge transformation into $C_0$ to simplify $C_2$. Once again, 
we note that analogous T-Duals along a $U(1)$ of an $S^2$ can lead to singularities in the dual description (see the discussion in \cite{Itsios:2013wd}).

 \end{itemize}

\subsubsection{Fixing $q\equiv \xi$}
We now investigate the solutions with $(p,b,u)=1,~(a,c,m)=0,~q\equiv \xi,s\equiv \zeta,v\equiv \gamma$, where the $U(1)_R$ component \eqref{eqn:U(1)} now becomes
  \begin{equation}
U(1)_R= (1+\zeta)\chi +(\xi+\gamma)\beta +\phi,
\end{equation}
where we again perform a dimensional reduction along $\chi$, followed by an ATD along $\beta$ and $\phi$ in turn. 

\begin{itemize}
\item \textbf{Performing an ATD long $\beta$}\\
In this case we calculate a background which can be derived from \eqref{eqn:ATD3}, by fixing $\xi\rightarrow 0,\gamma\rightarrow \gamma-\zeta\, \xi$ before applying the following gauge transformations
\beq
C_0\rightarrow C_0+\xi,~~~~~~~~~C_2\rightarrow C_2+\xi B_2.
\eeq

One then re-derives the existing $\mathcal{N}=1$ solution.

\item \textbf{Performing an ATD long $\phi$}\\
In this final case, one derives a solution which matches \eqref{eqn:chiredphiatd}  with $\xi=0$, followed by the following gauge transformations
\beq
C_2\rightarrow C_2-\xi d\phi\wedge d\beta,~~~~~~~~~B_2\rightarrow B_2 -\zeta\,\xi \,d\phi\wedge d\beta.
\eeq

\end{itemize}
\subsection{U(1)$\times$U(1) preserving $\mathcal{N}=1$ solution}
Fixing $(\zeta=-1,\gamma=0)$ in \eqref{eqn:ATD3} derives the following one parameter family of $\mathcal{N}=1$ solutions
     \begin{align}\label{eqn:IIBchiN1}
   &    ds_{10,st}^2=\frac{1}{X} f_1^{\frac{3}{2}}(f_5f_6^2+f_3)^{\frac{1}{2}} \sqrt{\Xi_2}\bigg[4ds^2(\text{AdS}_5)+f_4(d\sigma^2+d\eta^2)+ ds^2(M_3)\bigg] , \nn\\[2mm]
   & ds^2(M_3) = f_2 \bigg(d\theta^2+\frac{1}{\Pi_2} \sin^2\theta d\phi^2  \bigg)  +\frac{X^2}{f_1^3 f_3f_5\Pi_2 }D\beta^2,~~~~~D\beta=d\beta -\frac{1}{X}f_8\sin\theta \, d\theta,\nn\\[2mm]
 &     C_0=\frac{X f_5 (f_6+\xi) }{(f_5f_6^2+f_3) \Xi_2},~~~~~  C_2= \frac{ \sin\theta }{\Pi_2}\bigg[X\frac{f_2}{f_3 } \sin\theta\, d\beta + f_8 d\theta\bigg] \wedge d\phi ,~~~~~~~~e^{2\Phi_\mathcal{B}}=\frac{ (f_5f_6^2+f_3)^2}{X^2 f_3f_5}\frac{\Xi_2^2}{\Pi_2},   \nn\\[2mm]
&B_2=  \frac{\sin\theta}{\Pi_2} \bigg[ \frac{ f_2}{f_3 }(f_6+\xi)\sin\theta \,d\beta +\frac{1}{X} \bigg((f_7+\xi f_8)- \frac{ f_2}{f_3 }(f_6f_8-f_7)\sin^2\theta\bigg) d\theta\bigg] \wedge d\phi , \nn\\[2mm]
   &\Xi_2= \Delta_2+ \frac{f_2}{\big(f_5f_6^2+ f_3  \big)} \sin^2\theta,~~~~~~~~\Delta_2=1+\xi \frac{f_5(2f_6+\xi)}{\big(f_5f_6^2+ f_3 \big)},~~~~~~~~~~
 \Pi_2=1+\frac{f_2}{f_3}\sin^2\theta,
  \end{align}
which maps to \eqref{eqn:IIBN1} following the transformations in \eqref{eqn:N=1transforms2} (with $\xi\rightarrow 1/\xi$). We then focus on the G-structures for the (unique) $\xi=0$ case, corresponding to the supersymmetry preserving ATD along $\beta$ of the unique IIA $\mathcal{N}=1$ solution discussed in Section \ref{sec:chiN1}. 
\subsubsection{G-structure description}  
Here we follow the same approach as we did in Section \ref{sec:IIBN1one}, this time with $dy\equiv d\beta$ and $(u^\mathcal{A}, v^\mathcal{A}, j^\mathcal{A},\omega^\mathcal{A})$ given in \eqref{eqn:UnqIIAGs} - with $u^\mathcal{A}_{||}=v^\mathcal{A}_{||}=0$ and $u^\mathcal{A}_\perp=u^\mathcal{A},~v^\mathcal{A}_\perp=v^\mathcal{A}$.
Recall the following relations (see \eqref{eqn:A1Aval} with $\phi_1=\beta,\phi_2=\phi$)
\beq
 E^\beta_\mathcal{A}=  e^{C^\mathcal{A}}(d\beta +A_1^\mathcal{A}),~~~~~~~~~~~A_1^\mathcal{A}=\frac{1}{2}\frac{h_{\beta\phi}}{h_{\beta}}d\phi = \frac{f_2f_6\sin^2\theta}{f_3+f_2\sin^2\theta}d\phi,
 \eeq
using \eqref{eqn:chi-Gen} as appropriate. Then to derive the remaining elements we use the decompositions given in \eqref{eqn:jomegadecomps}, leading to
leading to
\beq
\hspace{-1.9cm}
\begin{aligned}
& j_{||}^\mathcal{A}= \frac{ e^{-C^\mathcal{A}} \kappa f_2^{\frac{1}{2}}}{2\big(1+\frac{f_2}{4} (\sin^2\theta-4)\big)^{\frac{1}{2}}}\bigg[f_5f_6\Big[e^{-3\rho}d(\sin^2\theta e^{3\rho}\dot{V}) -\frac{1}{2}d(\sin^2\theta \dot{V}) +\sin^2\theta d(\dot{V})\Big]- \frac{\Lambda f_3}{\dot{V}} d(V')-2\sin^2\theta d\eta \bigg],\\[2mm]
& j_\perp^\mathcal{A}=  \frac{  \kappa f_2^{\frac{1}{2}}}{2\big(1+\frac{f_2}{4} (\sin^2\theta-4)\big)^{\frac{1}{2}}}\Bigg[\bigg( (1-f_2) \Big(d(2\dot{V}\sin^2\theta)+e^{-6\rho}  \sin^2\theta d(2e^{6\rho} \dot{V})\Big) -2f_2 \sin^2\theta d(\dot{V})\bigg) \\[2mm]
&~~~~~- \frac{f_2f_6\sin^2\theta}{f_3+f_2\sin^2\theta} \bigg[f_5f_6\Big(e^{-3\rho}d(\sin^2\theta e^{3\rho}\dot{V}) -\frac{1}{2}d(\sin^2\theta \dot{V}) +\sin^2\theta d(\dot{V})\Big)- \frac{\Lambda f_3}{\dot{V}} d(V')-2\sin^2\theta d\eta \bigg] \Bigg]\wedge d\phi,\nn\\[2mm]
&\omega_{||}^\mathcal{A}= -2i\,e^{-C^\mathcal{A}}\kappa   \frac{\sqrt{f_2}}{\dot{V}}  e^{-3\rho} \, d(\sigma \dot{V}e^{3\rho}e^{i\phi}\sin\theta ) ,\\[2mm]
& \omega_\perp^\mathcal{A} = - 2\kappa   \frac{\sqrt{f_2}}{\dot{V}} \bigg( \sigma \dot{V} e^{-3\rho}  d(V')\wedge d(e^{3\rho}\sin\theta e^{i\phi})+   \tilde{\Delta} \Big(1-\frac{3 }{2}f_2\Big) e^{i\phi} \sin\theta d\sigma\wedge d\eta \\[2mm]
 &~~~~~~~~~~~~~~~~~~~~~~~~~~~~~~~~~~~~~~~~~~~~~~~~~~~~~~~~~~~~~~~~~~~~~ -i \frac{f_2f_6\sin^2\theta}{f_3+f_2\sin^2\theta}\, e^{-3\rho} \, d(\sigma \dot{V}e^{3\rho}e^{i\phi}\sin\theta )\wedge d\phi \bigg),
\end{aligned}
\eeq
and we note $(dy-B_1^\mathcal{A})=(d\beta -f_8\sin\theta d\theta)$. Notice that the $C^\mathcal{A}$ dependence drops out of the expressions. We then build the pure spinors
\begin{equation*}
\begin{aligned}
  \Psi_-^\mathcal{B} &=\frac{1}{8}e^{\Phi_\mathcal{B}-\Phi_\mathcal{A}} e^{\frac{1}{2}z^\mathcal{A}_{\perp}\wedge \overline{z}^\mathcal{A}_{\perp}}\wedge\Big( e^{C^\mathcal{A}}\omega^\mathcal{A}_{||}+   \omega^\mathcal{A}_{\perp}\wedge(d\beta -f_8\sin\theta d\theta)\Big),  \\
   \Psi_+^\mathcal{B}&=\frac{i }{8} e^{\Phi_\mathcal{B}-\Phi_\mathcal{A}}e^{-ij^\mathcal{A}_{\perp}}\wedge    z^\mathcal{A}_{\perp}\wedge \Big( (d\beta -f_8\sin\theta d\theta)-i    e^{C^\mathcal{A}} j^\mathcal{A}_{||}\Big),
\end{aligned}
\end{equation*}
which satisfy the supersymmetry conditions given in \eqref{eqn:IIBGconditions}.
\section{Following the IIA $\phi$ Reduction}\label{sec:followingphi}
We now investigate performing an abelian T-duality following a dimensional reduction along $\phi$, using the same procedure as the previous sections. This case largely mirrors the $\chi$ reduction procedure. 
\subsection{Three-Parameter Families}
Here we will perform an ATD on the two-parameter family of solutions given in \eqref{eqn:phieqmatch}. As in the $\chi$ reduction case, only an ATD along $\beta$ will lead to a SUSY preserved solution. We now investigate the ATD for the $\phi$ reduction with $c\equiv \xi$ and $v \equiv \xi$, in turn.

\subsubsection{Fixing $c\equiv \xi$}
We begin with the case where  $(p,b,u)=1,~(s,a,v)=0,~c\equiv \xi,m\equiv \zeta,q\equiv \gamma$, meaning \eqref{eqn:U(1)} becomes
\beq
U(1)_R= \chi + \gamma\beta +(\zeta+1)\phi.
\eeq
 We then perform a dimensional reduction along $\phi$, followed by an ATD to IIB. We will investigate an ATD along $\beta$ and $\chi$ in turn, beginning with the case which will give rise to an $\mathcal{N}=1$ solution.
   \begin{itemize}
 \item \textbf{Performing an ATD along $\beta$}\\
Using the T-dual formula \eqref{eqn:betaredchitdual}, we get the following three-parameter family of solutions
       \begin{align}\label{eqn:ATD6}
 &      ds_{10,st}^2=\frac{1}{X} f_1^{\frac{3}{2}}  f_2^{\frac{1}{2}} \sin\theta \sqrt{\Xi_3}\bigg[4ds^2(\text{AdS}_5)+f_4(d\sigma^2+d\eta^2)+ds^2(M_3)\bigg], \nn\\[2mm]
  & ds^2(M_3)= f_2d\theta^2 +\frac{1}{\Pi_3} \bigg(  f_3  d\chi^2  +\frac{X^2}{f_1^3 f_2f_5\sin^2\theta } D\beta^2\bigg) , ~~~~~~~D\beta =d\beta +\frac{1}{X}(f_8+\gamma f_7)\sin\theta\,  d\theta,   \nn\\[2mm]
 & C_0= \frac{X}{ f_2 \sin^2\theta \,\Xi_3} \Big(f_5(1+\gamma f_6)(\zeta f_6+\xi)+ \gamma\zeta f_3 \Big) , ~~~~~~~~~~~ e^{2\Phi_\mathcal{B}}=\frac{1}{X^2}\frac{f_2}{f_5}\frac{\Xi_3^2}{\Pi_3}  \sin^2\theta , \nn\\[2mm]
&B_2=\frac{1}{\Pi_3} \bigg[\bigg(f_6(1+\gamma f_6)+\gamma \frac{f_3}{f_5}+\xi(\gamma\xi-\zeta)\frac{f_3}{f_2\sin^2\theta} \bigg) d\beta~~~~~~~~~~~~~~~~~~~~~~~~~~~~~~~~~~~~~~~~~~~~~~~~~~~~~~~~~~    \nn\\[2mm]
&~~~~~~~~~- \sin\theta\frac{1}{X}\bigg( (f_7-f_6f_8)(1+\gamma f_6)-\gamma \frac{f_3f_8}{f_5} -(\gamma\xi-\zeta)(\zeta f_7+\xi f_8)\frac{f_3}{f_2\sin^2\theta} \bigg) d\theta\bigg] \wedge d\chi ,   \nn\\[2mm]
 &  C_2=(\gamma\xi-\zeta)\frac{f_3 }{f_2\sin^2\theta \,\Pi_3}\Big(X d\beta +(f_8+\gamma f_7) \sin\theta d\theta\Big) \wedge d\chi ,\nn\\[2mm]
     & \Xi_3= 1+\frac{f_5(\xi  +\zeta f_6)^2 + \zeta^2 f_3}{ f_2 \sin^2\theta}  \equiv 1+\frac{(\Delta_3+ \widehat{\Delta}_3)}{f_3}\bigg[f_5f_6^2 \bigg(1-\frac{2}{\Delta_3+ \widehat{\Delta}_3}\Pi_3\bigg)+f_3\bigg(1-\frac{2 }{\Delta_3+ \widehat{\Delta}_3}\bigg)  \bigg] ,\nn\\[2mm]
    &\Delta_3=1+\zeta^2 \frac{f_3}{f_2\,\sin^2\theta},~~~~~~\widehat{\Delta}_3= 1+\xi^2 \frac{f_3f_5}{(f_5f_6^2+f_3)f_2\sin^2\theta},
    \nn\\[2mm]
    &\Pi_3=1+\gamma\bigg(f_6(2+\gamma f_6)+\gamma \frac{f_3}{f_5}\bigg)+(\zeta-\gamma\,\xi)^2\frac{f_3}{f_2\sin^2\theta}.
  \end{align}
As in the $\chi$ reduction case discussed in the previous section, we can preserve the $U(1)_R$ component given in \eqref{eqn:chiredU(1)1} by fixing $\zeta=-1$ (to derive the $\mathcal{N}=1$ IIA solution) followed by fixing $\gamma=0$ under the T-duality. Of course, given that we have dimensionally reduced along $\phi$, we have no $S^2$ preserved solutions - see Figure \ref{fig:IIBtableplotphired}.

  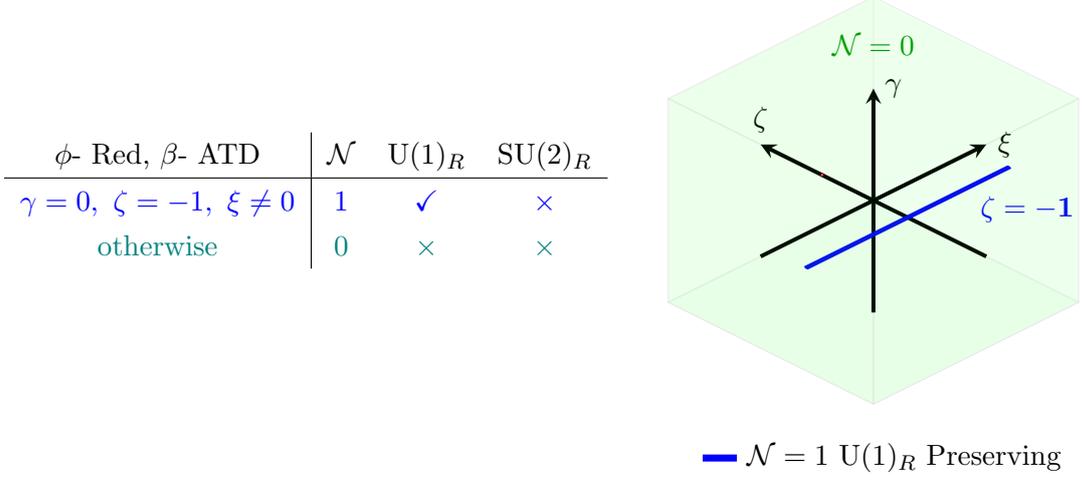
\begin{figure}[H]
\centering  
\subfigure
{
\centering
  \begin{minipage}{0.5\textwidth}
\begin{tabular}{c | c c c c  }
$\phi$- Red, $\beta$- ATD&$\mathcal{N}$&U(1)$_R$&SU(2)$_R$  \\
\hline
$\textcolor{blue}{ \gamma=0,~\zeta=-1,~\xi\neq 0}$&$ \textcolor{blue}{1}$ &$\textcolor{blue}{\checkmark}$&$\textcolor{blue}{\times}$  \\
\textcolor{teal}{otherwise}&$ \textcolor{teal}{0}$ &$\textcolor{teal}{\times}$ &$\textcolor{teal}{\times}$  
\end{tabular}
  \end{minipage}
     \begin{minipage}{.5\textwidth}
    \centering
\begin{tikzpicture}[scale=0.45,y={(1cm,0.5cm)},x={(-1cm,0.5cm)}, z={(0cm,1cm)}]
\draw[-stealth, line width=0.53mm] (0,0,-3.3)--(0,0,3.3) node[right ]{$\mathbf{\gamma}$};
\draw[green!60!black ] (2,2,2) node[above] {$\mathcal{N}=0$};
\begin{scope}[canvas is yx plane at z=0]
\draw[-stealth, line width=0.53mm] (-3.3,0)--(3.3,0) node[right ]{$\xi$};
\draw[-stealth, line width=0.53mm] (0,-3.3)--(0,3.3) node[above ]{$\zeta$};
\clip (-3,-3) rectangle (3,3);
\begin{scope}[cm={0.5,-0.5,  50,50,  (0,0)}]  
\end{scope}
\draw[line width=0.6mm,blue](-3,-1)--(3,-1);
\end{scope}
\begin{scope}[canvas is yx plane at z=-1]
 \node[blue ,rotate=0] at (3,-1.5) {$\mathbf{\zeta=-1}$};
\end{scope}


\draw[fill=green,opacity=0.025] (-3,3,-3) -- (3,3,-3) -- (3,3,3) -- (-3,3,3) -- cycle;
\draw[fill=green,opacity=0.035] (3,-3,-3) -- (3,3,-3) -- (3,3,3) -- (3,-3,3) -- cycle;
\draw[fill=green,opacity=0.05] (3,-3,-3) -- (3,3,-3) -- (-3,3,-3) -- (-3,-3,-3) -- cycle;

\draw[fill=green,opacity=0.05] (-3,-3,-3) -- (3,-3,-3) -- (3,-3,3) -- (-3,-3,3) -- cycle;
\draw[fill=green,opacity=0.05] (-3,-3,-3) -- (-3,-3,3) -- (-3,3,3) -- (-3,3,-3) -- cycle;
\draw[fill=green,opacity=0.05] (-3,-3,3) -- (-3,3,3) -- (3,3,3) -- (3,-3,3) -- cycle;
\begin{scope}[canvas is yz plane at x=1.5]
\coordinate (Origin) at (0,0);
\shade[ball color=red] (Origin) circle (0.05cm);
\end{scope}
\begin{scope}[canvas is yz plane at x=0]
\end{scope}
\end{tikzpicture}
  \end{minipage} 
}
\subfigure
{
\centering
  \begin{minipage}{\textwidth}
    \centering
\begin{tikzpicture}[scale=0.9]
\draw[ blue,line width=0.93mm]  (2,-4.5)--(2.5,-4.5);
\draw (2.5,-4.5) node[right]{$ \mathcal{N}=1$ U(1)$_R$ Preserving};
\draw  (-7,-4.5) node[right]{$~$};
\end{tikzpicture}
  \end{minipage}
}
\caption{In the general case, for arbitrary $(\xi,\zeta)$ (in green), the background breaks all SUSY. Along the $(\zeta=-1,\gamma=0)$ line (in blue), the $U(1)_R$-symmetry is preserved, leading to $\mathcal{N}=1$ solutions. No backgrounds preserves $SU(2)$ isometry, hence there are no $\mathcal{N}=2$ solutions here - as no solutions preserves the necessary $SU(2)_R\times U(1)_R$ R-symmetry.}
    \label{fig:IIBtableplotphired}
\end{figure}

As in previous cases, we once again find that we can map this solution to \eqref{eqn:ATD1} via the transformations given in \eqref{eqn:Transformation} (with $k_1=\frac{1}{\xi},k_2=\frac{\xi^{\frac{5}{2}}}{\zeta-\gamma \xi}$)
 \begin{align}\label{eqn:IIBtrans3}
&g_{MN}\rightarrow \frac{1}{\xi} g_{MN},~~~~~~~~~~~~~ e^{2\Phi_B}\rightarrow \frac{(\zeta-\gamma \xi)^2}{\xi^4} e^{2\Phi_B},~~~~~~~~~~~C_0\rightarrow \frac{\xi^2}{\zeta-\gamma\xi} C_0,\nn\\
&C_2\rightarrow \frac{\xi}{\zeta-\gamma\xi} C_2,~~~~~~~~~~~B_2\rightarrow \frac{1}{\xi}  B_2,\\
& C_0\rightarrow -C_0+\frac{\xi}{\zeta-\gamma\xi},~~~~~C_2\rightarrow -C_2 + \frac{\xi}{\zeta-\gamma\xi} B_2,\nn\\
& \chi\rightarrow -(\zeta-\gamma\xi)\phi ,~~~~~~~~~~\beta \rightarrow -(\zeta-\gamma\xi) \chi,~~~~~~\zeta \rightarrow \frac{\bar{\xi}}{\bar{\zeta}},~~~~\xi\rightarrow \frac{1}{\bar{\zeta}},~~~~\gamma\rightarrow- \frac{1}{\bar{\gamma}} (\bar{\zeta} - \bar{\gamma}\bar{\xi}).\nn
 \end{align}
 
 Here we see that $1/\xi,1/\zeta,1/\gamma$ and $1/(\zeta-\gamma\xi)$ terms are all required in the mapping. We therefore argue that fixing each of these denominators to zero, in turn, derives four separate new and unique solutions.  It does not appear possible to map this solution to \eqref{eqn:ATD2} or \eqref{eqn:ATD3}.

To derive the $\mathcal{N}=1$ solution we must fix $(\gamma=0,\zeta=-1)$, which re-derives \eqref{eqn:IIBchiN1} 
via the following transformations
  \begin{equation*}
  \begin{aligned}
  &C_2\rightarrow C_2-d\beta \wedge d\chi,~~~~~~~~~B_2\rightarrow-B_2+\xi d\beta \wedge d\chi,~~~~~~C_0\rightarrow-C_0,\\[2mm]
  &\chi\rightarrow \phi,~~~~~~~~~~~~~~~~~~~~~~~~~~\beta\rightarrow -\beta,~~~~~~~~~~~~~~~~~~~~~~~~~\xi\rightarrow-\xi.
  \end{aligned}
  \end{equation*}
  

 \item \textbf{Performing an ATD long $\chi$}\\
 In this case we find the following $\mathcal{N}=0$ solution
        \begin{align}\label{eqn:ATD2param6}
 &      ds_{10,st}^2= \frac{1}{X}f_1^{\frac{3}{2}} f_2^{\frac{1}{2}} \sin\theta \, \sqrt{\Xi_3}\bigg[4ds^2(\text{AdS}_5)+f_4(d\sigma^2+d\eta^2)+ds^2(M_3)\bigg], \nn\\[2mm]
  & ds^2(M_3) = f_2d\theta^2+ \frac{1}{(f_3+f_5f_6^2)\,\Pi_3}\bigg(  f_3 f_5d\beta^2 + \frac{X^2}{f_1^3f_2 \sin^2\theta  }D\chi^2\bigg),~~~~~~D\chi = d\chi +\frac{1}{X}f_7\sin\theta \, d\theta, \nn\\[2mm]
&  C_0= \frac{X}{ f_2 \sin^2\theta \, \Xi_3}\Big(f_5f_6 (\zeta f_6+\xi) + \zeta  f_3 \Big) ,~~~~~~~~~~~~~~~~~~~    e^{2\Phi^B}=\frac{1}{X^2}\frac{f_2 \sin^2\theta}{(f_3+f_5f_6^2)}\frac{ \Xi_3^2}{\Pi_3},   
 \nn\\[2mm]
&B_2=\frac{f_5}{(f_3+f_5f_6^2)\,\Pi_3}\bigg[ \bigg(  f_6 - \zeta \xi \frac{ f_3}{f_2\sin^2\theta}  \bigg) d\chi~~~~~~~~~~~~~~~~~~~~~~~~~~~~~~~~~~~~~~~~~~~~~~~~~~~~~~~~~~~~~~~~~~~~~~~~~~~~~~~~~~~~~~~~~~~~~~~~~   \nn\\[2mm]
&~~-\frac{1}{X}  \bigg( f_6(f_6f_8-f_7)+\frac{f_3f_8}{f_5}+\xi\frac{ f_3}{f_2\sin^2\theta} (\zeta f_7+\xi f_8)\bigg) \sin\theta\, d\theta\bigg] \wedge d\beta +\gamma d\chi\wedge d\beta,  \nn\\[2mm]
&  C_2=-\xi  \frac{f_3f_5  }{ f_1 f_2 \sin^2\theta \, \Xi_3}\Big(X d\chi +f_7 \sin\theta d\theta \Big)\wedge d\beta ,   \nn\\[2mm]
   & \Xi_3= 1+\frac{f_5(\xi  +\zeta f_6)^2 + \zeta^2 f_3}{ f_2 \sin^2\theta}  \equiv 1+\frac{(\Delta_3+ \widehat{\Delta}_3)}{f_3}\bigg[f_5f_6^2 \bigg(1-\frac{2}{\Delta_3+ \widehat{\Delta}_3}\Pi_3\bigg)+f_3\bigg(1-\frac{2 }{\Delta_3+ \widehat{\Delta}_3}\bigg)  \bigg] ,\nn\\[2mm]
    &\Delta_3=1+\zeta^2 \frac{f_3}{f_2\,\sin^2\theta},~~~~~\widehat{\Delta}_3= 1+\xi^2 \frac{f_3f_5}{(f_5f_6^2+f_3)f_2\sin^2\theta},~~~~~\Pi_3 =1+ \xi^2  \frac{f_3f_5}{f_2(f_3+f_5f_6^2)\sin^2\theta},
  \end{align}
 where, in a similar manner to \eqref{eqn:chiredphiatd}, we see that $\gamma$ only emerges as a large gauge transformation of $B_2$. 
 \end{itemize}
 
 \subsubsection{Fixing $v\equiv \xi$}
 We now investigate the solutions with $(p,b,u)=1,~(s,a,c)=0,~v\equiv \xi,m\equiv \zeta,q\equiv \gamma$, where the $U(1)_R$ component \eqref{eqn:U(1)} now becomes
\beq
U(1)_R = \chi +(\gamma+\xi)\beta +(\zeta+1)\phi,
\eeq
performing a dimensional reduction along $\phi$, followed by an ATD along $\beta$ and $\chi$ in turn. 
 \begin{itemize}
 \item \textbf{Performing an ATD along $\beta$}\\
In this case we compute a background which is derived via \eqref{eqn:ATD6} by fixing $\xi\rightarrow 0,\gamma\rightarrow \gamma-\zeta\, \xi$, before applying the following gauge transformations
 \beq
 C_0\rightarrow C_0+\xi,~~~~~~~~~~~~C_2\rightarrow C_2 +\xi B_2.
 \eeq

 Then, fixing $(\gamma=-\xi,~\zeta=-1)$ we derive another $\mathcal{N}=1$ solution
  which once again maps to the solutions already given, via
\beq
 \begin{aligned}
  &C_2\rightarrow C_2-d\beta\wedge d\chi,~~~~~~~B_2\rightarrow-B_2,~~~~~~~~C_0\rightarrow -C_0,\\[2mm]
 & \phi\rightarrow \chi,~~~~~~~\chi\rightarrow \phi,~~~~~~~\beta\rightarrow-\beta,~~~~~~~~~~~~\xi\rightarrow-\xi.
  \end{aligned}
  \eeq
 \item \textbf{Performing an ATD along $\chi$}\\
This final case derives a solution where $\gamma$ and $\xi$ are carried through the calculation as gauge transformations of $B_2$ and $C_2$. The background can be derived from \eqref{eqn:ATD2param6} by fixing $\xi=0$ 
before applying the following gauge transformations
  \beq
  B_2\rightarrow B_2-\zeta\, \xi \,d\chi\wedge d\beta,~~~~~~~~~~C_2\rightarrow C_2+\xi \,d\beta \wedge d\chi.
  \eeq

 \end{itemize}
\section{Summary so far}
\begin{itemize}
\item Following an ATD from type IIA, we constructed new three-parameter families of type IIB solutions. These solutions are in general $\mathcal{N}=0$ but contain one parameter families of $\mathcal{N}=1$ solutions, as well as a one-parameter family of $S^2$ preserved $\mathcal{N}=0$ backgrounds. Both families of $\mathcal{N}=1$ IIB solutions presented in this work have the interesting property of a zero five-form flux. We studied the supersymmetry preservation using the method of G-structures and investigated the boundary in an analogous manner to the IIA solutions, finding the presence of orbifold singularities in some cases. In fact, T-dualising within the spindle-like orbifold can lead to solutions for which the orbifold structure is completely broken, but the rational charge is still inherited.
\item The preservation of the holographic central charge under the deformation parameters leads to the conclusion that these solutions are dual to marginal deformations of the `parent' $\mathcal{N}=2$ SCFTs (dual to the GM class). 
\end{itemize}

\chapter{The $\gamma$-deformations of NRSZ}\label{chap:gamma}
\section{M-Theory
}
\begin{figure}[H]
\begin{center} 
 \begin{tikzpicture}

\draw (0,2.4) node[above] {\textbf{GM}};

\draw (0.2,1) node[right] {$\phi_3$};

\draw (0,-0.5) node[below] {\textbf{IIA}};

\draw (2.2,-0.5) node[above] {\textbf{T}};
\draw (2.2,-1) node[below] {ATD $\phi_1$};

\draw (4.5,-0.5) node[below] {\textbf{IIB}};

\draw (6.7,-0.5) node[above] {\textbf{s}};
\draw (6.7,-1) node[below] { $\phi_2\rightarrow \phi_2+\gamma\phi_1$};

\draw (9,-0.5) node[below] {\textbf{IIB}};

\draw (11.2,-0.5) node[above] {\textbf{T}};
\draw (11.2,-1) node[below] {ATD $\phi_1$};

\draw (13.5,-0.5) node[below] {\textbf{IIA}};

\draw (12.5,2.4) node[above] {\textbf{TsT deformed GM} };


\begin{scope}[ every node/.style={sloped,allow upside down}]

 \draw ((0,2)-- node {\midarrow} (0,0);
  \draw ((0.75,-0.75)-- node {\midarrow} (3.75,-0.75);
     \draw ((5.25,-0.75)-- node {\midarrow} (8.25,-0.75);
          \draw ((9.75,-0.75)-- node {\midarrow} (12.75,-0.75);
          
           \draw ((13.5,0)-- node {\midarrow} (13.5,2);
           \draw (13.3,1) node[left] {$\phi_3$};

\end{scope}


 
\end{tikzpicture}
\end{center}
\caption{Perform a dimensional reduction along $\phi_3$, followed by a TsT along $\phi_1$ and subsequent uplift along $\phi_3$.}
\label{fig:GMTSTtransformations }
\end{figure}
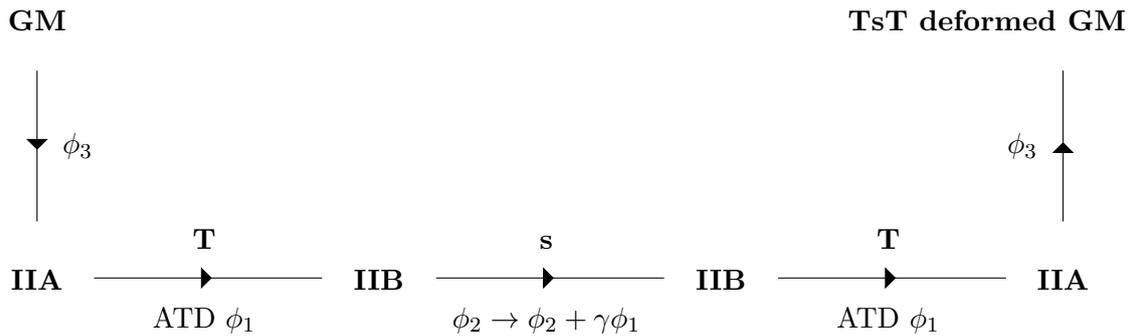

In this chapter we discuss a one parameter supersymmetry breaking deformation of the GM background in $d=11$, which was first derived in \cite{Nunez:2019gbg} (and later \cite{Merrikin:2024bmv}). As we recall, this solution has three $U(1)$ directions which we will for the moment label generically as $(\phi_1,\phi_2,\phi_3)$. The approach here is to dimensionally reduce along one of the $U(1)$ directions, $\phi_3$, before performing a TsT transformation down in type IIA by utilising the two remaining $U(1)$ directions $(\phi_1,\phi_2)$ and the formulas given in \eqref{eqn:TSTgenlgeneral}, \eqref{eqn:TSTgenl}, before uplifting along $\phi_3$. See Figure \ref{fig:GMTSTtransformations } for a graphical representation of these steps. This transformation will then pick up a non-trivial free parameter, which recovers the original GM background when fixed to zero, and breaks the $\mathcal{N}=2$ supersymmetry of the background to $\mathcal{N}=0$. 
\newpage
Using the reduction formula given in \eqref{eqn:reductionformula} (with $\psi\equiv \phi_3$), we uplift \eqref{eqn:TSTgenl} as follows
\begin{adjustwidth}{-1cm}{}
\vspace{-0.7cm}
       \begin{align}\label{eqn:TSTuplift}
 &      ds^2= \bigg(\frac{\delta_2 +\gamma_1^2 \delta_1}{\delta_2}\bigg)^{\frac{1}{3}}e^{-\frac{2}{3}\Phi_A}\Bigg[ds_8^2 +\frac{1}{\delta_2 + \gamma_1^2 \delta_1}\bigg[\delta_1 \delta_2(d\phi_2 -\gamma_1 B_{2,\phi_1}d\theta)^2~~~~~~~~~~~~~~~~~~~~~~~~~~~~~~~~~~~~~~~~~~~~~~~~~~~~~~\nn\\[2mm]
  & ~~~~~~~~~~~~~    + \bigg(d\phi_1 +\frac{1}{2}\frac{h_{\phi_1\phi_2}}{h_{\phi_1}}d\phi_2 +\gamma_1 \Big(B_{2,\phi_2}-\frac{1}{2}\frac{h_{\phi_1\phi_2}}{h_{\phi_1}}B_{2,\phi_1}\Big)d\theta\bigg)^2\,\bigg]\Bigg]\nn\\[2mm]
 &   ~~~~~  + \bigg(\frac{\delta_2}{\delta_2 +\gamma_1^2 \delta_1}\bigg)^{\frac{2}{3}}e^{\frac{4}{3}\Phi_A} \bigg(d\phi_3 + C_{1,\phi_2}d\phi_2 +C_{1,\phi_1}d\phi_1 +\gamma_1 (C_{3,\phi_1\phi_2} + C_{1,\phi_1}B_{2,\phi_2} -C_{1,\phi_2}B_{2,\phi_1})d\theta\bigg)^2,\nn\\[2mm]
&C_3=\frac{\delta_2}{\delta_2+\gamma_1^2 \delta_1}C_{3,\phi_1\phi_2}d\theta \wedge d\phi_2 \wedge d\phi_1 + \frac{\delta_2}{\delta_2+\gamma_1^2\delta_1}(B_{2,\phi_2}d\phi_2 +B_{2,\phi_1}d\phi_1)\wedge d\theta \wedge d\phi_3\nn\\[2mm]
&~~~~~~~~-\frac{\gamma_1 \delta_1}{\delta_2+\gamma_1^2\delta_1}d\phi_2 \wedge d\phi_1\wedge d\phi_3,\nn
 ~~~~~~~~~~~~~     \delta_1=\Gamma  \bigg(h_{\phi_2}-\frac{1}{4}\frac{h_{\phi_1\phi_2}^2}{h_{\phi_1}}\bigg),
       ~~~~~~~~~~\delta_2=\frac{1}{h_{\phi_1} \Gamma }.
  \end{align}
  \end{adjustwidth}
This result is sufficient for the GM background, which we will now investigate. Of course, the more general form given in \eqref{eqn:TSTgenlgeneral} follows in the same manner. 

\subsubsection{One-Parameter deformation of GM}
For the GM background, we have six iterations corresponding to $(\phi_1,\phi_2,\phi_3)=(\chi,\phi,\beta)$ etc. This is the choice which we will use in the following analysis, which in fact is sufficient, as it seems each of the six alternatives will lead to the same background (up to relabelling $\gamma\rightarrow-\gamma$). Using \eqref{eqn:TSTuplift} for the specific case in question, where  
\begin{align}
&ds_8^2= f_1 e^{\frac{2}{3}\Phi_A}  \Big[4ds^2(AdS_5)+f_2d\theta^2 +f_4(d\sigma^2+d\eta^2)\Big],\nn\\[2mm]
&e^{\frac{4}{3}\Phi_A} = f_1f_5,~~~~~~~~~~~~~~~~~~~~\Gamma = f_1 e^{\frac{2}{3}\Phi_A},~~~~~~~~~~~C_{1,\phi_1}=C_{1,\chi}=f_6,~~~~~~~~~~C_{1,\phi_2}=C_{1,\phi}=0,\nn\\[2mm]
&B_{2,\phi_1} = B_{2,\chi}=0  ,~~~~~~~~~~~~~~~
 B_{2,\phi_2} = B_{2,\phi}=-f_8\sin\theta,~~~~~~~~~~~~~C_{3,\phi_1\phi_2}=C_{3,\chi\phi} = f_7\sin\theta, \nn\\[2mm]
&h_{\phi_1} = h_\chi=f_3,~~~~~~~~~~~~~~~~~ h_{\phi_2} =h_\phi=f_2\sin^2\theta,~~~~~~~~ h_{\phi_1\phi_2} =h_{\chi\phi}=0,
\end{align}
we re-derive the following one-parameter deformation of the GM background
\beq\label{eqn:GMdeformed}
\hspace{-1.4cm}
 \begin{aligned}
    &   ds^2= f_1 Z^{\frac{1}{3}}  \Bigg[  4ds^2(AdS_5)+f_4(d\sigma^2+d\eta^2) +ds^2(M_4)\Bigg] , \quad\quad\quad\quad\quad\quad\quad\quad\quad\quad\quad\quad\quad\quad\quad \\[2mm]
 &ds^2(M_4)=f_2\bigg( d\theta^2+\frac{1}{Z }\sin^2\theta d\phi^2  \bigg)+ \frac{1}{Z } \bigg(f_3 (d\chi-\gamma f_8 \sin\theta d\theta )^2  +f_5\Big( d\beta +f_6 d\chi +\gamma (f_7-f_6f_8)\sin\theta d\theta \Big)^2 \bigg)  ,\\[2mm]
&A_3=\frac{1}{Z
}\bigg( f_7\sin\theta \,d\chi \wedge d\theta  + f_8\sin\theta  \,d\beta \wedge d\theta + \gamma   f_1^3f_2f_3f_5\sin^2\theta \,d\beta \wedge d\chi    \bigg)\wedge d\phi ,
\end{aligned}
\eeq
  where we have defined
  \beq\label{eqn:GMdeformedZval}
  Z=1+\gamma^2 f_1^3f_2f_3f_5\sin^2\theta.
  \eeq
It is easy to see that $Z\rightarrow 1$ when $\gamma=0$, recovering the GM background \eqref{eqn:GM}.  When keeping track of the $U(1)_R$ component through this calculation shows that, at best, it is necessarily broken by the TsT transformation prior to the uplift. In addition, it is clear to see from \eqref{eqn:GMdeformed} and \eqref{eqn:GMdeformedZval} that the $SU(2)_R$ component is broken by $\gamma$. Hence, $\gamma\neq 0$ breaks the $\mathcal{N}=2$ solution to $\mathcal{N}=0$. We will now demonstrate this more explicitly by investigating the G-structure description of this background.
\paragraph{Supersymmetry breaking}
We now investigate the G-structure description of this background. We find a simple deformation to the GM description, where 
\begin{equation}
\begin{aligned}
ds^2 &=e^{2\hat{A}}ds^2(\text{Mink}_4) +ds_7^2,~~~~~~~~~~~~~~~e^{2\hat{A}}=4f_1  Z^{\frac{1}{3}}  e^{2\rho},\\[2mm]
ds_7^2 &= \sum_{a=1}^{3}E^a \bar{E}^{\bar{a}}+K^2= f_1Z^{\frac{1}{3}} \Bigg[4d\rho^2 +f_2d\theta^2 +f_4(d\sigma^2+d\eta^2)  \quad\quad\quad\quad\quad\quad\quad\quad\quad\quad\quad\quad\quad\quad\quad \nn\\[2mm]
 &     ~~~~~~~~~~~ ~+ \frac{1}{Z} \bigg( f_2\sin^2\theta d\phi^2 +f_3 (d\chi-\gamma f_8 \sin\theta d\theta )^2  +f_5\Big( d\beta +f_6 d\chi +\gamma (f_7-f_6f_8)\sin\theta d\theta \Big)^2 \bigg)  \Bigg] ,\nn
\end{aligned}
\end{equation}
  which then leads to a simple deformation of the vielbeins \eqref{eqn:GstructureForms}, which become
  \beq\label{eqn:GstructureForms2}
     \begin{aligned}
&~~~~~~~~~~~~~~~~~~~~~~~~~~~~~~~~~~~~~~~~~~   \textbf{$\beta$ reduction frame} \\
&K= \frac{\kappa\, e^{-2\rho}}{f_1}Z^{\frac{1}{6}}d(\cos\theta e^{2\rho}\dot{V}),\\
&E_1 =  -Z^{\frac{1}{6}} \sqrt{f_1f_3} \bigg(\frac{1}{\sigma} d\sigma+ d\rho   +\frac{i}{\sqrt{Z}} ( d\chi -\gamma f_8 \sin\theta d\theta)\bigg),\\[2mm]
&E_2= \frac{\kappa \,e^{-2\rho} e^{i\phi}}{f_1} Z^{\frac{1}{6}} \bigg(d(\sin \theta e^{2\rho}\dot{V})+  \frac{i}{\sqrt{Z}} \, \dot{V} e^{2\rho} \sin \theta\, d\phi \bigg),\\[2mm]
&E_3 =-e^{i\chi}Z^{\frac{1}{6}}\sqrt{f_1f_5} \,\Bigg[-\frac{1}{4}f_3 \frac{\dot{V}'}{\sigma} d\sigma -V''d\eta +f_6 d\rho +\frac{ i }{\sqrt{Z}}\Big(d\beta +f_6d\chi +\gamma (f_7-f_6 f_8)\sin\theta d\theta \Big)\Bigg].
\end{aligned}
\eeq
One could now use the methods outlined in Section \ref{sec:framerotation}, or otherwise, to perform a rotation to an alternative reduction frame.

We can build the real two-form and complex three-form $(J,\Omega)$
     \begin{equation*}
\Omega= E^1\wedge E^2\wedge E^3  ,~~~~~~~~~~~~~~~ J=\frac{i}{2}(E^1\wedge \overline{E}^1+E^2\wedge \overline{E}^2+E^3\wedge \overline{E}^3),
\end{equation*}
and check the G-structure conditions given in \eqref{eqn:Geqns}. We find
\beq
\begin{aligned}
d(e^{2\hat{A}} K)&=\gamma^2 W_2,~~~~~~~~~~~~
d(e^{4\hat{A}} J)-e^{4\hat{A}} *_7G_4=\gamma W_3,\\[2mm]
d(e^{3\hat{A}} \Omega)&= \gamma W_4,~~~~~~~~~~~
d(e^{2\hat{A} }J\wedge J )+2 e^{2\hat{A} }G_4\wedge K=\gamma W_5,
\end{aligned}
\eeq
demonstrating that $\gamma$ does indeed break supersymmetry, with $W_k$ some $k$-form.

Investigating the boundary of this solution, we find the $\gamma$ does not lead to orbifold singularities but it does break the $S^2$ along the $\sigma=0,\eta\in (0,P)$ boundary (excluding the end points of $\eta$). 
\section{Type IIA}

Dimensionally reducing \eqref{eqn:GMdeformed}, using the reduction formula \eqref{eqn:reductionformula}
 leads to
\beq 
\hspace{-0.5cm}
 \begin{aligned}
    &   ds^2=  f_1^{\frac{3}{2}} f_5^{\frac{1}{2}}   \Bigg[  4ds^2(AdS_5)+f_4(d\sigma^2+d\eta^2) +f_2\Big( d\theta^2+\frac{1}{Z }\sin^2\theta d\phi^2  \Big)+ \frac{f_3 }{Z }  \Big(d\chi-\gamma f_8 \sin\theta d\theta \Big)^2 \Bigg] ,\\[2mm]
&C_1=f_6 d\chi +\gamma (f_7-f_6f_8)\sin\theta d\theta,~~~~~~~~~~~e^{\frac{4}{3}\Phi}= f_1f_5Z^{-\frac{2}{3}}  ,~~~~~~~~~  Z=1+\gamma^2 f_1^3f_2f_3f_5\sin^2\theta,\\[2mm]
&C_3= \frac{1}{Z }f_7\sin\theta \,d\chi \wedge d\theta\wedge d\phi,~~~~~~ B_2= \frac{1}{Z }\bigg(  f_8\sin\theta \, d\theta + \gamma   f_1^3f_2f_3f_5\sin^2\theta \,  d\chi    \bigg)\wedge d\phi ,
\end{aligned}
\eeq
where one can map to the form derived in \cite{Nunez:2019gbg} by the following mappings (which we label with a bar for clarity),  
  \begin{equation}\label{eqn:ftransformations}
  \begin{aligned}
 & f_3 =\frac{\bar{f}_4}{\bar{f}_1},~~~~~~~~~  f_2 =\frac{\bar{f}_3}{\bar{f}_1},~~~~~~~~~~~ f_5 =\frac{\bar{f}_8}{\bar{f}_1},~~~~~~~~~~~ f_4 =\frac{\bar{f}_2}{\bar{f}_1},\\
  & f_8=\bar{f}_5,  ~~~~~~~~~~  f_6=\bar{f}_6,~~~~~~~~~~~ f_7=\bar{f}_7,~~~~~~~~~~~  f_1^{\frac{3}{2}}  f_5^{\frac{1}{2}}=\kappa^2\Lambda^{\frac{1}{2}} = \kappa^2 \bar{f}_1,
   \end{aligned}
  \end{equation} 
with $(\theta =\bar{\chi},~\chi =\bar{\beta},~\phi =\bar{\xi})$ and appropriate rescaling.
\paragraph{Supersymmetry breaking}
We are now in a position to derive the G-structure description for this background, using \eqref{eqn:GstructureForms2} and the reduction formula \eqref{eqn:11Dto10Dforms}
\begin{equation} 
\hspace{-1.5cm}
\begin{aligned}
 v& =  \kappa\, e^{-2\rho}  f_1^{-\frac{3}{4} }f_5^{\frac{1}{4} } d(\cos\theta e^{2\rho}\dot{V}),~~~~~~~~~~
u= (f_1f_5)^{\frac{3}{4}}   \Big( \frac{f_3 \dot{V}'}{4\sigma} d\sigma +V''d\eta -f_6 d\rho\Big),\nn\\[2mm]
\omega& = \kappa \,e^{i(\chi+\phi)} e^{-2\rho}(f_3f_5)^{\frac{1}{2}} \bigg(\frac{1}{\sigma} d\sigma+ d\rho   +\frac{i}{\sqrt{Z}} ( d\chi -\gamma f_8 \sin\theta d\theta)\bigg)\wedge  \bigg(d(\sin \theta e^{2\rho}\dot{V})+  \frac{i}{\sqrt{Z}} \, \dot{V} e^{2\rho} \sin \theta\, d\phi \bigg),\\[2mm]
j&=\frac{1}{\sqrt{Z}} \bigg(\frac{f_1^{\frac{3}{2}}f_5^{\frac{1}{2}} f_3}{\sigma} e^{-\rho}d\left(e^{\rho}\sigma\right)\wedge  ( d\chi -\gamma f_8 \sin\theta d\theta)+\kappa\frac{\sigma f_2 f_3^{-\frac{1}{2}}e^{-4\rho}}{ \dot{V}^2}d(e^{4\rho}\sin^2\theta \dot{V}^2)\wedge d\phi \bigg) ,
\end{aligned}
\end{equation}
with
\begin{align}
\hat{E}^1 &= e^{\frac{1}{3}\Phi} E^1 = -   f_1^{\frac{3}{4}} f_5^{\frac{1}{4}} f_3^{\frac{1}{2}} \bigg(\frac{1}{\sigma} d\sigma+ d\rho   +\frac{i}{\sqrt{Z}} ( d\chi -\gamma f_8 \sin\theta d\theta)\bigg),\nn\\[2mm]
\hat{E}^2 &=e^{\frac{1}{3}\Phi} e^{-i\hat{\theta}_+} E^2 =- \kappa \,e^{i(\phi+\chi)}  e^{-2\rho} f_1^{-\frac{3}{4}} f_5^{\frac{1}{4}}  \bigg(d(\sin \theta e^{2\rho}\dot{V})+  \frac{i}{\sqrt{Z}} \, \dot{V} e^{2\rho} \sin \theta\, d\phi \bigg),
\end{align}
which re-derive the above $(j,\omega)$ via \eqref{eqn:IIAomegaj1}, and define the metric in the following manner
  \begin{align} 
&ds_{10}^2=e^{2A}ds^2(\text{Mink}_4)+
\hat{E}^1\bar{\hat{E}}^{1}+ \hat{E}^2 \bar{\hat{E}}^{2}+u^2+v^2,~~~~~~~~
e^{2A}=e^{2\hat{A}+\frac{2}{3}\Phi}=4  f_1^{\frac{3}{2}} f_5^{\frac{1}{2}} e^{2\rho} .\nn
\end{align}
 
By constructing the pure spinors \eqref{eqn:Psi}, we can check the conditions \eqref{eqn:IIAGconditions},  
\begin{align}
d_{H}(e^{3A-\Phi}\Psi_+)&=\gamma \,(Y_3+ Y_5 ), \nn\\
d_{H}(e^{2A-\Phi}\text{Re}\Psi_-)&=\gamma\,(\gamma\, Y_2+Y_4+\gamma \,Y_6), \\
d_{H}(e^{4A-\Phi}\text{Im}\Psi_-)-\frac{e^{4A}}{8}*_6\lambda(g)&=\gamma\,(\gamma \,\tilde{Y}_2+\tilde{Y}_4+\gamma \,\tilde{Y}_6), \nn
\end{align}
with the appropriate $k$-forms, $(Y_k,\tilde{Y}_k)$. We have now verified explicitly that $\gamma$ breaks supersymmetry by breaking all three conditions. 
 
\section{Type IIB}\label{sec:gammaIIB}
Let us now fix $\zeta=-\xi$ in \eqref{eqn:ATD1}, recalling that this is the first of the two necessary conditions to preserve $\mathcal{N}=1$ in the IIB theory (along with $\gamma=-1$). This of course derives a two parameter family of $\mathcal{N}=0$ backgrounds, enhancing to the $\mathcal{N}=1$ solution for $\gamma=-1$. However, let us for the moment keep $\gamma$ free and instead fix $\xi=0$ (with $X=1$), one then finds
\beq
\hspace{-0cm}
     \begin{aligned} 
&       ds_{10,st}^2= f_1^{\frac{3}{2}}f_5^{\frac{1}{2}}\bigg[4ds^2(\text{AdS}_5)+f_2d\theta^2+f_4(d\sigma^2+d\eta^2) + \frac{f_2f_3 \sin^2\theta d\phi^2 }{f_3 +\gamma^2 f_2\sin^2\theta}+\frac{(d\chi  -\gamma f_8  \sin\theta d\theta)^2}{f_1^3( f_3f_5  +\gamma^2\sin^2\theta f_2 f_5) }\bigg], \\[2mm]
&e^{2\Phi^B}=\frac{f_5}{ f_3 +\gamma^2 f_2\sin^2\theta },~~~~~~~~~~B_2=  \frac{  f_3  f_8 }{ f_3 +\gamma^2 f_2\sin^2\theta }     \sin\theta d\theta \wedge   d\phi  +\frac{\gamma f_2   }{f_3 +\gamma^2\sin^2\theta f_2} \sin^2\theta  d\chi  \wedge   d\phi  ,  \\[2mm]
&C_0=f_6 ,~~~~~~~~~~~~~~ 
  C_2=\frac{f_3f_7+\gamma^2f_2(f_7-f_6f_8)\sin^2\theta}{f_3 +\gamma^2 f_2\sin^2\theta} d\theta \wedge d\phi +\frac{\gamma f_2f_6}{f_3 +\gamma^2 f_2\sin^2\theta} \sin^2\theta   d\chi \wedge  d\phi .
  \end{aligned}
  \eeq
 Because $\xi=0$, the quantization of charge is automatically restored, allowing for $X=1$.
  We have in fact re-derived the TsT background presented in \cite{Nunez:2019gbg}. Rewriting the GM warp factors in terms of the definitions of \cite{Nunez:2019gbg}, by using \eqref{eqn:ftransformations},
   one arrives at 
            \begin{align}
 &       ds_{10,B}^2=\kappa^2\bigg[4\bar{f}_1ds^2(\text{AdS}_5)+ \bar{f}_3d\theta^2+ \bar{f}_2(d\sigma^2+d\eta^2)~~~~~~~~~~~~~~~~~~~~~~~~~~~~~~~~~~~~~~~~~ \nn\\[2mm]
 &  ~~~~~~~~~~~~~~~~~~~~~~~     +\frac{1}{\bar{f}_4+\gamma^2  \bar{f}_3 \sin^2\theta}\Big(\bar{f}_3 \bar{f}_4\sin^2\theta d\phi^2+\frac{1}{\kappa^4}(d\chi-\gamma\,\bar{f}_5 \sin\theta  d\theta )^2\Big)\bigg], \nn\\[2mm]
 &e^{2 \Phi^B}=\frac{\bar{f}_8}{\bar{f}_4+\gamma^2 \bar{f}_3\sin^2\theta},~~~~~~~~~~~~~
               B^B  =\frac{\gamma \bar{f}_3 \sin^2\theta}{\bar{f}_4+\gamma^2\bar{f}_3\sin^2\theta}( d\chi -\gamma \bar{f}_5\sin\theta d\theta)\wedge d\phi +\bar{f}_5\sin\theta   d\theta\wedge  d\phi  , \nn\\[2mm]
&C_0=\bar{f}_6,~~~~~~~~~~~
C_2=\bar{f}_7 \sin\theta d\theta \wedge d\phi + \frac{\gamma \bar{f}_6\bar{f}_3 \sin^2\theta}{\bar{f}_4+\gamma^2\bar{f}_3\sin^2\theta}( d\chi -\gamma \bar{f}_5\sin\theta d\theta)\wedge d\phi,
    \end{align} 
with $(\theta =\bar{\chi},~\chi =\bar{\beta},~\phi =\bar{\xi})$. Here, we have jumped in at the `s' stage of their TsT transformation, making a coordinate transformation in IIA before T-Dualising to IIB.
We can now say that the TsT solution of \cite{Nunez:2019gbg} is in fact an $\mathcal{N}=1$ background when $\gamma=-1$, and SUSY broken otherwise. It is also a specific example of the three parameter family of solutions given in \eqref{eqn:ATD1} (with $\zeta=-\xi=0$).

It is interesting to note this $\gamma=-1$ IIB solution of NRSZ \cite{Nunez:2019gbg} (corresponding to the $\zeta=-\xi=0,\gamma=-1$ case of \eqref{eqn:ATD1}) appears to be the only $\mathcal{N}=1$ solution presented in this work (and the work of \cite{Nunez:2019gbg}) which preserves integer quantization (and hence a dual Lagrangian description) in either IIA or IIB. This solution is then the most likely candidate for the holographic dual to the $\mathcal{N}=1$ marginal deformation of \cite{Nunez:2019gbg}. Hence, it would be additionally interesting to consider this $\mathcal{N}=1$  solution within the context of the soft-SUSY deformations reviewed in Section \ref{sec:softSUSY}.

\paragraph{Supersymmetry breaking}
The G-Structure analysis is a particular case of the more general results given in Section \ref{sec:GstructuresIIB} (after fixing $\zeta=-\xi=0$), where 
  \begin{subequations} 
\begin{align}
d_{H^\mathcal{B}_3}(e^{3A-\Phi_\mathcal{B}}\Psi^\mathcal{B}_-)&=(\gamma+1) \Big(U_2+U_4+U_6\Big),\\
d_{H^\mathcal{B}_3}(e^{2A-\Phi_\mathcal{B}}\text{Re}\Psi^\mathcal{B}_+)&=0,\\
d_{H^\mathcal{B}_3}(e^{4A-\Phi_\mathcal{B}}\text{Im}\Psi^\mathcal{B}_+)-\frac{e^{4A}}{8}*_6\lambda(g) &=0  ,
\end{align}
 \end{subequations}
for some $k$-form, $U_k$. Interestingly, only the domain-wall BPSness condition is broken by $\gamma$ (in a similar fashion to \cite{Lust:2008zd}), with $\mathcal{N}=1$ supersymmetry clearly recovered for $\gamma=-1$.

\addcontentsline{toc}{chapter}{Conclusions \& Future directions}
\chapter*{Conclusions \& Future directions}
\label{Conclusions}

We now give a brief summary of the results presented throughout this thesis.
\begin{itemize}
\item In Chapter \ref{chap:setup}, we derived the G-structure forms for the $\mathcal{N}=2$ GM class of solutions, for both ten and eleven dimensions. We perform an SL$(3,\mathds{R})$ transformation of the $d=11$ solution, and devise a method to rotate the G-structure forms in general. In addition, we analyse the ten-dimensional $\mathcal{N}=2$ solution at the boundary. We recover (the already known) brane set-up, consisting of stacks of D6 brane sources along the $\sigma=0$ boundary, at the positions of the kinks of the rank function (at $\eta=k\in\mathds{Z}$ with $k\in (0,P)$). In addition, there are $P$ NS5 branes at $\sigma\rightarrow\infty$, and $N_k$ D4 branes at $\sigma=0$ in each $\eta\in [k,k+1]$ interval, both considered pure flux.
\item  In Chapter \ref{chap:typeIIA}, we present three separate two-parameter families of type IIA solutions, corresponding to dimensional reductions along each of the three $U(1)$ directions, $(\beta,\chi,\phi)$, in turn.
\begin{itemize}
\item In Section \ref{sec:betaRed}, the $\beta$ reduction is considered. This solution recovers the $\mathcal{N}=2$ GM solution when both parameters are fixed to zero. Investigations at the boundary demonstrate the existence of orbifold singularities due to the additional parameters. In the $\zeta=0$ case, we find multiple neighbouring stacks of D6 branes, each orthogonal to its own spindle - with conical deficit angles defined by the slope of the rank function at that point. In the more general $\zeta\neq0$ case, we find higher dimensional analogues of the spindle, which take a more generalized form to previous examples found within supergravity. The relevant Euler characteristics are derived. These orbifolds then give rise to rational quantization of D6 charge (as a consequence of the rational form of the Euler characteristic of the corresponding spindle-like manifolds). Investigating the quantization further, along with the holographic central charge, leads to the conclusion that the two parameters $(\xi,\zeta)$ correspond to marginal deformations of the CFT. Because $\xi$ gives rise to rational quantization, we interpret these deformations as non-Lagrangian in the dual CFT description - breaking the Lagrangian nature of the theory. The $\zeta$ parameter appears to preserve the Lagrangian nature of the dual CFT, as integer quantization is unaffected. The D4 branes are only present for a preserved $S^2$ (with $\zeta=0$), however $P$ NS5 branes at $\sigma\rightarrow \infty$ remain in all solutions. 

We derive the G-structure forms for these solutions, finding the solutions are in general $\mathcal{N}=0$ - with a special $\mathcal{N}=1$ subclass for $\zeta=-\xi$. We find all three G-structure conditions are broken in general, the first explicit examples of the gauge BPSness condition being broken. This leads to the consideration of a new pair of pure spinors for which this condition is imposed. We interpret the supersymmetry breaking deformations as soft-SUSY breaking in the CFT description, and suggest the corresponding operators from the literature. 

The stability of D6 probes are considered in the $\zeta=0$ case, showing no sign of instability. Spin-2 analysis is also included for a subset of solutions, with a lower bound provided for the spectrum of dimensions.
\item  In Section \ref{sec:chiRed}, the $\chi$ reduction is considered. We demonstrate mappings to the $\beta$ reduction solutions, but due to the requirement that the parameters are integer, we propose that these solutions are physically different. We repeat the boundary analysis, showing very similar results to before - with neighbouring spindles and higher dimensional versions as in the $\beta$ reduction case, but defined by different conical deficit angles. The relevant Euler characteristics are derived. Switching both parameters off no longer recovers the GM solution, and in general, rational quantization seems a necessity for all solutions. A special $\mathcal{N}=1$ subclass is found, when $\zeta=-1$, with the G-structure forms provided. We again only find D4 branes for a preserved $S^2$, with $P$ NS5 branes at $\sigma\rightarrow \infty$ - although this limit is now a little more subtle, with D4 charge present for some solutions.
\item  In Section \ref{sec:phiRed}, the $\phi$ reduction is considered. In these solutions, the $S^2$ is broken in all cases. Again certain mappings to other solutions exist, but again we interpret these solutions as physically different. A special $\mathcal{N}=1$ subclass is again found for $\zeta=-1$. Investigations at the boundary are again repeated, however things behave differently to the previous two solutions. We no longer find the spindle (due to the broken $S^2$) but we do still find higher dimensional manifolds which include orbifold singularities - with the corresponding Euler characteristic derived. We still find rational quantization of D6 brane charge, however there now exists two separate integration cycles for D6 branes. NS5 branes are still found along the $\sigma\rightarrow\infty$ boundary, but no D4 branes exist along $\sigma=0$. 
\end{itemize}
\item In Chapter \ref{chap:typeIIB}, abelian T-duality from type IIA to type IIB is considered, with the G-structure forms and conditions presented. The three type IIA solutions are then T-dualised to type IIB along both remaining $U(1)$ directions, deriving three-parameter families of solutions. These solutions contain special $\mathcal{N}=1$ classes, which contain zero five-form flux. These infinite classes of solutions are potentially interesting in their own right.
 \begin{itemize}
 \item In Section \ref{sec:followingbeta}, the $\beta$ reduction solution of Section \ref{sec:betaRed} is T-dualised along both $\chi$ and $\phi$ (with a special $\mathcal{N}=1$ subclass existing in the first of these two solutions). The preservation of SUSY now requires two conditions, $\zeta=-\xi$ and $\gamma=-1$. We investigate the boundary of T-duality along $\chi$, finding stacks of $P$ NS5 branes at $\sigma\rightarrow\infty$ as in the IIA solution. This time, we find stacks of D7 branes at each kink of the rank function, with rational charge stemming from the IIA solution. The manifolds still contain orbifolds (in most cases), but are more complicated than in the IIA solutions due to T-dualising within the orbifold structure. As an additional consequence of this, we find solutions which contain no orbifold singularities but retain the rational quantization of charge. The general three-parameter G-structure forms are then derived, with explicit investigations into the breaking of supersymmetry conditions left for future study. Due to the integration cycles, we conclude that D5 branes are only present for a preserved $S^2$.
  \item In Section \ref{sec:followingchi}, the $\chi$ reduction solution of Section \ref{sec:chiRed} is T-dualised along both $\beta$ and $\phi$ (with a special $\mathcal{N}=1$ subclass existing in the first of these two solutions). Mappings exist between these solutions, but as in the IIA cases, it is likely that these solutions are still physically different. A more in depth study of these solutions is left for the future (such as analysis of the boundary), but some G-structure results are presented.
    \item In Section \ref{sec:followingphi}, the $\phi$ reduction solution of Section \ref{sec:chiRed} is T-dualised along both $\beta$ and $\chi$ (with a special $\mathcal{N}=1$ subclass existing in the first of these two solutions). Mappings once again exist between these solutions, but again it is likely they describe physically different backgrounds. A more in depth study of these solutions is left for the future.
 \end{itemize}
\item  In Chapter \ref{chap:gamma}, we provide the G-structure forms for the NRSZ $\gamma$-deformations of the GM class, for the M-Theory, IIA and IIB deformations in turn. The IIB solution is a subclass of the more general solution derived in Section \ref{sec:followingbeta}. We demonstrate that the only $\mathcal{N}=1$ preserved deformation is the IIB solution with $\gamma=-1$, which is in fact the only $\mathcal{N}=1$ solution to preserve integer quantization of charge - and hence the best candidate for the dual description of the $\mathcal{N}=1$ CFT presented in the NRSZ paper \cite{Nunez:2019gbg}. All other deformations are $\mathcal{N}=0$, and provide additional examples of non-supersymmetric solutions which break all three supersymmetry conditions (including the gauge BPSness condition).
\end{itemize}
\subsubsection{Future directions}
 This work remains unfinished -  we outline further analysis which should be conducted.
\begin{itemize}
\item 
Further derive and analyse the full $\mathcal{N}=0$ G-structure results within the context of generalized complex geometry and the formalism of \cite{Lust:2008zd,Legramandi:2019ulq,Menet:2023rnt,Menet:2023rml,Menet:2024afb,Held:2010az} - including the $\gamma$ deformations of NRSZ given in Chapter \ref{chap:gamma}.
\item 
Finish the investigation into an alternative set of pure spinors, for which the final supersymmetry condition is imposed. It remains to be seen whether such spinors exist. The language of generalized complex geometry may become useful.
\item 
Complete the analysis at the boundary for the remaining type IIB solutions within this work, investigating further their brane set-ups and charge quantization.
\item 
Perform a more in-depth investigation into explicit rank function examples of our solutions, including triangular, trapezium and Sfetsos-Thompson cases - in addition to the three potentials $(V_1,V_2,V_3)$ given in \cite{Couzens:2022yjl}. The stringent constraints imposed by the spindle on the allowed rank functions should definitely be investigated further. 
\item 
Conduct the spin-2 analysis more generally, for all the multi-parameter type IIA and type IIB solutions presented.
\item 
Gain a better understanding of the operators which trigger the marginal deformation and the soft-SUSY breaking.
\item 
Take a more in-depth look into the Lagrangian and non-Lagrangian deformations of the CFT, along the lines of \cite{Gaiotto:2009we,Gaiotto:2009gz}.
\item 
Investigate the rational quantization of D7 charge inherited purely from the IIA solution (without orbifold singularities present in the IIB metric). Do these general solutions have a physical interpretation (such as the rotating D-branes in Appendix E of \cite{Merrikin:2024bmv}), or should the system be restricted to rank functions such as Figure \ref{fig:Mathematica5}.
\item Further stability analysis should be conducted on the various $\mathcal{N}=0$ solutions presented throughout this work, with analysis along the lines of the paper \cite{Apruzzi:2021nle} particularly pertinent for our backgrounds.
In addition, the recent studies on the stability of non-SUSY backgrounds have been refined thanks to the application of Exceptional field theory techniques, see for example \cite{Eloy:2023acy,Giambrone:2021wsm,Malek:2020yue}. It would be interesting to apply this technology to our family of ${\cal N}=0$ backgrounds.
\item 
Study the SUSY probe-dynamics we encountered in the ${\cal N}=2$ and ${\cal N}=1$ systems.
\item 
Calculate more holographic observables for our multi-parameter families of solutions, including the calculations of \cite{Nunez:2018qcj,Pal:2023bjz,Roychowdhury:2023lxk,Roychowdhury:2024pvu}.
\item 
Place the ${\cal N}=1$ families of solutions within the context of the work \cite{Ashmore:2021mao,Ashmore:2016oug}.
\item 
Perform an S-duality on the three-parameter families of type IIB solutions, following the G-structure analysis through the calculation (the discussion in \cite{Legramandi:2020des} may prove useful here). This should derive further $\mathcal{N}=0$ families, and potentially new $\mathcal{N}=1$ solutions - for which this whole story could be repeated.
\item 
Further investigate the dual CFTs of the infinite families of $\mathcal{N}=1$ type IIB solutions (with zero five form flux) presented throughout this work, within the context of \cite{Macpherson:2014eza,Gauntlett:2005ww,Couzens:2016iot} and possible five-brane webs.
\item Consider non-abelian T-duality on the backgrounds with a preserved $SU(2)$ isometry.
\item 
Investigate these solutions in the context of non-AdS/non-CFT correspondence, such as the work of \cite{Aharony:2002up}. One could break the conformality of the dual field theories (as well as the supersymmetry for our $\mathcal{N}=0$ solutions), by replacing the AdS$_5$ factor with an asymptotically AdS$_5$ factor - introducing a non-zero temperature into the field theory description. Perhaps the effects of including a Black Hole within the solutions could also be investigated.
\item 
Consider whether soft-SUSY deformations of this type could one day have more phenomenological applications, for example within the context of a Minimally Supersymmetric Standard Model (MSSM). Particularly noteworthy are the naturally imposed and stringent restrains on the rank function (and hence the makeup of the dual field theory) coming from the presence of spindles in the supergravity.
\end{itemize}
Additionally, this work opens up some interesting topics to study:
\begin{itemize}
\item 
Investigate whether the two generalized forms of $\mathbb{WCP}^2$ which we uncover during our analysis of the type IIA solutions can appear in more standard wrapped brane scenarios, namely compactifications of AdS$_{n+4}$ to AdS$_n$ for $n=2,3$.
\item 
Repeat the analysis of this thesis for two qualitatively different but related systems: the Lin-Maldacena family \cite{Lin:2005nh} dual to different vacua of the BMN matrix model (see also \cite{vanAnders:2007ky,Shieh:2007xn,Donos:2010va,Lozano:2017ole}); and the system describing the 4d defect inside the $(0,2)$ 6d SCFT - see for example \cite{Capuozzo:2023fll}.
\item 
Consider whether these multi-parameter families could be applied within the context of 5d Minimal Gauged Supergravity, along the lines of \cite{Bah:2017wxp,Couzens:2022aki}. See also \cite{Cheung:2022wpg}.
\item Similar analysis to this work could be conducted on any supersymmetric M-Theory solution involving at least two U(1) directions.
\end{itemize}
I hope that at least some of this work will one day be conducted.


\appendix 

\chapter{Holographic Central Charge}
 
\section{A General $d$-dimensional form (with off-sets)}\label{sec:HCCdiversedim}
The following expression can be constructed from the results quoted in the literature for the holographic central charge in diverse Mink$_D$ dimensions
    \begin{equation}
\begin{aligned}
c_{hol_D}&=\text{coeff}(D)\sum_{n=1}^\infty \frac{P^{D-3}}{n^{D-4}}\mathcal{R}_n^2.
\end{aligned}
\end{equation}
Using the following Fourier decomposition for the rank function of a generic balanced quiver 
\begin{equation}\label{eqn:Rnval}
\begin{aligned}
\mathcal{R}_n&=\frac{2}{P}\int_0^P\mathcal{R}(\eta)\sin\bigg(\frac{n\pi\eta}{P}\bigg)d\eta =\frac{2}{P}\sum_{j=0}^{P-1}\int_j^{j+1}\big[N_j +(N_{j+1}-N_j)(\eta-j)\big]\sin\bigg(\frac{n\pi\eta}{P}\bigg)d\eta\\
&=\frac{2}{n^2\pi^2  }\sum_{j=0}^{P-1}\Bigg[n\pi \bigg[N_j \cos\bigg(\frac{n\pi j}{P}\bigg)-N_{j+1}\cos\bigg(\frac{n\pi(j+1)}{P}\bigg)\bigg]\\
&~~~~~~~~~~~~~~~~~~~~~+ P(N_{j+1}-N_j) \bigg[\sin\bigg(\frac{n\pi (j+1)}{P}\bigg)-\sin\bigg(\frac{n\pi j}{P}\bigg)\bigg]\Bigg]\\
&=\frac{2}{n\pi}\Big(N_0+(-1)^{n+1}N_P\Big)\bigg[1-\frac{P}{n\pi}\sin\bigg(\frac{n\pi}{P}\bigg)\bigg]+ \frac{2P}{(n\pi)^2}\sum_{j=1}^{P}b_j\,\sin\Big(\frac{n\pi}{P}j\Big),\\
&\equiv  \mathcal{R}_n\Big|_{N_0,\,N_P}+\mathcal{R}_n\Big|_{N_0=N_P=0},~~~~~~~~\text{with}~~~b_j=2N_j-N_{j+1}-N_{j-1}~~(\text{for}~N_0=N_P=0).
\end{aligned}
\end{equation}
The first term includes off-sets (namely, $N_0,N_P\neq0$) while the second term (including $b_j$) is defined with the requirement that $N_0=N_P=0$. 
\subsubsection{Without Off-sets}
We will first consider $\mathcal{R}_n|_{N_0=N_P=0}$, corresponding to the rank function given in \eqref{eqn:Rkwithderivs}, for which we recover the $\mathcal{R}_n$ given in \eqref{eqn:potential} (now replacing the variable $k$ with $j$).
It then follows, 
    \begin{equation}
    \begin{aligned}
    \mathcal{R}_n^2&=\frac{4P^2}{n^4\pi^4}\sum_{j=1}^{P}\sum_{l=1}^{P}b_jb_l\sin\bigg(\frac{n\pi}{P}j\bigg)\sin\bigg(\frac{n\pi}{P}l\bigg)\\
        &=-\frac{P^2}{n^4\pi^4}\sum_{j=1}^{P}\sum_{l=1}^{P}b_jb_l\,\,\Big(e^{\frac{in\pi}{P}(j+l)}+e^{-\frac{in\pi}{P}(j+l)}-e^{\frac{in\pi}{P}(j-l)}-e^{-\frac{in\pi}{P}(j-l)}\Big).
    \end{aligned}
    \end{equation}
Hence, the final result for $c_{hol_D}$ can be written as 
      \begin{equation}
      \hspace{-0.75cm}
\begin{aligned}
c_{hol_D}\Big|_{N_{0,P}=0}&=\text{coeff}(D)\sum_{n=1}^\infty \frac{P^{D-3}}{n^{D-4}}\mathcal{R}_n^2\\
&=-\frac{\text{coeff}(D)}{\pi^4}\sum_{j,l=1}^{P}b_jb_l\,P^{D-1}\sum_{n=1}^\infty\,\Bigg[\frac{\Big(e^{\frac{i\pi}{P}(j+l)}\Big)^n}{n^D}+\frac{\Big(e^{-\frac{i\pi}{P}(j+l)}\Big)^n}{n^D}-\frac{\Big(e^{\frac{i\pi}{P}(j-l)}\Big)^n}{n^D}-\frac{\Big(e^{-\frac{i\pi}{P}(j-l)}\Big)^n}{n^D}\Bigg]\\
&=\text{coeff}(D)\frac{2P^{D-1}}{\pi^4}\sum_{j,l=1}^{P}b_jb_l\,\,\text{Re}\bigg[\text{Li}_D\Big(e^{\frac{i\pi}{P}(j-l)}\Big) - \text{Li}_D\Big(e^{\frac{i\pi}{P}(j+l)}\Big)\bigg].
\end{aligned}
\end{equation}

\subsubsection{With Off-sets}
Including off-sets, we use the full expression in \eqref{eqn:Rnval}, recalling $b_j$ is defined with $N_0=N_P=0$.
Hence,
    \begin{equation}
     \begin{aligned}
c_{hol}&=\text{coeff}(D)\sum_{n=1}^\infty \frac{P^{D-3}}{n^{D-4}}\mathcal{R}_n^2. =c_{hol_D}\Big|_{N_{0,P}=0}+\text{coeff}(D)(T_2+T_3),
  \end{aligned}
    \end{equation} 
    with
    \begin{equation}
     \begin{aligned}
 T_2&=\frac{8P^{D-2}}{\pi^3}\sum_{j=1}^{P}b_j\sum_{n=1}^\infty \frac{1}{n^{D-1}}\Big(N_0+(-1)^{n+1}N_P\Big)\bigg[1-\frac{P}{n\pi}\sin\bigg(\frac{n\pi}{P}\bigg)\bigg] \sin\bigg(\frac{n\pi j}{P}\bigg) ,\\
T_3&=\frac{4P^{D-3}}{\pi^2}\sum_{n=1}^\infty \frac{1}{n^{D-2}}\Big(N_0+(-1)^{n+1}N_P\Big)^2\bigg[1-\frac{P}{n\pi}\sin\bigg(\frac{n\pi}{P}\bigg)\bigg]^2.
  \end{aligned}
    \end{equation} 
Now, evaluating $T_2$ gives
   \begin{equation}
     \begin{aligned}
 T_2
 &=\frac{8P^{D-2}}{\pi^3}\sum_{j=1}^{P}b_j\sum_{n=1}^\infty \frac{1}{n^{D-1}} \Big(N_0+(-1)^{n+1}N_P\Big)\bigg[\sin\bigg(\frac{n\pi j}{P}\bigg)-\frac{P}{n\pi}\sin\bigg(\frac{n\pi}{P}\bigg)\sin\bigg(\frac{n\pi j}{P}\bigg)\bigg] \\
  &=\frac{8P^{D-2}}{\pi^3}\sum_{j=1}^{P}b_j\sum_{n=1}^\infty  \Big(N_0+(-1)^{n+1}N_P\Big)\Bigg\{-\frac{1}{2}i\,\Bigg[\frac{\Big(e^{\frac{i\pi}{P}j}\Big)^n}{n^{D-1}}-\frac{\Big(e^{-\frac{i\pi}{P}j}\Big)^n}{n^{D-1}}\Bigg]\\
  &~~~~~~~~~~~~~~~~~~~~~~~~~~~~+\frac{P}{4\pi}\Bigg[\frac{\big(e^{\frac{i\pi}{P}(1+j)}\big)^n}{n^{D}}+\frac{\big(e^{-\frac{i\pi}{P}(1+j)}\big)^n}{n^{D}}-\frac{\big(e^{\frac{i\pi}{P}(1-j)}\big)^n}{n^{D}}-\frac{\big(e^{-\frac{i\pi}{P}(1-j)}\big)^n}{n^{D}}\Bigg]\Bigg\}\\
     &=\frac{8P^{D-2}}{\pi^3}\sum_{j=1}^{P}b_j\Bigg[ N_0\,\text{Im}\,\Big[\text{Li}_{D-1}\big(e^{\frac{i\pi j}{P}}\big)\Big]+\frac{P}{2\pi}N_0\,\,\,\text{Re}\bigg[\text{Li}_{D}\Big(e^{\frac{i\pi}{P}(j+1)}\Big)-\text{Li}_{D}\Big(e^{\frac{i\pi}{P}(j-1)}\Big)\bigg]\\
  &~~~~~~~~~~~        - N_P\,\text{Im}\,\Big[\text{Li}_{D-1}\big(-e^{\frac{i\pi j}{P}}\big)\Big]-\frac{P}{2\pi}N_P\,\,\,\text{Re}\bigg[\text{Li}_{D}\Big(-e^{\frac{i\pi}{P}(j+1)}\Big)-\text{Li}_{D}\Big(-e^{\frac{i\pi}{P}(j-1)}\Big)\bigg]\Bigg].\\
  \end{aligned}
    \end{equation} 
To evaluate $T_3$, we further sub-divide it
   \begin{equation}
     \begin{aligned}
T_3&= \frac{4P^{D-3}}{\pi^2}\sum_{n=1}^\infty \frac{1}{n^{D-2}}\Big(N_0+(-1)^{n+1}N_P\Big)^2\bigg[1-\frac{P}{n\pi}\sin\bigg(\frac{n\pi}{P}\bigg)\bigg]^2= \bar{T}_1+\bar{T}_2+\bar{T}_3,
  \end{aligned}
    \end{equation} 
with 
   \begin{equation}
     \begin{aligned}
     \bar{T}_1&=  \frac{4P^{D-1}}{\pi^4 }\sum_{n=1}^\infty \,\Big(N_0+(-1)^{n+1}N_P\Big)^2\frac{1}{n^D}\sin^2\bigg(\frac{n\pi}{P}\bigg),\\
\bar{T}_2&=\frac{4P^{D-3}}{\pi^2}\sum_{n=1}^\infty \,\Big(N_0+(-1)^{n+1}N_P\Big)^2 \, \frac{1}{n^{D-2}},\\
\bar{T}_3&=-\frac{8P^{D-2}}{\pi^3}\sum_{n=1}^\infty \,\Big(N_0+(-1)^{n+1}N_P\Big)^2 \frac{1}{n^{D-1}}\sin\bigg(\frac{n\pi}{P}\bigg).
  \end{aligned}
    \end{equation} 

The first term gives
  \begin{equation}
  \hspace{-1.8cm}
     \begin{aligned}
     \bar{T}_1&= -\frac{P^{D-1}}{\pi^4 }\sum_{n=1}^\infty \,\Big(N_0+(-1)^{n+1}N_P\Big)^2\Bigg(\frac{\big(e^{\frac{2i\pi}{P}}\big)^n}{n^D}+\frac{\big(e^{-\frac{2i\pi}{P}}\big)^n}{n^D}-\frac{2}{n^D}\Bigg)\\
      &=  -\frac{P^{D-1}}{\pi^4 }\sum_{n=1}^\infty \,\Big(N_0+(-1)^{n+1}N_P\Big)^2\Bigg(\frac{\big(e^{\frac{2i\pi}{P}}\big)^n}{n^D}+\frac{\big(e^{-\frac{2i\pi}{P}}\big)^n}{n^D}\Bigg) + \frac{2P^{D-1}}{\pi^4 }\sum_{n=1}^\infty \,\Big(N_0+(-1)^{n+1}N_P\Big)^2\frac{1}{n^D}\\
            &=  -\frac{2P^{D-1}}{\pi^4 }\bigg[\Big(N_0^2+N_P^2\Big)\,\,\text{Re}\big[\text{Li}_D\big(e^{\frac{2i\pi}{P}}\big)\big]-2N_0N_P\,\,\text{Re}\big[\text{Li}_D\big(-e^{\frac{2i\pi}{P}}\big)\big]-\sum_{n=1}^\infty \,\Big(N_0+(-1)^{n+1}N_P\Big)^2\frac{1}{n^D}\bigg],
  \end{aligned}
    \end{equation} 
where
   \begin{equation}
     \begin{aligned}
\sum_{n=1}^\infty \,\Big(N_0+(-1)^{n+1}N_P\Big)^2\frac{1}{n^D} &=\Big(N_0^2+N_P^2\Big) \sum_{n=1}^\infty \frac{1}{n^D}-2N_0N_P  \sum_{n=1}^\infty \frac{(-1)^n}{n^D}\\
 &=\zeta(D)\bigg[\Big(N_0^2+N_P^2\Big) -2\big(2^{1-D}-1\big)N_0N_P \bigg],\\
  \end{aligned}
    \end{equation} 
making use of the Riemann Zeta Function
\begin{equation*}
\zeta(s)=\sum_{n=1}^\infty \frac{1}{n^s},~~~~~~~~~~~~~~~~~~~~~~~~~~~~ \sum_{n=1}^\infty \frac{(-1)^n}{n^s}=\zeta(s)\big[2^{1-s}-1\big].
\end{equation*}
Hence,
\begin{align}
\bar{T}_1&=  -\frac{2P^{D-1}}{\pi^4 }\Bigg[\Big(N_0^2+N_P^2\Big)\,\,\text{Re}\big[\text{Li}_D\big(e^{\frac{2i\pi}{P}}\big)\big]-2N_0N_p\,\,\text{Re}\big[\text{Li}_D\big(-e^{\frac{2i\pi}{P}}\big)\big]\nn\\[2mm]
&~~~~~ -\zeta(D)\bigg[\Big(N_0^2+N_P^2\Big) -2\big(2^{1-D}-1\big)N_0N_P \bigg]\Bigg].
\end{align}
Now, for $\bar{T}_2$
  \begin{equation}
    \hspace{-0.7cm}
     \begin{aligned}
\bar{T}_2&= \frac{4P^{D-3}}{\pi^2}\sum_{n=1}^\infty \,\Big(N_0+(-1)^{n+1}N_P\Big)^2 \, \frac{1}{n^{D-2}}=\frac{4P^{D-3}}{\pi^2}\Bigg[\,\Big(N_0^2+N_P^2\Big)\sum_{n=1}^\infty  \,  \frac{1}{n^{D-2}}-2N_0N_P\sum_{n=1}^\infty (-1)^n \,  \frac{1}{n^{D-2}}\Bigg]\\
&=\frac{4P^{D-3}}{\pi^2} \zeta(D-2)\Bigg[\,\Big(N_0^2+N_P^2\Big)-2\big(2^{3-D}-1\big)N_0N_P\Bigg],\\
  \end{aligned}
    \end{equation} 
   and for $\bar{T}_3$
       \begin{equation}
     \begin{aligned}
\bar{T}_3&=-\frac{8P^{D-2}}{\pi^3}\sum_{n=1}^\infty \,\Big(N_0+(-1)^{n+1}N_P\Big)^2 \frac{1}{n^{D-1}}\sin\bigg(\frac{n\pi}{P}\bigg)\\
&=i\,\frac{4P^{D-2}}{\pi^3}\Bigg[\Big(N_0^2+N_P^2\Big)\sum_{n=1}^\infty \,\bigg(\frac{(e^{\frac{i\pi}{P}})^n}{n^{D-1}}-\frac{(e^{\frac{-i\pi}{P}})^n}{n^{D-1}}\bigg)-2N_0N_P\sum_{n=1}^\infty \,\bigg(\frac{(-e^{\frac{i\pi}{P}})^n}{n^{D-1}}-\frac{(-e^{\frac{-i\pi}{P}})^n}{n^{D-1}}\bigg)\Bigg]\\
&=-\frac{8P^{D-2}}{\pi^3}\,\text{Im}\bigg[\Big(N_0^2+N_P^2\Big)\,\text{Li}_{D-1}(e^{\frac{i\pi}{P}})-2N_0N_P\,\text{Li}_{D-1}(-e^{\frac{i\pi}{P}})\bigg].
  \end{aligned}
    \end{equation} 
    
    So, putting it all together, $T_3$ reads
     \begin{equation}
     \hspace{-0.4cm}
     \begin{aligned}
T_3&= -\frac{2P^{D-1}}{\pi^4 }\Bigg[\Big(N_0^2+N_P^2\Big)\,\,\text{Re}\big[\text{Li}_D\big(e^{\frac{2i\pi}{P}}\big)\big]-2N_0N_p\,\,\text{Re}\big[\text{Li}_D\big(-e^{\frac{2i\pi}{P}}\big)\big]\nn\\[2mm]
&-\zeta(D)\bigg(\Big(N_0^2+N_P^2\Big) -2\big(2^{1-D}-1\big)N_0N_P \bigg)\Bigg]+\frac{4P^{D-3}}{\pi^2} \zeta(D-2)\bigg(\,\Big(N_0^2+N_P^2\Big)-2\big(2^{3-D}-1\big)N_0N_P\bigg)\\
&~~-\frac{8P^{D-2}}{\pi^3}\,\text{Im}\Bigg[\Big(N_0^2+N_P^2\Big)\,\text{Li}_{D-1}(e^{\frac{i\pi}{P}})-2N_0N_P\,\text{Li}_{D-1}(-e^{\frac{i\pi}{P}})\Bigg].
  \end{aligned}
    \end{equation} 
    Hence, the final result for the holographic central charge with off-sets reads
                   \begin{equation}\label{eqn:HCCwithOffsets}
           \hspace{-1.75cm}
     \begin{aligned}
c_{hol}&=\frac{2P^{D-3}}{\pi^4}\text{coeff}(D)\Bigg\{ P^2 \,\text{Re}\Bigg[\sum_{j=1}^{P}\sum_{l=1}^{P}b_jb_l \,\,\,\bigg(\text{Li}_D\Big(e^{\frac{i\pi}{P}(j-l)}\Big)-\text{Li}_D\Big(e^{\frac{i\pi}{P}(j+l)}\Big)\bigg)\\
  &+2\sum_{j=1}^{P}b_j\bigg[N_0\bigg(\text{Li}_{D}\Big(e^{\frac{i\pi}{P}(j+1)}\Big)-\text{Li}_{D}\Big(e^{\frac{i\pi}{P}(j-1)}\Big)\bigg)       -N_P\bigg(\text{Li}_{D}\Big(-e^{\frac{i\pi}{P}(j+1)}\Big)-\text{Li}_{D}\Big(-e^{\frac{i\pi}{P}(j-1)}\Big)\bigg)\bigg]\\
  &+2N_0N_P\,\,\text{Li}_D\big(-e^{\frac{2i\pi}{P}}\big) - \Big(N_0^2+N_P^2\Big)\,\,\text{Li}_D\big(e^{\frac{2i\pi}{P}}\big)  \Bigg]\\
&+ 4\pi\,P\,\text{Im}\,\Bigg[\sum_{j=1}^{P}b_j\,\bigg[\,N_0\,\text{Li}_{D-1}\big(e^{\frac{i\pi j}{P}}\big)  -N_P\,\text{Li}_{D-1}\big(-e^{\frac{i\pi j}{P}}\big)\bigg]\nn\\[2mm]
&-\bigg[\Big(N_0^2+N_P^2\Big)\,\text{Li}_{D-1}(e^{\frac{i\pi}{P}})-2N_0N_P\,\text{Li}_{D-1}(-e^{\frac{i\pi}{P}})\bigg]\Bigg]\\
 &+P^2\zeta(D)\bigg(\Big(N_0^2+N_P^2\Big)\,\,-2\,\, \big(2^{1-D}-1\big)N_0N_p\bigg)+2 \pi^2 \zeta(D-2)\bigg(\,\Big(N_0^2+N_P^2\Big)-2\big(2^{3-D}-1\big)N_0N_P\bigg)\Bigg\}.
  \end{aligned}
    \end{equation} 
    
   \newpage
\subsection{Example rank functions}\label{sec:exampleRanks}    
\subsubsection{Triangle Rank Function}

 \begin{figure}[H]
\centering  
\subfigure
{
\centering
  \begin{minipage}{0.55\textwidth}
  \begin{equation*}
  \mathcal{R}(\eta) = \begin{cases}
      ~~~~~~~~N\eta & \eta\in[0,S]\\
     \frac{NS}{(P-S)}(P-\eta)  & \eta\in[S,P]
    \end{cases} ,
\end{equation*}
  \end{minipage}
     \begin{minipage}{.45\textwidth}
    \centering
\begin{tikzpicture}[scale=0.85, every node/.style={scale=0.95}]
\draw[black,line width=0.3mm] (1,0) --(5,2);
\draw[black,line width=0.3mm] (5,2)-- (7,0);

\draw (1,0) node[below] {$0$};
\draw[gray,dashed,line width=0.3mm] (5,0)-- (5,2);
\draw (5,0) node[below] {$S$};
\draw (7,0) node[below] {$P$};

\draw[gray,dashed,line width=0.3mm] (1,2)-- (5,2);
\draw (1,2) node[left] {$NS$};

\draw[-stealth, line width=0.53mm] (1,0)--(1,3) node[left ]{$\mathcal{R}(\eta)$};
\draw[-stealth, line width=0.53mm] (1,0) --(7.6,0) node[below ]{$\eta$};
\end{tikzpicture}
  \end{minipage}

}
\subfigure
{
   \begin{minipage}{\textwidth}
   \vspace{-0.4cm}
  \begin{equation}\label{eqn:triangle}
  \mathcal{R}'(\eta) = \begin{cases}
      ~~~~N & \eta\in[0,S]\\
    - \frac{NS}{P-S}  & \eta\in[S,P]
    \end{cases} 
    ,~~~~~~~~~~~~~~  \mathcal{R}''(\eta) =-\frac{NP}{P-S}\delta(\eta-S).
\end{equation}
  \end{minipage}
}
\subfigure
{
   \begin{minipage}{\textwidth}
\begin{center}
 \begin{tikzpicture}[square/.style={regular polygon,regular polygon sides=4},scale=1]
         
   \begin{scope}[xshift=-2cm,yshift=5cm]
\draw (8,1.3)  node[above] {$\frac{NP}{P-S}$ };
\node[label=above:] at (8,1){\LARGE $\otimes$};
\draw (1,-1.5) -- (1,2);
\draw (1,-2.1) node[above] {$1$};
\draw (3,-1.5) -- (3,2);
\draw (3,-2.1) node[above] {$2$};
\draw (5,-1.5) -- (5,2);
\draw (5,-2.1) node[above] {$3$};
\draw (7,-1.5) -- (7,2);
\draw (7,-2.1) node[above] {$S-1$};
\draw (9,-1.5) -- (9,2);
\draw (9,-2.1) node[above] {$S$};
\draw (11,-1.5) -- (11,2);
\draw (11,-2.1) node[above] {$S+1$};
\draw (13,-1.5) -- (13,2);
\draw (13,-2.1) node[above] {$P-1$};
\draw (15,-1.5) -- (15,2);
\draw (15,-2.1) node[above] {$P$};
\draw (1,0) -- (3,0);
\draw (3,0.2) -- (5,0.2);
\draw (5.5,0) node[right] {.~~.~~.};
\draw (7,0.2) -- (9,0.2);
\draw (9,0) -- (11,0);
\draw (13,0) -- (15,0);
\draw (11.5,0) node[right] {.~~.~~.};
\draw (2,-0.75) node[above] {$N$};
\draw (4,-.75) node[above] {$2N$};
\draw (8,-0.75)  node[above] {$NS$};
\draw (10,-0.75)  node[above] {\small{$\frac{NS(P-S-1)}{P-S}$}};
\draw (14,-0.75)  node[above] {$\frac{NS}{P-S}$};

\end{scope}
 
  \begin{scope}[xshift=0cm,yshift=0cm]
        \node at (0,0) [circle,inner sep=0.6em,draw] (c000) {$N$};
        \node at (6,1.75) [square,inner sep=0.8em,draw] (f300) {$   $};
        \node at (2,0) [circle,inner sep=0.5em,draw] (c100) {$2N$};
        \node at (6,0) [circle,inner sep=0.4em,draw] (c200) {$NS$};
        \node at (8,0) [circle,inner sep=0.9em,draw] (c300) {$   $};
        \node at (12,0) [circle,inner sep=0.9em,draw] (c400) {$   $};
        \draw (c100) -- (c000);
        \draw (c100) -- (3,0);
        \draw (5,0) -- (c200);
        \draw (c200) -- (f300);
        \draw (c200) -- (c300);
        \draw (c300) -- (9,0);
        \draw (11,0) -- (c400);
         \draw (3.5,0) node[right] {.~~.~~.};
        \draw (9.5,0) node[right] {.~~.~~.};
        \draw (6,1.35) node[above] {$\frac{NP}{P-S}$};
        \draw (8,0.5) node[above] {$\frac{NS(P-S-1)}{P-S}$};
         \draw (12,-0.4) node[above] {$\frac{NS}{P-S}$};
         
         \end{scope}
    \end{tikzpicture}
    \end{center}
       \end{minipage}
}
\caption{The Generic Triangular Rank Function}
    \label{fig:Triangular}
\end{figure}
In this case, one finds the following charges (following the conventions of \cite{Akhond:2021ffz})
\beq
Q_{NS5}=P,~~~~~~~~~Q_{D6}=\frac{NP}{P-S},~~~~~~~~~ Q_{D4}^{\text{Total}}=\int_0^P R(\eta) d\eta =\frac{1}{2}NSP .
\eeq
To calculate the holographic central charge, one finds using \eqref{eqn:potential} and integration by parts
\begin{equation}
\begin{aligned}
	\mathcal{R}_n&=\frac{1}{P}\int_{-P}^P\mathcal{R}(\eta)\sin\bigg(\frac{n\pi}{P}\eta\bigg)d\eta=\frac{2}{P}\int_0^P\mathcal{R}(\eta)\sin\bigg(\frac{n\pi }{P}\eta\bigg)d\eta\\
	&=\frac{2}{P}\int_0^SN\eta \sin\bigg(\frac{n\pi }{P}\eta\bigg)d\eta+\frac{2}{P}\int_S^P \frac{NS}{P-S}(P-\eta)\sin\bigg(\frac{n\pi }{P}\eta\bigg)d\eta=\frac{2}{n^2\pi^2}\frac{NP^2}{(P-S)} \sin\bigg(\frac{k\pi S}{P}\bigg),
\end{aligned}
\end{equation}
which then leads to
\beq\label{eqn:HCCTriang1}
\hspace{-2cm}
\begin{aligned}
c_{hol}&= \frac{ \kappa^3}{\pi^4}\sum_{k=1}^\infty P \, \mathcal{R}_n^2 = \frac{4 \kappa^3}{\pi^8} \frac{N^2P^5}{(P-S)^2}\, \sum_{n=1}^\infty \frac{1}{n^4} \sin^2\bigg(\frac{n\pi S}{P}\bigg)=- \frac{ \kappa^3}{\pi^8} \frac{N^2P^5}{(P-S)^2}\, \sum_{n=1}^\infty   \bigg[\frac{(e^{\frac{2i\pi S}{P}})^n}{n^4}+\frac{(e^{-\frac{2i\pi S}{P}})^n}{n^4}-\frac{2}{n^4}\bigg] \\[2mm]
&=- \frac{ \kappa^3}{\pi^8} \frac{N^2P^5}{(P-S)^2}\,  \bigg[\text{Li}_4(e^{i\frac{2\pi S}{P}})+\text{Li}_4(e^{-i\frac{2\pi S}{P}})-2 \,\text{Li}_4(1)\bigg]=- \frac{2 \kappa^3}{\pi^8} \frac{N^2P^5}{(P-S)^2}\, \text{Re}  \bigg[\text{Li}_4(e^{i\frac{2\pi S}{P}})- \,\text{Li}_4(1)\bigg]\\[2mm]
&=  \frac{1}{4\pi^5}\frac{N^2P^5}{(P-S)^2}\Big( \zeta(4)-  \text{Re}[\text{Li}_4(e^{i\frac{2\pi S}{P}})]\Big),~~~~~\text{with}~~~\kappa=\frac{\pi}{2},~~~~~\zeta(4)=\frac{\pi^4}{90}.
\end{aligned}
\eeq
One can now use \eqref{eqn:HCCnew} for a more convenient method of deriving this result. Here we have one D6 brane with charge given by $b_S=\frac{NP}{P-S}$. In this case, $j=l=S$, giving
          \begin{equation}\label{eqn:HCCTriang2}
\begin{aligned}
c_{hol_4}&=\text{coeff}(4)\frac{2P^{3}}{\pi^4}b_S^2\,\,\text{Re}\bigg[\text{Li}_4(1) - \text{Li}_4\Big(e^{\frac{i\pi}{P}2S}\Big)\bigg] =\frac{1}{4\pi^5}\frac{N^2P^5}{(P-S)^2}\,\,\bigg(\zeta(4) -\text{Re}\Big[ \text{Li}_4\Big(e^{\frac{i\pi}{P}2S}\Big)\Big]\bigg),
\end{aligned}
\end{equation}
matching \eqref{eqn:HCCTriang1}.
We now investigate the holographic limit $(P\rightarrow\infty)$ to leading order, as in \cite{Akhond:2021ffz}\footnote{noting in the $S=P/2$ case
\beq
\begin{aligned}
&\zeta(4)-  \text{Re}[\text{Li}_4(-1)] =   \zeta(4)-   \sum_{n=1}^\infty \frac{(-1)^n}{n^4}= \zeta(4)-   \zeta(4) (2^{1-4}-1) = \frac{15}{8}\zeta(4) =\frac{\pi^4}{48}\\
&\Rightarrow~~~~~~c_{hol}=  \frac{4}{\pi^5}\frac{N^2P^5}{(P)^2}\Big( \zeta(4)-  \text{Re}[\text{Li}_4(e^{i\pi})]\Big) = \frac{4}{\pi^5}N^2P^3\frac{\pi^4}{48} =  \frac{N^2P^3}{12\pi},\nn
\end{aligned}
\eeq}
\beq\label{eqn:trianglechol}
c_{hol}\Big|_{P\rightarrow\infty}=
\begin{cases}
\frac{N^2P^3}{12\pi}
&S=P-1\\
 \frac{N^2P^3}{48\pi}&S=\frac{P}{2}\\
 \frac{N^2S^2P}{12\pi}&S\in \mathds{Z}
\end{cases}.
\eeq
One can easily check that $N_f=2N_c$ at each node of the above linear quiver. The $N_{\text{v}}$ and $N_{\text{h}} $ were calculated in section F.2 of \cite{Nunez:2019gbg}, and read
\begin{equation}
\begin{aligned}
N_{\text{v}} &=\frac{1}{6(P-S)}\Big(2S^2P^2N^2+SP(N^2+6)-2S^3PN^2-6P(P-1)-6S\Big),\\
~~~~~
N_{\text{h}}&=\frac{N^2S}{3(P-S)}\Big(5P-S^2(P+3)+S(P^2+3P-3)\Big),
\end{aligned}
\end{equation}
which from \eqref{centralsN=2}, with $P\rightarrow\infty$, gives
\begin{equation}
c\,\Big|_{P\rightarrow\infty} \sim a\Big|_{P\rightarrow\infty}  =
\begin{cases}
\frac{N^2P^3}{12\pi} &S=P-1\\
 \frac{N^2P^3}{48\pi}&S=\frac{P}{2}\\
 \frac{N^2S^2P}{12\pi}&S\in \mathds{Z}
\end{cases},
\end{equation}
for large $NS$ in the last case, which matches the $c_{hol}$ given in \eqref{eqn:trianglechol}. 
\subsubsection{Trapezium Rank Function}

 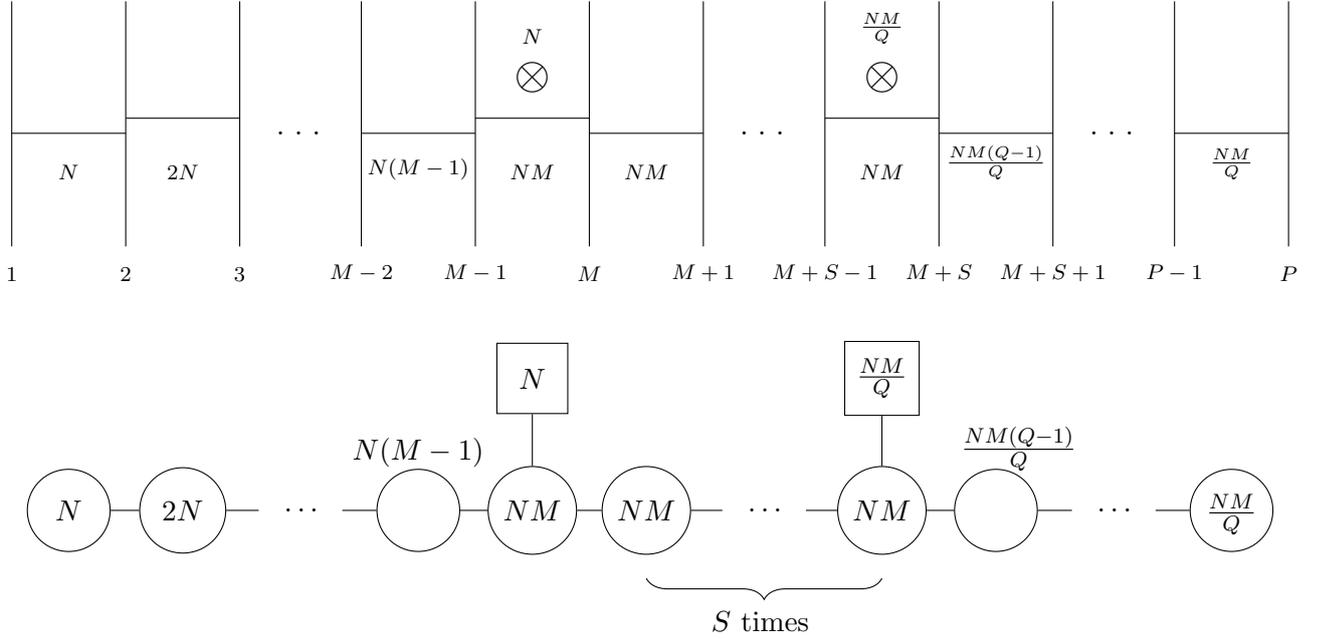
\begin{figure}[H]
\centering  
\subfigure
{
\centering
  \begin{minipage}{0.55\textwidth}
    \begin{equation*} 
  \mathcal{R}(\eta) =
        \begin{cases}
        ~~~~~  N\eta & \eta\in[0,M] \, ,\\
          ~~~~~N M   & \eta \in [M,M+S] \,  \\
        \frac{N M}{Q}(P-\eta) & \eta\in[M+S,P]\, 
        \end{cases}       ,
    \end{equation*}
  \end{minipage}
     \begin{minipage}{.45\textwidth}
    \centering
\begin{tikzpicture}[scale=0.85, every node/.style={scale=0.95}]
\draw[black,line width=0.3mm] (1,0) --(2.5,2);
\draw[black,line width=0.3mm] (2.5,2) --(5.5,2);
\draw[black,line width=0.3mm] (5.5,2)-- (7,0);

\draw (1,0) node[below] {$0$};
\draw[gray,dashed,line width=0.3mm] (5.5,0)-- (5.5,2);
\draw[gray,dashed,line width=0.3mm] (2.5,0)-- (2.5,2);
\draw (2.5,0) node[below] {$M$};
\draw (5.5,0) node[below] {$M+S$};
\draw (7,0) node[below] {$P$};

\draw[gray,dashed,line width=0.3mm] (1,2)-- (2.5,2);
\draw (1,2) node[left] {$NM$};

\draw[-stealth, line width=0.53mm] (1,0)--(1,3) node[left ]{$\mathcal{R}(\eta)$};
\draw[-stealth, line width=0.53mm] (1,0) --(7.6,0) node[below ]{$\eta$};

              \draw (6,2.4) node[above] {$Q\equiv P-S-M$};
                   \draw [decorate,decoration = {brace,amplitude=5pt}] (5.5,2.3) --  (7,2.3);
\end{tikzpicture}
  \end{minipage}

}
\subfigure
{
   \begin{minipage}{\textwidth}
      \vspace{-0.6cm}
  \begin{equation}\label{eqn:trapezium}
  \mathcal{R}'(\eta) =   \begin{cases}
        ~~~~~  N & \eta\in[0,M] \, \\
          ~~~~~0   & \eta \in [M,M+S] \,  \\
      -  \frac{N M}{Q} & \eta\in[M+S,P]\, 
        \end{cases}       
    ,~~~~~~~  \mathcal{R}''(\eta) =-N\delta(\eta-M) - \frac{N M}{Q}\delta(\eta-M-S).
\end{equation}
  \end{minipage}
}
\subfigure
{
   \hspace{-1cm}
   \begin{minipage}{\textwidth}
\begin{center}
 \begin{tikzpicture}[square/.style={regular polygon,regular polygon sides=4},scale=1]
         
   \begin{scope}[xshift=-3cm,yshift=5cm]
\draw (8.1,1.05)  node[above] {\scriptsize{$N$} };
\node[label=above:] at (8.1,0.75){\LARGE $\otimes$};
\node[label=above:] at (12.7,0.75){\LARGE $\otimes$};
\draw (12.7,1.05)  node[above] {\scriptsize{$\frac{NM}{Q}$ }};
\draw (1.25,-1.5) -- (1.25,1.75);
\draw (1.25,-2.1) node[above] {\scriptsize{$1$}};
\draw (2.75,-1.5) -- (2.75,1.75);
\draw (2.75,-2.1) node[above] {\scriptsize{$2$}};

\draw (4.25,-1.5) -- (4.25,1.75);
\draw (4.25,-2.1) node[above] {\scriptsize{$3$}};
\draw (5.85,-1.5) -- (5.85,1.75);
\draw (5.85,-2.1) node[above] {\scriptsize{$M-2$}};
\draw (7.35,-1.5) -- (7.35,1.75);
\draw (7.35,-2.1) node[above] {\scriptsize{$M-1$}};
\draw (8.85,-1.5) -- (8.85,1.75);
\draw (8.85,-2.1) node[above] {\scriptsize{$M$}};
\draw (10.35,-1.5) -- (10.35,1.75);
\draw (10.35,-2.1) node[above] {\scriptsize{$M+1$}};
\draw (11.95,-1.5) -- (11.95,1.75);
\draw (13.45,-1.5) -- (13.45,1.75);
\draw (11.95,-2.1) node[above] {\scriptsize{$M+S-1$}};
\draw (14.95,-1.5) -- (14.95,1.75);
\draw (13.45,-2.1) node[above] {\scriptsize{$M+S$}};
\draw (14.95,-2.1) node[above] {\scriptsize{$M+S+1$}};

\draw (16.55,-2.1) node[above] {\scriptsize{$P-1$}};

\draw (18.05,-2.1) node[above] {\scriptsize{$P$}};
\draw (16.55,-1.5) -- (16.55,1.75);
\draw (18.05,-1.5) -- (18.05,1.75);
\draw (1.25,0) -- (2.75,0);
\draw (2.75,0.2) -- (4.25,0.2);
\draw (4.6,0) node[right] {.~.~.};
\draw (15.3,0) node[right] {.~.~.};
\draw ( 5.85,0) -- (7.35,0);
\draw (7.35,0.2) -- (8.85,0.2);
\draw (8.85,0) -- (10.35,0);

\draw (11.95,0.2) -- (13.45,0.2);
\draw (13.45,0) -- (14.95,0);
\draw (16.55,0) -- (18.05,0);
\draw (10.7,0) node[right] {.~.~.};
\draw (6.6,-0.75) node[above] {\scriptsize{$N(M-1)$}};
\draw (2,-0.75) node[above] {\scriptsize{$N$}};
\draw (3.5,-.75) node[above] {\scriptsize{$2N$}};
\draw (8.1,-0.75)  node[above] {\scriptsize{$NM$}};
\draw (9.6,-0.75)  node[above] {\scriptsize{$NM$}};
\draw (12.7,-0.75)  node[above] {\scriptsize{$NM$}};
\draw (17.3,-0.75)  node[above] {\scriptsize{$\frac{NM}{Q}$}};
\draw (14.2,-0.75)  node[above] {\scriptsize{$\frac{NM(Q-1)}{Q}$}};

\end{scope}
 
  \begin{scope}[xshift=-1cm,yshift=0cm]

        \node at (0,0) [circle,inner sep=0.6em,draw] (c000) {$N$};
        \node at (1.5,0) [circle,inner sep=0.5em,draw] (c100) {$2N$};
        \node at (4.6,0) [circle,inner sep=1em,draw]   (c200) {$ $};
        \node at (6.1,0) [circle,inner sep=0.35em,draw] (c300) {$  NM $};
        \node at (7.6,0) [circle,inner sep=0.35em,draw] (c400) {$ N M $};
        \node at (10.7,0) [circle,inner sep=0.35em,draw] (c500) {$ N M $};
        \node at (12.2,0) [circle,inner sep=1em,draw] (c600) {$  $};
        \node at (15.3,0) [circle,inner sep=1em,draw] (c700) {$ $};

        \node at (6.1,1.75) [square,inner sep=0.4em,draw] (f300) {$ N  $};
        \node at (10.7,1.75) [square,inner sep=0.4em,draw] (f500) {$ ~~~ $};

        \draw (c000) -- (c100);
        \draw (c100) -- (2.5,0);
        \draw (2.7,0) node[right] {.\,.\,.};
        \draw (3.6,0) -- (c200);
        \draw (c300) -- (f300);
        \draw (c200) -- (c300);
        \draw (c300) -- (c400);
        \draw (c400) -- (8.6,0);
        \draw (8.8,0) node[right] {.\,.\,.};
        \draw (9.7,0) -- (c500);
        \draw (c500) -- (f500);
        \draw (c500) -- (c600);
        \draw (c600) -- (13.2,0);
        \draw (13.4,0) node[right] {.\,.\,.};
        \draw (14.3,0) -- (c700);

        \draw(4.6,1.1)  node[below] {$N(M-1)$};
        \draw [decorate,decoration = {brace,amplitude=8pt}] (10.7,-0.9) --  (7.6,-0.9);
        \draw (9.1,-1.2)  node[below] {$S$ times};
        \draw(12.5,1.25)  node[below] {$\frac{NM(Q-1)}{Q}$};
        
        \draw(15.3,0.35)  node[below] {$\frac{NM}{Q}$};
        \draw(10.7,1.35)  node[above] {$ \frac{NM}{Q}$};
         
         \end{scope}
    \end{tikzpicture}
    \end{center}
       \end{minipage}
}
\caption{The Generic Trapezium Rank Function, with $Q\equiv P-M-S$.}
    \label{fig:Trapezium}
\end{figure}

In this case, one finds the following charges (follow the conventions of \cite{Akhond:2021ffz})
\beq
Q_{NS5}=P,~~~~~~~~~Q_{D6}= N+\frac{NM}{Q}=\frac{N(P-S)}{P-S-M},~~~~~~~~~ Q_{D4}^{\text{Total}}=\int_0^P R(\eta) d\eta =\frac{NM}{2}(P+S) .
\eeq
We now have
\begin{equation}
\begin{aligned}
	\mathcal{R}_n=\frac{2}{P}\int_0^P\mathcal{R}(\eta)\sin\bigg(\frac{n\pi\eta}{P}\bigg)d\eta=\frac{2NP}{ n^2\pi^2Q} \Bigg[M\sin\bigg(\frac{n\pi (M+S)}{P}\bigg) +Q\sin\bigg(\frac{n\pi M}{P}\bigg)\Bigg].
\end{aligned}
\end{equation}
One can then calculate $c_{hol}$, either by performing the calculation long hand or by using \eqref{eqn:HCCnew}. Both approaches give the same result.  Here we will demonstrate the the second approach (with $\kappa=\pi/2$ again). In this case, we have two D6 branes, with charge given by $b_M=N$ and $b_{M+S}=\frac{NM}{Q}$. Here, $(j,l)$ will take the values $S$ and $M+S$ over the double sum, giving
          \begin{equation}\label{eqn:HCCTrap2}
          \hspace{-1.25cm}
\begin{aligned}
c_{hol_4}&=\text{coeff}(4)\frac{2P^{3}}{\pi^4} \Bigg[b_M\sum_{l=1}^{P} b_l\,\,\text{Re}\bigg[\text{Li}_4\Big(e^{\frac{i\pi}{P}(M-l)}\Big) - \text{Li}_4\Big(e^{\frac{i\pi}{P}(M+l)}\Big)\bigg]\\
&~~~~+b_{M+S}\,\sum_{l=1}^{P}b_l\,\text{Re}\bigg[\text{Li}_4\Big(e^{\frac{i\pi}{P}(M+S-l)}\Big) - \text{Li}_4\Big(e^{\frac{i\pi}{P}(M+S+l)}\Big)\bigg]\Bigg]\\
&= \frac{ \kappa^3}{\pi^4}\frac{2P^{3}}{\pi^4} \Bigg[b_M\bigg( b_M\,\,\text{Re}\bigg[\text{Li}_4(1) - \text{Li}_4\Big(e^{\frac{i\pi}{P}(2M)}\Big)\bigg]+ b_{M+S}\,\,\text{Re}\bigg[\text{Li}_4\Big(e^{\frac{i\pi}{P}(-S)}\Big) - \text{Li}_4\Big(e^{\frac{i\pi}{P}(2M+S)}\Big)\bigg]\bigg)\\
&~~~~+b_{M+S}\bigg( b_M\,\text{Re}\bigg[\text{Li}_4\Big(e^{\frac{i\pi}{P}(S)}\Big) - \text{Li}_4\Big(e^{\frac{i\pi}{P}(2M+S)}\Big)\bigg]+b_{M+S}\,\text{Re}\bigg[\text{Li}_4(1) - \text{Li}_4\Big(e^{\frac{i\pi}{P}(2(M+S))}\Big)\bigg]\bigg)\Bigg]\\
&= \frac{N^2P^3 }{ 4\pi^5Q^2} \bigg((M^2+Q^2)\zeta(4)\\
&~~~~~- \text{Re}\Big[M^2\text{Li}_4(e^{i\frac{2\pi (M+S)}{P}})  +Q^2 \text{Li}_4(e^{i\frac{2\pi M}{P}})+2M Q \Big(\text{Li}_4(e^{i\frac{\pi (2M+S)}{P}})-\text{Li}_4(e^{i\frac{\pi S}{P}})\Big)\Big]  \bigg).
\end{aligned}
\end{equation}
As before, investigating the holographic limit $(P\rightarrow \infty)$,
\beq\label{eqn:choltrap}
c_{hol}\Big|_{P\rightarrow\infty}=
\frac{N^2M^2P}{12\pi}~~~~~~ S,M\in \mathds{Z}.
\eeq
 From the linear quiver, using the results given in Section F.3 of \cite{Nunez:2019gbg}, one has (after correcting typos)
    \begin{equation}
    \begin{aligned}
     N_{\text{v}} &=\frac{1}{6}\left( 6+2P(M^2N^2-3)+MN^2\Big(1+4MS+\frac{M}{Q}\Big)\right), \\
    N_{\text{h}} &= \frac{MN^2 }{ 3Q}\left( MP^2 +P(5-M^2+MS)-S(5+2M^2+2MS)\right) \, .
    \end{aligned}
    \end{equation}
    Using \eqref{centralsN=2}, one then finds (for large $MN$)
    \beq
    c\Big|_{P\rightarrow\infty}\sim a\Big|_{P\rightarrow\infty} = \frac{N^2M^2P}{12\pi},
    \eeq
  matching the $c_{hol}$ given in \eqref{eqn:choltrap}.
\newpage
\section{Parametric deformations of GM}\label{sec:HCCcalc}
To calculate the general holographic central charge, $c_{hol}$, for a given background, one has
\begin{equation}
\begin{gathered}
ds^2=\alpha(\rho,\vv{\theta})\Big(dx_{1,\hat{d}}^2+\beta(\rho)d\rho^2\Big)+g_{ij}(\rho,\vv{\theta})d\theta^id\theta^j,   \\
c_{hol}=\frac{\hat{d}^{\hat{d}}}{G_N}\beta^{\hat{d}/2}\frac{H^{\frac{2\hat{d}+1}{2}}}{(H')^{\hat{d}}},~~~~H=V_{int}^2,~~~~~V_{int}=\int d\vv{\theta}\sqrt{\text{det}[g_{ij}]e^{-4\Phi}\alpha^{\hat{d}}},
\end{gathered}
\end{equation}
with $G_N=8\pi^6 \alpha'^4g_s^2=8\pi^6$. In the cases discussed in this work, $\hat{d}=3$.
\subsection{Type IIA}
All of the type IIA backgrounds encountered throughout the thesis have a metric which can be expressed in the same manner, with all dependence on the parameters dropping out of the following calculation. 
We have 
\begin{align}
    ds_{10,A}^2&=4 \rho^2e^{\frac{2}{3}\Phi_\mathcal{A}}f_1(\sigma,\eta)\bigg(dx_{1,3}^2+\frac{1}{\rho^4}d\rho^2\bigg)+  e^{\frac{2}{3}\Phi_\mathcal{A}}f_1(\sigma,\eta) \bigg[ f_4(\sigma,\eta)(d\sigma^2+d\eta^2)+ ds^2(M_3^{\mathcal{A}})\bigg].
\end{align}
It is now a straight forward calculation to derive
\begin{equation}\label{eqn:HCCIIA}
\sqrt{\text{det}[g_{ij}]e^{-4\Phi_\mathcal{A}}\alpha^3} = \frac{8}{X}\rho^3 f_1^{\frac{9}{2}}f_3^{\frac{1}{2}}f_5^{\frac{1}{2}} f_2f_4  \Vol(\text{S}^1)\Vol(\text{S}^2)  ,
\end{equation}
with all dilaton dependence, $\Phi_\mathcal{A}$, dropping out neatly, meaning that all the backgrounds take the above form for the holographic central charge!
\\\\Inserting the forms of the warp factors \eqref{eqn:fs}, gives
\begin{equation}
V_{int}=\frac{2^5\kappa^3}{X} \Vol(\text{S}^1)\Vol(\text{S}^2) \,\rho^3  \int \sigma \dot{V} V''  d\sigma d\eta    ,
\end{equation}
leading to (with $ \Vol(\text{S}^1)=2\pi,~\Vol(\text{S}^2) =4\pi$)

\begin{equation}
\begin{gathered}
c_{hol}=\frac{3^3}{8\pi^6}\bigg(\frac{1}{\rho^4}\bigg)^{3/2}\frac{V_{int}^7}{[(V_{int}^2)']^3}  = \frac{4\kappa^3L^8}{X \pi^4} \int \sigma \dot{V} V''  d\sigma d\eta,  
\end{gathered}
\end{equation}
where $(V_{int}^2)'$ is the derivative of $V_{int}^2$ with respect to $\rho$, 
\beq
(V_{int}^2)'=\frac{6}{\rho}V_{int}^2.
\eeq
One can calculate this via two slightly different approaches. First, we can integrate directly in $\sigma$
\beq
\begin{aligned}
\int \sigma \dot{V} V''  d\sigma d\eta   = -\int_0^P d\eta \int_0^\infty  \dot{V}\partial_\sigma(\dot{V})d\sigma =-\frac{1}{2}\int_0^P  \Big[\dot{V}^2\Big]_0^\infty d\eta.
\end{aligned}
\eeq
From the boundary conditions \eqref{eqn:BCs}, it is now easy to see the following result
 \beq
c_{hol} =\frac{2\kappa^3}{X\pi^4}\int_0^P\mathcal{R}(\eta)^2 d\eta ,
\eeq
where clearly the central charge is proportional to the area under $\mathcal{R}(\eta)^2$.

Now, using the following 
\beq
\hspace{-1cm}
\dot{V}(\sigma,\eta) = \frac{\pi}{P} \sum_{n=1}^\infty n\,\mathcal{R}_n \,\sigma\sin\bigg(\frac{n\pi}{P}\eta\bigg) K_1\bigg(\frac{n\pi}{P}\sigma\bigg),~~~~\lim_{x\rightarrow \infty} x K_1(x) \sim \sqrt{\frac{\pi}{2}}\sqrt{x}e^{-x} = 0,~~~~\lim_{x\rightarrow 0} x K_1(x)=1,\nn
\eeq
leads to
\begin{align}
\dot{V}^2\Big|_0^\infty &= \frac{\pi^2}{P^2}  \sum_{n=1}^\infty \sum_{m=1}^\infty n\,m\,\mathcal{R}_n\mathcal{R}_m \bigg[\sigma K_1\bigg(\frac{n\pi}{P}\sigma\bigg)\bigg]\bigg[\sigma K_1\bigg(\frac{m\pi}{P}\sigma\bigg)\bigg]\sin\bigg(\frac{n\pi}{P}\eta\bigg)\sin\bigg(\frac{m\pi}{P}\eta\bigg)\bigg|_{\sigma=0}^{\sigma=\infty}\nn\\[2mm]
 &=   \sum_{n=1}^\infty \sum_{m=1}^\infty \mathcal{R}_n\mathcal{R}_m  \bigg[\frac{n\,\pi \sigma}{P} K_1\bigg(\frac{n\pi}{P}\sigma\bigg)\bigg]\bigg[\frac{m\,\pi \sigma}{P} K_1\bigg(\frac{m\pi}{P}\sigma\bigg)\bigg]\sin\bigg(\frac{n\pi}{P}\eta\bigg)\sin\bigg(\frac{m\pi}{P}\eta\bigg)\bigg|_{\sigma=0}^{\sigma=\infty}\nn\\[2mm]
  &= -  \sum_{n=1}^\infty \sum_{m=1}^\infty \mathcal{R}_n\mathcal{R}_m \sin\bigg(\frac{n\pi}{P}\eta\bigg)\sin\bigg(\frac{m\pi}{P}\eta\bigg),
\end{align}
and using 
\begin{equation} 
  \int_0^P \sin\bigg(\frac{n\pi}{P}\eta\bigg)\sin\bigg(\frac{m\pi}{P}\eta\bigg) d\eta =\frac{P}{2}\delta_{n\,m},
    \end{equation}
one finds
\begin{align}
c_{hol} = \frac{4\kappa^3}{X\pi^4}  \int \sigma \dot{V} V''  d\sigma d\eta  &= \frac{2\kappa^3}{X\pi^4}  \int_0^P \sum_{n=1}^\infty \sum_{m=1}^\infty \mathcal{R}_n\mathcal{R}_m \sin\bigg(\frac{n\pi}{P}\eta\bigg)\sin\bigg(\frac{m\pi}{P}\eta\bigg)   d\eta\nn\\[2mm]
&=\frac{ \kappa^3}{X\pi^4}\sum_{n=1}^\infty P \, \mathcal{R}_n^2.
\end{align}
Alternatively, one can insert the following definitions into the initial form, $\int \sigma \dot{V} V''  d\sigma d\eta$,
\begin{equation}\label{eqn:VderivVals}
\hspace{-0.75cm}
\begin{aligned}
V(\sigma,\eta) &=-\sum_{n=1}^\infty \mathcal{R}_n\,  \sin\bigg(\frac{n\pi}{P}\eta\bigg) K_0\bigg(\frac{n\pi}{P}\sigma\bigg),\\
\dot{V}(\sigma,\eta) &=  \sum_{n=1}^\infty \frac{n\pi}{P}\mathcal{R}_n\sigma  \sin\bigg(\frac{n\pi}{P}\eta\bigg) K_1\bigg(\frac{n\pi}{P}\sigma\bigg),\\
V''(\sigma,\eta) &=\sum_{n=1}^\infty \bigg(\frac{n\pi}{P}\bigg)^2\mathcal{R}_n \sin\bigg(\frac{n\pi}{P}\eta\bigg) K_0\bigg(\frac{n\pi}{P}\sigma\bigg),\\
\sigma \dot{V}(\sigma,\eta)V''(\sigma,\eta) &=\sum_{n=1}^\infty\sum_{m=1}^\infty \,  \bigg(\frac{n\pi}{P}\bigg)^2\frac{m\pi}{P}  \mathcal{R}_n  \mathcal{R}_m \,\sigma^2 \sin\bigg(\frac{n\pi}{P}\eta\bigg) \sin\bigg(\frac{m\pi}{P}\eta\bigg) K_0\bigg(\frac{n\pi}{P}\sigma\bigg) K_1\bigg(\frac{m\pi}{P}\sigma\bigg),
\end{aligned}
\end{equation}
using the following results 
  \begin{equation}\label{eqn:standard}
    \int_0^P \sin\bigg(\frac{n\pi}{P}\eta\bigg)\sin\bigg(\frac{m\pi}{P}\eta\bigg) d\eta =\frac{P}{2}\delta_{n\,m},~~~  \int_{0}^\infty \sigma^2 K_0\bigg(\frac{n\pi}{P}\sigma\bigg)K_1\bigg(\frac{n\pi}{P}\sigma\bigg) d\sigma = \frac{P^3}{2\pi^3n^3} ,
    \end{equation}
which leads to the same result as the first approach 
\begin{equation}
\begin{aligned}
c_{hol} &= \frac{4\kappa^3}{X\pi^4} \int \sigma \dot{V} V''  d\sigma d\eta 
=\frac{ \kappa^3}{X\pi^4}\sum_{k=1}^\infty P \, \mathcal{R}_k^2.
\end{aligned}
\end{equation}

\subsection{Type IIB}
We now calculate this quantity for the form of the IIB metrics given throughout this thesis, which reads in general 
\begin{align}
    ds_{10,B}^2&=4 \rho^2e^{\frac{2}{3}\Phi_\mathcal{A}}f_1(\sigma,\eta)\bigg(dx_{1,3}^2+\frac{1}{\rho^4}d\rho^2\bigg)+   e^{\frac{2}{3}\Phi_\mathcal{A}}f_1(\sigma,\eta) \bigg[ f_4(\sigma,\eta)(d\sigma^2+d\eta^2)+ ds^2(M_3^{\mathcal{B}})\bigg],\nn\\[2mm]
    ds^2(M_3^{\mathcal{B}}) &=f_2d\theta^2+ \frac{1}{\hat{\Xi} }\bigg(f_2f_3f_5\sin^2\theta\, d\phi_1^2+\frac{X^2}{f_1^3}\,\big(d\phi_2+ h(\eta,\sigma) \sin\theta  d\theta\big)^2\bigg),~~~~~~ e^{2\Phi_\mathcal{B}}=X^2\frac{e^{\frac{8}{3}\Phi_\mathcal{A}} }{\hat{\Xi} f_1^2},
\end{align}
for some functions $(\hat{\Xi},h)$.
 We now calculate the following quantities
\beq
\begin{gathered}
\text{det}[g_{ij}] =X^2\frac{e^{\frac{10}{3}\Phi_\mathcal{A}}}{\hat{\Xi}^2}f_1^2f_4^2f_2^2f_3f_5\sin^2\theta,~~~~~~~~~~\alpha =4\rho^2 e^{\frac{2}{3}\Phi_\mathcal{A}}f_1,\\
\Rightarrow \sqrt{\text{det}[g_{ij}]e^{-4\Phi_\mathcal{B}}\alpha^3} = \frac{8}{X}\rho^3 f_1^{\frac{9}{2}}f_3^{\frac{1}{2}}f_5^{\frac{1}{2}}f_4f_2\text{Vol}(S^1)\text{Vol}(S^2),
\end{gathered}
\eeq
with all dilaton and $\hat{\Xi}$ dependence dropping out neatly. This matches exactly the IIA result \eqref{eqn:HCCIIA}, and hence leads to the same holographic central charge. Here we note that det$(M_3^{\mathcal{B}})=\delta_2^{2}\,\text{det}(M_3^{\mathcal{A}})$, and hence det$(g_{ij}^{\mathcal{B}})=\delta_2^{2}\,\text{det}(g_{ij}^{\mathcal{A}})$, with $e^{4\Phi_\mathcal{B}} = \delta_2^{2}\,e^{4\Phi_\mathcal{A}}$ and $\delta_2$ defined in \eqref{eqn:betaredchitdual}. Consequently, all $\delta_2$ contributions simply cancel, recovering the IIA result. This is as expected, as these arguments match those summarised in \eqref{eqn:HCCtransform}, from the original presentation in Section 4.3 of \cite{Macpherson:2014eza}. 
\chapter{Deformations of GM}

\section{Transformation from LLM to GM}\label{sec:Transformation}
We present below a step-by-step transformation from the LLM class of backgrounds to the GM class, via the `B$\ddot{\text{a}}$cklund' transformation, demonstrating explicitly the following relation (as described in \cite{Bah:2019jts,Bah:2022yjf,Couzens:2022yjl})
\begin{equation}\label{eqn:pickup}
\tilde{\chi}\rightarrow\chi+ \beta,~~~~~~\tilde{\beta}\rightarrow -\beta.
\end{equation}
We begin with the LLM background with the additional $U(1)$ isometry, $\tilde{\beta}$, picked up via the coordinate transformation \eqref{eqn:x1x2transformation}, and given in \eqref{eqn:LLM}. We re-write the solution here for convenience 
\begin{adjustwidth}{-1.75cm}{}
\vspace{-0.7cm}
\begin{align}
\frac{ds^2}{\kappa^{\frac{2}{3}}}&= e^{2\lambda}\bigg[4ds^2(\text{AdS}_5)+y^2 e^{-6\lambda} ds^2(\text{S}^2)+\frac{4}{1-y\partial_yD}\Big(d\tilde{\chi}-\frac{r}{2}\partial_rD d\tilde{\beta}\Big)^2-\frac{\partial_yD}{y}\Big(dy^2+e^D (dr^2+r^2d\tilde{\beta}^2)\Big)\bigg],\nn\\[2mm]
G_4&= \kappa \bigg[-d(2y^3 e^{-6\lambda}) \wedge d\tilde{\chi} +  \bigg(d\bigg(e^{-6\lambda}\frac{y^2}{\partial_yD}r\partial_rD\bigg) - \partial_y(e^D)r\,dr + r\partial_rD\,dy\bigg) \wedge d\tilde{\beta} \bigg]\wedge \text{vol}(\text{S}^2), \nn
\end{align}
\end{adjustwidth}
with $D(r,y)$ satisfying
\begin{equation}
\frac{1}{r}\partial_r(r \partial_r D) +\partial_y^2e^D=0,~~~~e^{-6\lambda} = \frac{-\partial_yD}{y(1-y\partial_yD)}.
\end{equation}
By direct comparison with the GM case \eqref{eqn:GM}, re-written below for clarity
\begin{align}
ds^2&=f_1 \Bigg[4ds^2(\text{AdS}_5) +f_2 ds^2(\text{S}^2)+f_3 d\chi^2 +f_4 \big(d\sigma^2 +d\eta^2\big)+f_5 \Big(d\beta +f_6 d\chi\Big)^2\Bigg],\nn\\[2mm]
A_3&=\Big(f_7 d\chi +f_8 d\beta \Big)\wedge \text{vol}(\text{S}^2),\nn
\end{align}
it is immediately clear that 
\begin{equation}
\kappa^{\frac{2}{3}}e^{2\lambda} = f_1,~~~~y^2 e^{-6\lambda} =f_2.
\end{equation}
After the `B$\ddot{\text{a}}$cklund' transformation, which replaces $(r,y)$ with $(\sigma,\eta)$
\begin{equation}\label{eqn:Backlund}
r^2e^D=\sigma^2,~~~~y=\dot{V},~~~~\text{log}(r)=V',
\end{equation}
along with the given relation for $\lambda$, 
\begin{equation}
e^{-6\lambda}= \frac{-\partial_yD}{y(1-y\partial_yD)} ,~~~~\Rightarrow~~~~ \frac{2V''}{\tilde{\Delta}} = \frac{-\partial_yD}{(1-y\partial_yD)} ,
\end{equation}
one arrives at
\begin{equation}\label{eqn:box3}
\partial_yD =\frac{2V''}{2V''\dot{V}-\tilde{\Delta}}=\frac{2V''}{\ddot{V}V''-(\dot{V}')^2},
\end{equation}
which matches the form given in \cite{Reid-Edwards:2010vpm} after inserting the Laplace equation \eqref{eqn:laplace}.\\
\\Using $r=e^{V'}$ and $\sigma^2=r^2e^D$ from the B$\ddot{\text{a}}$cklund transformation, one finds
\begin{equation}\label{eqn:box4}
e^{\frac{1}{2}D}dr=  \sigma V''d\eta +\dot{V}'d\sigma.
\end{equation}
Hence, 
\begin{equation}
dy^2+e^Ddr^2 =(d\sigma^2+d\eta^2)\bigg(-V''\ddot{V}+(\dot{V}')^2\bigg),
\end{equation}
meaning, from the LLM metric, we see
\begin{align}
-\frac{\partial_yD}{y}(dy^2+e^Ddr^2 )&= -\frac{1}{\dot{V}}\frac{2V''}{(\ddot{V}V''-(\dot{V}')^2)}\big(-V''\ddot{V}+(\dot{V}')^2\big)(d\sigma^2+d\eta^2)\nn\\[2mm]
&=f_4 (d\sigma^2+d\eta^2).
\end{align}
Thus, the only part of the LLM metric still left to transform is 
 \begin{equation}
\frac{4}{1-y\partial_yD}\Big(d\tilde{\chi}-\frac{r}{2}\partial_rD d\tilde{\beta}\Big)^2-\frac{\partial_yD}{y}e^D r^2d\tilde{\beta}^2.
\end{equation}
Before doing so, we note the following relation from the transformation $\sigma(r,y)$ (as in \cite{Reid-Edwards:2010vpm})
\begin{equation}
\frac{d\sigma}{d\eta}=\frac{\partial \sigma}{\partial r} \frac{\partial r}{\partial \eta}+\frac{\partial \sigma}{\partial y} \frac{\partial y}{\partial \eta} = \partial_r\sigma \partial_\eta r+\partial_y \rho \partial_\eta y=0,
\end{equation}
giving 
\begin{equation}
\partial_r\sigma=-\frac{\partial_y \sigma \partial_\eta y}{ \partial_\eta r}.
\end{equation}
Now we use the forms of the transformation to get (assuming $\sigma =re^{\frac{D}{2}}$)
\begin{equation}
\partial_r\sigma= \frac{1}{2}e^{\frac{D}{2}}(2+r\partial_r D),~~~~\partial_y\sigma= \frac{\sigma}{2}\partial_y D,~~~~\partial_\eta y = \dot{V}',~~~~\partial_\eta r = V'' r.
\end{equation}
Then, by substituting these forms into the previous equation, we find
\begin{equation}
\begin{aligned}
& e^{\frac{D}{2}}(2+r\partial_r D)=-\frac{\sigma\partial_y D \dot{V}'}{V'' r}\\
\Rightarrow~~~~~ & r\partial_r D=- \frac{\partial_y D \dot{V}'}{V'' }-2 = - \frac{2 \dot{V}'}{\ddot{V}V''-(\dot{V}')^2}-2 = -2\Big(g(\eta,\sigma)+1\Big).
\end{aligned}
\end{equation}
Inserting this result into the remaining part of the LLM metric, it becomes clear that this additional (plus one) contribution to $g(\eta,\sigma)$ introduces a $\tilde{\beta}$ term into the definition of $\chi$ for the GM class
 \begin{equation}
\frac{4}{1-y\partial_yD}\Big(d\tilde{\chi}-\frac{r}{2}\partial_rD d\tilde{\beta}\Big)^2-\frac{\partial_yD}{y}e^D r^2d\tilde{\beta}^2 
= \frac{4}{1-y\partial_yD}\bigg(d\chi-g(\eta,\sigma)d\beta\bigg)^2-\frac{\partial_yD}{y}e^D r^2d\beta^2,
\end{equation}
where 
\begin{equation}
\chi = \tilde{\chi} + \tilde{\beta},~~~~\beta=-\tilde{\beta},~~~~g(\eta,\sigma) = \frac{ \dot{V}'}{\ddot{V}V''-(\dot{V}')^2}.
\end{equation}
Let us now we re-arrange this metric component such that the roles of $\chi$ and $\beta$ are switched
\begin{align}
\frac{4}{1-y\partial_yD}&\bigg(d\chi-g(\eta,\sigma)d\beta\bigg)^2-\frac{\partial_yD}{y}e^D r^2d\beta^2= -\frac{4r^2e^D \partial_yD}{4y\,g(\eta,\sigma)^2 -e^D r^2\partial_yD (1-y\partial_yD)}d\chi^2\nn\\[2mm]
+& \bigg(\frac{4y g(\eta,\sigma)^2 -r^2 e^D \partial_yD (1-y\partial_yD)}{y(1-y\partial_yD)}\bigg)\bigg[d\beta -\frac{4y\, g(\eta,\sigma)}{4y\,g(\eta,\sigma)^2 -e^D r^2 \partial_yD (1-y\partial_yD)}d\chi\bigg]^2 .
\end{align}
Noting the following relations
\begin{equation}
4y g(\eta,\sigma)^2 -r^2 e^D \partial_yD (1-y\partial_yD) = -\frac{2\Lambda V''}{\ddot{V}V''-(\dot{V}')^2},~~~~~~~~~~y (1-y\partial_yD) = -\frac{\dot{V}\tilde{\Delta}}{\ddot{V}V''-(\dot{V}')^2},
\end{equation}
we now transform each of the above elements  
\begin{align}
\frac{4y g(\eta,\sigma)^2 -r^2 e^D \partial_yD (1-y\partial_yD)}{y(1-y\partial_yD)} = \frac{2\Lambda V''}{\dot{V} \tilde{\Delta}}&=f_5,\nn\\[2mm]
\frac{4y\, g(\eta,\sigma)}{4y\,g(\eta,\sigma)^2 -e^D r^2 \partial_yD (1-y\partial_yD)}= -\frac{2\dot{V}\dot{V}'}{\Lambda V''} &=-  f_6,\nn\\[2mm]
\frac{4r^2e^D \partial_yD}{4y\,g(\eta,\sigma)^2 -e^D r^2\partial_yD (1-y\partial_yD)} = -\frac{4\sigma^2}{\Lambda}&=-f_3.
\end{align}
Hence, our metric component transforms as follows
\begin{align}
\frac{4}{1-y\partial_yD}\bigg(d\chi&-g(\eta,\sigma)d\beta\bigg)^2-\frac{\partial_yD}{y}e^D r^2d\beta^2= f_5\big(d\beta +f_6d\chi\big)^2+f_3d\chi^2. 
\end{align}
For the transformation of $A_3$, we first note the relations
\begin{equation}
\hspace{-0.5cm}
2\kappa\, y^3 e^{-6\lambda} = -f_7,~~~~\kappa\, e^{-6\lambda}\frac{y^2}{\partial_yD}(2y\partial_yD-2g(\eta,\sigma)) = -f_7 - \kappa \frac{2\dot{V}\dot{V}'}{\tilde{\Delta}},~~~~\kappa\bigg(2e^{-6\lambda}\frac{y^2}{\partial_yD}+2y\bigg) = -f_7,\nn
\end{equation}
and 
\begin{equation}
- \partial_y(e^D)r\,dr +(2+r\partial_rD)\,dy =2d\eta,
\end{equation}
and after taking the positive square-root for $\sigma = \pm r e^{\frac{D}{2}}$, we find
\begin{align}
&G_4= \kappa \bigg[-d(2y^3 e^{-6\lambda}) \wedge d\tilde{\chi} -d(2y^3 e^{-6\lambda}) \wedge d\tilde{\beta} \nn \\[2mm]
&~~~~~~+  \bigg(d\bigg(e^{-6\lambda}\frac{y^2}{\partial_yD}(2y\partial_yD+r\partial_rD)\bigg) - \partial_y(e^D)r\,dr + r\partial_rD\,dy\bigg) \wedge d\tilde{\beta} \bigg]\wedge \text{vol}(\text{S}^2) \nn\\[2mm]
&=  \bigg[d(f_7) \wedge d\tilde{\chi} +d(f_7) \wedge d\tilde{\beta} \nn\\[2mm]
&~~~~~+ \kappa \bigg(d\bigg(e^{-6\lambda}\frac{y^2}{\partial_yD}(2y\partial_yD-2g(\eta,\sigma)-2)\bigg) - \partial_y(e^D)r\,dr + r\partial_rD\,dy\bigg) \wedge d\tilde{\beta} \bigg]\wedge \text{vol}(\text{S}^2) \nn\\[2mm]
&=  \bigg[d(f_7) \wedge d\tilde{\chi} + d(f_7) \wedge d\tilde{\beta}+  \bigg(d\bigg( -f_7 -\kappa \frac{2\dot{V}\dot{V}'}{\tilde{\Delta}}\bigg) + 2\kappa\, d\eta -\kappa\, d\bigg(2e^{-6\lambda}\frac{y^2}{\partial_yD}+2y\bigg) \bigg) \wedge d\tilde{\beta}  \bigg]\wedge \text{vol}(\text{S}^2) \nn\\[2mm]
&=  \bigg[d(f_7) \wedge (d\tilde{\chi} +d\tilde{\beta}) - d(f_8)  \wedge d\tilde{\beta}  \bigg]\wedge \text{vol}(\text{S}^2)\nn\\[2mm]
&=  \bigg(d(f_7) \wedge d\chi + d(f_8)  \wedge d\beta  \bigg)\wedge \text{vol}(\text{S}^2).
\end{align}
We finally arrive at the form of the GM solution written in \eqref{eqn:GM}. We have demonstrated explicitly that the $\chi$ in the GM class has the form $\chi = \tilde{\chi} + \tilde{\beta}$ when written in terms of the LLM form (as stated in \cite{Bah:2019jts,Bah:2022yjf,Couzens:2022yjl}), and $\beta= -\tilde{\beta}$.

\section{Behaviour of $f_i$ at the boundaries}\label{sec:fs}
Here we simply quote the values of the functions $f_i$ \eqref{eqn:fs} at each $\sigma\in [0,\infty),\eta\in[0,P]$ boundary appearing in our various deformed GM solutions. 

\subsubsection*{At $\sigma\rightarrow \infty$}
To leading order
  \beq\label{eqn:Vlimit}
  V=-\mathcal{R}_1 e^{-\frac{\pi}{P}\sigma}\sqrt{\frac{P}{2\sigma}}\sin\bigg(\frac{\pi}{P}\eta\bigg)+...,
  \eeq
      with 
\begin{align}\label{eqn:finf}
&f_1=\left(\frac{\pi^3 {\cal R}^2_1 \kappa^2 \sigma^2}{4 P^2}e^{-\frac{2\pi}{P}\sigma}\right)^{\frac{1}{3}},~~~f_2=\frac{2P}{\pi\sigma}\sin^2\left(\frac{\pi\eta}{P}\right),~~~f_3=4,~~~~f_4=\frac{2\pi}{P\sigma},~~~f_5=\frac{4P^2}{\pi^3 {\cal R}_1^2}e^{\frac{2\pi}{P}\sigma},\\[2mm]
&f_6=\sqrt{\frac{2}{P\sigma}}\pi {\cal R}_1 e^{-\frac{\pi}{P}\sigma}\cos\left(\frac{\pi\eta}{P}\right),~~f_7=-2\kappa{\cal R}_1 \sqrt{\frac{2 P}{\sigma}}e^{-\frac{\pi}{P}\sigma}\sin^3\left(\frac{\pi \eta}{P}\right),~~f_8=\kappa\left(-2 \eta+\frac{P}{\pi}\sin\left(\frac{2\pi\eta}{P}\right)\right)\nn.
\end{align}

\subsubsection*{At $\eta=0$ with $\sigma\neq 0$}
        \begin{align}\label{eqn:feq}
 &   \dot{V}'=f,~~~\dot{V}=f\eta,~~~V''=-\frac{\eta}{\sigma^2}\dot{f},~~~\ddot{V}=\eta \dot{f},~~~f(\sigma)=\frac{\pi^2}{P^2}\sum_{n=1}^\infty \mathcal{R}_n\,\sigma\,n^2K_1\Big(\frac{n\pi}{P}\sigma\Big),\nn\\[2mm]
&f_1=\left(\frac{\kappa^2 \sigma^2 f^3}{-2\dot{f}}\right)^{\frac{1}{3}},~~~f_2=\frac{-2\eta^2\dot{f}}{\sigma^2 f},~~~f_3=\frac{-4 \dot{f}}{2f- \dot{f}},~~~f_4=-\frac{2 \dot{f}}{f\sigma^2},~~~f_5=\frac{2(2f -\dot{f})}{f^3},\\[2mm]
&f_6=\frac{2 f^2}{2f-\dot{f}},~~~f_7=\frac{4\kappa\eta^3 \dot{f}}{\sigma^2},~~~f_8=2\kappa \eta^3 \frac{\dot{f}}{ f \sigma^2}\bigg(2-\frac{\dot{f}}{f}\bigg),
~~~~~~~~~~  \text{where}~~ |\dot{f}|=-\dot{f}. \nn
\end{align}

\subsubsection*{At $\sigma=0,~\eta\in (k,k+1)$}
     Along the $\sigma=0$ boundary, $\ddot{V}=0$ to leading order, hence (using the boundary condition given in \eqref{eqn:BCs} in the final step), we first note
       \begin{equation}
  \frac{f_2}{f_5}\Big|_{\sigma\rightarrow0}=\frac{\dot{V}^2V''}{2\dot{V}-\ddot{V}} \Big|_{\sigma\rightarrow0} = \frac{1}{2}\dot{V}V''=\frac{1}{2}\mathcal{R}(\eta)V'',
  \end{equation}
and using \eqref{eq:alternative}, in this limit
\beq \label{eqPdef2}
\hspace{-1.5cm}
V''= P_k= \sum_{j=k+1}^P\frac{b_j}{j-\eta}+\frac{1}{2P}\sum_{j=1}^Pb_j\left[\psi\left(\frac{\eta+j}{2P}\right)-\psi\left(\frac{\eta-j}{2P}\right)+\frac{\pi}{2}\left(\cot\left(\frac{\pi(\eta+j)}{2P}\right)-\cot\left(\frac{\pi(\eta-j)}{2P}\right)\right)\right],
\eeq
with $\psi$ the digamma function. This doesn't vanish or blow up between these bounds. One finds
\begin{align}\label{eqn:fsat0}
f_1&=\left(\frac{\kappa^{2}{\cal R}(2 {\cal R}P_k +({\cal R}')^2)}{2 P_k}\right)^{\frac{1}{3}},~~~f_2=\frac{2 {\cal R} P_k}{2 {\cal R} P_k+({\cal R}')^2},~~~f_3= \frac {2 \sigma^2 P_k}{\cal R},~~~f_4=\frac{ 2 P_k}{{\cal R}},\nn\\[2mm]
f_5&=\frac{4}{2{\cal R}P_k+({\cal R}')^2},~~~f_6={\cal R}',~~~f_7=-\frac{4\kappa {\cal R}^2 P_k}{2{\cal R} P_k+({\cal R}')^2},~~~f_8=2\kappa\left(-\eta+\frac{{\cal R}{\cal R}'}{2{\cal R}P_k+({\cal R}')^2}\right), 
\end{align}
recalling ${\cal R}=N_k+(N_{k+1}-N_{k})(\eta-k)$.

           \subsubsection*{At $\sigma=0,~\eta=0$} To approach this boundary, adopt the coordinate change $(\eta=r \cos\alpha,~\sigma=r \sin\alpha)$, expanding about $r=0$. To leading order
  \begin{align}\label{eqQdef}
&\dot{V}=N_1 r \cos\alpha,~~~~~~\dot{V}'= N_1,~~~V''=\frac{1}{4P^2}r\cos\alpha\sum_{j=1}^Pb_j\left(2 \psi^1\left(\frac{j}{2P}\right)-\pi^2\csc^2\left(\frac{j\pi}{2P}\right)\right),\nn\\[2mm]
&f_1=\left(\frac{\kappa^2 N_1^3 }{2 Q}\right)^{\frac{1}{3}},~~~f_2=\frac{2 r^2 Q \cos^2\alpha }{N_1},~~~f_3= \frac{2 r^2 Q \sin^2\alpha }{N_1},~~~f_4=\frac{2 Q}{N_1},\\[2mm]
&f_5= \frac{4}{N_1^2},~~~f_6= N_1,~~~f_7=-4\kappa Q r^3 \cos^3\alpha,~~~f_8=0,\nn
\end{align}
   where $Q$ is extracted via $V''=r Q\,\cos\alpha$. Notice that $f_5,f_6$ remain finite whereas $f_3$ and $f_2$ vanish.
     
          \subsubsection*{At $\sigma=0,~\eta=k$}
           Now one should make the following coordinate change $(\eta=k-r \cos\alpha,~\sigma=r \sin\alpha)$ (where $0<k<P$ and $k\in \mathds{Z}$). To leading order  
\begin{align}\label{eqn:fsfork}
&f_1=(\kappa N_k)^{\frac{2}{3}},~~~f_2=1,~~~f_3=\frac{r^2\sin^2\alpha}{N_k}\frac{b_k}{r},~~~f_4=\frac{1}{N_k}\frac{b_k}{r},~~~f_5=\frac{4}{N_k}\frac{r}{b_k},~~f_7=-2\kappa N_k,~~f_8=-2\kappa k.\nn\\[2mm]
&f_6=\frac{b_k}{2}(1+\cos\alpha)+N_{k+1}-N_k=\cos^2\Big(\frac{\alpha}{2}\Big) (N_k-N_{k-1})+\sin^2\Big(\frac{\alpha}{2}\Big) (N_{k+1}-N_{k}) \equiv g(\alpha),
\end{align} 
noting the use of $b_k=2N_k-N_{k+1}-N_{k-1}$ to re-write $f_6$.

\section{$GL(3,\mathds{R})$ Reductions}\label{sec:GL(3R)}
\subsubsection{ $\beta$ reduction}
Here we present the full type IIA solution, with all nine $GL(3,\mathds{R})$ transformation parameters intact, following a dimensional reduction along $\beta$
  \begin{align}\label{eqn:beta-Gen}
 &       ds_{10,st}^2=e^{\frac{2}{3}\Phi}f_1\bigg[4ds^2(\text{AdS}_5)+f_2d\theta^2+f_4(d\sigma^2+d\eta^2)\bigg]+\frac{1}{X^2}f_1^2 e^{-\frac{2}{3}\Phi} ds^2_2,\nn\\[2mm]
  &e^{\frac{4}{3}\Phi}=  \frac{1}{X^2}f_1 \Big[q^2(f_3+f_5f_6^2)+b^2 f_5 +2bqf_5f_6 + v^2 \sin^2\theta f_2\Big] ,~~~~~~~~~  B_2 =\frac{1}{X}\Big(B_{2,\chi}d\chi  +B_{2,\phi}d\phi \Big) \wedge d\theta,\nn\\[2mm]
&C_1= \frac{1}{X}\Big(C_{1,\chi} d\chi  + C_{1,\phi} d\phi\Big)  ,~~~~~~~~~~~~~~~~~
       C_3= C_{3,\chi\phi} d\chi \wedge d\theta \wedge d\phi,\nn\\[2mm]
    &   B_{2,\chi} =   \sin\theta \Big((vp -sq )f_7+(va -sb)f_8 \Big),~~~~~~~~B_{2,\phi} = \sin\theta \Big((vm - uq)f_7 +(vc-ub)f_8 \Big),\nn\\[2mm]
     &  C_{1,\chi} =   f_1 e^{-\frac{4}{3}\Phi}\Big(bf_5 (a+pf_6) +p\,q f_3 + qf_5f_6(a+pf_6) +s\,v \sin^2\theta f_2\Big), \nn\\[2mm]
    &   C_{1,\phi} =  f_1 e^{-\frac{4}{3}\Phi} \Big(bf_5 (c+mf_6) +m\,q f_3 + qf_5f_6(c+mf_6) +u\,v \sin^2\theta f_2\Big),\nn\\[2mm]
    &   C_{3,\chi\phi}=\sin\theta \,\bigg[u (p f_7+a f_8) - s(mf_7 +cf_8)  \bigg], \nn\\[2mm]
   &    ds^2_2 = h_\chi(\eta,\sigma,\theta)d\chi^2 + h_\phi(\eta,\sigma,\theta)d\phi^2 + h_{\chi\phi}(\eta,\sigma,\theta)d\chi d\phi\nn\\[2mm]
   &    h_\chi(\eta,\sigma,\theta) = f_3f_5 (bp-aq)^2 +\sin^2\theta f_2\Big[b^2s^2 f_5 +(pv-qs)^2f_3 - 2bsf_5\big((pv-qs)f_6 +av\big) \nn\\[2mm]
   &   ~~~~~~~~~~~~~~~~~~~~~~~~~~~~~~ +f_5\big((pv-qs)f_6 +av\big)^2\Big],\nn\\[2mm]
    &          h_\phi(\eta,\sigma,\theta) = f_3f_5 (bm-cq)^2 +\sin^2\theta f_2\Big[c^2v^2 f_5 +(qu-mv)^2f_3 - 2cv f_5\big((qu-mv)f_6 +bu\big)\nn\\[2mm]
    &          ~~~~~~~~~~~~~~~~~~~~~~~~~~~~~~~~~ +f_5\big((qu-mv)f_6 +bu\big)^2\Big],\nn\\[2mm]
    &                        h_{\chi\phi}(\eta,\sigma,\theta) =2\bigg[ f_3f_5 (bp-aq)(bm-cq)  +\sin^2\theta f_2 h_4(\eta,\sigma)\bigg],\nn\\[2mm]
   &                       h_4(\eta,\sigma) =  cv f_5\Big((pv-qs)f_6+av\Big) +b^2suf_5 -(qu-mv) \Big[(pv-qs)f_3+f_5f_6 \Big((pv-qs)f_6+av\Big)\Big]\nn\\[2mm]
   &~~~~~~~~~~~~~~~~~~ - bf_5\bigg(sv(c+mf_6)+u\Big(av+(pv-2qs)f_6\Big)\bigg),
    \end{align}
with $X$ the generalised reduction parameter, from \eqref{eqn:reductionformulaGEN}.

\subsubsection{ $\chi$ reduction}
The full type IIA solution corresponding to a dimensional reduction along $\chi$, with all nine $GL(3,\mathds{R})$ transformation parameters intact, reads
  \begin{align}\label{eqn:chi-Gen}
  &      ds_{10,st}^2=e^{\frac{2}{3}\Phi}f_1\bigg[4ds^2(\text{AdS}_5)+f_2d\theta^2+f_4(d\sigma^2+d\eta^2)\bigg]+\frac{1}{X^2}f_1^2 e^{-\frac{2}{3}\Phi} ds^2_2,\nn\\[2mm]
    &    e^{\frac{4}{3}\Phi}=\frac{1}{X^2} f_1 \Big[p^2(f_3+f_5f_6^2)+a^2 f_5 +2apf_5f_6 + s^2 \sin^2\theta f_2\Big] ,~~~~C_3 = C_{3,\beta\phi} d\beta \wedge d\theta \wedge d\phi,
      \nn\\[2mm]
    &   C_1=\frac{1}{X}(C_{1,\beta}d\beta+ C_{1,\phi}d\phi),~~~~~~~~~~  B_2=\frac{1}{X}(B_{2,\beta}d\beta +B_{2,\phi}d\phi)\wedge d\theta, \nn\\[2mm]
   &    B_{2,\beta} = \sin\theta\Big((sq-vp)f_7+(s b-va)f_8\Big),~~~~~~~B_{2,\phi}= \sin\theta\Big( (s m-up )f_7 +(s c-ua) f_8\Big), \nn\\[2mm]
   &    C_{1,\beta}=  f_1 e^{-\frac{4}{3}\Phi}\Big(bf_5 (a+pf_6) +p\,q f_3 + qf_5f_6(a+pf_6) +s\,v \sin^2\theta f_2\Big),  \nn\\[2mm]
   &    C_{1,\phi} =    f_1 e^{-\frac{4}{3}\Phi}\Big(af_5 (c+mf_6) +m\,p f_3 + pf_5f_6(c+mf_6) +u\,s \sin^2\theta f_2\Big)   ,\nn\\[2mm]
     &   C_{3,\beta\phi} =\sin\theta\Big[u (qf_7+bf_8) - v(mf_7 +cf_8)  \Big],  \nn\\[2mm]
   &    ds^2_2 = h_\beta(\eta,\sigma,\theta)d\beta^2 + h_\phi(\eta,\sigma,\theta)d\phi^2 + h_{\beta\phi}(\eta,\sigma,\theta)d\beta d\phi\nn\\[2mm]
   &    h_\beta(\eta,\sigma,\theta) = f_3f_5 (bp-aq)^2 +\sin^2\theta f_2\Big[b^2s^2 f_5 +(pv-qs)^2f_3 - 2bsf_5\big((pv-qs)f_6 +av\big) \nn\\[2mm]
   &    ~~~~~~~~~~~~~~~~~~~~~~~~~~~~~~+f_5\big((pv-qs)f_6 +av\big)^2\Big],\nn\\[2mm]
  &            h_\phi(\eta,\sigma,\theta) = f_3f_5 (pc - am)^2 +\sin^2\theta f_2\Big[c^2s^2 f_5 +(pu-ms)^2f_3 - 2cs f_5\big((pu-ms)f_6 +au\big) \nn\\[2mm]
    &          ~~~~~~~~~~~~~~~~~~~~~~~~~~~~~~~+f_5\big((pu-ms)f_6 +au\big)^2\Big],\nn\\[2mm]
     &                       h_{\beta\phi}(\eta,\sigma,\theta) =2\bigg[ f_3f_5 (bp-aq)(pc-am)  +\sin^2\theta f_2 h_4(\eta,\sigma)\bigg],\nn\\[2mm]
    &                      h_4(\eta,\sigma) =-  sb f_5\Big((pu-ms)f_6-cs\Big) +a^2vuf_5 +(pv-qs) \Big[(pu-ms)f_3\nn\\[2mm]
    &~~~~~~~+f_5f_6 \Big((pu-ms)f_6-cs\Big)\Big] - af_5\bigg(sv(c+mf_6)+u\Big(bs+(qs-2pv)f_6\Big)\bigg),
    \end{align}
    with $X$ the reduction parameter from \eqref{eqn:reductionformulaGEN}.

\subsubsection{ $\phi$ reduction}

The dimensional reduction along $\phi$, with all all nine $GL(3,\mathds{R})$ transformation parameters intact, reads
   \begin{align}\label{eqn:phi-Gen}
  &      ds_{10,st}^2=e^{\frac{2}{3}\Phi}f_1\bigg[4ds^2(\text{AdS}_5)+f_2d\theta^2+f_4(d\sigma^2+d\eta^2)\bigg]+\frac{1}{X^2}f_1^2 e^{-\frac{2}{3}\Phi} ds^2_2,\nn\\[2mm]
 &       e^{\frac{4}{3}\Phi}=\frac{1}{X^2} f_1 \Big[m^2(f_3+f_5f_6^2)+c^2 f_5 +2cmf_5f_6 + u^2 \sin^2\theta f_2\Big] ,~~~~~~~~~B_2=\frac{1}{X}(B_{2,\chi}d\chi+B_{2,\beta}d\beta)\wedge d\theta,\nn\\[2mm]
&        C_1=\frac{1}{X}(C_{1,\chi}d\chi+C_{1,\beta}d\beta),~~~~~~~~~~C_3 = C_{3,\chi\beta}d\chi\wedge d\theta \wedge d\beta,\nn\\[2mm]
   &    B_{2,\chi} = \sin\theta\Big((up -sm) f_7+(ua-sc) f_8\Big),~~~~~~~~~~~~~~ B_{2,\beta} =\sin\theta \Big((uq-vm)f_7+(ub-vc)f_8\Big)  ,\nn\\[2mm]
 &      C_{1,\chi}=    f_1 e^{-\frac{4}{3}\Phi}\Big(af_5 (c+mf_6) +p\,m f_3 + pf_5f_6(c+mf_6) +s\,u \sin^2\theta f_2\Big), \nn\\[2mm]
   &    C_{1,\beta}=  f_1 e^{-\frac{4}{3}\Phi}\Big(bf_5 (c+mf_6) +m\,q f_3 + qf_5f_6(c+mf_6) +u\,v \sin^2\theta f_2\Big)   ,\nn\\[2mm]
  &      C_{3,\chi\beta}=\sin\theta \bigg[v(p f_7+a f_8)-s (qf_7+bf_8) \bigg] ,\nn\\[2mm]
 &      ds^2_2 = h_\chi(\eta,\sigma,\theta)d\chi^2 + h_\beta(\eta,\sigma,\theta)d\beta^2 + h_{\chi\beta}(\eta,\sigma,\theta)d\chi d\beta, \nn\\[2mm]
 &      h_\chi(\eta,\sigma,\theta) =    f_3f_5 (am-pc)^2 +\sin^2\theta f_2\Big[c^2s^2 f_5 +(pu-ms)^2f_3 - 2cs f_5\big((pu-ms)f_6 +au\big) \nn\\[2mm]
  &     ~~~~~~~~~~~~~~~~~~~~~~~~~~~+f_5\big((pu-ms)f_6 +au\big)^2\Big],\nn\\[2mm]
   &           h_\beta(\eta,\sigma,\theta) = f_3f_5 (bm-cq)^2 +\sin^2\theta f_2\Big[c^2v^2 f_5 +(qu-mv)^2f_3 - 2cv f_5\big((qu-mv)f_6 +bu\big) \nn\\[2mm]
 &             ~~~~~~~~~~~~~~~~~~~~~~~~~~~~+f_5\big((qu-mv)f_6 +bu\big)^2\Big], \nn\\[2mm]
     &                       h_{\chi\beta}(\eta,\sigma,\theta) =2\bigg[ f_3f_5 (am-pc)(bm-cq)  +\sin^2\theta f_2 h_4(\eta,\sigma)\bigg],\nn\\[2mm]
     &                     h_4(\eta,\sigma) =  ub f_5\Big((pu-ms)f_6+au\Big) +c^2sv f_5 +(qu-mv) \Big[(pu-ms)f_3+f_5f_6 \Big((pu-ms)f_6+au\Big)\Big]\nn\\[2mm]
     &~~~~~~~~~~~~~~~~~~ - cf_5\bigg(uv(a+pf_6)+s\Big(ub+(qu-2mv)f_6\Big)\bigg),
    \end{align}
    with $X$ the reduction parameter from \eqref{eqn:reductionformulaGEN}.
\newpage

\section{T-dualising the G-structure description}\label{sec:ATDGstructure}
 Here we present in full detail the derivation of the IIB G-Structure description, following an ATD from IIA. 
Note that in this calculation the convention $d_{H_3}=d+H_3\wedge$ is required - in order to use the minus sign convention (used throughout the rest of the thesis) we would need to appropriately flip the sign of the $B$ field in the T-Dual rules given in \eqref{eqn:TD2} (such that $E_\mathcal{B}^y=e^{-C^\mathcal{A}}(dy+B_1^\mathcal{A})$). 
 
As discussed in Section \ref{sec:Mink4} and \ref{sec:ATDGssumarry}, the pure spinors transform in the same manner as the Ramond fields under the T-duality. We can then use the T-dual rules \eqref{eqn:TD2} to make the following decomposition 
\begin{equation*}
 \Psi_\pm^\mathcal{A}=\Psi_{\pm_{\perp}}^\mathcal{A} + \Psi_{\pm_{||}}^\mathcal{A}\wedge E^y_\mathcal{A},
\end{equation*}
with the following initial ansatz
\beq\label{eqn:PsiAnsatz}
\Psi_\mp^\mathcal{B}=e^{C^\mathcal{A}}\Psi_{\pm_{||}}^\mathcal{A} + \Psi_{\pm_{\perp}}^\mathcal{A}\wedge (dy-B_1^\mathcal{A}),
\eeq
 where  $E^y_\mathcal{A}=  e^{C^\mathcal{A}}(dy+A_1^\mathcal{A})$ and $E^y_\mathcal{B}=  e^{C^\mathcal{B}}(dy+A_1^\mathcal{B}) =  e^{-C^\mathcal{A}}(dy-B_1^\mathcal{A}) $. We will see shortly that this ansatz for $\Psi_\mp^\mathcal{B}$ will need some minor adjusting. 
 
 We recall that the roles of $\Psi_{\pm}$ need to swap for the type IIB G-structure description, to account for the condition \eqref{eqn:Gconditionrepeated} and the fact the Ramond fields switch from even in IIA to odd in IIB.

\subsection{The G-structure conditions}\label{sec:ATDGstructureConds}

We begin by transforming the G-Structure condition under abelian T-Duality from IIA to IIB. For convenience, we will summarise the left-hand side of the following IIA G-Structure conditions simply as $d_{H^\mathcal{A}_3}(e^{\alpha A-\Phi_\mathcal{A}}\Psi_\pm^\mathcal{A}) $,
\begin{equation*}
\begin{aligned}
d_{H^\mathcal{A}_3}(e^{3A-\Phi_\mathcal{A}}\Psi^\mathcal{A}_+)&=0,\\
d_{H^\mathcal{A}_3}(e^{2A-\Phi_\mathcal{A}}\text{Re}\Psi^\mathcal{A}_-)&=0,\\
d_{H^\mathcal{A}_3}(e^{4A-\Phi_\mathcal{A}}\text{Im}\Psi^\mathcal{A}_-)&=\frac{e^{4A}}{8}*_6\lambda(g),
\end{aligned}
\end{equation*}
 with the choice of $\alpha\in (2,3,4)$ depending on the specific condition.
Now, the condition which will transform in the same manner as the Ramond fields is
\beq
  \text{Vol}_4\wedge d_{H^\mathcal{A}_3}(e^{\alpha A-\Phi_\mathcal{A}}\Psi_\pm^\mathcal{A}) =   \text{Vol}_4\wedge\Big[d(e^{\alpha A-\Phi_\mathcal{A}}\Psi_\pm^\mathcal{A})  + dB^\mathcal{A} \wedge e^{\alpha A-\Phi_\mathcal{A}}\Psi_\pm^\mathcal{A}\Big].
\eeq
From the T-Dual rules \eqref{eqn:TD2}, we have
\beq
\begin{aligned}
dB^\mathcal{A}&=dB_2^\mathcal{A}+dB_1^\mathcal{A}\wedge dy\\
&=dB_2^\mathcal{A} +e^{-C^\mathcal{A}}dB_1^\mathcal{A}\wedge e^{C^\mathcal{A}}(dy+A_1^\mathcal{A})-dB_1^\mathcal{A}\wedge A_1^\mathcal{A}\\
&=(dB_2^\mathcal{A} -dB_1^\mathcal{A}\wedge A_1^\mathcal{A})+ e^{-C^\mathcal{A}}dB_1^\mathcal{A}\wedge E^y_\mathcal{A},
\end{aligned}
\eeq
and using the decomposition for $\Psi_\pm^\mathcal{A}$, we get
\beq
\begin{aligned}
 dB^\mathcal{A} \wedge e^{\alpha A-\Phi_\mathcal{A}}\Psi_\pm^\mathcal{A}&=[(dB_2^\mathcal{A} -dB_1^\mathcal{A}\wedge A_1^\mathcal{A})+ e^{-C^\mathcal{A}}dB_1^\mathcal{A}\wedge E^y_\mathcal{A}]\wedge [e^{\alpha A-\Phi_\mathcal{A}}\Psi_{\pm_{\perp}}^\mathcal{A} +e^{\alpha A-\Phi_\mathcal{A}} \Psi_{\pm_{||}}^\mathcal{A}\wedge E^y_\mathcal{A}]\\
 &=(dB_2^\mathcal{A} -dB_1^\mathcal{A}\wedge A_1^\mathcal{A})\wedge e^{\alpha A-\Phi_\mathcal{A}}\Psi_{\pm_{\perp}}^\mathcal{A} \\
 &~~~~+\Big[(dB_2^\mathcal{A} -dB_1^\mathcal{A}\wedge A_1^\mathcal{A}) \wedge e^{\alpha A-\Phi_\mathcal{A}} \Psi_{\pm_{||}}^\mathcal{A} -   e^{\alpha A-\Phi_\mathcal{A}}\Psi_{\pm_{\perp}}^\mathcal{A} \wedge e^{-C^\mathcal{A}}dB_1^\mathcal{A} \Big]\wedge E^y_\mathcal{A}.
\end{aligned}
\eeq
In addition,
\beq
\begin{aligned}
 d(e^{\alpha A-\Phi_\mathcal{A}}\Psi_\pm^\mathcal{A})  &=  d(e^{\alpha A-\Phi_\mathcal{A}}\Psi_{\pm_{\perp}}^\mathcal{A} + e^{\alpha A-\Phi_\mathcal{A}}\Psi_{\pm_{||}}^\mathcal{A}\wedge E^y_\mathcal{A}) \\
 &=  d(e^{\alpha A-\Phi_\mathcal{A}}\Psi_{\pm_{\perp}}^\mathcal{A})+d( e^{\alpha A-\Phi_\mathcal{A}}\Psi_{\pm_{||}}^\mathcal{A})\wedge E^y_\mathcal{A}+ e^{\alpha A-\Phi_\mathcal{A}}\Psi_{\pm_{||}}^\mathcal{A} \wedge \Big(d(e^{C^\mathcal{A}})\wedge dy +d(e^{C^\mathcal{A}}A_1^\mathcal{A})\Big)\\
     &=  d(e^{\alpha A-\Phi_\mathcal{A}}\Psi_{\pm_{\perp}}^\mathcal{A})+e^{\alpha A-\Phi_\mathcal{A}}\Psi_{\pm_{||}}^\mathcal{A} \wedge e^{C^\mathcal{A}}d(A_1^\mathcal{A})\\
  &~~~~~+\Big[d( e^{\alpha A-\Phi_\mathcal{A}}\Psi_{\pm_{||}}^\mathcal{A}) + e^{\alpha A-\Phi_\mathcal{A}}\Psi_{\pm_{||}}^\mathcal{A} \wedge e^{-C^\mathcal{A}}d(e^{C^\mathcal{A}})\Big]\wedge E^y_\mathcal{A} .
 \end{aligned}
\eeq
Hence, we have
\beq\label{eqn:Gammas}
\begin{gathered}
  \text{Vol}_4\wedge d_{H^\mathcal{A}_3}(e^{\alpha A-\Phi_\mathcal{A}}\Psi_\pm^\mathcal{A}) =  \text{Vol}_4\wedge(\Gamma_{\perp}^\mathcal{A}+\Gamma_{||}^\mathcal{A}\wedge E^y_\mathcal{A}),\\\\
  \Gamma_{\perp}^\mathcal{A} =  d(e^{\alpha A-\Phi_\mathcal{A}}\Psi_{\pm_{\perp}}^\mathcal{A})+e^{\alpha A-\Phi_\mathcal{A}}\Psi_{\pm_{||}}^\mathcal{A} \wedge e^{C^\mathcal{A}}d(A_1^\mathcal{A})+ (dB_2^\mathcal{A} -dB_1^\mathcal{A}\wedge A_1^\mathcal{A})\wedge e^{\alpha A-\Phi_\mathcal{A}}\Psi_{\pm_{\perp}}^\mathcal{A} ,\\
  \Gamma_{||}^\mathcal{A} = d( e^{\alpha A-\Phi_\mathcal{A}}\Psi_{\pm_{||}}^\mathcal{A}) + e^{\alpha A-\Phi_\mathcal{A}}\Psi_{\pm_{||}}^\mathcal{A} \wedge e^{-C^\mathcal{A}}d(e^{C^\mathcal{A}})~~~~~~~~~~~~~~~~~~~~~~~~~~~~~~~~~~~~~~~~~~~~~~~~~~~~~~~~~~~ \\
  +(dB_2^\mathcal{A} -dB_1^\mathcal{A}\wedge A_1^\mathcal{A}) \wedge e^{\alpha A-\Phi_\mathcal{A}} \Psi_{\pm_{||}}^\mathcal{A} -e^{\alpha A-\Phi_\mathcal{A}}\Psi_{\pm_{\perp}}^\mathcal{A}\wedge e^{-C^\mathcal{A}}dB_1^\mathcal{A}  .
 \end{gathered}
\eeq
Now, applying the T-Dual rules, one gets for the IIB equations
\beq
\text{Vol}_4\wedge\Big(e^{C^\mathcal{A}}\Gamma_{||}^\mathcal{A} + \Gamma_{\perp}^\mathcal{A}\wedge (dy-B_1^\mathcal{A})\Big).
\eeq
Before proceeding, from the transformation rules
\beq
B^\mathcal{B}=B_2^\mathcal{B}+B_1^\mathcal{B}\wedge dy =B_2^\mathcal{A} -A_1^\mathcal{A}\wedge (dy-B_1^\mathcal{A})  ,
\eeq
the following result will be required
\beq
dB_2^\mathcal{A} -dB_1^\mathcal{A}\wedge A_1^\mathcal{A} = dB^\mathcal{B} +dA_1^\mathcal{A} \wedge (dy-B_1^\mathcal{A}),
\eeq
which we substitute directly into \eqref{eqn:Gammas}, giving

\beq 
\hspace{-0.85cm}
\begin{aligned}
  \Gamma_{\perp}^\mathcal{A} &=  d(e^{\alpha A-\Phi_\mathcal{A}}\Psi_{\pm_{\perp}}^\mathcal{A})+  dB^\mathcal{B} \wedge e^{\alpha A-\Phi_\mathcal{A}}\Psi_{\pm_{\perp}}^\mathcal{A}+e^{\alpha A-\Phi_\mathcal{A}}\Psi_{\pm_{||}}^\mathcal{A} \wedge e^{C^\mathcal{A}}d(A_1^\mathcal{A})-dA_1^\mathcal{A}\wedge e^{\alpha A-\Phi_\mathcal{A}}\Psi_{\pm_{\perp}}^\mathcal{A} \wedge (dy-B_1^\mathcal{A})  ,\\
  \Gamma_{||}^\mathcal{A} &= d( e^{\alpha A-\Phi_\mathcal{A}}\Psi_{\pm_{||}}^\mathcal{A}) + e^{\alpha A-\Phi_\mathcal{A}}\Psi_{\pm_{||}}^\mathcal{A} \wedge e^{-C^\mathcal{A}}d(e^{C^\mathcal{A}})~~~~~~~~~~~~~~~~~~~~~~~~~~~~~~~~~~~~~~~~~~~~~~~~~~~~~~~~~~~ \\
 & ~~~+\Big( dB^\mathcal{B} +dA_1^\mathcal{A} \wedge (dy-B_1^\mathcal{A})\Big) \wedge e^{\alpha A-\Phi_\mathcal{A}} \Psi_{\pm_{||}}^\mathcal{A} - e^{\alpha A-\Phi_\mathcal{A}}\Psi_{\pm_{\perp}}^\mathcal{A}\wedge e^{-C^\mathcal{A}}dB_1^\mathcal{A} ,
 \end{aligned}
\eeq
hence
\beq
\hspace{-1.75cm}
\begin{aligned}
e^{C^\mathcal{A}}\Gamma_{||}^\mathcal{A} + \Gamma_{\perp}^\mathcal{A}\wedge (dy-B_1^\mathcal{A})&=e^{C^\mathcal{A}}d( e^{\alpha A-\Phi_\mathcal{A}}\Psi_{\pm_{||}}^\mathcal{A}) + e^{\alpha A-\Phi_\mathcal{A}}\Psi_{\pm_{||}}^\mathcal{A} \wedge d(e^{C^\mathcal{A}})+e^{C^\mathcal{A}} dB^\mathcal{B} \wedge e^{\alpha A-\Phi_\mathcal{A}} \Psi_{\pm_{||}}^\mathcal{A}\\
&~~~~~~~ -e^{C^\mathcal{A}} dA_1^\mathcal{A}\wedge e^{\alpha A-\Phi_\mathcal{A}} \Psi_{\pm_{||}}^\mathcal{A} \wedge (dy-B_1^\mathcal{A})  - e^{\alpha A-\Phi_\mathcal{A}}\Psi_{\pm_{\perp}}^\mathcal{A}\wedge dB_1^\mathcal{A}\\
+\bigg[ &d(e^{\alpha A-\Phi_\mathcal{A}}\Psi_{\pm_{\perp}}^\mathcal{A})+  dB^\mathcal{B} \wedge e^{\alpha A-\Phi_\mathcal{A}}\Psi_{\pm_{\perp}}^\mathcal{A}+e^{\alpha A-\Phi_\mathcal{A}}\Psi_{\pm_{||}}^\mathcal{A} \wedge e^{C^\mathcal{A}}d(A_1^\mathcal{A})\bigg]\wedge (dy-B_1^\mathcal{A})\\
&=d( e^{\alpha A-\Phi_\mathcal{A}}e^{C^\mathcal{A}}\Psi_{\pm_{||}}^\mathcal{A}) + dB^\mathcal{B} \wedge e^{\alpha A-\Phi_\mathcal{A}}e^{C^\mathcal{A}} \Psi_{\pm_{||}}^\mathcal{A}\\
&~~~~~~+  d\Big(e^{\alpha A-\Phi_\mathcal{A}}\Psi_{\pm_{\perp}}^\mathcal{A}\wedge (dy-B_1^\mathcal{A})\Big)+  dB^\mathcal{B} \wedge e^{\alpha A-\Phi_\mathcal{A}}\Psi_{\pm_{\perp}}^\mathcal{A}\wedge (dy-B_1^\mathcal{A})\\
&= d_{H^\mathcal{B}_3}\Big[ e^{\alpha A-\Phi_\mathcal{A}} \Big(e^{C^\mathcal{A}}\Psi_{\pm_{||}}^\mathcal{A} +\Psi_{\pm_{\perp}}^\mathcal{A}\wedge (dy-B_1^\mathcal{A}) \Big)\Big]\\
&= d_{H^\mathcal{B}_3}( e^{\alpha A-\Phi_\mathcal{A}} \Psi_\mp^\mathcal{B} ).
\end{aligned}
\eeq

The IIB G-Structure equations now read
\begin{equation*}
\begin{aligned}
d_{H^\mathcal{B}_3}(e^{3A-\Phi_\mathcal{A}}\Psi^\mathcal{B}_-)&=0,\\
d_{H^\mathcal{B}_3}(e^{2A-\Phi_\mathcal{A}}\text{Re}\Psi^\mathcal{B}_+)&=0,\\
d_{H^\mathcal{B}_3}(e^{4A-\Phi_\mathcal{A}}\text{Im}\Psi^\mathcal{B}_+)&=\frac{e^{4A}}{8}*_6\lambda(g). 
\end{aligned}
\end{equation*}
Of course, by adjusting the transformation as follows
\begin{equation*}
e^{-\Phi_\mathcal{B}}  \Psi_\mp^\mathcal{B}=e^{-\Phi_\mathcal{A}}\Big[e^{C^\mathcal{A}}\Psi_{\pm_{||}}^\mathcal{A} + \Psi_{\pm_{\perp}}^\mathcal{A}\wedge (dy-B_1^\mathcal{A})\Big],
\end{equation*}
one can re-write the above G-Structure conditions in terms of $\Phi_\mathcal{B}$, 
\begin{equation*}
\begin{aligned}
d_{H^\mathcal{B}_3}(e^{3A-\Phi_\mathcal{B}}\Psi^\mathcal{B}_-)&=0,\\
d_{H^\mathcal{B}_3}(e^{2A-\Phi_\mathcal{B}}\text{Re}\Psi^\mathcal{B}_+)&=0,\\
d_{H^\mathcal{B}_3}(e^{4A-\Phi_\mathcal{B}}\text{Im}\Psi^\mathcal{B}_+)&=\frac{e^{4A}}{8}*_6\lambda(g). 
\end{aligned}
\end{equation*}
\subsection{The G-structure forms}
      Now we wish to calculate the IIB pure spinors which we need for the G-Structure conditions just derived. The $SU(2)$ pure spinors for IIA are given in \eqref{eqn:Psi}, and re-written here for clarity
      
      \begin{equation*}
\Psi^\mathcal{A}_+=\frac{1}{8} e^{\frac{1}{2}z^\mathcal{A}\wedge \overline{z}^\mathcal{A}}\wedge \omega^\mathcal{A},~~~~~\Psi^\mathcal{A}_-=\frac{i}{8}  e^{-i j^\mathcal{A}}\wedge z^\mathcal{A}.
\end{equation*}
 In what follows, it will prove useful to make the following decompositions
\beq\label{eqn:decomps}
\begin{gathered}
\omega^\mathcal{A}=\omega^\mathcal{A}_{\perp}+\omega^\mathcal{A}_{||}\wedge E^y_\mathcal{A},~~~~~~~~~~~~~~~~~
j^\mathcal{A}=j^\mathcal{A}_{\perp}+j^\mathcal{A}_{||}\wedge E^y_\mathcal{A},~~~~~~~~~~~~~
z^\mathcal{A}=z^\mathcal{A}_{\perp}+z^\mathcal{A}_{||}\wedge E^y_\mathcal{A},\\
z^\mathcal{A}_{\perp}=u^\mathcal{A}_{\perp}+i\,v^\mathcal{A}_{\perp},~~~~~~~~~~~~~~~~~~z^\mathcal{A}_{||}=u^\mathcal{A}_{||}+i\,v^\mathcal{A}_{||}.
\end{gathered}
\eeq  
 We then note
\beq
\begin{aligned}
e^{\frac{1}{2}z^\mathcal{A}\wedge \overline{z}^\mathcal{A}} &= 1+\frac{1}{2}z^\mathcal{A}\wedge \overline{z}^\mathcal{A}\\
&= \Big(1+\frac{1}{2} z^\mathcal{A}_{\perp}\wedge \overline{z}^\mathcal{A}_{\perp}\Big) +\frac{1}{2}( z^\mathcal{A}_{\perp}\wedge \overline{z}^\mathcal{A}_{||} -z^\mathcal{A}_{||}\wedge \overline{z}^\mathcal{A}_{\perp})\wedge E^y_\mathcal{A},\\
\end{aligned}
\eeq
as $( z^\mathcal{A}\wedge \overline{z}^\mathcal{A})\wedge ( z^\mathcal{A}\wedge \overline{z}^\mathcal{A})  =0$, and 

\beq
\hspace{-1cm}
\begin{aligned}
e^{-ij^\mathcal{A}} &= 1+(-ij^\mathcal{A})+\frac{1}{2}(-ij^\mathcal{A})\wedge (-ij^\mathcal{A}) +\frac{1}{3!}(-ij^\mathcal{A})\wedge (-ij^\mathcal{A})\wedge (-ij^\mathcal{A})+....\\
&=1-ij^\mathcal{A} -\frac{1}{2}j^\mathcal{A}\wedge j^\mathcal{A} +\frac{i}{3!}j^\mathcal{A}\wedge j^\mathcal{A} \wedge j^\mathcal{A}+....\\
&=1-i(j^\mathcal{A}_{\perp}+j^\mathcal{A}_{||}\wedge E^y_\mathcal{A}) -\frac{1}{2}(j^\mathcal{A}_{\perp} \wedge j^\mathcal{A}_{\perp} +2j^\mathcal{A}_{\perp}\wedge j^\mathcal{A}_{||}\wedge E^y_\mathcal{A}) +\frac{i}{3!}(j^\mathcal{A}_{\perp}\wedge j^\mathcal{A}_{\perp}\wedge j^\mathcal{A}_{\perp} + 3j^\mathcal{A}_{\perp}\wedge j^\mathcal{A}_{\perp}\wedge j^\mathcal{A}_{||}\wedge E^y_\mathcal{A})+ ..\\
&=\Big(1- i j^\mathcal{A}_{\perp} - \frac{1}{2}j^\mathcal{A}_{\perp} \wedge j^\mathcal{A}_{\perp}+\frac{i}{3!}j^\mathcal{A}_{\perp}\wedge j^\mathcal{A}_{\perp}\wedge j^\mathcal{A}_{\perp}+... \Big) +\Big(-i j^\mathcal{A}_{||} -j^\mathcal{A}_{\perp}\wedge j^\mathcal{A}_{||} +\frac{i}{2}j^\mathcal{A}_{\perp}\wedge j^\mathcal{A}_{\perp}\wedge j^\mathcal{A}_{||}+... \Big)\wedge E^y_\mathcal{A}\\
&=e^{-ij^\mathcal{A}_{\perp}}\wedge (1-i j^\mathcal{A}_{||}\wedge E^y_\mathcal{A}).
\end{aligned}
\eeq

Hence, we get
\beq
\begin{gathered}
 \Psi_\pm^\mathcal{A}=\Psi_{\pm_{\perp}}^\mathcal{A} + \Psi_{\pm_{||}}^\mathcal{A}\wedge E^y_\mathcal{A},~~~~~~~~\\
 \Psi_{+_{\perp}}^\mathcal{A}=\frac{1}{8}\Big(1+\frac{1}{2}z^\mathcal{A}_{\perp}\wedge \overline{z}^\mathcal{A}_{\perp}\Big)\wedge \omega^\mathcal{A}_{\perp},~~~~~\Psi_{+_{||}}^\mathcal{A}=\frac{1}{8}\bigg[\Big(1+\frac{1}{2}z^\mathcal{A}_{\perp}\wedge \overline{z}^\mathcal{A}_{\perp}\Big)\wedge \omega^\mathcal{A}_{||}+\frac{1}{2}( z^\mathcal{A}_{\perp}\wedge \overline{z}^\mathcal{A}_{||} -z^\mathcal{A}_{||}\wedge \overline{z}^\mathcal{A}_{\perp})\wedge \omega^\mathcal{A}_{\perp}\bigg],\\
 \Psi_{-_{\perp}}^\mathcal{A}=\frac{i}{8}e^{-ij^\mathcal{A}_{\perp}}\wedge z^\mathcal{A}_{\perp},~~~~~~~~~~~~~~~~~~~\Psi_{-_{||}}^\mathcal{A}=\frac{i}{8}e^{-ij^\mathcal{A}_{\perp}}\wedge (z^\mathcal{A}_{||}+ij^\mathcal{A}_{||}\wedge z^\mathcal{A}_{\perp}).
\end{gathered}
\eeq
Now, recalling
\begin{equation*}
\Psi_\mp^\mathcal{B}=e^{\Phi_\mathcal{B}-\Phi_\mathcal{A}}\Big[e^{C^\mathcal{A}}\Psi_{\pm_{||}}^\mathcal{A} + \Psi_{\pm_{\perp}}^\mathcal{A}\wedge (dy-B_1^\mathcal{A})\Big],
\end{equation*}
we finally arrive at the following results for the IIB pure spinors, written in terms of the IIA G-structure forms
\beq
\begin{aligned}\label{eqn:IIBforms}
  \Psi_-^\mathcal{B}
  &=\frac{1}{8}e^{\Phi_\mathcal{B}-\Phi_\mathcal{A}}\bigg[e^{\frac{1}{2}z^\mathcal{A}_{\perp}\wedge \overline{z}^\mathcal{A}_{\perp}}\wedge\Big( e^{C^\mathcal{A}}\omega^\mathcal{A}_{||}+   \omega^\mathcal{A}_{\perp}\wedge (dy-B_1^\mathcal{A})\Big)+e^{C^\mathcal{A}}\frac{1}{2}( z^\mathcal{A}_{\perp}\wedge \overline{z}^\mathcal{A}_{||} -z^\mathcal{A}_{||}\wedge \overline{z}^\mathcal{A}_{\perp})\wedge \omega^\mathcal{A}_{\perp}\bigg], \\
   \Psi_+^\mathcal{B}&=\frac{i }{8} e^{\Phi_\mathcal{B}-\Phi_\mathcal{A}}e^{-ij^\mathcal{A}_{\perp}}\wedge \bigg[ (e^{C^\mathcal{A}}z^\mathcal{A}_{||}+ z^\mathcal{A}_{\perp}\wedge (dy-B_1^\mathcal{A}))+i  e^{C^\mathcal{A}} j^\mathcal{A}_{||}\wedge z^\mathcal{A}_{\perp}\bigg].
\end{aligned}
\eeq


\bibliographystyle{JHEP}

\bibliography{ref}

\end{document}